\pdfoutput=1
\documentclass[cernpreprint, atlasdraft=false, UKenglish,british,orcidlogo, texmf, orcidlogo]{atlasdoc}
\usepackage[subfig]{atlaspackage}
\usepackage{multirow}
\usepackage{adjustbox}
\usepackage{graphicx}
\usepackage{grffile}
\usepackage{atlasbiblatex}
 
\usepackage[process=true,hion=true]{atlasphysics}

\newglossarystyle{mylist}{

}
 
\setacronymstyle{long-short}
 
\makenoidxglossaries

\graphicspath{{logos/}{figures/}}
 
\usepackage{ANA-GENR-2019-02-PAPER-defs}
\usepackage{atlasmisc} 
 
\usepackage{chngcntr}
 
\counterwithin{figure}{section}
\counterwithin{table}{section}
 
\AtlasTitle{The ATLAS Experiment at the CERN Large Hadron Collider: A Description of the Detector Configuration for Run~3}
 
\PreprintIdNumber{CERN-EP-2022-259}
\AtlasAbstract{
The ATLAS detector is installed in its experimental cavern at Point~1 of the Large Hadron Collider (LHC) at CERN. It was originally designed for proton-proton collisions at a centre-of mass energy of $\sqrt{s}=14$\,TeV with an instantaneous luminosity of $\mathcal{L}=10^{34}\mathrm{cm}^{-2}\mathrm{s}^{-1}$, and heavy ion collisions at $\sqrt{s}=5.5$\,TeV per nucleon pair and 
$\mathcal{L}=10^{27}\mathrm{cm}^{-2}\mathrm{s}^{-1}$.
During Run~2 of the LHC, the energy was $\sqrt{s}=13$\,TeV,
with $\mathcal{L}=2\times 10^{34}\mathrm{cm}^{-2}\mathrm{s}^{-1}$ routinely achieved at the start of fills.
For Run~3, a small energy increase to $\sqrt{s}=13.6$\,TeV accompanies accelerator improvements, notably luminosity levelling allowing sustained running at an instantaneous luminosity of $\mathcal{L}=2\times 10^{34}\mathrm{cm}^{-2}\mathrm{s}^{-1}$, with an average of up to 60 interactions per bunch crossing.
The ATLAS detector has been upgraded to recover Run~1 single-lepton trigger thresholds while operating comfortably under Run~3 sustained pileup conditions. A fourth pixel layer 3.3~cm from the beam axis was added before
Run~2 to improve vertex reconstruction and $b$-tagging performance. New Liquid Argon Calorimeter digital trigger electronics, with corresponding upgrades to the Trigger and Data Acquisition system, take advantage of a factor of 10 finer granularity to improve triggering on electrons, photons, taus, and hadronic signatures through increased pileup rejection. The inner muon endcap wheels 
were replaced by New Small Wheels with Micromegas and small-strip Thin Gap Chamber detectors, providing both precision tracking and Level-1 Muon trigger functionality.
Trigger coverage of the inner barrel muon layer near one endcap region was augmented with modules integrating new thin-gap resistive plate chambers and smaller-diameter drift-tube chambers.
Tile Calorimeter scintillation counters were added to improve electron energy resolution and background rejection. Upgrades to Minimum Bias Trigger Scintillators and Forward Detectors improve luminosity monitoring and enable total proton-proton cross section, diffractive physics, and heavy ion measurements. These upgrades are all compatible with operation in the much harsher environment anticipated after the High-Luminosity upgrade of the LHC and are the first steps towards preparing ATLAS for the High-Luminosity upgrade of the LHC.
The Run~3 configuration of the ATLAS detector is described in this paper.
}

\author{The ATLAS Collaboration}
 
\AtlasRefCode{GENR-2019-02}

\AtlasJournalRef{JINST 19 (2024) P05063}
\AtlasDOI{10.1088/1748-0221/19/05/P05063}

\hypersetup{pdftitle={ATLAS document},pdfauthor={The ATLAS Collaboration}}
 
\begin{document}
 
\maketitle
 
\tableofcontents
 
\clearpage
\newpage
 
\section{Overview}

 
The \gls{LHC} at CERN and its detectors have ushered in a new era in particle physics, colliding protons at a centre-of-mass energy up to $\rts = \SI{13}{\TeV}$ and a peak instantaneous luminosity of \lumiruntwopeak. The ATLAS detector, which was installed at the \gls{LHC} \gls{IP}~1 during the period $2000-2008$, has performed extremely well during \RunOne and \RunTwo of the \gls{LHC}, recording \SIlist{5;21;147}{\ifb} of proton-proton collision data at $\rts = $\SIlist{7;8;13}{\TeV}, respectively. The highlight of this period was the discovery of the Higgs boson at a mass of \SI{125}{\GeV}, which was announced by the ATLAS and \acrshort{CMS} collaborations (each comprising nearly \num{3000} scientific authors) on July 4\textsuperscript{th}, 2012~\cite{HIGG-2012-27,CMS-HIG-12-028}. The excellent performance of the ATLAS detector and of the \gls{LHC} has enabled over \num{1000} publications exploring the nature of the Higgs boson, searching for new phenomena, and continuing to probe the Standard Model.
 
The ATLAS detector was designed for a peak instantaneous luminosity of $\luminominal$ at a proton-proton (\pp) centre-of mass energy of $\rts=\SI{14}{\TeV}$ with about \num{25} interactions per bunch crossing, and of $\lumiHInominal$ of heavy ion (lead-lead: \PbPb, and proton-lead: \pPb) collisions at \SI{5.5}{\TeV} per nucleon pair. The original configuration of the detector, as it was built for the start of the \gls{LHC}, is described in Ref~\cite{PERF-2007-01}. However, the \gls{LHC} operating conditions differed from expectations; in particular, the instantaneous luminosity surpassed design, reaching a maximum value of \lumiruntwopeak\ during \RunTwo. There were on average \num{33.7} interactions per bunch crossing during \RunTwo, with a peak value of over \num{60} interactions per bunch crossing recorded in 2017 and 2018.
The ATLAS detector performed well despite these harsher conditions; performance and analysis techniques were adapted and improved to maintain the experiment's excellent physics reach during \RunTwo. A substantial upgrade to the ATLAS detector, the ``Phase-I Upgrade'', has consisted of improvements to the detector subsystems and their electronics in order to withstand the expected \RunThr conditions of
\murunthree (and at maximum \num{60})
interactions per bunch crossing, and maintain the lowest achievable trigger thresholds, enabling the continued broad physics program planned for \RunThr.
The ATLAS detector configuration during \RunThr of the \gls{LHC} is described in this paper.

\subsection{Brief history of the ATLAS detector performance and LHC roadmap}
 
The ATLAS detector has been in operation since autumn 2009. During this period, the \gls{LHC} and ATLAS have alternated between distinct running periods and long shutdown periods:
\begin{description}
\item[\RunOne, $2009-2013$:] ATLAS recorded \SI{5}{\ifb} of \pp collision data at $\rts = \SI{7}{\TeV}$ and \SI{21}{\ifb} at $\rts = \SI{8}{\TeV}$; \intlumiPbPbrunone\ of \PbPb collisions were recorded in 2010 and 2011; \intlumipPbrunone\ of \pPb collisions were recorded in 2013.
 
\item[\gls{LS1}, $2013-2015$:] This shutdown was used to consolidate the \gls{LHC} machine elements (repairing the magnet splices and upgrading the collimation scheme) to increase the centre-of-mass energy and reach the design luminosity. 
Upgrades to ATLAS installed and commissioned during \gls{LS1} are referred to as ``Phase-0'' Upgrades, and are described in this document; the most significant Phase-0 upgrade was the addition of the innermost layer of silicon pixel detectors described in Section~\ref{sec:ID-IBL}.
 
\item[\RunTwo, $2015-2018$:] ATLAS recorded \SI{147}{\ifb} of proton-proton collision data at $\rts = \SI{13}{\TeV}$; \intlumipPbruntwo\ of \pPb collisions were recorded in 2016 and \intlumiPbPbruntwo\ of \PbPb collisions were recorded in 2018. The \gls{LHC} reached a peak instantaneous luminosity of \lumiruntwopeak\ and, on average, delivered \num{33.7} interactions per bunch crossing.
 
\item[\gls{LS2}, $2019-2022$:] During this shutdown, a new linear accelerator (\gls{LINAC-4}) was connected into the injector complex~\cite{LHCcontributionInThisJournal}, and
the injection beam energy of the \gls{PS} Booster was upgraded in order to reduce the beam emittance. New cryogenics plants were installed to separate the cooling circuits of the superconducting \gls{RF} cavities from those of the superconducting magnets. The ATLAS Phase-I Upgrade was installed and commissioned during \gls{LS2} and its description constitutes the bulk of this document.
 
\item[\RunThr, $2022-2025$:] The \gls{LHC} is expected to deliver a peak instantaneous luminosity of approximately $\lumirunthree$ and an integrated luminosity of \SI{250}{\ifb} of proton-proton collision data at a centre-of-mass energy of $\rts = \SI{13.6}{\TeV}$.
 
\item[\gls{LS3}, $2026-2028$:] The \gls{LHC} will undergo a major upgrade of its components (e.g. low-$\beta$ quadrupole triplets, crab cavities at the interaction regions). The ATLAS Phase-II Upgrade will be installed and commissioned.
 
\item[\gls{HL-LHC}, $2029$ and beyond:] The \gls{LHC} complex is expected to deliver a levelled instantaneous luminosity of $\lumihllhc$ and an annual integrated luminosity of approximately \SI{250}{\ifb} to reach a total dataset of \SI{3000}{\ifb}.
 
\end{description}
 
\begin{figure}[htpb]
\begin{center}
\subfloat[]{\label{figOv:pplumi}\includegraphics[width=0.49\textwidth]{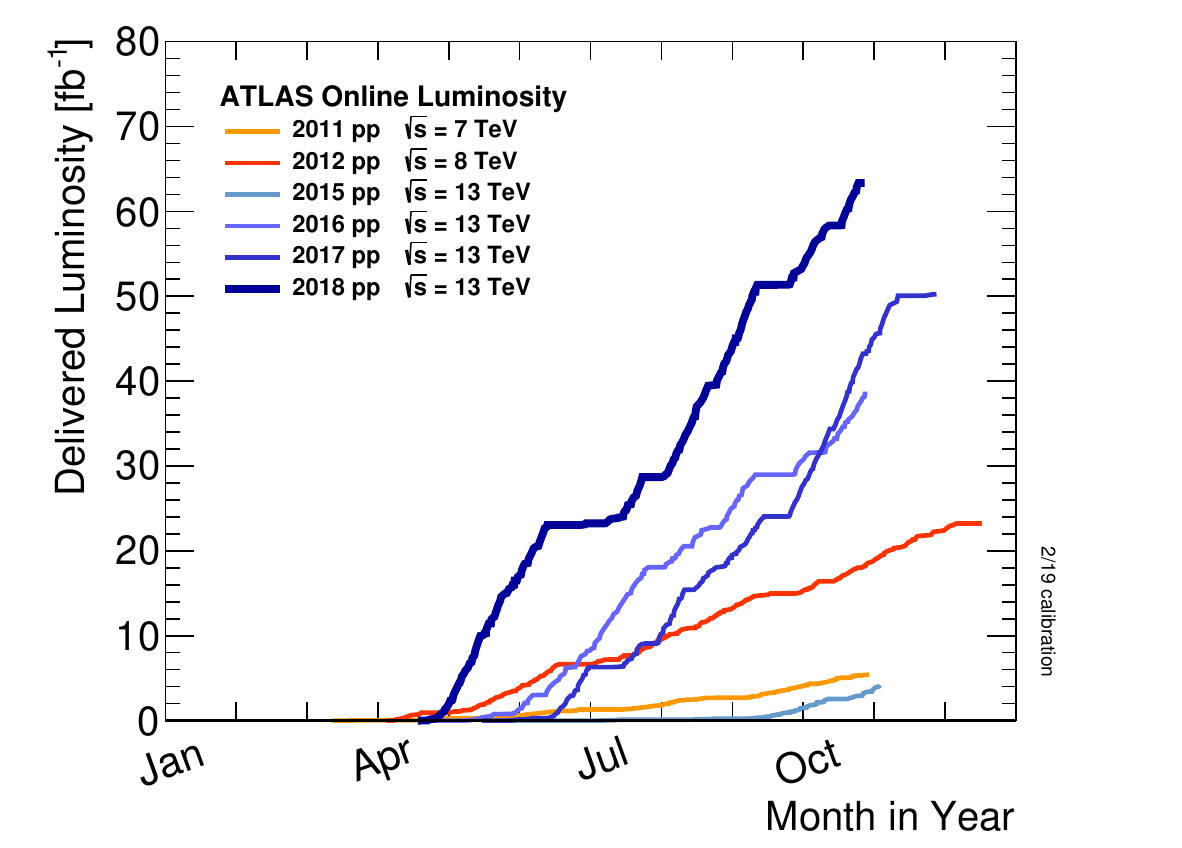}}
\subfloat[]{\label{figOv:PbPblumi}\includegraphics[width=0.49\textwidth]{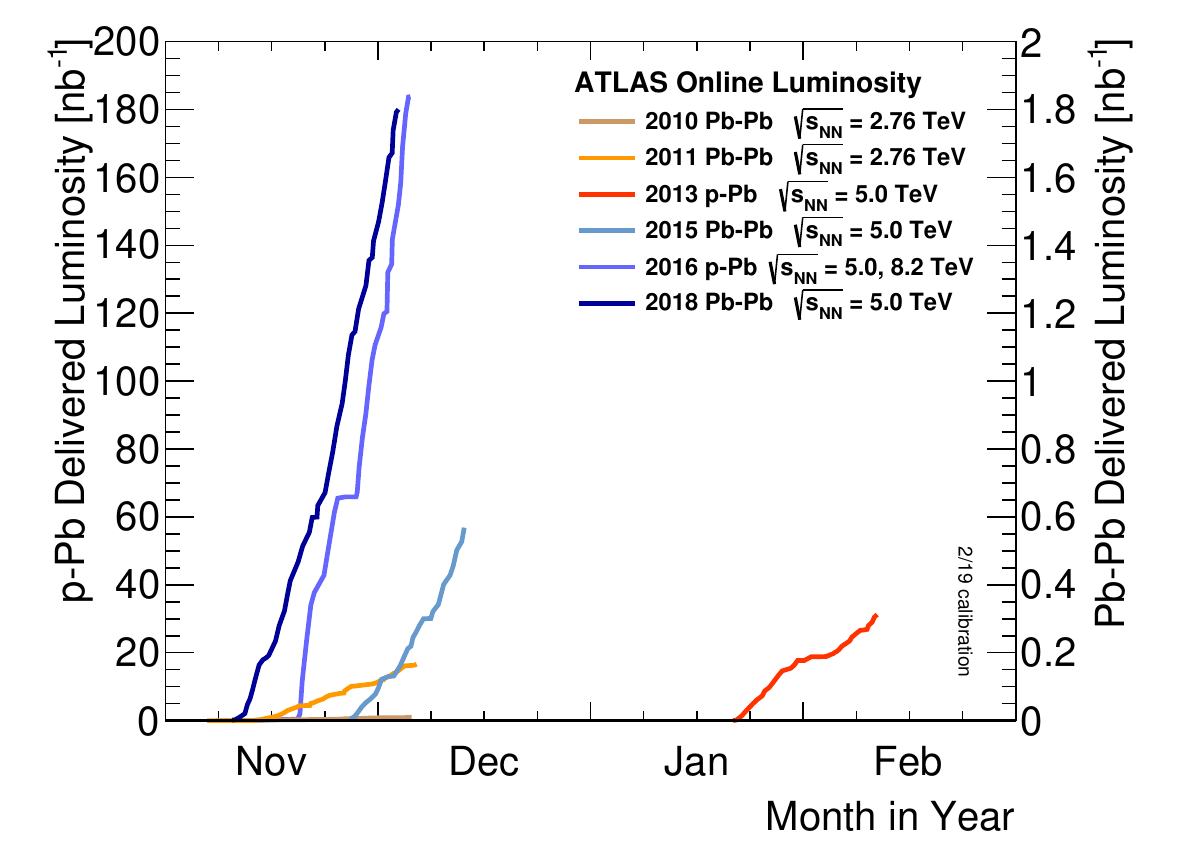}}\\
\subfloat[]{\label{figOv:muRun2}\includegraphics[width=0.49\textwidth]{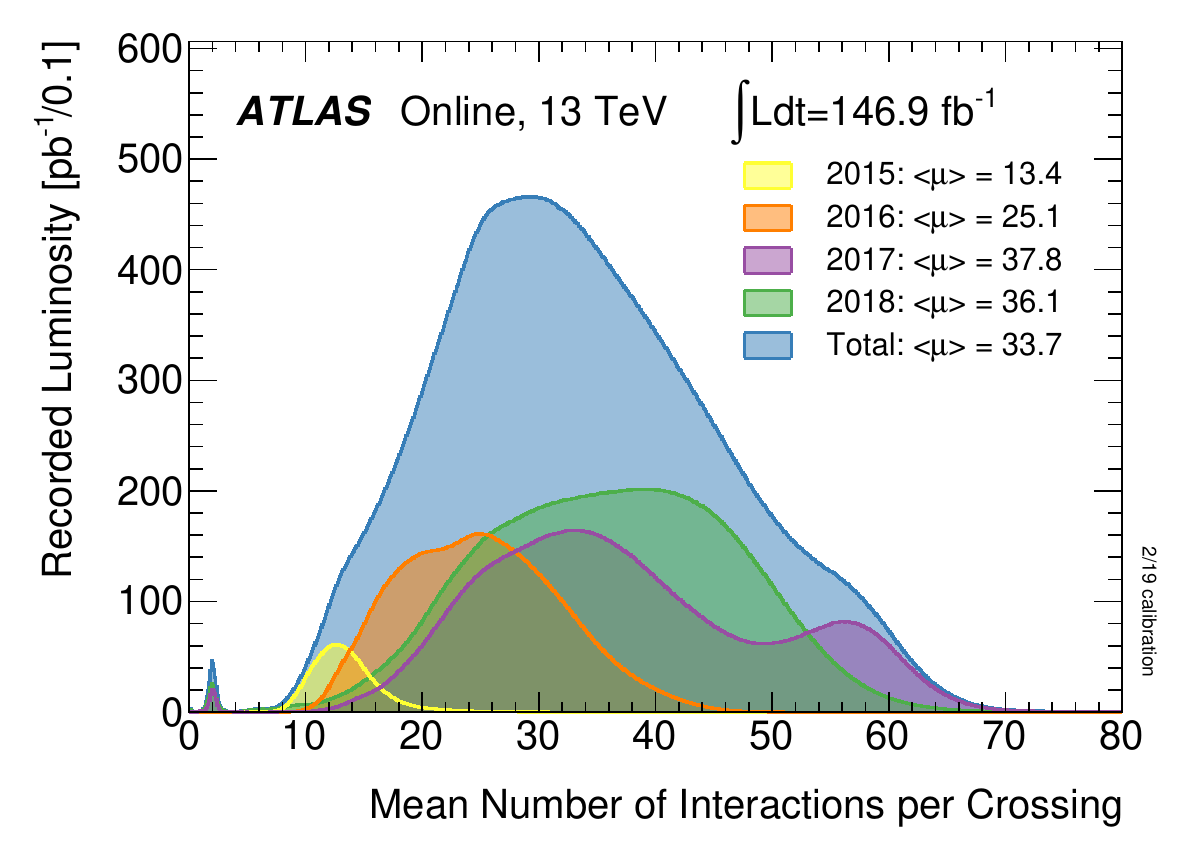}}
\caption{\protect\subref{figOv:pplumi} Cumulative luminosity versus day delivered to \ATLAS during stable beams for high energy \pp collisions. \protect\subref{figOv:PbPblumi} Cumulative luminosity versus day delivered to \ATLAS during stable beams for \pPb and \PbPb collisions (indicated on the left and right vertical axes, respectively). \protect\subref{figOv:muRun2} Integrated luminosity as a function of the mean number of interactions per crossing, $\langle\mu\rangle$, for the 2015--2018 \pp collision data at $\rts=\SI{13}{TeV}$. All data recorded by ATLAS during stable beams are
shown, including the integrated luminosity and the mean $\mu$ value for each year.
The mean number of interactions per bunch crossing, $\langle\mu\rangle$, corresponds to the calculated mean of the Poisson distribution
of the number of inelastic interactions per bunch crossing.
\label{fig:Overview:lumi-pileup}}
\end{center}
\end{figure}

Figure~\ref{fig:Overview:lumi-pileup} presents the distributions of the delivered luminosity for each production year between 2011 and 2018 and the distribution of the \pileup ($\mu$, the number of \pp collisions per \gls{LHC} bunch crossing) for the four data-taking years of the \gls{LHC} \RunTwo.
 
Over the ten years since the start of \gls{LHC}, the ATLAS detector has operated reliably under the challenging conditions provided by the excellent performance of the \gls{LHC}. Three indicators illustrate the detector performance:
the fraction of operational channels, the data taking and the data quality efficiencies. The fraction of operational channels at the start of \gls{LS2} was above 99.5\% for calorimeters and about 95\% for tracking detectors. By the start of \RunThr, some faulty channels had been repaired, while some detectors that had undergone extensive modifications were still being commissioned. The \glspl{RPC} required flushing with argon to remove debris that had accumulated during \gls{LS2}, after which the operational fraction improved.
The full details are shown in Table~\ref{tab:livechannels}. ATLAS data-taking efficiency (the fraction of the time ATLAS collects data while
\gls{LHC} delivers collisions in stable conditions) improved from 90 to 95\% in the course of the eight years of data taking. Over the same period, the fraction of collected data declared good for physics analysis has increased from 88.8 to 97.5\%~\cite{DAPR-2018-01}.

\begin{table}
\begin{center}
\caption{The number of channels vs.\ the approximate operational fraction for ATLAS subdetectors with more than \num{100} channels as of the end of \RunTwo (March 2019) compared with the start of \RunThr (May 2022). The \gls{RPC} operational fraction is expected to improve as commissioning continues,
and compares favourably with the operational fraction at the start of \RunTwo. None of the 8704 \gls{RPC} channels of the \gls{BIS78} were operational at the start of \RunThr.}
\label{tab:livechannels}
\begin{tabular}{l|c|c|c|c}
& \multicolumn{2}{c|}{{\textbf{\RunTwo}}} & \multicolumn{2}{c}{{\textbf{\RunThr}}} \\ \cline{2-5}
\textbf{Subdetector} & \textbf{Number of} & \textbf{Operational} &  \textbf{Number of} & \textbf{Operational} \\
& \textbf{Channels} & \textbf{Fraction} &  \textbf{Channels} & \textbf{Fraction} \\ \hline
Pixel \glstext{IBL} & 12M & $99.3\%$ & 12M & 96.7\%\\
Pixel outer 3 layers & 80M & $94.8\%$ & 80M & 96.7\% \\
\glstext{SCT} & 6.3M & $98.6\%$ & 6.3M & 98.3\% \\
\glstext{TRT} & 350k & $97.2\%$ & 350k & 96.6\% \\
\glstext{LAr} Calorimeter (\glstext{EM}) & 170k & $100\%$ & 170k & 100\% \\
Tile Calorimeter & \num{5200} & $99.5\%$ & \num{5200} & 99.2\% \\
\glstext{LAr} Calorimeter (\glstext{HEC}) & \num{5600} & $99.7\%$ & \num{5600} & 99.9\% \\
\glstext{LAr} Calorimeter (\glstext{FCAL}) & \num{3500} & $99.8\%$ & \num{3500} & 99.8\% \\
\glstext{L1Calo} Legacy Trigger & \num{7160} & $99.9\%$ & \num{7160} & 99.9\%\\
\glstext{L1Calo} \glstext{SC} Trigger & not present & -- & 34k  & 100\% \\ 
\glstext{L1Muon} \glstext{RPC} Trigger & 383k & $100\%$ & 392k & 97.5\% \\
\glstext{L1Muon} \glstext{TGC} Trigger & 320k & $99.9\%$ & 312k & 100\% \\
\glstext{MDT} & 367k & $99.7\%$ & 344k & $99.7\%$ \\ 
\glstext{CSC} & 31k & $93.0\%$ & not present & -- \\
\glstext{RPC} & 383k & $93.3\%$ & 392k & 85.8\%\\
\glstext{TGC} & 320k & $98.9\%$ & 312k & 99.4\%\\
\glstext{sTGC} & not present & -- & 358k & 99.2\%\\
\glstext{MM} & not present & -- & 2.1M & 98.0\% \\
\glstext{ALFA} & 10k & $98.9\%$ & 10k & 100\% \\
\glstext{AFP} & 430k & $97.0\%$ & 430k & 100\%\\
\end{tabular}
\end{center}
\end{table}
 
\subsection{LHC Performance and Status}
\label{ss:LHC}

 
The \gls{LHC} and its injector chain have undergone multiple major
repairs, consolidations and upgrades since \RunOne as detailed in Ref.~\cite{LHCcontributionInThisJournal}.
These changes facilitated the increase in collision energy and luminosity during \RunTwo
and are expected to provide further increases for \RunThr and beyond.
 
During \gls{LS1} the more than \num{10000} high current splices between the
LHC superconducting magnets were repaired and consolidated, 18 dipole
magnets replaced and new safety systems added in order to safely
increase the beam energy for \RunTwo. The \gls{LHC} collimation
system, injection kicker magnets and injection protection system were
upgraded to handle long trains of bunches with \SI{25}{\ns} bunch
spacing and higher bunch brightness. Additional improvements to the
injection and beam dump systems were done in winter shutdowns during
\RunTwo to be able to further increase the beam intensity.
 
High-luminosity proton-proton collisions in \RunTwo were delivered at
$\rts = \SI{13}{\TeV}$, while Pb-Pb collisions were delivered at $\rts
= \SI{5.02}{\TeV}$ per nucleon-pair. Multiple improvements were
deployed during the running period to increase the instantaneous
luminosity a factor two beyond the increases brought by the higher
beam energy and the larger number of bunches from shorter bunch
spacing. As the beam optics and the understanding of collimation
tolerances were improved, the interaction optics (\betastar), which determines the transverse beam size at the interaction point, was gradually squeezed more and more. In 2016 brighter bunches were introduced
through a change to the bunch-production scheme in the \gls{PS} called Batch Compression Merging and Splitting 
which gave smaller transverse beam size~\cite{Steerenberg:2259071}.  In 2017 crossing angle
anti-levelling was introduced, where the crossing angle in the
\gls{IP} was gradually reduced during each fill as bunch
intensity decreased. In 2018 this was augmented with a \betastar
anti-levelling
where a \SI{15}{\percent} additional squeeze at the \gls{IP}
was done towards the end of each fill. Both anti-levelling schemes
increased the integrated luminosity without increasing the peak
instantaneous luminosity, by running near the peak instantaneous luminosity for a much larger fraction of the total running time.
 
In \gls{LS2} a major consolidation campaign was carried out for the
bypass diodes of the \gls{LHC} superconducting dipole magnets in order
to more safely condition the magnets for highest beam energies. A
large set of upgrades was carried out in the injector chain, including
the connection of the new \gls{LINAC-4} to the injector complex and an
increase of the \gls{PS} Booster beam energy. This will
allow very low emittance, high intensity bunches to be produced for
the \gls{HL-LHC}. The full intensity beams of the injector will only
be usable after \gls{LS3}, but upgrades and modifications during
\gls{LS2} to the \gls{LHC} collimators, injection kicker magnets and
the beam dump will allow the \gls{LHC} to use up to 60\% more intense
beams in \RunThr. In the \gls{SPS}, upgrades to the RF
system will be used to produce ion beams with \SI{50}{ns} bunch
spacing using a technique called slip-stacking~\cite{slipstacking}, thus enabling a
larger beam intensity to be injected in the \gls{LHC} for Pb-Pb collisions.
 
In \RunThr, the proton-proton collision energy will increase to
\SI{13.6}{TeV}, but the peak instantaneous luminosity will remain limited
to about \lumirunthree in ATLAS. The main
limitation is luminosity-induced heating of the inner triplet
magnets that provide the final focus before the beams reach the \gls{IP}. The higher beam brightness from the \gls{LS2} injector
upgrade will instead be used to provide extended periods of luminosity
leveled at \lumirunthree for as long as 10\,hours per \gls{LHC} fill. The corresponding peak pileup
level is expected to be between 52 and 57 collisions per crossing
during the luminosity levelling period. This is expected to give a
yearly integrated luminosity of more than \SI{80}{\ifb} in the latter
part of \RunThr and potentially more than \intlumirunthree for the full
run. For
ions, more than \SI{6}{\nb^{-1}} of Pb-Pb collisions is expected in \RunThr.


\subsection{Brief overview of the ATLAS detector configuration for \RunThr}
 
The ATLAS detector~\cite{PERF-2007-01} at the \gls{LHC} covers nearly the entire solid angle around the collision point\footnote{ATLAS uses a right-handed coordinate system with its origin at the nominal \gls{IP} in the centre of the detector and the $z$-axis along the \beampipe. The $x$-axis points from the \gls{IP} to the centre of the \gls{LHC} ring, and the $y$-axis points upward. Cylindrical coordinates $(r,\phi)$ are used in the transverse plane, $\phi$ being the azimuthal angle around the $z$-axis. The pseudorapidity is defined in terms of the polar angle $\theta$ as $\eta=-\ln\tan(\theta/2)$. The two ends of the detector are labelled A for $+z$ and C for $-z$, with B used for elements at $\eta=0$.}.
 
It consists of an inner tracking detector surrounded by a thin superconducting solenoid, electromagnetic and hadronic calorimeters,
and a muon spectrometer incorporating three large superconducting air-core toroidal magnets. The dimensions of the detector are \SI{25}{\m} in height and \SI{44}{\m} in length; the overall weight of the detector is approximately \SI{7000}{\tonne}.
 
The \gls{ID} system is immersed in a \SI{2}{\tesla} axial magnetic field
and provides charged-particle tracking in the range \(|\eta| < 2.5\).
The high-granularity silicon \gls{Pixel} detector covers the vertex region and typically provides four measurements per track,
the first hit normally being in the \gls{IBL} installed before \RunTwo~\cite{ATLAS-TDR-19,PIX-2018-001}.
It is followed by the \gls{SCT}, a silicon microstrip tracker that usually provides eight measurements per track.
These silicon detectors are complemented by the \gls{TRT},
which enables radially extended track reconstruction up to \(|\eta| = 2.0\).
The \gls{TRT} also provides electron identification information
based on the fraction of hits above a higher energy-deposit threshold corresponding to transition radiation.
 
The calorimeter system covers the pseudorapidity range \(\abseta < 4.9\).
Within the region \(\abseta< 3.2\), electromagnetic calorimetry is provided by barrel and
endcap high-granularity lead/\gls{LAr} calorimeters,
with an additional thin \gls{LAr} presampler covering \(\abseta < 1.8\)
to correct for energy loss in material upstream of the calorimeters.
Hadron calorimetry is provided by the steel/scintillator-tile calorimeter,
segmented into three barrel structures within \(\abseta < 1.7\), and two copper/\gls{LAr} hadron endcap calorimeters.
The solid angle coverage is completed with forward copper/\gls{LAr} and tungsten/\gls{LAr} calorimeter modules
optimised for electromagnetic and hadronic energy measurements respectively.
 
The \gls{MS} comprises separate trigger and
high-precision tracking chambers measuring the deflection of muons in a magnetic field generated by the superconducting air-core toroidal magnets.
The field integral of the toroids ranges between \num{2.0} and \SI{6.0}{\tesla\metre}
across most of the detector.
Three stations of precision chambers, each consisting of layers of \glspl{MDT}, cover the region \(\abseta < 2.7\),
except in the innermost station of the endcaps, in the range $\abseta>1.3$, where the \gls{NSW} detectors, described below, have replaced the detectors used in \RunOneTwo. 
The muon trigger system covers the range \(\abseta < 2.4\) with \glspl{RPC} in the barrel  \(\abseta < 1.0\) , \glspl{TGC} in the endcap  \(\abseta > 1.0\) regions outside the toroids, and the \glspl{NSW} between the cryostats of the endcap calorimeters and the endcap toroids. 
 
Interesting events are selected by the first-level trigger system implemented in custom hardware,
followed by selections made by algorithms implemented in software in the high-level trigger~\cite{TRIG-2016-01}.
The first-level trigger accepts events from the \SI{40}{\MHz} bunch crossings at a rate below \SI{100}{\kHz},
which the high-level trigger further reduces in order to record events to disk at about \SI{3}{\kHz}.
 
An extensive software suite~\cite{ATL-SOFT-PUB-2021-001} is used in the reconstruction and analysis of real
and simulated data, in detector operations, and in the trigger and data acquisition systems of the experiment.
 
\begin{figure}[t]
\centerline{\includegraphics[width=0.95\textwidth]{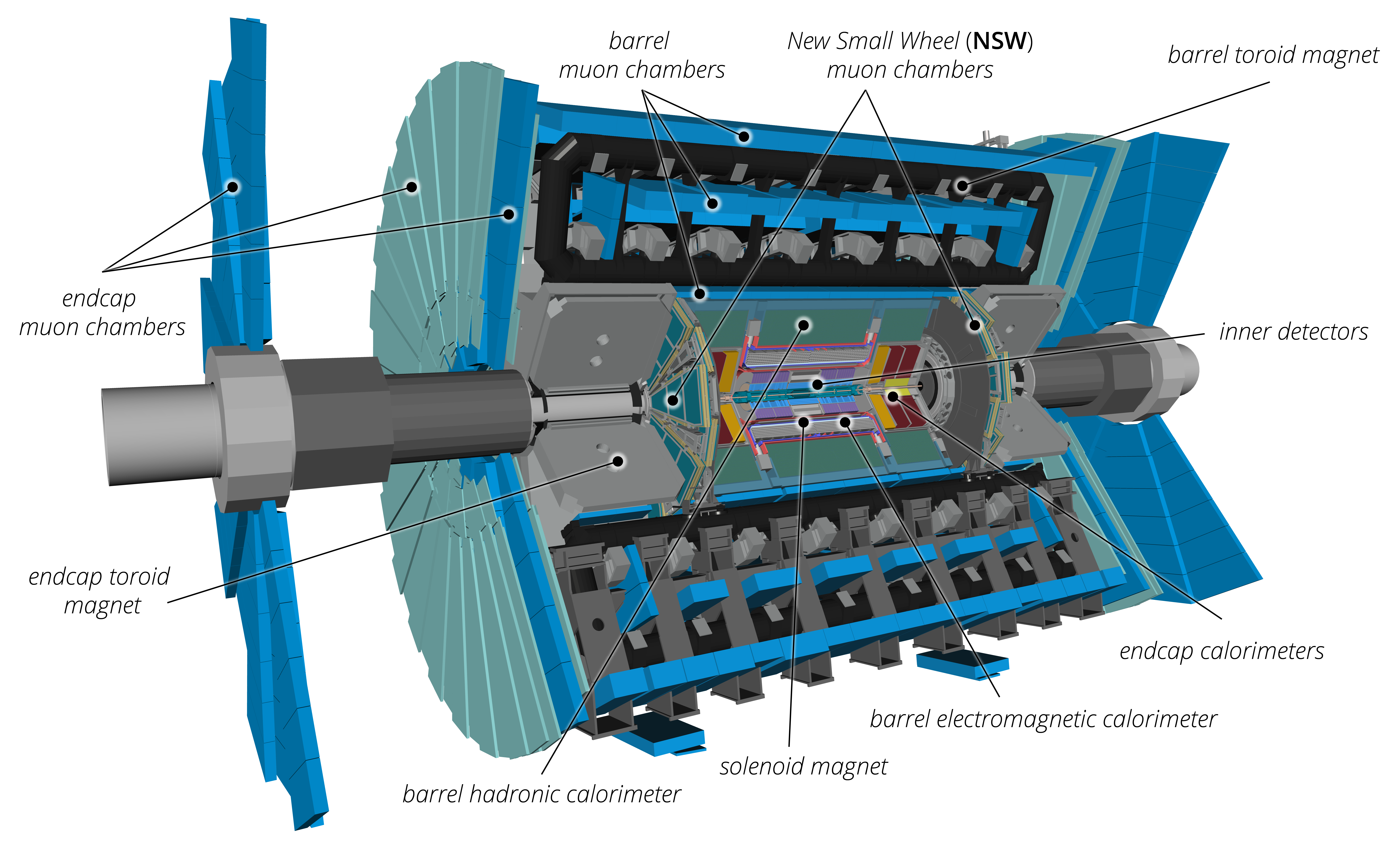}}
\caption{Cut-away view of the \RunThr configuration of the \ATLAS detector indicating the locations of the larger detector sub-systems.
\label{fig:Overview:ATLAS}}
\end{figure}

Figure~\ref{fig:Overview:ATLAS} illustrates the \RunThr\ configuration of the ATLAS detector; the main modifications implemented to the detector, its electronics, and the trigger and data acquisition system as well as their expected impact on the radiation levels are summarised here and will be described in detail in the following sections of this document.
 
\subsubsection{Radiation and shielding}
 
At the \gls{LHC}, the primary source of radiation at full luminosity
comes from collisions at the \gls{IP}. In the \gls{ID}, charged hadron
secondaries from inelastic proton-proton interactions dominate the radiation
backgrounds at small radii while at larger radii other sources,
such as neutrons, become more important for the calorimeters and the \gls{MS}. The expected background radiation levels for \RunThr are presented in Section~\ref{sec:BkgRadiation}, based on measurements from the \gls{LHC} \RunOne and \RunTwo and the modified and improved detector geometry.
 
In ATLAS, most of the energy from collisions at the \gls{IP} is dumped into two regions: the
collimators protecting \gls{LHC} quadrupoles and the \glspl{FCAL}.
The beam vacuum system spans the length
of the detector, and in the forward region it is a major source of radiation backgrounds.
Primary particles from the \gls{IP} strike the \beampipe\ at very shallow angles,
such that the projected material depth is large. Studies have shown that the beam-line
material contributes more than half of the radiation backgrounds in the muon system.
 
Details of the predicted \RunThr radiation levels and methods for monitoring and simulating the radiation in the ATLAS cavern are described in Section~\ref{sec:BkgRadiation}.

\subsubsection{Tracking}
\label{subsec:OverviewTracking}
\begin{figure}[t]
\centerline{\includegraphics[width=0.95\textwidth]{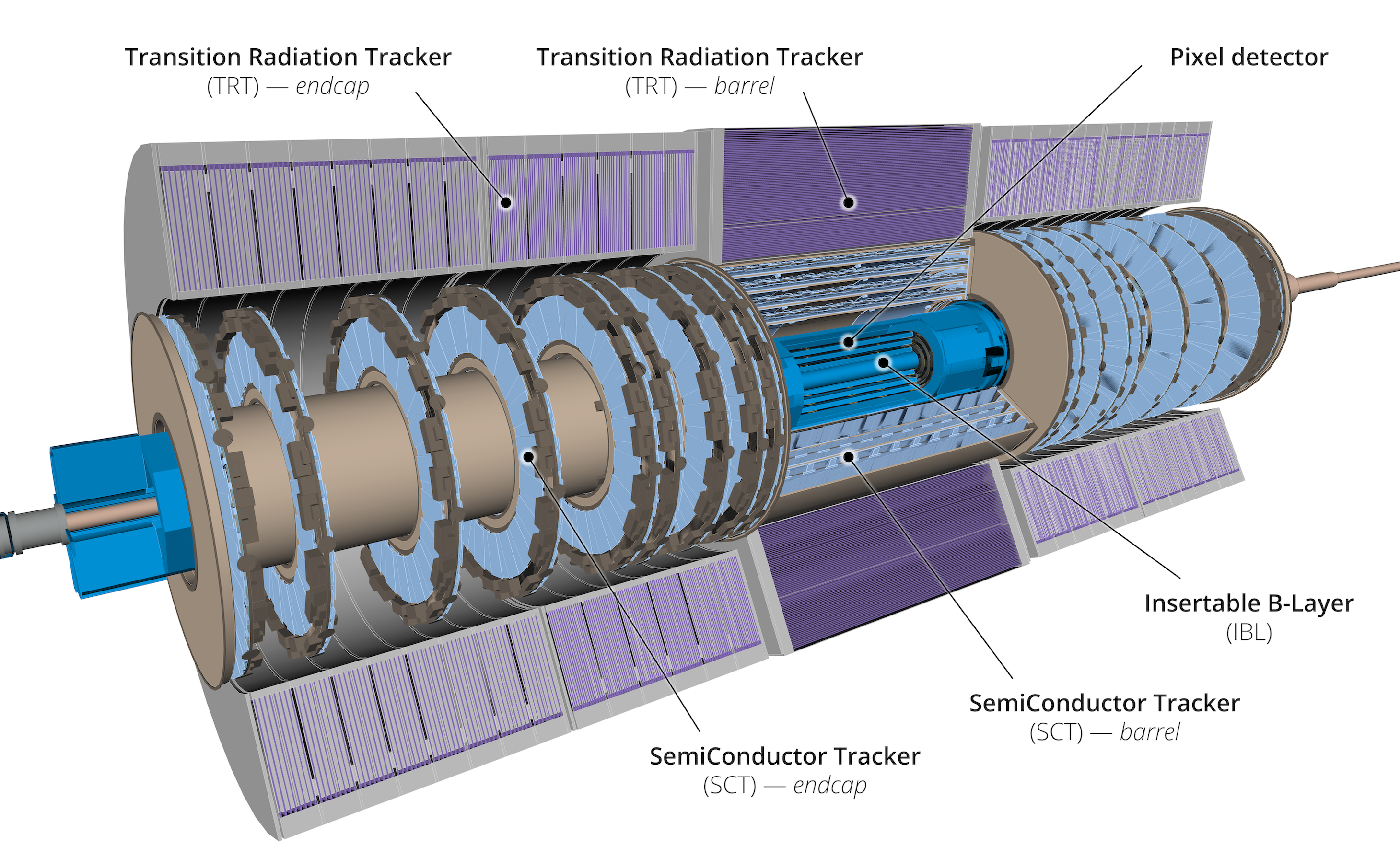}}
\caption{Cut-away view of the \ATLAS \gls{ID}, which is designed to provide a high-precision reconstruction of charged-particle trajectories. The \gls{ID} covers the pseudorapidity range $|\eta| < 2.5$ and has full coverage in $\phi$. It consists of a silicon \gls{Pixel} detector at the innermost radii surrounded by a silicon microstrip detector (\gls{SCT}) and a straw-tube detector, the \gls{TRT}, that combines continuous tracking capabilities with particle identification based on transition radiation.
\label{fig:Overview:ID}}
\end{figure}

The layout of the \gls{ID} is illustrated in Figure~\ref{fig:Overview:ID} and detailed in
Chapter~\ref{sec:ID}. Its
basic parameters  are summarised in Table~\ref{tab:IDpara}.
The \gls{ID} is immersed in a \SI{2}{\tesla} magnetic field generated by the
central solenoid, which extends over a length of \SI{5.3}{\m} with
a diameter of \SI{2.5}{\m}.
 
The precision tracking detectors (\gls{Pixel} and \gls{SCT}) cover
the region $\abseta<2.5$. In the barrel region, they are arranged on concentric
cylinders around the beam axis while in the endcap regions they are located
on discs perpendicular to the beam axis.
The highest granularity is achieved around the vertex
region using silicon \pixel\  detectors. The \ATLAS \gls{Pixel} detector consists of three barrel layers and three discs on each side, and has
approximately 80~million readout channels.  The \pixel\ layers are segmented in $r\phi$ and $z$ with typically
three \pixel\  layers crossed by each track. For these three outer layers, all pixel sensors are identical, with a pixel size of $\SI{50}\micron
\times \SI{400}{\micron}$. The intrinsic accuracies in the barrel
are \SI{10}{\micron} ($r\phi$) and \SI{115}{\micron} ($z$) and in the discs
are \SI{10}{\micron} ($r\phi$) and \SI{115}{\micron} ($r$). A fourth inner layer, the Insertable $b$-layer or \gls{IBL}, was installed during \gls{LS1} and started to be operational at the start of \RunTwo data taking. The \gls{IBL} sensors have $\SI{50}{\micron} \times \SI{250}{\micron}$ pixels and are at an average radius of \SI{33.4}{\mm}, adding an additional 12 million readout channels to the system (for a total of 92~million pixel channels).
For the \gls{SCT}, eight strip layers (four space points) are crossed by each track.
In the barrel region, this detector uses small-angle (\SI{40}{\milli\radian}) stereo strips
to measure both $r\phi$ and $z$, with one set of strips in each layer parallel
to the beam direction, measuring $r\phi$.
They consist of two \SI{6.4}{\cm} long daisy-chained sensors with a strip pitch
of \SI{80}{\micron}. In the endcap region, the detectors have a set of strips running radially and a
set of stereo strips at an angle of \SI{40}{\milli\radian}.  The mean pitch of the strips is also
approximately \SI{80}{\micron}. The intrinsic accuracies per module in the barrel
are \SI{17}{\micron} ($r\phi$) and \SI{580}{\micron} ($z$) and in the discs are
\SI{17}{\micron} ($r\phi$) and \SI{580}{\micron} ($r$).
There are approximately 6.3~million readout channels in the \gls{SCT}.
 
The \gls{TRT} is the outermost of the three tracking subsystems of the \gls{ID}, and comprises several layers of gas-filled straw tubes interleaved with transition radiation material. The \num{300000} thin-walled proportional-mode drift tubes provide on average \num{30} $(r,\phi)$ points with \SI{130}{\micron} resolution for charged particle tracks with $|\eta| < 2$ and $\pT > \SI{0.5}{\GeV}$, contributing to the combined tracking system \pT\ resolution. Along with continuous tracking, the \gls{TRT} provides electron identification capability through the detection of transition radiation X-ray photons.
 
\begin{table}[htbp]
 
\begin{center}
\caption{Main parameters of the Inner Detector system.}
\label{tab:IDpara}
\begin{tabular}{| l l | c | c |}
\hline
\textbf{Item} & & \textbf{Radial extension (\si{\mm})} &\textbf{Length (\si{\mm})}\\[2pt] \hline
& & & \\
\textbf{Overall \gls{ID} envelope}  &                      & $0<r<1150$           & $0<|z|<3512$ \\
\textbf{Beampipe}                &                       & $ 23.5 <r<30$        & \\
& & & \\ \hline
\textbf{\gls{Pixel} (\gls{IBL} included) }               & Overall envelope    & $ 31 <r<242$            & $0<|z|<3092$\\
4 cylindrical layers       & Sensitive barrel      & $33.5<r<122.5$        & $0<|z|<400.5$\\
$2\times 3$ discs         & Sensitive endcap   & $88.8<r<149.6$            & $495<|z|<650$\\
& & & \\
\textbf{\gls{SCT}}                & Overall envelope   &$255<r<549$ (barrel)  &$0<|z|<805$\\
&                    &$251<r<610$ (endcap )&$810<|z|<2797$\\
4 cylindrical layers       &Sensitive barrel    &$299<r<514$           &$0<|z|<749$\\
$2 \times 9$ discs         &Sensitive endcap  &$275<r<560$           &$839<|z|<2735$\\
& & & \\
\textbf{\gls{TRT}}                  & Overall envelope  & $554<r<1082$ (barrel)    & $0<|z|<780$\\
&                              & $617<r<1106$ (endcap ) & $827<|z|<2744$ \\
73 straw planes            &Sensitive barrel      & $563<r<1066$                & $0<|z|<712$\\
160 straw planes           &Sensitive endcap  & $644<r<1004$                & $848<|z|<2710$\\
\hline
\end{tabular}
\end{center}
\end{table}
 
\begin{figure}[t]
\centerline{\includegraphics[width=0.95\textwidth]{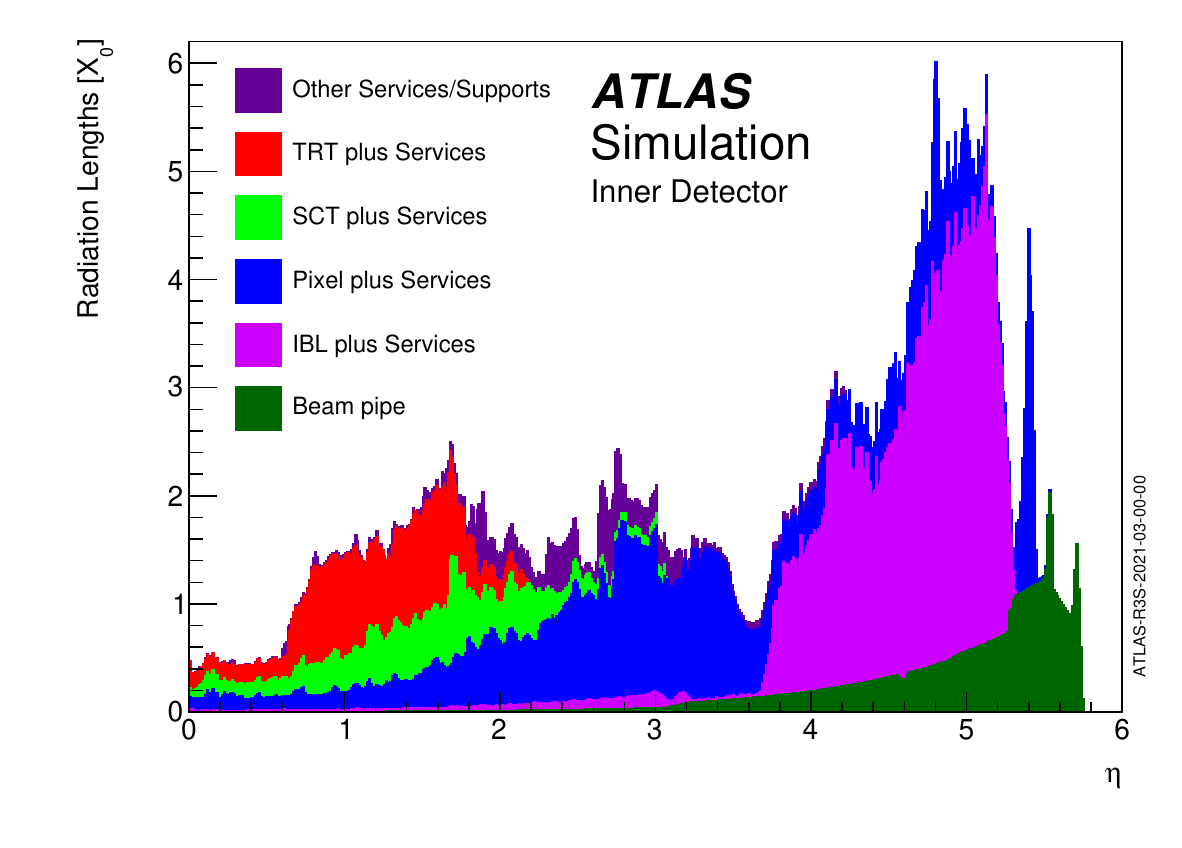}}
\caption{Radiation length ($X_0$) as a function of $|\eta|$ (averaged over $\phi$) for the \beampipe and different \gls{ID} components as implemented in the ATLAS geometry model describing the \RunTwo and \RunThr configuration. ``Services'' refers to supports, cooling infrastructure and cabling.
\label{fig:Overview:material}}      
\end{figure}

The updated \gls{ID} material distribution as a function of pseudorapidity is presented in Figure~\ref{fig:Overview:material}.
Compared with the equivalent distribution for the \RunOne detector, there is considerably more material at \(\abseta > 3.5\), due to the additional services for the \gls{IBL}, but almost no change in the region covered by tracking.
 
\subsubsection{Calorimetry}
 
The ATLAS calorimeters measure the energies and positions of charged and neutral electromagnetically or strongly interacting particles. They are designed to absorb most of the particles coming from a collision, forcing them to deposit all of their energy and stop within the detector. The ATLAS calorimeters are sampling calorimeters, in that they consist of layers of ``absorbing" high-density materials that stop incoming particles, interleaved with layers of ``active" media that measure the particle energies.
 
\begin{figure}[t]
\centerline{\includegraphics[width=0.95\textwidth]{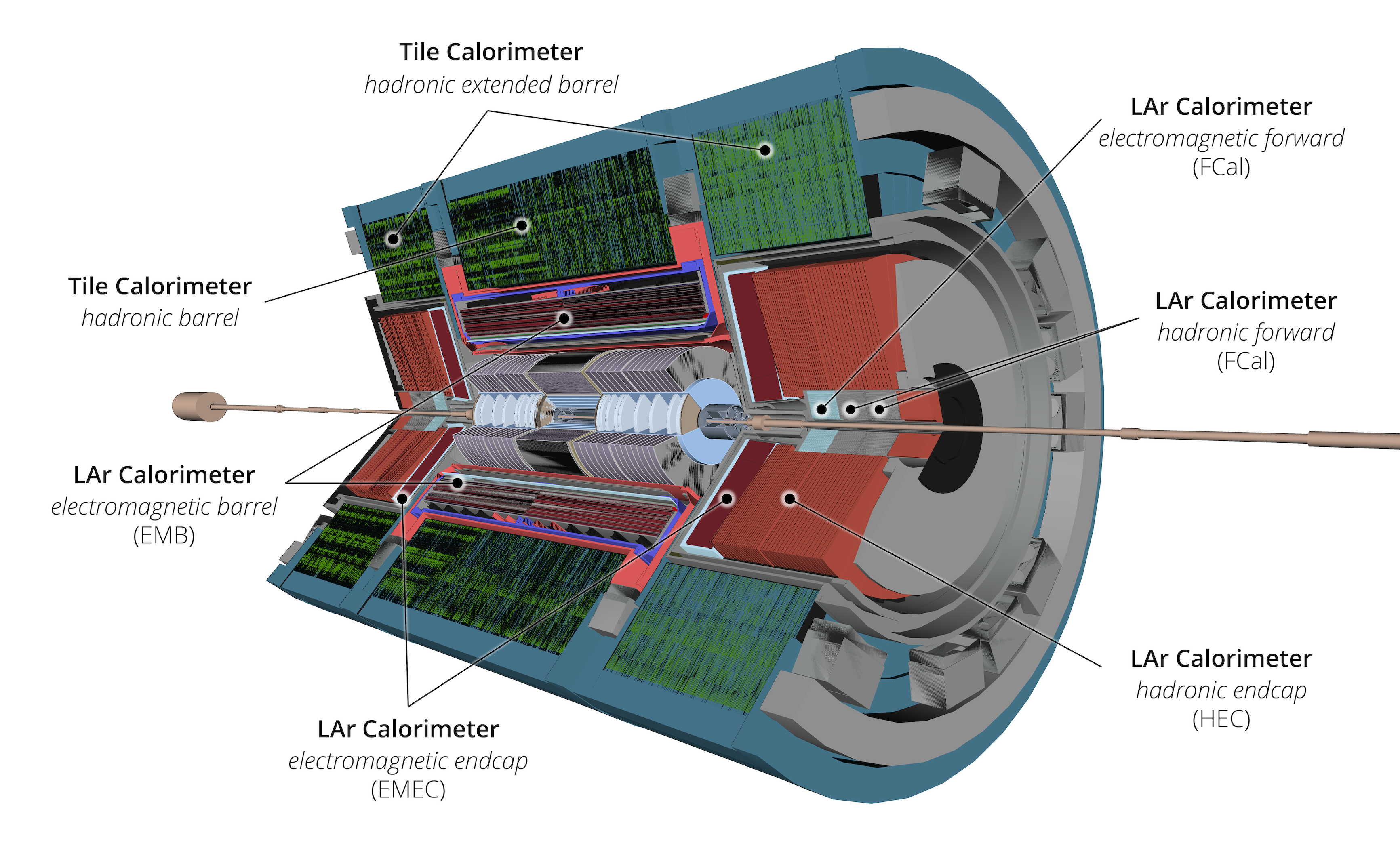}}
\caption{Cut-away view of the \ATLAS\ calorimeter system that measures the energies and positions of charged and neutral particles through interleaved absorber and active layers out to $|\eta| < 4.9$. \gls{LAr} is used as the active medium for the electromagnetic calorimeters and all of the endcap and forward calorimeters; scintillating Tiles are used for hadron calorimetry in the central region.\label{fig:Overview:calos} }
\end{figure}

ATLAS uses two sampling calorimeter technologies: \gls{LAr}~\cite{ATLAS-TDR-02} for the electromagnetic calorimeters and all of the endcap and forward calorimeters, and scintillating Tiles~\cite{ATLAS-TDR-03} for hadron calorimetry in the central region. The calorimeters are highlighted in Figure~\ref{fig:Overview:calos}.
 
The \gls{LAr} Calorimeter system consists of several subsystems, namely the
\gls{EMB}, the \gls{EMEC}, the \gls{HEC},
and the \gls{FCAL}. The ATLAS Tile Calorimeter covers the pseudorapidity region $\abseta<1.7$ using pseudo-projective calorimeter towers composed of scintillating tiles in a steel matrix, read out by wavelength-shifting fibers. The pseudorapidity coverage, granularity, and longitudinal
segmentation of the calorimeters are summarised in Table~\ref{tab:overview-calo-parameters}.
These calorimeters cover the range  $\abseta<4.9$, using different
geometries and absorber materials suited to the widely varying requirements of the physics processes of interest
and of the radiation environment over this large $\eta$-range.
Over the $\eta$ region for which the inner detector provides tracking, the fine granularity of
the \gls{EM} calorimeter is ideally suited for precision measurements of electrons and photons.
The coarser granularity of the rest of the calorimeter is sufficient to satisfy the
physics requirements for jet reconstruction and missing transverse momentum (\MET) measurements.
 
The ATLAS calorimeter detectors~\cite{LARG-2009-01} require very few changes to run at luminosities substantially higher than the original design, and are expected to last the entire lifetime of the \gls{LHC} and \gls{HL-LHC}. The main calorimeter upgrades were performed on the \gls{LAr} calorimeter electronics. A new digital trigger path provides finer granularity inputs to the upgraded trigger system and aims to better control the trigger rates by improving the selectivity of electron, photon, and tau lepton objects, the resolution of jets and \MET\ trigger signatures, and the discrimination power against background emerging from both out-of-time and in-time pileup. These upgrades as well as improvements to Tile calorimeter cryostat scintillation counters and Minimum Bias Trigger Scintillators are described in Chapter~\ref{sec:Calorimeters}.
 
\begin{table}
\scriptsize
\begin{center}
\caption{Main parameters of the calorimeter system.}
\label{tab:overview-calo-parameters}
\begin{tabular}{|c|l r|l l|}
\hline
&  \multicolumn{2}{c|}{\textbf{Barrel}}  & \multicolumn{2}{c|}{\textbf{Endcap}}  \\
\hline
\hline
\multicolumn{5}{|c|}{{\textbf{\gls{EM} calorimeter }}} \\
\hline
\multicolumn{5}{|c|}{Number of layers and $\abseta$ coverage} \\
\hline
Presampler &  1 &  $\abseta<1.52$ & 1 & $1.5<\abseta<1.8$ \\
\hline
Calorimeter & 3 & $\abseta<1.35$           & 2 & $1.375<\abseta<1.5$ \\
& 2 & $1.35<\abseta<1.475$     & 3 & $1.5<\abseta<2.5$ \\
&   &                         & 2 & $2.5<\abseta<3.2$ \\
\hline
\multicolumn{5}{|c|}{Granularity $\Delta\eta\times\Delta\phi$ versus $\abseta$}
\\
\hline
Presampler &  $0.025\times 0.1$ & $\abseta<1.52$ &  $0.025\times 0.1$ &
$1.5<\abseta<1.8$\\
\hline
Calorimeter 1st layer &  $0.025/8\times 0.1$ & $\abseta<1.40$
& $0.050\times 0.1$ & $1.375<\abseta<1.425$   \\
&  $0.025\times 0.025$ & $1.40<\abseta<1.475$
& $0.025\times 0.1$ & $1.425<\abseta<1.5$ \\
&    &                          & $0.025/8\times 0.1$ & $1.5<\abseta<1.8$ \\
&    &                          & $0.025/6\times 0.1$ & $1.8<\abseta<2.0$ \\
&    &                          & $0.025/4\times 0.1$ & $2.0<\abseta<2.4$ \\
&    &                          & $0.025\times 0.1$ & $2.4<\abseta<2.5$ \\
&   &                          & $0.1\times 0.1$ & $2.5<\abseta<3.2$ \\
\hline
Calorimeter 2nd layer   &  $0.025\times 0.025$ & $\abseta<1.40$
& $0.050\times 0.025$ & $1.375<\abseta<1.425$ \\
& $0.075\times 0.025$ & $1.40<\abseta<1.475$
& $0.025\times 0.025$ & $1.425<\abseta<2.5$ \\
&     &                         & $0.1\times 0.1$ & $2.5<\abseta<3.2$ \\
\hline
Calorimeter 3rd layer   &
$0.050\times 0.025$ & $\abseta<1.35$  & $0.050\times 0.025$
& $1.5<\abseta<2.5$ \\
\hline
\multicolumn{5}{|c|}{Number of readout channels} \\
\hline
Presampler & \num{7808} & & \num{1536} (both sides) & \\
Calorimeter & \num{101760} & & \num{62208} (both sides)& \\
\hline
\hline
\multicolumn{5}{|c|}{{\textbf{\gls{LAr} hadronic endcap }}} \\
\hline
$\abseta$ coverage & & & $1.5<\abseta<3.2$ & \\
Number of layers & & & 4 & \\
\hline
Granularity ${\Delta\eta\times\Delta\phi}$ & & &
$0.1\times 0.1$ & $1.5<\abseta<2.5$ \\
&  & &  $0.2\times 0.2$ & $2.5<\abseta<3.2$  \\
\hline
Readout channels & & & 5632 (both sides) & \\
\hline
\hline
\multicolumn{5}{|c|}{{\textbf{\gls{LAr} forward calorimeter }}} \\
\hline
$\abseta$ coverage & & & $3.1<\abseta<4.9$ & \\
Number of layers & & & 3 & \\
\hline
Granularity ${\Delta x\times\Delta y }$ (cm)& & & \gls{FCAL}1:  $3.0 \times 2.6$ &  $3.15<\abseta<4.30$  \\
& & & \gls{FCAL}1: $\sim$ four times finer &  $3.10<\abseta<3.15$,\\
& & &                          &  $4.30<\abseta<4.83$ \\
& & & \gls{FCAL}2:  $3.3 \times 4.2$ &  $3.24<\abseta<4.50$\\
& & & \gls{FCAL}2:  $\sim$ four times finer &  $3.20<\abseta<3.24$,\\
& & &         &  $4.50<\abseta<4.81$\\
& & & \gls{FCAL}3: $5.4 \times 4.7$ &  $3.32<\abseta<4.60$\\
& & & \gls{FCAL}3: $\sim$ four times finer &  $3.29<\abseta<3.32$,\\
& & &                         &  $4.60<\abseta<4.75$\\
\hline
Readout channels & & & \num{3524} (both sides) &  \\
\hline
\hline
\multicolumn{5}{|c|}{{\textbf{Scintillator tile calorimeter}}} \\
\hline
& Barrel & & Extended barrel & \\
\hline
$\abseta$ coverage & $\abseta<1.0$ & & $0.8<\abseta<1.7$ & \\
Number of layers & 3 & & 3 & \\
\hline
Granularity ${\Delta\eta\times\Delta\phi}$ & $0.1\times 0.1$ & & $0.1\times 0.1$ & \\
Last layer                       & $0.2\times 0.1$ & & $0.2\times 0.1$ & \\
\hline
Readout channels &  \num{5760} & &  \num{4092} (both sides) &  \\
\hline
\end{tabular}
\end{center}
\end{table}

\subsubsection{Muon system}
 
\begin{figure}[t]
\centerline{\includegraphics[width=0.95\textwidth]{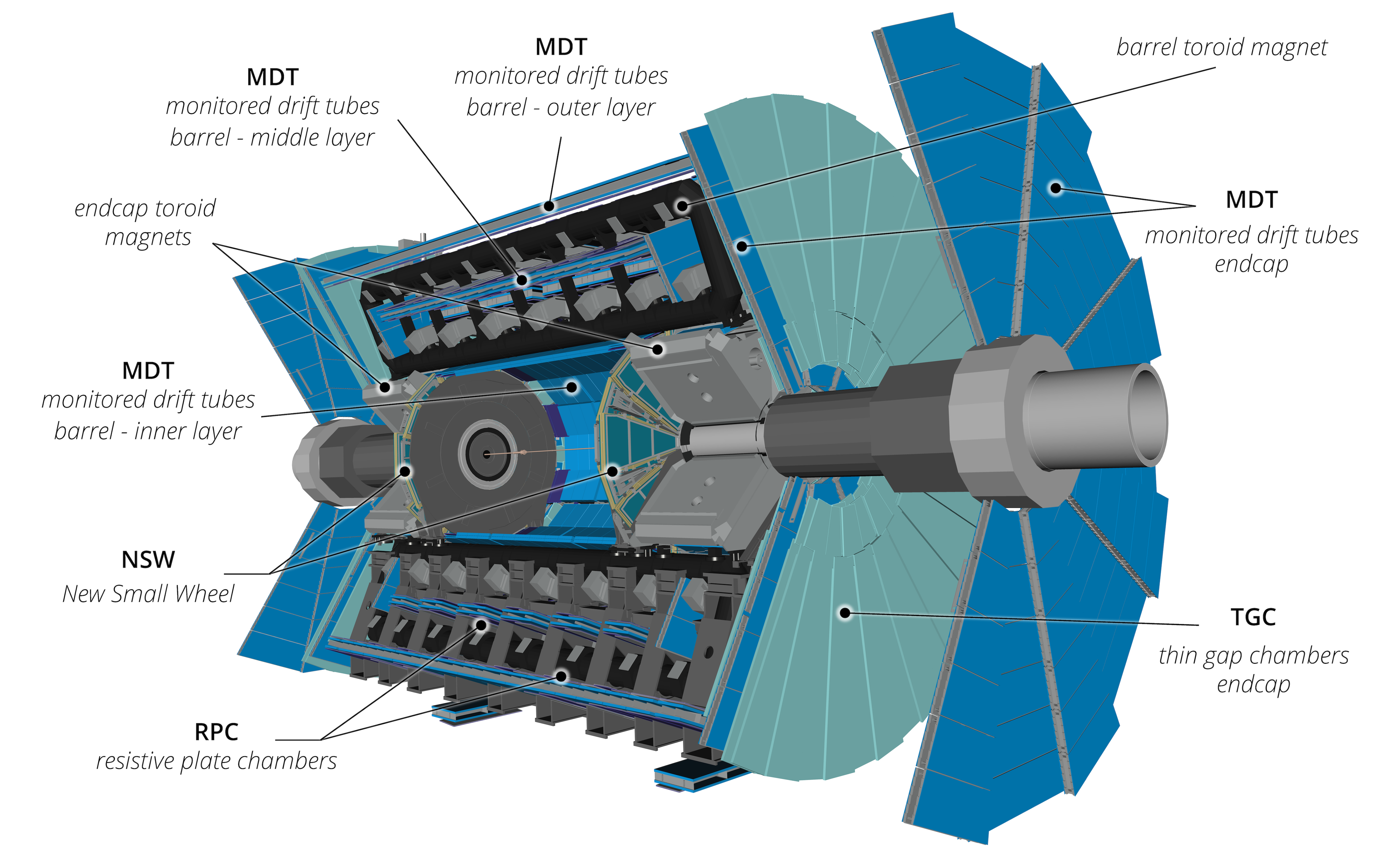}}
\caption{Cut-away view of the \ATLAS\ muon system that detects charged particles exiting the barrel and endcap
calorimeters and measures their momentum in the pseudorapidity range
$\abseta<2.7$. The muon system is built around an air-core toroidal magnet system. The main muon detector components upgraded for \RunThr\ (e.g., the \glspl{NSW}) are visible. \label{fig:Overview:muons}  }
\end{figure}

The muon spectrometer forms the large outer part of the ATLAS detector and
detects charged particles exiting the barrel and endcap
calorimeters, measuring their momentum in the pseudorapidity range
$\abseta<2.7$; the layout of the muon spectrometer is shown
in Figure~\ref{fig:Overview:muons}. The muon system is based on the
magnetic deflection of muon tracks in the large superconducting
air-core toroid magnets, instrumented with separate trigger and
high-precision tracking chambers. Over the range $\abseta<1.4$,
magnetic bending is provided by the large barrel toroid. For
$1.6<\abseta<2.7$, muon tracks are bent by two smaller endcap toroid magnets
inserted into both ends of the barrel toroid. In between these two regions, $1.4<\abseta<1.6$, magnetic deflection is provided by a combination of barrel and endcap fields. This magnet configuration provides a field which is mostly
orthogonal to the muon trajectories, while minimising the
degradation of resolution due to multiple scattering. The muon detector is included in the trigger system in the region $|\eta| < 2.4$.
 
The \glsfirst{MS} comprises a ``barrel'', consisting of three concentric, roughly cylindrical, stations (the inner, middle and outer barrels), and two endcaps, each consisting of three discs, referred to as the inner, middle, and outer endcap ``wheels'', supplemented by an ``extended'' endcap ring of detectors positioned outside the radius of each endcap toroid cryostat.
The anticipated high level of particle flux has had a major impact on
the choice and design of the spectrometer instrumentation and its upgrade, affecting
performance parameters such as rate capability, granularity, ageing properties, and radiation hardness.
 
The majority of the barrel detectors are unchanged from the original \RunOne configuration described in Ref.~\cite{ATLAS-TDR-20}:
all three stations use multilayered \gls{MDT} chambers for the precision measurements in the bending coordinate,
and the outer and middle stations are also equipped with \glspl{RPC} for triggering and to measure the azimuthal coordinate of the tracks. The middle and outer wheels are unchanged from \RunOne. The middle wheels are located on the far side of the endcap toroid cryostats and the outer wheels are mounted on scaffolding attached to the end-walls of the ATLAS cavern. The ``extended'' endcap rings provide a third measurement station between the inner and middle wheels for tracks with $1.05 < \abseta < 1.3$, which are outside the acceptance of the outer wheel.
The outer wheels contain only \glspl{MDT}, while the middle wheels have both \glspl{MDT} for precision tracking in the bending coordinate, and \glspl{TGC} for triggering and for measuring the azimuthal coordinate. The inner wheels (often nicknamed ``small wheels''), which are the main focus of the Phase-I upgrade, sit between the calorimeters and the endcap toroid cryostats, inside the barrel toroids. These inner wheels have been completely replaced by \glspl{NSW} occupying the same position and providing tracking over the same polar angle range: $1.3 < \abseta < 2.7$. The \glspl{NSW} use two chamber technologies: \glspl{sTGC} and \gls{MM} detectors, both fast enough for Level-1 trigger functionality, and both designed for precision tracking in the bending direction, as well as improved resolution in the azimuthal coordinate. These \glspl{NSW} allow for improved \pT\ resolution in the trigger and increased background rejection, allowing for a low muon \pT threshold and manageable Level-1 trigger rate, thus maintaining the acceptance for many interesting physics processes.

The main parameters of the muon chambers are listed in Table~\ref{tab:muon_parameters}; the upgrade of the muon system for \RunThr is described in detail in Chapter~\ref{chapter:muons}. The asymmetry of the \glstext{RPC} coverage is due to the installation of \gls{BIS78} detectors on only the $\eta>0$ side of ATLAS for \RunThr.
 
\begin{table}[t]
\begin{center}
\caption{Main parameters of the  muon spectrometer.\label{tab:muon_parameters}}
\begin{tabular}{|l|c|}  \hline
\multicolumn{2}{|c|}{\textbf{\gls{MDT}}}              \\ \hline
$\eta$ coverage                 & $\abseta<2.7$ (innermost layer: $\abseta<1.3$)     \\ 
Number of modules       & \num{1098}             \\
Number of channels       &
355k \\
Function                 & Precision tracking \\ \hline
\multicolumn{2}{|c|}{\textbf{\gls{sTGC}}}                 \\ \hline
$\eta$ coverage                 & $1.3<\abseta<2.7$ (2.4 for trigger) \\
Number of quadruplets    & \num{192}                 \\
Number of gas volumes         & \num{768}\\
Number of channels       &
357k \\
Function                 & Trigger, precision tracking, 2nd coordinate \\ \hline
\multicolumn{2}{|c|}{\textbf{\gls{MM}}}                \\ \hline
$\eta$ coverage                 & $1.3<\abseta<2.7$ (2.4 for trigger)\\
Number of quadruplets    & \num{128}                \\
Number of gas volumes         & \num{512}                \\
Number of channels       &
2.05M \\
Function                 & Precision tracking, trigger, 2nd coordinate \\ \hline
\multicolumn{2}{|c|}{\textbf{\gls{RPC}}}                 \\ \hline 
$\eta$ coverage                 & $-1.05<\eta<1.3$     \\  
Number of modules       & \num{652}              \\
Number of channels       &
389k\\
Function                 & Trigger, 2nd coordinate \\ \hline
\multicolumn{2}{|c|}{\textbf{\gls{TGC}}}                \\ \hline
$\eta$ coverage                 & $1.05<\abseta<2.7$ (2.4 for trigger)\\
Number of modules       & \num{1530}              \\
Number of gas volumes       & \num{3492}              \\
Number of channels       &
312k \\
Function                 & Trigger, 2nd coordinate  \\ \hline
\end{tabular}
\end{center}
\end{table}

\subsubsection{Forward detectors}
Four smaller detector systems cover the \ATLAS\
forward region (see Chapter~\ref{sec:Forward}).
At $\pm\SI{17}{\m}$ from the \gls{IP} lies \gls{LUCID}, which detects inelastic proton-proton scattering in the forward direction, and is the main online and offline luminosity monitor for \ATLAS. The second detector is \gls{ALFA}. Located at
$\pm\SI{240}{\m}$, it consists of scintillating fibre trackers housed inside Roman pots which are designed to approach as
close as \SI{1}{\mm} to the beam. \gls{ALFA} is used in dedicated low luminosity and high \betastar running of the \gls{LHC} and can provide a measurement of the total cross section for proton-proton interactions.
The third system is the \gls{ZDC}, which plays
a key role in determining the centrality of heavy-ion collisions. It is located at
$\pm\SI{140}{\m}$ from the \gls{IP}, just beyond the point  where the common straight-section
vacuum-pipe divides back into two independent beampipes. The \gls{ZDC} modules consist of layers of alternating quartz
rods and tungsten plates which measure neutral particles at pseudorapidities $\abseta \ge 8.2$.
The fourth system, which was installed in 2015 and 2016 is the \gls{AFP}, which consists of two arms, each with two stations, that are $\pm\SI{210}{\m}$ from the \gls{IP}. The \gls{AFP} is designed to study soft, hard, and central (exclusive) diffractive events at low luminosities using a silicon-based tracker for momentum measurements and a time-of-flight system to match protons from the two arms to a single inner detector vertex, thereby reducing the background from multiple proton-proton collisions.
 
\subsubsection{Trigger and Data Acquisition System}
\label{subsec:OverviewTDAQ}
 
\begin{figure}[b]
\centerline{\includegraphics[width=0.95\textwidth]{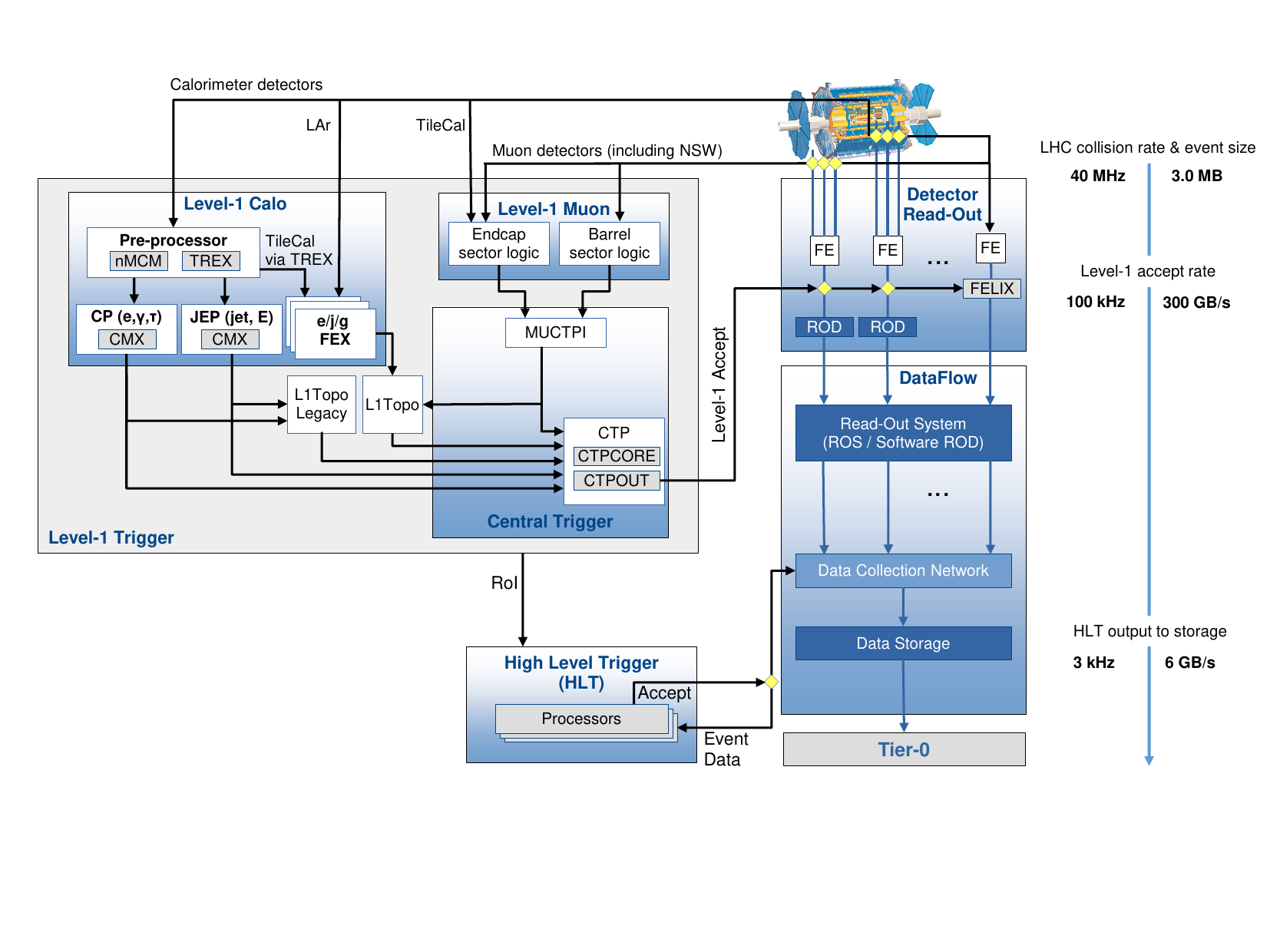}}
\caption{Schematic overview of the Trigger and \gls{DAQ} system in \RunThr. }
\label{fig:Overview:TDAQ}       
\end{figure}

The ATLAS \gls{TDAQ} system selects events with distinguishing characteristics (such as the presence of energetic leptons, photons, hadronic jets, $\tau$ leptons, or large missing energy) that make them interesting for physics analyses, and reads them out for further offline processing. It is based on a two-level event selection system: the \gls{L1} system,
which consists of custom-built electronics, and the \gls{HLT}, which is a software-based system implemented on commercial computers. Interwoven with these levels is the \gls{DAQ} system, which transports data from custom subdetector electronics through to offline processing, according to the decisions made by the trigger. A diagram of the complete \gls{TDAQ} system in \RunThr is shown in Figure~\ref{fig:Overview:TDAQ}.
 
The \gls{L1} trigger uses reduced-granularity information from the calorimeters and muon system to search for signatures of these events.
The maximum \gls{L1} accept rate supported by the detector readout systems is \SI{100}{\kHz}, and a share of this rate budget is allocated to each underlying physics object according to the physics goals of ATLAS. All processing for an event must be completed within the time window (latency) permitted by the detector electronics. This latency is \SI{2.5}{\micro\s} per event.
 
The \gls{HLT} software is designed to reproduce the offline selection as closely as possible, a philosophy that will be taken one step further in \RunThr through the use of AthenaMT (see additional details in Section~\ref{subsubsec:tdaq_daq_hlt_athenamt}). On average, the event processing time at the \gls{HLT} in 2018 was approximately \SI{400}{\ms} for runs with a peak luminosity of \SI{2e34}{\per\cm\squared\per\s}. The mean \gls{HLT} event processing time in \RunThr\ is expected to be larger due to widespread use of full-detector \gls{HLT} track reconstruction for hadronic signatures. Upgrades to the \gls{HLT} farm, and algorithmic improvements to tracking will keep the \gls{HLT} resource usage within the limits of the deployed system. The \gls{HLT} reduces the event rate from \SI{100}{\kHz} after the \gls{L1} selection to approximately \SI{3}{\kHz} (averaged over the course of an \gls{LHC} fill), after which the events are stored for offline analysis.
 
The \gls{DAQ} system has also undergone an upgrade in order to read out the full detector during \RunThr. The \gls{ROS} must support the \gls{L1} accept rate of \SI{100}{\kHz} plus $20\%$ contingency. The throughput of the \gls{ROS} will increase in \RunThr\ due to the increased request rate at which the \gls{HLT} will request data and a $30\%$ increase in the average event size (the event size at $\langle\mu\rangle \approx 60$ is \SI{2.1}{MB}).
The system must accommodate an average rate of \SI{3}{\kHz} of physics events to mass storage with the flexibility to handle variations in such rate. The maximum throughput is \SI{8}{GB/s}, which offers sufficient margin for these variations in rate. For example, at $\langle\mu\rangle \approx 60$, assuming a maximum physics rate to disk of \SI{3.3}{\kHz} and the above event size, the system throughput is \SI{6.9}{GB/s}. This represents nearly a factor of two increase in the required performance compared to \RunTwo.  In order to store the raw data volume for continuous operation at \SI{3}{\kHz} average output, guaranteeing 24 hours of storage in case of downtime of the CERN mass storage services, the contribution of file transfer and deletion latencies, and the margin needed to guarantee the required file system throughput characteristics, at least \SI{1.4}{PB} of effective storage volume is available at Point~1.
 
\subsection{Physics and performance goals for \RunThr}
\label{sec:Overview:PhysPerfGoals}
 
The ATLAS physics goals for \RunThr will build on the successful discovery of the Higgs boson and take advantage of the \SI{250}{\ifb} proton-proton collisions planned by the \gls{LHC}. The exploration of the mechanism of electroweak symmetry breaking includes an exploration of the phenomenology of the Higgs boson, precisely measuring its mass and width, the couplings to both fermions and bosons, and the observation of rare decay modes. Furthermore, ATLAS will continue to exploit the unique access to the energy frontier offered by the \gls{LHC} through the study of rare Standard Model processes, flavour physics, and searches for new phenomena such as \gls{SUSY} and exotic \gls{BSM} scenarios that may reveal the nature of dark matter.
 
The Phase-I upgrades of the ATLAS detector have focused on the following improvements to the detector and trigger system, which are required to remain effective not only for \RunThr, but throughout the lifetime of ATLAS. The first overall objective is the preservation, and in some cases improvement, of the low transverse-momentum electron and muon trigger thresholds that enabled the successful \RunOne\ and \RunTwo\ physics program through the collection of a rich dataset of electroweak boson ($H, \Wboson$, and $\Zboson$) decays. The second objective is to maintain sensitivity to electroweak-scale particles that produce hadronically-decaying tau leptons, jets and missing transverse momentum, the magnitude of which is referred to as \MET. These objectives are accomplished through the detector upgrades described in this paper: calorimeter electronics upgrades that provide finer granularity and higher energy resolution to the trigger system,  a new endcap muon detector that can tolerate the high background radiation environment that will be present in the \gls{HL-LHC} era, and \gls{TDAQ} system upgrades that take advantage of the calorimeter and muon upgrades while rejecting background in a high pileup environment.
 
In addition to the Phase-I upgrades, the physics performance of the ATLAS detector in \RunThr critically depends on its ability to measure charged particle tracks in order to reconstruct primary and secondary vertices; this performance was significantly enhanced by the installation of the \gls{IBL} described above.
 
The techniques for the measurement of the instantaneous and integrated luminosities developed in \RunTwo will be refined further in \RunThr to ensure a continuously high-performant luminosity measurement. \RunThr provides, moreover, a testing ground for any new luminometer envisaged for the \gls{HL-LHC} era to gain experience at moderate number of simultaneous particle interactions, before that number increases further for the \gls{HL-LHC}. The interplay of different independent luminometers, carefully calibrated and together covering the full range in the number of simultaneous interactions over up to four orders of magnitude, is of key importance to achieve the targeted precision in the integrated luminosity in \RunThr of $1\%$ or below.
 
The final \RunThr\ ATLAS detector configuration will enable the broad physics goals of the experiment described above through the enhanced selection of the signatures relevant to the ATLAS physics programme, including electrons, photons, muons, $\tau$-leptons, jets, $b$-jets, $B$ mesons, and \MET. The trigger menu developed for \RunThr translates the physics priorities of the experiment into allocations of the total \gls{L1} and \gls{HLT} rates. At \lumirunthree\ these rates are about \SI{95}{\kHz} and \SI{3}{\kHz}, respectively\footnote{These include $10\%$ ($13\%$) of the rate at \gls{L1} (\gls{HLT}) for support triggers, as was done in \RunTwo.}. The trigger menu comprises a list of trigger chains used for data-taking, where a chain consists of a \gls{L1} trigger item and a series of \gls{HLT} algorithms that reconstruct physics objects and apply kinematic selections to them. Roughly equal shares of the overall rate are given to electron and muon trigger chains, with a large share of bandwidth reserved for jets, \MET, taus, and multi-object triggers; rates for single electrons and muons are each limited to approximately \SI{25}{\kHz} or less. Improvements in the trigger performance allow  for lower single-lepton  \pT\ thresholds, yielding increased physics acceptance for a given trigger rate.
 
\begin{figure}[!hp]
 
\centering
\subfloat[]{\label{fig:ElectronTriggerPerformance}\includegraphics[width=0.6\textwidth]{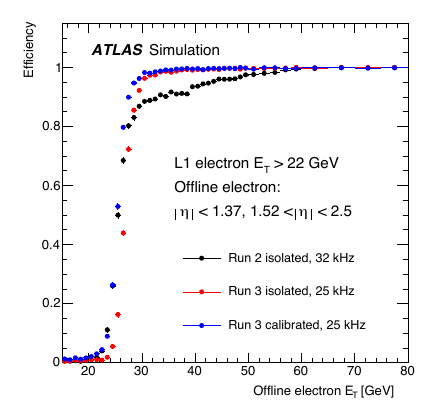}}
\\
\centering
\subfloat[]{\label{fig:MuonTriggerPerformance}\includegraphics[width=0.6\textwidth, height=0.6\textwidth, trim=100 0 100 0]{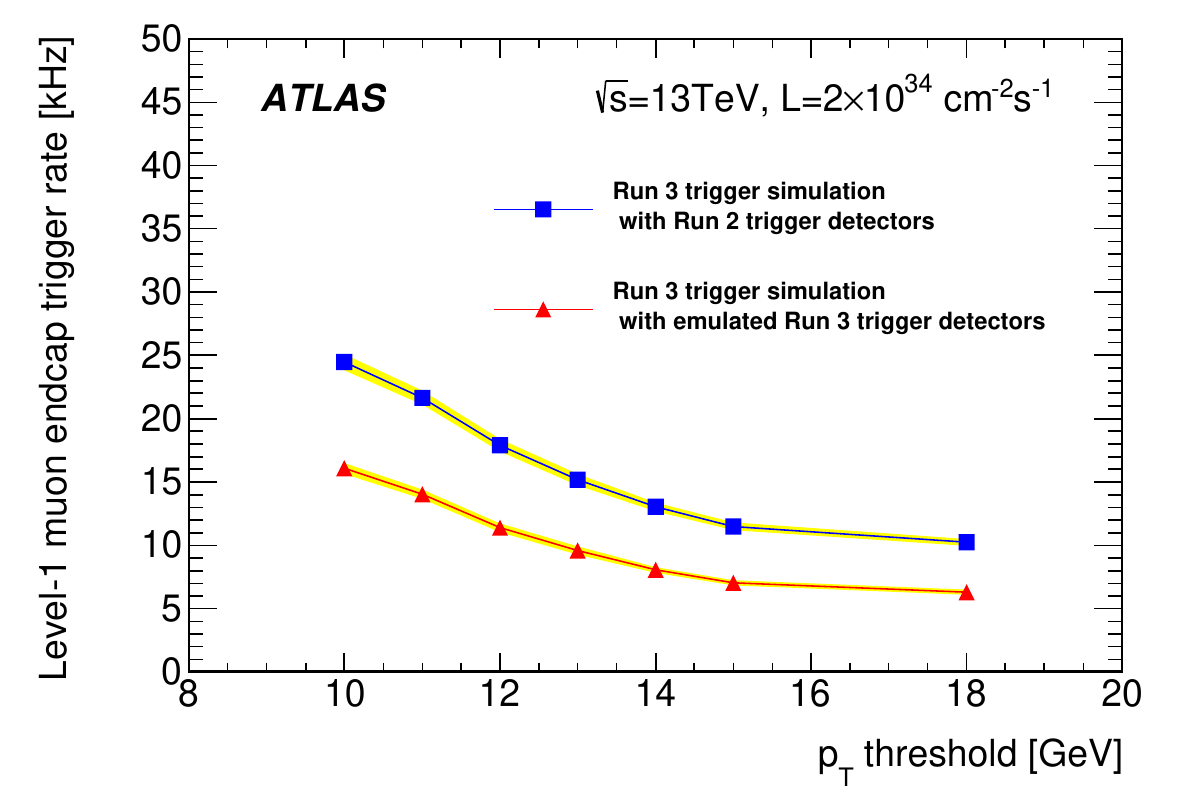}}
 
\caption{Expected \protect\subref{fig:ElectronTriggerPerformance} electron trigger efficiencies and \protect\subref{fig:MuonTriggerPerformance} muon trigger rate reduction after the Phase-I upgrades.}
\label{fig:LeptonTriggerPerformance}
\end{figure}

The expected single-lepton trigger performance is demonstrated in Figure~\ref{fig:LeptonTriggerPerformance}. Figure~\ref{fig:ElectronTriggerPerformance} demonstrates the expected performance of the \RunThr\ single-electron trigger using $\Zee$ Monte Carlo simulation: a lower rate and improved efficiency is achieved compared to the \RunTwo\ electron trigger. A threshold of \SI{22}{\GeV} is used for the \RunTwo\ (black) and uncalibrated \RunThr\ (red) electron and photon trigger. The \RunThr\ isolation thresholds were tuned to give the lowest rate while introducing only a $2\%$ inefficiency for electrons passing the \gls{L1} energy threshold; this isolation requirement is not applied for clusters with $\ET > 50 (60)\,\GeV$ in the \RunTwo\ (\RunThr) trigger. A layer- and $\eta$-dependent calibration is introduced (blue) to compensate for the varying detector response. The threshold on calibrated cluster energy is chosen to produce the same rate as the uncalibrated trigger, resulting in an improved efficiency. For muon triggers, the \glspl{NSW} reduce the trigger rate through the use of trigger chambers that provide up to \num{16} space points for muons, allowing a constraint on both the position and direction of the candidate muon at the location of the \gls{NSW} through a pointing segment matched to the primary trigger from the \glspl{TGC} of the middle station of the \gls{MS} (see also Section~\ref{sec:Muons}); this rate reduction is illustrated in Figure~\ref{fig:MuonTriggerPerformance}. In this figure, the expected \RunThr muon trigger performance is emulated by using the offline muon segments reconstructed by the \glspl{MDT} and \glspl{CSC} in the old Small Wheels on so-called ``enhanced bias'' data\footnote{To assure statistical sensitivity in the most relevant kinematic regions, a mix of events is selected by the \gls{L1} trigger system that emphasises higher energies and object multiplicities. This sample, which is taken from Run~360026, is constructed in such a way that the selection bias is removable with event weights.}. The \glspl{NSW} will provide even more important rate reductions in the \gls{HL-LHC} era.
 
\subsection{Outline of this Paper}
 
This paper describes the \RunThr\ configuration of the ATLAS detector. The background radiation environment and shielding are described in Section~\ref{sec:BkgRadiation}. Then the \glsfirst{ID}, \gls{LAr} and Tile calorimeters, and Muon systems are described in Sections~\ref{sec:ID}, \ref{sec:Calorimeters}, and~\ref{sec:Muons}, respectively. The upgraded Forward Detectors are presented in Section~\ref{sec:Forward}, followed by a description of the \glsfirst{TDAQ} System in Section~\ref{sec:TDAQ}. Finally, an outlook to the Phase-II upgrades planned to prepare ATLAS for the \gls{HL-LHC} are provided in Section~\ref{sec:Outlook}.


\clearpage
\newpage

\section{Background Radiation and Shielding} 
\label{sec:BkgRadiation}

 
The radiation field within and around the ATLAS detector is almost entirely created by the \pp collisions,
occurring at a rate of more than \SI{e9}{\per\s} 
at the \gls{IP}. The secondary particles produced in
these collisions distribute the total collision energy into different parts of the detector
and the \gls{LHC} ring.
 
About \SI{50}{\percent} of the particles produced are emitted into the acceptance of the central calorimeters
($|\eta| < 3$) but, on average, they carry only about \SI{1}{\percent} of the total energy. Around \SI{5}{\percent} of the collision energy
is deposited in the forward calorimeters ($3 < \abseta < 5$). Roughly a third is dissipated in the \gls{LHC} machine elements within the
experimental cavern, carried by \SI{20}{\percent} of the particles produced.
The remainder escapes into the \gls{LHC} ring and is of no consequence for the
detector.
 
Without any shielding the secondary particles interacting with the beam-line
elements and other material in the experimental area would result in prohibitively high radiation
levels in the ATLAS cavern and detectors. Even with dedicated and carefully optimised shielding
the radiation damage to sensors, electronics and power supplies is a serious concern and necessitates
a careful evaluation of radiation tolerance and in many cases recourse to dedicated radiation-hard
components. The margin between radiation exposure accumulated over the lifetime of the experiment and
the maximum tolerated by the components is often narrow. Accurate predictions of radiation
levels were therefore imperative for the original design of ATLAS and are ever more important for future upgrades.

\subsection{LHC luminosity evolution and expectation}
 
Figure~\ref{fig:Overview:lumi-pileup} shows the luminosity which the \gls{LHC} has delivered to ATLAS
during proton-proton operation in the years \numrange{2011}{2018}. The \SI{28}{\ifb} (at \sqrts{8}) from \RunOne and \SI{157}{\ifb} (at \sqrts{13})
from \RunTwo add up to \SI{185}{\ifb}. This was sufficient to cause non-negligible damage in
detectors, especially the silicon sensors of the \gls{ID}.
 
In Section\,\ref{sect:radsimcomp} the observed radiation damage, accumulated between 2010 and 2018, is compared with simulation results
in order to estimate the uncertainty of the simulation. For this the annealing (recovery) of the damage over
the 7-year period, including the end-of-year technical stops and the long shutdown between \RunOne and \RunTwo,
has to be corrected for.
 
The luminosity to be delivered during \RunThr might be as high as \intlumirunthree, but since this value
is uncertain, the radiation values in this section will be kept generic by normalising them to \SI{1}{\ifb}.
These values can be converted to instantaneous rates at, e.g. \SI{e34}{\instLumiUnit}, by multiplying them by
\SI{e-5}{\fb/\s}.

\subsection{Radiation shielding}
\label{ss:RadShielding}
\begin{figure}
\begin{center}
\includegraphics[width=0.9\textwidth]{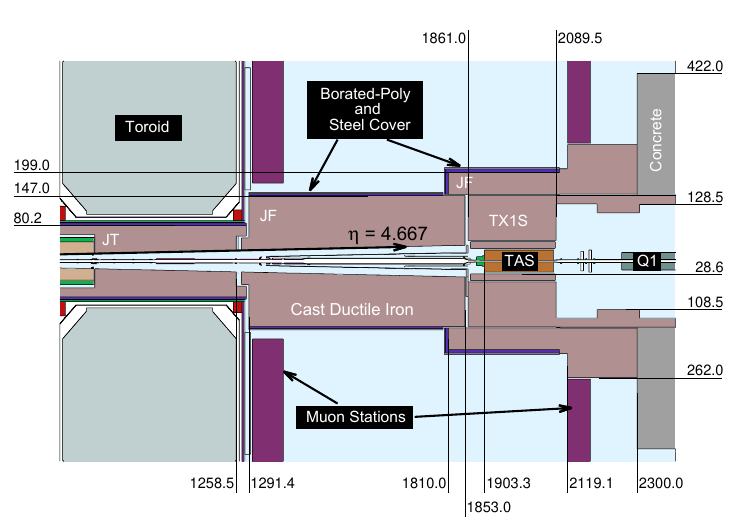}
\end{center}
\caption{ATLAS forward radiation shielding, as described in the $\phi$-symmetric \fluka geometry. All dimensions are shown in \si{\cm}. The $z$ dimensions (horizontal axis) are measured from the nominal \gls{IP}, which is to the left of the left-hand side of the drawing. The $r$ dimensions (vertical axis) are measured from the centre of the \beampipe (which is shown in the horizontal centre of the drawing). The role of the \glsentrytext{TAS} absorber
is to protect the focusing magnets, e.g. Q1, against particle debris from collisions at the \gls{IP}. The shielding around the \glsentrytext{TAS}, here labelled ``TX1S'', is also referred to as ``JN'' in some chapters of this document. The \gls{newJD} shielding that supports the \gls{NSW} of the \gls{MS} is located to the left of the toroid (outside the region shown in the drawing); the tip of its central hub can be seen projecting into the central opening in the toroid at the far left. }
\label{fig:fw_shielding}
\end{figure}
 
The high energy particles emitted in the forward direction cross the \gls{LHC} \beampipe at very shallow
angles and a significant fraction of them interact in the thin \beampipe wall. The particle showers
resulting from these interactions dissipate some of the energy locally while most continue into
the dedicated absorbers (\gls{TAS}) at $|z|=\SI{19}{\m}$ from the \gls{IP}. These absorbers protect the superconducting final focus
quadrupoles, e.g. Q1, of the \gls{LHC} against the heat load due to the collision secondaries emerging from the \gls{IP}.
These \gls{TAS} absorbers, two \SI{1.8}{\m} long copper cylinders with $r=\SI{17}{\mm}$ apertures for the beam, are
the most intense sources of background radiation in the ATLAS cavern.
They are therefore surrounded by shielding made of cast ductile iron of up to \SI{1.8}{\m} thickness, lined
with a layer of boron-loaded polyethylene, as shown in
Figure~\ref{fig:fw_shielding}.
In order to suppress the \SI{470}{\keV} photons from thermal neutron capture on boron, but also for fire protection, the polyethylene is
enclosed in a steel cover. Because boron-10 (which makes up \SI{20}{\percent} of natural boron) has a thermal neutron capture cross-section
of \SI{3800}{barn}, a boron concentration of as little as a few percent is sufficient to remove thermal neutrons and to prevent them from being captured on other elements that would emit more energetic capture photons.
 
The \beampipe within ATLAS represents a distributed source of radiation. Since calorimeters and
muon stations are extended to high pseudorapidities, space left for dedicated shielding around the
\beampipe is very limited. In addition, any shielding has to be placed such that it does not intercept particles
directed towards the \gls{TAS} absorbers. Since the latter are embedded deep inside the shielding they are the preferred
element to deposit waste energy. This is achieved by a conical inner bore of the shielding, as shown in
Figure~\ref{fig:fw_shielding}.
 
The purpose of the calorimeters is to measure the energy of particles,
so the radiation load on them cannot be reduced by shielding and sufficiently
radiation hard technologies must be applied. The liquid argon used in the ATLAS electromagnetic calorimeters,
the hadronic endcaps and the forward calorimeters is intrinsically radiation hard. The scintillators in the Tile calorimeter
are subject to radiation damage, but their dose is reduced because they are shielded by the electromagnetic calorimeter.
The electronics and power supplies are the most vulnerable parts and have to be protected
by shielding, and by optimising their location.
 
The albedo from the calorimeters, thermal neutrons emitted back towards the \gls{IP}, is a significant
contribution to the radiation load of the inner tracking detector, adding to the irreducible
particle flux emerging from collisions at the \gls{IP}. The absorber material
used in the ATLAS electromagnetic calorimeters is lead, which results in copious neutron emission
upon hadronic interactions. The neutron albedo is, however, reduced significantly by
polyethylene moderators lining the sides of the calorimeters which face the \gls{IP}.
 
\subsection{Characterisation of the radiation environment}
\label{sect:radDamMech}
 
The radiation environment, being composed of several particle types, all with a wide energy spectrum,
cannot be described by a single quantity. On the other hand it is not generally feasible -- and usually not
necessary -- to consider detailed particle spectra at all locations of interest.
In most cases the characterisation of the radiation environment can be based on a few quantities that have proven to
be related to the effects caused in various types of detectors or microelectronic circuits.
The three generic damage mechanisms and the associated radiation quantities are:
 
\begin{itemize}
\item Bulk damage in silicon is usually assumed to be proportional to the \SI{1}{MeV} neutron equivalent
fluence (\phieqv)~\cite{summers87}, where fluence is defined as total track length per unit volume.
For silicon this equivalence is defined through energy-dependent hardness factors\,\cite{moll-TNS65:1561}, which depend on particle type and are not applicable to any
other material\footnote{Hardness factor compilations also exist for other semiconductor materials, such as GaAs and diamond.}.
These factors have been experimentally
determined only over limited energy ranges for neutrons, protons and pions. For other particles
they are extrapolated by theoretical calculations\,\cite{huhtinen-nim335:580, huhtinen-nim491:194}.
In addition, different electrical properties, e.g. increase of leakage current and changes of the effective doping concentration, exhibit differences in particle type and material
dependence\,\cite{moll-TNS65:1561}.
For typical particle spectra encountered in ATLAS a systematic uncertainty of about \SI{30}{\percent} has to
be assigned to the characterisation of the leakage current increase in silicon through \phieqv.
For estimates of the change of effective doping concentration,
where the particle type and silicon properties (e.g. doping level and impurities) play a non-negligible role, it is better to consider
\phieqv divided into two components:  neutrons and other particles.
\item \gls{TID}, measured as the amount of energy deposited via ionising processes
per unit mass of material, leads to damage in electronics through charge trapped in oxide layers.
For instance in transistors, the accumulation of this trapped charge can induce a shift of the threshold voltage needed to switch the state
or increase lateral leakage current in the oxide layer. Scintillating materials and optical
fibres also suffer from damage that to good approximation is proportional to the \gls{TID}~\cite{ZHU1998297}.
The damage manifests itself mainly as a reduction of the light transmission, but sometimes also as induced
phosphorescence or changes of the scintillating properties.
\item \gls{SEE} in electronics circuits are caused by large energy depositions
close to sensitive regions of the chips. The released charge can be sufficient to flip the logic state
of a transistor or, in the worst case, permanently damage the component.
The amount of ionisation needed to cause a \gls{SEE} can only be deposited by slow heavy ions.
At the \gls{LHC} such slow ions are nuclear fragments from hadronic interactions within the chip itself. The
generally adopted characterisation of the radiation field for \gls{SEE} rate estimates is to use the total
flux of hadrons with energies above \SI{20}{\MeV}~\cite{huhtinen-nim450:155} (\phiSEU). Contrary to the two first
mechanisms, which depend on the cumulative radiation exposure, the probability of a \gls{SEE} occurring in
any given time interval can, to a good approximation, be assumed to be independent of the irradiation
history.
\end{itemize}
 
In addition to the characterisation by the damage potential of the radiation a further quantity
is needed to describe the potential impact on detector performance: the instantaneous rate of charged
particles at peak luminosity. These charged particles create hits in detectors and lead to an increase of occupancy and possibly spurious
triggers which can compromise the available bandwidth and increase the dead-time. The muon system, which
is designed for much more modest hit rates than the inner detector, is particularly sensitive to
the detailed kinematics of the background which gives rise to fake muon trigger signals (see Section~\ref{section:muonOverview}).
Since low-energy charged particles have short ranges in matter, an accurate
estimation of the hit rate in the sensitive volumes of the muon detectors requires a very detailed description
of the detector and its shielding. General fluence simulations, as presented here, give only an indication of the
hit rates to be expected.
 
\subsection{Radiation monitoring}
\label{sect:radmon}
 
In view of the significant role which radiation damage was expected to play for ATLAS components, a monitoring
of the radiation environment was foreseen early on. The RadMon system~\cite{RadMonRef} consists of sensors at 14 locations
inside and around the ATLAS detector. Each RadMon station comprises several \gls{RadFet} with varying oxide thicknesses such that they
together cover a range from few \si{\milli\gray} up to \SI{100}{\kilo\gray} of \gls{TID}.
In order to account for the fact that the radiation field around ATLAS is composed of many particle types and energies,
the \glspl{RadFet} are calibrated at several facilities using different particle types: protons of different energies, X-rays,
$\gamma$-rays and neutrons, allowing for an assessment of the differences in response for \gls{TID} from different particle types.
The uncertainty of the dose measurement with the \glspl{RadFet} was estimated at \SI{20}{\percent}~\cite{RadMonRef}.
 
The RadMon system also includes two different kinds of \gls{PIN}-diodes which provide two ranges of sensitivity to measure \phieqv: through
an increase of the leakage current under reverse bias or through the resistivity change by monitoring the voltage under forward current.
The diodes have different sensitivity ranges, such that the high-sensitivity ones provide a linear response
from about \SI{e9}{\per\cm\squared} up to few times \SI{e12}{\per\cm\squared}, where the
low-sensitivity ones pick up and can measure  up to around \SI{e15}{\per\cm\squared}. The diodes are also calibrated at various facilities,
using protons, pions and neutrons, and no significant particle type dependence was uncovered. They are thus
well suited for the mixed radiation field present in ATLAS.
 
In addition to the dedicated RadMon system, the radiation damage in the detectors themselves can be used to determine their total exposure.
In particular the silicon sensors of the \gls{ID} are ideal for such a study. Their irradiation and temperature (annealing)
histories are recorded accurately and their leakage currents are constantly monitored. Since sensors are distributed over the entire
Pixel and \gls{SCT} volumes, they provide an excellent means to determine the spatial variation of \phieqv in the \gls{ID}~\cite{IDET-2017-10};
however, these sensors are not suited to measure the \gls{TID}.
 
The scintillators of the Tile calorimeter have also suffered from radiation damage that has been compared with the simulated \gls{TID}~\cite{TileDamage}.
However, the radiation-induced light loss at the end of \RunTwo was still very small, with large uncertainties, so
the Tile scintillator data are not yet sufficient to constrain the simulation.
 
\subsection{Radiation simulation codes and methods}
 
The radiation environment in ATLAS has formerly been studied with the \fluka\,\cite{flukaref1, flukaref2, Ferrari:300336}
and \gcalor\,\cite{gcalorref} Monte Carlo simulation programs.
Recent simulations, however, have been done with \fluka and \GEANTV{4}\,v10.4~\cite{Agostinelli:2002hh}. While the
first has a long history as a dedicated radiation simulation program, \GEANTV{4} has only recently been
employed for such tasks. The advantage of the \GEANTV{4} simulation is that it benefits from a very
detailed three-dimensional model of the ATLAS detector, built for physics performance simulations.
Having two different simulation packages, utilising independent geometry models, provides a
means to assess the systematic uncertainties due to geometry description and accuracy of the
physics modelling.
 
The \fluka description of the ATLAS detector is much simpler than that of \GEANTV{4} and, with a few
exceptions, $\phi$-symmetric.
Since, however, the \fluka simulations are specifically targeted at
estimating the radiation levels, great attention has been paid to the modelling details and material
composition of the shielding. This is especially true for the \gls{ID}, while for the calorimeter and cavern
regions, where the \fluka geometry lacks many details, the \GEANTV{4} results are considered more reliable.
 
During the simulations several generic maps are generated, which illustrate the spatial variation of the intensity of the
most relevant radiation quantities.
All these maps are averaged over $\phi$ even if the underlying geometry is not perfectly
$\phi$-symmetric\footnote{In a few situations, where the $\phi$-structure has a significant
impact on the radiation levels, the \GEANTV{4} results have been averaged over restricted $\phi$-ranges.}
and mirror-symmetry with respect to $z=0$ is assumed.
The entire ATLAS cavern is covered by maps with $\SI{10}{\cm} \times \SI{10}{\cm}$ 
cell size in $r$ and $|z|$.
For the calorimeter regions a finer binning with $\SI{4}{\cm}\times\SI{4}{\cm}$ 
cell size is used. In the \gls{ID}
the radial gradient is steep and detector elements are thin. Here the radial bin size is \SI{2}{\mm} up to $r=\SI{20}{\cm}$ and
\SI{2}{\cm} beyond that. In $|z|$ the bin size in the \gls{ID} region is \SI{2}{\cm}.
 
It is expected that the \gls{LHC} will operate at a collision energy of \SI{13.6}{\TeV} throughout \RunThr. Since an
increase to \SI{14}{\TeV} is anticipated after \gls{LS3}, the radiation simulations have been performed assuming the ultimate design energy of \SI{14}{\TeV}.
The effect of this difference in the radiation levels is about \SI{2}{\percent}.
The primary \pp collisions were
simulated with \PYTHIAV{8}\,\cite{Sjostrand:2014zea} and results were normalised assuming an
inelastic cross section of \SI{79.3}{mb}.
 
\subsection{Comparison of observed radiation levels with simulations}
\label{sect:radsimcomp}
 
Prior to \gls{LHC} operation very few comparisons of radiation damage and simulations were available, so
the initial radiation predictions\,\cite{PERF-2007-01} relied heavily on simulations and only
rather vague estimations of the uncertainties could be given.
 
During the ten years of \gls{LHC} operation the radiation field in and around ATLAS has been measured and monitored by
the methods described in section\,\ref{sect:radmon}. A comparison of these measurements allows for better estimates of the uncertainties associated with the simulations.
The simulation models themselves have been further refined over the past decade and, most importantly, the
available computing power has increased dramatically. These developments have made it possible to
reduce the statistical uncertainties of the simulations to an extent that makes them negligible in almost
all cases. The only exceptions are regions with very low radiation levels where the statistical fluctuations from individual energy depositions
may be very large in a fully \analog simulation.
 
\subsubsection{Inner detector}
 
\begin{figure}
\begin{center}
\subfloat[]{\includegraphics[width=0.49\textwidth]{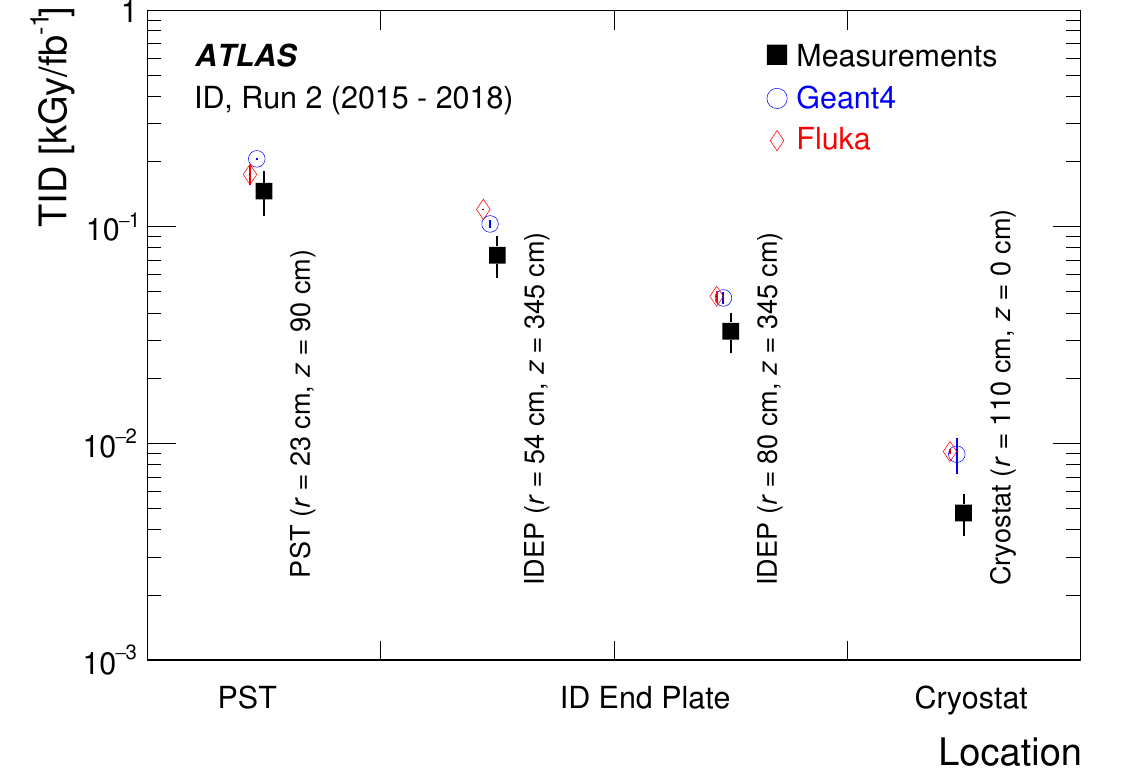}
\label{fig:Radmon_ID_dose}} \subfloat[]{\includegraphics[width=0.49\textwidth]{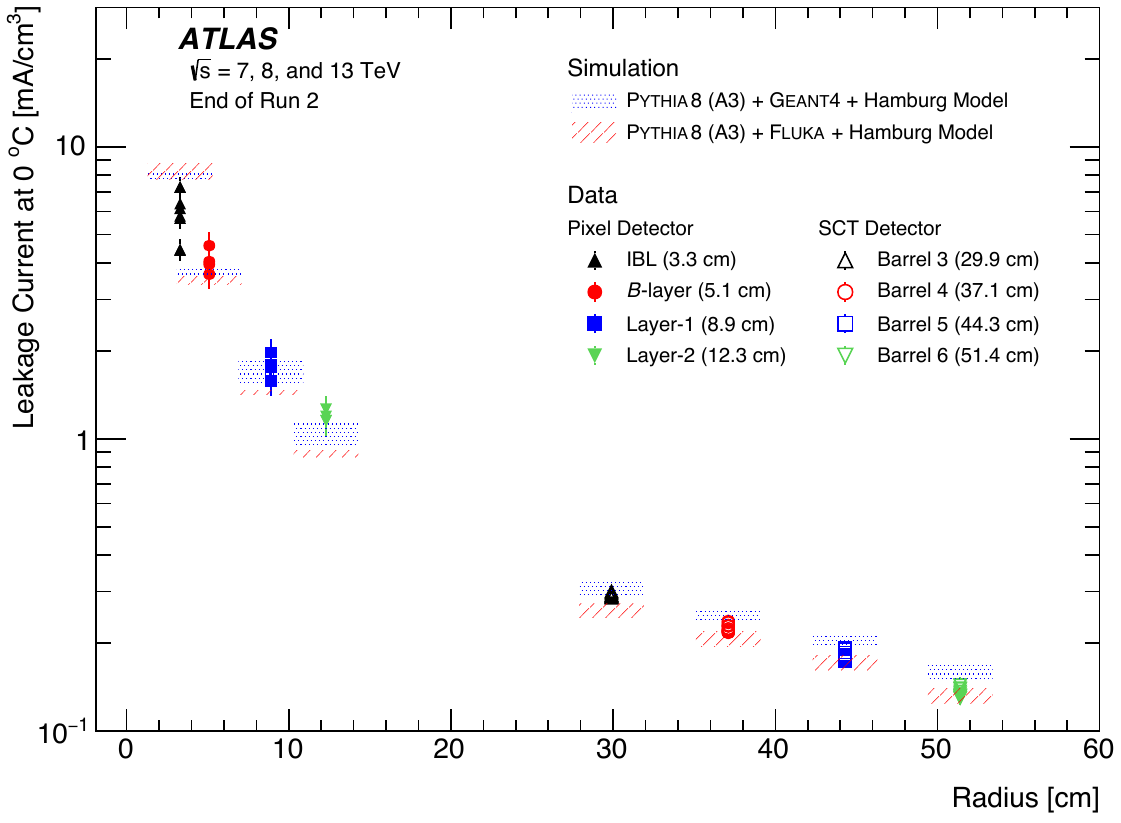}
\label{fig:SCT_sidam}}
\end{center}
\caption{\protect\subref{fig:Radmon_ID_dose} Comparison of measured and simulated \gls{TID} at several
locations inside the \gls{ID} volume. One set of monitors is close to the \gls{IP}, fixed on the \gls{PST}.
Two sets are at different radii on the \gls{IDEP} next to the endcap calorimeter, and a fourth set on the
wall of the cryostat of the solenoid. The error bars reflect the statistical uncertainty of the simulations and
the variation related to a position uncertainty of the RadMon sensors.
\protect\subref{fig:SCT_sidam} Comparison of measured and simulated leakage current in \gls{ID} sensors as a function of radius\,\protect\cite{IDET-2020-01}.
The error bars on the symbols include various uncertainties affecting the measurement while the widths of the simulated bands
reflect the statistical uncertainty and the variation as a function of sensor radius. Uncertainties of the silicon damage
factors, discussed in Section\,\ref{sect:radDamMech}, are not included.
\label{fig:damageComparisonID}
}
\end{figure}

Figure~\ref{fig:Radmon_ID_dose} shows the comparison of the RadMon measurements  with \fluka and \GEANTV{4}
simulations of the \gls{TID} inside the \gls{ID} volume. The contribution of \RunOne has been
subtracted from the measurements in order to compare only the \RunTwo dose, received at \sqrts{13}.
The simulations have a tendency to overestimate the \gls{TID}. The difference with respect to measurements is
\SIrange{30}{100}{\percent} with the largest deviation on the cryostat wall where the dose is lowest.
 
Figure~\ref{fig:SCT_sidam} shows a comparison of the leakage current measured in \gls{SCT} sensors with \fluka
and \GEANTV{4} predictions.
The simulated \phieqv values are converted to a prediction of the leakage current using the Hamburg annealing model
\cite{hamburgmodel}, which takes into account the detailed irradiation and temperature histories of the silicon
modules. The agreement is better than about \SI{20}{\percent}, i.e. within the uncertainty assumed for the silicon hardness
factors.
 
\subsubsection{Calorimeter and muon regions}
\label{sss:bkgCaloMuon}
\begin{figure}
\begin{center}
\subfloat[]{\includegraphics[width=0.49\textwidth]{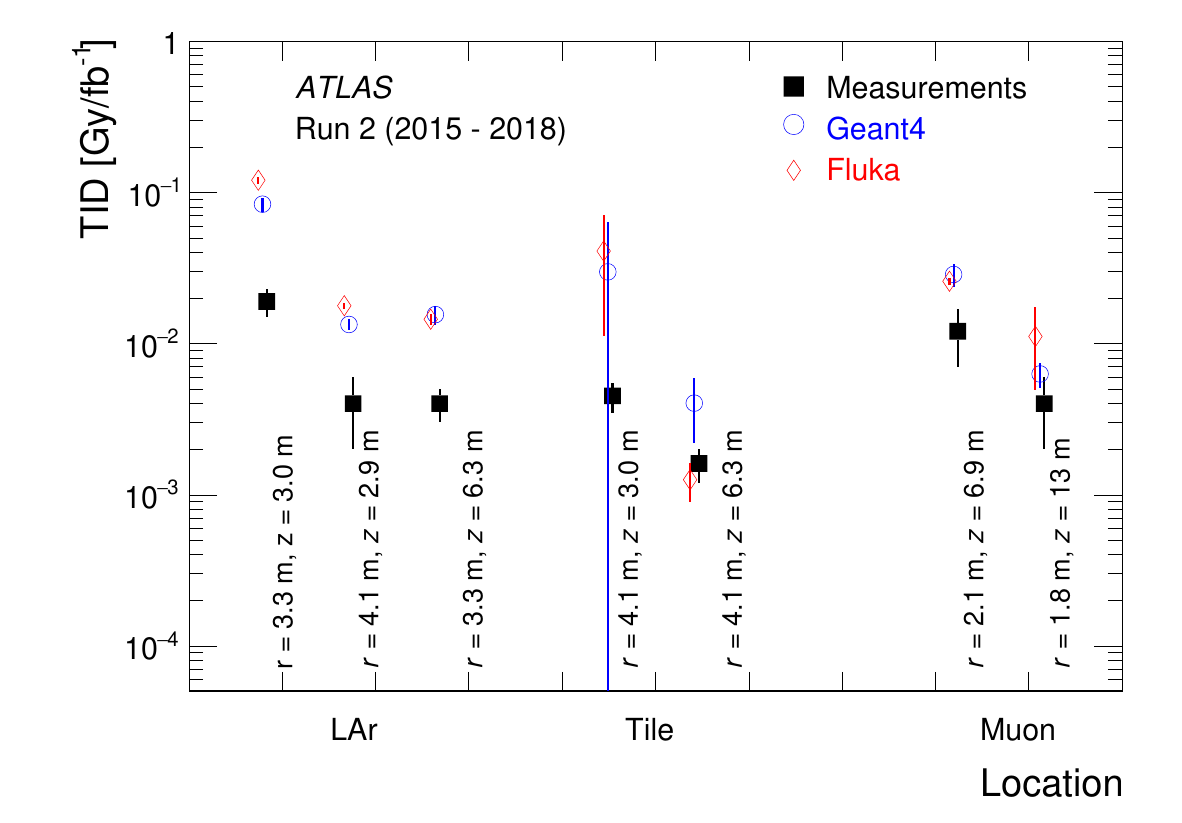}
\label{fig:Radmon_LAr_dose}}
\subfloat[]{\includegraphics[width=0.49\textwidth]{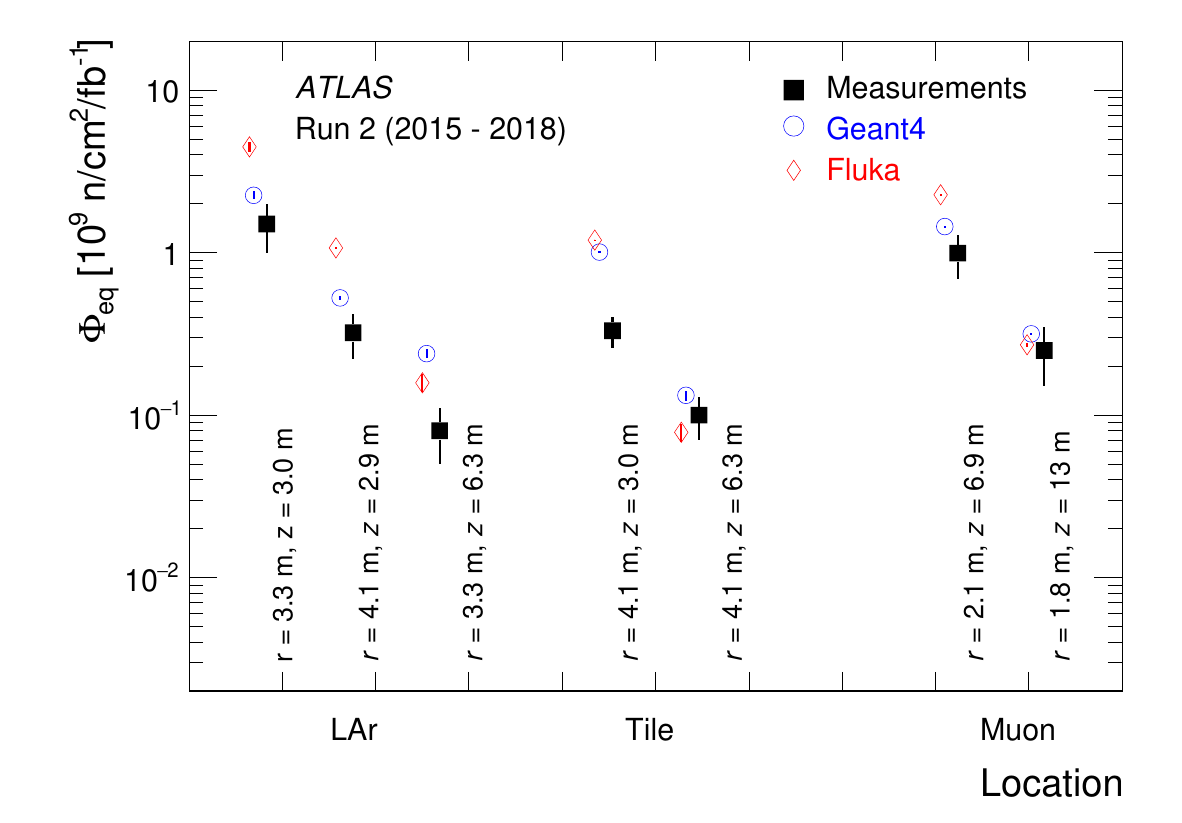}
\label{fig:Radmon_LAr_sidam}}
\end{center}
\caption{Comparison of measured \protect\subref{fig:Radmon_LAr_dose} \gls{TID}  and \protect\subref{fig:Radmon_LAr_sidam} \phieqv
in different locations around the calorimeters and the ATLAS muon system. The error bars reflect the statistical uncertainty of
the simulations and the variation related to a position uncertainty of the RadMon sensors. Some of the very large error bars for TID
are due to the fact that the quantities are recorded in a small well-shielded volume. Here a single deposition by an energetic
electron can dominate and cause a large fluctuation.
}
\label{fig:damageComparisonCalo}
\end{figure}

Figures~\ref{fig:Radmon_LAr_dose} and~\ref{fig:Radmon_LAr_sidam} show the comparison of the RadMon measurements
with \gls{TID} and \phieqv predictions of \fluka and \GEANTV{4} at the locations of calorimeter front-end electronics
and power supplies.
The agreement between the simulated and measured \gls{TID} is worse and less consistent than for the \gls{ID} regions. At some
locations the simulations are in satisfactory agreement with the measurements, but there are regions where the
predictions are five times higher than the RadMon devices indicate. All locations shown in
Figure~\ref{fig:damageComparisonCalo} are behind a significant amount of material with respect
to the \gls{IP}. Some are close to, or even within, service channels where the exact amount of material is difficult to
describe accurately in the simulation models. It is therefore very likely that the large deviations, which are all
overestimations by the simulations, are caused by an underestimate of cables and other services.
However, 
in many cases, \gls{TID} predictions of \fluka and \GEANTV{4} are more consistent mutually than with data. It is possible that
this indicates an issue with the calibration transfer of the RadFet devices, i.e. a systematic effect between the
spectra at the calibration facilities and the wide particle spectrum at the location where the sensors are installed in ATLAS.
This possibility needs further investigation and cross checks with additional monitors in \RunThr.
 
For \phieqv the agreement is better with maximum discrepancies not exceeding a factor of two. This is quite plausible
since the \gls{TID} is more sensitive to the accurate description of material in the immediate vicinity of the RadMon
detectors. Even small inaccuracies in the description of these immediate surroundings can result in a large deviation
of the \gls{TID} estimate while the other radiation quantities are less affected.
 
\subsubsection{Simulation safety factors}
 
Figures\,\ref{fig:damageComparisonID} and, especially, \ref{fig:damageComparisonCalo} indicate a that the simulations can deviate from
measurements by up to a factor of five. The larger deviations, however, are systematically overestimates of the simulations. All of these
appear in regions behind a significant amount of material, and most likely indicate that some material is missing in the simulation models.
In some regions of the inner detector the simulations underestimate the measurements, but by not more than 20\%.
 
These comparisons are crucial input for defining the safety factors to be applied on top of the simulations, when designing the detectors and
their electronics. Since the largest deviations systematically correspond to an overestimation of the simulations, ATLAS has adopted a
unique simulation safety factor of 1.5 for all radiation quantities and detector regions.

\subsection{Predicted \RunThr radiation levels}
The radiation levels are mapped separately for each of the three main detector subsystems, as the very different sizes of the \gls{ID}, the Calorimeters and
the \gls{MS} make different scales appropriate.
 
\subsubsection{Radiation levels in the Inner detector}
 
In the inner detector region the background originates from two main sources, of those discussed in Section~\ref{ss:RadShielding}:
\begin{itemize}
\item particles produced in the proton-proton collisions at the \gls{IP} and secondaries from their
interactions in the \beampipe walls or the material of the \gls{ID}.
\item albedo from the electromagnetic calorimeters.
\end{itemize}
The first comprises a mix of all particle types, but for silicon bulk damage, charged pions are
the most significant component.
While the fluence of charged hadrons decreases rapidly with radius, the neutron fluence, which is
dominated by calorimeter albedo, is more uniform throughout the Pixel and \gls{SCT} volumes.
 
\begin{figure}
\begin{center} 
\subfloat[]{\includegraphics[width=0.49\textwidth]{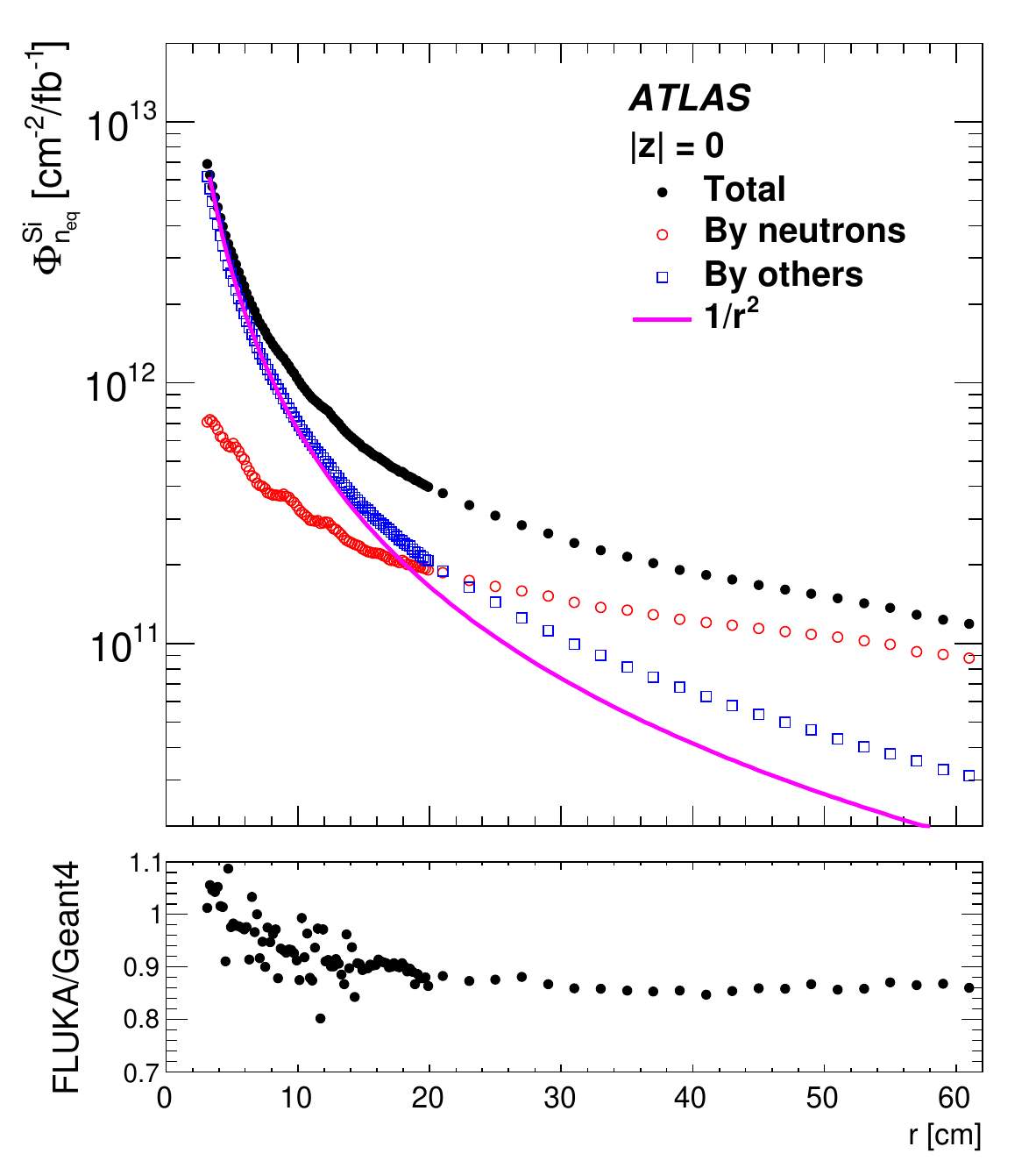}
\label{fig:rDepIDz0}}
\subfloat[]{\includegraphics[width=0.49\textwidth]{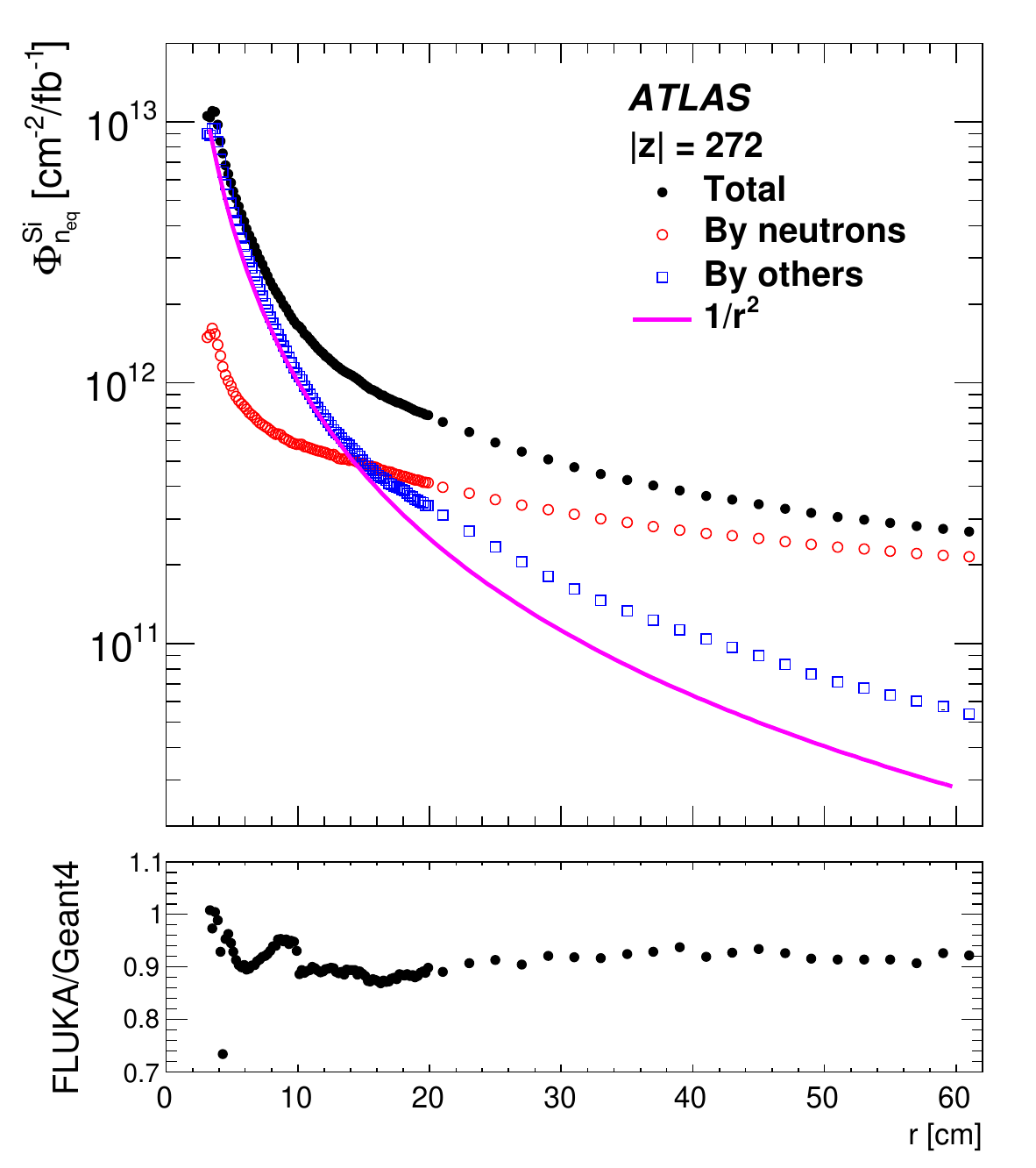}
\label{fig:rDepIDz272}}
\end{center}
\caption{Radial dependence of \phieqv at \protect\subref{fig:rDepIDz0} $|z|=0$ and \protect\subref{fig:rDepIDz272} $|z|=\SI{272}{\cm}$ in the \gls{ID}, as obtained from \fluka
simulations. The total \phieqv is divided into the contribution by neutrons and by all other particles.
The $r^{-2}$ dependence is normalised at $r=\SI{3.1}{\cm}$ to the `By others' value. The lower panels show the ratio
of \fluka and \GEANTV{4} results for the total.}
\label{fig:rDepID}
\end{figure}

Figure~\ref{fig:rDepID} shows that, in terms of \phieqv, the contributions of both generalised particle categories (neutrons, and all other particles)
are equal at radii of  \SI{20}{\cm} and \SI{15}{\cm} at $z=0$ (in the central transverse plane of the detector) and at
$|z|=\SI{272}{\cm}$ (at the extremities of the \gls{SCT} envelope), close to the endcap calorimeters,
respectively.
The bulk damage in the pixel detector is dominated by charged pions, but in the \gls{SCT} by neutrons. Figure\,\ref{fig:rDepID} shows that the neutron
component is more significant closer to the endcap calorimeters. This is due to the intense neutron albedo from the lead absorbers, which cannot be
entirely suppressed by the neutron moderators on the calorimeter face.
While for older technologies, including all those
installed for \RunThr, \phieqv provides a sufficiently accurate parametrisation of the bulk damage in silicon, an additional
particle-type dependence has been observed in some more recent devices. 
 
In the absence of scattering and a magnetic field, it can be shown that particles emerging from the \gls{IP} with a flat $|\eta|$-distribution
would result in a fluence independent of $|z|$ and dropping as $r^{-2}$.
Figure~\ref{fig:rDepID} shows that this
radial dependence models the data well up to $r\sim$\SI{20}{\cm} but in the \gls{SCT} region the fluence drops somewhat more slowly
due to particle production in inelastic interactions and possibly the curling up of tracks in the \SI{2}{\tesla} axial field.
Comparing Figures~\ref{fig:rDepIDz0} and~\ref{fig:rDepIDz272} also reveals a small $|z|$-dependence of the
\SI{1}{\MeV} neutron-equivalent fluence.
Besides the fact that the $|\eta|$-distribution is not perfectly flat, this increase of \phieqv with
$|z|$ may also be due to secondary production and the magnetic field.
The increase of the neutron contribution closer to the endcap calorimeter, clearly seen when comparing the two plots,
shows the impact of the calorimeter albedo, which is the main source of neutrons.
 
\begin{figure}
\begin{center}
\includegraphics[width=0.99\textwidth]{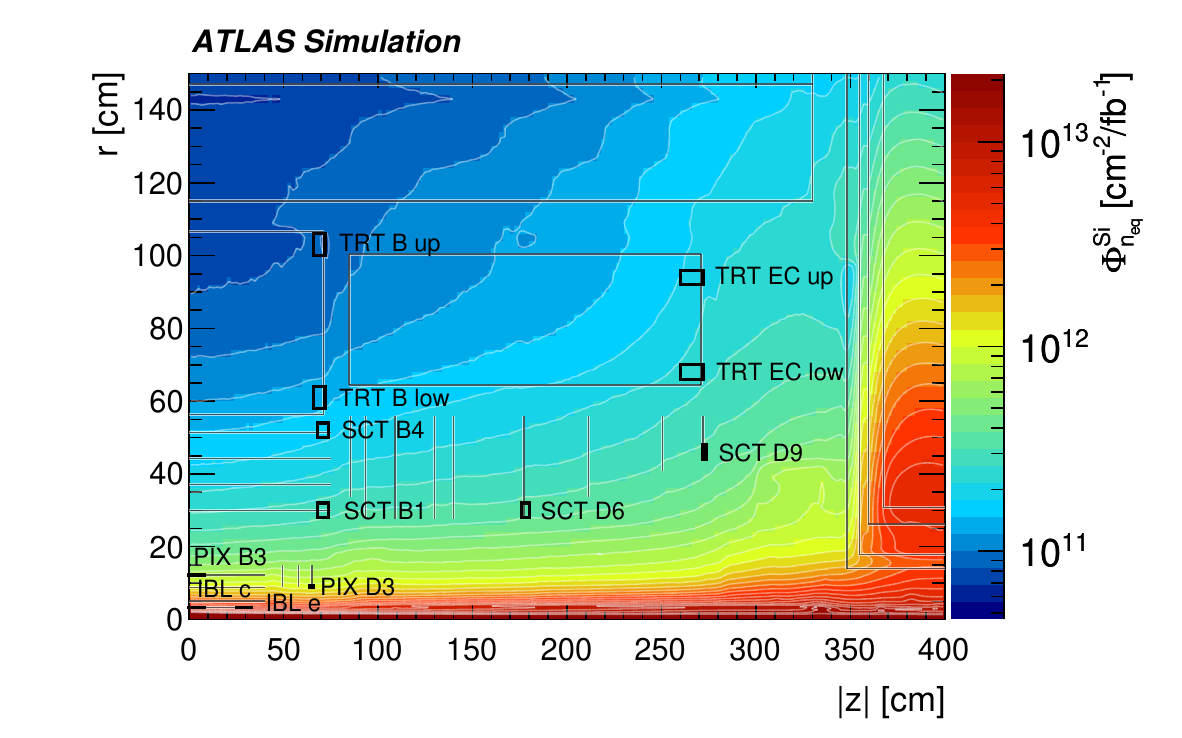}
\end{center}
\caption{Radiation levels in the inner detector region. The contours show \phieqv obtained from \fluka
simulations. The radiation levels, averaged over the labelled volumes are given in Table~\protect\ref{tab:radValuesID}
The Pixel and \gls{SCT} layers are indicated by grey lines; the \gls{TRT} and adjacent calorimeter and associated neutron moderator
volumes are indicated by boxes.
}
\label{fig:radLevelID}
\captionof{table}{\gls{TID} and characteristic particle fluences in the inner detector volume, obtained from \fluka
simulations at the locations indicated in Figure~\protect\ref{fig:radLevelID}.
The statistical uncertainty is in the last digit given, or smaller.
\label{tab:radValuesID}
}
\begin{center}

\begin{tabular}{c|r|r |c |c |c }
Region   &   \multicolumn{1}{c|}{|z| [cm]}   &   \multicolumn{1}{c|}{r [cm]}   &  TID [Gy/fb$^{-1}$]  &  \phieqv[cm$^{-2}$/fb$^{-1}$]  &  $\Phi_\mathrm{20}^\textrm{had}$ [cm$^{-2}$/fb$^{-1}$] \\ \hline
IBL e     &   29     &  3 & \expfor{3.42}{3} & \expfor{5.37}{12} & \expfor{9.08}{12} \\
IBL c     &   4     &  3 & \expfor{3.12}{3} & \expfor{6.19}{12} & \expfor{8.49}{12} \\
PIX D3     &   65     &  9 & $821$ & \expfor{1.13}{12} & \expfor{1.45}{12} \\
PIX B3     &   4     &  12 & $377$ & \expfor{7.85}{11} & \expfor{7.22}{11} \\
SCT D6     &   178     &  30 & $171$ & \expfor{3.58}{11} & \expfor{2.49}{11} \\
SCT B1     &   71     &  30 & $118$ & \expfor{2.93}{11} & \expfor{2.00}{11} \\
SCT D9     &   273     &  46 & $99.1$ & \expfor{3.36}{11} & \expfor{1.58}{11} \\
TRT EC low     &   266     &  68 & $65.0$ & \expfor{2.34}{11} & \expfor{8.50}{10} \\
SCT B4     &   71     &  52 & $46.0$ & \expfor{1.68}{11} & \expfor{8.12}{10} \\
TRT B low     &   69     &  61 & $36.7$ & \expfor{1.34}{11} & \expfor{6.09}{10} \\
TRT EC up     &   266     &  94 & $35.2$ & \expfor{1.86}{11} & \expfor{5.10}{10} \\
TRT B up     &   69     &  103 & $12.9$ & \expfor{8.27}{10} & \expfor{2.37}{10} \\ \hline
\end{tabular}

 
\end{center}
\end{figure}

This $r$- and $|z|$-dependence of \phieqv can also be appreciated qualitatively from Figure~\ref{fig:radLevelID}.
Table~\ref{tab:radValuesID} lists the three main radiation quantities averaged over the regions indicated by
rectangles and labelled with letters in Figure~\ref{fig:radLevelID}. These areas are selected to cover
representative regions of the Pixel, \gls{SCT} and \gls{TRT} detectors. The $r$ and $|z|$ positions given in Table~\ref{tab:radValuesID}
correspond to the centre of the rectangle. The statistical uncertainties of the simulations are typically less
than \SI{1}{\percent} but if the last digit given in Table~\ref{tab:radValuesID} is affected, this is indicated.
Systematic uncertainties are not included and, as discussed before, can amount to several
tens of percent.
 
\subsubsection{Radiation levels around the calorimeters}
 
\begin{figure}
\begin{center}     \includegraphics[width=0.99\textwidth]{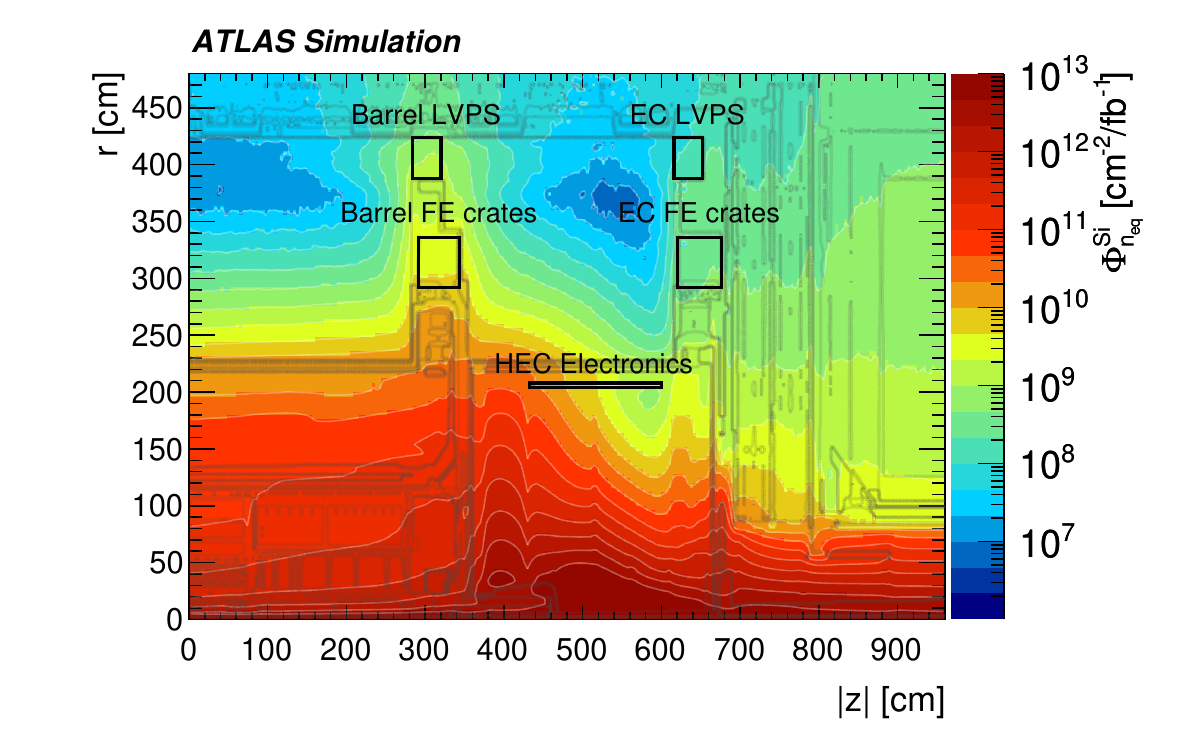}
\end{center}
\caption{Radiation levels at the locations of calorimeter electronics. The contours show \phieqv obtained from
\GEANTV{4} simulations. The radiation levels in the indicated volumes are given in Table~\protect\ref{tab:radValuesCalo}. The faint grey lines are
density contours from the \GEANTV{4} geometry model and only serve to give an indication of the detector geometry.
}
\label{fig:radLevelCalo}
\captionof{table}{\gls{TID} and characteristic particle fluences obtained from \GEANTV{4} simulations at the locations of calorimeter
electronics and power supplies, i.e the regions indicated in Figure~\protect\ref{fig:radLevelCalo}.
\label{tab:radValuesCalo}
}
\begin{center}

\begin{tabular}{c|r|r |c |c |c }
Region   &   \multicolumn{1}{c|}{|z| [cm]}   &   \multicolumn{1}{c|}{r [cm]}   &  TID [Gy/fb$^{-1}$]  &  \phieqv[cm$^{-2}$/fb$^{-1}$]  &  $\Phi_\mathrm{20}^\textrm{had}$ [cm$^{-2}$/fb$^{-1}$] \\ \hline
Barrel FE crates     &   318     &  314 & \expfor{1.42}{-1} & \expfor{3.80}{9} & \expfor{6.50}{8} \\
HEC Electronics     &   516     &  206 & \expfor{8.00}{-2} & \expfor{7.15}{9} & \expfor{4.44}{8} \\
Barrel LVPS     &   302     &  406 & \expfor{1.97}{-2} & \expfor{1.00}{9} & \expfor{1.45}{8} \\
EC FE crates     &   648     &  314 & \expfor{1.88}{-2} & \expfor{3.40}{8} & \expfor{4.14}{7} \\
EC LVPS     &   634     &  406 & \expfor{5.9}{-3} & \expfor{1.31}{8} & \expfor{1.33}{7} \\ \hline
\end{tabular}


\end{center}
\end{figure}

Figure~\ref{fig:radLevelCalo} and the associated Table~\ref{tab:radValuesCalo} show the \phieqv distribution in the
calorimeter regions and the volume-averaged main radiation quantities, respectively. The regions for which the averages
have been calculated correspond to the location of front-end and power supply electronics of the calorimeters, i.e. to
equipment susceptible to suffer from radiation damage.
In Figure~\ref{fig:radLevelCalo} the streaming of radiation through the gaps between the calorimeter parts, especially at
$|z|\approx 320$\,cm, is clearly visible.
In reality these gaps are filled with electrical cables, optical fibres and cooling pipes following complicated routing.
These can be described only in an approximate and average way in the simulation codes, which is the possible cause for the
overestimation of the radiation levels, seen in Figure\,\ref{fig:damageComparisonCalo}.
Although no direct measurements are available to support the predictions of \phiSEU, high-energy hadrons are less sensitive
to material than the other two radiation components. Therefore it can be assumed that the uncertainties are bounded by the
uncertainty on \phieqv, seen in Figure\,\ref{fig:Radmon_LAr_sidam}.
 
\subsubsection{Radiation levels in the muon system and cavern}
 
\begin{figure}
\begin{center}
\includegraphics[width=0.99\textwidth]{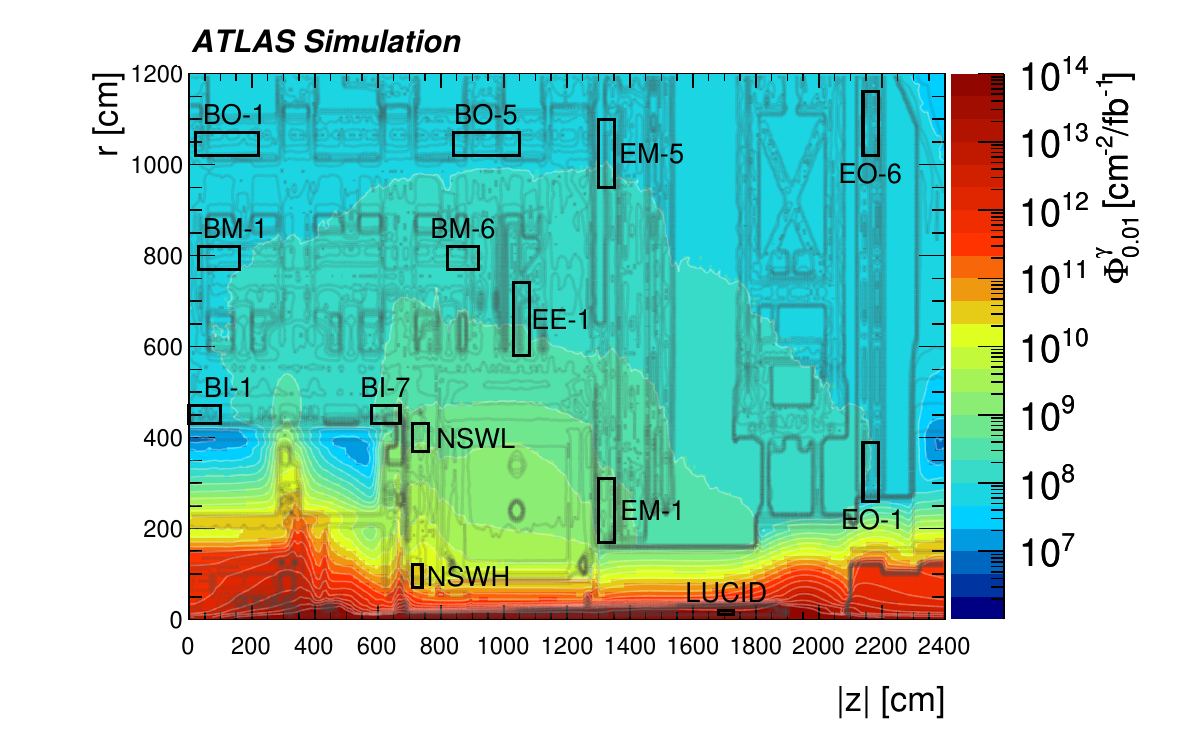}
\end{center}
\caption{Contours of constant fluence for photons with $E>\SI{10}{\keV}$ ($\Phi^\gamma_\textrm{0.01}$) in the ATLAS cavern and muon spectrometer regions.
The black boxes indicate regions over which the main radiation quantities given in Table~\ref{tab:radValuesCavern} are averaged. More detailed explanations of the
detectors in these regions can be found in Section~\ref{chapter:muons}.
The faint grey lines are density contours from the \GEANTV{4} geometry model and only serve to give an indication of the detector geometry.
\label{fig:radLevelsCavern}
}
\captionof{table}{\gls{TID} and \phieqv obtained from \GEANTV{4} simulations in the ATLAS cavern and muon spectrometer
as averages over the volumes indicated in Figure~\protect\ref{fig:radLevelsCavern}. The rightmost column gives the
fluence of photons with $E>\SI{10}{\keV}$ (\phiGamma).
$r$ and $|z|$ refer to the centre of the given region.
The statistical uncertainty is in the last digit given, or smaller.
\label{tab:radValuesCavern}
}
\begin{center}

\begin{tabular}{c|r|r |c |c |c |c }
Region   &   \multicolumn{1}{c|}{|z| [cm]}   &   \multicolumn{1}{c|}{r [cm]}   &  TID [Gy/fb$^{-1}$]  &  \phieqv[cm$^{-2}$/fb$^{-1}$]  &  $\Phi_\mathrm{20}^\textrm{had}$ [cm$^{-2}$/fb$^{-1}$]  &  $\Phi^\gamma_\mathrm{0.01}$ [cm$^{-2}$/fb$^{-1}$] \\ \hline
LUCID     &   1705     &  15 & \expfor{2.91}{3} & \expfor{2.57}{12} & \expfor{4.83}{11} & \expfor{5.78}{13} \\
NSWH     &   725     &  95 & \expfor{4.40}{-1} & \expfor{1.79}{10} & \expfor{1.86}{9} & \expfor{3.09}{10} \\
EM-1     &   1325     &  240 & \expfor{7.6}{-3} & \expfor{3.03}{8} & \expfor{6.45}{7} & \expfor{6.17}{8} \\
NSWL     &   735     &  400 & \expfor{6.9}{-3} & \expfor{2.02}{8} & \expfor{4.81}{7} & \expfor{9.03}{8} \\
BI-7     &   625     &  450 & \expfor{4.4}{-3} & \expfor{9.05}{7} & \expfor{1.13}{7} & \expfor{3.42}{8} \\
EE-1     &   1055     &  660 & \expfor{2.43}{-3} & \expfor{1.10}{8} & \expfor{2.30}{7} & \expfor{2.05}{8} \\
BM-6     &   870     &  795 & \expfor{1.63}{-3} & \expfor{4.63}{7} & \expfor{1.13}{7} & \expfor{1.56}{8} \\
EO-1     &   2165     &  325 & \expfor{1.3}{-3} & \expfor{3.19}{7} & \expfor{3.99}{6} & \expfor{9.58}{7} \\
EM-5     &   1325     &  1025 & \expfor{1.15}{-3} & \expfor{3.21}{7} & \expfor{8.34}{6} & \expfor{9.32}{7} \\
BO-5     &   945     &  1045 & \expfor{1.07}{-3} & \expfor{2.36}{7} & \expfor{6.94}{6} & \expfor{7.98}{7} \\
BM-1     &   95     &  795 & \expfor{8.1}{-4} & \expfor{1.52}{7} & \expfor{2.43}{6} & \expfor{8.87}{7} \\
BI-1     &   50     &  450 & \expfor{7.3}{-4} & \expfor{2.70}{7} & \expfor{7.0}{5} & \expfor{8.55}{7} \\
BO-1     &   120     &  1045 & \expfor{6.4}{-4} & \expfor{1.03}{7} & \expfor{2.59}{6} & \expfor{5.69}{7} \\
EO-6     &   2165     &  1090 & \expfor{5.9}{-4} & \expfor{1.36}{7} & \expfor{3.23}{6} & \expfor{5.82}{7} \\ \hline
\end{tabular}


\end{center}
\end{figure}

Since the cavern is efficiently shielded with respect to the beam-line, the available simulation samples have too much
statistical fluctuation between bins to produce a contour plot of \gls{TID} or charged particle fluences in the entire ATLAS
experimental cavern. Instead Figure~\ref{fig:radLevelsCavern} shows the distribution of the photon fluence.
This quantity is chosen for the illustration, since the photons are generating the e$^{\pm}$, which dominate the
charge particle fluence and TID in the muon spectrometer region.
 
All major radiation quantities  are collected in Table~\ref{tab:radValuesCavern} as averages over representative volumes.
The dose rates in the muon system vary from \SI{0.59}{\milli\gray}\perinversfb\ in the central barrel to \SI{44}{\milli\gray}\perinversfb\ at the innermost radius of the \gls{NSW}
region. At an instantaneous luminosity of \SI{2e34}{\instLumiUnit} these correspond to \SI{12}{\nano\gray/\s} and \SI{0.9}{\micro\gray/\s}, respectively. Assuming
all particles to be minimum ionising, these doses can be converted to rough charged particle flux estimates of
about \SI{50}{\instLumiUnit} 
and \SI{5000}{\instLumiUnit}, 
respectively. In reality the \gls{TID} has contributions from
particles with higher ionisation potential, which means that those estimates serve only as rough upper limits.
 
As shown in Figure~\ref{fig:radLevelsCavern}, the \gls{LUCID} detector is situated in the immediate proximity of the \beampipe where it is
also subject to intense albedo from the \gls{TAS} regions. A comparison of the radiation levels in that region with those of the cavern
gives a good indication of the performance of the forward shielding of ATLAS: the particle fluences are reduced by about four
orders of magnitude and the \gls{TID} by almost 6 orders of magnitude (e.g. comparing regions `LUCID' and `EM-1'). This larger shielding efficiency
for \gls{TID} is due to the fact that charged particle showers, except for muons, developing around the beam-line all get
suppressed and only neutrons with an associated photon component penetrate the shielding.
Thus the radiation field in the cavern is almost entirely due to neutrons, photons from neutron capture and e$^\pm$ from
interactions of the photons.
One notable exception in the radiation exposure of the muon system is the high-$\eta$ edge of the \gls{NSW}
(NSWH in Table~\ref{tab:radValuesCavern}); here the space available for shielding is limited by the required acceptance and
the clearance needed with respect to the \beampipe. Even though dedicated shielding materials are employed, the radiation levels at the
smallest radii of the \gls{NSW} are almost two orders of magnitude higher than anywhere else in the muon system.


\clearpage
\newpage
 
\section{Inner Detector} 
\label{sec:ID}



The \glsfirst{ID} is the primary tracking device for measuring the paths of all charged particles in ATLAS. It has been designed to provide hermetic and robust pattern recognition, excellent momentum resolution and both primary and secondary vertex measurements for charged-particle tracks within the pseudorapidity range $\abseta<2.5$. It is contained in a cylindrical envelope 
\SI{7024}{\mm}
long and \SI{1150}{\mm} in radius, immersed in the \SI{2}{\tesla} field of a solenoidal magnet.
 
\begin{figure}[h!]
\centering
\includegraphics[width=0.99\textwidth]{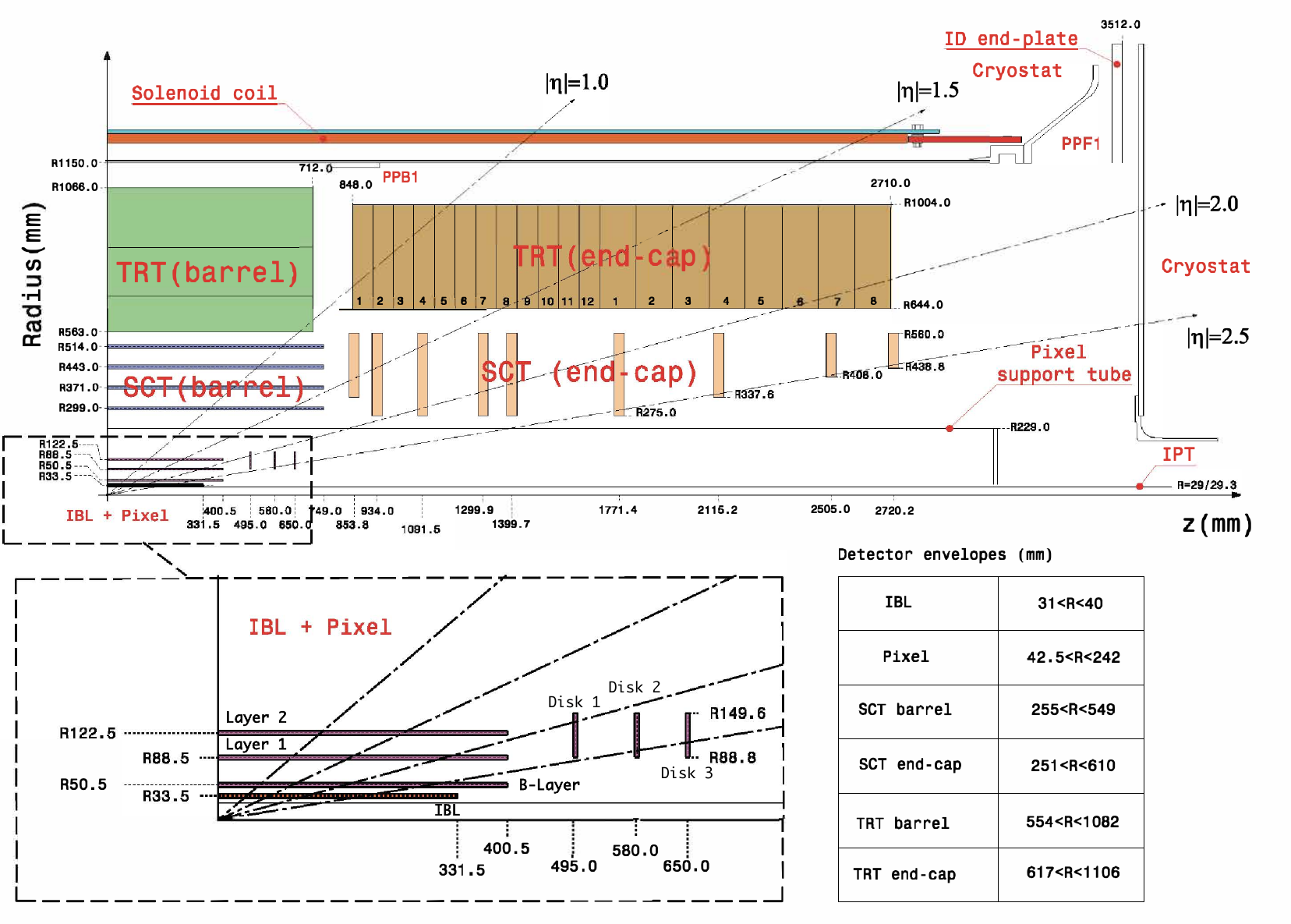}
\caption{The layout of the ATLAS \gls{ID}, including the \gls{IBL} detector~\cite{PIX-2018-001}.}
\label{fig:ID_Run2Layout}
\end{figure}

As described in Section~\ref{subsec:OverviewTracking}, the \gls{ID} consists of three complementary sub-detectors arranged coaxially around the beam line (see Figure~\ref{fig:ID_Run2Layout}): a high-resolution silicon Pixel detector~\cite{Pixel} ($r< \SI{122.5}{\mm}$), the \glsfirst{SCT}~\cite{IDET-2013-01} relying on stereo micro-strips ($299<  r< \SI{514}{\mm}$) and the \glsfirst{TRT}~\cite{TRT} comprising several layers of gaseous straw tubes interleaved with transition radiation material ($563<  r< \SI{1066}{\mm}$).
 
\begin{figure}[h!]
\centering
\includegraphics[width=0.70\textwidth]{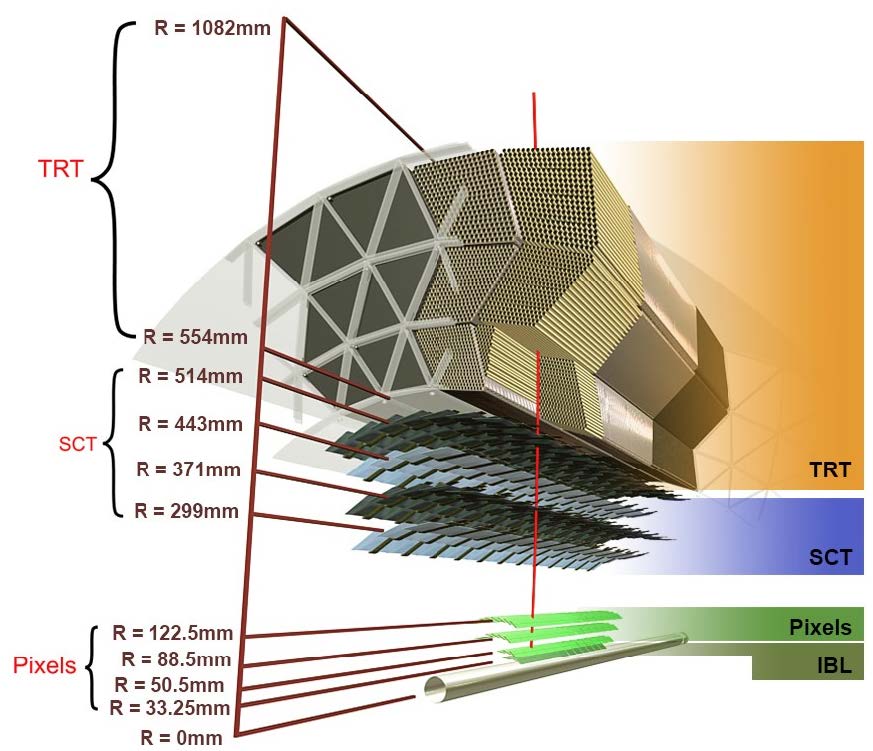}
\caption{Transverse view of the ATLAS \gls{ID} including the \gls{IBL} detector.}
\label{fig:IDRun2}
\end{figure}

For many physics channels, particularly those involving relatively long-lived particles such as $B$-hadrons that decay and produce a secondary vertex inside the \beampipe, the performance of the ATLAS experiment depends critically on the innermost layer of the Pixel detector. For this reason,
during \gls{LS1}, the detector underwent a major upgrade. During this period a fourth pixel layer, the \glsfirst{IBL}~\cite{ATLAS-TDR-19,ATLAS-TDR-19-addm,PIX-2018-001}, 
was added to the Pixel detector between a new, narrower beryllium \beampipe and the previously existing innermost pixel layer (Pixel $B$-Layer). Figure~\ref{fig:IDRun2} shows the transversal layout of the upgraded ATLAS \gls{ID} during \RunTwo and \RunThr. A description of the main concepts and characteristics of the \gls{IBL} is given in Section~\ref{sec:ID-IBL}.
 
At the same time, part of the Pixel services was replaced by the \glsfirstplural{nSQP}, allowing for the optical-to-electrical converters to be moved into an area accessible for service by extending the 
electrical readout cables
.
A brief overview of the \glspl{nSQP}  installation together with a description of the Pixel off-detector readout upgrades in \RunTwo is given in Section~\ref{sec:Pixel}.
 
The readout systems of the \gls{SCT} and \gls{TRT} were also upgraded during \RunTwo, to cope with the increasing requirements coming from the \gls{LHC} performance. A brief description of these upgrades is presented in Sections~\ref{sec:ID-SCT} and~\ref{sec:ID-TRT}.


 
\subsection{Insertable $B$-layer}
\label{sec:ID-IBL}
 
The driving motivation for the \glsfirst{IBL} detector was the consolidation and enhancement of the \gls{ID} tracking performance in high luminosity scenarios. The reduced distance of the \gls{IBL} from the beam axis (\SI{3.3}{\cm}, compared to \SI{5.0}{\cm} for the Pixel $B$-Layer) increases the resolution of track impact parameters and thus enhances the vertex reconstruction and flavour-tagging performance of the tracking system.  An additional layer of highly segmented pixel sensors helps to mitigate inefficiencies of the Pixel detector caused by overall radiation damage and irreparable module failures, and reduces the occurrence of fake tracks due to the combinatorics of high pileup backgrounds.
 
The reduced distance from the beam axis, combined with the increased luminosity of the machine, set stricter radiation hardness requirements.
Simulations~\cite{ATLAS-TDR-19} predicted that \gls{IBL} sensors  and electronics would be subject to an overall fluence \phieqv of \SI{3.3e15}{\per\cm\squared}
for an integrated luminosity of \SI{550}{\ifb}, which was the original design requirement for \RunThr.
A planar technology adapted according to the \RunOne Pixel experience was used for the sensors in the central part while a new 3D technology was developed for the forward region. Section~\ref{sec:ID-IBL-Sensors} describes these two technologies.
For the electronics, a new front-end chip was developed on the basis of the  \RunOne Pixel experience to cope with the more stringent
requirements foreseen in future data taking scenarios.
 
\begin{figure}
\centering
\includegraphics[width=0.70\textwidth]{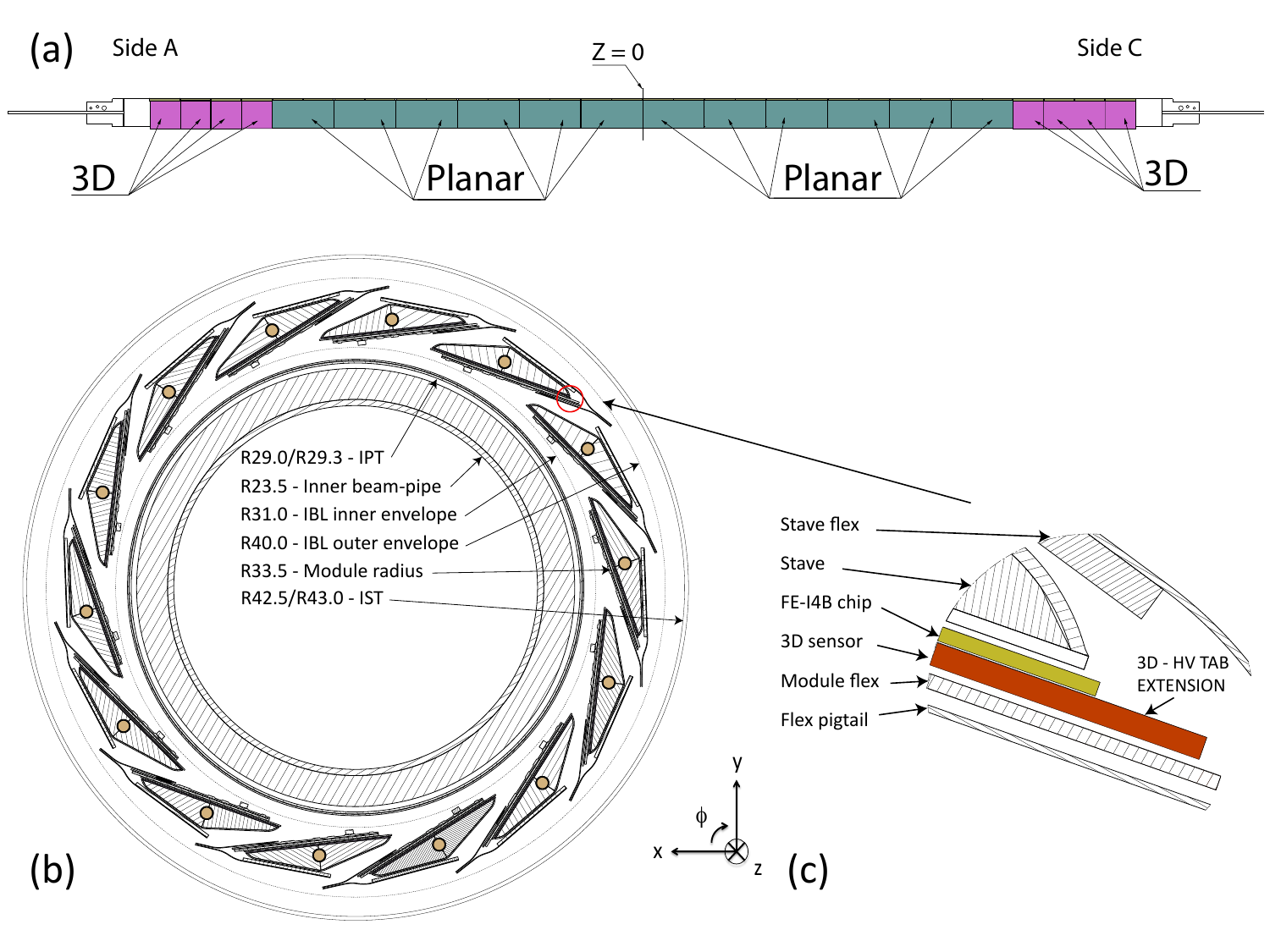}
\caption{\gls{IBL} layout~\cite{PIX-2018-001}: (a) Longitudinal layout of planar and 3D modules on a stave. (b) An $r-\phi$ section showing the \beampipe, the \glsfirst{IPT}, the staves of the \gls{IBL} detector and the \glsfirst{IST}, as viewed from the C-side. (c)  An expanded $r-\phi$ view of the corner of a 3D module fixed to the stave.}
\label{fig:ID-IBL_Layouts}
\end{figure}
 
The \gls{IBL} layout is shown in Figure~\ref{fig:ID-IBL_Layouts}. The \gls{IBL} consists of a single layer of pixel sensors assembled with their readout chips (modules) and arranged on \num{14} longitudinal supports (staves) fastened onto a high precision carbon fibre tubular structure, the \glsfirst{IPT} (outer radius of \SI{29.3}{\mm}), via the stave support ring. The staves surround a \SI{23.5}{\mm}-radius beryllium \beampipe. The outer envelope is defined by a second carbon fibre cylinder, the \glsfirst{IST} (inner radius of \SI{42.5}{\mm}), fastened to the Pixel detector support structure.
 
The staves are tilted by \ang{14} in a classic turbine design in order to ensure complete azimuthal coverage for high $\pT$ tracks and to partially compensate for the Lorentz angle affecting the trajectory of the charges drifting in the planar sensors. With this arrangement, the average distance of the innermost pixel sensors from the beam-axis was reduced from \SI{5.05}{\cm} of the Pixel $B$-Layer down to \SI{3.35}{\cm}. Each stave hosts \num{20} modules aligned along the beam axis (z-direction): \num{12} two-chip modules covering the central part and eight single-chip modules dedicated to the high-\abseta regions extending to $\abseta=3.0$. The full detector coverage is effectively $\abseta=2.58$ if the \SI{122}{\mm} two-standard-deviation spread of the primary vertex distribution is taken into account. The main layout parameters of the IBL are presented in Table\,\ref{tab:ID-IBL_TableLayout}.
 
\begin{table}
\caption{Main layout parameters of the \gls{IBL}. The nominal $\eta$ coverage of the detector (assuming no spread on the longitudinal position of the \gls{IP})  is $\abseta < 3.0$ but a spread of the vertex position on the $z$-coordinate reduces the detector coverage down to $\abseta < 2.58$.
\label{tab:ID-IBL_TableLayout}
}
\centering
\begin{tabular}{lc}
\hline\hline
Item & Value \\
\hline
Number of staves & 14 \\
Number of modules per stave & 12 planar $+$ 8 3D \\
Number of front-end chips per stave & 32 \\
Coverage in $\eta$ -- no vertex spread & $\abseta<3.0$ \\
Coverage in $\eta$ -- $2\sigma$ (\SI{122}{\mm}) vertex spread
& $\abseta<2.58$ \\
Active $|z|$ stave length  & \SI{330.15}{\mm} \\
Stave tilt in $\phi$ & \ang{14} \\
Overlap in $\phi$ & \ang{1.82} \\
Center of the sensor radius  & \SI{33.5}{\mm} \\
\hline\hline
\end{tabular}
\end{table}
 
Limited \gls{IBL} radial clearance prevents sensor shingling along a stave; for this reason, in order to maximise the coverage, thin-edge sensors were developed for the detector. An effective inactive edge width of \SI{215}{\micron} (\SI{175}{\micron}) was measured for planar (3D) sensors, considerably reduced with respect to the \SI{1100}{\micron}  of the Pixel detector in \RunOne. An air gap of \SI{205}{\micron} is maintained between contiguous modules to provide adequate electrical insulation.
 
Cooling is achieved by means of a ${\text{CO}}_2$ two-phase system where the coolant is circulated within titanium pipes embedded in the stave structure; in order to optimise the thermal contact between the active components and the pipes, staves are filled with carbon foam, which contributes to the global stiffness of the mechanical supports. A detailed description of the cooling system is given in Section\,\ref{sec:IBL-Cooling}.
 
\subsubsection{New \beampipe}
 
The ATLAS \beampipe had to be replaced to allow sufficient radial clearance for the insertion of the \gls{IBL} detector and its mechanical structure. The new \beampipe, characterised by an inner radius of \SI{23.5}{\mm} (reduced from the previous \SI{29}{\mm}), consists of a \SI{7100}{\mm}-long beryllium section with an average wall thickness of approximately \SI{870}{\micron}; its extremities are welded to aluminium flanges, the size of which allow its insertion through the \gls{IPT}.
 
The inner surface of the \beampipe is treated with a \gls{NEG} coating to improve the vacuum quality by bonding to gas molecules remaining within the partial vacuum;
this getter coating had to be activated with a bake-out procedure using heaters wrapped around the \beampipe surface. To mitigate the effect of extreme heat on the silicon sensors, a layer of aluminium was interleaved between the \beampipe and the \gls{IBL} to reduce the infra-red emissivity of the \beampipe. The heater temperature was carefully monitored during the process.
 
\subsubsection{Sensors}
\label{sec:ID-IBL-Sensors}
 
Two sensor technologies are used in the \gls{IBL}: an improved version of the ATLAS \RunOne Pixel planar sensors~\cite{Pixel}, able to
cope with unprecedented operational conditions, and an innovative design of 3D devices~\cite{3D} a first-ever application in high energy physics.
 
The two sensor implementations share a common footprint in order to be interfaced to the same readout chip, with two-chip planar sensor tiles having an equal longitudinal dimension of two single-chip 3D sensors in order to fit a common mechanical layout. Both technologies are characterised by the same granularity, with $\SI{250}{\micron}\times\SI{50}{\micron}$
pixels organised in $80 \times 336$ arrays (to be compared with the $18 \times 160$ matrix configuration of $\SI{400}{\micron}\times\SI{50}{\micron}$ pixels of the three outermost Pixel layers). Pixels of sizes different from the nominal are also present in order to fill the gaps between front-ends. A comparison of the main characteristics of the Pixel and \gls{IBL} detectors is shown in Table~\ref{tab:ID-IBL_TablePixelComparison}.
 
\begin{table}
\caption{Comparison of the main characteristics of the IBL pixels with the original Pixel detector layers.
\label{tab:ID-IBL_TablePixelComparison}
}
\centering
\begin{tabular}{lcc}
\hline\hline
Technical Characteristic & Pixel & IBL \\
\hline
Active surface (\si{\m\squared}) & 1.73 & 0.15 \\
Number of channels ( $\times \num{e6}$) & 80.26 & 12.04 \\
\hline
Pixel size (\si{\micron\squared}) - nominal/long
& $50 \times 400/600$ & $50 \times 250/500$ \\
Pixel array (columns$\times$rows)
& $160 \times 18$ & $336 \times 80$ \\
Front-end chip size (\si{\mm\squared})
& $7.6 \times 10.8$ & $20.2 \times 19.2$ \\
Active surface fraction ($\%$) & 74 & 89 \\
\Analog current (\si{\SIUnitSymbolMicro{A}}) & 26 & 10 \\
Digital current (\si{\SIUnitSymbolMicro{A}}) & 17 & 10 \\
\Analog voltage (\si{\V}) & 1.6 & 1.4 \\
Digital voltage (\si{\V}) & 2.0 & 1.2 \\
Data out transmission (Mbps)
& $40-160$ & 160 \\
\hline
Sensor type & planar & planar/3D \\
Sensor thickness (\si{\micron}) & 250 & 200/230 \\
Layer thickness ($X_0$) & 2.8 & 1.9 \\
Cooling fluid & ${\text{C}}_3{\text{F}}_8$ & ${\text{CO}}_2$ \\
\hline\hline
\end{tabular}
\end{table}
 
\gls{IBL} planar sensors are fabricated with a n-in-n design on a \SI{200}{\micron}-thick substrate, thinner than the \SI{250}{\micron} devices used in the rest of the Pixel detector.
Slim inactive edges (\SI{200}{\micron} wide) are achieved by shifting the guard rings on the p-side underneath the pixel implantations.
Planar sensors are produced in two-chip tiles of an overall dimension of
$\SI{18.59}{\mm}\times\SI{41.32}{\mm}$;
the operational voltage is expected to evolve during the device lifetime from \SI{80}{\V} up to \SI{1000}{\V}.
 
\begin{figure}
\centering
\includegraphics[width=0.68\textwidth]{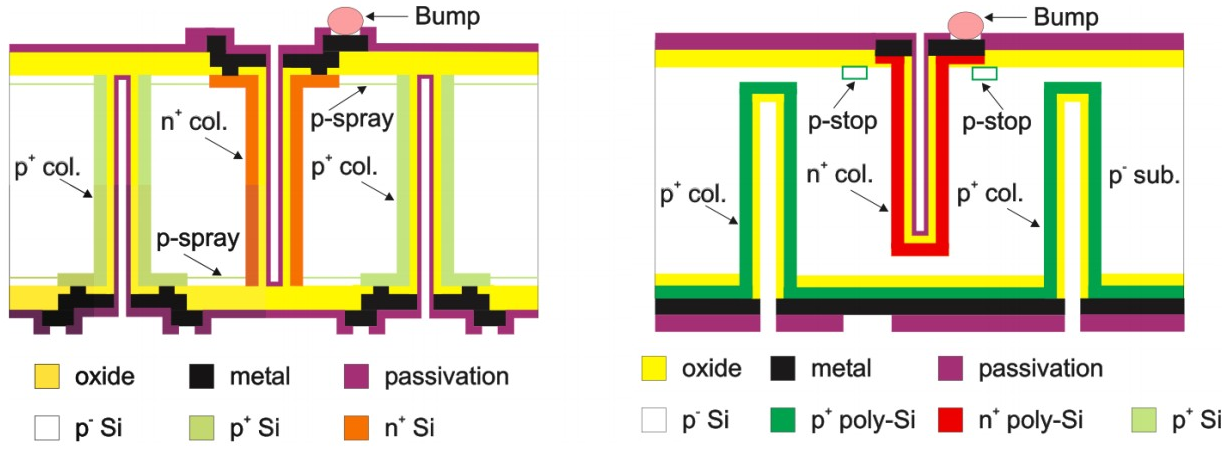}
\caption{Design of the columns of (a) FBK and (b) CNM 3D sensors, including the location of the bump used to bond the sensor to the front-end chip~\cite{PIX-2018-001}. This sketch is for illustration only and is not to scale. }
\label{fig:ID-IBL_3D}
\end{figure}

\gls{IBL} 3D sensors rely on a \SI{230}{\micron}-thick p-type substrate and are fabricated with a process in which \SI{10}{\micron}-diameter columnar electrodes are implanted by double-sided \gls{DRIE} -- i.e. n-type and p-type columns penetrate the substrate from opposite sides.
Columnar electrodes are either passing-through or \SI{210}{\micron} deep according to the different designs implemented by the vendors (FBK\footnote{Fondazione Bruno Kessler, Trento, Italy} and CNM\footnote{Centro Nacional de Microelectronica, Barcelona, Spain}) at the time of production; see Figure \ref{fig:ID-IBL_3D} for a visual comparison of the two designs.
The pixel layout consists of two n-type readout electrodes connected at the wafer surface and surrounded by six p-type ohmic electrodes  which are shared with the neighbouring cells. Edge isolation is achieved by a fence of ohmic electrodes corresponding to an inactive area of \SI{200}{\micron} deep.
Due to the characteristic design, which decouples the sensor thickness from the drift distance, 3D sensors are expected to be particularly radiation tolerant and for this reason suitable to equip detectors exposed to high irradiation levels.
The operational bias of these sensors started with \SI{20}{\V} at the beginning of \RunTwo and is expected not to exceed \SI{200}{\V} during the device lifetime.
 
\subsubsection{Modules and staves}
\label{sec:ID-IBL-ModulesStaves}
 
The connection between sensor and front-end chip is achieved by means of fine-pitch bump-bonding and flip-chip technology.
Due to more stringent radiation and occupancy requirements, the front-end electronics of the \gls{IBL} has been completely redesigned with respect to the integrated circuit used in the rest of the Pixel detector in \RunOne (the FE-I3).
The new front-end chip, the FE-I4B~\cite{FE-I4}, was manufactured in \SI{130}{\nm} \gls{CMOS} technology, featuring a large footprint
($\SI{20.3}{\mm}\times\SI{19.2}{\mm}$)
and \num{26880} readout channels replicating the pixel matrix. Each cell comprises an independent amplifier with adjustable shaping and a discriminator with an individually adjustable threshold. The \analog threshold (typically 1500-2500 electrons) sets the minimum collected charge to be processed (and converted to \gls{ToT}) by the front-end chip; the digital threshold (in units of \gls{ToT})
sets the minimum digital collected charge amplitude to be transmitted by the module and is typically $2$\gls{BC}. The collected charge amplitude is measured as \gls{ToT} in units of the \SI{25}{\ns} \gls{LHC} bunch crossing period with a four-bit resolution.
The FE-I4B chips (see Section~\ref{sec:IBL-Readout} for details about the readout architecture) are slimmed to \SI{150}{\micron} in order to reduce the material budget.
 
\begin{figure}
\centering
\includegraphics[width=0.70\textwidth]{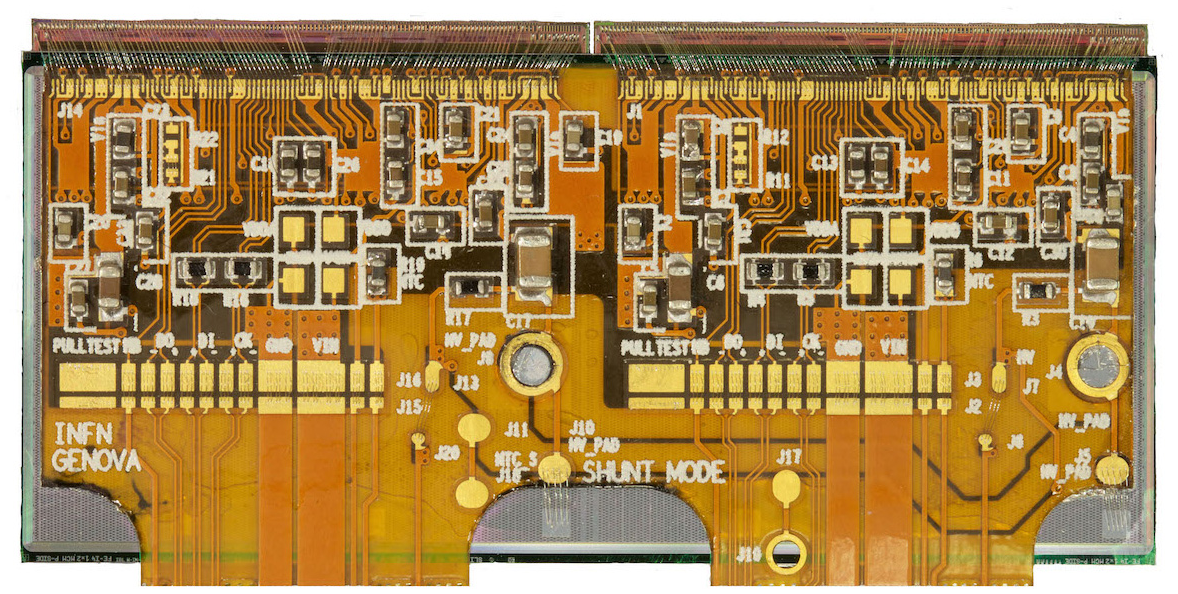}
\caption{Photograph of an IBL planar module~\cite{PIX-2018-001}. The full size of the silicon sensor is $\SI{18.8}{\mm}\times\SI{41.3}{\mm}$.}
\label{fig:ID-IBL_PlanarModule}
\end{figure}

A double-sided, flexible printed circuit (known as the module flex hybrid) is glued to the back side of the sensor (see Figure~\ref{fig:ID-IBL_PlanarModule}) and connects the module to external electrical services.
Read-out and power lines connections between the module flex hybrid, the FE-I4B chips and the sensors are wire-bonded.
The module temperature monitoring and interlock is achieved with a \gls{NTC} thermistor mounted on the module flex hybrid.
 
\gls{IBL} modules are fastened by two glue dots to their supporting staves. Staves are carbon foam structures glued to V-shaped carbon fiber laminates; the carbon foam provides efficient thermal coupling to the titanium cooling pipes that are embedded within the staves.
Cooling is obtained by a ${\text{CO}}_2$ evaporative system, capable of maintaining the silicon sensors at an operating temperature of \SI{-15}{\degreeCelsius}; the minimum value achievable for future operation is \SI{-35}{\degreeCelsius}. Power, data acquisition and module configuration are routed through a multi-layer bus (the stave flex hybrid) laminated on the back of the staves, from which they are distributed to modules by means of the module flex hybrids.
 
During the first year of the \gls{IBL} operation in 2015, a significant increase of the low voltage current of the front-end chip and a detuning of its parameters (threshold and \gls{ToT}) were observed in relation to the received \gls{TID}. The increase of the low voltage current  of the FE-I4B chip and the drifting of its tuning parameters were traced back to the generation of a leakage current in \gls{NMOS} transistors induced by radiation. The radiation induces positive charges that are quickly trapped into the shallow-trench-isolation oxide at the edge of the transistor. Their accumulation builds up an electric field sufficient to open a source-drain channel where the leakage current flows. If the accumulation of positive charges is relatively fast, the formation of interface states is a slower process. The negative charges trapped into interface states start to compete with the oxide-trapped charges with a delay. This is what gives origin to the \gls{TID} effect at low dose~\cite{CYP}.
 
In order to study the dependence of the low voltage current increase on temperature and dose rate, several irradiation tests were performed by setting one of those variables and changing the other.
Given the results of these tests, it was decided to increase the \gls{IBL} operational set point from \SI{-10}{\degreeCelsius} to \SI[retain-explicit-plus]{+15}{\degreeCelsius} for a few months in early 2016 to limit the TID effects during the first radiation exposure, before reducing it to \SI{-20}{\degreeCelsius} for the remainder of \RunTwo.
In addition, the digital supply voltage (V$_{\mathrm{D}}$) was lowered from \SI{1.2}{\V} to \SI{1}{\V} to decrease the low voltage current.
 
\subsubsection{Integration and installation of the \glstext*{IBL} in the \glstext*{ID}}
\label{sec:IBL-Integration}
 
A total of \num{20} staves were assembled during the production phase. An issue concerning wire-bond corrosion~\cite{PIX-2018-001} was identified on most staves mid-way through the production.
This resulted from a combination of extreme susceptibility to corrosion due to chemical contamination on the flex and accidental exposure to humidity during the temperature cycling tests after stave loading; all but two staves were fully repaired.
Because the wire bonds of the \gls{IBL} were left unprotected (potting the bond foot or the use of spray coatings such as polyurethane were initially considered as options to protect them), it was crucial that the temperature be maintained well above the dew point at all stages of the integration and installation, and it remains crucial for operation.
 
Of the \num{18} staves available for the integration, \num{14} were selected on the basis of quality criteria and installed in the detector.
These criteria rely on a weighted pixel failure fraction aimed at minimising the geometrical inefficiencies and clusterisation of defects.
The average bad pixel fraction for the integrated \gls{IBL} staves at the time of installation amounted to \SI{0.09}{\percent} (\SI{0.07}{\percent} for $\abseta<2.5$).
 
The \gls{IBL} was integrated as a single package that was lowered into the ATLAS experimental cavern and finally inserted inside the Pixel detector.
Once all staves were loaded, an annular support was installed and clipped to each stave center in order to provide additional radial stiffness.
This constraint did not eliminate other degrees of freedom, most notably a rotation around the beam axis arising from the mismatch of thermal expansion coefficients of the stave and stave flex and their asymmetric assembly; this resulted in a distortion of the staves of the order of a few \SI{}{\micron} / \SI{}{\degreeCelsius} that required an enhanced temperature stability at the level of \SI{0.2}{\degreeCelsius} and the development of in-run alignment correction procedures~\cite{IDTR-2019-05}.
 
\subsubsection{Readout of the \glstext*{IBL}}
\label{sec:IBL-Readout}
 
The \gls{IBL} readout system was designed to be fully efficient at a few times the nominal \gls{LHC} peak luminosity
(up to \SIrange[range-phrase=--]{2}{3e34}{\per\cm\squared\per\second})  with a \gls{L1} trigger rate of \SI{100}{kHz}.
The readout can be naturally divided into two parts, the on-detector and the off-detector systems, which communicate via optical fibres in both directions.
 
The on-detector system consists of the FE-I4B front-end readout chips and the service electronics, which include the electro-optical converter boards (optoboards).  The optoboards are connected to the front-end chips by the FE-I4B input and output lines.
 
The building block of the readout system, the \gls{IBL} \gls{DAQ} module, is formed by two neighbouring front-end chips.
Each \gls{IBL} \gls{DAQ} module shares the same clock and command lines (including the \gls{L1} trigger signals) that are distributed to the pair of front-end chips.
 
The FE-I4B input lines run at \SI{40}{Mbps}. When charge deposition is detected by the discriminator in each pixel, the hit and its timestamp are briefly buffered in the pixel cells. The FE-I4B pixel array is organised in double-columns like the FE-I3 chip, but the readout architecture is very different. Instead of moving all hits from the pixel array to a global shared memory structure for later trigger processing, the FE-I4B double-columns are further divided in $2\times 2$ pixel regions where hits are stored locally. This results in an enhanced capability to cope with high hit rates and reduced inefficiency. Each region contains four identical \analog pixels and one shared memory and logic block called the Pixel Digital Region. This memory can store up to five events.
 
When the \gls{L1A} arrives, any event for which the timestamp matches is read out via a serial \gls{LVDS} output
that operates at \SI{160}{Mbps}. There is one \gls{LVDS} output line per front-end.
 
The optoboards are responsible for translating the optical signals received from the off-detector electronics into electrical signals for the FE-I4B inputs; vice versa they translate the electrical signals received from the FE-I4B outputs to optical signals before transmitting them to the off-detector electronics.
\gls{PIN} diodes are used to convert optical into electrical signals. \gls{VCSEL} arrays convert electrical into optical signals.

\begin{figure}
\centering
\includegraphics[width=0.70\textwidth]{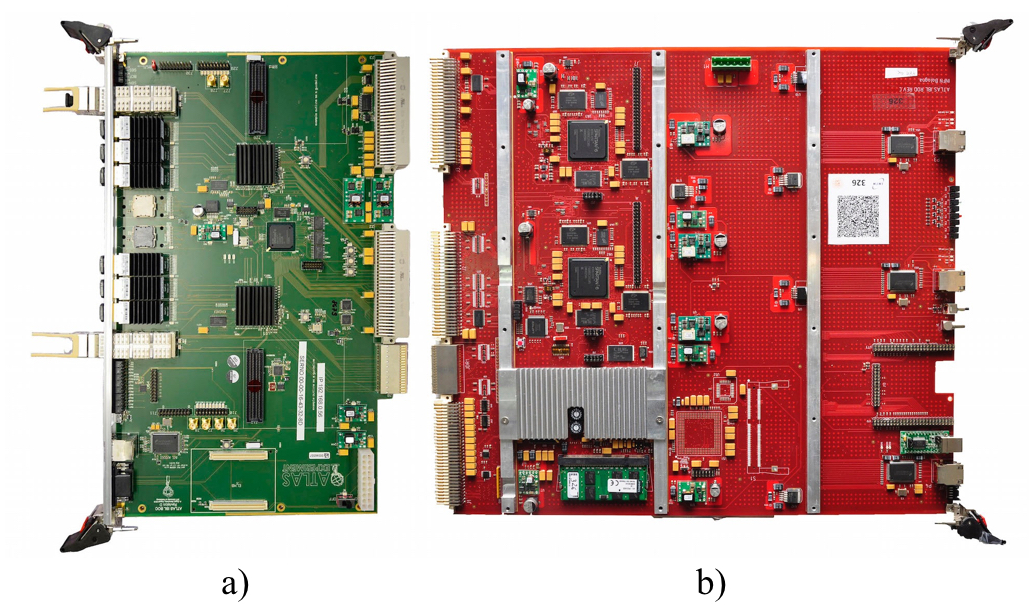}
\caption{The \gls{IBL} \gls{ROD}/\gls{BOC} readout cards~\cite{PIX-2018-001} (used also for the Pixel readout upgrades during three consecutive winter shutdowns) are located at the off-detector side of the optical link. (a) The \gls{BOC} card and (b) the \gls{ROD} card are paired in a \gls{VME} crate via its back-plane. }
\label{fig:IBL_Readout}
\end{figure}

\begin{figure}
\centering
\includegraphics[width=0.90\textwidth]{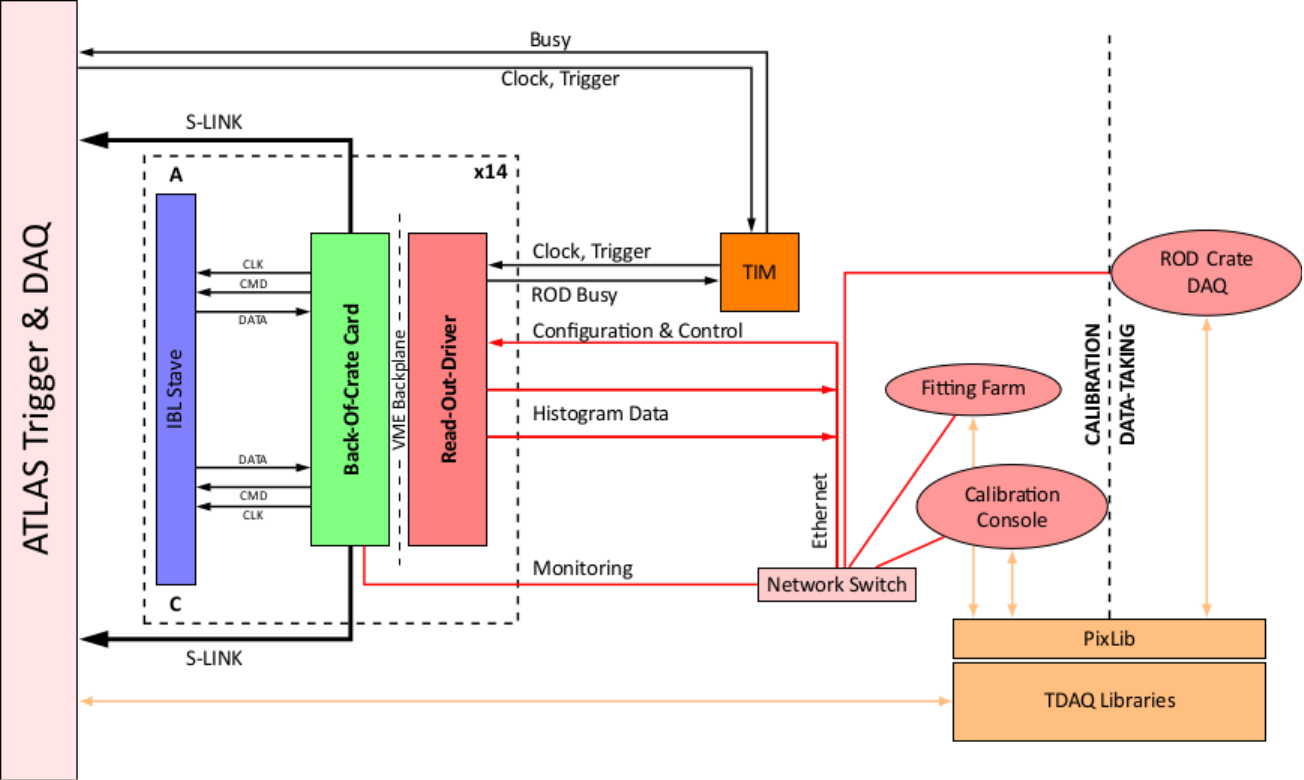} 
\caption{Schematic of the \gls{IBL} read-out system~\cite{PIX-2018-001}. The \gls{IBL} Stave is the only on-detector component in this diagram.}
\label{fig:IBL_DAQ}
\end{figure}

The off-detector electronics consists of commercial optical \gls{Rx} plugins and \gls{Tx} plugins, used to interface the readout hardware with the optical fibres. The plugins are mounted in the \gls{BOC} cards, used to transmit  the clock, triggers and commands to the modules after \gls{BPM} encoding, and to receive and decode 8b/10b-encoded data from the front-ends. The \gls{BOC} communicates through the \gls{VME} backplane to the \gls{ROD} cards. The \gls{ROD} has several tasks: it distributes the \gls{L1} trigger upstream, along with signals received from the ATLAS \gls{TIM}, reformats the data received from the front-ends (via the \gls{BOC}), and finally transmits the generated event fragments to the ATLAS \gls{ROS}~\cite{ROS} via CERN \glspl{s-link}~\cite{slink}.
This transmission is done by forwarding the reformatted data back through the backplane to dedicated optical transceivers (\gls{QSFP}~\cite{QSFP}) on the \gls{BOC}.
Each \gls{s-link} has a bandwidth of \SI{1.28}{Gbps}. As shown in Figure~\ref{fig:IBL_Readout}, a single \gls{IBL} \gls{ROD}/\gls{BOC} pair hosts four \glspl{s-link} 
for a total output bandwidth towards the ATLAS \gls{ROS} units of \SI{5.12}{Gbps}.
The \gls{ROD} is also used to calibrate and tune the \gls{IBL} detector.
A schematic of the entire readout chain is shown in Figure~\ref{fig:IBL_DAQ}.
 
Each \gls{IBL} \gls{ROD}/\gls{BOC} pair reads out an entire \gls{IBL} stave via 32 FE-I4B chips. Two \gls{Tx} plugins are used to communicate with \num{16} double-chip modules via 16 independent serial lines, eight lines for each \gls{Tx} plugin. Four \gls{Rx} plugins read back the data from each of the \num{32} front-ends individually; each \gls{Rx} plugin receives data from eight front-ends. A total of \num{14} \gls{IBL} \gls{ROD}/\gls{BOC} pairs have been equipped in a 9U \gls{VME} crate located in the ATLAS service cavern (\gls{USA15}).
 
The same type of readout board was used to upgrade the outer Pixel layers during \RunTwo.  A detailed description of these upgrades will be given in Section~\ref{sec:Pixel-Readout}.

\subsubsection{Off-detector services and detector power supplies}
 
\begin{figure}
\centering
\includegraphics[width=1.\textwidth]{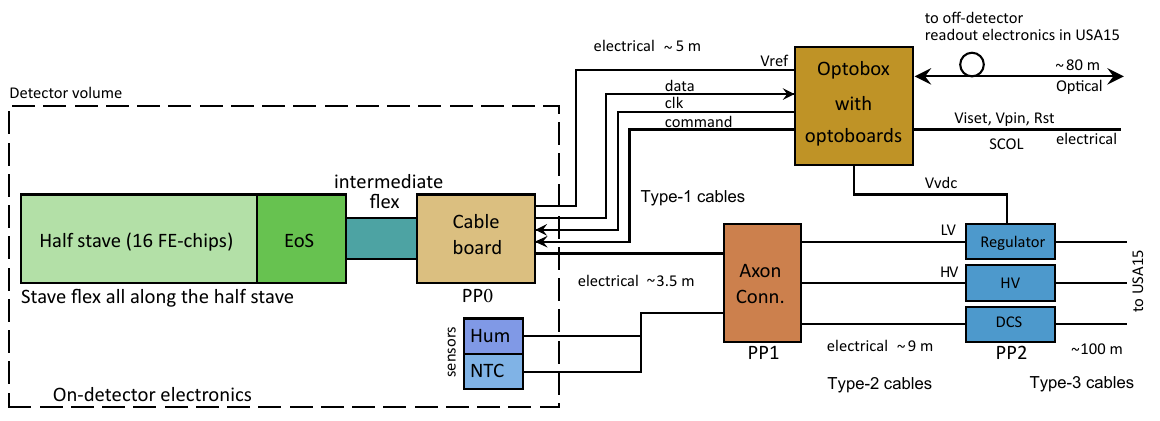}
\caption{Block diagram of on-detector and off-detector electrical services for one half-stave of the IBL detector~\cite{PIX-2018-001}.
}
\label{fig:IBL_Services}
\end{figure}

The basic element for the segmentation of the \gls{IBL} electrical services is the half stave.
The end of each half-stave is connected to a set of corrugated flex \glspl{PCB} which on the other side connect to a \gls{PCB} called the cable board as is schematically shown in Figure~\ref{fig:IBL_Services}. The cable board is part of a cable harness, called a type-1 cable, which makes the connection out to the \gls{ID} endplate region. Each harness is divided into two individual sub-cables, one for data transmission and one for power and the \gls{DCS}.

The type-1 cable transmission bundle consists of 24 polyimide-clad copper twisted pairs.
For the data from the front-ends \SI{28}{\gls{AWG}} wire is used due to the higher rate of \SI{160}{Mbps}, while for
the \SI{40}{\MHz} clock and the \SI{40}{Mbps} command transmission \SI{36}{\gls{AWG}} wire is used. Each command and clock line
is shared by two frontend chips. The standard for all signals is \gls{LVDS}. Four \SI{36}{\gls{AWG}} ground wires in each
bundle ensure a common ground level between transmitters and receivers. The transmission
cable is approximately \SI{5}{\m} long and terminates at the optobox in the outer part of the \gls{ID} endplate region.
 
The second cable in the type-1 cable harness contains 61 polyimide-clad copper wires: 32 low voltage \SI{24}{\gls{AWG}} wires,
eight wires (24/28\gls{AWG}) for low-voltage sensing, 16 \SI{32}{\gls{AWG}} \gls{DCS} wires, four \SI{32}{\gls{AWG}} wires for high voltage and
one \SI{26}{\gls{AWG}} drain wire. The cable is \SI{3.5}{\m} long and terminates in a custom 67-pin AXON33 connector of \SI{21}{\mm}
diameter.
 
In the \gls{ID} endplate area the type-1 power sub-cables connect to \SI{9}{\m}-long type-2 cables which end in the \gls{PP2}
situated inside the \gls{MS} volume. For the entire \gls{IBL} there are four \gls{PP2}
crates, two per detector side.
The \gls{PP2} contains regulators for the low voltage and is passive for \gls{HV} and \gls{DCS}. Type-3 cables, \SI{75}{\m} long, connect the \gls{PP2} with the \gls{HV} and \gls{LV} power supplies and the \gls{DCS} electronics inside the counting room.
 
Commercial power supplies provide high and low voltage for the detector. The maximum high voltage for planar sensors is \SI{1000}{\volt} and for 3D sensors \SI{500}{\volt}, at a maximum current of \SI{8}{\milli\ampere} (planar) and \SI{10}{\milli\ampere} (3D). The low voltage from the power supplies is regulated down to around \SI{2}{\volt} by the \gls{PP2} regulator boards (see Table\,\ref{tab:ID-IBL-PS}). Four front-ends chips share one \gls{HV} channel, as well as one \gls{LV} channel on a regulator board. Each \gls{PP2} regulator board, supplying one half-stave, is fed by a single primary voltage supply channel.
 
The optoboard \gls{LV} supply chain replicates the frontend \gls{LV} chain design from power supply to the detector-side end of
the type-2 cable at which point a direct connection to the optobox is made.
In addition to \gls{LV}, the optoboards require a bias voltage for the \gls{PIN} diodes of \SI{5}{\V} and a control voltage for the VCSELs of about \SI{0.8}{\V} (see Table~\ref{tab:ID-IBL-PS}).
All voltages, as well as a reset signal, are provided by custom \gls{SCOL} modules in the counting room.
Each optoboard which reads out half a stave is powered separately.
 
The \gls{DCS} lines coming from each half stave are used to monitor the temperature on every fourth frontend chip, on the cable board,
on the cooling pipe, and for a fraction of the staves on the type-1 bundle while for the rest of the staves this temperature sensor is replaced by a humidity sensor on the cable board.
The temperature monitoring uses \gls{NTC} sensors. The humidity sensors were not radiation hard; for this reason they were only used during the installation phase.
In the counting room the \gls{DCS} lines are connected to a custom crate that processes the signals.
 
Table~\ref{tab:ID-IBL-PS} summarises the power requirements for Pixel and \gls{IBL}. The power consumption of the \gls{IBL} was about \SI{1.6}{\kW} for the low voltage at the end of \RunTwo,  while for the high voltage it was about \SI{15}{\W}. The power consumption is expected to increase further towards the end of life due to a rise in leakage current (\gls{HV}) and \gls{LV} current caused by radiation damage.
 
\begin{table}
\caption{Summary of the silicon sensor bias voltage and front-end electronics requirements and granularity. The different front-end electronics supply voltages are
described in the text. The power consumption expected during initial operation is also shown. Following irradiation, the power consumption increases significantly in the sensors, front-end electronics and cables.
\label{tab:ID-IBL-PS}
}
\centering
\begin{adjustbox}{max width=0.9\textwidth}
\begin{tabular}{l|l|c|c}
\hline\hline
& & Pixel & IBL Planar (3D) \\
\hline\hline
\multirow{5}{*}{\textbf{Bias voltage supplies}} & Voltage (maximum)      & \SI{600}{\V} & \SI{1000}{\V} (\SI{500}{\V}) \\
& Voltage (start of \RunTwo) &   \SI{250}{\V} & \SI{80}{\V} (\SI{20}{\V})              \\
& Voltage (end of \RunTwo)  &   \SI{400}{\V} &  \SI{400}{\V} (\SI{40}{\V})            \\
& Current (maximum)       &  \SI{4}{\milli\ampere} &   \SI{8}{\milli\ampere}  (\SI{10}{\milli\ampere})   \\
& Segmentation     &  one per module & one per 4 FE-I4B\\    \hline
\multirow{6}{0.3\textwidth}{\textbf{Front-end electronics low-voltage supplies}} & Voltages & \SIrange{1.7}{2.1}{\volt} \analog          & \SIrange{1.2}{2.2}{\volt} feeding both \\
&                & \SIrange{2.1}{2.5}{\volt} digital                &  \analog and digital \\
&                & \SIlist{0.8;2.5;5}{\volt} opto-device  & \SIlist{0.8; 2.5;5}{\volt} opto-device \\
& Current  & \SI{3.7}{\kilo\ampere} & \SI{209}{\ampere} \\
& Segmentation of bulk supply & One per \numrange{6}{7} modules & One per 4 FE-I4B \\
& Segmentation of regulated supplies & One regulator per module & One regulator per 4 FE-I4B\\
\hline
\multirow{4}{*}{\textbf{Power}} & Front-end electronics power
& $\sim\SI{6}{\kW}$     &  $\sim\SI{0.3}{\kW}$ \\
& Cables plus regulators    & $\sim\SI{18}{\kW}$   &  $\sim\SI{0.8}{\kW}$ \\
& Total power (start of \RunTwo)  & $\sim\SI{24}{\kW}$  &  $\sim\SI{1.1}{\kW}$ \\
& Total power (end of \RunTwo) & $\sim\SI{32}{\kW}$  &  $\sim\SI{1.6}{\kW}$\\
\hline\hline
\end{tabular}
\end{adjustbox}
\end{table}
 
\subsubsection{Detector control and interlock systems}
 
\begin{figure}
\centering
\includegraphics[width=0.70\textwidth]{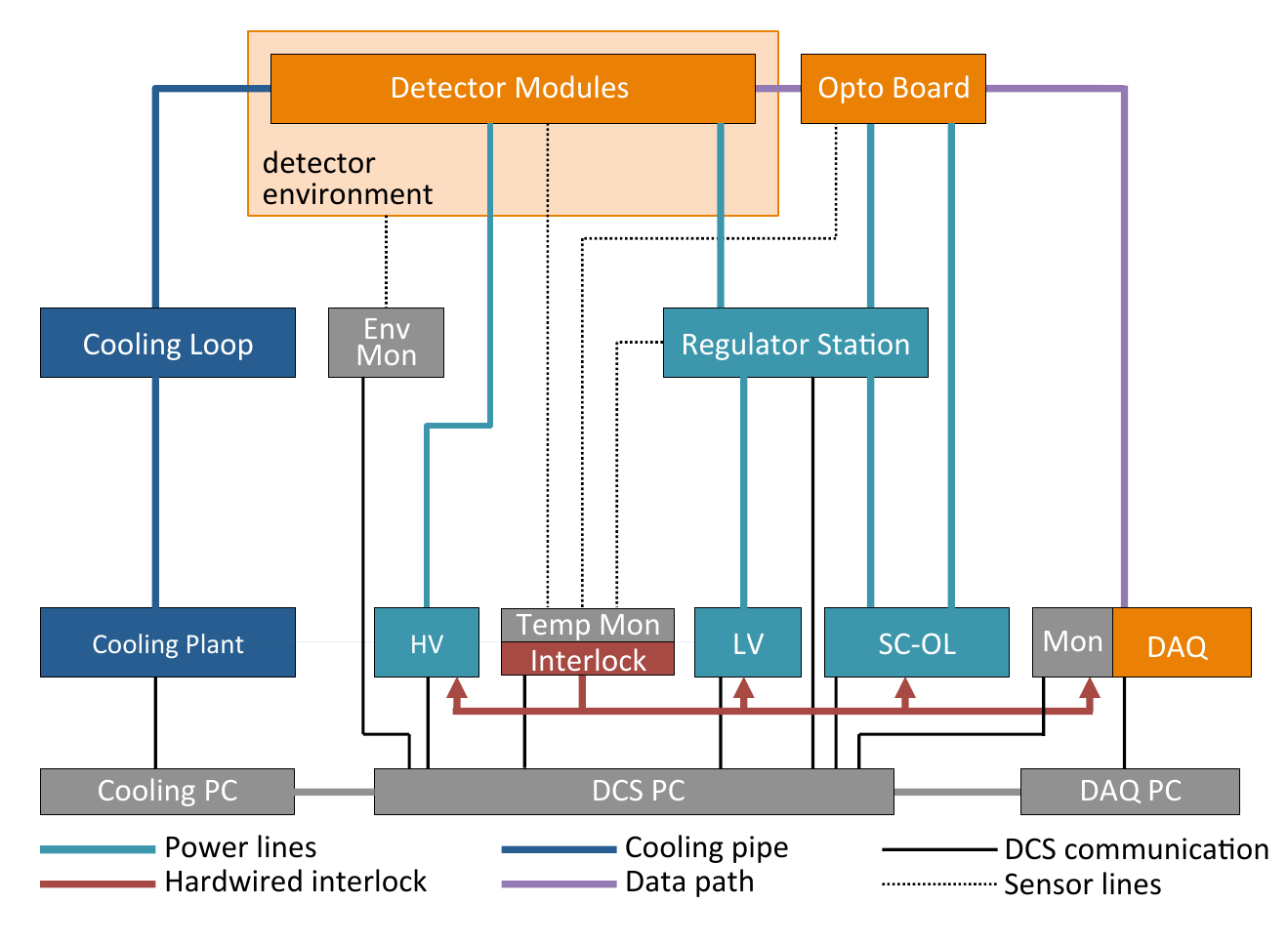}
\caption{Schematic~\cite{PIX-2018-001} of the \gls{IBL} \gls{DCS} control and monitoring functions on the sensor (\gls{HV}), front-end electronics (\gls{LV}) and optoboard (\gls{SCOL}) power supplies, as well as temperature and humidity monitoring. The schematic also indicates an independent hardware interlock system used for detector and operational security. }
\label{fig:IBL_DCS}
\end{figure}

The \gls{IBL} \gls{DCS} provides monitoring, control, and safety for the \gls{IBL}. Its hardware encompasses the sensors (humidity and temperature), the monitoring crates, and the interlock crates. From an operational point of view, the power supplies and the regulator stations at \gls{PP2} also belong to the \gls{DCS}. Table~\ref{tab:IBL_DCS} gives an overview of the \gls{DCS} hardware components while their interconnections and functional dependencies are shown in Figure~\ref{fig:IBL_DCS}. The heart of all custom built crates is an \gls{ELMB}128~\cite{ELMB}.
 
\begin{table}
\caption{Main \gls{IBL} \gls{DCS} hardware components.
\label{tab:IBL_DCS}
}
\centering
\begin{adjustbox}{max width=1.0\textwidth}
\begin{tabular}{|l|l|l|l|l|}
\hline\hline
Crate & Location & Task & Communication & Vendor \\
\hline\hline
Beam-building module & PP3, UX15 & Monitoring of environmental temperature and humidity & CAN & custom built\\
\hline
\gls{HV} & USA15 &Power Supply: sensor depletion voltage for planar and 3D modules& CAN &ISEG\\
\hline
\gls{LV} & USA15 &Power Supply: Front-end Electronics & TCP/IP&Wiener\\
\hline
LV-PP4 & USA15 & Distribution of \gls{LV} and current monitoring per module & CAN &custom\\
\hline
\gls{SCOL} & USA15 & PS: Optical Link& CAN &custom\\
\hline
Regulator Station & PP2, UX15 & LV regulation & CAN &custom\\
\hline
IMC & USA15 & HW Interlock. Temperature monitoring of interlock controlled devices & CAN &custom\\
\hline
\end{tabular}
\end{adjustbox}
\end{table}
 
The \gls{DCS} software is based on the \gls{WinCC OA} \gls{SCADA} system~\cite{PVSS} which is used throughout the \gls{LHC} experiments. The data transmission and communication between \gls{DCS} hardware and software is mainly done via \gls{CAN} bus and an \gls{OPC} client-server architecture, where the servers are provided by the vendor, and the client provided by \gls{WinCC OA}. External information (for example, for the ${\text{CO}}_2$ cooling system) can be accessed via distributed systems such as the ATLAS-wide \gls{DSS}~\cite{DSS} or the information published by the \gls{LHC} regarding the machine and beam state.
 
The \gls{IBL}-specific \gls{DCS} is organised in three layers: the lowest level implements the control at a hardware based layer and provides the necessary setup and configuration. The middle level maps the hardware channels to the detector units. The smallest units that can be controlled independently are groups of four front-ends, or a \gls{DCS} module, and the optoboards each serving four \gls{DCS} modules. At the top level, detector oriented control and monitoring is provided by a \gls{FSM} hierarchy implemented in a framework provided by the CERN \gls{JCOP} and ATLAS groups. Thanks to this tree structure, shifters can easily monitor the detector state by observing the changes of the upper level of the tree; the detector can be controlled via \gls{FSM} commands by non-\gls{DCS}-experts. The lowest level of the \gls{FSM} hierarchy is based on modules and optoboards and follows the detector geometry of readout units and staves building the innermost Pixel layer.
 
For detector safety, a dedicated hardware interlock system is installed. It is fully independent from the \gls{DCS} software,
but its input and output signals are monitored by the \gls{DCS}. The interlock system protects the detector and certain electronics against overheating by generating interlock signals from \analog temperature monitoring with a high granularity. In addition, it receives signals from \gls{DSS} and distributes the corresponding interlocks, and protects the \gls{IBL} from dangerous beam conditions by receiving the Stable Beams signal from the \gls{LHC} and using it to provide a software-generated beam injection permit signal.
 
\subsubsection{\glstext*{IBL} ${\text{CO}}_2$ cooling system}
\label{sec:IBL-Cooling}
 
\begin{figure}
\centering
\includegraphics[width=1.0\textwidth]{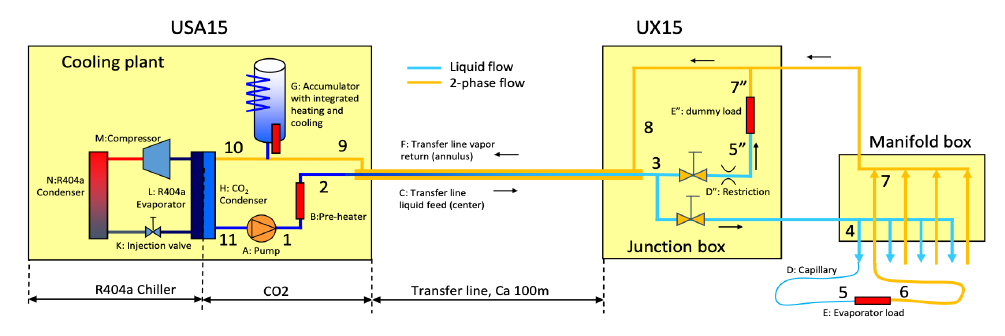}
\caption{Schematic representation of the \gls{IBL} ${\text{CO}}_2$ cooling system~\cite{PIX-2018-001}. The numbers indicate the loop direction.}
\label{fig:IBL-CO2Cooling}
\end{figure}

The \gls{IBL} is cooled by a two-phase ${\text{CO}}_2$ system which guarantees an operational temperature in the range between \SIrange[retain-explicit-plus]{-35}{+15}{\degreeCelsius}, thanks to a total design cooling power of about \SI{3}{\kW}. The cooling prevents thermal runaway of the irradiated sensors. Figure~\ref{fig:IBL-CO2Cooling} shows a simplified schematic view of the system.
 
\begin{figure}[htbp]
\centering
\subfloat{\label{fig:iblCoolLines}\includegraphics[width=.5\textwidth]{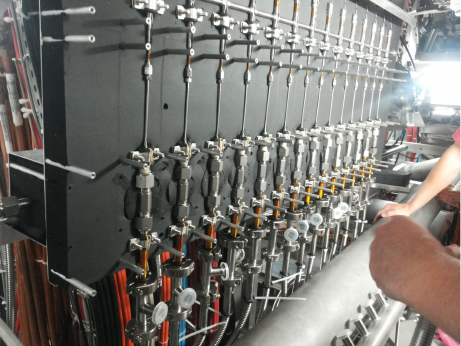}}\hspace{0.5 cm}
\subfloat{\label{fig:IDendPlateCool}\includegraphics[height=6.1cm]{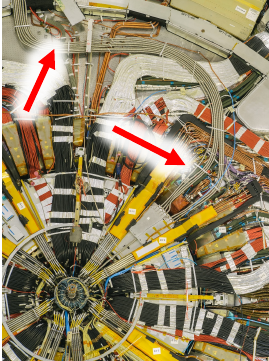}} \\
\caption{\protect\subref{fig:iblCoolLines} \gls{IBL} cooling lines and \protect\subref{fig:IDendPlateCool} the \gls{ID} end-plate region, with the path of the cooling lines highlighted.}
\label{fig:IBL_CO2CoolingLines}
\end{figure}

The ${\text{CO}}_2$ is cooled by commercial chillers based on a hydrofluorocarbon mixture (R404A) and pumped to the experimental cavern through a concentric transfer line of a length of about \SI{100}{\m}.  The liquid supply line is inside the two-phase return line to pre-heat the sub-cooled liquid from the plant to approximately the same temperature as the returning two-phase fluid.
This also protects the supply liquid from possible ambient heating. The transfer line is vacuum insulated with a stationary vacuum shield of \SI{63.5}{\mm}.
At the end of the transfer line the fluid is distributed to the detector staves through a manifold box (see the left side of Figure~\ref{fig:IBL_CO2CoolingLines}) where \num{14} concentric vacuum flex lines (seven for each side) bring the fluid to the detector through the same routing as the cables on the \gls{ID} end plate (see the red arrows in Figure~\ref{fig:IBL_CO2CoolingLines} on the right).
The flex line vacuum is obtained thanks to an active vacuum pump system installed in \gls{UX15} and monitored by a dedicated
\gls{PLC}.
Next to the manifold box there is a junction box where temperature and pressure sensors are installed to monitor the
${\text{CO}}_2$ conditions closest to the detector. Also a \SI{3}{\kW} heater is present in order to simulate the detector load running in bypass mode for
commissioning or test of the system.
 
The evaporation temperature in the detector is regulated by the pressure in the accumulator (${\text{CO}}_2$ tank providing coolant reservoir).
The return pressure (saturation temperature) is controlled by heating or cooling of the two-phase mixture inside the accumulator.
By controlling the accumulator it is possible to control in a very precise way the evaporation temperature in the detector and to provide the required cooling conditions and stability.
The diameter of the return lines is chosen to be slightly larger than that of the \gls{IBL} cooling pipes, to ensure that the pressure drop along these lines remains sufficiently low that the accumulator pressure directly determines the evaporator pressure.
 
In \gls{USA15}, two redundant cooling units are installed and running, one providing cooling to the detector
with the other running cold in stand-by mode through a by-pass. The purpose of the stand-by mode is to keep the pump
of the inactive system cold to allow for a fast start in case it is required.
One unit is running with normal electrical power while the other is also supplied by \gls{UPS} batteries:
in case of a failure of the normal power, there is an automatic swap between the two units which is
transparent for the detector, and only a very small and short temperature variation is visible.
The logic of installing two redundant units allows uninterrupted operation in case of
failure or servicing of one line. The two cooling units are controlled by two independent \glspl{PLC}.
 
Since the start of operation, the \gls{IBL} cooling has operated at three different temperatures. In May~2015 the \gls{IBL} was operated at \SI{-10}{\degreeCelsius}, until the start of the study of the high current drawn by the readout chips discussed in Section~\ref{sec:ID-IBL-ModulesStaves},
when the cooling system was run at the much higher temperature of (\SI[retain-explicit-plus]{+15}{\degreeCelsius}) until June~2016. From then on, the operational set point of the plant was reduced to \SI{-20}{\degreeCelsius}.
 
The \gls{IBL} cooling system has been shown to operate stably over the full range between room temperature and its lowest verified limit, which is \SI{-35}{\degreeCelsius}.
Its reliability was demonstrated during external events like power cuts, glitches and primary cooling failures. No downtime occurred during data taking from the beginning of \RunTwo and, on the occasion of the only system failure, the stand-by unit kept the detector in the required conditions.
 
The temperature stability, needed to limit the detector bowing effect described in Section~\ref{sec:IBL-Integration},
was better than \SI{0.2}{\degreeCelsius}  (with less than \SI{0.05}{\degreeCelsius} RMS) and made it possible to correct for bowing through alignment corrections.


\subsection{The ATLAS Pixel upgrade} 
\label{sec:Pixel}
 
\subsubsection{Installation of new Service Quarter Panels}
 
The installation of the \glspl{nSQP} during \gls{LS1} enabled the relocation of the optoboards outside the \gls{ID} volume to an accessible area, making future repairs possible without extraction of the Pixel detector. All defects originating from broken data transmission lines and faulty optoboards were therefore repaired during \gls{LS1}.
Additionally, all faulty connections outside the active Pixel detector volume were repaired during the process of reconnection after the \glspl{nSQP} installation. Faulty connections within the active volume were not accessible and thus could not be repaired.
 
The full detector package was removed, taken to the surface and tested there, before being re-installed in ATLAS in December 2013.
The refurbished three-layer Pixel detector was then reconnected and tested.
The number of disabled modules was decreased to 33, 
$1.9\%$ of the total. The biggest improvement was achieved in the $B$-Layer, where the disabled fraction was reduced from $6.3\%$ to $1.4\%$, and Layer~2, where the $7.0\%$ faulty fraction was reduced to $1.9\%$.
 
The \glspl{nSQP} included additional data fibres dedicated to the readout of Layer~1 (and part of the Discs); this eventually resulted in doubling the bandwidth of the transmitted data once new \gls{IBL} \gls{ROD}/\gls{BOC} card pairs were deployed at a later stage of the project (see Section \ref{sec:Pixel-Readout}).
For Layer~1 and part of the Discs, the readout upgrade increased the bandwidth from \SIrange{80}{160}{Mbps}, and for Layer~2 from \SIrange{40}{80}{Mbps}.
This increase of the bandwidth by a factor two would allow the detector to run without bandwidth limitation up to an instantaneous luminosity of \lumirunthreedesign. Further details of the readout upgrade are given in the Section~\ref{sec:Pixel-Readout}.
 
Because of the easier maintenance afforded by the installation of the \glspl{nSQP}, another optoboard replacement campaign was conducted during \gls{LS2}. The detector status at the start of \RunThr was as follows: 10 modules ($3.5\%$) disabled in the $B$-layer, 14 modules ($2.8\%$) disabled in Layer~1, 32 modules ($4.7\%$) disabled in Layer~2, and 10 modules ($3.5\%$) disabled in the Discs. In addition, three \gls{IBL} front ends were disabled.

\subsubsection{Readout upgrades during the winter shutdowns}
\label{sec:Pixel-Readout}
 
New readout components developed for the \gls{IBL} (see Section\,\ref{sec:IBL-Readout} for details) were used to upgrade the off-detector readout hardware of the outer Pixel layers, making use of the new fibre installation provided with the \glspl{nSQP}. The \gls{nSQP} project was an opportunity to increase the bandwidth for some of the Pixel layers and remove some limitations that would have arisen with the expected increase of luminosity.
The strategy was to gradually upgrade the various layers during three consecutive winter shutdowns between 2015 and 2018. The timeline of these upgrades was as follows:
 
\begin{description}
 
\item[2015/16] Layer 2 + 12.5\% of Layer 1, a total of \textbf{22 \gls{ROD}/\gls{BOC}} card pairs
 
\item[2016/17] Remaining part of Layer 1 (87.5\%), a total of \textbf{42 \gls{ROD}/\gls{BOC}} card pairs
 
\item[2017/18] $B$-Layer and Disc, a total of \textbf{38 \gls{ROD}/\gls{BOC}} card pairs
 
\end{description}
 
By the end of \RunTwo, in 2018, both Pixel and \gls{IBL} finally shared the same type of readout system, making it easier to operate the detector and offering more powerful debugging capabilities. No further changes to the off-detector readout system are foreseen during \RunThr operation.


\subsection{Semiconductor Tracker readout upgrade}
\label{sec:ID-SCT}
 
\begin{figure}
\centering
\includegraphics[width=0.7\textwidth]{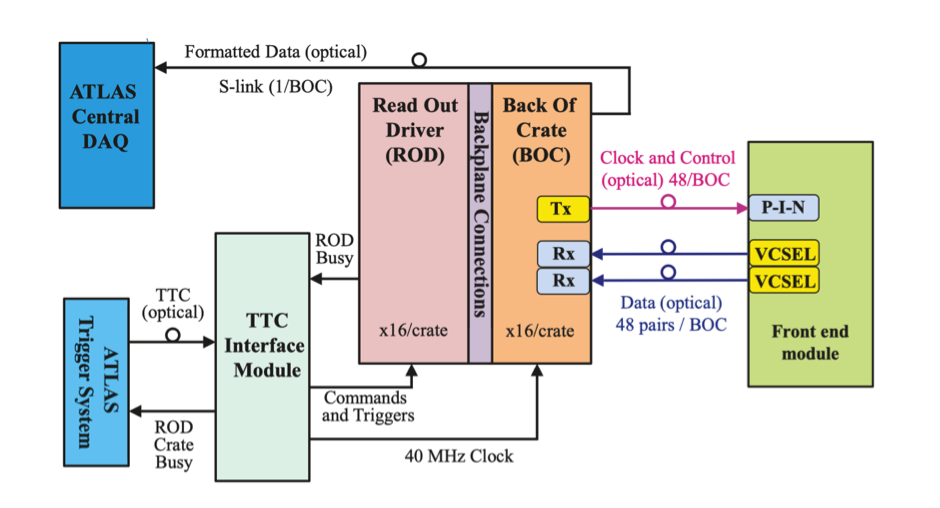}
\caption{Schematic of the \gls{SCT} Data Acquisition System.}
\label{fig:SCT_DAQ}
\end{figure}

The \gls{SCT} data acquisition system in \RunOne comprised \num{90} \gls{ROD} and \gls{BOC} cards.  The \gls{BOC} provides the optical interface between a \gls{ROD} and up to \num{48} \gls{SCT} modules; for each module the \gls{BOC} transmits the clock and trigger via a single command  stream and receives data back via two optical links from the six \glspl{ASIC} on each side of the module. The \gls{ROD} processes incoming data from the \num{96} data links at the \gls{L1} trigger rate, and combines those data into a single event fragment which is broadcast on a single optical \gls{s-link} to the ATLAS \gls{DAQ}. Figure~\ref{fig:SCT_DAQ} shows a schematic of the \gls{SCT} \gls{DAQ}.
 
The \gls{SCT} was designed to operate with \SIrange{0.2}{0.5}{\percent} occupancy from its 6.3 million sampled strips with an \gls{LHC} pileup of up to about $\mu=23$ \pp collisions per \gls{BC}. The fundamental bottlenecks arising from higher pileup that restrict increasing occupancy are the bandwidth limitations of the data links from the modules (which transmit bit streams at \SI{40}{Mbps}) and the \glspl{s-link} (which transmit 32-bit words at \SI{40}{\MHz}, a throughput of \SI{1.28}{Mbps}). Extrapolations of occupancy with increasing pileup during \RunOne suggested that the data flow in the data links and \glspl{s-link} would exceed bandwidth limitations at pileups of around $\mu=87$ and \num{33} \pp collisions per \gls{BC} respectively, at a \gls{L1} trigger rate of \SI{100}{\kHz}. These projections used the optimum on-chip data compression mode (where only hits exceeding a threshold from the in-time \gls{BC} are read out, while simultaneously vetoing hits from the preceding \gls{BC}), and the standard data packing schema of the \gls{ROD} which included readout of the hits from each strip over three consecutive \gls{BC}.
 
A number of mitigation steps were deployed to prepare the \gls{SCT} for the anticipated higher pileup during \RunTwo. Furthermore, it was discovered early in \RunTwo that \gls{SCT} hit occupancy had increased significantly compared to \RunOne data, which is now attributed to secondary interactions with the newly-installed \gls{IBL} services. The increased occupancy led to a revision of the pileup limits for \gls{SCT} operation, leading to further mitigation steps during \RunTwo in order to achieve a comfortable margin for operation with $\mu\sim 60$. These mitigation steps
were:

\begin{description}
 
\item[Expanded \gls{DAQ}]
To address the pileup limit imposed by the \glspl{s-link}, the numbers of \glspl{ROD} and \glspl{BOC} were increased in 2014, during \gls{LS1}, thereby reducing the number of modules serviced by each \gls{ROD} from 48 to 36. During \gls{LS2} the \gls{DAQ} system was expanded from 90 to 128 cards (the full complement of cards possible in each \gls{ROD} crate), increasing the total number of \glspl{s-link} from 90 to 128.
 
\item[Cable remapping]
In 2015, the cabling of the data links to the \glspl{ROD} was rearranged to harmonize the data load on each \gls{ROD}. This resulted in a flatter occupancy distribution across the \glspl{s-link}, removing spikes from the highest occupancy \glspl{s-link} and thereby improving the pileup limit imposed by the \glspl{s-link}.
 
\item[Data compression]
A new highly efficient data packing scheme with compression on the \gls{ROD} (``supercondensed mode'') was developed during \gls{LS1} and was deployed routinely from 2016.  The new scheme resulted in a roughly 25\% data size reduction, at the expense of losing the 3-bit timing information.
 
\item[Chip Masking]
From 2017, the noisiest chips were masked to reduce data throughput on the data links. This worked by using a reference table of the noisiest chips as a function of pileup, and masking the \num{128} channels of those chips. The mechanism worked ``on the fly'', gradually reducing to zero the number of masked chips as instantaneous luminosity decreased during a run. Masking was flagged in the data to avoid an arbitrary loss in hit efficiency. Typically, even at the highest pileup of $\langle\mu\rangle\sim 60$, well under 1\% of chips were masked.
 
\end{description}
 
\begin{figure}
\centering
\includegraphics[width=3.5in]{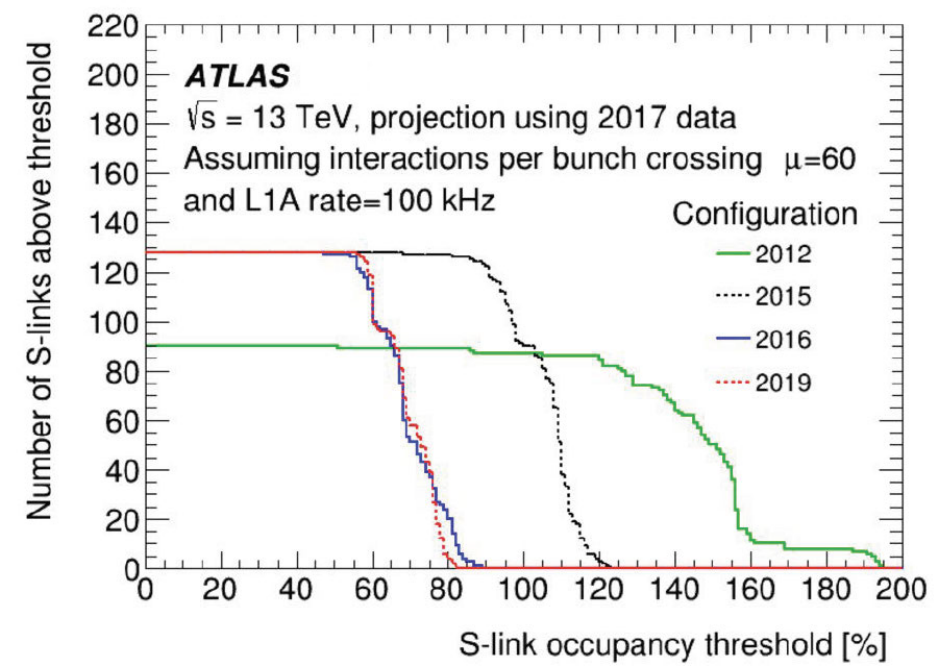}
\caption{Number of \glspl{s-link} that would be operating above the bandwidth occupancy threshold as a function of that threshold. The 2012 configuration was the original 90-\gls{ROD} system, the 2015 configuration used 128 \glspl{ROD}. Neither of these configurations was suitable for \RunThr conditions; however, the 2016 configuration incorporated supercondensed mode and the fibre reshuffling, and this allows all the \glspl{s-link} to operate below the bandwidth occupancy threshold. A further minor cable remapping was performed in 2019, improving matters even more.}
\label{fig:SCT_SLINK}
\end{figure}

Figure~\ref{fig:SCT_SLINK} shows the impact of the different mitigation steps related to \glspl{s-link}, showing the number of \glspl{s-link} which operate above a defined fraction of the available bandwidth (occupancy threshold)
as a function of that threshold. The plot illustrates, for example, that no \glspl{s-link} used more than about $90\%$ of the bandwidth from 2016.
 
Following these mitigation steps, pileup limits at \SI{100}{\kHz} trigger rate are around $\langle\mu\rangle = 70$ for both the data links and the \glspl{s-link}, which is sufficient for \RunTwo and \RunThr operation. Hot chip masking is not deployed routinely, but remains an option for future operation.


\subsection{Transition Radiation Tracker upgrades}
\label{sec:ID-TRT}
 
\subsubsection{\glstext*{TRT} \glstext*{DAQ} upgrades in \RunTwo}
The \gls{TRT}~\cite{TRT,TRTBarrel,TRTEndcap} \gls{DAQ} consists of custom \analog/digital front-end electronics and the back-end \glspl{ROD}~\cite{TRTElectronics}.
An \analog component of the front-end, the \gls{ASDBLR}, detects signals from two drift straws using two different thresholds: a tracking threshold of about \SI{200}{\eV}, the \gls{LT}, and the transition radiation threshold of about \SI{6}{\kilo\eV}, the \gls{HT}.
During each \SI{25}{\nano\second} bunch-crossing, the digital component of the electronics, the \gls{DTMROC}, samples the output of the \glspl{ASDBLR} eight times, creating one bit for each \SI{3.125}{\ns} indicating whether the straw signal exceeds the low threshold, and one additional bit to record whether the signal crosses the high threshold at any time during the 25~ns.
Upon receiving a \gls{L1A} signal, the data from three bunch-crossings are sent to the \glspl{ROD} on the detector back-end.
These 27~bits of information constitute a ``straw word'', which is the fundamental block of data from a readout and compression perspective.
Each \gls{ROD} validates and formats the data from up to 120 \glspl{DTMROC}, building a ``data fragment'' by serialising the straw words.
The fragment is then compressed with an entropy-based Huffman algorithm~\cite{Huffman} implemented in the \gls{ROD} firmware and is sent to the ATLAS central \gls{ROS} system via an \gls{s-link}.
 
With this setup, the \gls{TRT} \gls{DAQ} system was capable of operating at \gls{L1} input rates up to \SI{80}{\kHz} at the average pileup of $\langle \mu \rangle = 35$ in \RunOne; however,
as the \gls{LHC} exceeded its initial design luminosity, the goal for the final years of \RunTwo (2017-18) and for \RunThr became running at an \gls{L1} input rate of \SI{100}{\kilo\Hz} with pileup up to $\langle \mu \rangle = 60 - 70$ \pp collisions per \gls{BC}.
This pileup corresponds to an average per-event straw hit occupancy of $75 - 85\%$ in the high-$z$ slices of the \gls{TRT} endcaps, which have the highest occupancies due to their high-\abseta positions.
Under these more strenuous conditions, bottlenecks were observed when reading data both into and out of the \glspl{ROD}.
Together, these bandwidth saturations limited the system to a \gls{L1} rate of \SI{90}{\kHz} and a maximum straw hit occupancy of 50\%, for which the following \gls{DAQ} upgrades were implemented between the end of \RunOne and early \RunTwo (2015--16).
 
To alleviate the \gls{ROD} input bandwidth saturation, the four trailing bits in the 27-bit straw word were truncated
off in the \glspl{DTMROC}. This corresponds to discarding hits with the longest drift time where the ionizing particles barely cross the edge of the straws, which are rather rare and are less important to tracking. This truncation reduces the straw word size sent to the \glspl{ROD} by 15\% with no impact on tracking performance, enabling the \gls{TRT} to run without saturating the \gls{ROD} input bandwidth at up to \SI{104}{\kHz} with evenly spaced triggers, and up to \SI{102}{\kHz} in physics data taking where the rate can instantaneously exceed \SI{104}{\kHz}.
 
To cope with the \gls{s-link} bandwidth saturation when reading data out of the \glspl{ROD} at high rates, a series of \gls{ROD} firmware updates and hardware upgrades were performed. The firmware updates include the introduction of a validity gate to remove words leaving no \gls{LT} hits within the \SIrange{18.75}{56.25}{\nano\second} time window.
Such hits are most likely to originate from a neighbouring bunch-crossing, and
only about 3\% of the hits from the bunch of interest are lost to this validity gate.
Similarly, the \gls{HT} bits from the first and third bunch-crossings were masked to zero to allow more efficient compression, which keeps 94\% of \gls{HT} hits from the bunch of interest.
Additionally, non-vital information was suppressed in the error blocks attached to the \gls{ROD} data fragments.
An auxiliary word-caching hash table was also implemented to relieve the heavy use of the Network Search Engine \glspl{ASIC} on the \glspl{ROD}~\cite{TRTElectronics} during the Huffman compression.
Finally, the Huffman compression table was updated to maintain optimal compression performance in the harsher run conditions of late \RunTwo and \RunThr.
Together, these firmware updates resulted in a 40\% reduction of the event fragment size sent off the \glspl{ROD}, without significant loss of performance and operational stability.
On the hardware side, the \gls{HOLA} \gls{s-link} interface cards~\cite{HOLA} used to transfer data from the \gls{TRT} \glspl{ROD} to the \glspl{ROS} were upgraded with a faster clock crystal (\SI{60}{\MHz} vs \SI{40}{\MHz}) and a faster \gls{FPGA} (\SI{2.5}{Gbps} vs \SI{2.0}{Gbps}).
This created 50\% additional bandwidth for the \glspl{s-link}.
 
\begin{figure}[htbp]
\centering
\includegraphics[width=.6\textwidth]{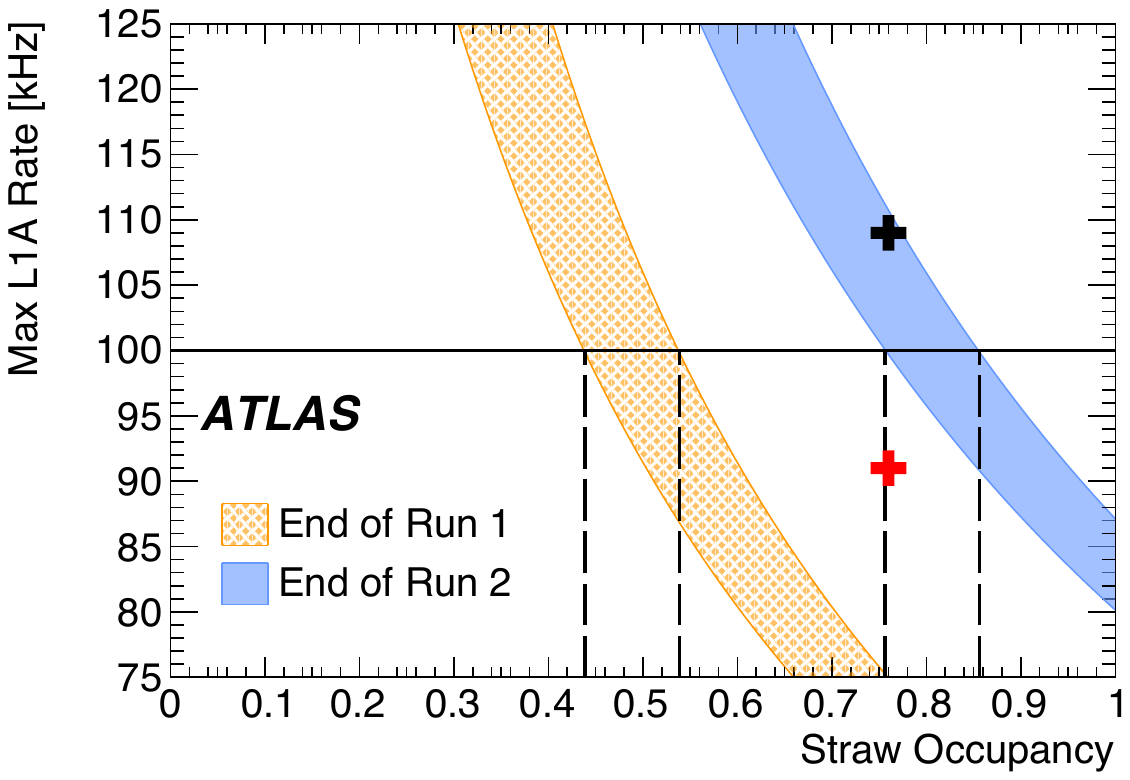}
\caption{
Projected maximum achievable \gls{L1} input rate as a function of \gls{TRT} straw occupancy (dotted orange) before and (blue) after increasing the \gls{s-link} bandwidth and reoptimising the compression table.
Only the high-$z$ endcap \glspl{ROD} are considered in the calculation, as they experience the highest occupancies.
The bands represent the uncertainty due to the variable size of the error blocks in the event fragments; the upper (lower) edge of the area corresponds to the minimum (maximum) error block.
Two data points are shown for the \gls{ROD} closest to the \gls{s-link} bandwidth limit. These data points are taken from the beginning of a 2018 run with $\langle \mu \rangle = 57$, resulting in an occupancy of 76\% in this \gls{ROD}. The red cross marks the \gls{L1} rate at which this \gls{ROD}'s data was collected, while the black cross shows the maximum rate at which this data could have been collected before saturating the \gls{s-link}, extrapolated based on the average word length after the compression.
}
\label{fig:maxL1}
\end{figure}

Figure~\ref{fig:maxL1} summarises the effect of the updated compression table and \gls{HOLA} cards.
The projected maximum \gls{L1} rate at which the \gls{TRT} can read out its data fragments without saturating the \gls{s-link} is shown as a function of the detector occupancy.
These limits are calculated from the average Huffman codeword length for the hit straws, the number of empty straws, and the \gls{s-link} bandwidth.
Only the high-$z$ endcap \glspl{ROD} are used in this calculation, as they have the highest occupancies and are closest to the bandwidth limit.
The bands represent the uncertainty in the fragment size due to the variable length of each fragment's error block, which can range from 0 to 60 32-bit words; the upper (lower) edge of the shaded area corresponds to the minimum (maximum) error block lengths.
Under the final \RunTwo conditions, the error block sizes were typically around 15 words or less.
After all the upgrades, the system can handle occupancies up to 76-86\% at the target \gls{L1} rate of \SI{100}{\kHz}, depending on the error block size.

\subsubsection{\glstext*{TRT} Gas Configuration Update}
 
By the end of \RunOne, several leaks developed in flexible active gas exhaust pipes made of \gls{PEEK}. The leaks originated from cracks occurring in places of local stress on the \gls{PEEK} pipes due to a reaction with ozone produced in the active gas during \gls{TRT} operation. An attempt to repair the leaking pipes during \gls{LS1} in 2013 was partially successful for the \gls{TRT} endcaps. However, the leaks in the \gls{TRT} barrel and some of those in the \gls{TRT} endcaps are located in inaccessible areas and therefore repairs are not possible. Since the number of leaks is expected to grow with increasing \gls{LHC} luminosity, it became unaffordable to operate the entire detector with the baseline xenon-based gas mixture. In order to understand the \gls{TRT} performance with a lower-cost argon-based gas mixture, dedicated studies were performed during the proton-lead collision running at the end of \RunOne. In that study, leaking modules in the barrel and endcaps were operated with the argon-based gas mixture. The study found that the transition radiation based \gls{PID} performance of the detector was significantly reduced but that the tracking properties were preserved~\cite{IDET-2015-01}.
 
\begin{figure}[htbp]
\centering
\subfloat[End of \RunTwo setup]{\includegraphics[width=.48\textwidth]{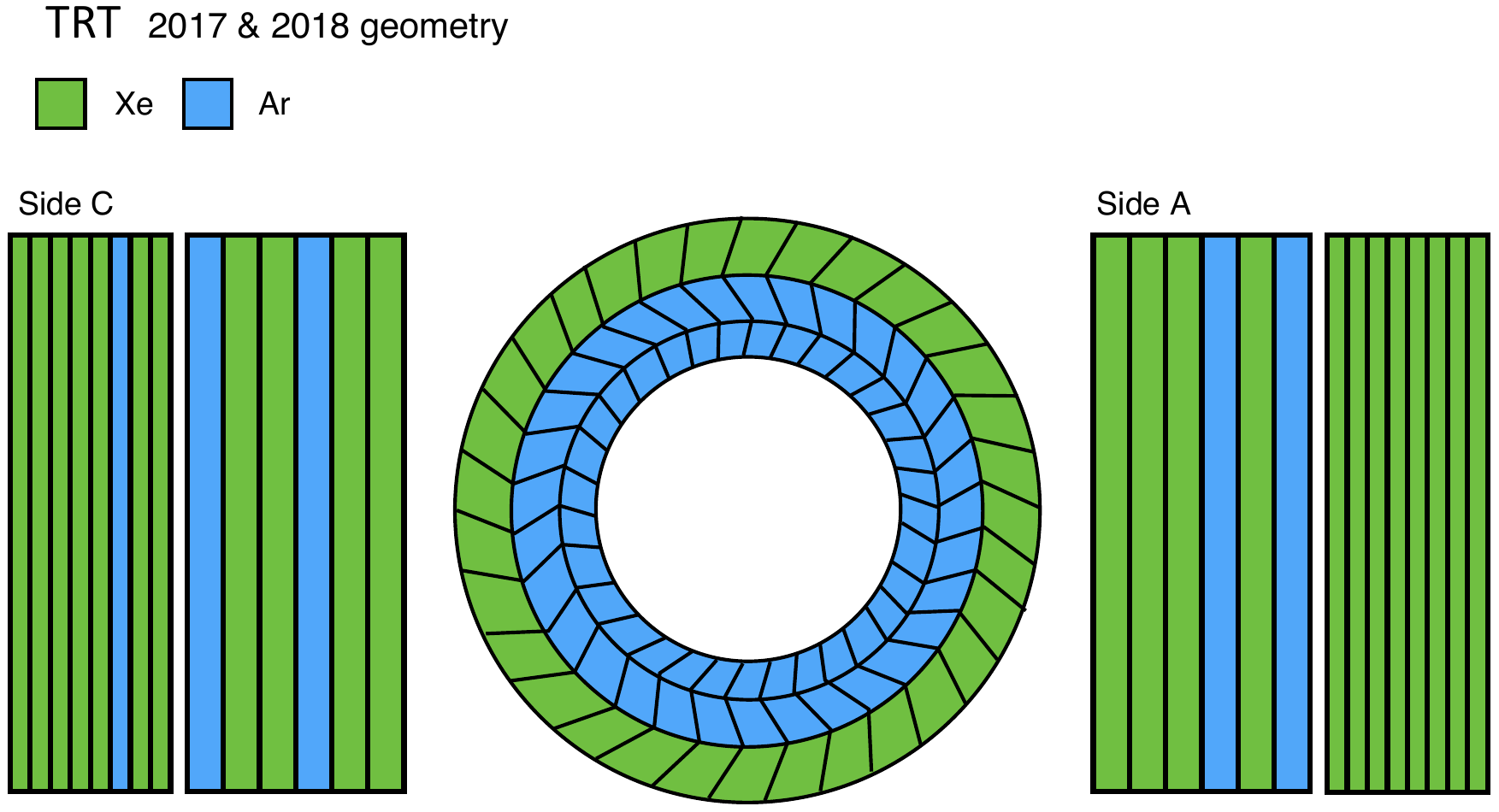} \label{fig:TRT_gas_Run2}}\\
\subfloat[Setup at start of \RunThr]{\includegraphics[width=.48\textwidth]{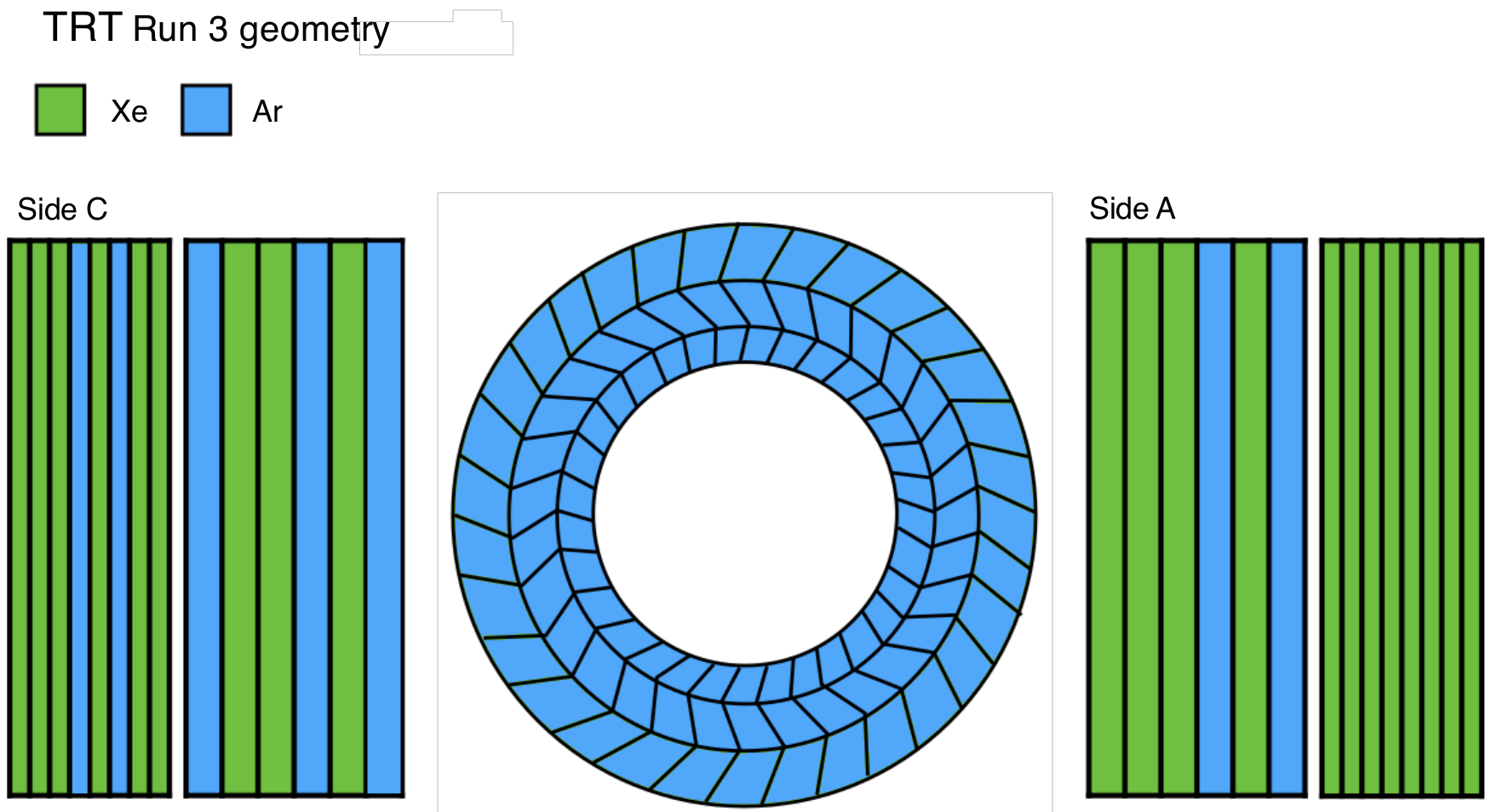} \label{fig:TRT_gas_Run3}}
\caption{
\gls{TRT} gas geometries \protect\subref{fig:TRT_gas_Run2} at the end of \RunTwo and \protect\subref{fig:TRT_gas_Run3} at the start of \RunThr. Blue represents \gls{TRT} sections supplied with the Ar-based gas mixture and green represents the sections supplied with the Xe-based gas mixture.
}
\label{fig:TRTgeo}
\end{figure}

During \RunTwo many leaking modules were changed to the argon-based gas mixture in stages. In 2015 all modules of the barrel inner layer and one wheel from each endcap were operated with the argon-based mixture. Starting in 2016 additional leaks led to a configuration where the two inner barrel module layers were supplied with the argon-based gas mixture and only the outer module layer was operated with the xenon-based gas mixture. This configuration still left half of the straws in the \gls{TRT} barrel volume with full transition radiation information. In the endcaps, two wheels on side~A (4 and 6) and three wheels on side~C (3, 6, 9) were supplied with the argon-based gas mixture, as can be seen in Figure~\ref{fig:TRTgeo}\subref{fig:TRT_gas_Run2}. In this configuration, the \gls{PID} properties of the \gls{TRT} endcaps were not significantly affected.
 
For \RunThr it is planned to use the argon-based mixture for the entire barrel and for a few more endcap wheels on the C side, as illustrated in Figure~\ref{fig:TRTgeo}\subref{fig:TRT_gas_Run3}.
This will significantly reduce the loss of xenon-based gas mixture while maintaining a stable gas configuration hopefully for the entire \RunThr.
In the \RunThr gas configuration the \gls{PID} performance of the endcaps is largely preserved.
In the barrel the \gls{PID} function is significantly reduced because of a poor absorption of transition radiation photons by the argon gas
but in a combination with $\text{d}E/\text{d}x$ measurements it still contributes to the ATLAS electron identification particularly at particle energies below \SI{10}{\GeV}~\cite{IDET-2015-01}.
 
\subsection{New \glstext*{ID} thermosiphon system}
 
\begin{figure}
\centering
\includegraphics[width=0.4\textwidth]{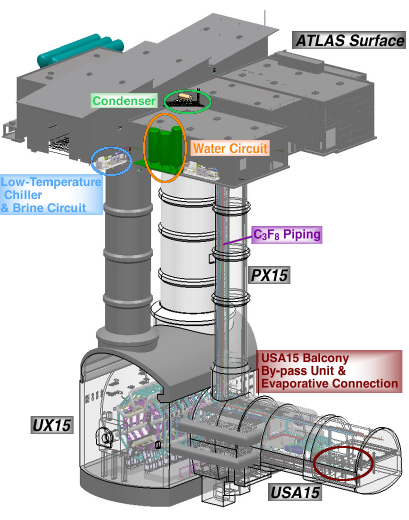}
\caption{
General layout of the \gls{ID} thermosiphon system distributed between the ATLAS Surface area, the ATLAS service cavern (\gls{USA15}) and the ATLAS detector cavern (\gls{UX15}).
Colour-coded labels identify the main elements of the system, which are highlighted in the drawing. The \SI{92}{\m} column of coolant is contained in a pipe extending the full depth of the PX15 service shaft.
}
\label{fig:ID_Thermo_Layout}
\end{figure}

\begin{figure}
\centering
\includegraphics[width=0.9\textwidth]{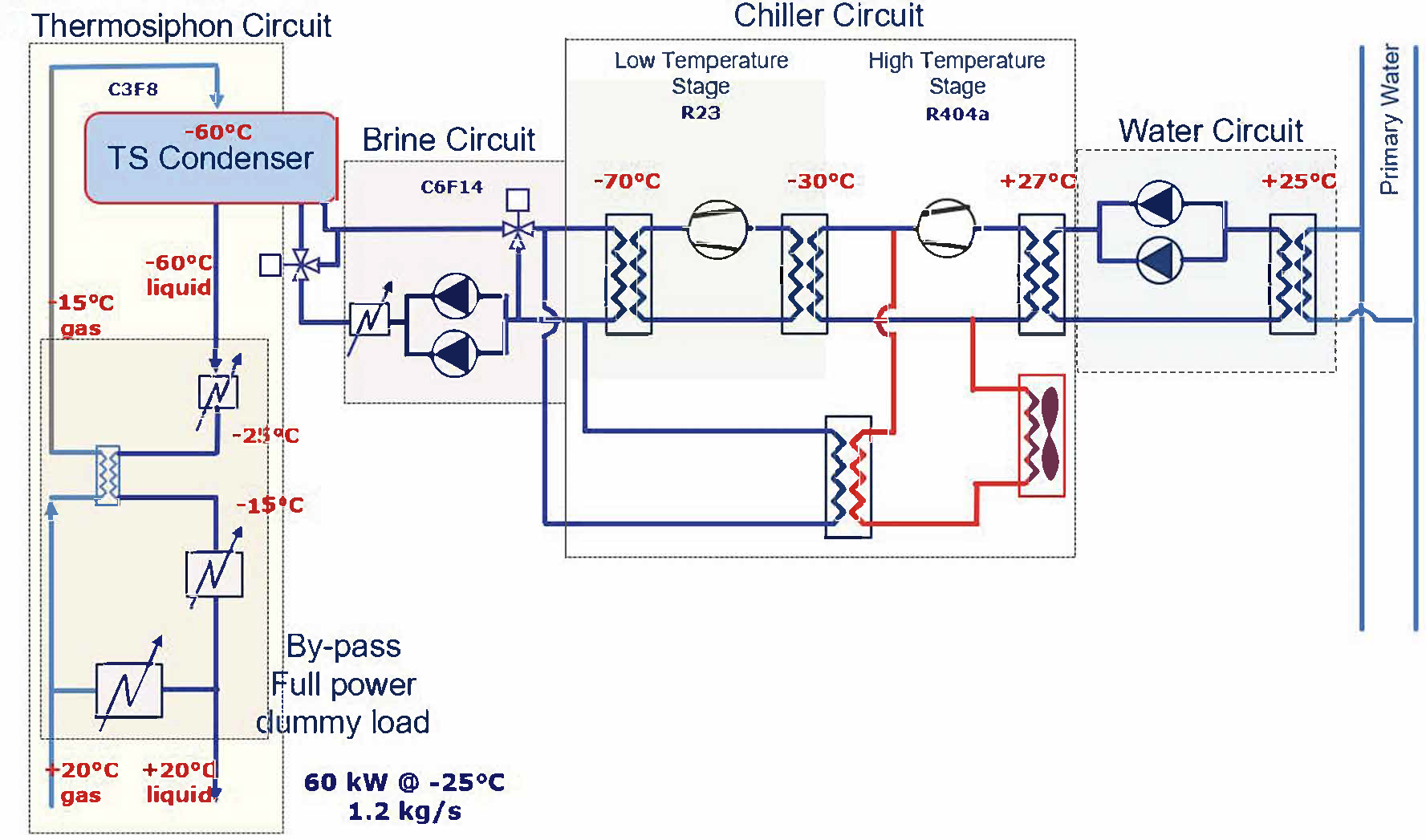}
\caption{Full schematic of the thermosiphon circuit showing, from left to right, the primary thermosiphon circuit using the full depth of the shaft to cool through the natural pressure differential, the brine circuit used to condense the coolant in the primary circuit, the chiller circuit that cools the brine, and the water circuit that provides cool water from the ATLAS cooling towers to the condenser of the chiller first stage. }
\label{fig:ID_Thermo_Circuit}
\end{figure}

From 2008 to 2017, the ATLAS \gls{ID} was cooled using an oil-free compressor plant. This system requires maintenance every year and has a suction pressure limitation of \SI{1}{\bar}. It was observed that below this limit the failure rate of one of the compressors increased drastically. In parallel, between 2012 and 2016, a thermosiphon system was developed using the height difference between the surface and the experimental cavern (see Figure~\ref{fig:ID_Thermo_Layout} for the general layout), with the advantage of having all active elements on the surface and therefore easily accessible for maintenance. This system is composed of four major units (see Figure\,\ref{fig:ID_Thermo_Circuit} for the full circuit schematic):
 
\begin{description}
\item[Water circuit] providing cold water to the condenser of
the chiller first stage, sourced from the ATLAS
cooling towers at approximately \SI{25}{\degreeCelsius}.
\item[Chiller circuit] using a two-stage vapour compression cycle
to cool down a brine (${\text{C}}_6{\text{F}}_{14}$) to
\SI{-70}{\degreeCelsius}; the chiller operates in a cascade of two refrigerant fluids, using
R404A at the first stage and R23 for the second one.
\item[Brine circuit] using a ${\text{C}}_6{\text{F}}_{14}$ closed
loop to condense the ${\text{C}}_3{\text{F}}_{8}$ of the primary circuit through heat exchange across the tubes of the thermosiphon condenser. This
loop is able to run at about \SI{40}{\kg\per\second} to provide
enough flux to condense the \SI{1.2}{\kg\per\second} of
${\text{C}}_3{\text{F}}_{8}$ corresponding to the \SI{60}{\kilo\watt}
of heat dissipated by the ATLAS inner detector.
\item[Thermosiphon primary circuit] condensing the
${\text{C}}_3{\text{F}}_{8}$ at the surface of ATLAS to produce a
natural hydrostatic liquid column over the \SI{92}{\metre} of height
difference from the surface to the cavern. The liquid evaporates
in the (unchanged) on-detector cooling channels and returns to the
surface as vapour by differential pressure. The thermosiphon
circuit was developed to provide the same performance as the
existing compressor plant.
\end{description}
 
The compressor plant and the thermosiphon are interconnected; since 2018 the thermosiphon is used as the main cooling system and the compressor plant as a back-up in case of failure. The two systems run with the same pressure parameters, allowing for a fully transparent swap from one to the other.


\subsection{Updated material description of the \glstext*{ID}}

Obtaining an accurate description of the material is essential to understand the performance of the detector.
The \gls{IBL} insertion during \gls{LS1} and the modifications to the layout of the cables and support structures of the existing pixel detector (\gls{nSQP} upgrade) required a new analysis of the \gls{ID} material in \RunTwo.
 
Three complementary techniques were applied to measure the material in the \gls{ID}:
\begin{itemize}
\item \textbf{photon conversion vertex reconstruction}, taking advantage of the precise theoretical understanding of the electromagnetic interaction processes.
\item \textbf{hadronic interaction vertex reconstruction}, sensitive to the material through nuclear interactions and offering much better resolution in the radial position of the vertex compared to the photon conversion. However, the description of hadronic interactions is quite complex and only phenomenologically modelled in the simulation.
\item \textbf{track-extension efficiency method}, complementary approach applicable to the full tracking acceptance to measure the nuclear interaction rate of charged hadrons through hadronic interactions by matching track segments reconstructed in the Pixel detector with tracks that are also reconstructed in the \gls{SCT} and \gls{TRT} detectors, where the unmatched Pixel segments are assumed to be associated with charged hadron that interact hadronically while traversing the region between the Pixel and \gls{SCT} detectors.
\end{itemize}
All were studied in a dedicated set of analyses using a low-luminosity $\sqrt{s}$ = \SI{13}{\TeV} \pp collision sample corresponding to around \SI{2.0}{\per\nb} collected in 2015~\cite{PERF-2015-07}.
 
While the first two methods probe the barrel region of the inner detector, in particular the new detector components installed before \RunTwo (the \beampipe, the \gls{IBL} and the supporting tubes of \gls{IPT} and \gls{IST}), the track-extension efficiency method is also sensitive in the endcap regions of  $1.0 < \abseta < 2.5$ where most of the refurbished pixel services reside.

The precision of each measurement varies depending on the detector region.
All of these approaches are used together to measure a large part of the inner detector's volume and cross-check individual measurements.
The description of the geometry model was examined in detail both in radial and longitudinal distributions of the rate of reconstructed hadronic interaction and photon conversion vertices.
 
\begin{figure}
\begin{center}
\subfloat[]{\label{sfig:IDmaterialData}\includegraphics[width=0.40\textwidth]{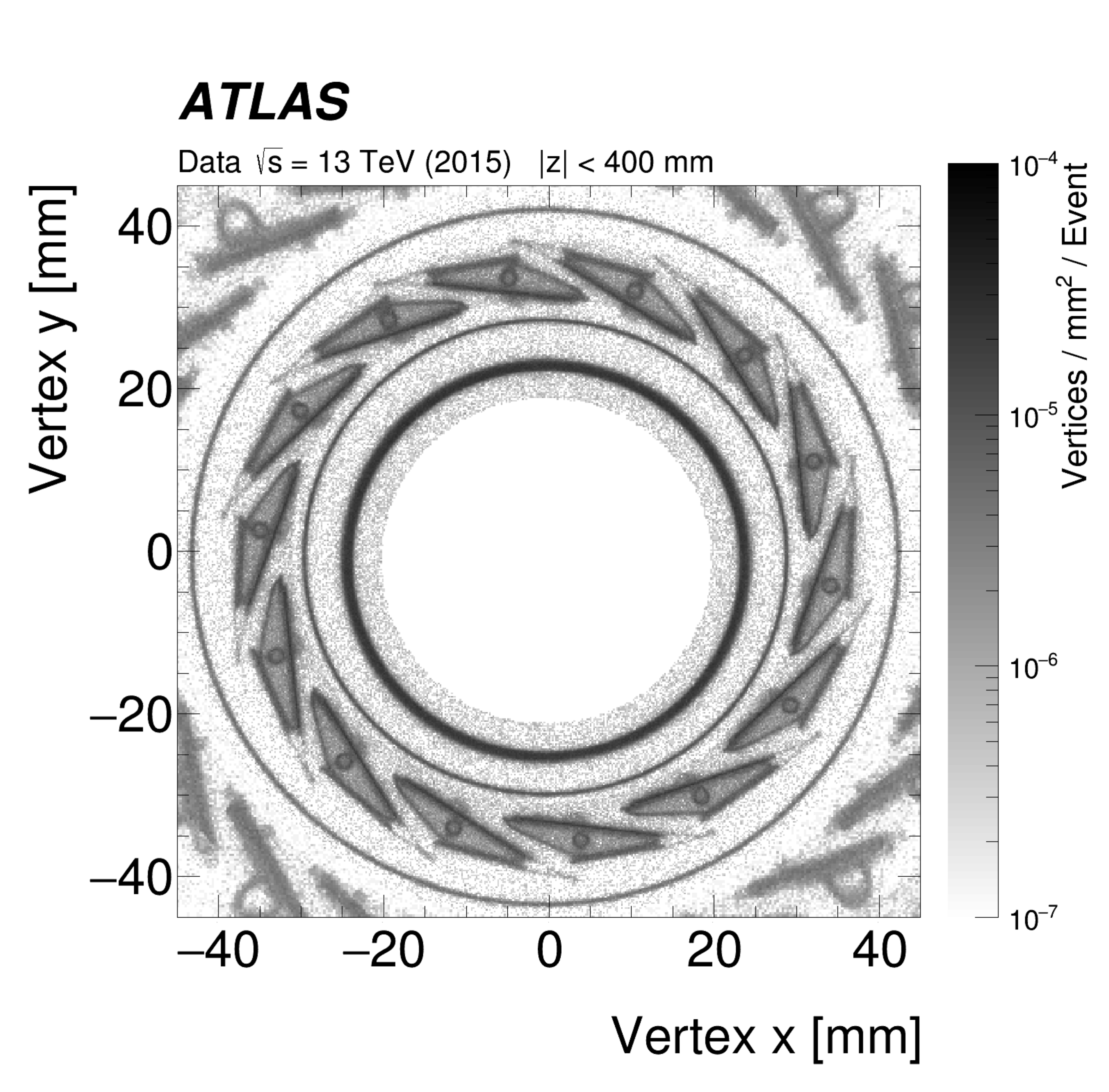} }
\subfloat[]{\label{sfig:IDmaterialMC}\includegraphics[width=0.40\textwidth]{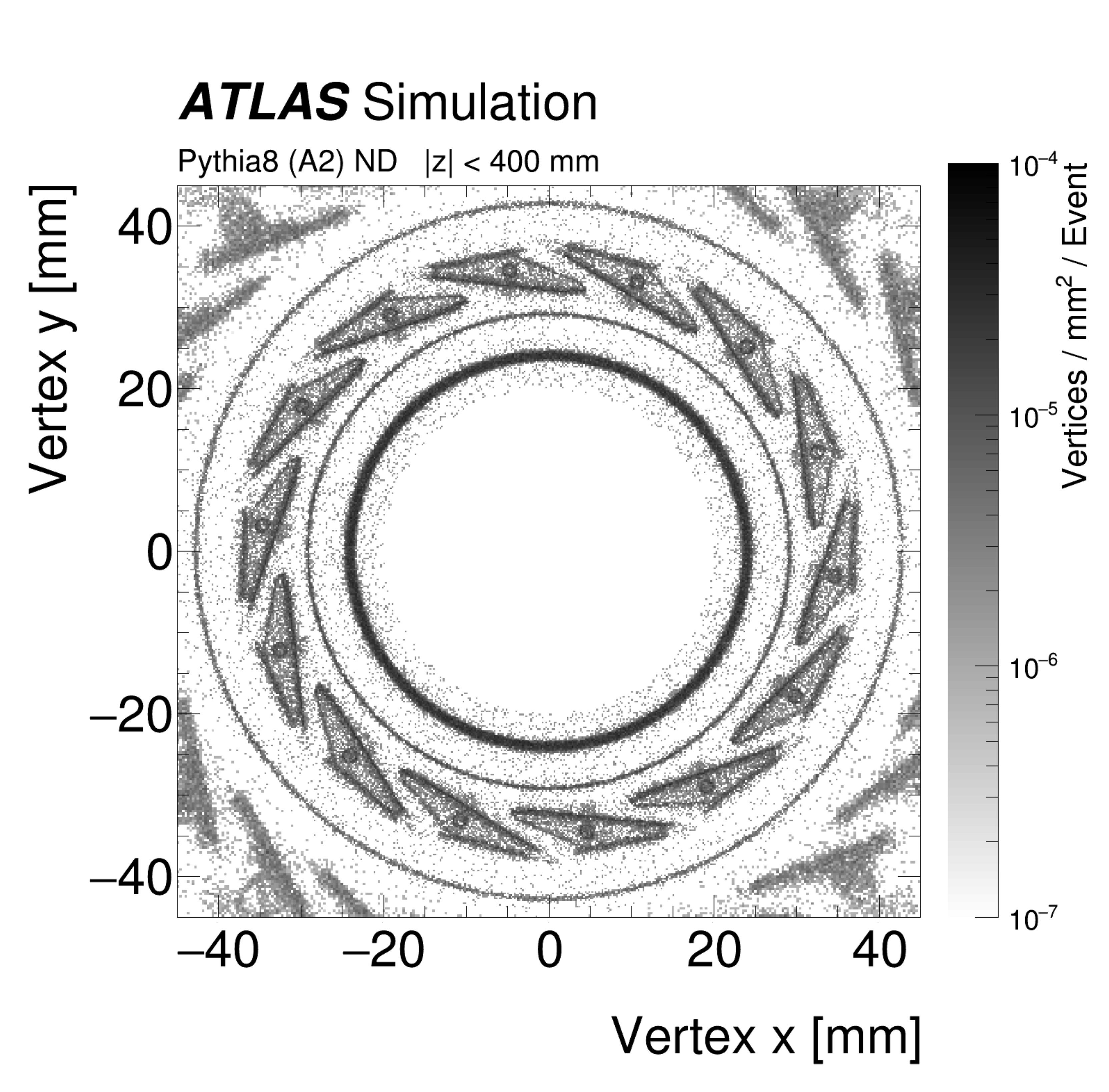} }
\end{center}
\caption{ Distribution of hadronic-interaction vertex candidates with $\abseta<2.4$ and $|z|<\SI{400}{\mm}$ \protect\subref{sfig:IDmaterialData} for data and \protect\subref{sfig:IDmaterialMC} for the \textsc{Pythia8} MC simulation with the updated geometry model~\cite{PERF-2015-07}. 
}
\label{fig:ID2Dmap}
\end{figure}

In the central barrel region, a significant amount of missing material in the \gls{IBL} front-end electronics for the flex bus, surface mounted
devices on the front-end chips and the \gls{IPT} and \gls{IST} was identified in the original geometry model that was used for ATLAS \gls{MC} simulation in 2015. Figure~\ref{fig:ID2Dmap} compares the updated geometry model in simulation with the 2015 data, showing good agreement.
 
The results of these studies have been taken into account in an improved description of the material in the ATLAS inner detector simulation, resulting in a reduction in the uncertainties associated with the charged-particle reconstruction efficiency determined from simulation.
The updated geometry model, which was created to resolve the above discrepancies, provides a much better description of the material in the ATLAS \gls{ID} simulation and is used in analyses~\cite{IDTR-2019-05}.

\begin{figure}
\begin{center}
\includegraphics[width=0.70\columnwidth]{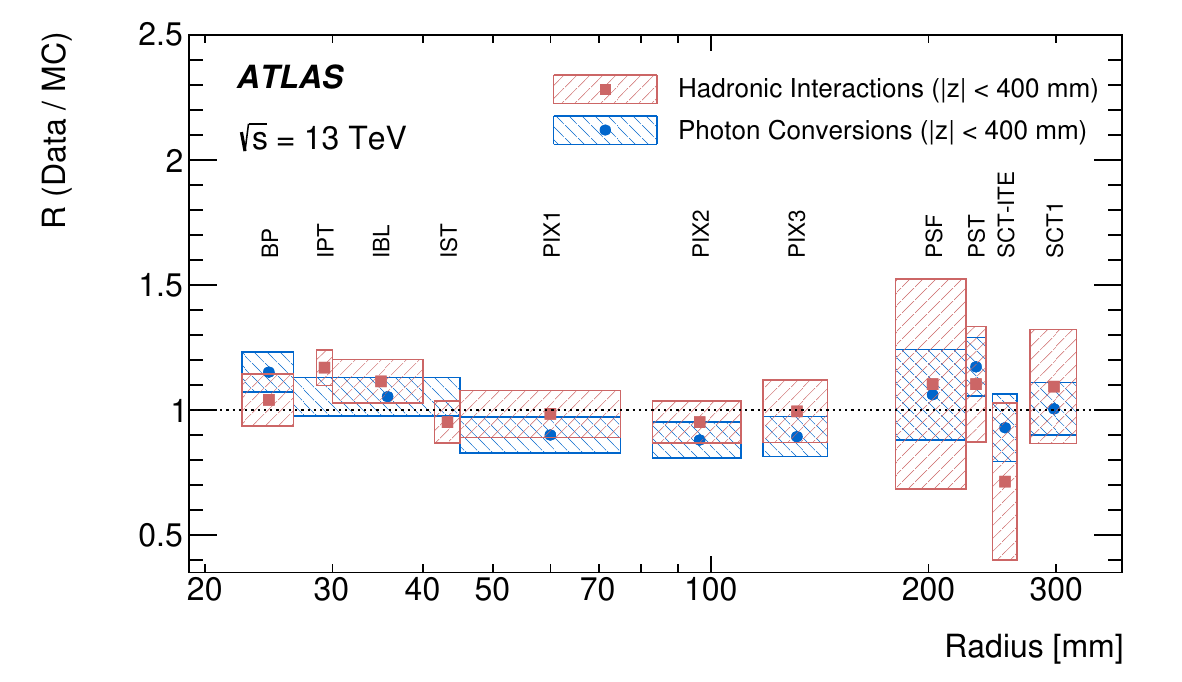}
\end{center}
\caption{Comparison of the rate ratio, denoted $R$, between data and \gls{MC} simulation, for hadronic interactions and photon conversions as a function of radius~\cite{PERF-2015-07}.  The horizontal range of each marker represents the radial range of vertices used in each measurement, while the vertical range represents the total uncertainty.
The radial regions used for comparing data to \gls{MC} simulation are the following: BP (\beampipe), \gls{IPT}, IBL (\gls{IBL} staves), \gls{IST}, PBX1-2-3 (Pixel Barrel layers 1-2-3), PSF (Pixel Support Frame), \gls{PST}, SCT-ITE (SCT Inner Thermal Enclosure), SCT1 (first
\gls{SCT} barrel layer).
}
\label{fig:hadroconv}
\end{figure}

The \beampipe is found to be very accurately described except the central region ($|z| < \SI{40}{\mm}$).
The simulated material in the \gls{IBL} within the updated geometry model is found to be consistent with that observed in data, within less than 10\%, mainly due to uncertainties of the hadronic interaction and conversion measurements (see Figure~\ref{fig:hadroconv}).
 
The Pixel barrel layers are found to be described well, and the results from the analyses using the hadronic interactions and photon conversions agree within the systematic uncertainties.
They confirm the results of the previous hadronic interaction analysis obtained with the \RunOne data set.
 
The updated geometry model provides reasonable agreement with the data in the ratio of the rate measurements of hadronic interactions and photon conversions within the uncertainties of the measurements. The measured rates of photon conversions and hadronic interactions reconstructed in data are found to agree to within 7\% -- 18\% with those predicted by simulation, based on the updated geometry model, out to the outer envelope of the Pixel detector. This is also supported by a study of the transverse impact parameter resolution below $\pT=\SI{1}{\GeV}$, where multiple scattering is dominant (see also Ref.~\cite{PERF-2015-07}).

In the forward region, the material in the pixel service region is found to be underestimated in the geometry model by up to $\delta N_{\lambda j} = (3.7 \pm 0.9)\%$ at some values of $\eta$. This corresponds to roughly 10\% of the material in the pixel services in the corresponding regions.
Furthermore, in the very forward region ($3.1 < | \eta | < 5 $), outside of the tracking acceptance and corresponding to the \gls{FCAL} acceptance, an extra contribution of material was identified.
The mismatch concerns the \gls{IBL} Type-1 low voltage cable; the original idea to use copper-clad aluminium wires was dropped in favour of a more robust option of all-copper wires, offering better connectability and routing flexibility, and very similar resistance, but higher material density.


 
\subsection{Performance of the \glstext*{ID} at the end of \RunTwo and projections for \RunThr}
The performance of the Pixel and \gls{IBL} detectors during \RunTwo (discussed in Section~\ref{sss:pixIblPerf}) is indicative of the performance expected during \RunThr, as these detectors were designed for high instantaneous luminosity and pileup conditions. The \gls{SCT} and \gls{TRT} required more substantial changes to cope with sustained periods of peak instantaneous luminosity; the expected impact of these changes is discussed in Sections~\ref{sss:sctPerf} and~\ref{sss:trtPerf}.
 
\subsubsection{Pixel and \glstext*{IBL} detector performance}
\label{sss:pixIblPerf}
\begin{figure}[ht!]
\begin{center}
\subfloat[]{\label{figID:transpT}\includegraphics[width=.40\textwidth]{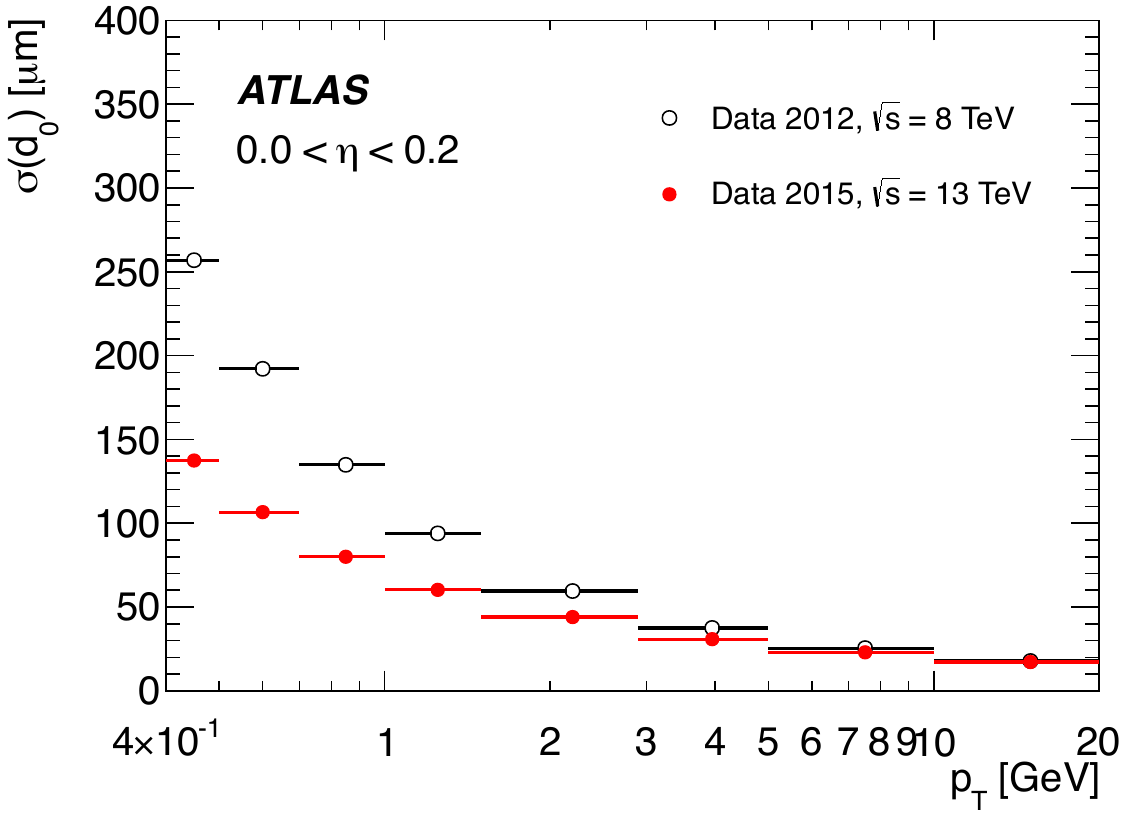}}
\subfloat[]{\label{figID:longpT}\includegraphics[width=.40\textwidth]{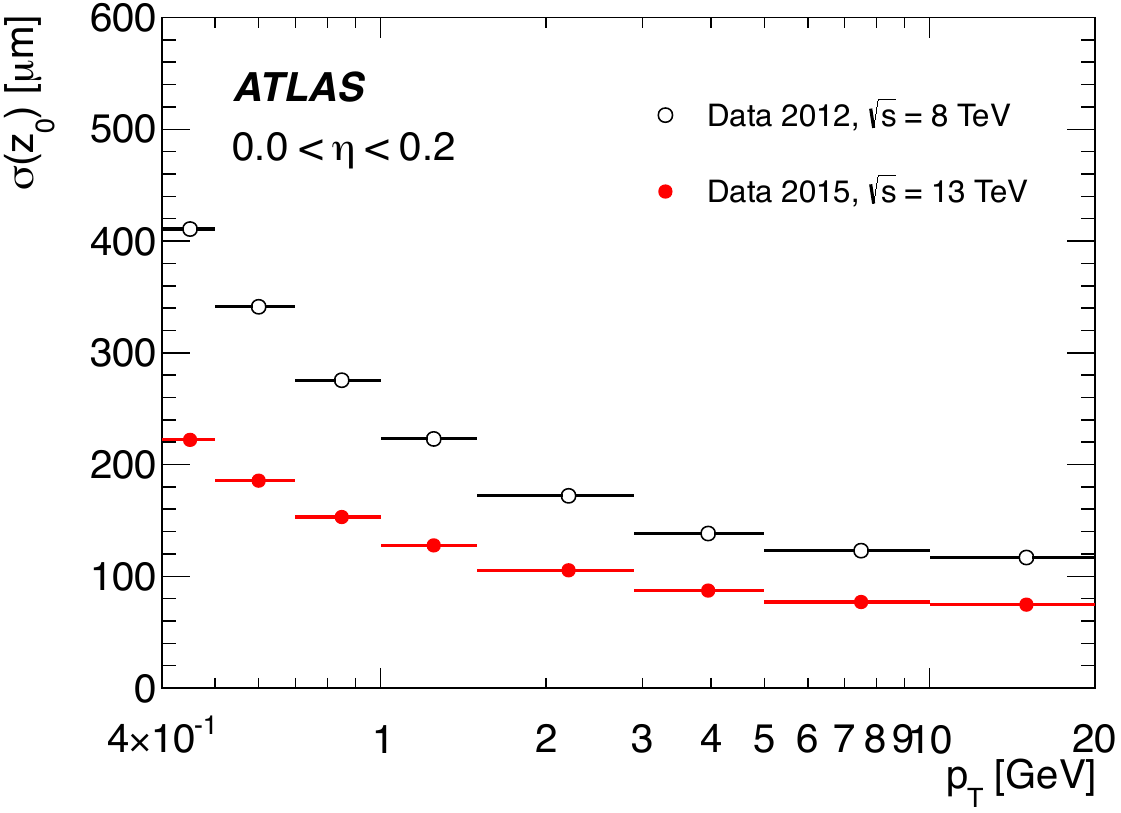}}\\
\subfloat[]{\label{figID:transEta}\includegraphics[width=.40\textwidth]{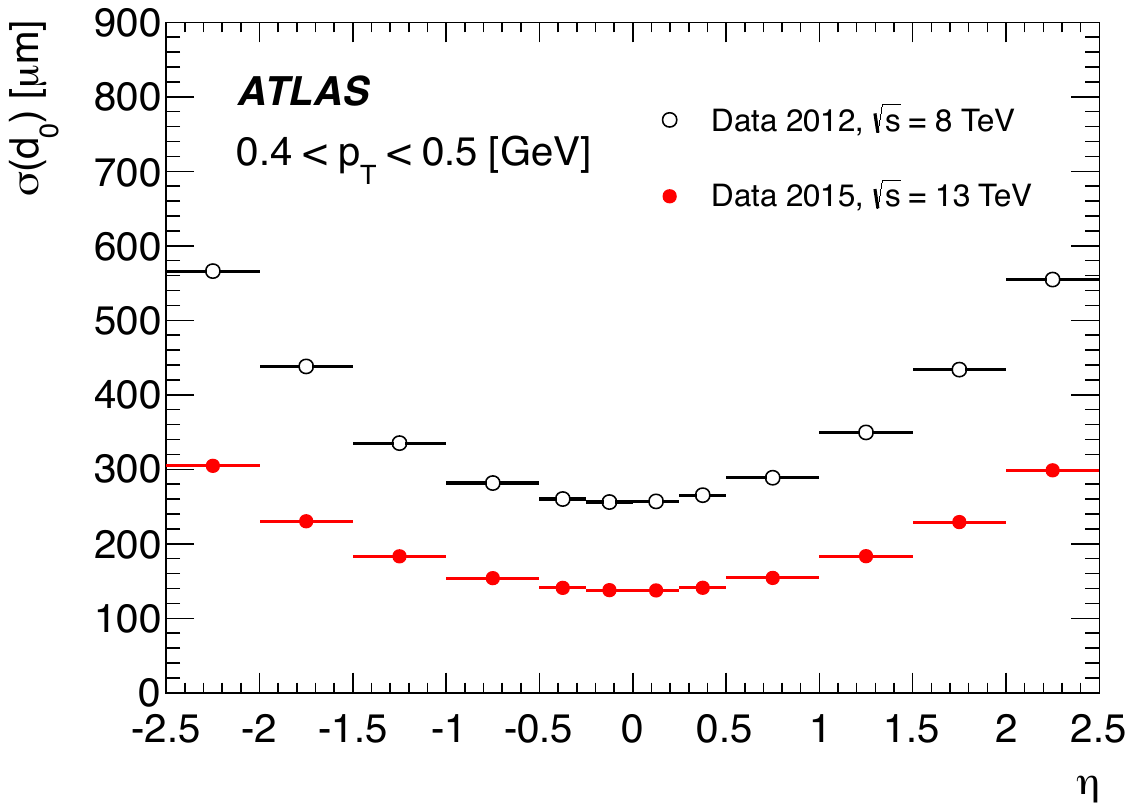}}
\subfloat[]{\label{figID:longEta}\includegraphics[width=.40\textwidth]{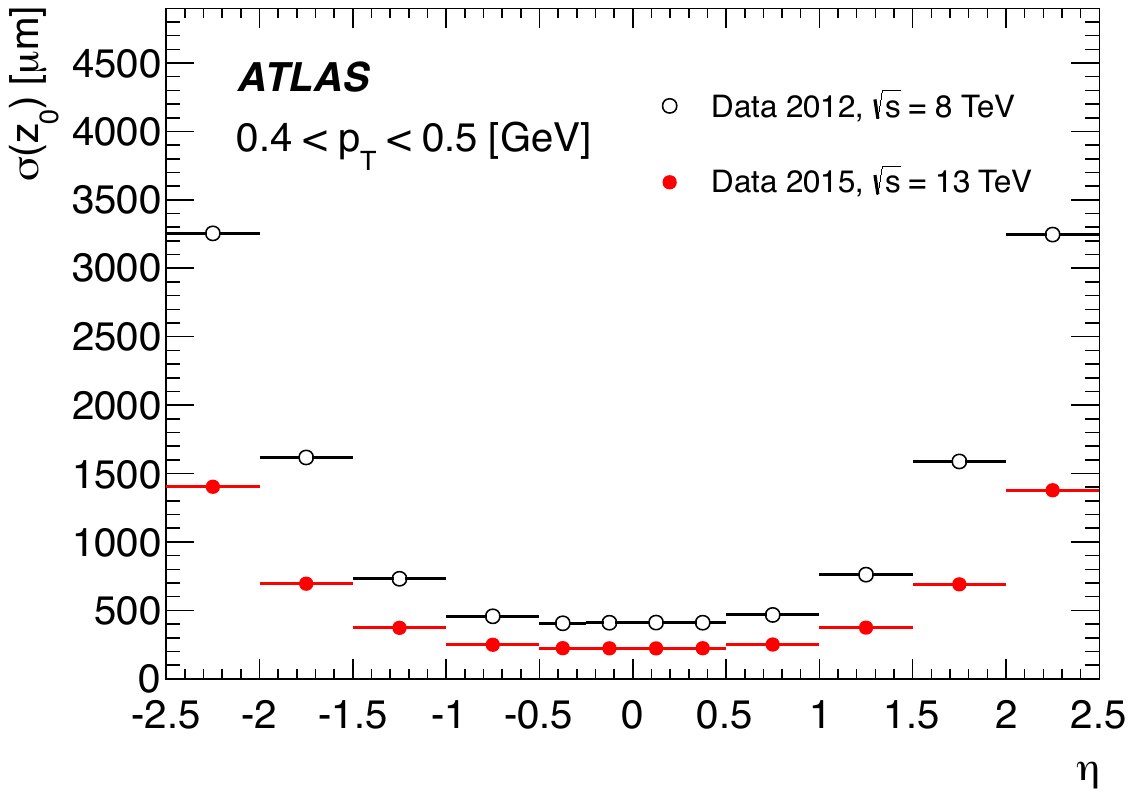}}
\caption{Comparison between 2012 \RunOne and 2015 \RunTwo detector
configuration for
(left) the transverse impact parameter as functions of \protect\subref{figID:transpT} \pT\ and \protect\subref{figID:transEta} $\eta$, and (right) the longitudinal impact parameter as functions of \protect\subref{figID:longpT} \pT\ and \protect\subref{figID:longEta} $\eta$.
}
\label{fig:tracking}
\end{center}
\end{figure}

The inclusion of the \gls{IBL}, adding a new point measurement with higher precision at a radius just outside the \beampipe, has significantly improved the tracking performance over the whole acceptance, and this improvement is seen most clearly in the impact parameter resolution.
Figure~\ref{fig:tracking} shows this benefit in terms of impact parameter resolution by comparing two measurements
performed in early \RunTwo and at the end of \RunOne.
 
The global enhancement, which is in line with expectations, can be explained in terms of proximity to the \gls{IP} of the innermost layer (which drives the performance at low track $\pT$) and smaller pixel pitch in the $z$ direction.
Furthermore, these results demonstrate the robustness of the in-run alignment procedure that was put in place in order to mitigate the distortion of the \gls{IBL} described in Section~\ref{sec:IBL-Integration}.
 
\begin{figure}[ht!]
\begin{center}
\includegraphics[width=0.70\columnwidth]{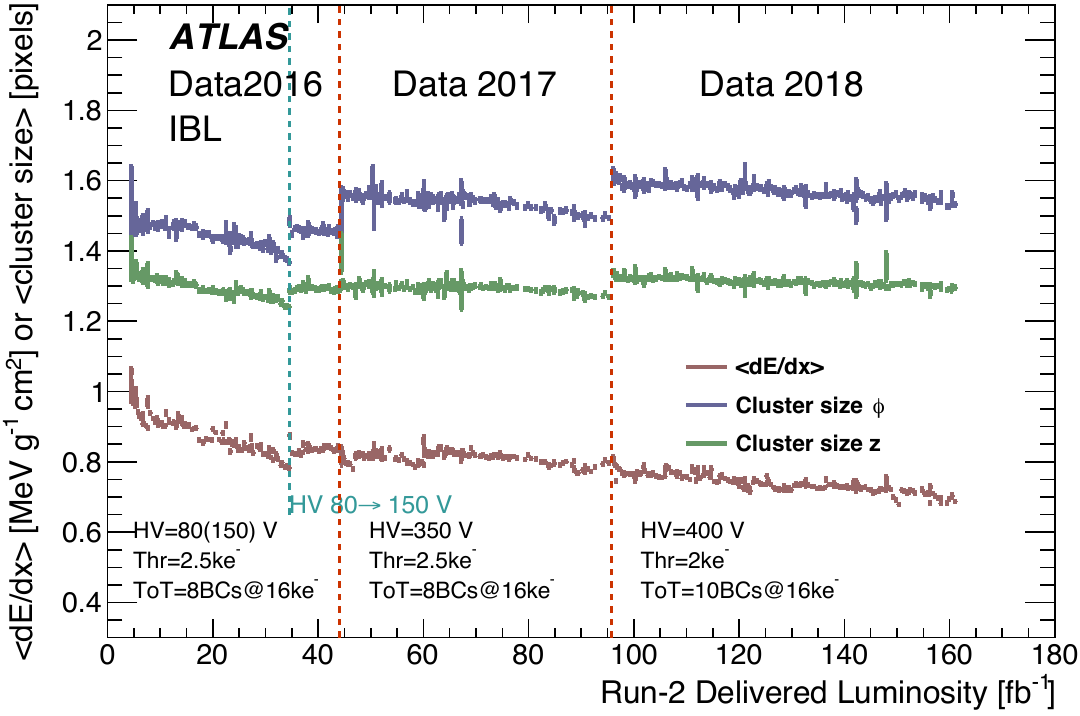}
\end{center}
\caption{The dependence on the delivered luminosity of the average cluster size and the ${\text{d}}E/{\text{d}}x$ measured with the \gls{IBL}. Each point represents a single run, and only runs recorded in 2016, 2017 and 2018 are shown. Clusters are selected which match exactly one reconstructed charged track with $\pT>\SI{10}{\GeV}$ and $\abseta<1.4$, associated to jets with $\pT>\SI{200}{\GeV}$ by $0.1<\Delta R\text{(track,jet)}<0.4$. The lower cut is to reduce contamination from two particle clusters. The impact of changing the bias voltage in the \gls{IBL} is clearly visible. The gradual decrease of the measured ${\text{d}}E/{\text{d}}x$ is due to the reduced charge collection fraction due to radiation damage. Red dotted lines indicate the different data-taking years.
\label{fig:IBLdEdx}}
\end{figure}

During \RunTwo, the pixel detector sensors accumulated a significant fluence that reached
\phieqv $\sim \SI{e15}{\per\cm\squared}$
at the location of the \gls{IBL}, as predicted by the simulations shown in Table~\ref{tab:radValuesID} for the \SI{147}{\ifb} of \pp collision data taken during \RunTwo.
The resulting radiation damage required a careful tuning of the detector operating parameters in order to minimise its impact on the pixel performance in the reconstruction of charged particle tracks. The observed decrease of the charge collection has been countered with the increase of the bias voltage; this can be seen in Figure~\ref{fig:IBLdEdx}, where the cluster size and the ${\text{d}}E/{\text{d}}x$ decreasing trends for increasing integrated luminosity were mitigated during the run.
 
\begin{figure}[ht!]
\begin{center}
\includegraphics[width=0.70\columnwidth]{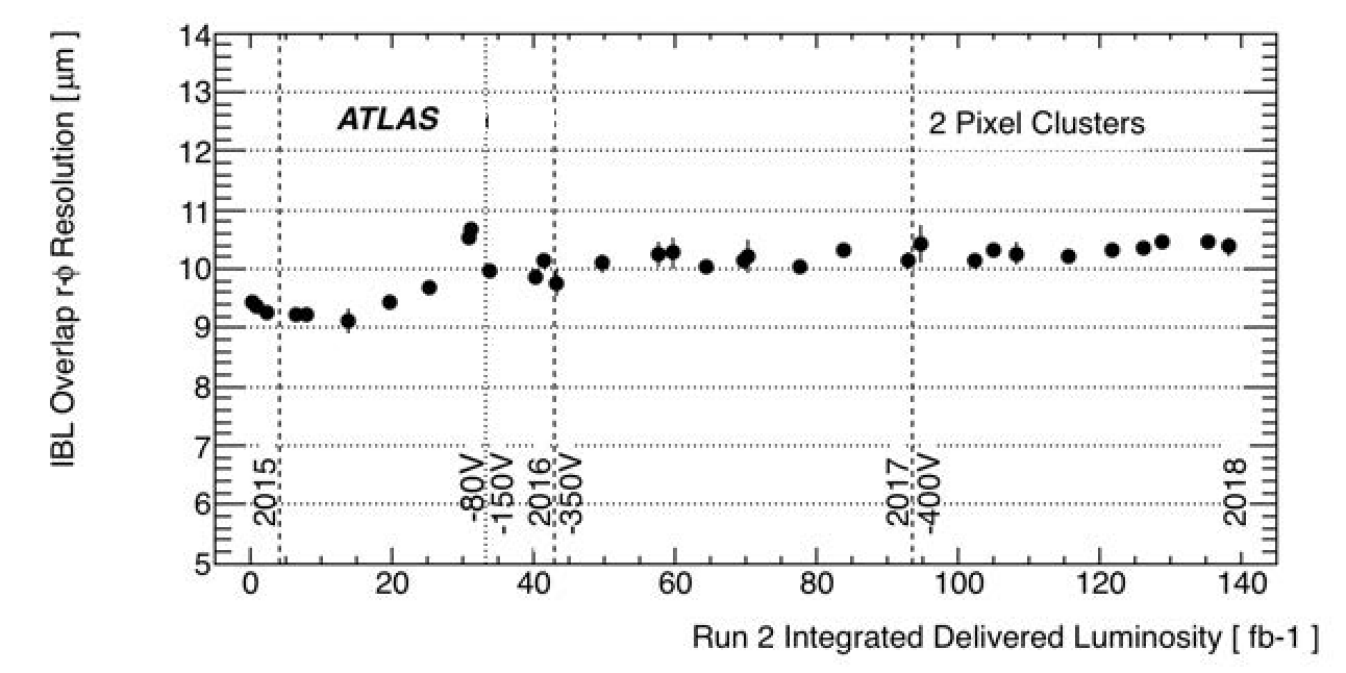}
\end{center}
\caption{\gls{IBL} spatial resolution in the transverse coordinate (small pixel pitch) as a function of the integrated luminosity in \RunTwo. The resolution is determined from the corrected residuals of pairs of reconstructed \gls{IBL} clusters associated to tracks from charged particles traversing the detector in the region of the module overlaps. Only clusters with two pixels are considered here to remove the dependence of the resolution on the cluster shape. The dashed lines indicate the different data-taking years and the dotted lines the change of depletion voltage during \gls{LHC} operation. \label{fig:IBLResX}}
\end{figure}

The effect of this countermeasure of increasing the bias voltage can be observed in Figure~\ref{fig:IBLResX}, where the \gls{IBL} resolution in the transverse coordinate ($r\phi$) shows a nearly constant behaviour as a function of the integrated luminosity.
 
\begin{figure}[ht!]
\begin{center}
\includegraphics[width=0.50\columnwidth]{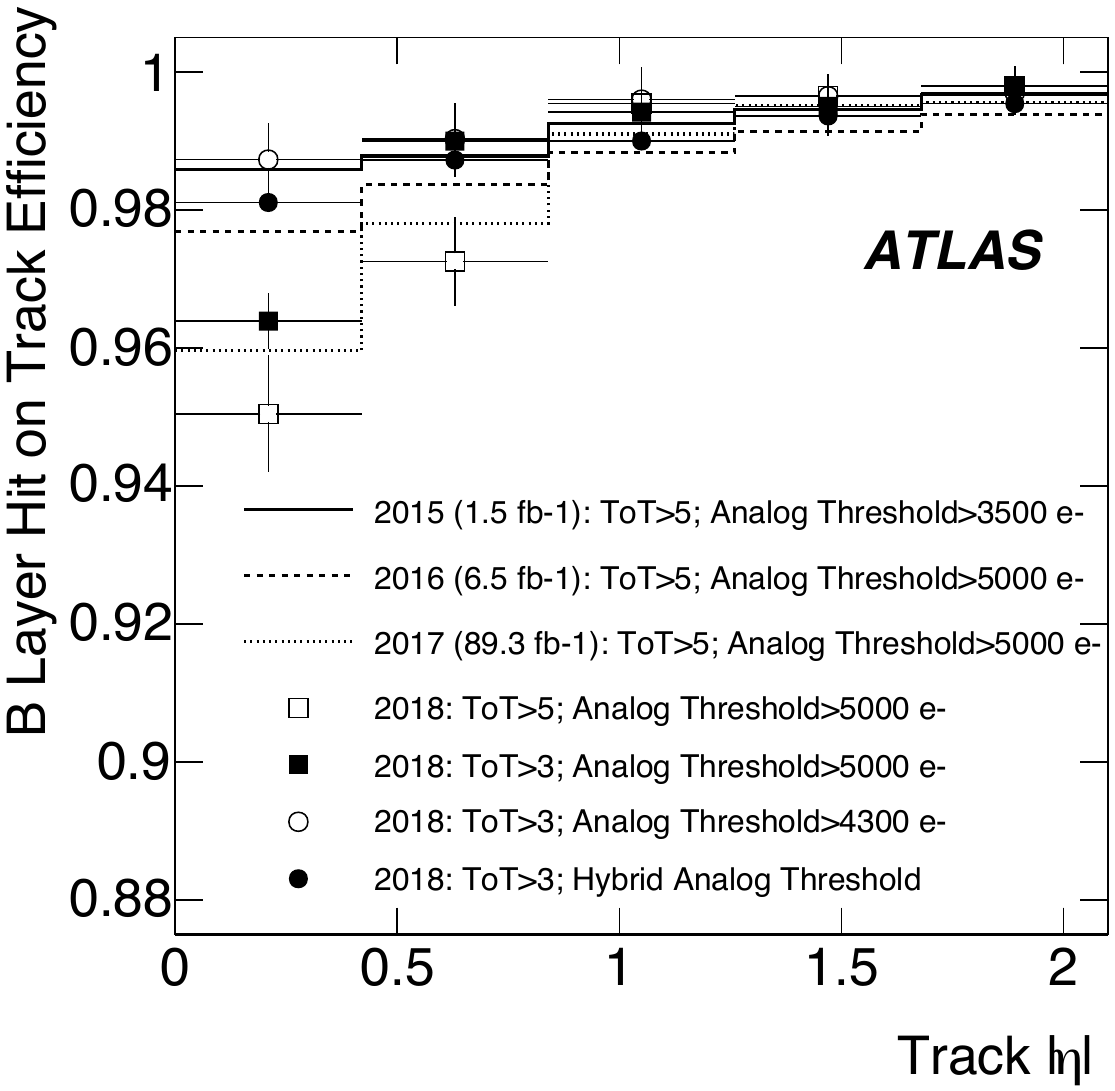}
\end{center}
\caption{Efficiency for $B$-Layer clusters associated to a reconstructed particle track as a function of the track $|\eta|$ for different years in \RunTwo and different threshold settings at the beginning of 2018. The use of a hybrid setting with lower thresholds for the central modules exposed to higher particle fluence and higher thresholds for forward modules affected by larger occupancy recovers the efficiency obtained in operations at the beginning of \RunTwo.
\label{fig:BLEff}}
\end{figure}

In the $B$-Layer, the decrease of the efficiency has been minimised by suitable combinations of \analog (applied on the charge that is deposited in each pixel by its individual discriminator) and digital (or \gls{ToT}) thresholds (see Figure~\ref{fig:BLEff}). Different threshold schemes along the $\eta$ coordinate have been applied to take into account the uneven distribution of the radiation fluence and readout bandwidth occupation, leading to a hybrid threshold scenario. Despite a decrease of the collected charge of up to 35\% on the central $B$-Layer modules, these mitigation actions resulted in no change of the pixel efficiency by the end of \RunTwo (see Figure~\ref{fig:BLEff}).
 
A similar strategy is foreseen for \RunThr.  The planned operating conditions in \RunThr follow from those adopted towards the end of \RunTwo. The bias voltage will be set at up to \SI{500}{\volt} for the \gls{IBL} planar sensors, at up to \SI{100}{\volt} for the \gls{IBL} 3D sensors and at up to \SI{600}{\volt} for the other Pixel sensors. At these large values of the depletion voltage the Lorentz angle is small (from \SI{250}{\mrad} at \SI{80}{\volt} at the beginning of \RunTwo to \SI{100}{\mrad} at \SI{400}{\volt} in 2018 on the \gls{IBL}) but charge sharing between neighbouring pixels is still ensured by the non-zero angle of incidence of particle tracks. This ensures stable spatial resolution. The tuning to lower and, if needed, $\eta$-dependent thresholds, successfully adopted for the $B$-layer at the beginning of 2018, may also be extended to the \gls{IBL} if the response of the detector efficiency along $z$ does not otherwise remain uniform.
 
\subsubsection{\glstext*{SCT} tracking performance}
\label{sss:sctPerf}
In \RunTwo the \gls{LHC} delivered high instantaneous luminosity and pileup conditions that were far in excess of the original \gls{SCT} design goals. The \gls{SCT} \gls{DAQ} system had to be re-optimised to mitigate bandwidth limitations as described in Section~\ref{sec:ID-SCT}.
As a result of these changes, the \gls{SCT} can now provide efficient tracking with a pileup $\mu$ of up to 70 \pp interactions per \gls{BC} and a \gls{L1} trigger rate up to \SI{100}{\kHz}. 
During \RunTwo, the first significant operational impacts arising from radiation damage to the sensors and to the on-detector electronics in the \gls{SCT} were observed. All \gls{SCT} $p^+$-on-$n$ silicon sensors underwent type inversion, followed by a continuous
increase in depletion voltage. Consequently, higher operating voltages are being applied progressively to ensure full depletion of the sensors and to maintain hit efficiency~\cite{SCTD-2019-01}.
 
\begin{figure}[p]
\begin{center}
\includegraphics[width=\columnwidth]{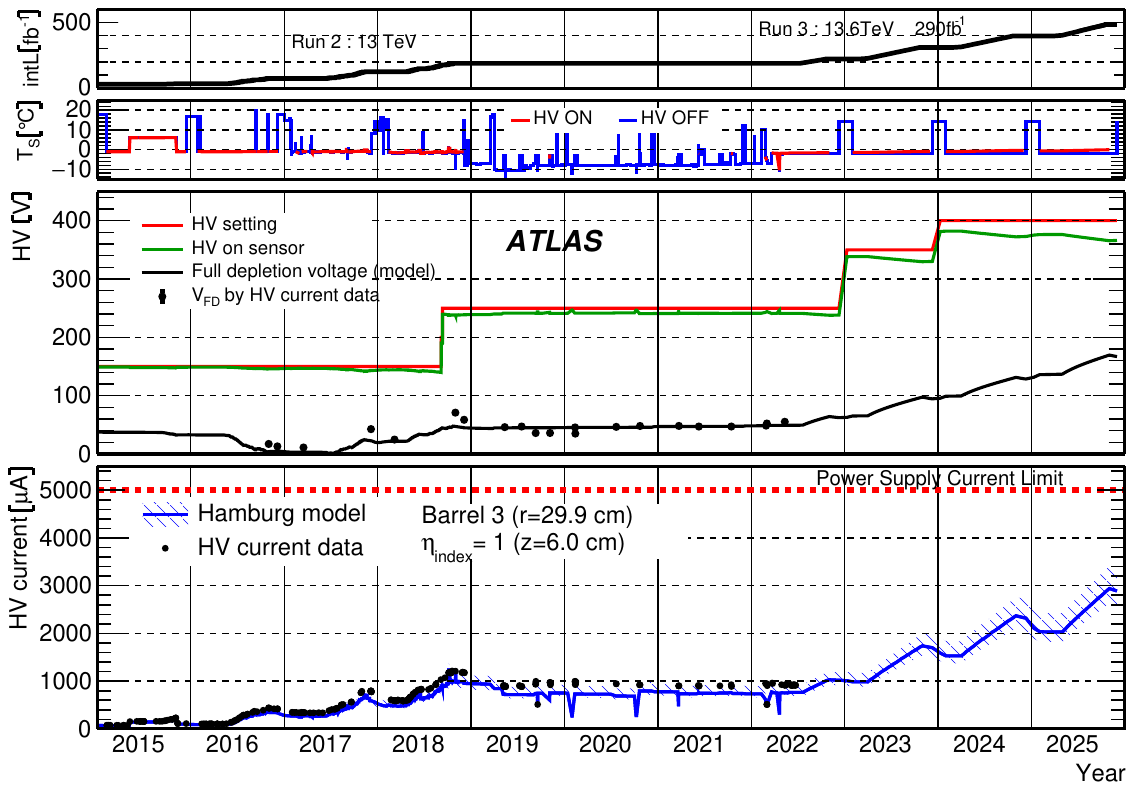}
\caption{Projection of the module \gls{HV} for the central Barrel~3 modules from 2015 to the end of 2025. Top plot: The integrated luminosity vs.\ time; in \RunThr\ the integrated luminosity is assumed to be \SI{33}{\ifb} in 2022 and \SI{86}{\ifb} per year in 2023-2025 at \SI{13.6}{\TeV}.
The second plot shows the expected time profile of the sensor temperature with the \gls{HV} on (red) and off (blue). The red line indicates the expected temperature rise due to bulk-heating by leakage current. The third plot shows the \gls{HV} setting (red), the actual \gls{HV} values on the sensor (green), and the full depletion voltage (black) estimated by the Hamburg model. The black points show measurements obtained from I-V scans, which show a kink that is an indicator for the full depletion voltage. The bottom plot shows the evolution of leakage current using the Hamburg model (blue line) compared to data (black points). Hashed areas indicated model uncertainties which do not include (unknown) errors of the luminosity to fluence conversion factors.}
\label{SCT001}
\end{center}
\end{figure}

\begin{figure}[htbp]
\centering
\subfloat[]{\includegraphics[width=.50\textwidth]{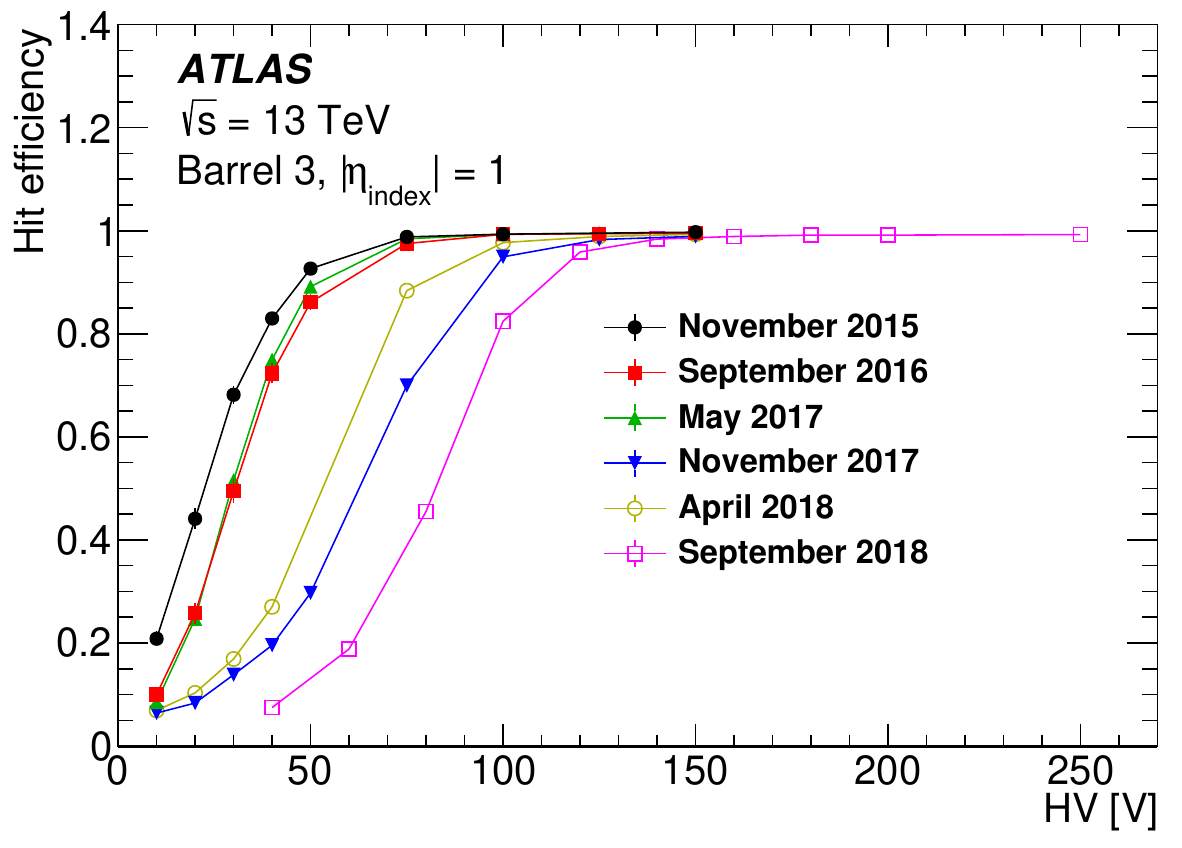} \label{fig:SCT002}}
\subfloat[]{\includegraphics[width=.50\textwidth]{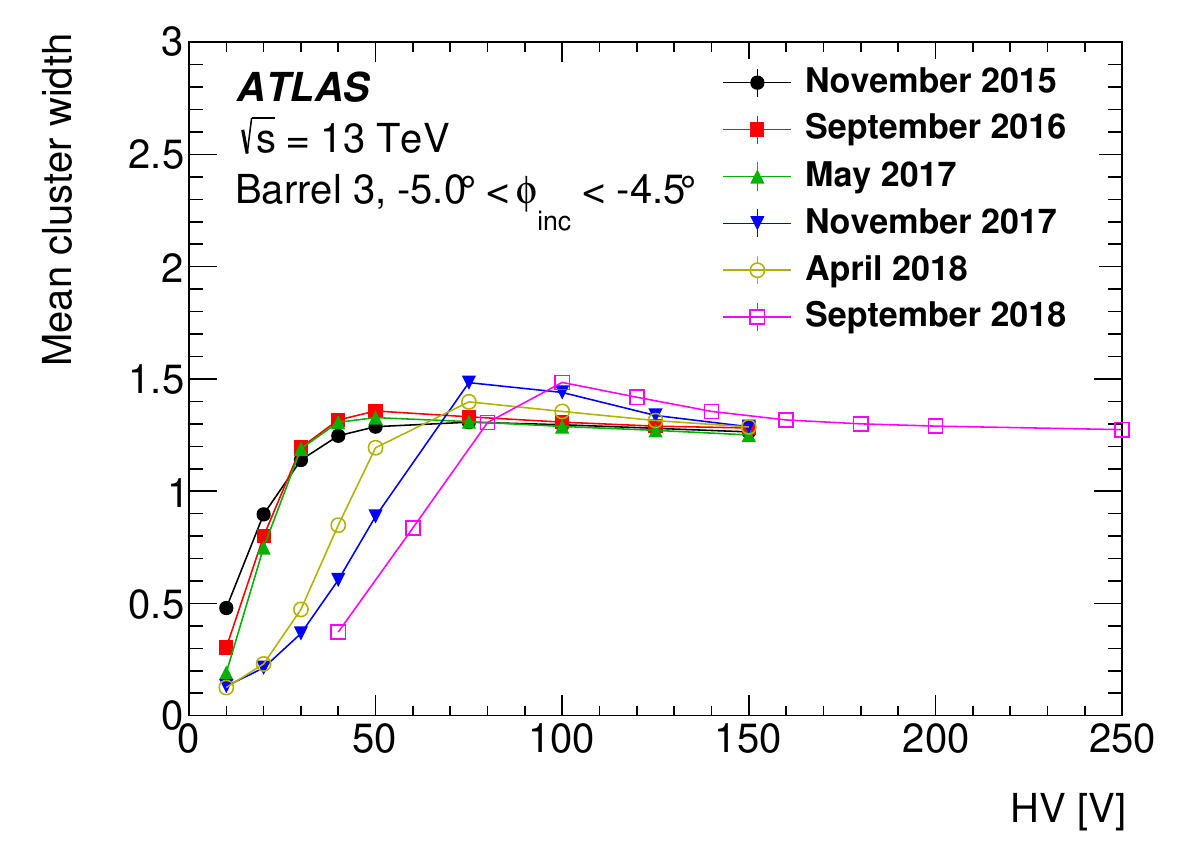} \label{fig:SCT003}}\\
 
\caption{\protect\subref{fig:SCT002} Hit efficiency as a function of \gls{HV} for central-most (designated $|\eta_\text{index}|=1$) modules in \gls{SCT} Barrel~3, measured from November 2015 to September 2018~\cite{SCTD-2019-01}. \protect\subref{fig:SCT003} Evolution of mean cluster width as a function of \gls{HV} for Barrel~3 during \RunTwo~\cite{SCTD-2019-01}. }
\end{figure}

\begin{figure}[htbp]
\begin{center}
 
\includegraphics[width=.60\columnwidth]{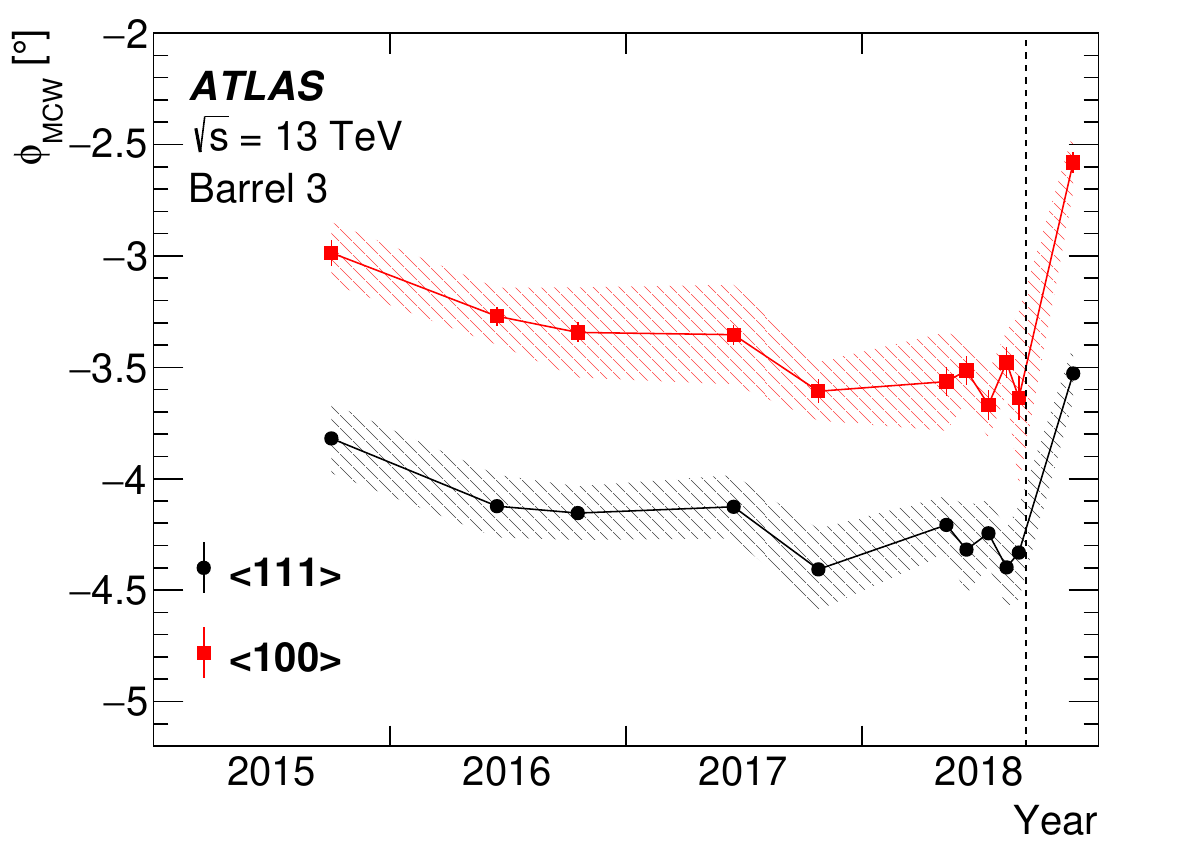}
 
\caption{Time dependence of $\phi_\text{MCW}$ between 2015 and 2018 in Barrel~3. Only the sides with no stereo angle are shown. The error bars are statistical, while the shaded bands show the systematic uncertainties. The nominal operational \gls{HV} was raised from \SIrange{150}{250}{\volt} at the time in 2018 indicated by the vertical dashed line~\cite{SCTD-2019-01}. $\langle111\rangle$ and $\langle100\rangle$ are Miller indices, indicating the crystal lattice orientation of the silicon wafers.
}
\label{SCT004}
\end{center}
\end{figure}

The maximum $\mu$ in \RunThr will be regulated by $\beta$ levelling so as not to exceed the values seen at the end of \RunTwo and the maximum trigger rate is expected to be \SI{100}{\kHz}~\cite{SCTD-2019-01}. Given the optimisation to address the \RunTwo conditions, the \gls{SCT} \gls{DAQ} is expected to operate smoothly in \RunThr.
Radiation damage in \gls{SCT} will continue to evolve in \RunThr, primarily in the form of increases of the full depletion voltage ($V_{FD}$) and the leakage current in the sensor. Figure~\ref{SCT001} shows $V_{FD}$ and the leakage current observed in \RunTwo for \gls{SCT} Barrel Layer~3 (the innermost \gls{SCT} layer, henceforth abbreviated to Barrel~3). The data points in the figure match well with the predictions of the Hamburg model~\cite{hamburgmodel}. The projection for $V_{FD}$ and leakage current has been extended for \RunThr with the proposed luminosity delivery plan, showing the anticipated increase of $V_{FD}$ and leakage current. Figure~\ref{fig:SCT002} shows measured hit efficiencies as a function of \gls{HV} during \RunTwo for modules close to $z=0$ on Barrel~3 (designated $|\eta_\text{index}|=1$). As this position corresponds to the maximum level of accumulated radiation, it required higher \gls{HV} to maintain hit efficiency. Figure~\ref{fig:SCT003} shows the mean cluster width as a function of \gls{HV} for tracks passing through Barrel~3 modules for  $\ang{-5.0}<\phi_{inc}<\ang{-4.5}$, where $\phi_{inc}$ is the incident angle with respect to the sensor surface. The required \gls{HV} to obtain a good cluster width increased toward the end of \RunTwo. Figure~\ref{SCT004} shows the time evolution of the incident angle with minimum cluster width, $\phi_\text{MCW}$ (a good estimator for the Lorentz angle)~\cite{SCTD-2019-01}, in Barrel~3 modules: the change of the electrical field in the sensor due to the increase of $V_{FD}$ affected the measurement. The big jumps in 2018 are due to increases of \gls{HV} from \SI{150}{\V} to \SI{200}{\V}. Those measurements will be repeated periodically and if necessary \gls{HV} will continue to be increased to maintain optimal tracking conditions.
 
\subsubsection{\glstext*{TRT} tracking performance}
\label{sss:trtPerf}
\begin{figure}[htbp]
\centering
\subfloat[Position accuracy]{\includegraphics[width=.65\textwidth]{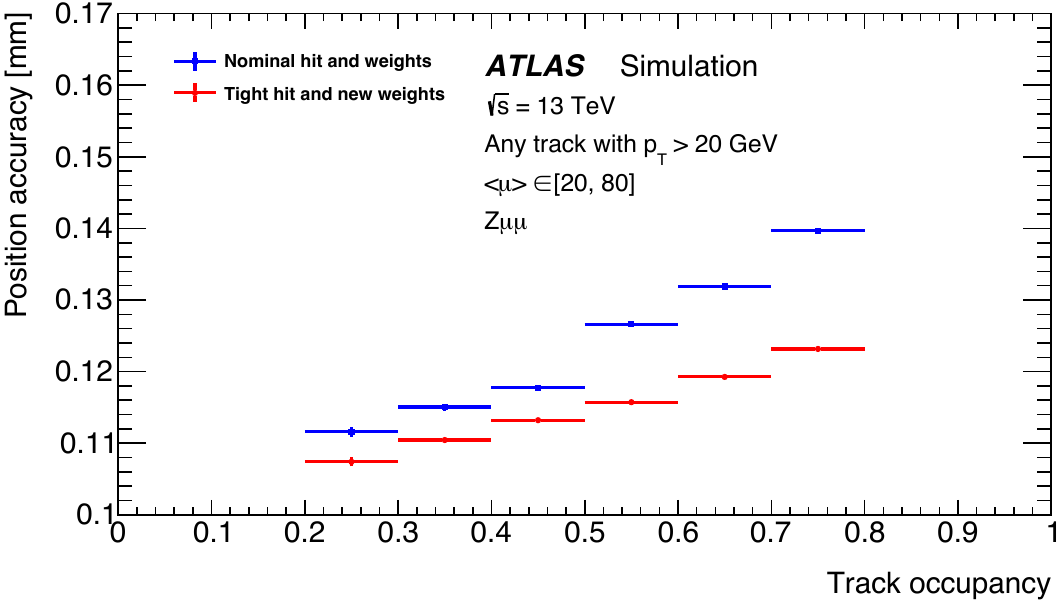}\label{fig:hitResoPosition}}\\
\subfloat[\pt resolution]{\includegraphics[width=.65\textwidth]{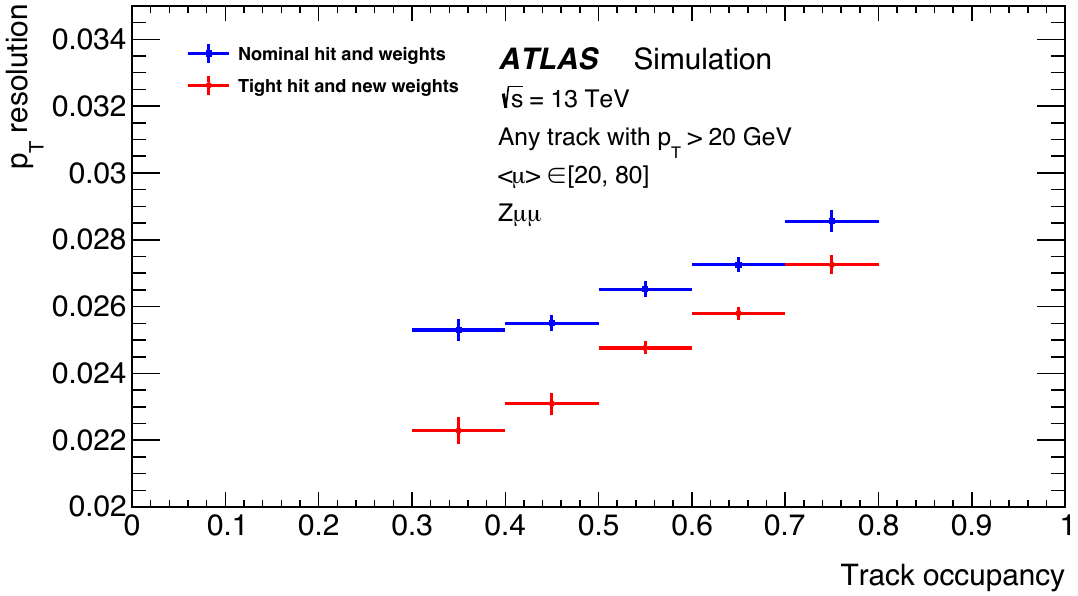}\label{fig:hitResoPT}}
\caption{
\protect\subref{fig:hitResoPosition} The \gls{TRT} hit position measurement accuracy and \protect\subref{fig:hitResoPT} the relative transverse momentum resolution of the \gls{ID} for tracks with $\pt > \SI{20}{\GeV}$ in simulated $\Zmm$ events as a function of \gls{TRT} track occupancy for both the \RunTwo hit weights (``nominal'') and the expected \RunThr \gls{TRT} hit weights (``new'').
}
\label{fig:hitReso}
\end{figure}

The expected \gls{LHC} operation mode in \RunThr is to use luminosity levelling at \lumirunthree. Under these conditions the average hit occupancy of the \gls{TRT} straws will be significantly higher than in previous runs. The track occupancy in the \gls{TRT}, defined as the hit occupancy in straws in the path of a track of interest, is expected to be up to 0.75 in \RunThr, while it was typically less than 0.5 during \RunTwo. In order to ensure optimal \gls{TRT} performance, some modifications to the \gls{TRT} reconstruction software were implemented. For the \gls{TRT} track reconstruction these modifications have been implemented, and include a tighter requirement for hits to have a significant weight in the track fit. Figure~\ref{fig:hitReso} shows the resulting hit position measurement accuracy in the \gls{TRT} straws and the relative transverse momentum resolution in the ATLAS \gls{ID} for tracks with $\pt > \SI{20}{\GeV}$ as a function of the \gls{TRT} track occupancy for simulated $\Zmm$ events. The dependencies in Figure~\ref{fig:hitReso} are shown for both the \gls{TRT} hit weights used in \RunTwo and for recalculated weights intended for use during \RunThr. The new hit weights preserve excellent track reconstruction up to the highest occupancy expected in \RunThr.
 
\clearpage
\newpage
 
\section{Calorimeters} 
\label{sec:Calorimeters}

ATLAS uses two sampling calorimeter technologies: Liquid Argon~\cite{ATLAS-TDR-02} for the electromagnetic calorimeters and all the endcap and forward calorimeters, and scintillating Tiles~\cite{ATLAS-TDR-03} for hadron calorimetry in the central region.
The ATLAS calorimeters~\cite{LARG-2009-01} were designed to last for the entire lifetime of the \gls{LHC}, and the detectors themselves require very few modifications to run at higher luminosity.
This section contains a discussion of the changes in the Liquid Argon (Section~\ref{sec:LAr}) and Tile (Section~\ref{sec:tile}) Calorimeters for \RunThr.
 
\subsection{Liquid Argon Calorimeters}
\label{sec:LAr}
The \glsfirst{LAr} system consists of several subsystems, namely the
\glsfirst{EMB}, the \glsfirst{EMEC}, the \glsfirst{HEC},
and the \glsfirst{FCAL}.
Using the full granularity of the calorimeters, read out with precision readout electronics,
the \gls{LAr} Calorimeter system
measures the energy
of electrons, photons, $\tau$ leptons, and jets
as they are slowed by the dense calorimeter material,
and contributes to the identification of these physics objects.
The system
also participates in the calculation of missing transverse energy (\met). In addition, it provides
lower-granularity information to the \gls{L1} calorimeter trigger system in order to select
events potentially containing electrons, photons, $\tau$ leptons, jets or \met.
 
During \RunOneTwo, these
signals to the trigger system for most of the calorimeter
consisted of    $\Delta\eta\times\Delta\phi=0.1\times 0.1$  \glspl{TT}, groups of elementary calorimeter cells
for which the readout signals are \analog sums of the signals in the longitudinal layers of the calorimeters.
For \RunThr and beyond, a new digital trigger readout path was implemented~\cite{ATLAS-TDR-22,LArPhaseIPaper}, increasing the granularity
of the summed signals by up to a factor of ten: \glspl{SC} group together elementary calorimeter cells to
provide information for the energy deposition in each individual electromagnetic calorimeter layer
and with finer granularity for the front and middle layers for $\abseta<2.5$.
For the \gls{HEC}, the granularity of the  signals for the digital trigger is the same as the \RunOneTwo trigger towers.
For the \gls{FCAL}, where the division in $\eta$--$\phi$ trigger cells is less regular,
the increase in granularity depends on the layer.
With
the increased granularity of the new trigger signals, the efficiency in selecting
events with interesting signatures and the discrimination power against jets are expected to improve.
The legacy (\RunOneTwo) \analog trigger path~\cite{ATLAS-TDR-22} will be kept operating in parallel to the new digital trigger
path at least until the new system has been fully commissioned with the first collision data in \RunThr
and its performance is equal to or exceeds the performance of the old trigger system.
 
The modifications and improvements
to the original \gls{LAr} calorimeter system that
are not directly related to the new digital trigger path are described in \Sect{\ref{sec:LArMods}}.
The implementation of the new digital
trigger path is then presented in \Sect{\ref{sec:LArDigitalTrigger}}.

\subsubsection{Modifications to the Original LAr System}
\label{sec:LArMods}

\begin{figure}[htbp]
\begin{center}
\includegraphics[width=0.9\textwidth]{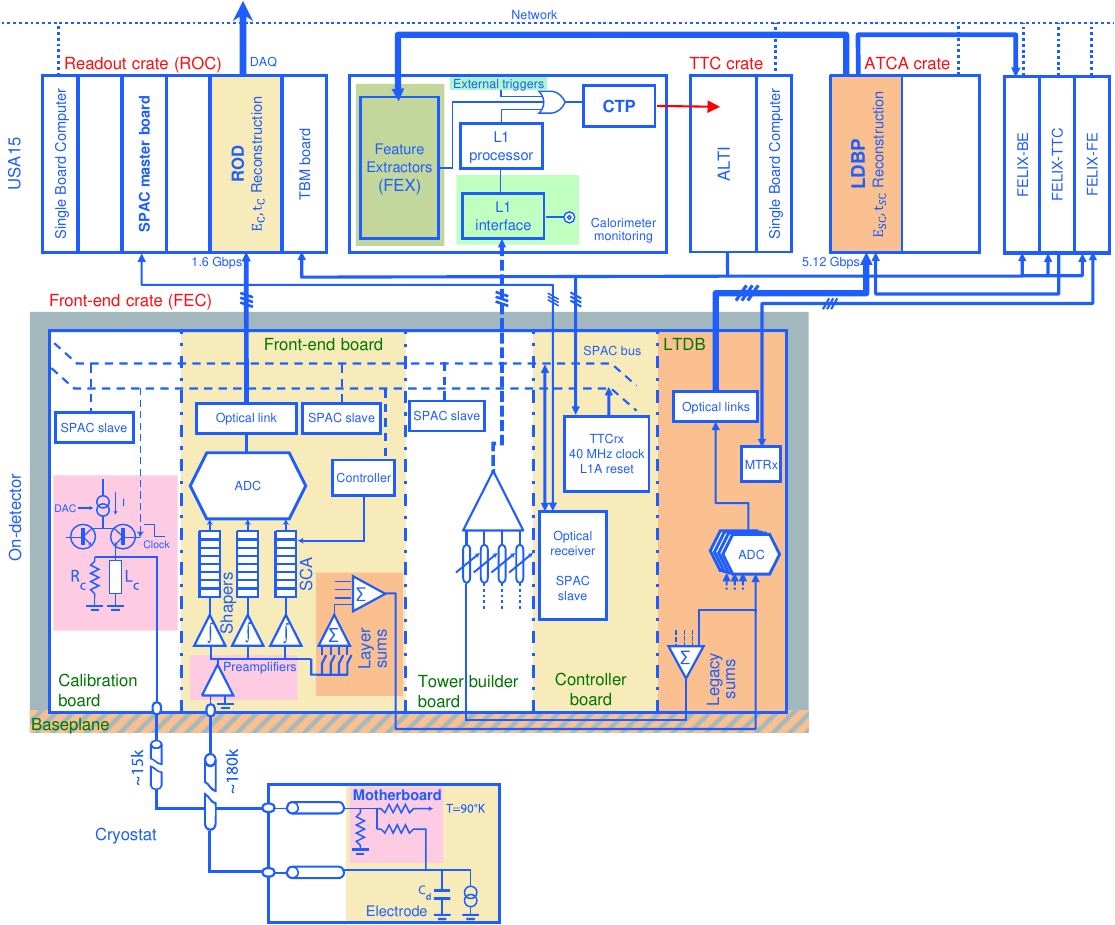}
\end{center}
\caption{The schematic block diagram of the \RunThr \gls{LAr} readout electronics architecture. In this
representation, the
\gls{LAr} ionisation signal proceeds upwards, through the front-end crates mounted on the detector to the off-
detector electronics. The elements added during \gls{LS2} are highlighted in orange.
This diagram is valid for the electromagnetic calorimeters, where only the very minimal boards shown at the bottom of the diagram are located inside the cryostats; the \gls{HEC} and \gls{FCAL} have additional electronics inside the endcap cryostats.
}
\label{fig:lar-run3-arch}
\end{figure}

 
The readout of the \gls{LAr} Calorimeter
remains largely unchanged; an independent digital
trigger path has been added which operates in parallel to the legacy electronics.
An updated schematic block diagram of the upgraded \gls{LAr} readout electronics architecture for \RunThr is shown in \Fig{\ref{fig:lar-run3-arch}}. The legacy electronics include the precision main readout, the calibration
system, and the \analog trigger path providing lower-granularity energy sums to the \gls{L1Calo} system. As can be
seen on the diagram, the new digital trigger can be configured and read out independently of the legacy system.

In the main readout, the \glspl{FEB} installed in \glspl{FEC} on the detector
shape, amplify, and sample the \gls{LAr} ionisation
signals in three overlapping gain scales at \SI{40}{MHz}. Upon a \gls{L1A},
the \glspl{FEB} typically digitise four samples corresponding to the signal
of the triggered collision
and transmit them to the off-detector \glspl{ROD} which employ \glspl{DSP} to calculate the energy deposited in each cell and the
peaking time of the ionisation signal using the \gls{OF} technique~\cite{cleland}.
For every \gls{L1A}, the calculated energies for each calorimeter cell are
sent to the \gls{ROS} for eventual integration with the ATLAS event.
Under certain configurable conditions (such as for cells above a given (programmable) energy threshold), the peaking time, a pulse quality factor, and/or the raw \gls{ADC} samples are also transmitted.
 
\begin{figure}[htbp]
\begin{center}
\includegraphics[width=0.9\textwidth]{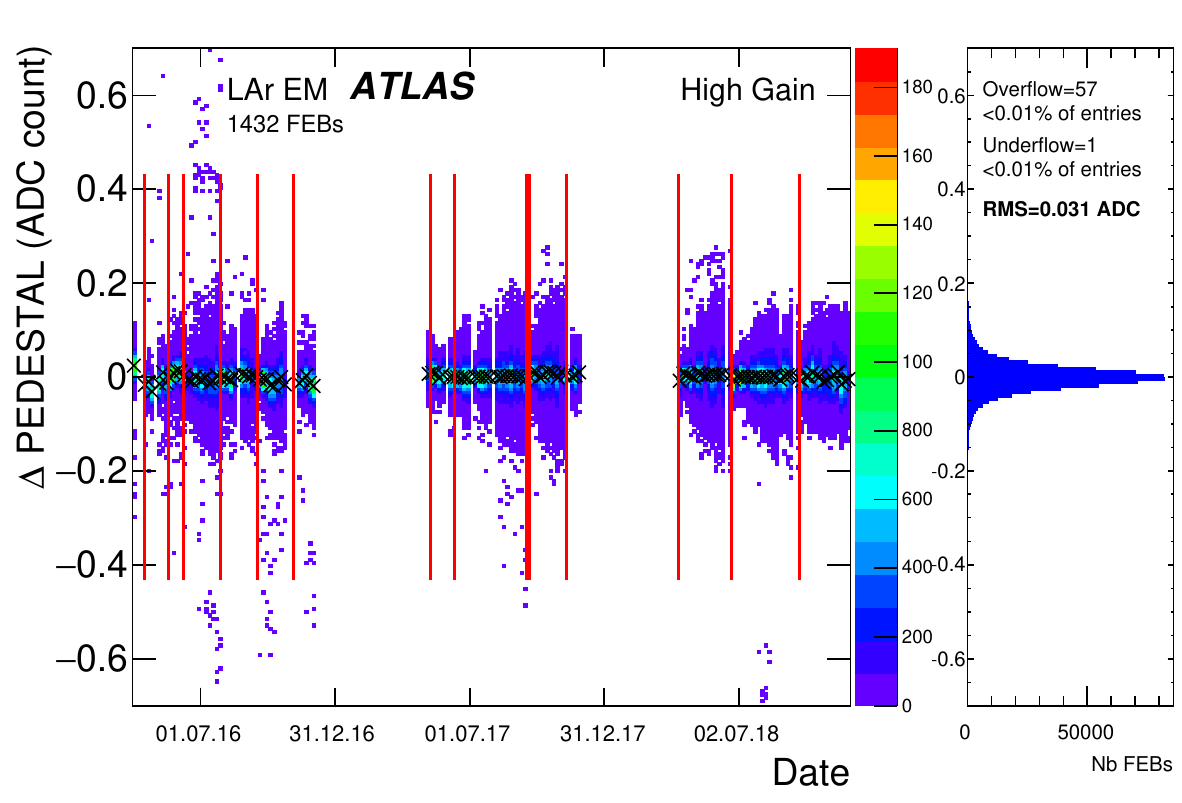}
\end{center}
\caption{The absolute deviation of the pedestal per-FEB average from running
reference values as a function of time for the LAr EM calorimeters
in High Gain. The dates on which the running reference values were obtained
are marked with red lines. The right panel is a projection plot of the same data; the RMS values shown include all entries.
The overflow and underflow entries reported are due to a few problematic FEBs which were replaced.
}
\label{fig:LArCalibStability}
\end{figure}

The calibration
system relies on the front-end calibration boards, to inject pulses of
known amplitude and of known shape directly
into the signal path, close to the calorimeter cells. The calibration pulse
then propagates to the electronics through the same path as the
signals from the ionisation of the \gls{LAr}, from which a prediction of the physics pulse shape,
required for the calculation of the \glspl{OFC},
and electronics calibration constants are calculated. Calibration campaigns are
typically performed several times per week during periods between \gls{LHC} physics fills and the
constants are updated periodically to ensure calorimeter response
stability better than 1\%.
As an example, \Fig{\ref{fig:LArCalibStability}} shows the
evolution of the pedestal measurements in calibration runs  over the course of \RunTwo,
for the electromagnetic calorimeter and for the highest gain scale.

Finally, the \analog trigger path begins with the Trigger \glspl{TBB}, also installed in the \glspl{FEC}, which create fixed-size projective energy sums for each \gls{TT}
by adding together the signal sums from each of the up to four \gls{LAr} calorimeter layers.
The layer sum signals are aligned in time using delays applied to each \gls{TBB} input.
The \gls{TT} sums
are routed with twisted pair copper cables to receivers in the off-detector electronics cavern and are
digitised and processed by dedicated \gls{L1Calo} trigger electronics.

The \gls{LAr} calorimeters operated very reliably during \RunOneTwo with a data-taking
efficiency
ranging from 99.2\% in 2011~\cite{LARG-2013-01} to better than 99.7\% in 2018~\cite{DAPR-2018-01}. The improved performance
can be partly attributed to optimised monitoring and recovery procedures.
Examples include the more efficient identification and vetoing of events with large-scale coherent noise
(noise bursts)~\cite{LARG-2013-01,DAPR-2018-01} and the online identification
of noisy calorimeter cells contributing to high \gls{L1} trigger rates and their
automatic removal from the \gls{TT} energy sums.
Improvements of some components in the \gls{LAr} system have also contributed to the performance.
One such improvement was the upgrade of the original high-voltage units with
more robust high-voltage modules during \gls{LS2}. The new modules can operate in current-control (current-limiting) mode
avoiding high-voltage trips and the needed recovery time.

An additional modification to the original \gls{LAr} system was the replacement of the modules for \gls{TTC}, discussed in more detail in Section~\ref{sec:TDAQ_MUCTPI}.
The legacy off-detector \gls{TTC} distribution system consisted of several \gls{VME} modules
chained together with flat-ribbon cable connections to receive the \gls{TTC} signals from the ATLAS \gls{CTP}
and propagate them to the \gls{LAr} front-end and off-detector electronics.
The original \gls{TTC} chain was replaced during \gls{LS2} with the new \gls{ALTI}, a single-board \gls{VME} module developed by ATLAS. The \gls{ALTI} modules
are less susceptible to electromagnetic interference and eliminate multiple potential points of failure in the \gls{TTC} chain.
 
Finally, during the course of the installation of the new digital trigger path, the \gls{FEC} water cooling
system was refurbished. In particular, it had been observed that
the material of the cooling hoses circulating water to the \glspl{FEB}
and to the \gls{FEC} \glspl{LVPS} was ageing.
The cooling hoses were therefore replaced, to mitigate the risk of leaks during \RunThr.


\subsubsection{New digital trigger path}
\label{sec:LArDigitalTrigger}

 
Several new hardware components are required both on- and off-detector to implement the new digital trigger path:
a new trigger front-end board, the \gls{LTDB} was designed and constructed to transmit the higher-granularity trigger signals
off the detector, where they are read out and processed by a completely new readout system. The detailed \gls{SC} granularity is presented in Table~\ref{tab:LArSuperCells}.
Some modifications to the on-detector electronics in the \glspl{FEC} and off-detector electronics were also required.
These modifications and new components, all of which were installed during \gls{LS2}, are briefly described below.
 
\begin{table}[p]
\caption{Size of the \glspl{SC} in the \gls{LAr} Calorimeter digital trigger path, in terms of both elementary cells and $\Delta \eta$ and $\Delta \phi$. The $|\eta|$ ranges correspond to the \gls{SC} granularity changes. The numbers of elementary cells in a \gls{SC} is given by $n_\eta$ and $n_\phi$. The notation $(0.05)0.025$ stands for a \gls{SC} composed of 1 cell of $\Delta \eta =0.05$ and 3 cells of $\Delta \eta= 0.025$. In the \gls{HEC}, the \gls{SC} granularity is the same as in the legacy system; the layers are summed. The \gls{FCAL} modules are built with a non-pointing $x-y$ geometry; therefore, the \gls{SC} geometry is somewhat irregular in shape and size and only approximate constant $\eta-\phi$ regions can be defined.}
\label{tab:LArSuperCells}
\begin{center}
 
\begin{tabular}{|c|c||c||c|c|}
\hline
 
& &Elementary cell&\multicolumn{2}{c|}{Super Cell} \\ \cline{3-5}
$|\eta|$-range &Layer & $\Delta\eta \times \Delta\phi$ & $n_\eta \times n_\phi$ &$\Delta\eta \times \Delta\phi$ \\
\hline \hline
\multicolumn{5}{|l|}{\gls{EM} Barrel} \\ \hline
0--1.4   & Presampler & 0.025 $\times$ 0.1    & 4 $\times$ 1 & 0.1 $\times$ 0.1 \\
& Front      & 0.003125 $\times$ 0.1 & 8 $\times$ 1 & 0.025 $\times$ 0.1  \\
& Middle     & 0.025 $\times$ 0.025  & 1 $\times$ 4 & 0.025 $\times$ 0.1\\
& Back       & 0.05  $\times$ 0.025  & 2 $\times$ 4 & 0.1 $\times$ 0.1 \\
\hline
1.4--1.5   & Presampler & 0.025 $\times$ 0.1    & 4 $\times$ 1 & 0.1 $\times$ 0.1 \\
&Front       & 0.025 $\times$ 0.025 & 1 $\times$ 4 & 0.025 $\times$ 0.1   \\
& Middle    & 0.075 $\times$ 0.025  & 1 $\times$ 4 & 0.075 $\times$ 0.1  \\
\hline
 
\multicolumn{5}{|l|}{\gls{EMEC}: Outer Wheel} \\
\hline
1.375--1.5   & Front      & (0.05)0.025 $\times$ 0.1 & 4 $\times$ 1 & 0.125 $\times$ 0.1 \\
& Middle     & (0.05)0.025 $\times$ 0.025  & 1 $\times$ 4 &(0.05)0.025 $\times$ 0.1 \\
\hline
1.5--1.8      & Presampler & 0.025 $\times$ 0.1    & 4 $\times$ 1 & 0.1 $\times$ 0.1 \\
& Front      & 0.003125 $\times$ 0.1 & 8 $\times$ 1 & 0.025 $\times$ 0.1 \\
& Middle     & 0.025 $\times$ 0.025  & 1 $\times$ 4 & 0.025 $\times$ 0.1 \\
& Back       & 0.05  $\times$ 0.025  & 2 $\times$ 4 & 0.1 $\times$ 0.1 \\
\hline
1.8--2.0              & Front      & 0.004167 $\times$ 0.1 & 4 $\times$ 1 & 0.0167 $\times$ 0.1 \\
& Middle     & 0.025 $\times$ 0.025  & 1 $\times$ 4 & 0.025 $\times$ 0.1 \\
& Back       & 0.05  $\times$ 0.025  & 2 $\times$ 4 & 0.1 $\times$ 0.1  \\
\hline
2.0--2.4              & Front      & 0.00625 $\times$ 0.1 & 4 $\times$ 1 & 0.025 $\times$ 0.1  \\
& Middle     & 0.025 $\times$ 0.025  & 1 $\times$ 4 & 0.025 $\times$ 0.1 \\
& Back       & 0.05  $\times$ 0.025  & 2 $\times$ 4 & 0.1 $\times$ 0.1 \\
\hline
 
2.4--2.5             & Front      & 0.025 $\times$ 0.1 & 4 $\times$ 1 & 0.1 $\times$ 0.1  \\
& Middle     & 0.025 $\times$ 0.025  & 1 $\times$ 4 & 0.025 $\times$ 0.1 \\
& Back       & 0.05  $\times$ 0.025  & 2 $\times$ 4 & 0.1 $\times$ 0.1 \\
\hline
\multicolumn{5}{|l|}{\gls{EMEC}: Inner Wheel} \\
\hline
2.5--3.1             & Front      & 0.1 $\times$ 0.1 & 2 $\times$ 2 & 0.2 $\times$ 0.2  \\
& Middle     & 0.1 $\times$ 0.1  & 2 $\times$ 2 & 0.2 $\times$ 0.2  \\
 
\hline
3.1--3.2             & Front      & 0.1 $\times$ 0.1 & 1 $\times$ 2 & 0.1 $\times$ 0.2  \\
& Middle     & 0.1 $\times$ 0.1  & 1 $\times$ 2 & 0.1 $\times$ 0.2  \\
 
\hline
\multicolumn{5}{|l|}{\gls{HEC}} \\ \hline
1.5--2.5  & Summed &  0.1 $\times$ 0.1 & 1 $\times$ 1 & 0.1 $\times$ 0.1\\
2.5--3.2 & Summed & 0.2 $\times$ 0.2 & 1 $\times$ 1 & 0.2 $\times$ 0.2  \\
\hline
\multicolumn{5}{|l|}{\gls{FCAL}} \\ \hline
3.1--3.5  & 0 & x,y-various & various & $\approx$0.1 $\times$ 0.4\\
3.5--4.0  & 0 & x,y-various & various & $\approx$0.1-0.15 $\times$ 0.4\\
4.0--4.9  & 0 & x,y-various & various & $\approx$0.15-0.2 $\times$ 0.4\\
3.1--4.9 & 1 & x,y-various & various & 0.1-0.3 $\times$ 0.4  \\
3.1--4.9 & 2 & x,y-various & various & 0.4-0.5 $\times$ 0.4  \\

\hline
\end{tabular}
\end{center}
\end{table}

\begin{figure}[t]
\begin{center}
\includegraphics[height=5.9cm]{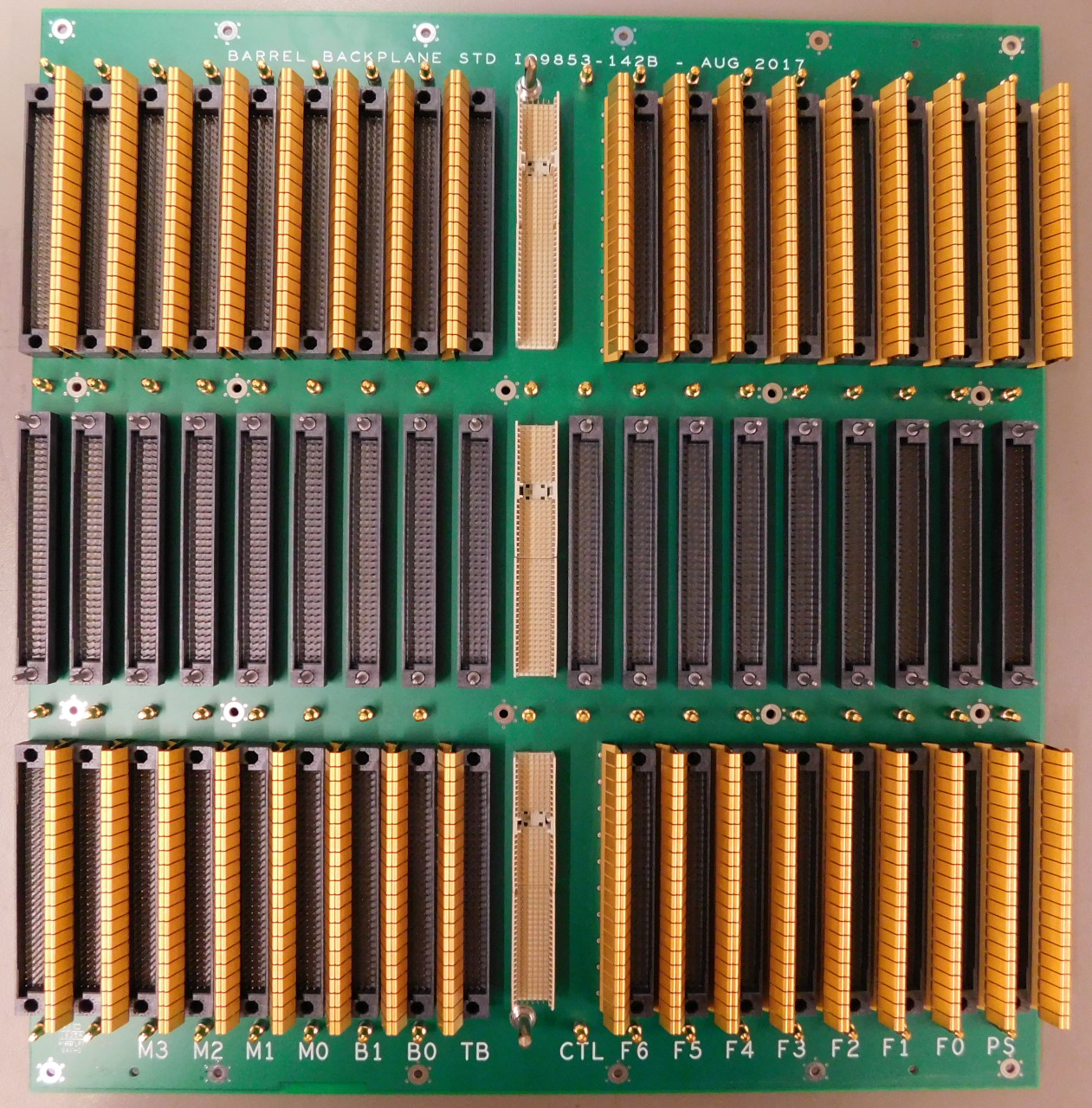}
\includegraphics[height=5.9cm]{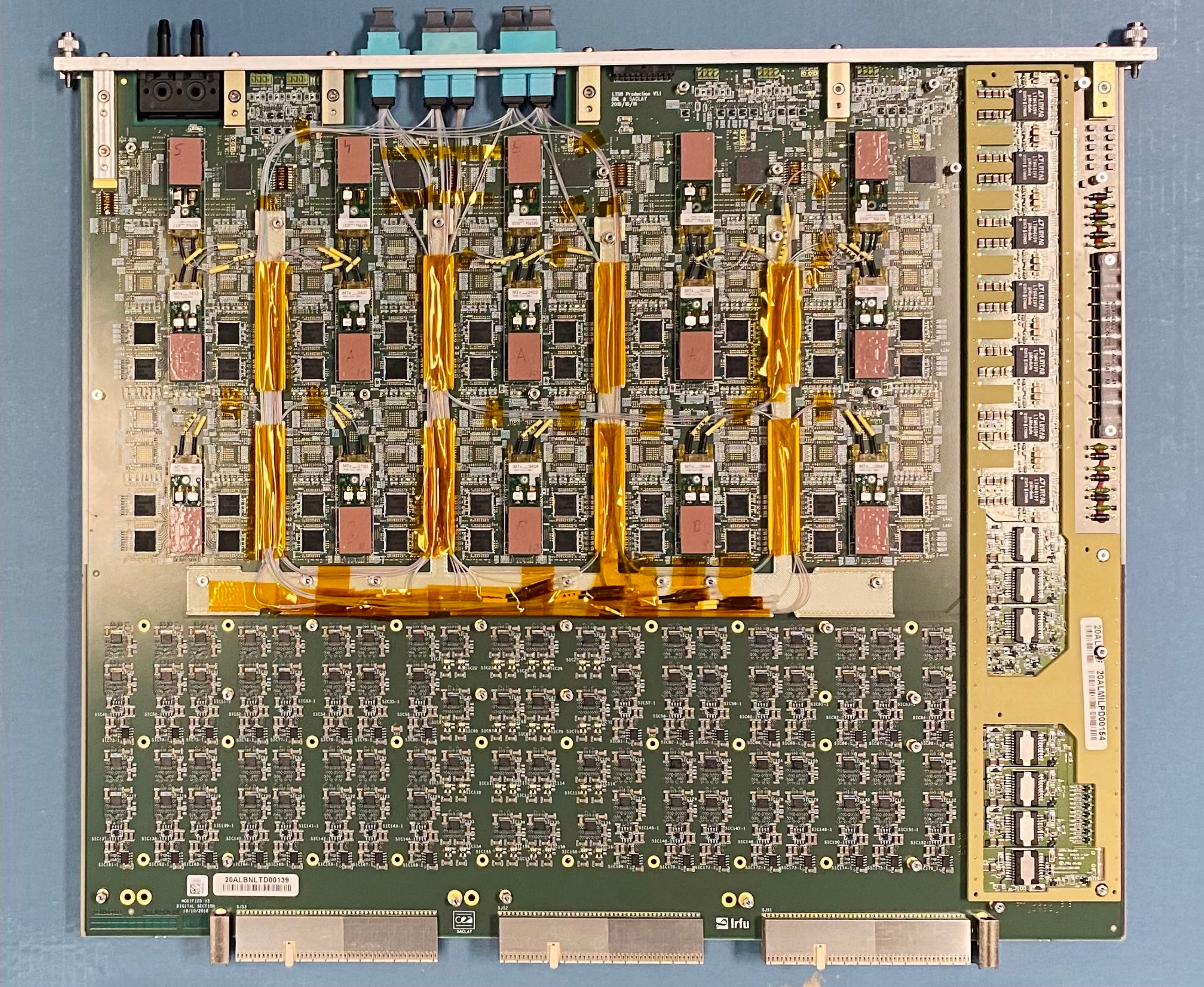}
\end{center}
\caption{Photographs of on-detector electronics components for the new \gls{LAr} digital trigger path: a new baseplane (left) of the type installed in the \gls{LAr} calorimeter front-end crates during \gls{LS2} to accommodate the installation of the \gls{LTDB} (right).}
\label{fig:larLTDB}
\end{figure}
 
\paragraph{New Baseplanes}
To support the new \glspl{LTDB}, the baseplanes in all the \glspl{FEC} had to be replaced.
The purpose of the baseplanes is to support the boards in the \gls{FEC} and to deliver the
signals arriving from the detector at the back of the baseplane to the readout \glspl{FEB}. In addition, the baseplanes route
the \analog sums of the readout signals provided by the \glspl{FEB} and the \glspl{LTDB} to the appropriate boards.
The new baseplanes (\Fig{\ref{fig:larLTDB}} left), have slots to receive the \glspl{LTDB} in addition to the slots in
which the legacy boards are seated. The intricate, multilayered design of the new baseplanes enables boards with the same
footprint as the original baseplanes to provide the many additional traces required to route the finer granularity
\analog sums to the \glspl{LTDB} and subsequently, in most cases, after an additional sum by the \gls{LTDB} (recreating the legacy \analog sums)
to the legacy Trigger \glspl{TBB}. \Analog sums that have the same granularity as the legacy sums (typically the back layer sums,
the presampler sums and in some regions first layer sums) are delivered to the \glspl{TBB} independently of the \glspl{LTDB},
while the rest of the sums are not available if the \gls{LTDB} is not installed.
In total, 114 new baseplanes of six different types are needed to equip all the \glspl{FEC} of the \gls{LAr} Calorimeter system.

\paragraph{Modifications of the FEB with new Layer Sum Boards} The \glspl{FEB} perform the first stage of \analog summing
of the signals to reduce the granularity used in the trigger system.
During \RunOneTwo, this first sum was performed by a daughter board, the \gls{LSB},
which summed the signals from all cells in the same calorimeter layer belonging to the same trigger tower. A series
of new \glspl{LSB} was produced in order to provide sums of the signals from cells in the same \gls{SC},
thereby increasing the granularity of the sums by up to a factor of four.
All 1524 \glspl{FEB} installed on the detector were removed and refurbished with the new \glspl{LSB}.
In addition, their cooling plates and cooling hoses were replaced and tested thoroughly for leaks.
After the baseplanes were replaced, the \glspl{FEB} were reinstalled
in the \glspl{FEC} together with the new \glspl{LTDB} and the rest of the legacy boards.

\paragraph{LAr Trigger Digitizer Board} The \glsfirst{LTDB} transmits \gls{LAr}
pulse samples for \glspl{SC} in four layers at \SI{40}{MHz} to the off-detector electronics.
The \gls{LTDB} is also responsible for the second stage of
\analog summing in order to recreate the legacy trigger sums and provide them
to the legacy \glspl{TBB} through the baseplane.
The board is designed so that the summing can be performed independently of
the configuration state of its digital part.
To cover the entire \gls{LAr} calorimeter system, 124 boards of seven different
types are needed. 
The assembled printed circuit board
for one type of \gls{LTDB} is shown in \Fig{\ref{fig:larLTDB}}.
On the \gls{LTDB}, the pulse for each \gls{SC} is sampled every \SI{25}{ns} and the
samples are digitised with a custom-designed 12-bit, 4-channel \gls{ADC}. The ADC is
implemented in a \SI{130}{nm} CMOS technology and achieves an \gls{ENOB} of 11 with a dynamic
range of 11.7 bits while consuming less than \SI{50}{\mW} per channel.
Each \gls{LTDB} can handle up to 320 \glspl{SC}, so 80 \gls{ADC} chips are installed on each \gls{LTDB}.
The digitised samples are serialised and transmitted with custom-designed optoelectronics
with \glspl{ASIC} implemented in a \SI{250}{nm} Silicon-on-Sapphire (SoS) process.
Up to 40 optical fibres operating at \SI{5.12}{Gbps} transfer the samples to the off-detector electronics.

\begin{figure}[t]
\begin{center}
\includegraphics[width=0.9\textwidth]{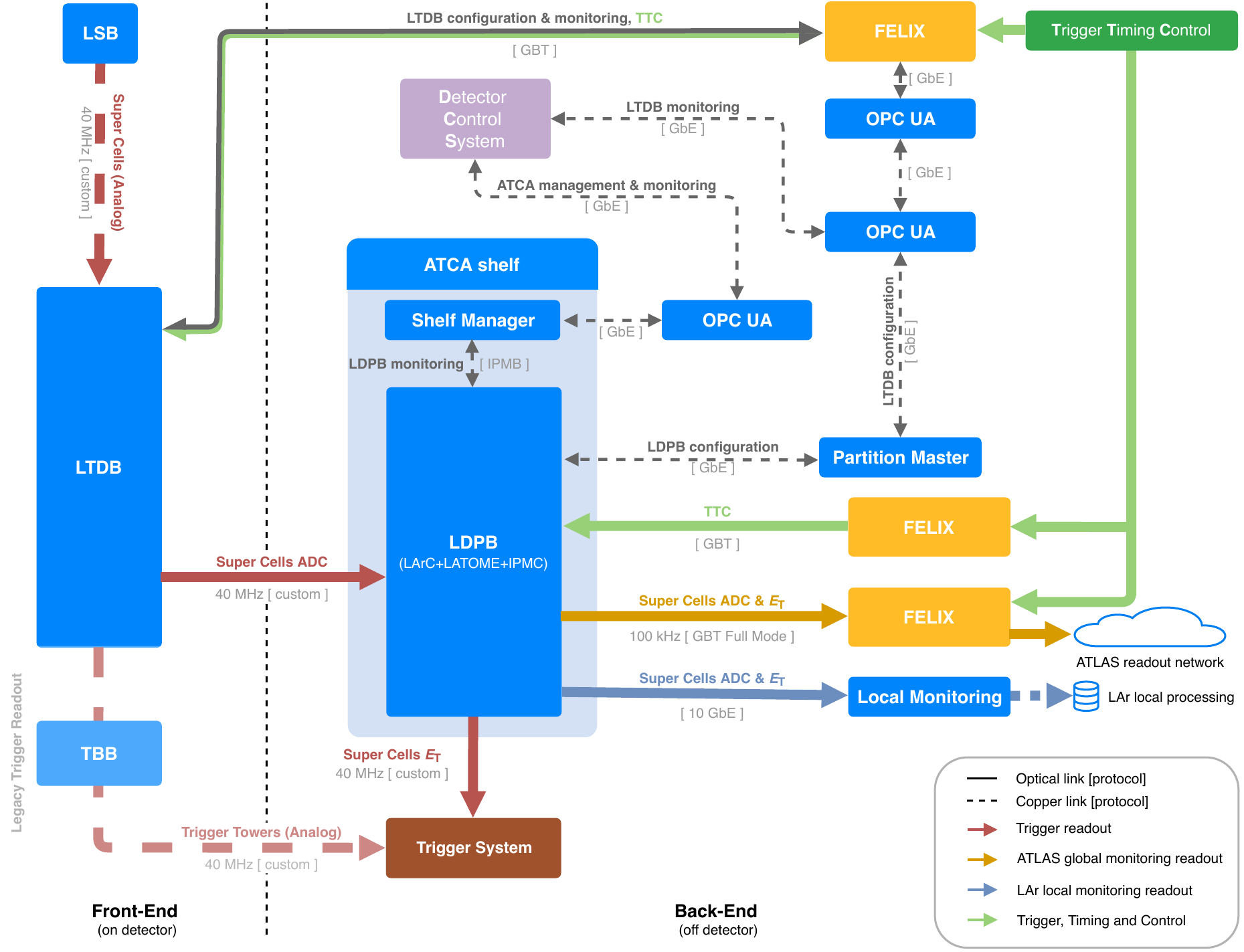}
\end{center}
\caption{The \gls{LAr} calorimeter digital trigger readout system installed during \gls{LS2}.}
\label{fig:lar-phase-I-BE-arch}
\end{figure}

\paragraph{New off-detector electronics} The configuration, monitoring, and readout of the \gls{LTDB} are performed
with new off-detector electronics installed in 
Underground Service Area
\gls{USA15} in 2020. A diagram of the architecture of the new off-detector electronics,
operated independently of the legacy electronics,
is shown in \Fig{\ref{fig:lar-phase-I-BE-arch}}.
The configuration and monitoring of the \glspl{LTDB}
is performed using the \gls{FELIX} system over optical links
interfaced to \gls{GBT-SCA} devices (see Section~\ref{TDAQ_DCS_GBT_SCA} for further details).
For the readout, the \gls{LDPS} comprises
\num{30} \glspl{LDPB} implemented in \gls{ATCA} technology and installed in three shelves.
 
\begin{figure}[t]
\begin{center}
\subfloat[]{\label{latomeFront}\includegraphics[height=6cm]{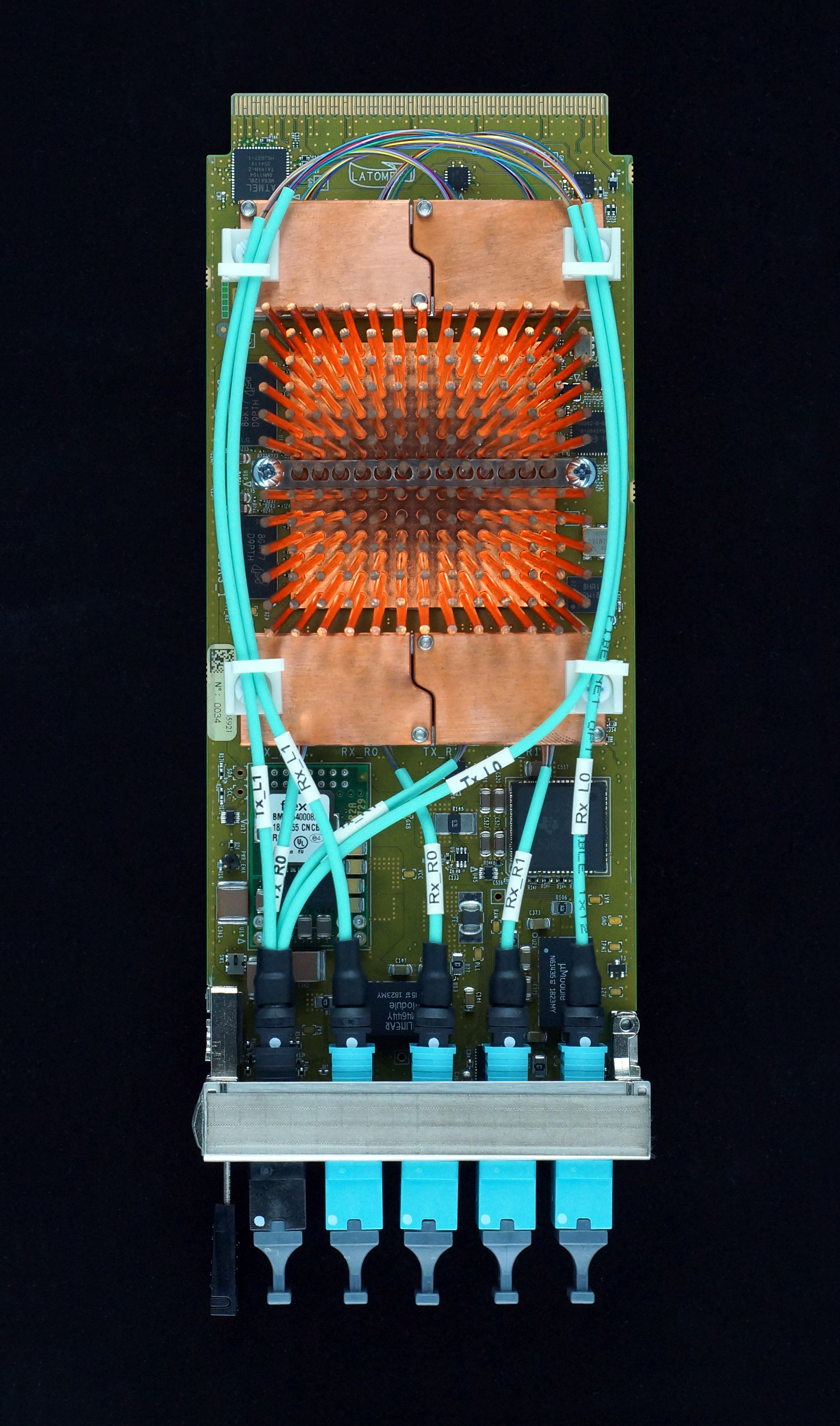}}
\subfloat[]{\label{latomeBack}\includegraphics[height=6cm]{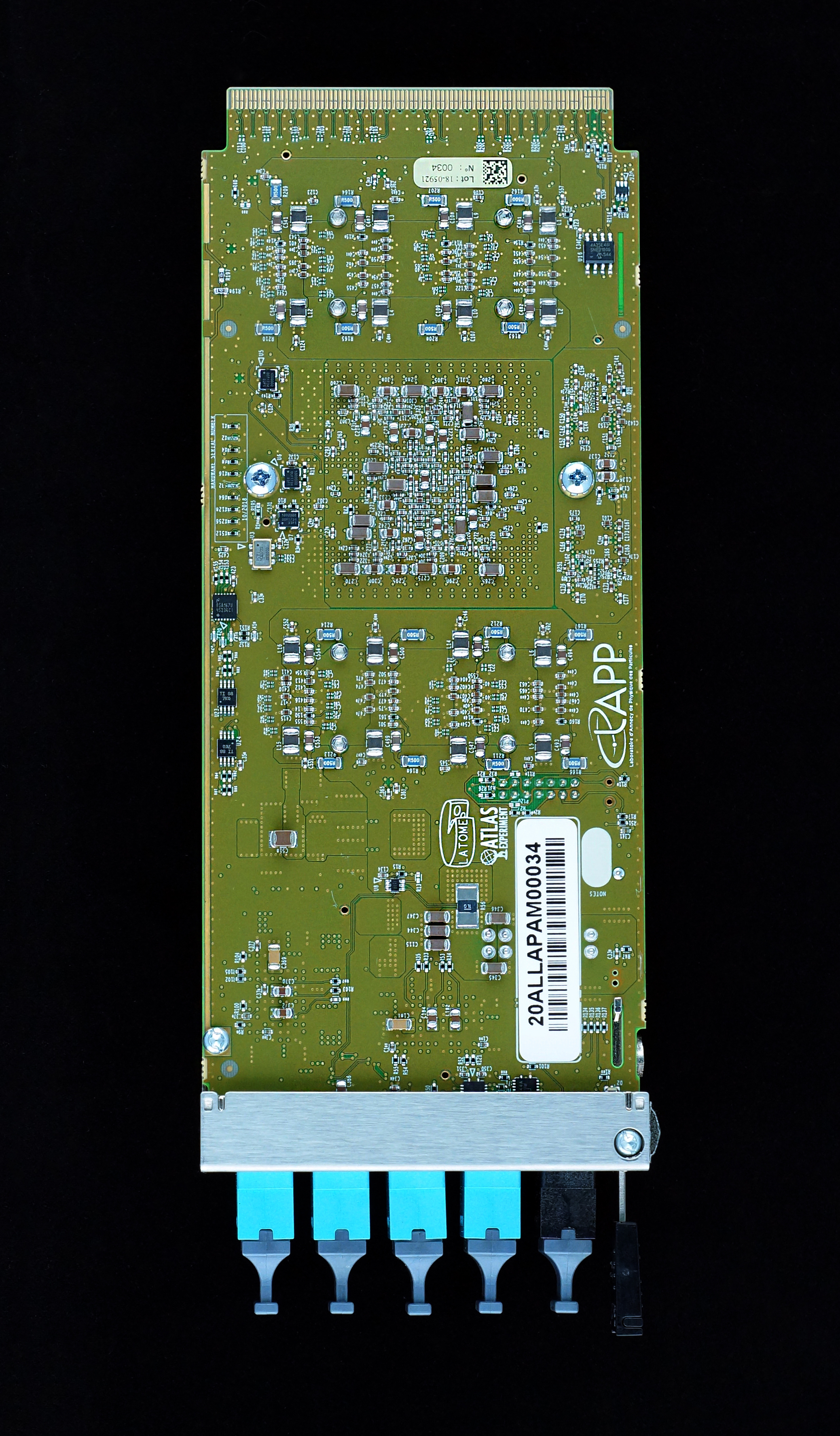}}
\subfloat[]{\label{latomeCarrier}\includegraphics[height=6cm]{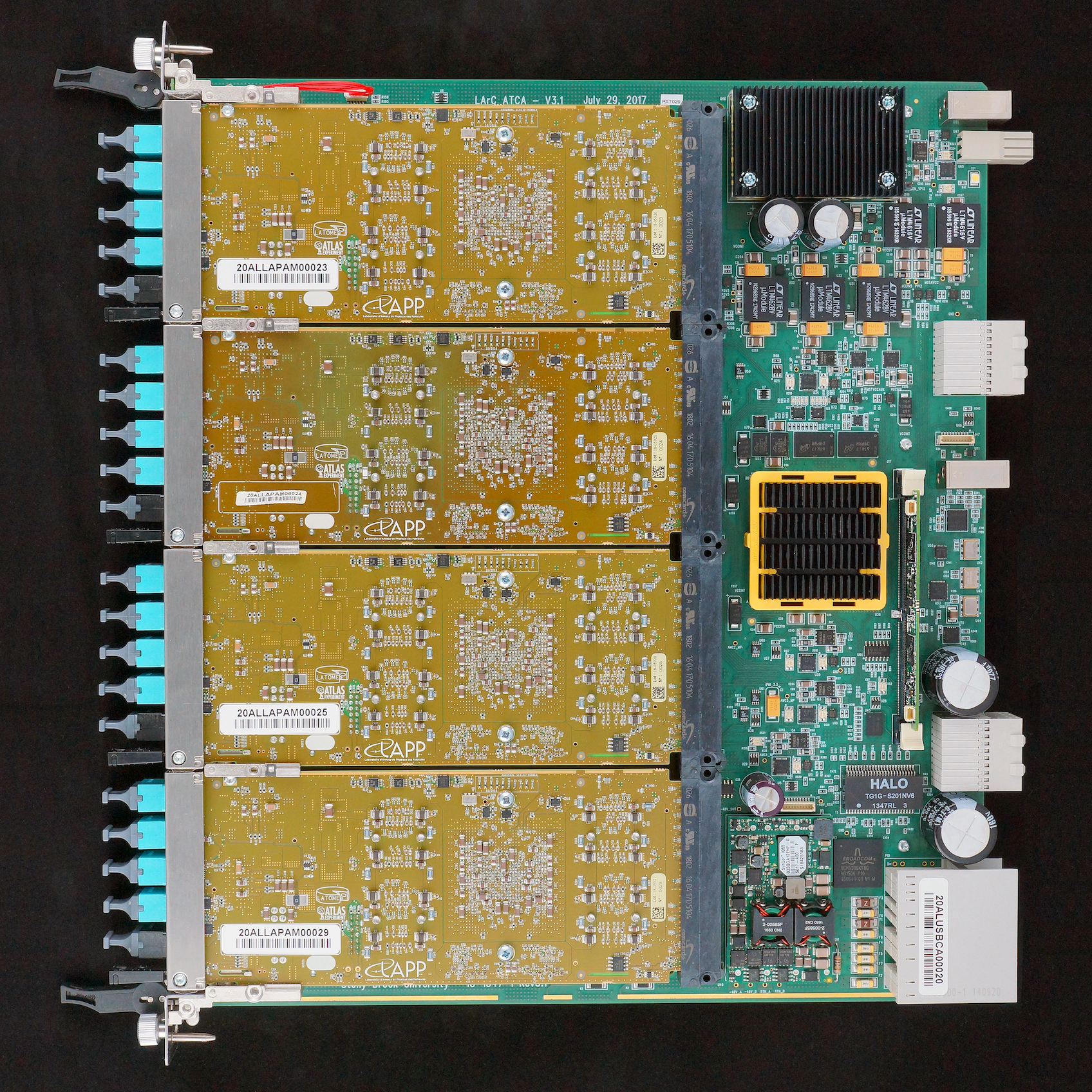}}
 
\end{center}
\caption{Photographs of off-detector electronics components for the new \gls{LAr} digital trigger path.
\protect\subref{latomeFront} and \protect\subref{latomeBack} show the front and back faces of a \gls{LATOME}, respectively, while \protect\subref{latomeCarrier} shows a \gls{LArC} fully equipped with four \glspl{LATOME}.}
\label{fig:larLDPB}
\end{figure}
 
\paragraph{LAr Digital Processing Blades} Each \glsfirst{LDPB} consists of a \gls{LArC} motherboard that houses up to four \glsfirst{LATOME} \glsfirst{AMC} daughterboards as shown in
\Fig{\ref{fig:larLDPB}}. Up to 48 input fibres can be connected to each \gls{LATOME} routing the
digitised pulse samples from the \glspl{LTDB} at \SI{5.12}{Gbps}, receiving 12-bit \gls{ADC} samples
from up to 320 \glspl{SC} at \SI{40}{MHz}. The samples are processed on the \gls{LATOME} in real time by
custom-designed firmware running on an Arria10 \gls{FPGA}. An \glsfirst{OF} procedure,
similar to the one used for the precision main-readout, is used to
calculate the energy deposited in each \gls{SC} at \SI{40}{MHz} for each bunch crossing.
The resulting calibrated transverse energy values for each \gls{SC} (\etsc) are routed to the
\gls{FEX} processors (see Section~\ref{sec:TDAQ_L1Calo}) directly from the \glspl{LATOME} over 48 output fibres
each operating at \SI{11.2}{Gbps}. In some cases, several copies of these values are transmitted to the different \glspl{FEX}.
In total, the \gls{LDPS} receives \SI{25.2}{Tbps} of input from the front-end and
streams up to \SI{41.1}{Tbps} to the \gls{L1} system.  The \gls{LArC} facilitates the control,
configuration, and monitoring of the blade and the hosted \glspl{LATOME}; it connects to
the \gls{ATCA} shelf, distributes the power to the various elements of the \gls{LDPB}, and
provides a connection to the \gls{ATCA} shelf manager via an \gls{IPMB}
for the low-level monitoring of the status of the blade, including voltage and temperature measurements.
Additional direct \gls{GbE} connections allow the reprogramming, loading of configuration constants, and
further monitoring of the \gls{LDPB}. The hardware control and monitoring of each \gls{LArC} blade proceeds via an \gls{IPMC}.
The \gls{LDPB} also implements two additional paths for reading out
processed data from the \glspl{LATOME}: the ``\gls{TDAQ} path'' transmits
on \gls{L1A} (therefore with a rate of \SI{100}{kHz}) for each \gls{SC} the \etsc and the \gls{ADC} values of the five
time slices used in its calculation. The data are transmitted over FULL mode links
(see Section~\ref{subsubsec:tdaq_daqhlt_felixswrod}) to
\gls{FELIX} to be integrated with the ATLAS event for recording.
The additional data throughput to \gls{TDAQ} is of the order of \SI{400}{Gbps} and can be used
for debugging the trigger decision, comparisons to the main readout, and calorimeter noise studies.
The second additional data path implemented on the \gls{LArC} is the  ``monitoring path'' which provides
an independent data flow of order of a few \SI{10}{Gbps} for local monitoring of the data, transmitted to a local computer farm over 10~\gls{GbE} links
using the \gls{ATCA} infrastructure.
Finally, the \gls{LArC} receives and distributes the ATLAS \gls{TTC} signals
to the \glspl{LATOME}.


Further technical details of the new LAr digital trigger system including
its integration and commissioning,
as well as the procedures used to validate its proper functioning and
the performance achieved during the commissioning of the system, are available
in Ref.~\cite{LArPhaseIPaper}.

\subsection{Tile Calorimeter}
\label{sec:tile}
 
The ATLAS \gls{Tile} covers the region $\abseta < 1.7$ using pseudo-projective calorimeter towers composed of scintillating tiles in a steel matrix, read out by wavelength-shifting fibres.
The \gls{Tile} comprises a cylindrical Barrel section that surrounds the \gls{LAr} \gls{EMB} cryostat and two Extended Barrels that surround the \gls{LAr} endcap cryostats.
The cryostat scintillation counters sit between the central and endcap cryostats. They are mounted to the \gls{Tile} Extended Barrel, and are read out through the \gls{Tile} electronic system. They are used for luminosity monitoring and to correct for energy losses in the outer wall of the barrel cryostat and \gls{ID} services in this region where there is a large amount of passive material.
The counters present in \RunOneTwo covered
the region $1.2 < \abseta < 1.6$, while for \RunThr this coverage has been extended to $1.2 < \abseta < 1.72$.
 
The \gls{MBTS} cover $2.0 < \abseta < 4.0$
and are used for triggering and for luminosity monitoring. As they are read out through the \gls{Tile} electronic system, they have been historically considered as part of the \gls{Tile}.
 
The only upgrades in \RunThr for the \gls{Tile} are in these two systems: the cryostat and \gls{MBTS} scintillation counters. The upgrades have been motivated primarily by the radiation damage that affected these systems in \RunTwo. Additionally, there has been development of a \gls{Tile} electronics demonstrator in preparation for \RunFour, which is also described in this section.

\subsubsection{Radiation environment}
The $1.2 < |\eta| < 4.0$ region in the ATLAS detector is a high-radiation environment for plastic scintillators. At a distance of about $z=\pm\SI{3500}{\mm}$ from the \gls{IP}, the average ionisation dose in the pseudorapidity region from \numrange[range-phrase=--]{3.0}{4.0} is of the order of \SI{830}{\gray}\perinvfb
while the section from pseudorapidity \numrange[range-phrase=--]{2.0}{3.0} sees an average dose of the order of \SI{95}{\gray}\perinvfb, with the dose in each rapidity interval dropping rapidly with increasing distance from the beam-line (for example from \SI{2700}{\gray}\perinvfb at $|\eta|=4$, to \SI{213}{\gray}\perinvfb at $|\eta|=3$).  For a projected total luminosity in \RunThr of \intlumirunthree, the dose is expected to be of the order of \SI{200}{\kilo\gray} for $3 < |\eta| < 4$,  and \SI{20}{\kilo\gray} for $2 < |\eta| < 3$. 
The dose at a pseudorapidity near 1.7 (E4) is of the order of \SI{10}{\gray}\perinvfb, or a total of \SI{2.5}{\kilo\gray} for \SI{250}{\ifb}, again dropping rapidly with increasing radius. Significant light loss can be expected in plastic scintillators for radiation doses greater than \SI{5}{\kilo\gray}~\cite{Jivan:2015mqk,Liao:2015zsa}. To give an indication of the impact of the light loss in \gls{MBTS}, at the start of \RunTwo in 2015, the two-track efficiency was $98\%$; in 2018 when the light loss reached $95\%$ in the outer counters, this two-track efficiency dropped to $75\%$.
 
\subsubsection{Cryostat counters}
\begin{figure}[h!]
\centering
\subfloat[]{\label{fig:TileE1-E4Counters}\includegraphics[width=0.65\textwidth,trim=100 -50 0 0,clip]{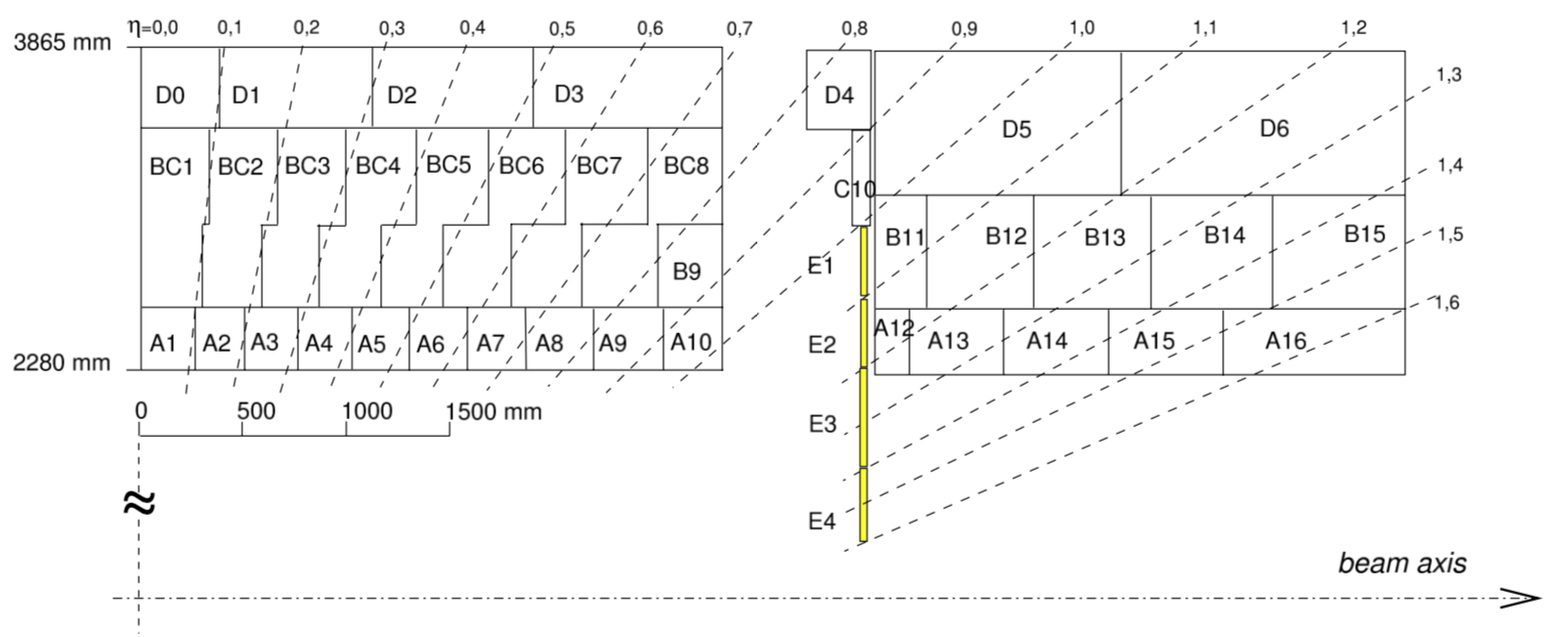}}\\
\subfloat[]{\label{fig:TileMaterial}\includegraphics[width=0.65\textwidth,trim=30 30 0 10,clip]{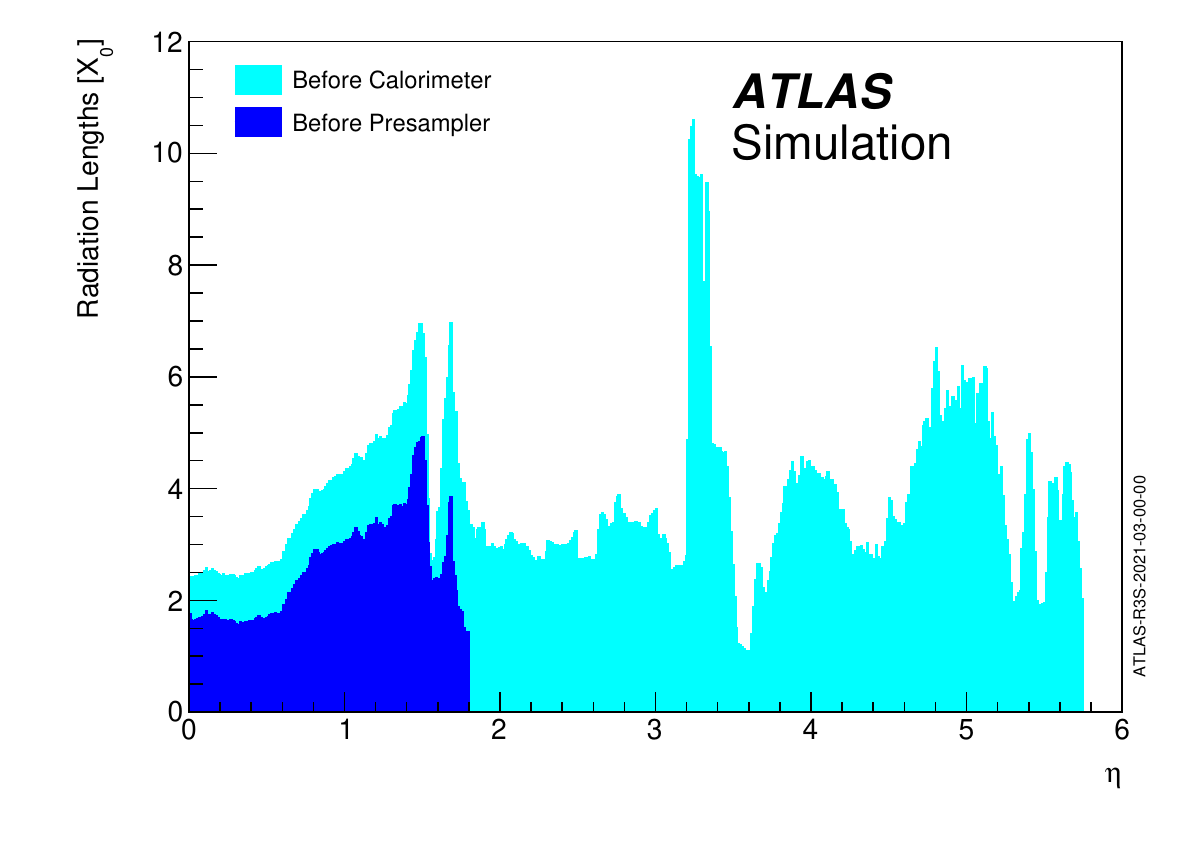}}
\caption{\protect\subref{fig:TileE1-E4Counters} Tile E1-E4 counter location (in yellow) and \protect\subref{fig:TileMaterial} amount of material (in radiation lengths) in front of the cryostat counters.
}
\label{fig_atlas_crack}
\end{figure}

In the region occupied by the cryostat counters, the
fraction of passive material along the particle path is high
(as shown in Figure~\ref{fig_atlas_crack}), primarily due to the cryostat walls. In the region of maximum passive material, there is up to \num{10} radiation lengths of material in front of the cryostat counters.
The electromagnetic showers are insufficiently sampled in this region, with the first active layer located close to the shower maximum.
Using the
cryostat counters
to correct for the energy loss in the passive material by adding the weighted energy deposit in the scintillators
leads to a partial recovery of the electron and photon energy resolution in the affected areas~\cite{ATL-PHYS-PUB-2016-015}.
 
In \RunOneTwo, each counter was divided into two segments, covering $1.2 < |\eta| < 1.4$ (E3)  and $1.4 < |\eta| < 1.6$ (E4). However, a significant energy resolution degradation is also present in the $\eta$ region \numrange[range-phrase=--]{1.6}{1.75}.
Before the start of \RunTwo, two of the 128 cryostat scintillators were re-built with an extension to cover the $\eta$ range from \numrange[range-phrase=--]{1.6}{1.75}. The energy resolution in the region covered by this extension was significantly improved with the corrections provided by the energy deposited in this scintillator. The replacement of the cryostat scintillators for \RunThr allows for an extension of the $|\eta|$ coverage to 1.72; an extension to 1.75 is not possible  due to interference with the liquid argon cryostat heaters in some $\phi$ regions.  The number of electronics channels for the cryostat scintillators is fixed, so the segmentation for E3 was modified to an $\eta$ range of \numrange[range-phrase=--]{1.2}{1.6}, and E4 to \numrange[range-phrase=--]{1.6}{1.72}. Simulations indicated that the energy resolution in the $\eta$ range from \numrange[range-phrase=--]{1.2}{1.6} would not be significantly degraded by joining the two segments, and that the new region \numrange[range-phrase=--]{1.6}{1.72} would survive the radiation  expected in \RunThr.
 
There are 64 cryostat counters corresponding to the 64 \gls{Tile} modules, each counter containing an E3 and an E4 channel. The cryostat counters are mounted at their outer radius to the extended barrel \gls{Tile} modules using specially designed brackets. Neighbouring counters are  attached at the inner radius using connectors designed for the purpose.  During \RunOneTwo, the counters were
aligned (in $\phi$) at the outer radius, where they are connected to the brackets. Due to imperfections in the circularity of the extended
barrels, this led to the appearance of gaps between some of the counters at the inner radius, resulting in additional uninstrumented areas, distributed non-uniformly. The alignment scheme was redesigned for \RunThr, with the alignment carried out at the inner radius (with each counter maintaining a separation of \SI{1}{\mm} from its neighbours), with the impact of any non-circularity taken into account in the brackets at the outer radius. The old E3/E4 counters were removed, and the new ones installed, in 2019--2020.
 
The cryostat counters use \SI{6}{\mm}-thick scintillators (EJ-208 from Eljen, a scintillator designed with an improved radiation hardness) sandwiched between two trapezoidal aluminium shells. Care was taken during installation to avoid any electrical contact between the aluminium shells and the endcap cryostat wall. The scintillating plates are wrapped in Tyvek sheets,
to prevent damage to the surface of the scintillator and also to reflect back some of the light that escapes from the top/bottom of the scintillator. Opaque spacers separate the two pieces of scintillator comprising E3 and E4. For light-tightness, the counters are sealed with both black electrical tape and aluminium tape.
 
\begin{figure}[h!]
\centering
\includegraphics[scale=0.70]{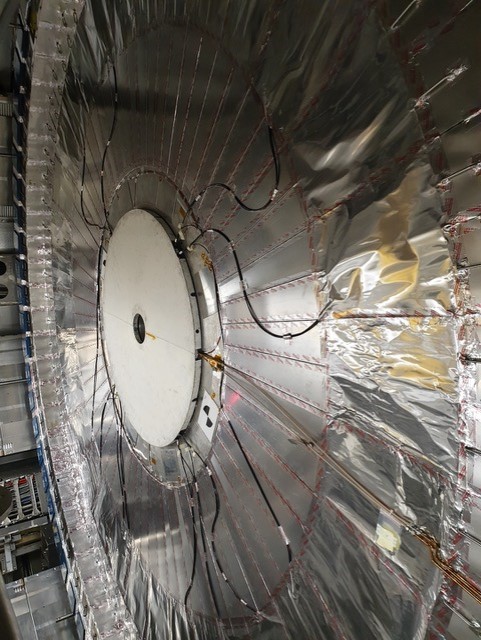}
\caption{The surface of one of the ATLAS \gls{Tile} extended barrels and the liquid argon endcap cryostat where the \gls{MBTS} counters (protected by a white cover) and cryostat scintillation counters can be seen.
Also shown are the aluminium foil panels used for light-tightness. The black tubing contains clear fibres which are used to transmit the light signals of the \gls{MBTS} counters.}
\label{fig_atlas_crack_MBTS}
\end{figure}

The EJ-208 scintillator emits light in the blue wavelength region. Y-11(200)MSJ \gls{WLS} fibres from Kuraray coupled to one side of each scintillator capture the blue light and then re-emit light in the green portion of the spectrum, a fraction of which is captured and then transmitted  to optical connectors glued into the tops of the counters. From there,  optical cables using BCF098 clear fibres from St.~Gobain transmit the light to photo-tubes in the extended barrel \gls{Tile} modules, where the signals are read out through the standard \gls{Tile} data acquisition chain. The area above the optical cables is covered with aluminium foil panels in order to improve the light-tightness.
Before installation, each assembled counter was checked for light yield and uniformity using a strontium $\beta$ source.
Figure~\ref{fig_atlas_crack_MBTS} shows the end surface of one of the ATLAS \gls{Tile} extended barrels and the \gls{LAr} endcap cryostat,
where the \gls{MBTS} counters, the cryostat scintillators, and the aluminium foil panels can be seen.
 
In summary, the E3/E4 cryostat scintillators are very useful in improving the lepton energy resolution and luminosity monitoring in the region $1.2 < \abseta < 1.72$. New counters have been constructed for \RunThr, extending the rapidity range and using the more radiation-hard scintillator described above. The \RunThr\ degradation is estimated to be in the range $28-43\%$ (for E3$-$E4, respectively) after \SI{250}{\ifb}.
 
\subsubsection{Minimum Bias Trigger Scintillator counters}
 
\begin{figure}[h!]
\centering
\includegraphics[scale=0.15]{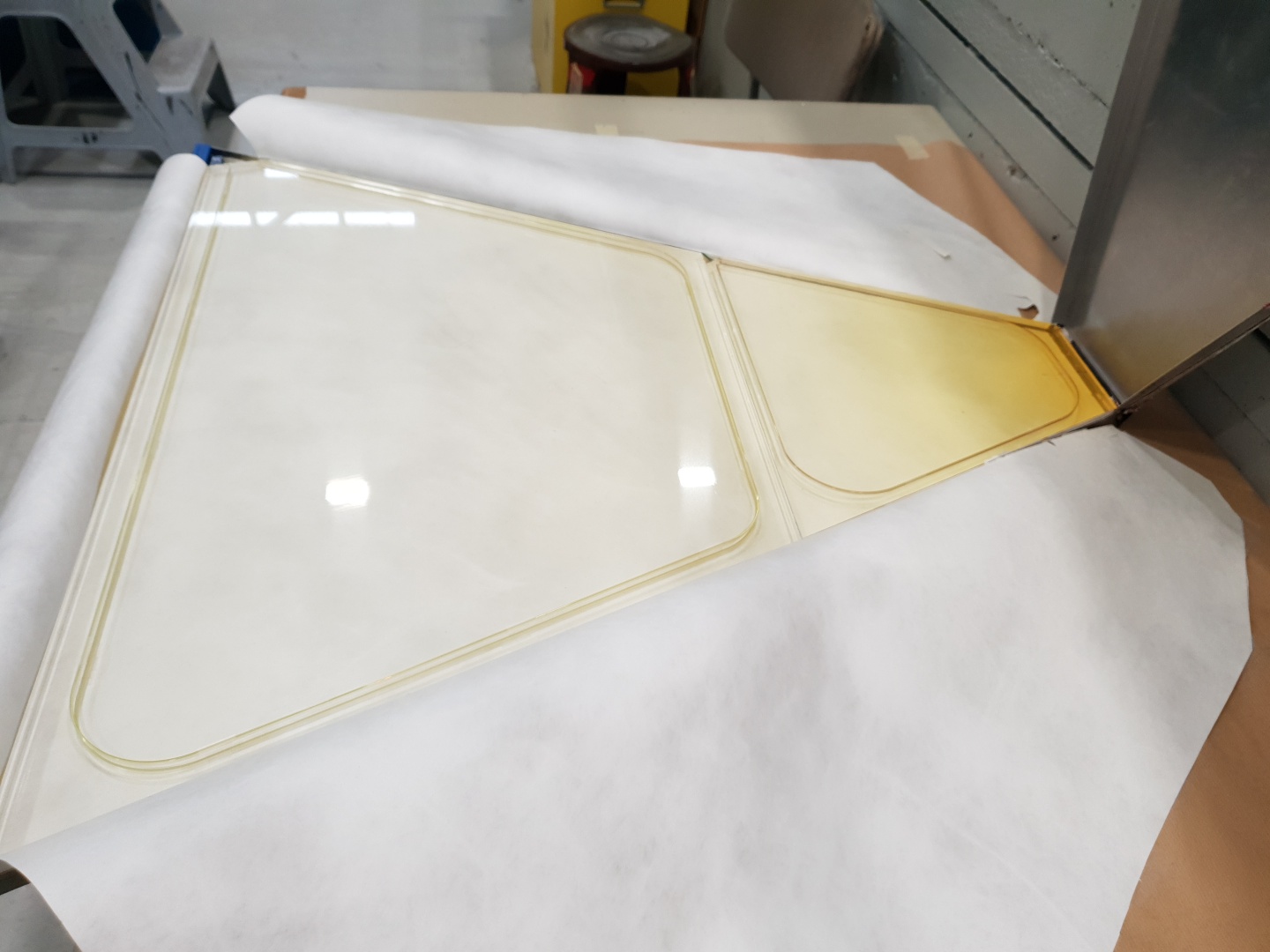}
\caption{The two scintillator segments from one of the \gls{MBTS} counters present in ATLAS during \RunTwo. The effects of the radiation exposure in \RunTwo can be seen as yellowing at the small end.   }
\label{fig_atlas_MBTS_rad_damage}
\end{figure}
 
The \glsfirst{MBTS} counters cover the high-rapidity interval from \numrange[range-phrase=--]{2.0}{4.0} on both sides of the ATLAS detector. The \gls{MBTS} counters are divided into eight $\phi$ segments per endcap, with each $\phi$ segment having two radial subdivisions in $\eta$, \numrange[range-phrase=--]{2.0}{3.0} and \numrange[range-phrase=--]{3.0}{4.0}. The \gls{MBTS} channels are read out through the standard \gls{Tile} data acquisition chain. In addition, a high-gain trigger signal is available from each channel, allowing the \gls{MBTS} signals to be used in the trigger. Initially, the \gls{MBTS} counters were intended to be used only in the early running of ATLAS in order to determine whether a particular beam crossing contained a proton-proton collision, needed in the extremely low luminosity collisions. They were not expected to be needed, nor functioning, in high luminosity conditions. However, they did maintain functionality, and proved to be useful for luminosity monitoring, van der Meer scans, and for triggering during the heavy ion running.
 
The high radiation environment in this rapidity region required the \gls{MBTS} counters to be replaced at the end of \RunOne, with the expectation that they would also need to be replaced at the end of \RunTwo. The rapidity interval covered by the inner \gls{MBTS} counters has some of the highest radiation levels for which plastic scintillators have ever been used. At the end of \RunTwo, the inner section of the \gls{MBTS} counters had lost 99\% of the original light yield, and the outer section had lost 95\% of its original light yield,  so care was taken in the design for their replacement in \RunThr, to improve the radiation hardness.
 
Light loss in polystyrene-based scintillators from radiation exposure results primarily from damage to the polystyrene matrix, resulting in the formation of radicals and other color centres, which absorb the light produced by the secondary fluors. A picture of one of the counters from \RunTwo is shown in Figure~\ref{fig_atlas_MBTS_rad_damage}, showing the coloration pattern due to the damage from radiation. The bonds can be re-formed over time (annealing). The annealing can be accelerated by the presence of oxygen. For \RunThr, polystyrene-based scintillator is used for both the inner and outer segments, but different dopants were used in the two areas. In particular, the dopants paraterphenyl (PTP) (primary) and  BBQ (secondary) were used in the inner region. With these dopants, the output scintillation light of the inner segment is in the green region of the spectrum, and is less sensitive to absorption by the color centres produced by radiation exposure.
The outer segment used the standard PTP and POPOP ((1,4-bis-(2-(5-phenyloxazolyl))-benzene) dopants, producing blue scintillator light.
 
Wavelength-shifting fibres are placed in $\sigma$-shaped grooves cut into the scintillator, with the \gls{WLS} fibres glued into optical connectors at the outer radius of the counter. The same type of \gls{WLS} fibres are used in the cryostat scintillators and at the outer radius of the \gls{MBTS} counters. Green scintillation light needs to be shifted to orange in the \gls{WLS} fibres used in the inner segment \gls{MBTS} counters. This requires a different type of \gls{WLS} fibre than, for example, in the cryostat scintillators or in the outer segment of the \gls{MBTS} counters. O2 fibres from Kuraray, shifting the green light to orange are used. Optical cables transmit the light from the \gls{MBTS} counters to the standard optical path of the \gls{Tile} readout, with the difference that red-sensitive photo-multipliers (Hamamatsu R7600-20 ERMA) are used.
 
\begin{figure}[hp]
\centering
\subfloat[]{\label{fig:MBTSModule}\includegraphics[width=0.45\textwidth,trim=0 0 -20 0]{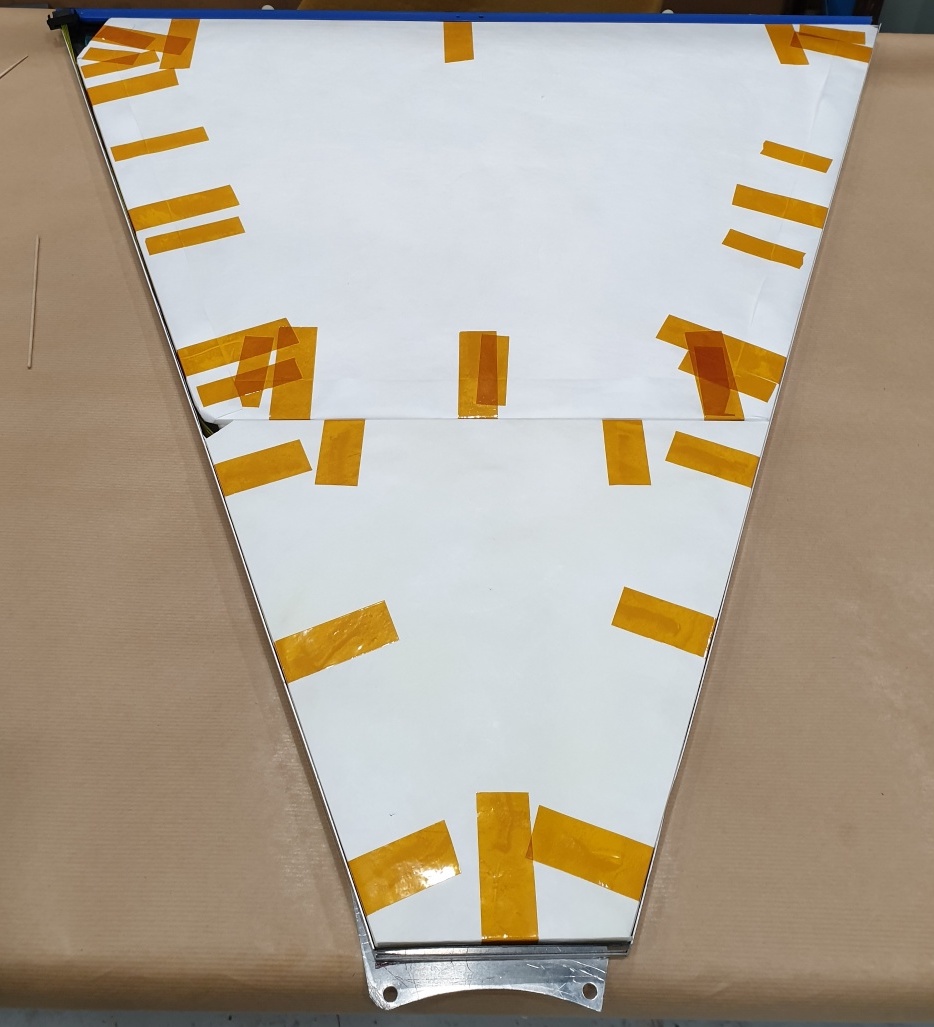}}
\subfloat[]{\label{fig:MBTSfibers}\includegraphics[width=0.45\textwidth,trim=-20 0 0 0]{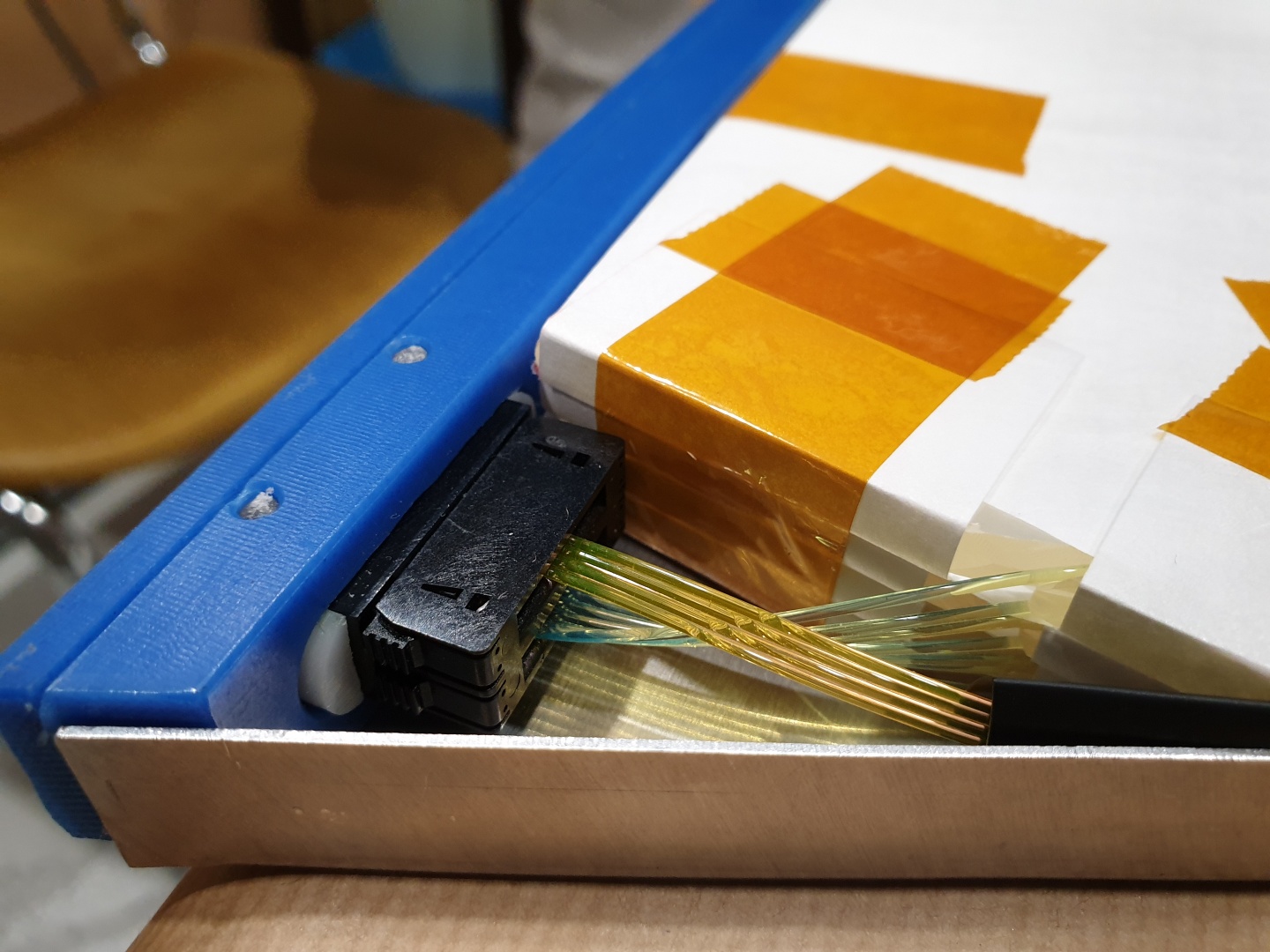}}
\caption{\protect\subref{fig:MBTSModule} An assembled \RunThr\ \gls{MBTS} counter without the top cover. \protect\subref{fig:MBTSfibers} Arrangement of wavelength-shifting fibres from the scintillator sandwich to the optical connector.
}
\label{fig:MBTS_Run3}
\end{figure}

For both the inner and outer \gls{MBTS} regions, the scintillator is divided into four layers, each \SI{5}{\mm} thick (rather than the single layer of \SI{2}{\cm} used in \RunOneTwo), individually wrapped with Tyvek sheets.
This layered structure presents more surface area to allow for the diffusion of oxygen, accelerating the annealing.
Each $\phi$-segment is placed inside its own aluminium can, with the edges sealed with both black electrical and aluminium tape. The aluminium cans are bolted onto a boronated polyethylene moderator attached to the front of the endcap cryostat, as shown in Figure~\ref{fig:MBTS_Run3}.
The outside of the aluminium can is protected by a white cover, as seen in Figure~\ref{fig_atlas_crack_MBTS} .
 
The new \gls{MBTS} scintillator, along with the new \gls{WLS} fibre readout used in the inner segment, has been shown to be less sensitive to radiation effects. Using four separate \SI{5}{\mm} thick sections of scintillator instead of a solid \SI{2}{\cm} thick piece of scintillator should allow for more oxygen diffusion into the scintillator, and thus a faster annealing of the color centres formed upon radiation exposure.
 
\subsubsection{TileCal Demonstrator for \RunFour}
The readout electronics for the Tile Calorimeter for the \gls{HL-LHC} running will have to be upgraded to deal with the increased radiation levels and the increased out-of-time pileup. Substantial progress towards this Phase~II upgrade was already made by the start of \RunThr. In order to gain experience with the new readout scheme prior to the high-luminosity running, a hybrid demonstrator module, combining the new \gls{Tile} module read-out scheme, but still compatible with the present (to be used in \RunThr) readout system, was constructed to be used for one barrel module in \RunThr. The demonstrator electronics were extensively tested in test beams, and were inserted into one barrel module on the ATLAS detector in 2019. The plan is for this demonstrator to remain in place for at least part of \RunThr. The goal is to carry out further studies under realistic ATLAS running conditions.
 
The demonstrator consists of a super-drawer that partitions a legacy \gls{Tile} drawer into four mini-drawers, each servicing up to 12 \gls{PMT} channels. The super-drawer continuously digitises \gls{PMT} signals using two gains, resulting in an effective 17-bit dynamic range, and sends the sampled data to off-detector systems at a rate of \SI{40}{MHz}.
 
The current \gls{L1} trigger system (\RunThr) is \analog, however. Compatibility with the \gls{Tile} legacy trigger system is achieved by the use of an adder-based board that groups the \gls{PMT} \analog trigger signals in pseudo-projective towers, and sends the \analog sums to the legacy \gls{L1} trigger system. The in-situ tests (since 2019) have been successful and have resulted in useful information for the future installation of the full Phase~II upgrade.


\clearpage
\newpage
 
\section{Muon Spectrometer} 
\label{sec:Muons}

\label{chapter:muons}
\subsection{Overview of Muon Spectrometer Upgrades \label{section:muonOverview}} 
The \glsfirst{MS} forms the large outer part of the ATLAS detector and
detects charged particles exiting the barrel and endcap
calorimeters, measuring their momentum in the pseudorapidity range
$\abseta<2.7$. It also triggers on these particles in the
region $\abseta<2.4$. The driving performance goal is a stand-alone
transverse momentum resolution better than 15\% 
for \SI{1}{\TeV} tracks.
This translates into a requirement to measure an effective sagitta
of about \SI{500}{\um}
with a resolution
of about \SI{75}{\um}.
In the \gls{MS}, this sagitta must be determined from tracks of particles passing through an inhomogeneous field, and measured at just three stations spanning a distance ranging from about \SI{6}{\m} in the central barrel region to more than \SI{15}{\m} in the endcaps.
 
Since the volume of the \gls{MS} is very large, it is impossible to provide continuous tracking; instead, tracks are reconstructed from straight segments reconstructed at (usually) three points: an inner station close to where the tracks exit the calorimeters and upstream of the toroid magnetic field, a middle station inside or immediately downstream from the toroid field, and an outer station well outside the magnetic field of ATLAS. The detectors in each station are multilayered, and provide at least six points along the muon trajectory that can be reconstructed as a straight track segment with a well-measured orientation in space, especially in the bending direction of the toroid magnets. Tracks are then constructed by fitting curved paths to the three ``pointing segments'' and matching them to tracks from the \gls{ID}.
 
\begin{figure}[!hp]
\centering
\subfloat[Cut through Small sectors]{\label{fig:MuonSmallQuadrant}\includegraphics[width=0.9\textwidth,trim=0 3cm 0 4cm,clip]{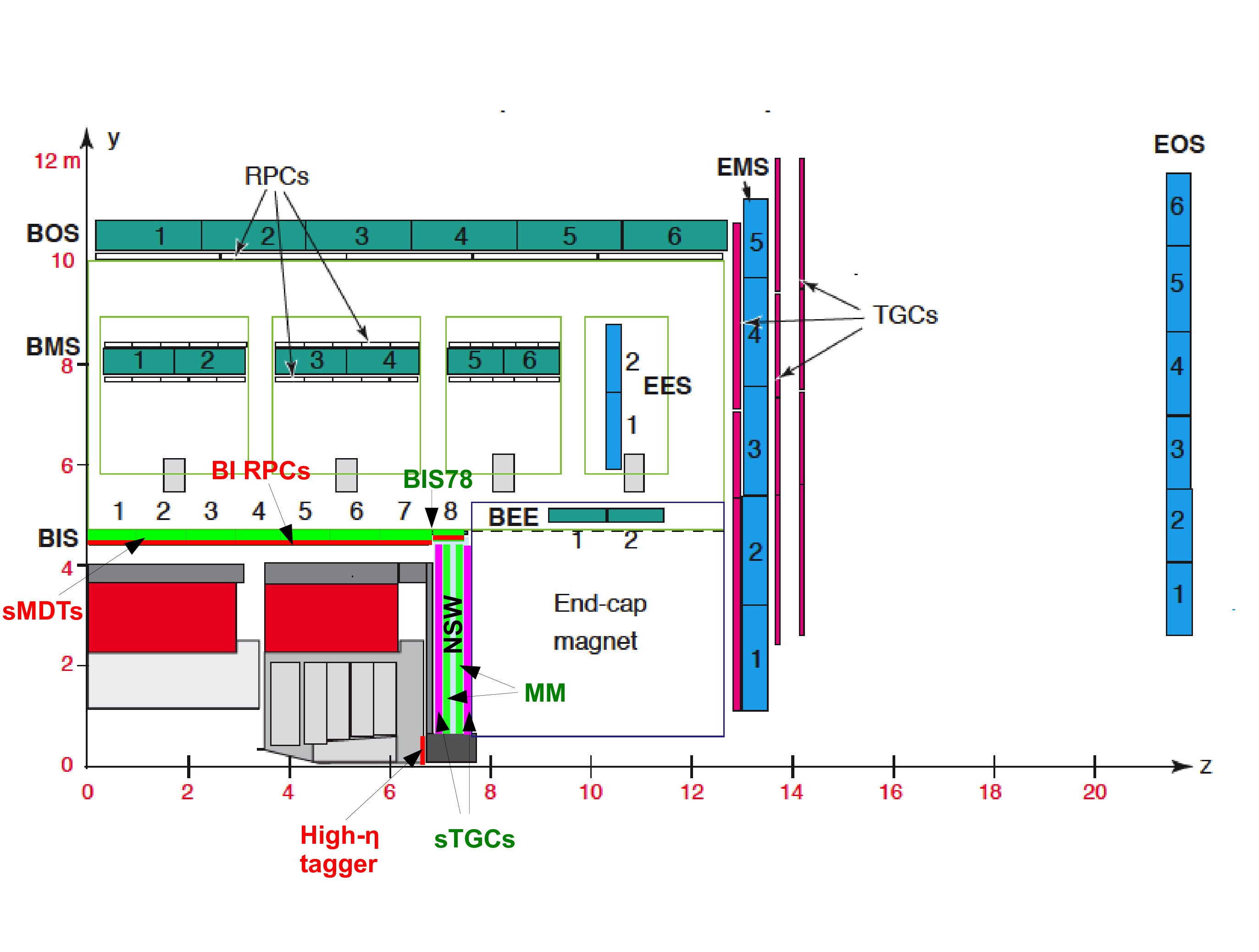}} \\
 
\subfloat[Cut through Large sectors]{\label{fig:MuonLargeQuadrant}\includegraphics[width=0.9\textwidth,trim=0 3cm 0 4cm,clip]{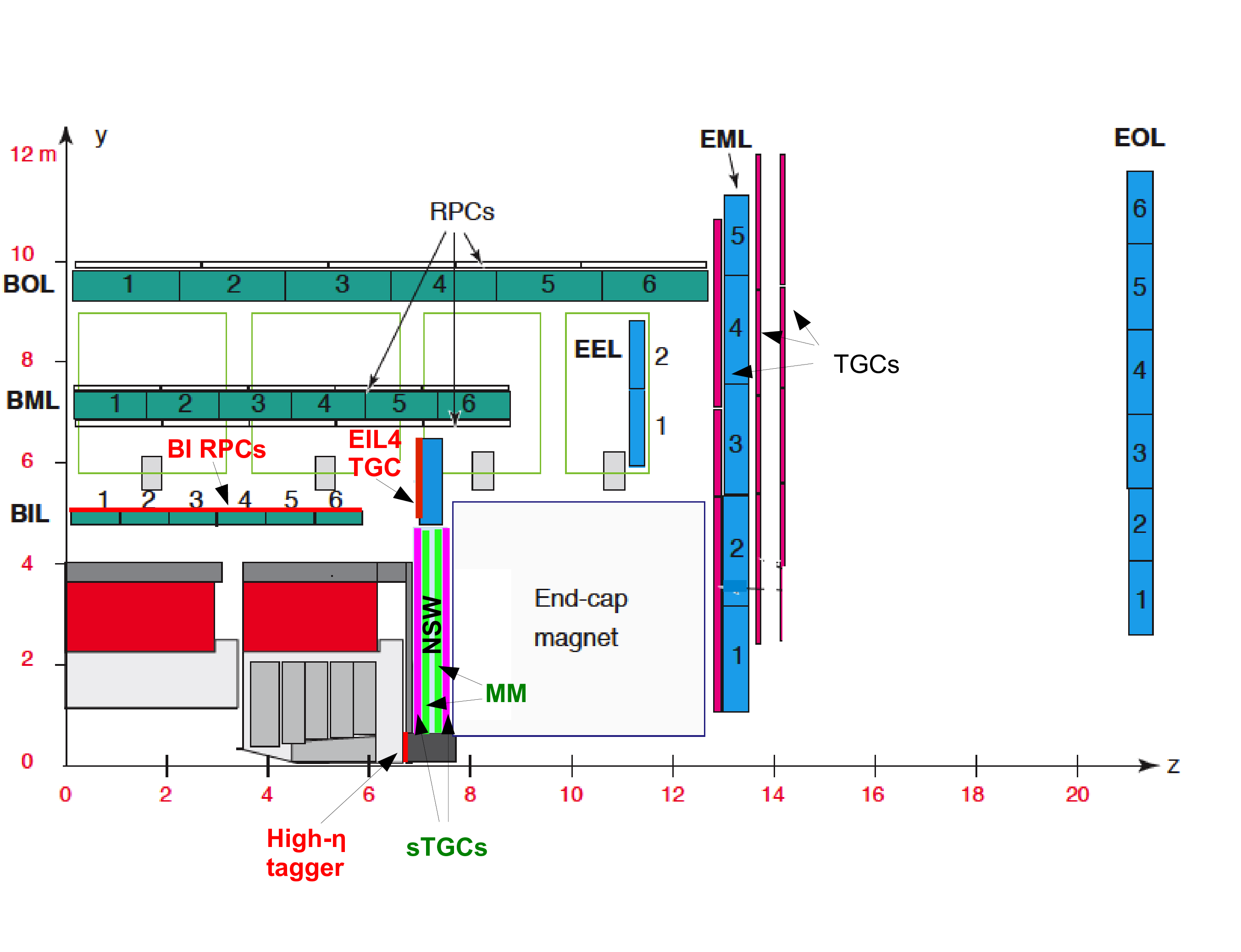}}
 
\caption{Schematic quadrant view showing locations of the main components of the \gls{MS} for \RunThr. \protect\subref{fig:MuonSmallQuadrant} shows a cut through a Small sector and \protect\subref{fig:MuonLargeQuadrant} a cut through a Large sector. The new detectors installed in \gls{LS2} are the \gls{NSW} (the new EI wheel), visible on both cuts in lime green and magenta, and the BIS chambers in positions 7 and 8 on Side~A, labelled in green as \gls{BIS78}. The blue and dark green rectangles are \glspl{MDT}. The large red rectangles represent the Tile Calorimeter. Components indicated with red labels will not be installed until \RunFour~\cite{ATLAS-TDR-26}.
}
\label{fig:Muon_quadrant}
\end{figure}

The \gls{MS} comprises a Barrel, consisting of three concentric, roughly cylindrical, stations (Barrel Outer: BO, Barrel Middle: BM and Barrel Inner: BI), and two Endcaps, each consisting of three discs, referred to as Wheels (Endcap Outer, Middle and Inner: EO, EM, and EI), as well as an Extended Endcap Ring (EE) positioned outside the radius of each endcap toroid cryostat.
The EI wheels, also called the Small Wheels, sit between the calorimeters and the endcap toroid cryostats, inside the barrel toroids (see Figures~\ref{fig:Muon_quadrant} and~\ref{fig:Overview:muons}).
The EM wheels (or Big Wheels), are located on the far side of the endcap toroid cryostats. The EO wheels, similar in size to the EM wheels, are mounted on metal scaffolding structures attached to the end-walls of the ATLAS cavern.
The EE rings provide a third measurement station in planes between the EI and EM wheels for tracks in the intermediate pseudorapidity region, $1.05<\abseta < 1.3$, not covered by the EO wheel.
The outermost (\gls{EIL4}) stations of the EI region covering the range $1.0<\abseta<1.25$ are not mounted on the EI wheels; they are instead permanently fixed between the barrel toroid coils.
The \gls{MS} has an eight-fold symmetry: eight Small (S) sectors are aligned with the eight coils of the barrel toroid magnet, and eight Large (L) sectors cover the regions between the coils.
\footnote{
Sectors are numbered from \numrange{1}{16}, starting with the Large sector at the $x$-axis of the ATLAS coordinate system and proceeding azimuthally (thus, Sector~5 is at the top of the detector, and Sector~13 at the bottom).
Modules within each sector are numbered sequentially according to their distance from the \gls{IP}: thus, BIL1 chambers sit near the centre ($z=0$) of the Barrel Inner layer in the Large sectors, while EOS6 chambers are at the outer radius ($r\sim\SI{11}{\m}$) of Small sectors in the Endcap Outer wheel, as shown in Figure~\ref{fig:Muon_quadrant}.
The only chambers at $z=0$ are the BOL0 chambers in Sectors~12 and 14. Other sectors have a gap at $z=0$ to allow services to exit from the calorimeters.
}
 
The majority of the barrel detectors are unchanged from the original \RunOne configuration described in Ref.~\cite{ATLAS-TDR-20}: all three stations use multilayered
\glsfirst{MDT}
chambers for the precision measurements in the bending coordinate, and the Outer (BO) and Middle (BM) stations are also equipped with
\glsfirstplural{RPC}
for triggering and to measure the azimuthal coordinate of the tracks.
Additional barrel chambers added since \RunOne are described in Section~\ref{muonSS:BMEBOEBMG}.
In response to the increasing number of gas leaks due to cracks in the gas inlets of the \gls{RPC} system that developed over time during \RunTwo,
significant work was undertaken during \gls{LS2} to reinforce the \gls{RPC} gas inlets and recover a large number of channels that had become inactive. Inlets were repaired and no-return valves installed. The gas inlets on \num{100} service boxes (two out of the four on the corners of each of \num{50} chambers) were reinforced by foam injection in 2022. Preliminary results indicate that this prevents new leaks from developing, and it will be done for all the accessible service boxes during winter shutdowns over the course of \RunThr. 
 
During \RunOneTwo, the endcap \gls{MS} comprised three technologies: \glspl{MDT}, as in the barrel, \glsfirstplural{CSC} in the innermost region of the EI wheel, and \glsfirstplural{TGC}, used for triggering and to provide the azimuthal coordinate of muons in the endcaps of the ATLAS \gls{MS}~\cite{PERF-2007-01}.
The EO and EM wheels and the EE rings are unchanged from \RunOne.
The EO wheels contain only \glspl{MDT}, and therefore measure only in the (precision) bending coordinate, while the EM wheels
each comprise four distinct wheels: one of \gls{MDT} chambers for precision tracking in the bending coordinate,
sandwiched between one of triplet \glspl{TGC} and two of doublet \glspl{TGC} for triggering and for measuring the azimuthal coordinate.
The \num{528} \gls{TGC} doublet modules and \num{216} triplet modules in the three EM \gls{TGC} Wheels
of each endcap continue to play the same essential roles in the \gls{L1} muon trigger and in offline tracking as they did in \RunOneTwo.
During \gls{LS2}, \num{24} 
faulty \gls{TGC} modules in the EM wheels were replaced by spares, increasing the number of operational channels.
The EE rings consist only of \gls{MDT} detectors for tracking, and do not participate in the trigger.
In the inner (EI) wheels of the endcap \gls{MS} from \RunOneTwo~\cite{PERF-2007-01}, however, the original \gls{TGC} doublets did not have sufficient resolution in the bending direction, nor enough detector layers, to form pointing segments for the \gls{L1} trigger.
For the \RunThr trigger, the EI wheels are required to provide pointing segments of sufficient angular precision to test for matches with muon candidates in the EM wheels and for consistency with muon tracks originating from the \gls{IP}.
The \glspl{SW} have therefore been completely replaced by the
\glsfirstplural{NSW}, and are the main subject of this chapter.
The sixteen \gls{EIL4} assemblies, comprising eight \gls{MDT} detectors and \num{21} \gls{TGC} doublets on each side of ATLAS, fixed between the barrel toroid magnets, are retained for \RunThr.
EM-wheel trigger candidates without corresponding \gls{NSW} segments satisfying these matching criteria can be dropped, and this is expected to reduce the false-trigger rate to an acceptable level.

The main objective of the Phase-I ATLAS upgrades~\cite{ATLAS-Phase-I-LOI}
is to sharpen the \gls{L1} trigger threshold
turn-ons and discriminate against background
while maintaining
the \gls{L1} rate at a manageable level.
This must be achieved without pre-scaling or raising \pT thresholds for the single-muon \gls{L1} triggers, which would result in a significant loss of
acceptance for many interesting physics processes.
 
When the design of the Phase-I upgrades began, the estimated total rate of the \gls{L1} trigger for single muons with $p_T >$\SI{20}{\GeV} was expected to rise above \SI{50}{\kHz} in \RunThr
if no measures were taken, while ATLAS can allocate only \SI{25}{\kHz}
for muon triggers out of a total \gls{L1} bandwidth of \SI{100}{\kHz}~\cite{ATLAS-TDR-23}.
This scenario will be relevant for the \gls{HL-LHC} but, because of luminosity-levelling, will probably not be experienced during \RunThr.

\begin{figure}[htbp!]
\centerline{\includegraphics[width=0.5\textwidth]{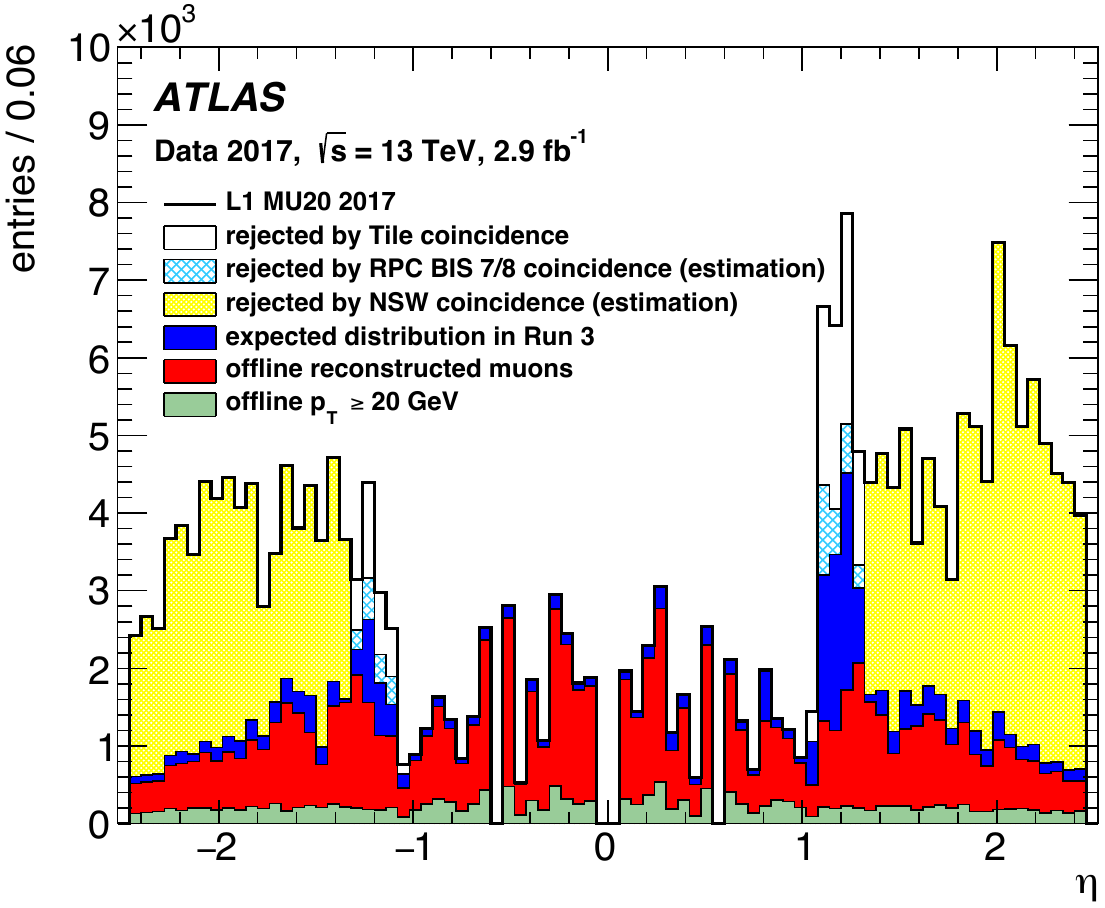}}
\caption{The expected muon trigger rate reduction from the \gls{NSW}, \gls{BIS78} and \glsfirst{L1Muon} endcap upgrades; however, the \gls{BIS78} on the negative-$\eta$ side, where the fake trigger rate is lower, will only be installed after \RunThr.
}
\label{fig:TDAQL1MuonEndcapPerformance}
\end{figure}

The \gls{MS} upgrades focus chiefly on the endcap regions.
The \RunOne endcap muon trigger relied only on the \glspl{TGC} of the EM wheels.
There, a substantial background arises (see Figure~\ref{fig:TDAQL1MuonEndcapPerformance}), proportional to the instantaneous luminosity, due to relatively low-energy charged particles emerging
from hadronic showers in the forward shielding, that enter the endcap toroid cryostats without passing through the EI muon wheel. The paths of these particles are then bent in the toroids, and a fraction of them enter the EM wheels on trajectories that closely mimic those of muons coming from the \gls{IP}. This led to a large fake rate when only the EM wheels were used to trigger~\cite{ATLAS-TDR-20}.
The number of \glspl{RoI} identified by the \gls{L1} single muon trigger thus increases sharply for $\abseta > 1$, while the $\eta$ distribution of reconstructed muons is almost flat. Most of the \glspl{RoI} for $\abseta > 1$ in \RunOne were due to this shielding-induced background.
The asymmetry seen in the trigger between the two ends of the detector is due to the opposite bending directions of the positively charged protons originating from the shielding interactions.
 
Since the background rate in the endcap region (including the transition region between the barrel and endcaps) increases with luminosity, it will eventually exceed the allocated \gls{L1Muon} trigger rate.
Failure to suppress this background would result in the need to increase the \pT\ threshold of the lowest unprescaled muon trigger.
The first steps to reduce the charged-particle background were taken in \RunTwo, when coincidences with the \gls{TGC} doublets in the \glspl{SW}, or with the Tile Calorimeter, were required in the \gls{L1} trigger to ensure that particles causing \gls{L1Muon} triggers in the endcap regions were on trajectories consistent with an origin at the \gls{IP}.
However, the background rates in the region of the EI stations are much higher than in the EM stations (as discussed in Section~\ref{sss:bkgCaloMuon}), so the relatively coarse granularity of the legacy EI
\glspl{TGC} implied that at high luminosity, nearly all EM triggers would have had corresponding background hits in the \glspl{SW}, rendering the coincidence requirement less effective.
Moreover, the \SI{30}{mm}-diameter \glspl{MDT} used in the \glspl{SW} are limited by space charge build-up effects~\cite{4436512} at high background rates, and their precision tracking capability was expected to degrade at the \gls{HL-LHC}.
The goal of the Phase~I upgrade was therefore to replace the EI wheels with detectors able to provide very fast and precise track-segment matching, while maintaining the offline tracking performance of the \glspl{SW} from \RunOneTwo.
 
To accomplish this, the \glspl{SW} have been completely replaced by \glsfirstplural{NSW} occupying the same envelopes and providing tracking over the same polar angle range: $1.3 < \abseta < 2.7$. These are described in Section~\ref{MuonSS:NSW}.
The large-radius region of the inner station
$1.0 < |\eta| < 1.3$ remains covered by the \gls{EIL4} \gls{MDT} and doublet \gls{TGC} detectors of the
original muon endcap system, which are mounted between the toroid coils and not attached to the EI wheels. The \gls{TGC} doublets in \gls{EIL4} will continue to provide adequate confirmation under \RunThr conditions that a particle has traversed the endcap
toroid zone, reducing the fake endcap triggers in this region; they are scheduled for replacement by triplet \glspl{TGC} during \gls{LS3} to improve their efficiency in rejecting fake triggers.
 
The
\gls{EIL4} chambers, however, cover only about 70\%~\cite{ATLAS-TDR-26} of the full azimuthal
angle in a ring surrounding the EI wheels (now the \glspl{NSW}), with gaps for the barrel toroid coils in the Small sectors.
To fill these azimuthal gaps in trigger coverage, the original BIS7 and BIS8 \gls{MDT} chambers (those farthest from the \gls{IP}) on Side~A of ATLAS were replaced with new, smaller, \gls{BIS78} chambers using
\glspl{sMDT}, leaving enough space to add \glspl{sRPC}
to complete the trigger coverage in the Small sectors of this transition region (described in Section~\ref{muonSS:BIS78}).
 
The \gls{sMDT} technology used in the \gls{BIS78} upgrade was first introduced in chambers used to fill several small holes in the original coverage of the \gls{MS} during winter shutdowns in the course of \RunTwo. These upgrades are briefly described in Section~\ref{muonSS:BMEBOEBMG}, along with some additional (original technology) \gls{MDT} chambers added during the first \gls{LHC} long shutdown.
 
\subsection{Endcap Upgrades: The New Small Wheels \label{MuonSS:NSW}}
The \gls{NSW} detectors operate in a high-background
radiation region (where detected hit rates could potentially increase to as much as \SI{20}{\kHz/\cm\squared} in the small region closest to the beamline, at the upper-limit estimate of \gls{HL-LHC} luminosity of \lumihllhchigh),
reconstructing muon tracks with high precision and furnishing
information for the \gls{L1} trigger~\cite{ATLAS-TDR-20}.
 
The \glspl{NSW} use two chamber technologies: \glsfirst{sTGC} detectors, and  \glsfirst{MM} detectors.
Both can be used for triggering: the single \gls{BC} identification capability (more than \SI{95}{\percent} of signals are  collected within one \gls{BC}) of the \glspl{sTGC} makes them particularly suitable for the primary
triggering role, and the small drift gap in the \gls{MM} chambers means that they are also very fast, with very little dead time.
Both technologies have precision tracking capabilities, with ultimate resolutions in the bending direction of the order of \SI{100}{\micron}.
In the \gls{MM} detectors the resolution is due in part to their very fine strip pitch (less than \SI{0.5}{\mm}), while the \glspl{sTGC} (with a strip pitch of \SI{3.2}{\mm}) rely more heavily on charge sharing.
Such precision is crucial to maintain the
current muon momentum resolution in the high background
environment of the upgraded \gls{LHC} when the full track reconstruction is run.
Second-coordinate resolution (in the azimuthal direction) is greatly improved with respect to the \glspl{SW}, with good trigger-level second-coordinate resolution from \gls{sTGC} pads and \gls{MM} stereo roads, and excellent resolution from the \gls{sTGC} wire groups and full \gls{MM} stereo strips in the final offline reconstruction.
The \gls{sTGC}--\gls{MM} chamber technology combination thus forms a fully redundant detector system for tracking, both for the trigger and
offline. This detector combination is designed
to provide robust, fast, high-resolution performance for the \gls{HL-LHC}, as well as meeting the immediate requirements of \RunThr.
 
The demanding performance requirements are summarised in Section~{\ref{muonSS:performance}}. A brief description of the layout of the \gls{NSW} follows in Section~{\ref{muonSS:layout}}. The \gls{MM} and \gls{sTGC} technologies are described in Sections~{\ref{muonSS:MM}} and~{\ref{muonSS:sTGC}} respectively. The mechanics of the supporting disc, spokes and shielding hub are summarised in Section~{\ref{muonSS:shielding}}, and Section~{\ref{muonSS:alignment}} covers the extensions to the endcap optical alignment system. The trigger hardware mounted on the detector (along with a broad outline of the logic of its firmware) is discussed briefly in Section~{\ref{muonSS:trigger}}, while the off-detector elements of the trigger are described later, in Section~{\ref{sec:TDAQ_L1Muon}}. Changes to the services for the EI wheel to accommodate the new detectors are described in Section~{\ref{muonSS:services}}.
 
\subsubsection{Performance requirements for the NSW \label{muonSS:performance}}
The \gls{NSW} has the same precision requirements as the \gls{SW}: it must provide track segments that, in combination with measurements from the \gls{ID} and the EM and EO wheels, allow the endcap \gls{MS} to measure the transverse momentum (\pT) of passing muons with a precision of better than \SI{15}{\percent} for \SI{1}{\TeV} muons in the full pseudorapidity coverage $\abseta < 2.7$.
To satisfy this requirement, the \gls{NSW} must reconstruct track segments with a position resolution of about \SI{50}{\micron} in the bending plane, irrespective of background conditions, and even if some detector planes are not operational. With sixteen detector planes, of which twelve measure the precision coordinate (the radial, or bending, direction), this implies a resolution of the order of \SIrange{150}{175}{\micron} is required for each plane of detector strips.
 
The \gls{L1} trigger track
segments must be reconstructed online with an angular resolution of
approximately \SI{1}{\mrad}, matching the ultimate (\RunFour) angular resolution of the EM trigger segments. The angular resolution requirement for \RunThr comes from the $\eta$-resolution of the EM-wheel \glspl{TGC}, and is about \SI{3}{\mrad}, but new electronics for the \gls{MDT} chambers of the EM Wheels in the Phase~II upgrade~\cite{ATLAS-TDR-26} will allow the \glspl{MDT} to participate in the \gls{L1} trigger at the \gls{HL-LHC}, improving the resolution of the EM-wheel segments.
 
Segment finding efficiencies must be better than 97\% for muons with \pT greater than \SI{10}{\GeV} to equal the performance of the \glspl{SW}.
Efficiencies and resolutions are required not to degrade for high-momentum muons that emit $\delta$-rays or showers.
 
The muon \gls{L1} trigger rate must be kept below \SI{25}{\kHz} and fit within the overall ATLAS \gls{L1} fixed latency budget for \RunThr of approximately \SI{2}{\us}.
The trigger design is also required to meet the ATLAS \gls{HL-LHC} \gls{L0}~\footnote{During \RunThr, there is no difference between the \gls{L0} and \gls{L1} trigger rates, but the readout is designed to function at the \gls{HL-LHC}, where a two-level trigger with \gls{L0} running at \SI{1}{\MHz} with up to \SI{10}{\SIUnitSymbolMicro\s} latency will be used.}
latency budget of \SI{10}{\us}, for a single-muon \gls{L0} trigger rate expected to be around \SI{45}{\kHz}~\cite{ATLAS-TDR-29}.

\subsubsection{Layout of the NSW \label{muonSS:layout}}
\begin{figure}[!h]
\subfloat[]{\label{fig:Muon_NSW_vp1}\includegraphics[width=0.59\textwidth]{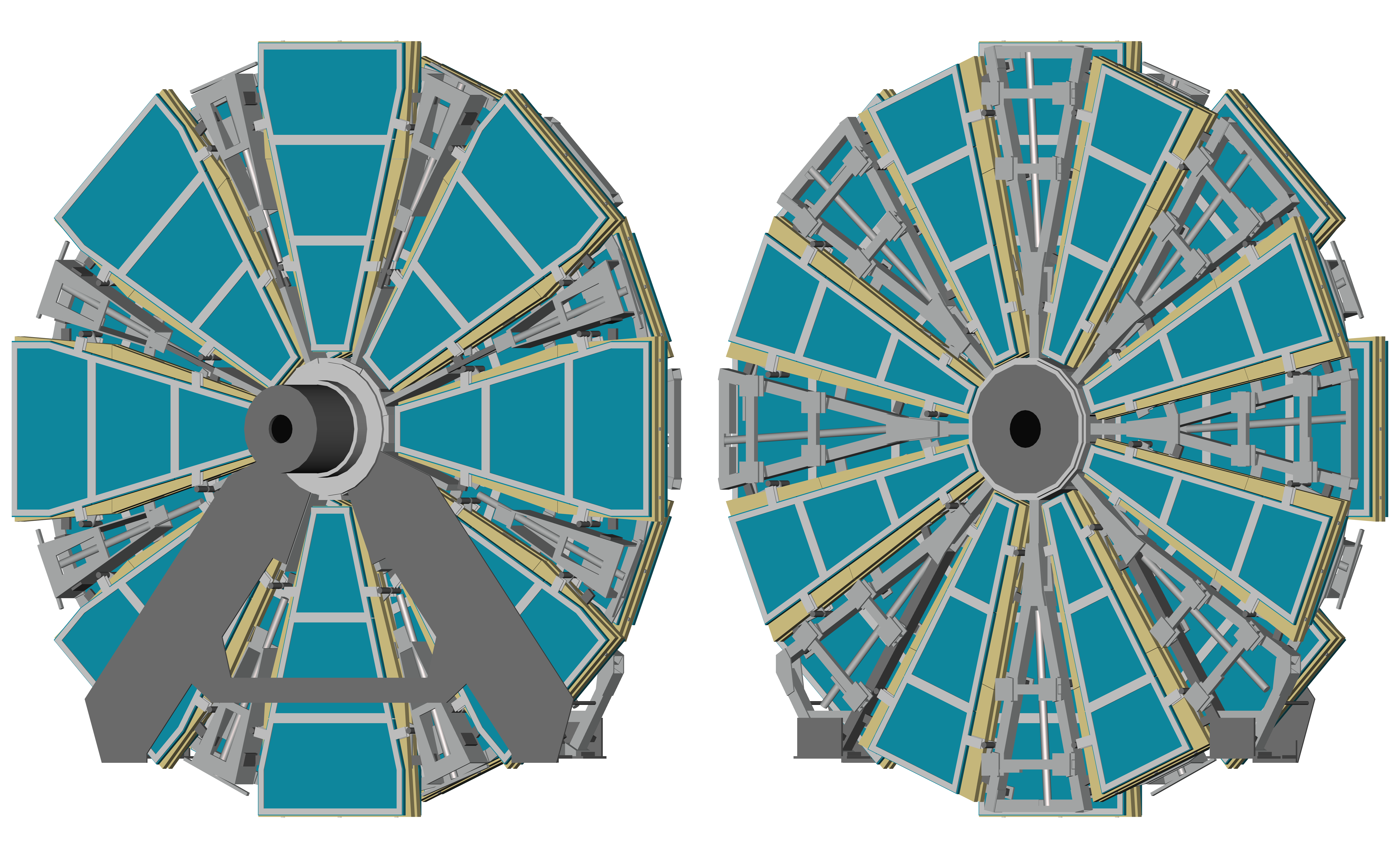}}
\hfil
\subfloat[]{\label{fig:Muon_NSW_photo}\includegraphics[width=0.4\textwidth]{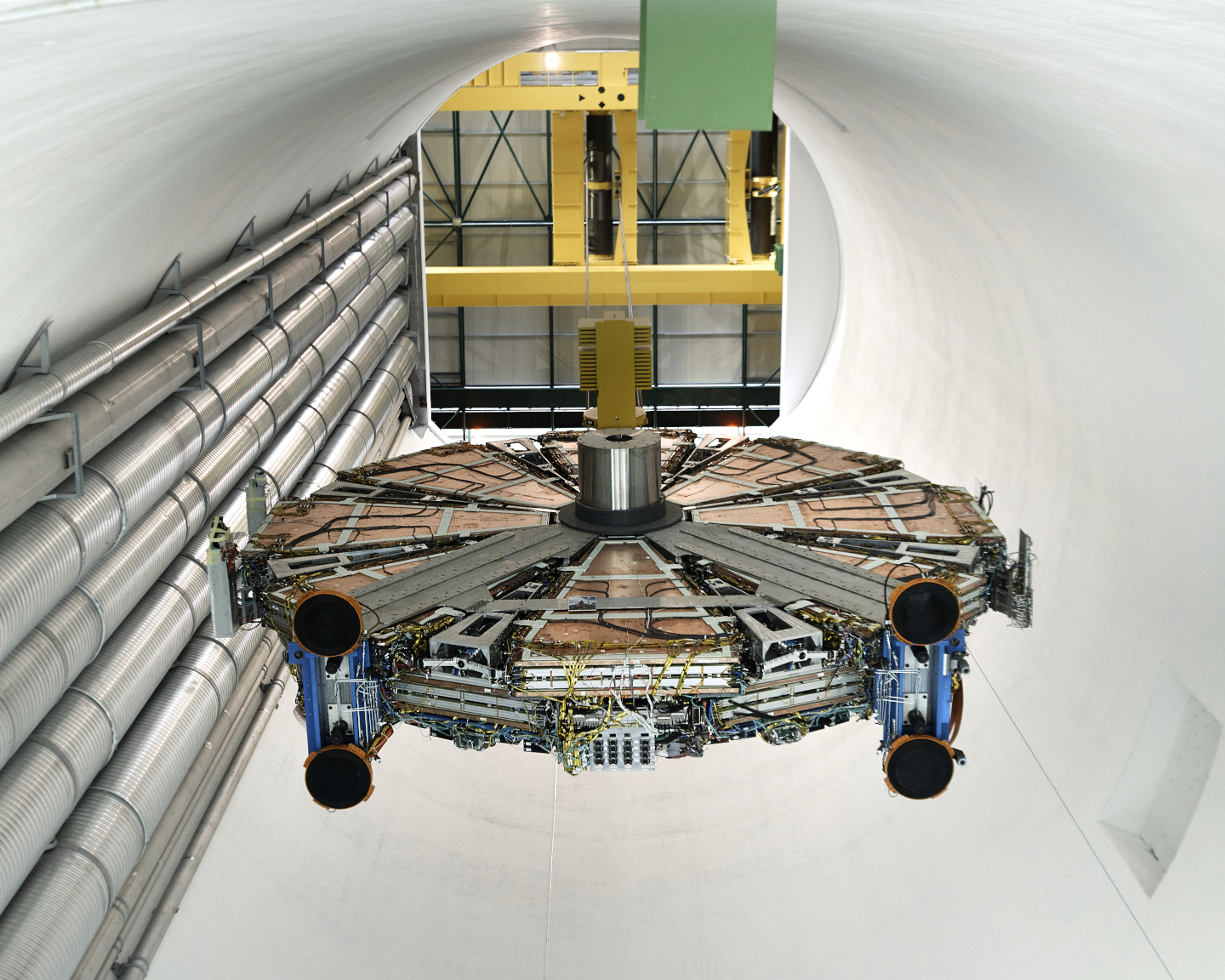}}
\caption[New Small Wheels]{New Small Wheels: \subref{fig:Muon_NSW_vp1} shows the structure: the wheel on the left is seen from the cavern end-wall, with the A-frame and shielding hub visible, along with the eight Large sectors and the eight spokes supporting them; the wheel on the right is seen from the \gls{IP}, and since the \gls{newJD} shielding disc is not shown, the eight Small sectors, aligned with the barrel toroid coils, are visible, along with the spokes supporting them. Some of the structural elements can be seem more clearly in Figure~\protect\ref{fig:Muon_NSWPassive}.
\subref{fig:Muon_NSW_photo} shows the first \gls{NSW} being lowered into the ATLAS cavern; it is shown from the side with the Large sectors and the A-frame.  \label{fig:Muon_NSW}}
\end{figure}

The geometry of the \gls{NSW} retains the eight-fold symmetry of most of the legacy \gls{MS}. Each \gls{NSW} consists of 16 sectors: eight Large, and eight Small (see Figure~\ref{fig:Muon_NSW} and Table~\ref{table:Muon_modularity}). The Small sectors are aligned with the barrel toroid coils, and form a plane close against the shielding disc (the \glstext{newJD} in Figure~\ref{fig:Muon_NSWPassive}). The Large sectors form a second plane, slightly farther from the \gls{IP}. The Large and Small sectors overlap mechanically, to avoid uninstrumented regions between the sectors; the instrumented areas of the \gls{MM} overlap between adjacent Large and Small sectors, but the instrumented \gls{sTGC} areas of adjacent sectors do not overlap.
 
\begin{table}[h!]
\caption[Modularity of NSW components]{Modularity of the \gls{sTGC} and \gls{MM} in the \gls{NSW}. There are two \gls{MM} wedges and two \gls{sTGC} wedges in each sector (Large or Small), and there are eight Large sectors and eight Small sectors in each wheel.
\label{table:Muon_modularity}
}
\centering
\begin{tabular}{llll}
\hline
Module & Per Quadruplet & Per Wedge & Full ATLAS (2 wheels)\\[0.5ex]
\hline
Large sectors (\ang{28} azimuthal) &  &  & 16 \\
Small sectors (\ang{17} azimuthal) &  &  & 16 \\
Total sectors & & & 32 \\
\gls{sTGC} Wedges & & & 64 \\
\gls{sTGC} Quadruplets & & 3 & 192 \\
\gls{sTGC} Gas Volumes & 4 & 12 & 768\\
\gls{MM} Wedges & & & 64\\
\gls{MM} Quadruplets & & 2 & 128\\
\gls{MM} Gas Volumes & 4 & 8 & 512\\[1ex]
\hline
\end{tabular}
\end{table}
 
\begin{figure}[!h]
\subfloat[]{\label{figMuon:WedgesNoFrames}\includegraphics[width=0.74\textwidth]{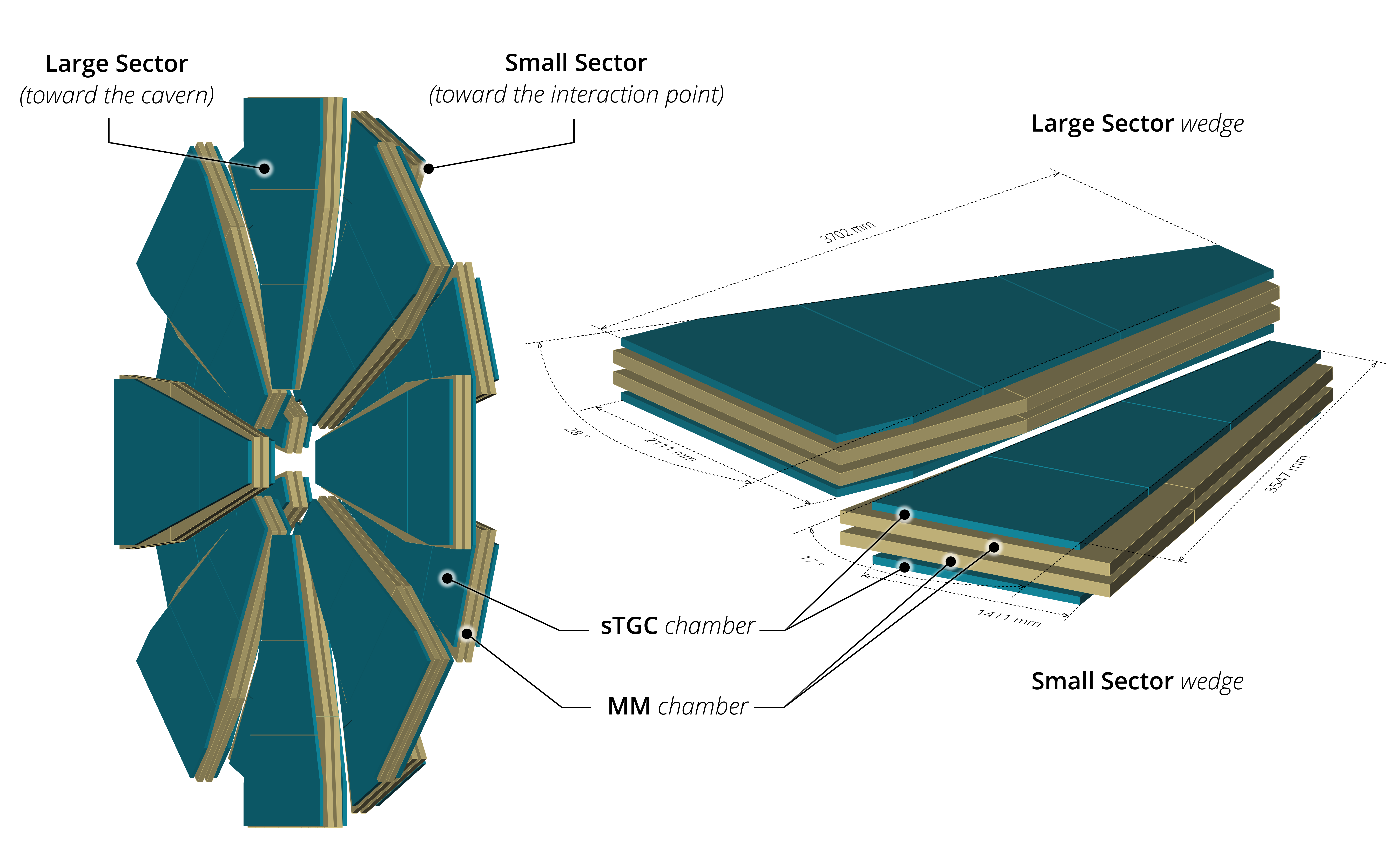}}
\subfloat[]{\label{figMuon:WedgesSide}\includegraphics[trim=1300 0 3900 0,clip,width=0.25\textwidth]{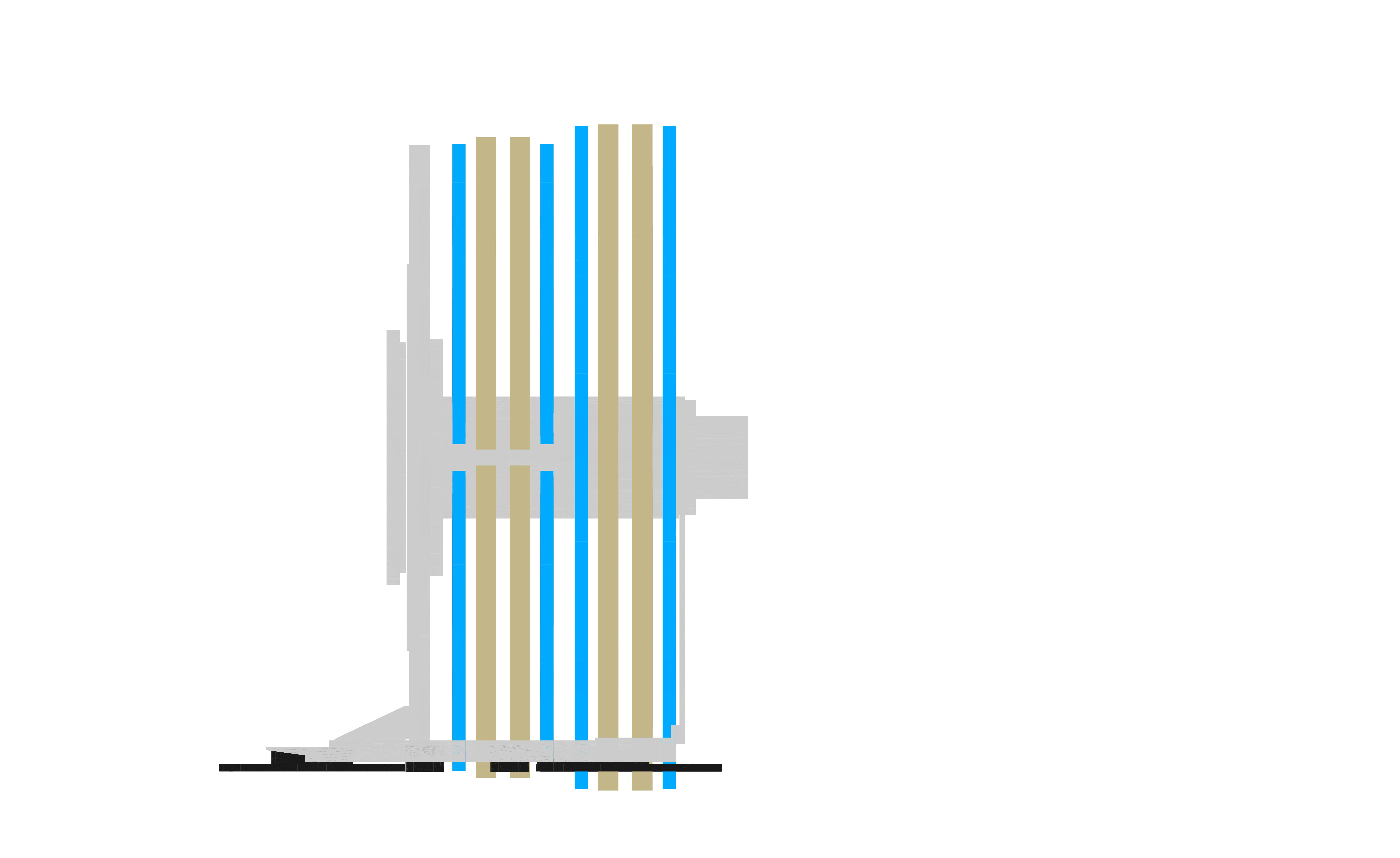}}
\caption{\protect\subref{figMuon:WedgesNoFrames} Structure of Small and Large \gls{NSW} sectors. Each sector consists of two wedges of \gls{sTGC} (green) hung kinematically on either side of a central spacer frame, to which \gls{MM} wedges (brown) are directly bolted on each side. Each \gls{sTGC} wedge consists of three quadruplets, and each \gls{MM} wedge consists of two quadruplets. \protect\subref{figMuon:WedgesSide} $rz$-plane view of an \gls{NSW} with $z$-scale exaggerated, showing from left to right (\gls{IP} to cavern wall): the \gls{newJD} in grey, the Small sectors, and finally the Large sectors.  Each sector contains \num{16} active layers: eight of \gls{MM} and eight of \gls{sTGC}. The \gls{sTGC} wedges are shown in blue and the \gls{MM} wedges in brown.}
\label{figMuon:Wedges}
\end{figure}

Both the \gls{MM} and the \gls{sTGC} detectors are built in (mostly) trapezoidal modules with four layers of gas gaps corresponding to active detector planes. These are referred to as Quadruplets. Quadruplets of various sizes are assembled radially into Large and Small \gls{sTGC} wedges and \gls{MM} double wedges, and these are assembled to build Large and Small sectors (as shown in Figure~\ref{figMuon:WedgesNoFrames}). A sector thus comprises \num{16} active detector layers in total: eight from the \glspl{sTGC} and eight from the \gls{MM}.
 
Each sector is built on a central spacer frame that is kinematically mounted to the spokes of the wheel (see Section~\ref{muonSS:shielding}).
Each spacer frame has a wedge comprising two \gls{MM} quadruplets secured to each side using in-plane sliding attachments, and this structure is the \gls{MM} double wedge.
A wedge built from three \gls{sTGC} quadruplets, positioned precisely relative to each other with glued fibreglass frames, is kinematically mounted on each side of the spacer frame, sandwiching the \gls{MM} assembly.
This arrangement maximises the distance between the \gls{sTGC} wedges (see Figure~\ref{figMuon:WedgesSide}), which provide the primary trigger.
The relative alignment of the \gls{MM} and \gls{sTGC} quadruplet modules is monitored with an optical alignment monitoring system. This system takes readings approximately every two hours to determine the displacements, rotations, and certain deformation modes of each individual quadruplet. The extension of the optical endcap alignment monitoring system to the \gls{NSW} is explained in more detail in Section~\ref{muonSS:alignment}.
 
\subsubsection{Mechanics and Shielding \label{muonSS:shielding}} 
\begin{figure}[!h]
\centerline{\includegraphics[width=0.75\textwidth]{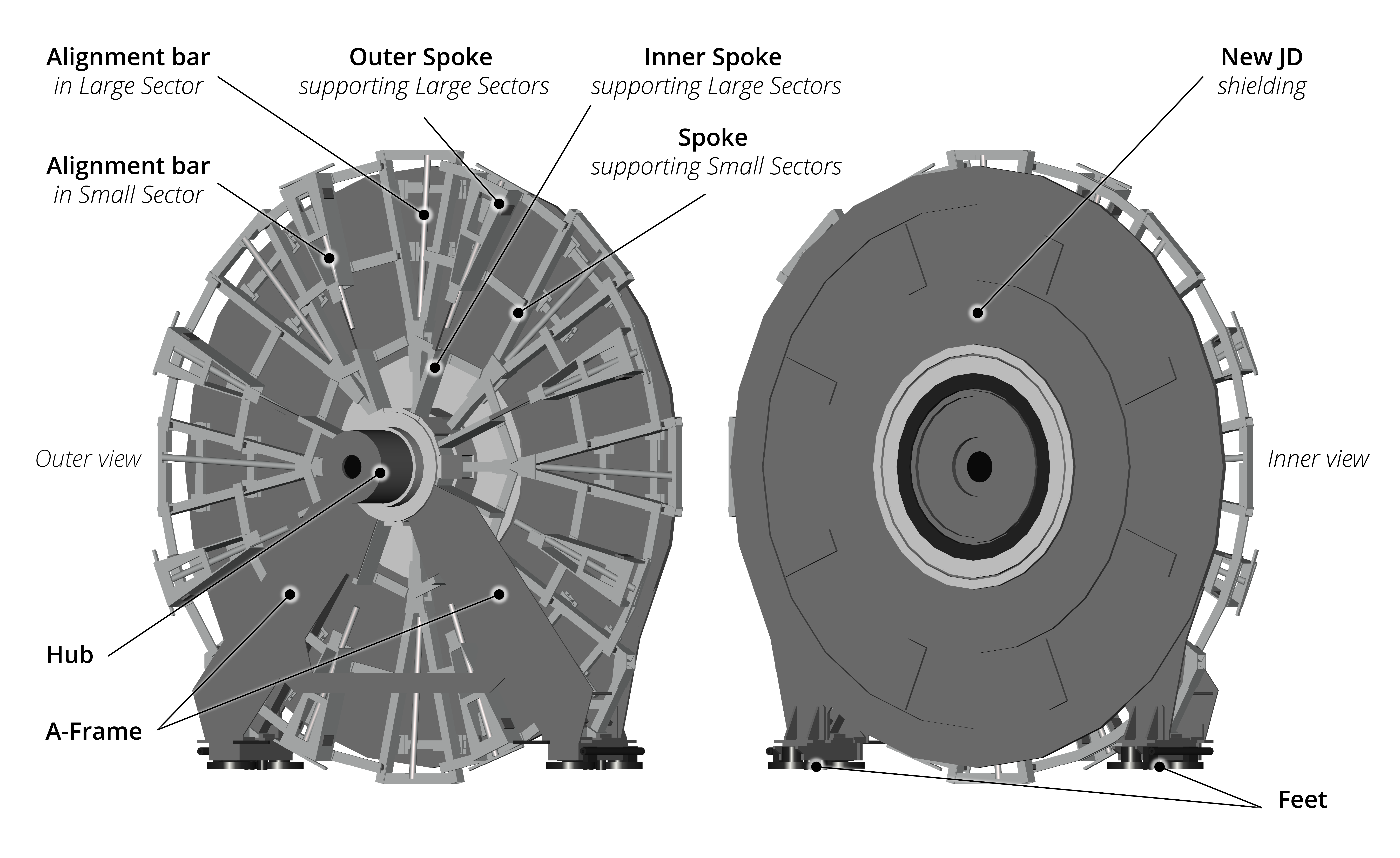}}
\caption{Support structures for the \glspl{NSW}, including the \gls{newJD} discs (which are mounted on the side closer to the \gls{IP}).}
\label{fig:Muon_NSWPassive}
\end{figure}

The increased mass and different geometry of the \glspl{NSW} with respect to the \glspl{SW} of \RunOneTwo required an entirely new mounting structure; moreover, the increased background rates anticipated at the \gls{HL-LHC} required a different shielding configuration.
The shielding disc and central hub of the EI Wheels were therefore entirely replaced, and the \glspl{NSW} are mounted on the \gls{newJD} shielding discs. 
These are made of structural steel, up to \SI{90}{\mm} thick.
The central structural axle supporting the detector wheel is the \gls{NSW} shielding hub, made of an external shell of stainless steel containing several interlocking pieces of forged copper, with no interior voids.
The \gls{NSW} supporting structure (see Figure~\ref{fig:Muon_NSWPassive}) provides the mechanical network from which the Large and Small Sectors of the \gls{NSW} are suspended.
It is built from many separate bolted and welded elements, made from either aluminium profiles and plates, or from austenitic steel.
The main elements of each \gls{NSW} structure are two ``foot spoke'' assemblies, which support the weight of the \gls{NSW}, six ``standard'' Small-sector spoke assemblies, and eight Large-sector spoke assemblies, the latter comprising both ``inner'' and ``outer'' parts.
Each spoke contains adjustable supports for an alignment bar (see Section~\ref{muonSS:alignment}) and adjustable systems for the kinematic mounts that hold the sectors.
Since the radiation hub constitutes the majority of the mass of the \gls{NSW}, large mechanical deformations are avoided by supporting the structure by a bridge-like assembly, of which the shielding disc forms one side, the other side consisting of a bolted A-frame connecting the two foot-spoke anchorages to the hub and to the aluminium structure. 
 
Three kinematic mounts secure each sector. Two of them suspend the sector from the two neighbouring spokes, one 
allowing only rotation in the plane of the sector about a fixed axis, and the second 
allowing only translation in a single direction in the plane of the detector.
The third kinematic mount 
allows only small translations in the plane.
The distribution of the kinematic mounts on the spokes is driven by the position of the maximum force applied on the fixed kinematic support.
 
Since the \glspl{NSW} are required to move out of their running position inside the barrel toroids during maintenance periods to allow access to the calorimeters, services including power, gas and cooling water are routed through four flexible cable trays (the ``flexible chains'') attached to each \gls{NSW} that allow the wheel to move without being uncabled.
 
\subsubsection{Micromegas Technology \label{muonSS:MM}} 
\begin{figure}[!h] \centerline{\includegraphics[width=0.95\textwidth]{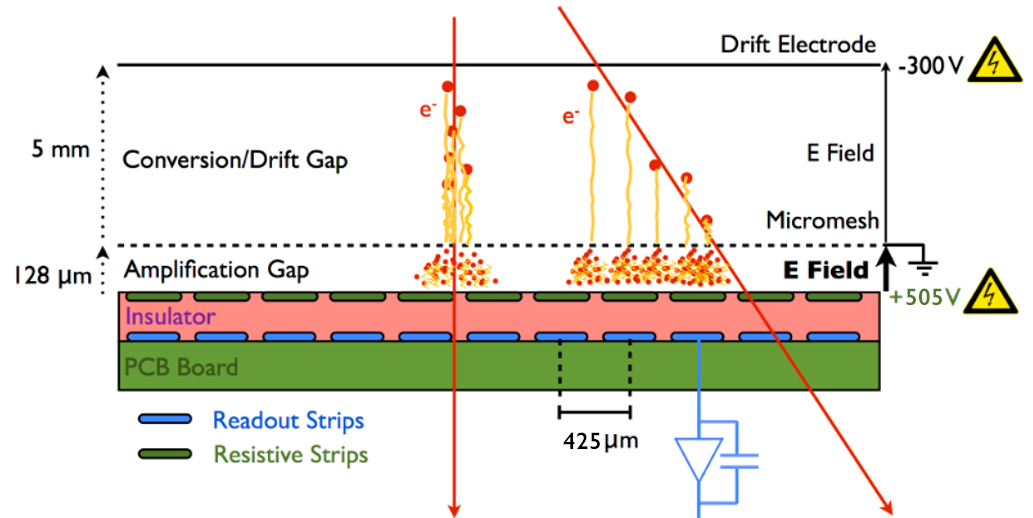}}
\caption{Layout and operating principle of the ATLAS \gls{NSW} \gls{MM} detectors. The micro-mesh is integrated into the drift panel, and stretched over the grid of support pillars on the readout panel when the two panels are pressed together. The red arrows indicate trajectories of incoming muons, the large red dots show the positions of primary ionization clusters, the diffusion paths of the ionization electrons are shown in yellow, and the small red dots in the amplification gap indicate the gain stages of the avalanche.
Parameters shown are for the inner quadruplets and baseline gas mixture; see Table~\ref{table:Muon_MMparams} for additional operating parameters.
\label{fig:Muon_MMprinciple}}
\end{figure}

\gls{MM} technology was developed in the mid-1990s~\cite{ref21TDR}. It permits the construction of thin planar
gaseous particle detectors where the traditional planes of \gls{HV} wires (which can be fragile) are replaced by a thin metallic micro-mesh.
\gls{MM} detectors consist of a planar (drift) electrode, a gas gap of a few millimetres acting as both the conversion and the drift region, and the micro-mesh at a distance of \SIrange{120}{130}{\micron} from the readout electrode, creating the amplification region. The \gls{MM} operating principle, as implemented in ATLAS, is illustrated in Figure~\ref{fig:Muon_MMprinciple}.
The \gls{HV} potentials are chosen such that the electric field is a few hundred \si{\volt/\cm} in the drift region, and \SIrange{40}{50}{\kilo\volt/\cm} in the amplification region.
These are adjusted to set the drift velocity close to its local maximum and plateau, while minimising longitudinal diffusion~\cite{ALEXOPOULOS2019125} and ion back-diffusion.
The field in the amplification region is adjusted, according to the gas mixture used, to get a gas gain close to \num{10000}.
Charged particles traversing the drift space ionise the gas; the electrons liberated by the ionisation process drift towards the micro-mesh. With an electric field in the amplification region 50 to 100 times stronger than the drift field, the micro-mesh is transparent to more than 95\% of the electrons. The electron avalanche takes place in the thin amplification region, immediately above the readout electrode. The drift of the electrons in the conversion gap to reach the micro-mesh is a comparatively slow process (though still very fast compared with most other detector techniques): it depends on the drift gas, drift distance, and electric field, and in the configurations used in ATLAS can take up to around \SI{100}{\ns}.
The amplification process, however, happens in a fraction of a nanosecond, resulting in a fast pulse of electrons on the readout strip. The ions produced in the avalanche move in the opposite direction to the electrons, back to the micro-mesh. Most of the ions are produced in the last avalanche step, close to the readout strip. It is the fast evacuation of the positive ions that particularly suits the \gls{MM} to operation at very high particle fluxes.
 
The main challenge addressed in the ATLAS \gls{MM} design is electrical discharging~\cite{ALEXOPOULOS2019125} in large-area detectors. Electrical discharges occur
when the total number of electrons in the avalanche reaches a few tens of millions
(the Raether limit~\cite{ref26TDR}). These discharges may damage the  detector and readout electronics, and the voltage breakdown can lead to large dead times.
High detection efficiency for minimum ionising muons requires gas amplification factors of the order of \num{10000}, meaning that ionisation processes producing more than \num{1000} electrons over distances comparable to the typical lateral extent of an avalanche (a few hundred microns) carry the risk of discharging (see, for example, Ref.~\cite{ref27TDR}). Such ionisation levels are easily reached by low-energy alpha particles or slowly moving charged debris from neutron (or other) interactions in the detector gas or detector materials.
 
The electrical discharge protection system developed for the \gls{NSW} \gls{MM} detectors consists of adding a layer of resistive strips on top of the readout strips, separated by a thin insulator, as shown in Figure~\ref{fig:Muon_MMprinciple}. The readout electrode is no longer directly exposed to the charge created in the amplification region, but capacitively coupled to the signals~\cite{ATLAS-TDR-20}. Some fraction of the signal height is lost, but the chamber can be operated at higher gas gain, with electrical discharge intensities reduced by about three orders of magnitude.
 
To create the amplification and drift potentials, positive \gls{HV} is applied to the resistive strips, and the micro-mesh is connected to ground, while the drift cathode is held at a negative potential~\cite{ALEXOPOULOS2019125}.
This scheme helps to reduce electrical discharges and allows for more stable operation of the detectors, as voltage breakdowns remain local and charge can be evacuated very quickly to ground through the micro-mesh, and the micro-mesh potential does not change. This \gls{HV} scheme
allows for easy segmentation of the \gls{HV} connection scheme, which simplifies detector fabrication.
 
The readout strips are etched on \SI{0.5}{\mm} thick \glspl{PCB} covered by a \SI{64}{\micron}-thick layer of insulator, and are overlaid with the resistive strips.
The micro-mesh support pillars are deposited on top of the resistive strips in a $\SI{7}{\mm}\SI{7}{\mm}\SI{7}{\mm}$ triangular grid.
Unlike most \gls{MM} detectors now in operation, in the ATLAS implementation the micro-mesh is not integrated into the readout structure, but rather connected to the drift panel, and stretched across the support pillars of the readout panel by mechanical tension~\cite{Farina_LaRochelle,ALLARD2022166143,AGARWALA2022167285}.
The precision direction of the \gls{MM} coincides with the bending direction of the toroidal magnetic field, while the second coordinate is determined using strips with a small stereo angle.
The main detector and operating parameters of the \gls{MM} detectors for ATLAS are summarised in Table~\ref{table:Muon_MMparams}.
 
The \gls{HV} stability of the \gls{MM} is found to be strongly correlated with the actual resistance of the resistive strips of the readout anode.
Lower resistance values are often found very close to the panel border, giving rise to specific discharge points in those areas.
To combat this, a process of edge passivation~\cite{INST20_MM,ALLARD2022166143,AGARWALA2022167285} was applied to the readout panels prior to their assembly.
A thin film of epoxy was applied
to the active region around the border of readout panels if the measured resistance was below a defined threshold. The breadths of the passivated border regions were up to a few centimetres.
This procedure improved the \gls{HV} stability of the chambers at the expense of a very small reduction in their active area. 
 
Two gas mixtures have been extensively tested.
The originally proposed and tested gas mixture, now considered the backup mixture, is \SI{93}{\percent} argon and \SI{7}{\percent} CO$_2$; this is the same mixture used for the ATLAS \glspl{MDT}, except that the \gls{MDT} gas requires active humidification up to the level of a few hundred ppm of water, while \gls{MM} need a very dry gas.
Superior \gls{HV} stability is, however, obtained by using an argon mixture in which \SI{2}{\percent} of CO$_2$ is replaced by 2\% of
isobutane, as the chambers reach the plateau of full efficiency for a lower \gls{HV} than without the isobutane. The mixture remains non-flammable.
This isobutane mixture, now considered the baseline, will be used for \RunThr, and if ageing tests confirm that it does not give rise to deposits on the electrodes, it will remain the default gas for the lifetime of the experiment.
 
The \gls{MM} require a positive potential of about \SIrange{500}{600}{\volt} on the resistive strips, and a negative potential of
about \SI{250}{\volt} 
on the drift cathode, producing an amplification field of about \SIrange{40}{45}{\kilo\volt/\cm} and a drift field of around 
\SI{500}{\volt/\cm},
respectively, as illustrated in Figure~\ref{fig:Muon_MMprinciple}.
The precise choices for the amplification potential depend on the gas choice, and are given in Table~\ref{table:Muon_MMparams}.
One of the great advantages of adding isobutane to the gas mixture is that it allows the chambers to run at full efficiency with lower amplification potentials, improving stability and reducing the number and intensity of electrical discharges.
The inner quadruplets, which are subject to higher background rates, may be run at a lower potential than the outer ones.
All the drift electrodes of each quadruplet can be supplied by a single \gls{HV} channel,
so each wedge requires just two channels for the negative drift voltage.
Since the resistive strips are split in the middle, positive voltage is supplied separately to each side of each readout \gls{PCB}. Each of these half-\glspl{PCB} is isolated by its own \gls{HV} capacitor.
As each layer of a wedge has eight \glspl{PCB}, this results in \num{16} readout \gls{HV} sections per wedge layer; however, the half-\glspl{PCB} of one layer of one quadruplet share a common \gls{HV} line, so there are two independent \gls{HV} inputs on each side of each wedge layer (one to each quadruplet). A splitter box distributes the input \gls{HV} to the groups of readout \gls{HV} sections.

\begin{table}[h!]
\centering
\caption{Main \gls{MM} detector and operating parameters. }
\label{table:Muon_MMparams}
\begin{tabular}{lll}
\hline
Item/Parameter & Characteristics & Value \\ [0.5ex]
\hline
Micro-mesh & Stainless steel; mesh separate from readout board  &  \\
Micro-mesh & Wire diameter & \SI{30}{\micron}\\
Micro-mesh & Gap between wires  & \SI{71}{\micron} \\
& Micro-mesh is separate from readout board. & \\
Amplification gap & & \SIrange{120}{130}{\micron} \\
Drift/conversion gap & & \SI{5}{\mm}\\
Resistive strips & Interconnected & R=\SIrange{10}{20}{\mega\ohm/\cm}\\
Readout strip width & & \SI{0.3}{\mm} \\
Readout strip pitch & Inner / Outer Modules & 0.425 / \SI{0.45}{\mm}\\
Stereo angle & one $+$ and one $-$ slope layer per quadruplet & $\pm$ \ang{1.5} \\ 
Total number of strips & & 2.1M \\ 
\hline
Baseline Gas & Ar:CO$_2$:iC$_4$H$_{10}$ & 93:5:2\\
Backup Gas & Ar:CO$_2$ & 93:7\\
\gls{HV} on resistive strips & Baseline gas & \SI{500}{\volt}  \\
\gls{HV} on resistive strips & Backup gas & \SI{570}{\volt} \\
\gls{HV} on drift panel &
& \SI{-240}{\volt} \\
Amplification field & & $\sim\SI{40}{\kilo\volt/\cm}$ \\ 
Drift field & & \SI{480}{\volt/\cm}\\ 
\hline
Bending Coordinate Resolution & single-plane, $\eta$-strips centroid fit & \SIrange{100}{200}{\micron}\\
Second Coordinate Resolution & $\phi$, single-plane, stereo strips centroid fit & $\SI{2.7}{\mm}$ \\
\hline
\gls{NSW-TP} Bending Coord. Res. & from \gls{NSW-TP} fitter & \SI{300}{\micron}\\
\gls{NSW-TP} Second Coord. Res. & $r\phi$, from \gls{NSW-TP} fitter & \SIrange{11}{12}{\mm} 
\\ [1ex]
\hline
\end{tabular}
\end{table}

\paragraph{\glstext{MM} Quadruplet and Wedge Structure}
 
Each sector of the \gls{NSW}, Large or Small, contains two \gls{MM} wedges attached to the faces of the central aluminium spacer frame as described in Section~\ref{muonSS:shielding}. The spacer frame is hollow, with integrated water cooling channels, and most of the \gls{MM} services are routed through this central volume.
 
\begin{figure}[!h]
\centerline{\includegraphics[width=0.85\textwidth]{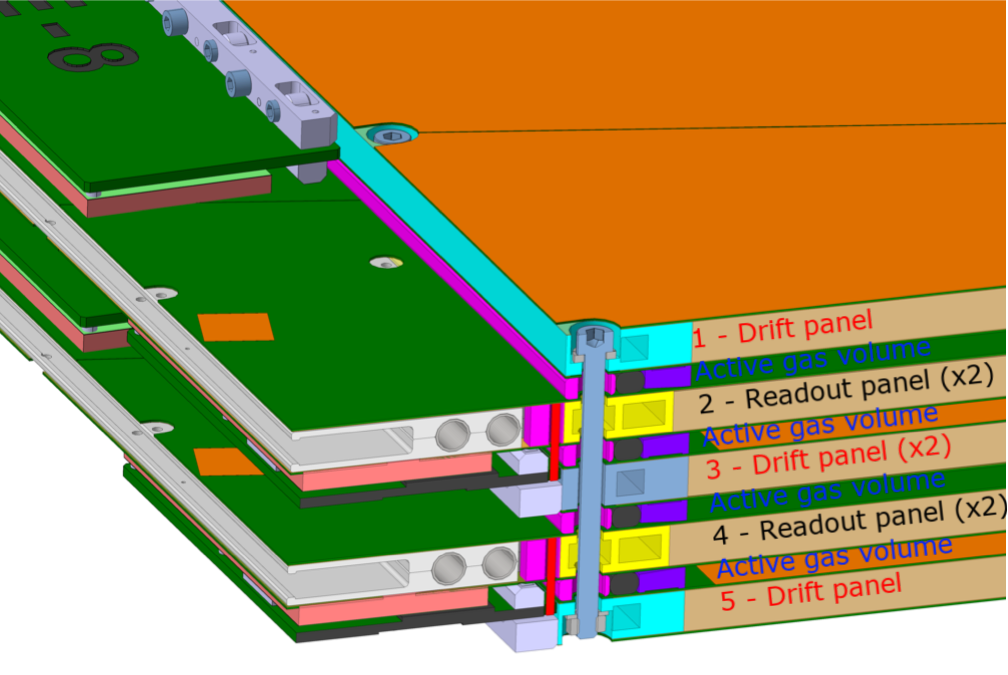}}
\caption{Arrangement of readout and drift panels in a \gls{MM} quadruplet. An interconnect is shown in cross-section at the front of the drawing; these interconnects limit bulging of the panels due to the slight (\SI{3}{\milli\bar}) overpressure of the gas volumes.}
\label{fig:Muon_MMquad}
\end{figure}

Each \gls{MM} wedge is assembled from two separate quadruplet modules, as shown in Figure~\ref{figMuon:Wedges}. There are thus four sizes of \gls{MM} quadruplets: Inner and Outer (radially) for the Small wedges, and similarly for the Large wedges.
Each quadruplet consists of five stiff panels, \SI{1}{\cm} thick.
Three are drift panels (the outer, single-sided, and the central, double-sided).
The other two are the double-sided readout panels.
Between the panels are the
four active layers, corresponding to independent gas volumes \SI{5}{\mm} thick, as shown in Figure~\ref{fig:Muon_MMquad}.
A particle traversing a sector of the \gls{NSW} thus passes through eight \gls{MM} drift regions (four in each of the two wedges, as shown in Figure~\ref{figMuon:WedgesSide}), inducing charges on readout strips on the corresponding readout panels. Two of these layers on each wedge have readout strips perpendicular to the radial centre-line of the wedge, (``eta strips'')
and two have strips at a small ``stereo'' angle of $\pm$\ang{1.5} (one positive and one negative) with respect to the eta strips.
 
The double-sided drift panel forms the centre of each quadruplet.
Double-sided readout panels mounted on either side of the central drift panel form the two central gas volumes.
Single-sided drift panels mounted on the outside complete the two outer gas volumes.
The four active layers of the quadruplet are grouped in two pairs, each pair sharing one double-sided readout panel in a back-to-back configuration.
This ensures that background tracks not synchronous with the bunch crossing will not be collinear in the two neighbouring planes, as the drift in the two adjacent detectors is in opposite directions, and the offsets of the track segments due to assuming the wrong start time for the drift go in opposite directions. Out-of-time background can thus be rejected. The back-to-back structure also ensures that systematic shifts of reconstructed particle positions due to the deviation of the electrons' drift path in a magnetic field (the Lorentz angle) cancel out.
 
The readout panels are assembled from separate \glspl{PCB} spanning the full azimuthal width of the sector, but with radial dimensions in the range \SIrange{435}{475}{\mm} to allow for production on standard \gls{PCB} machines.
The inner quadruplet in each wedge has five radial \gls{PCB} segments in each readout panel, and the outer quadruplet has three.
Readout strips must be precisely aligned from one face of each readout panel to the other, and from one readout panel to the next.
Several complementary methods were used to locate and align the readout strips with precision~\cite{Kuger_2016}, including a mechanical hole-and-slot pair, a pair of targets included in the copper strip pattern and aligned with the board axis and strips, and four \gls{RASNIK}~\cite{Beker_2019}
masks etched on the edge of each readout \gls{PCB}. \gls{RASNIK} masks consist of a small ``chessboard'' pattern of dark and light squares, with absolute positions encoded by inverting the colours of a few squares, to be read out with a CCD.
\gls{RASNIK} masks etched on the PCBs on both faces of the readout panels can be read out simultaneously with a dedicated calibrated device called a RasFork~\cite{ALLARD2022166143}.
The \gls{RASNIK} mask positions are read out during various stages of chamber assembly, with an accuracy of \SIrange{3}{5}{\micron}, such that the strip positions on the \glspl{PCB} are known with an accuracy better than \SI{50}{\micron} at the end of the construction process.
Subsequent deformations of the \gls{MM} quadruplets due to temperature and humidity can affect the positions of the strips, increasing this uncertainty to around \SI{100}{\micron}.
Alignment source platforms, described in more detail in Section~\ref{muonSS:alignment}, are glued to the outer surface of each \gls{MM} wedge.
 
Each \gls{MM} wedge is read out by \num{64}  \glspl{MMFE8}~\cite{Iakovidis_LaRochelle} (described in Section~\ref{MuSP:MMFE})
connected directly to the ends of the readout strips with flexible pressure connectors. 
Also mounted on the wedge are eight \glspl{L1DDC} for data readout (described in Section~\ref{MuSP:L1DDC}), each connected to eight of the \glspl{MMFE8}, and eight \glspl{ADDC} for the \gls{MM} trigger (described in Section~\ref{MuSP:ADDC}), also each connected to eight \glspl{MMFE8}.
 
The single-plane spatial resolution obtainable in the precision (``eta'') direction has been estimated using small \gls{MM} bulk prototype chambers (with different electronics) in a test beam to be about \SIrange{90}{100}{\micron} per plane; by using the ``$\upmu$TPC method''~\cite{ALEXOPOULOS2019125} to reconstruct the position from inclined tracks, it is expected to be possible to obtain a single-plane resolution within that range for all inclinations between \ang{0} and \ang{40}.
A second-coordinate resolution of about \SI{2.7}{\mm} can be obtained using the stereo strips~\cite{ATL-MUON-PUB-2015-001},
and this has been demonstrated using the centroid method in a test beam
~\cite{KNtekas}
using prototype detectors ($\SI{1}{\m}\SI{0.5}{\m}$) with a slightly smaller strip pitch (\SI{415}{\micron}) than the ATLAS \gls{MM}, positioned perpendicular to the beam axis.
Similar results for the spatial resolution of perpendicularly incident \SIrange{120}{150}{\GeV} muons and pions were achieved in the precision direction and second coordinate in \gls{SPS} experiments using three Module~0 chambers  and one series production chamber which was read out using the \gls{VMM} electronics. The results of the first test with a Module~0 chamber appear in Refs.~\cite{Iodice:2017Kt} and ~\cite{ALEXOPOULOS2020162086}; the others have not yet been published.
\RunThr ATLAS data will be required to determine the ultimate resolution obtainable with the full-sized detectors and non-perpendicular tracks.
 
\subsubsection{Small-Strip Thin Gap Chamber Technology \label{muonSS:sTGC}} 
The \glsfirst{sTGC} is a new development of the \gls{TGC} technology first developed in the early 1980s~\cite{MAJEWSKI1983265}, to allow for very fast on-line tracking that can be used in the \gls{L1} (hardware) trigger, with sufficient precision for the offline muon tracking as well.
Wherever possible, the \gls{sTGC} retain the characteristics of the \gls{TGC} already used in ATLAS: two \gls{FR4} cathodes coated with a resistive graphite-resin mixture span a gap \SI{2.8}{\mm} thick.
The gap is filled with a mixture of \SI{55}{\percent} CO$_2$ and \SI{45}{\percent} n-pentane.
The anode plane in the middle of the gap is strung with \SI{50}{\micron} gold-plated tungsten wires at a pitch of \SI{1.8}{\mm}.
The wire direction is parallel to the central radial axis of the chamber (contrary to the existing \gls{TGC} in ATLAS where the wires are strung azimuthally).
The wires are ganged together, with groups of typically \num{20} sharing a high-voltage capacitor.
 
The resistivity of the cathode coating in the \gls{sTGC} is significantly lower than for the other \gls{TGC} in ATLAS: about \SI{150}{\kilo\ohm}$/\square$~\footnote{Surface resistivity is the resistance between two opposite sides of a square of a thin-film resistor, and is independent of the size of the square; it is measured in \si{\ohm}, but in order to distinguish it from resistance and emphasise its two-dimensional property, it is usually written as $\si{\ohm}/\square$ or \si{\ohm/square}.} 
for the chambers closest to the \gls{IP}, and \SI{200}{\kilo\ohm}$/\square$ for the others.
This low resistivity allows for rapid clearing of charge on the cathode surface.
Signals are capacitively induced on copper readout strips (running perpendicular to the wires) on one side of the gap, and on large copper pads on the other side. The signal layer is under a thin (\num{150} or \SI{200}{\micron}) insulating pre-preg layer directly under the graphite cathode, increasing the capacitance compared with the other ATLAS \gls{TGC} (where the signal strips are on the back of the \gls{FR4}), to keep the same transparency for fast signals.
This readout layer is supported by a \SIrange{1.3}{1.4}{\mm}-thick \gls{PCB} backed by a copper grounding skin to limit cross-talk. The pad board has an additional pre-preg layer between the core of the \gls{PCB} and the ground skin to support the readout traces, which are connected by conductive vias to the pads.
 
\begin{figure}[!h]
\subfloat[]{\label{figMuon:struct}\includegraphics[width=0.7\linewidth]{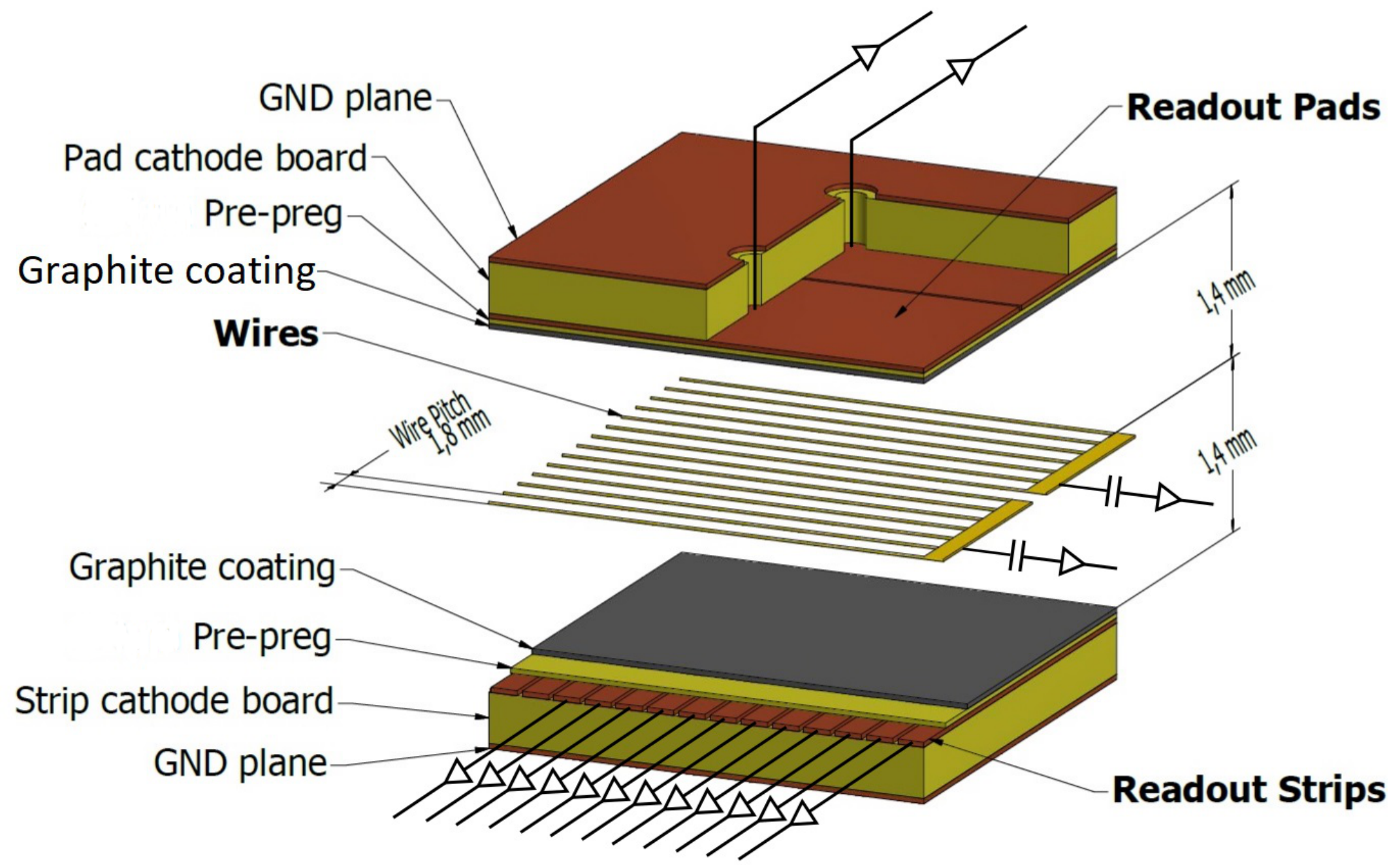}}
\hfil
\subfloat[]{\label{figMuon:quad}\includegraphics[width=0.25\linewidth]{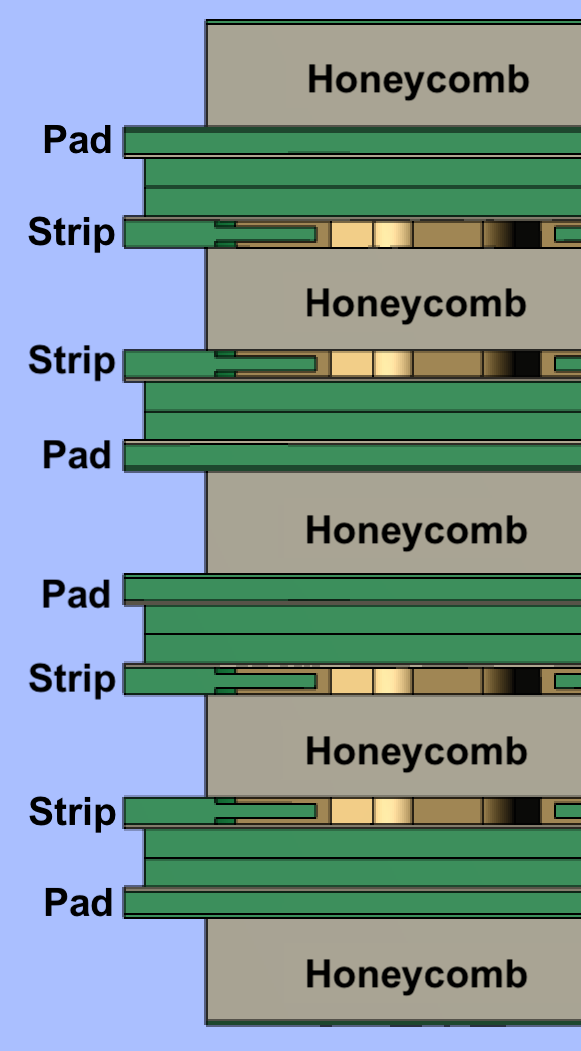}}
\caption[\gls{sTGC} detector operating principle]{The \gls{sTGC} \protect\subref{figMuon:struct} detector operating principle and \protect\subref{figMuon:quad} arrangement of gas gaps (in green, each comprising a strip cathode and a pad cathode separated by two frames clamping the wire plane) to form a quadruplet, shown to scale at the corner of a quadruplet, with brass alignment features inset in the strip cathodes. The pre-preg thickness is \SI{150}{\micron} for the innermost quadruplets and \SI{200}{\micron} for the middle and outer ones.
\label{fig:Muon_sTGCprinciple}}
\end{figure}

The main detector and operating parameters of the \gls{sTGC} detectors for ATLAS are summarised in Table~\ref{table:Muon_sTGCparams} and illustrated in Figure~\ref{fig:Muon_sTGCprinciple}.
 
\begin{table}[h!]
\caption{Main \gls{sTGC} detector and operating parameters. Total numbers of channels are for both wheels together. ``Outer Quadruplets'' refers to both Middle and Outer Quadruplets; the Inner Quadruplets are exposed to substantially higher background rates, and therefore have finer pad granularity and less resistive cathodes for faster charge evacuation. The wire groups in the innermost part of the Inner Quadruplets are not read out.
\label{table:Muon_sTGCparams}}
\centering
\begin{tabular}{lll}
\hline
Item/Parameter & Characteristics & Value \\ [0.5ex]
\hline
Readout strip pitch & & \SI{3.2}{\mm}\\
Readout strip width & & \SI{2.7}{\mm}\\
Total number of strips & & \num[fixed-exponent=3, round-mode = figures, round-precision = 3]{282.240e3}\\ 
Typical pad azimuth & Inner / Outer Quadruplets & \ang{5} / \ang{7.5} \\
Typical pad radial height &  & \SI{80}{\mm} \\
Range of ``full'' pad areas & & \SIrange{61}{519}{\cm\squared} \\
Total number of pads & & \num{46656}\\ 
Anode-cathode gap & & 1.4 \si{\mm}\\
Wire pitch & & \SI{1.8}{\mm}\\
Wire diameter & Gold-plated tungsten & \SI{50}{\micron}\\
Total number of wires & Ganged in groups of \num{20}, & \num{6390296} \\ 
& group boundaries offset & \\
& by \num{5} wires between layers &\\
Number of wire groups & Total / Read out & \num{31776} / \num{28704} \\ 
\gls{HV} on wires & Positive polarity & \SI{2.8}{\kilo\volt} \\
Cathode resistivity & Inner / Outer Quadruplets & 150 / \SI{200}{\kilo\ohm}$/\square$\\
Pre-preg thickness between readout and cathode & Inner / Outer Quadruplets & 150 / \SI{200}{\micron}\\
Gas & n-pentane:CO$_2$ & 45:55 \\
\hline
Bending Coordinate Resolution & single-plane $\eta$ from strips & \SIrange{100}{200}{\micron} \\ 
Azimuthal Resolution & single-plane $r\phi$ (from wire groups) & \SI{2.6}{\mm} \\ 
\hline
\gls{NSW-TP} Bending Coordinate Res. & \gls{L1} $\eta$ from centroid fit to strips in band & $<\SI{1}{\milli \radian}$ \\
Pad Trigger Azimuthal Res.  & $\phi$ at \gls{L1} from pad towers (Inner/Outer) &
\SI{7}{\milli \radian} / \SI{10}{\milli \radian}  \\
& $r\phi$ (at \gls{L1} from pad towers) &
\SIrange{7}{38}{\mm} 
\\ [1ex]
\hline
\end{tabular}
\end{table}
 
The readout strips are \SI{2.7}{\mm} wide, with a pitch of \SI{3.2}{\mm}, much finer than the \gls{TGC} strips.
The strip width and pitch required for the trigger and tracking functions were both optimised
by studying the position resolution of small detectors built with a range of strip widths and separations,
to reduce capacitive coupling between channels and non-linear corrections as much as possible for trigger purposes,
while maintaining the smallest possible number of readout channels.
In combination with a reduction of the HV and ground
decoupling resistors, this allows the \gls{sTGC} to remain efficient for minimum ionising particles even in large-surface detectors subject to detected rates of up to \SI{20}{\kHz/\cm\squared} over the full surface~\cite{ATLAS-TDR-20}.
The intrinsic position resolution achievable with the final strip configuration was measured using a quadruplet made with full-sized prototype boards, in conjunction with a silicon pixel telescope (used to determine independently the individual particle trajectories), in a \SI{32}{\GeV} pion testbeam at Fermilab~\cite{ABUSLEME201685}, and was found to be around \SI{45}{\micron} for particles incident perpendicular to the plane of the quadruplet.
The resolution obtainable in practice is limited by the precision with which the as-built positions of the strips are known.

\paragraph{sTGC Quadruplet and Wedge Structure \label{par:sTGCwedgestruct}}
In order to provide a track segment of sufficiently good pointing resolution, as well as for trigger efficiency, the \gls{sTGC} gas gaps are assembled in quadruplets, with four gas gaps separated from each other by \SI{5}{\mm}-thick fibreglass frames and paper honeycomb spacers, as shown in Figure~\ref{figMuon:quad}.
The dimensions of the quadruplets were dictated by the maximum size of cathode boards that could be produced industrially: the largest quadruplets are about \SI{2.2}{\m} wide (the full width of the Large sector at its outer radius).
Three trapezoidal\footnote{In the case of the outermost quadruplets in the Large sectors, the trapezoids are modified, with a rectangular section at the outer radius (see Figure~\ref{figMuon:WedgesNoFrames}).} quadruplets of different sizes are precisely glued together with a pair of fibreglass frames (visible in Figure~\ref{fig:Muon_NSW_vp1}) to form the wedges shown in Figure~\ref{figMuon:Wedges}. The fibreglass frames incorporate three pairs of projecting fibreglass bars that are used to support the mount points from which the wedges are hung on the sectors.  The positions of these projections were mechanically constrained during the gluing of the wedges,
relative to alignment features~\footnote{Each strip board has two machined brass alignment features on one edge, allowing for alignment against a pair of precisely positioned cylindrical pins: a ``V''-shaped feature allowing only rotation in the plane about the pin defines a fixed origin, and a flat feature parallel to the edge of the board defines a direction with respect to the first pin while avoiding over-constraining the board in case of small scaling variations due to temperature or humidity changes.} on the cathode strip boards of all four detector planes, using precisely machined cylindrical pins, placed with an accuracy of a few tens of \si{\micron} on the surface plates used for wedge assembly.
The same brass alignment features were used to position the alignment source platforms  (described in Section~\ref{muonSS:alignment}) on the surface of the \gls{sTGC} wedge that faces the \gls{MM} double-wedge. The accuracy with which the positions of the brass features are known with respect to the platform positions (combined with the generally lower accuracy with which the strip positions are known relative to the brass features) limits the precision with which the absolute strip positions can be determined.
The kinematic mounts are designed to allow for adjustment between the two \gls{sTGC} wedges to achieve parallelism of their strips, as defined by the alignment platforms, to better than \SI{1}{\mrad}.
 
There are six sizes of \gls{sTGC} quadruplets: three for the Small-sector wedges and three for the Large.
Each sector has two \gls{sTGC} wedges, hung kinematically from the sector spacer frames, outside the \gls{MM} wedges. This maximises the lever arm between the track segments provided by the two \gls{sTGC} wedges (refer to Figure~\ref{figMuon:WedgesSide}).
 
The \gls{HV} wires of the innermost quadruplets in each wedge (those closest to the \beampipe) are split in two, creating two \gls{HV} regions within a common gas volume.
This separates the high-background region ($2.4<\abseta <2.7$) and allows it to operate at the same effective amplification as the rest of the detector, by means of the \gls{HV} divider network.
Only the higher-radius portion of the wires ($\abseta\lesssim 2.4$) is read out for azimuthal coordinate measurements;
since the azimuthal granularity of the pad cathode boards is comparable to that of the wire groups for $\abseta > 2.4$, the pads provide the azimuthal coordinate in the region with no wire readout.
 
No active components are mounted on the cathode boards. All pad and strip readout channels are connected by soldered jumper wires to long multilayer adapter boards attached to both sides of each gas gap. Wire groups are read out through narrow adapter boards soldered to the longer parallel edge of each quadruplet, and their signals routed through the pad adapter boards by means of a ribbon cable.
A pad or strip front-end board, described in Section~\ref{muonSS:trigger}, is connected to each pad or strip adapter board respectively, using a high-density low-profile connector.

As for all \glspl{TGC} in ATLAS, each \gls{sTGC} wedge is enclosed around its periphery by a gas-tight envelope that is continuously flushed with CO$_2$. This is done to maintain a dry atmosphere around the \gls{HV} elements, as well as to dilute any possible leak of the operating gas. If traces of n-pentane are detected in the CO$_2$ stream at the output of a wedge, \gls{HV} and \gls{LV} as well as gas supplies are automatically switched off, and an alarm is activated.
The gas channel also serves (in conjunction with the copper ground skins of the detectors themselves) to complete the Faraday cage surrounding the detectors, adapter boards and front-end electronics.
The connectors for the front-end boards protrude through sealed slots in the Faraday cage; the front-end boards thus sit outside the gas channel.
The Faraday cage is completed by electrical shielding mounted around the front-end boards, to which the cooling pipes (described in Section~\ref{muonSS:services}) are connected by a system of thermally conductive clips.
 
\subsubsection{Extending the Endcap Alignment System \label{muonSS:alignment}} 
The ATLAS muon endcap alignment system\cite{Aefsky_2008} is based on the concept of calibrated alignment bars connected by optical lines.
Each alignment bar is a hollow aluminium tube, very precisely measured, containing precisely calibrated instrumentation to monitor its own length and deformations.
 
The smallest alignable units of the \gls{NSW} are the \gls{MM} and \gls{sTGC} quadruplets.
The alignment system monitors the rotations and translations of the quadruplets, as well as a limited number of deformation modes.
Within the individual quadruplets, it is assumed that the internal geometry does not change (apart from the allowed deformations); however, it is not assumed to be ideal: data taken during the construction of the quadruplets allow the as-built position of each individual strip to be reconstructed with respect to the coordinate system of the quadruplet.
 
\begin{figure}[!h]
\subfloat[]{\label{figMuon:azimuthal}\includegraphics[width=0.45\textwidth]{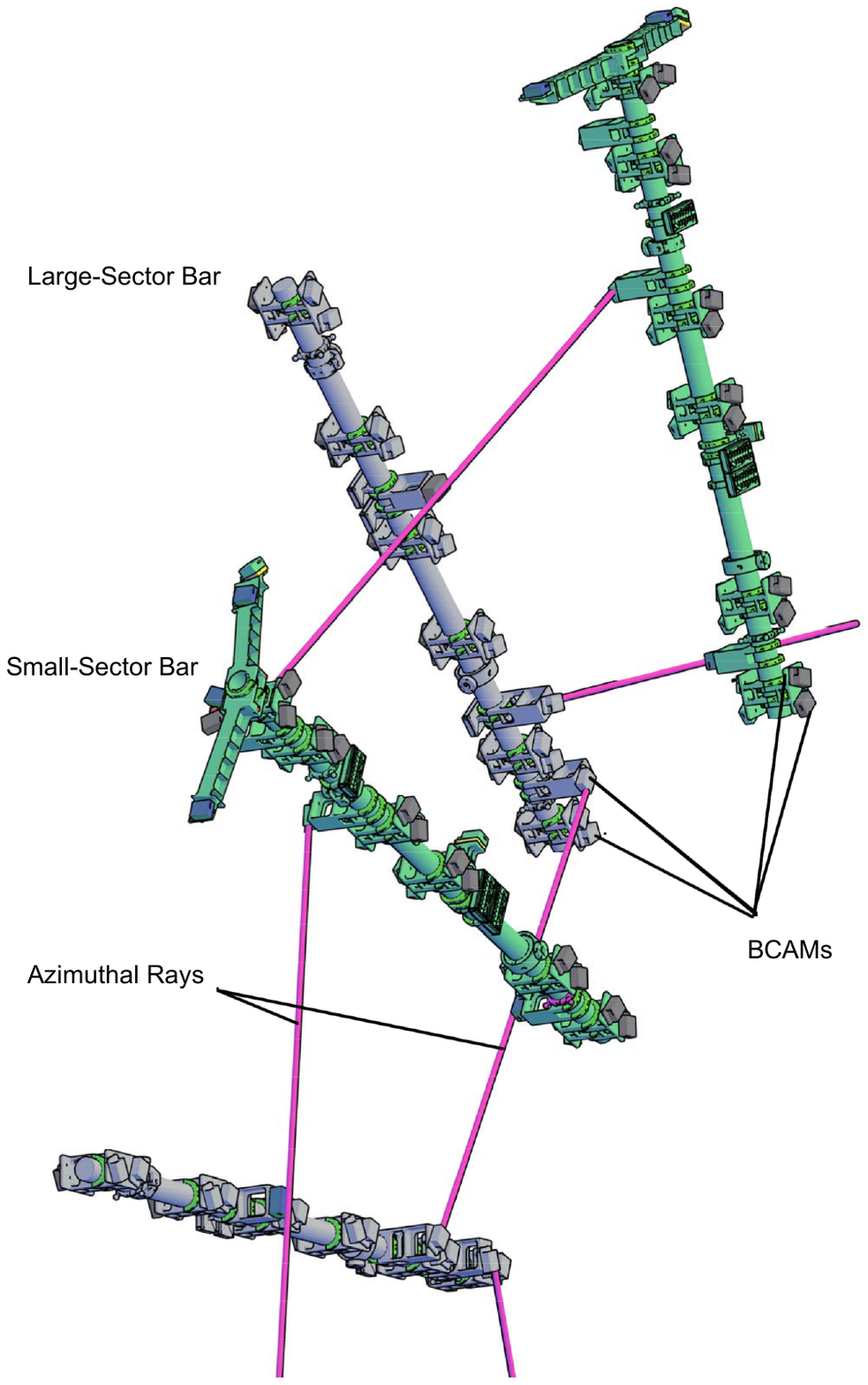}}
\subfloat[]{\label{figMuon:proximity}\includegraphics[width=0.45\textwidth]{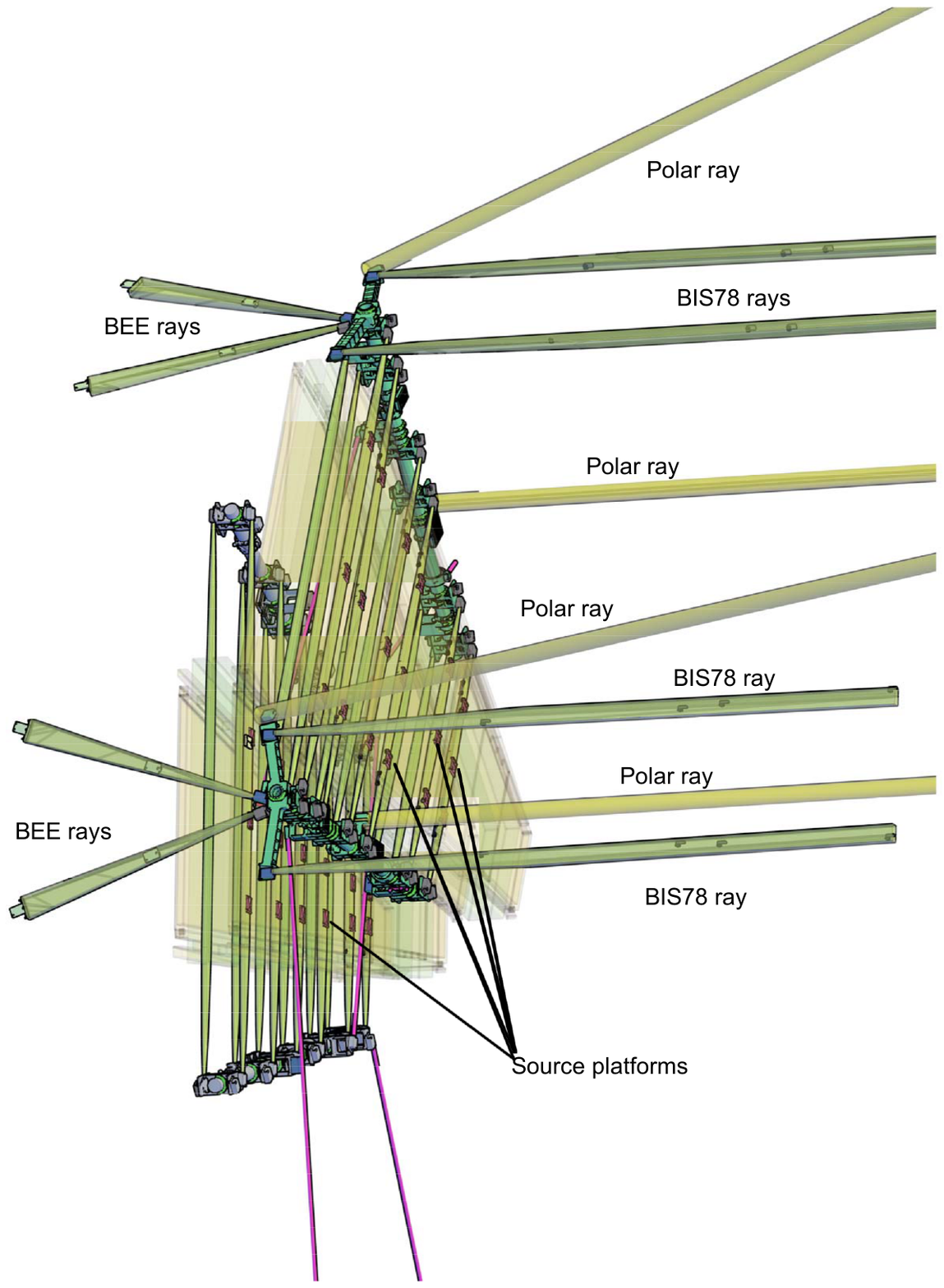}}
\caption{Alignment rays in the \gls{NSW}.  \protect\subref{figMuon:azimuthal} Alignment bars (green) in the spokes  supporting the Large sectors occupy the positions previously filled by the old EI alignment bars.
They are called the Small-sector bars because the spokes supporting Large sectors sit in the Small sectors on either side of each Large sector.
The Small-sector bars are connected by polar rays to the EM alignment bars exactly as the original alignment bars were.
They are connected to each other and to the Large-sector alignment bars (grey, located in the spokes supporting the Small sectors) by azimuthal rays (magenta) connecting \glspl{BCAM} (grey boxes) with integrated light sources to each other. \protect\subref{figMuon:proximity} The Small-sector bars are connected to the Large-sector \gls{sTGC} and \gls{MM} quadruplets (shown transparent), and the Large-sector bars are connected to the Small-sector \gls{sTGC} and \gls{MM} quadruplets, by proximity rays (yellow) connecting \glspl{BCAM} on the bars to light sources on platforms mounted on the quadruplet surfaces (small magenta rectangles). The Small-sector bars also have \glspl{BCAM} that connect them to EM (Big Wheel) alignment bars via the polar alignment rays, shown in yellow and pointing toward the right, and to sources on the \gls{BIS78} chambers (grey-yellow rays pointing to the right), and to sources on the BEE chambers (grey-yellow rays pointing to the left).
\label{figMuon:Alignment_principle}}
\end{figure}

Each \gls{NSW} has \num{16} alignment bars, one in each sector, unlike the EM and EO Wheels (and the \glspl{SW}), which have only eight.
This doubling allows \glspl{BCAM} on the alignment bars to see both \gls{sTGC} wedges and both \gls{MM}
wedges in each sector.
The \gls{NSW} alignment bars are mounted in the Large-sector and Small-sector spokes (see Section~\ref{muonSS:shielding}), and connected to each other and to the adjacent chambers by a network of optical lines, where CCD cameras in \glspl{BCAM}~\cite{Aefsky_2008} mounted on the alignment bars monitor the positions of light sources mounted on the chambers and on \glspl{BCAM} on the neighbouring alignment bars.
The \glspl{NSW} are aligned with respect to the other wheels of the endcap \gls{MS} using the existing
{\em Polar} alignment corridors, which traverse tightly constrained unobstructed paths through holes and gaps in detectors and support structures, and through dedicated hollow tubes passing through the endcap toroid cryostats.
Polar \glspl{BCAM} are therefore positioned on the eight Large-sector alignment bars, in the same positions as the former EI polar \glspl{BCAM}.
{\em Azimuthal} alignment corridors connect all the alignment bars to their neighbours, as shown in Figure~\ref{figMuon:azimuthal}, with \glspl{BCAM} on the alignment bars pointing at light sources on the nearest neighbour \glspl{BCAM} in both the adjacent Large and Small sectors (Large-to-Large and Large-to-Small, or Small-to-Small and Small-To-Large).
The optical connections among the \num{16} alignment bars and the \num{64} wedges in each endcap are illustrated in Figure~\ref{figMuon:Alignment_principle}.
 
Quadruplets on each \gls{sTGC} or \gls{MM} wedge must be aligned with respect to the adjacent alignment bars.
{\em Proximity} lines connect the \glspl{BCAM} on the alignment bars to light sources on the \gls{sTGC} and \gls{MM} wedge surfaces inside the \SI{27}{\mm} gaps between the \gls{sTGC} and \gls{MM} wedges in each sector.
Light injectors located at the periphery (``rim'') of the \gls{NSW} illuminate optical fibres.
The ends of these fibres are held in precisely determined positions on the alignment source platforms (mentioned in Sections~\ref{muonSS:MM} and \ref{muonSS:sTGC} above) installed on the \gls{sTGC} and \gls{MM} wedge surfaces.
Each wedge has \num{18} platforms glued to its surface. In the \gls{sTGC} wedges, six of these (in two lines of three) are glued to the outermost quadruplet in the wedge, nine (in three lines) to the middle quadruplet, and three (in one line) to the innermost quadruplet. In the \gls{MM} wedges nine platforms (in three lines) are glued to the outer quadruplet and nine (in three lines) to the inner quadruplet. Each platform holds two or four optical fibre ends. The \gls{MM} and \gls{sTGC} platforms are sufficiently close to each other that the \glspl{BCAM} on the alignment bars can each see ten light sources: two each from two \gls{MM} platforms and two from each of the corresponding \gls{sTGC} platforms, and two from its partner proximity \gls{BCAM} on the opposite side of the sector.
The positions of the source platforms are precisely measured using photogrammetry with a resolution of about \SI{50}{\micron}.
Their positions are measured with respect to the \gls{RASNIK} masks that locate the strip positions on the \gls{MM} readout boards, and to the brass edge-features (see Figure~\ref{figMuon:quad}) that determine the strip-pattern alignment of the \gls{sTGC} cathodes; thus, for both \gls{MM} and \gls{sTGC} the source platform positions are directly related to the positions of the readout strips.
The positions of the illuminated fibre ends on the source platforms are continuously monitored by \glspl{BCAM} on the alignment bars.
The alignment system can detect and measure chamber movements as small as \SI{40}{\micron}~\cite{Aefsky_2008}.
 
\subsubsection{NSW Electronics \label{muonSS:electronics}}
\begin{figure}[!h]
\centerline{\includegraphics[width=0.95\textwidth]{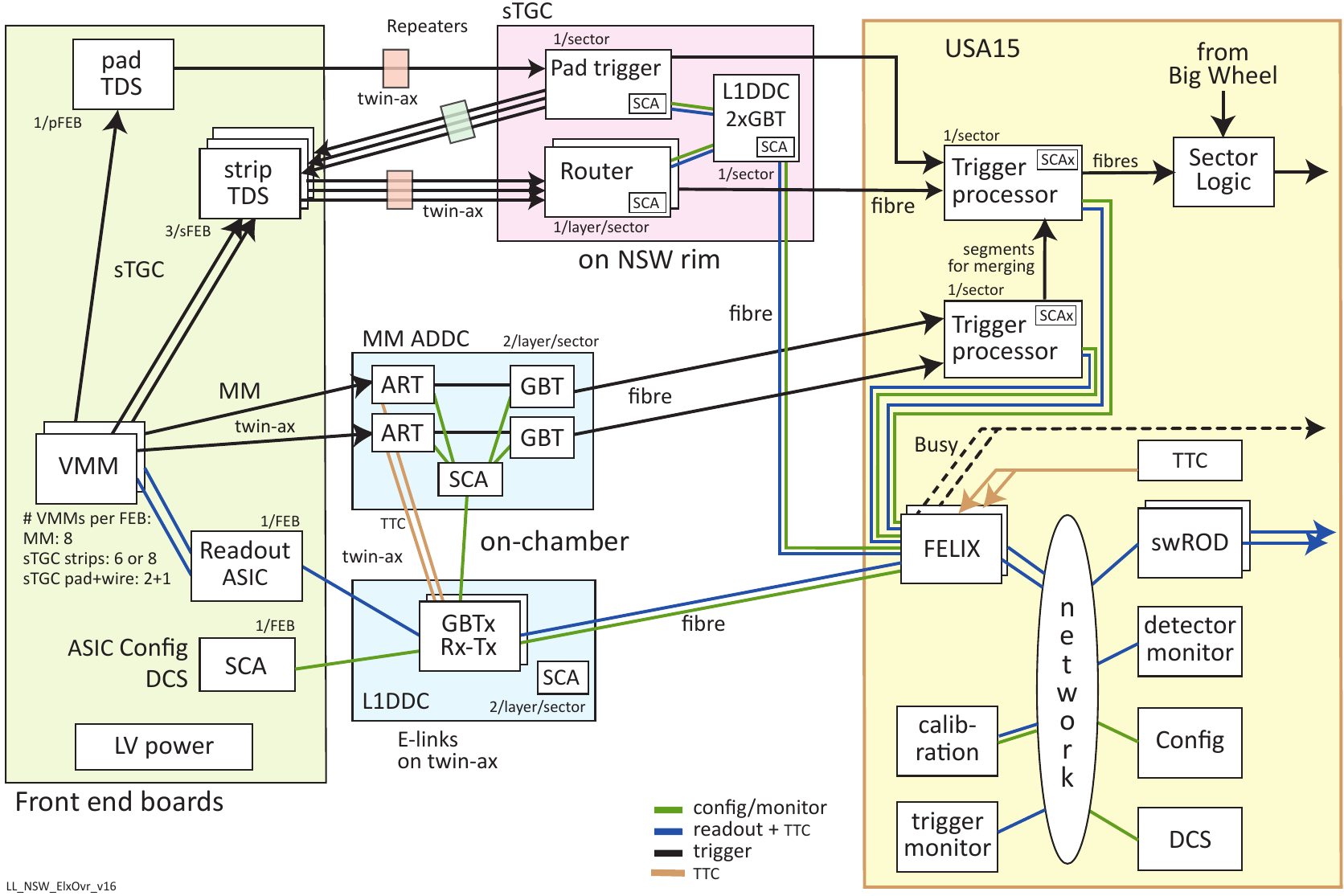}}
\caption{Overview of the \gls{NSW} electronics on- and off-detector
electronics and their connectivity.
The front-end boards are connected directly to the detectors (left box).
Also mounted directly on the wedges are the \gls{ADDC} boards for the \gls{MM} trigger (upper blue box) and the \gls{L1DDC} used for readout by both detectors (lower blue box).
The boards for the \gls{sTGC} Pad Trigger are located in the Rim Crates of the \gls{NSW} (central pink box), which also contain the Routers for the \gls{sTGC} trigger, the Rim-\gls{L1DDC} and an \glstext{LVDB}. Every \gls{NSW} electronics board installed in the ATLAS cavern has a \gls{GBT-SCA} for configuring and monitoring its components.
Off the detectors in \gls{USA15} (right box) are
the \gls{NSW-TP} and the \gls{FELIX} aggregator that
deliver the data to the ATLAS trigger and
data acquisition systems, and are described in Section~\ref{sec:TDAQ}.}
\label{fig:Muon_NSWelectronics}
\end{figure}

The trigger and readout electronics for the \gls{NSW} are closely intertwined.
Three interconnected pathways are illustrated in Figure~\ref{fig:Muon_NSWelectronics}: the Trigger and Timing path, described in Section~\ref{muonSS:trigger}, the Data Acquisition and Readout path, described in Section~\ref{muonSS:readout}, and the Configuration and Monitoring Path (see Section~\ref{sec:DCS}).
Each of these pathways has elements located directly on the \gls{sTGC} and \gls{MM} wedges, in the ``Rim Crates'' mounted at the outer radius of the \gls{NSW}, and in the service cavern \gls{USA15}.
This chapter describes only what is located on the \glspl{NSW}: the front-end electronics, the \gls{sTGC} Pad Trigger, and the additional elements that provide inputs to the \gls{NSW-TP} for the \gls{sTGC} and \gls{MM} strip triggers, and to the data acquisition system.
The \gls{NSW-TP} and \gls{FELIX} aggregator that deliver the data to the ATLAS trigger and
data acquisition systems are described in Sections~\ref{sec:NSWTP} and~\ref{subsubsec:tdaq_daqhlt_felixswrod}.
More detailed descriptions of the architecture, design considerations, and functionality of all the electronics described in this section are provided in Ref.~\cite{NSWelx}.
 
\paragraph{CERN ASICs.}
The following \glspl{ASIC} developed at CERN and used extensively in other experiments as well as ATLAS are fundamental elements of the \gls{NSW} electronics design:
\begin{itemize}
\item \gls{GBTx}: the Gigabit Transceiver~\cite{GBTx} aggregates many slow serial \glspl{e-link} into a single serial link running at \SI{4.8}{Gb/\s}
\item \gls{GBT-SCA}: every \gls{NSW} board includes a uniquely identified \gls{GBT-SCA}~\cite{GBTSCA,GBTSCA2} (see Section~\ref{sec:DCS})
for configuration, calibration and monitoring using the \gls{GBT-SCA} \glstext{OPC UA} Server.
Each \gls{GBT-SCA} includes a 31-channel 12-bit \gls{ADC} for monitoring the temperature sensors, on-chip temperatures and voltage power levels of the front-end boards.
It is configured via the \gls{FELIX} (Section~\ref{subsubsec:tdaq_daqhlt_felixswrod}).
\item \gls{FEAST}: The \gls{FEAST}~\cite{FEAST} is an integrated DC-to-DC convertor providing up to \SI{10}{\watt} of power at the voltages required by the \glspl{ASIC} on the \gls{NSW} boards, in the range \SIrange{1.2}{3.3}{\volt}, from an input supply in the range of \SIrange{5}{12}{\volt} DC. All the \gls{NSW} boards are powered by the \gls{FEAST}.
\item Versatile Links: the \gls{VTRx} and \gls{VTTx} are radiation tolerant optical link interfaces, transmitting data between on-detector and off-detector electronics at up to \SI{5}{Gb/\s}~\cite{GBT}.
\end{itemize}
 
\paragraph{NSW Custom ASICs.}
The \gls{NSW} electronics rely on a number of custom \glspl{ASIC}, described in this section. The \glstext{VMM} and \glstext{ROC} are designed to work in both \gls{NSW} subsystems, while the \glstext{TDS} and \glstext{ART} are specific to the \gls{sTGC} and \gls{MM}, respectively, and supply the inputs required by their respective \gls{NSW-TP} algorithms.
 
\subparagraph{VMM.} \label{MuSP:VMM}
A new, highly configurable custom \gls{ASD} front-end \gls{ASIC}, the \gls{VMM}3a~\cite{VMMref},
was developed for both the \gls{sTGC} and \gls{MM} front-end boards.
The \gls{VMM} can read out, amplify and shape, and provide peak finding and digitisation for up to 64 channels from the negative anode strip signals of the \gls{MM}, or the negative wire-group signals, or positive cathode strip or cathode pad signals of the \glspl{sTGC}.
Thresholds can be set independently for each channel, and a configurable option can also read out the peak charge of the nearest neighbours of strips with signals above threshold, so that a cluster of at least three strips can always be used for centroid-finding.
The \gls{VMM} also provides trigger outputs from the \gls{sTGC} pads or strips, or from the \gls{MM}.
 
The \gls{VMM} has four independent output paths, of which three are used in the \gls{NSW}: 
\begin{itemize}
\item Precision (10-bit) amplitude and (effective) 20-bit timestamp readout at \gls{L1A}~\footnote{\gls{L0A} after Phase-II upgrades.} with \SI{250}{\ns} deadtime per channel and a 64-deep FIFO per channel guaranteeing no data loss for a trigger latency up to \SI{16.0}{\us}; 
\item Serial out \gls{ART} synchronised to a \SI{160}{\MHz} clock, used to provide a 6-bit address of the first strip above threshold for the \gls{MM} trigger;
\item Parallel prompt outputs from all \num{64} channels in a variety of selectable formats for the \gls{sTGC} trigger, including a 6-bit \gls{ADC}.  When a peak is found in the \gls{sTGC} strips, its amplitude is digitised in 6-bits and immediately sent serially, one line per channel, to the strip-\gls{TDS}.
\end{itemize}

\subparagraph{Readout Controller.} 
Present on all the \glspl{FEB}, the \gls{ROC}~\cite{ROC} receives \gls{L0A} and \gls{L1A} bits from the trigger (see Section~\ref{sec:TDAQ_L1Muon}).
On receipt of a \gls{L1A}~\footnote{\gls{L0A} for \gls{HL-LHC}.}, the \gls{ROC} reads out data from the \glspl{VMM}.
The \gls{VMM} writes to the \gls{ROC} with an output bandwidth of \SI{640}{Mb/\s} (\SI{512}{Mb/\s} before 8b/10b encoding).
Each \gls{ROC} receives and processes data from up to eight \glspl{VMM}, and transmits the output to the \gls{L1DDC}.
 
\subparagraph{sTGC Trigger Data Serialiser.} 
The \gls{TDS}~\cite{TDSref} is a single \gls{ASIC} designed with two operation modes to handle pad and strip signals from the \glspl{VMM} on the \gls{sTGC} \glspl{FEB}, respectively denoted ``pad-\gls{TDS}'' and ``strip-\gls{TDS}''.
In either mode, the \gls{TDS} consists of 64-channel interfaces to two \glspl{VMM}, as well as a Preprocessor and a Serialiser.
A pad-\gls{TDS} reads out up to 104 pads, and a strip-\gls{TDS} up to 128 strips.

\subparagraph{\glstext*{ART}. \label{MuSP:ART}}
The \gls{ART} chips~\cite{ARTref} on the \gls{MM} \gls{ADDC} boards receive the six-bit address of the strip that fired first in a \gls{BC} from each of \num{32} \glspl{VMM}, and aggregate the addresses (and only the addresses) of the fired strips that are the inputs to the \gls{MM} strip trigger algorithm in the \gls{NSW-TP}.
 
\paragraph{NSW Front-End Boards.}
As far as possible, the \gls{MM} and \gls{sTGC} share front-end electronics components: the highly configurable \gls{VMM} was designed to meet the requirements of both technologies, and all the front-end boards use the same \gls{ROC} to buffer and process signals from the \glspl{VMM} for readout and transmit them to the \glspl{L1DDC}, as described in Section~\ref{muonSS:readout}. All use the \gls{GBT-SCA} for slow controls. The \glspl{FEB} are all described in detail in Ref.~\cite{NSWelx}.
The left-hand box of Figure~\ref{fig:Muon_NSWelectronics} illustrates the major similarities and differences between the three types of \gls{FEB}.
The \glspl{FEB} all receive power from \glspl{LVDB}, described in Section~\ref{MuSP:LVDB}.
 
\subparagraph{sTGC Front-End Boards.}  
The \gls{sTGC}s have two types of front-end boards: the \gls{sFEB} to read out the strip signals, and the \gls{pFEB} to read out the pad and wire signals.
Each \gls{sFEB} has six (outer and middle quadruplets) or eight (inner quadruplets) \glspl{VMM}, depending on the number of strips -- one \gls{VMM} can read out up to \num{64} strip channels.
Each \gls{pFEB} has three \glspl{VMM}: two for the pads and one for the wires.
Both types of \gls{FEB} use the \gls{ROC} to prepare \gls{VMM} signals for readout via an \gls{L1DDC} (Section~\ref{MuSP:L1DDC}).
Both use \gls{TDS} \glspl{ASIC} to digitise signals for the \gls{NSW-TP}, as described in Section~\ref{muonSS:trigger}.
There is one \gls{TDS} on the \gls{pFEB}, connected to two \glspl{VMM} (the third \gls{VMM} on the \gls{pFEB} reads out the wires, which are not used in the trigger, and therefore requires no \gls{TDS}).
The pad-\gls{TDS} transmits serialised pad hits to the Pad Trigger.
Every \gls{sFEB} has a \gls{TDS} connected to each pair of \glspl{VMM} that tags all active strip charges with the \gls{BCID} and saves them in its ring buffer, awaiting possible selection by the Pad Trigger for transmission to Routers in the Rim Crates.
Communications between the \gls{ROC} and the \gls{L1DDC}, and between the pad-\gls{TDS}, the strip-\gls{TDS} and the Pad Trigger and Router, all use miniSAS twin-axial ribbon cables.
 
\subparagraph{Micromegas Front-End Board. \label{MuSP:MMFE}}
The \gls{MM} have a single type of front-end board, the \gls{MMFE8}, that acts as an interface between the \gls{MM} detectors and the trigger (\gls{ADDC}) and data acquisition (\gls{L1DDC}) electronics cards.
Each \gls{MMFE8} has eight \gls{VMM} chips to read out the strips.
Each \gls{VMM} sends signals for the trigger to an \gls{ART} \gls{ASIC} on an \gls{ADDC} board.
A \gls{ROC} formats and buffers \gls{VMM} strip signals for transmission to an \gls{L1DDC}.
All signals to the \gls{L1DDC} and \gls{ADDC} are sent over miniSAS twin-axial ribbon cables.

\paragraph{Trigger, Readout and Low voltage On-Detector Boards.}
The \gls{sTGC} and \gls{MM} provide different types of data to the trigger and to the \gls{DAQ} readout.
The \glsfirstplural{L1DDC} aggregate and transmit the data acquisition readout for both technologies.
The \gls{MM} have an additional set of cards mounted directly on the detectors, the \glspl{ADDC}, which transmit the addresses of the strips with the first hits to the \gls{NSW-TP} (see Section~\ref{muonSS:MMtrigger}).
This section also describes the \glspl{LVDB} used to distribute power, the Router cards used by the \gls{sTGC} strip trigger, the \gls{sTGC} Pad Trigger Boards, and the Serial and \gls{LVDS} Repeaters.
 
\subparagraph{Level-1 Data Driver Card (L1DDC). \label{MuSP:L1DDC}}
The \gls{L1DDC}~\cite{Gkountoumis:2018usk} is an intermediate element in the \gls{DAQ} system for both the \gls{MM} and \gls{sTGC} detectors.
It is transparent to the type of data being transmitted or received.
It combines the three pathways listed above (Trigger and Timing, \gls{DAQ} and Readout, Configuration and Monitoring) into one or more bidirectional optical links (\gls{VTRx}).
Each \gls{L1DDC} collects detector signals and monitoring data through \glspl{e-link} to a \gls{GBTx}~\cite{GBT}, which then transmits the data to the relevant back-end systems through bidirectional (\gls{VTRx}) or unidirectional (\gls{VTTx}) optical fibre links at \SI{4.8}{Gbps} each.
The links are powered by \gls{FEAST} DC-DC converters.
The \gls{L1DDC} also distributes synchronous clocks, and trigger and configuration data from the back-end systems, to the \glspl{FEB}.
 
The \gls{NSW} uses three different types of \gls{L1DDC} boards, all made with the same custom radiation-tolerant \glspl{ASIC}, including \gls{GBTx} for high-speed serialisation and deserialisation,
Gigabit \gls{VTRx} optical transceiver and transmitter modules, and a \gls{GBT-SCA} to monitor voltages and temperatures and to configure the \gls{ROC} on the \glspl{FEB}.
The \gls{MM} and \gls{sTGC} detectors have slightly different \gls{L1DDC} boards mounted on the wedges to read out the strips, as well as the \gls{sTGC} pads and wire groups. When a \gls{L1A} initiates the transfer of data buffered on a \gls{ROC}, the data are transferred to a \gls{GBTx} that, in turn, drives the output to \gls{FELIX} over the optical links.
The \gls{MM} \gls{L1DDC} contains three \gls{GBTx} \glspl{ASIC} so it can read out eight \glspl{MMFE8} within one detector layer.
A common \gls{L1DDC} board design, with room for two \gls{GBTx} chips, is used for \gls{sTGC} strips and for \gls{sTGC} pads and wires, although the \glspl{L1DDC} for pads and wires have only one of the two \gls{GBTx} chips installed.
Each \gls{sTGC} \gls{L1DDC} reads out the three \glspl{FEB} on one side of one detector layer within a wedge.
The Rim-\gls{L1DDC} is mounted in the \gls{NSW} Rim Crates (see Figure~\ref{fig:Muon_NSWelectronics}) and provides readout of the \gls{sTGC} Pad Trigger input hits and its output decisions.
The \gls{GBT-SCA} on the Rim-Crate \gls{L1DDC} configures the Pad Trigger and the eight Router cards in the \gls{sTGC} trigger path.
Low-jitter synchronous clock signals for the Pad Trigger and Routers can be provided either from the \gls{GBTx} on the Rim-Crate \gls{L1DDC} or from a direct fibre from \gls{USA15}.
 
Each \gls{MM} wedge layer requires two \glspl{L1DDC} (eight in total per wedge) mounted at intervals along the outside edges of the wedge, on both edges.
The \gls{sTGC} on-wedge \glspl{L1DDC} are also mounted on both edges.
There are eight on each wedge in total, one each for the \glspl{pFEB} and \glspl{sFEB} of each layer.

\subparagraph{ART Data Driver Card (ADDC). \label{MuSP:ADDC}}
The \gls{MM} \gls{ADDC} contains two \gls{ART} \glspl{ASIC},
each receiving data as described above in Section~\ref{MuSP:ART} from four \glspl{MMFE8}, and thus from \num{32} \glspl{VMM}, all treated as independent data streams by the \gls{ART}.
Each \gls{ART} transmits its data to a \gls{GBTx}, which in turn transmits it to the \gls{NSW-TP} through one of the two transmission channels of a \gls{VTTx}. 
 
\subparagraph{sTGC Pad Trigger Board.}
In order to place the track finding and extrapolation logic in a less harsh and more accessible location, a coincidence, typically requiring at most a three-out-of-four coincidence of pads in each of the four-layer quadruplets, is used to choose the relevant bands of strips to be sent off-detector (as described in Section~\ref{muonSS:stgcPadTrigger}).
This substantially reduces both the required bandwidth and the amount of centroid and track-finding logic.
The board responsible for this decision-making is the \gls{sTGC} Pad Trigger~\cite{NSWelx},
located in the Rim Crate for each sector, and implemented in radiation- and magnetic field-tolerant electronics.
Placing the boards at the rim of the wheel, where radiation is less intense, allows the flexibility of forming the pad tower coincidence in programmable logic on \glspl{FPGA}.
Each Pad Trigger board receives \num{24} input links (three per layer) at \SI{4.8}{Gb/\s} from the pad-\gls{TDS} \glspl{ASIC} on the \glspl{pFEB} of the sector.
The Pad Trigger sends outputs of its logic to the strip-\glspl{TDS} on the \glspl{sFEB} via MiniSAS ribbon cables, to the \gls{NSW-TP} via optical \gls{VTTx}, as well as sending the pad hit readout data directly to \gls{FELIX} via \gls{e-link}.
 
\subparagraph{sTGC Router.}
The \gls{sTGC} Router~\cite{router_7287804} serves as a packet switch for routing strip charge information from the strip-\gls{TDS} on the \glspl{sFEB} to the \gls{sTGC} \gls{TP}~\cite{NSWelx}. There is one per layer for each sector. They are implemented in radiation- and magnetic field-tolerant electronics and sit in the sector's Rim Crate with the Pad Trigger board.
 
\subparagraph{sTGC Serial and LVDS Repeaters.}
Signals on the fast copper TwinAx connecting the \glspl{FEB} to the Pad Trigger and the Router suffer excessive attenuation if the cable length exceeds about \SI{5}{\m}.
The pad-\gls{TDS} to Pad Trigger and strip-\gls{TDS} to Router links operate at a speed of \SI{4.8}{Gb/\s}; the Pad Trigger to strip-\gls{TDS} at \SI{640}{Mb/\s}, based on \gls{LVDS} signals.
Serial and  \gls{LVDS} Repeaters~\cite{ATL-MUON-PUB-2022-003} (as appropriate) are therefore placed on all these lines to mitigate the attenuation.
 
\subparagraph{Low Voltage Distributor Board.}
\label{MuSP:LVDB} 
The input power to the \glspl{FEAST} on all the \gls{NSW} on-detector electronics is supplied by \glspl{NGPS} located in \gls{US15}. The \glspl{NGPS} provide \SI{280}{\volt} to the \gls{ICS}~\cite{NSWelx} modules (see Section~\ref{muonParLV}), which convert it to \SI{11}{\volt} and supply the electronics through the \glspl{LVDB}. One \gls{LVDB} can supply up to eight front-end boards (\analog section) and up to four digital boards (digital section).
Each \gls{MM} wedge has eight \glspl{LVDB} which act as splitters for the channels from the \gls{ICS}. Every input channel is connected to eight \glspl{MMFE8}, with an individual fuse for each \gls{MMFE8}.
There are two types of \gls{sTGC} \gls{LVDB}: one mounted on the Faraday cage of each wedge (\num{64} in total), and one in each Rim-Crate (of which there are \num{32}, one per sector).
The on-wedge \glspl{LVDB} have fuses for each \gls{FEB}, and each of the four \gls{LV} power supply channels corresponds to all the \glspl{FEB} on one detector layer (three \glspl{pFEB} and three \glspl{sFEB}); they also power the on-wedge \glspl{L1DDC}.
The Rim-Crate \glspl{LVDB} aggregate two power supply channels each, and power all the Rim-Crate electronics.
The Serial Repeaters typically draw power from a nearby \gls{L1DDC} or, in a few cases, from a nearby \gls{LVDS} Repeater.

\subsubsection{Trigger \label{muonSS:trigger}} 
The overall requirements for the trigger electronics (and for the data acquisition --- Section~\ref{muonSS:readout}) are stringent: the trigger decision has to arrive at the \Gls{SL} within \SI{1075}{\ns} (\num{43} \glspl{BCID}) of the collision,
and nearly half this time is required for fibre and cable delays.
The on-detector electronics must be radiation-tolerant to
\SI{3000}{\gray},
and operate in a highly inhomogeneous magnetic field, exceeeding \SI{0.5}{\tesla} in places.
Once the detectors are assembled, access to their electronics is extremely limited: most of the \glspl{FEB} are completely inaccessible, and most of the Rim Crates are only accessible during shutdowns long enough for the \gls{NSW} to be moved out of its running position. 
The trigger is therefore designed so that it still functions efficiently even if only a subset of the detector layers are operational, and some key components are redundant.
Despite the fact that the \gls{NSW} consists of two very different detector technologies, many components were designed to be used by both.
A defining aspect of the architecture is the use of \gls{FELIX} (see  Section~\ref{subsubsec:tdaq_daqhlt_felixswrod}) for the Readout, Configuration and \gls{TTC} distribution paths.
 
The \gls{L1} Muon trigger for \RunThr is described in Section~{\ref{sec:TDAQ_L1Muon}}. As in \RunOneTwo, forward muon triggers require track segments in the \glspl{TGC} of the \glspl{EM-TGC}; however, for \RunThr, in order to reduce the fake rate, the requirement of a matching segment in the \gls{NSW} is added to the \gls{L1} Muon \gls{SL} trigger decision.
 
This section focuses on relevant features of the detectors designed for \gls{NSW} triggering, and on the trigger functionality of the front-end electronics located on the detector, and on the rim of the \gls{NSW}.
Three distinct trigger pathways identify potential signals from the \gls{NSW}: the \gls{sTGC} Pad Trigger, the \gls{sTGC} Strip Trigger and the \gls{MM} Trigger.
The Pad Trigger (described in Section~\ref{muonSS:stgcPadTrigger}) provides an initial \gls{RoI} defined by an eight-layer coincidence ``tower'' of pads, defining an azimuthal range (identified by a ``$\phi$-ID'') that is used directly by the \gls{TP}, and a radial range (the ``Band-ID'') that uniquely determines which group of \gls{sTGC} strips should be read out by the \gls{sTGC} Strip Trigger.
The \gls{sTGC} Strip Trigger performs fast cluster-finding using only strips within the range of the Band-ID indicated by the Pad Trigger, as described in Section~\ref{muonSS:stgcStripTrigger}. The \gls{MM} Trigger runs independently of the \gls{sTGC} triggers, looking for particle ``roads'' based around the first \gls{MM} strip channel to fire for each \gls{VMM}; this is explained in Section~\ref{muonSS:MMtrigger}.
 
The Pad Trigger runs in the Rim Crates on the \glspl{NSW}, while the \gls{sTGC} Strip Trigger and the \gls{MM} Trigger algorithms run in the \gls{NSW-TP}, located off the detector in the Service Cavern \gls{USA15}.
At each bunch crossing, the \gls{NSW-TP} looks for local track segments that point to the \gls{IP}, and sends them to the \gls{SL} 
to corroborate muon triggers from the \gls{EM-TGC}.
The track segments are inputs to the \gls{L1}-Muon trigger decision, described in Section~\ref{sec:TDAQ_L1Muon}.
 
\paragraph{sTGC Pad Trigger \label{muonSS:stgcPadTrigger}}
The \gls{VMM} sends a discriminated 
pulse to the pad-\gls{TDS} from its 
threshold-crossing circuit
as described in Section~\ref{MuSP:VMM}.
 
The pad signals  are captured every \gls{BC}. The pad-\gls{TDS} has per-channel programmable delays that compensate for the different time-of-flight and pad trace lengths to the \gls{VMM}. At the end of the \gls{BC}, the firing status of all the channels is multiplexed and serialised by the pad-\gls{TDS}, and the vector of pad ``hit'' bits in a gas gap is sent within one \gls{BC} to the Pad Trigger Board in the Rim Crate for the sector.
 
To reduce the number of strip channels to be read out, while keeping the number of physical pads small, pad patterns are staggered across the four layers of each detector wedge by half a pad in both directions
to make logical towers corresponding to \glspl{RoI} built from virtual pads one quarter the area of a physical pad (about \SI{4}{\cm} in radial extension), typically corresponding to about 13 strips). The geometry of the pad offsets between the two wedges in each sector ensures that these logical towers point toward the \gls{IP}.
 
Because a pad-\gls{TDS} reads out at most \num{104} pads per cathode board, this is the maximum that can be used in the trigger. Some pad cathode boards of the innermost quadruplets have up to \num{112} pads, and in these cases the pads at the smallest radii are not used in the trigger; however these pads are not in the region $\abseta <2.4$ covered by the \gls{EM-TGC} trigger that the \gls{NSW} trigger is required to confirm.
 
The Pad Trigger uses the inputs from the pad-\gls{TDS} on each of the four layers of the three quadruplets in each of the two \gls{sTGC} wedges of its sector to tag a bunch crossing and identify the band of strips passing through the triggered tower in each layer that must be selected for readout from the strip-\gls{TDS}.
Every pattern corresponding to a logical pad tower is checked to see if at least three out of four layers in both quadruplets have hits. 
There are approximately \num{4000} possible trigger ``candidates'' (logical pad towers corresponding to tracks from the \gls{IP}) in a Large sector and about \num{1800} in a Small sector, which are stored in \glspl{LUT}.
 
The $\phi$-ID of towers satisfying these criteria is sent to the \gls{NSW-TP}, as well as directly to the \gls{FELIX} for readout.
The Band-ID is sent to the strip-\gls{TDS} on the \glspl{sFEB} to request the corresponding band of strips in each layer for their associated \gls{BCID}.
 
\paragraph{sTGC Strip Trigger Inputs. \label{muonSS:stgcStripTrigger}}
Each layer of each quadruplet has three (four for the inner quadruplets) strip-\gls{TDS} chips on its \gls{sFEB}.
At each \gls{BC}, the Pad Trigger can request up to four strip-\gls{TDS} chips per wedge-layer, and at most one candidate per strip-\gls{TDS}, to select and send data for a band of strips.
The strip-\gls{TDS} in each layer that holds a selected band transmits its digitised strip charges for that \gls{BCID} to a Router in the Rim Crate.
The data transmitted include the \gls{BCID}, band-ID, $\phi$-ID and the strip charges.
Although data from 17 strips (the 13 in the \gls{RoI} and two neighbouring strips on either side) are serialised, there is only enough time to send 120 bits of data from the strip-\gls{TDS} in one \gls{BC}.
Since the size of a muon cluster is typically around four to five active strips, it is possible to reduce the number of strips transmitted from 17 to 14 and add one bit indicating whether the highest or lowest 14 are selected.
The Router has inputs from nine strip-\gls{TDS} chips; however, only four \gls{TDS} inputs are active for any given \gls{BCID}.
For each \gls{BCID}, the Router selects the four active strip-\gls{TDS} inputs requested by the Pad Trigger and sends their data to the centroid-finding and track-extrapolation logic in the \gls{NSW-TP} via four optical fibres per layer.
The \gls{sTGC} strip trigger algorithm in the \gls{NSW-TP} is described in Section~\ref{tdaq:stgcStripTrigger}.
 
\paragraph{Micromegas Trigger Inputs. \label{muonSS:MMtrigger}}
The address of the first strip channel in a \gls{BC} to fire for each \gls{VMM} is transmitted to the \gls{ADDC}, which aggregates and forwards the address data (as described in Section~\ref{MuSP:ADDC}) to the \gls{NSW-TP}.
At most eight hits per \gls{ART} are forwarded to the \gls{NSW-TP} per \gls{BCID} (see Section~\ref{MuSP:ADDC}).
The \gls{MM} trigger algorithm in the \gls{NSW-TP} is described in Section~\ref{tdaq:MMtrigger}.

\paragraph{NSW Trigger Processor. }
The right-hand box of Figure~\ref{fig:Muon_NSWelectronics} illustrates the connectivity of the \gls{NSW-TP}, which is located off the detector, and described in detail in Section~\ref{sec:NSWTP}.
The \gls{sTGC} strip trigger and the \gls{MM} trigger algorithms run independently on separate \glspl{FPGA} in the \gls{NSW-TP}, which then merges the \gls{sTGC} and \gls{MM} candidate segments and sends at most eight merged \gls{NSW} candidate segments for each \gls{BCID} to the \gls{SL} to be compared with \gls{EM-TGC} muon candidates. Inputs and outputs are also sent to the \gls{FELIX}.
 
\subsubsection{Data Acquisition and Readout \label{muonSS:readout}} 
\gls{MM} strips are read out~\cite{NSWelx} 
through the \gls{MMFE8}; \gls{sTGC} Strips are read out through the \glspl{sFEB}, and pads and wire groups through the \glspl{pFEB}.
All data from all \gls{MM} readout strips and \gls{sTGC} pads, wire groups and strips with signals above threshold (along with the signals of the neighbouring channels in the case of the strips) are digitised using the 10-bit \gls{ADC} of the \gls{VMM} \glspl{ASIC} on the respective \glspl{FEB}.
The digitised signals are buffered on the \glspl{VMM} until a \gls{L1A}\footnote{From \RunFour, this will be a \gls{L0A}.} is received from the \gls{CTP}.
Zero data loss is guaranteed for a maximum latency of \SI{16}{\us}, but if the \gls{VMM} readout is delayed for longer, new data are lost; however, this far exceeds the latency requirements for \RunThr, unchanged from \RunOneTwo.
The \gls{ROC} \gls{ASIC} on all of these \glspl{FEB} (described above in Section~\ref{muonSS:electronics}), receives \gls{L0A} and \gls{L1A} bits from the \gls{CTP} via \gls{FELIX} (see Section~\ref{sec:TDAQ_L1Muon}).
On receipt of a \gls{L1A}~\footnote{\gls{L0A} at \gls{HL-LHC}.}, the \gls{ROC} reads out data from the \glspl{VMM} and transmits it to the \gls{FELIX} via the \gls{GBTx} on the \glspl{L1DDC} mounted on the wedge.

\subsubsection{Temperature and Magnetic Field Sensors \label{sss:TandBsensors}} 
The magnetic field is complicated and highly non-uniform in the region occupied by the \gls{NSW}.
To measure it {\em in situ}, twelve 3D~Hall probe cards per Large sector, similar to those that were mounted on the \gls{MDT} chambers of the \glspl{SW}~\cite{PERF-2007-01}, are mounted on the inner (\gls{IP}-facing) faces of each of the Large-sector \gls{sTGC} wedges in the plane farthest from the \gls{IP} (four per quadruplet).
 
All \gls{MM} wedges have nine temperature sensors (\gls{NTC} 10K thermistors) mounted on the same surface as the alignment source platforms (the surface facing toward the corresponding \gls{sTGC} wedge).
All \gls{sTGC} wedges also have nine of these temperature sensors, epoxied
in locations spread over the full surface of the non-alignment (outward) face.
All \gls{MM} and \gls{sTGC} wedges have four additional \gls{NTC} sensors, mounted on the cooling water inlet and outlet pipes on either edge of the wedge (Section~\ref{muonSS:services}).
 
The B-field sensors and temperature sensors are all read out through \gls{ELMB} modules~\cite{ELMB} in crates at the rim of the \gls{NSW}. \gls{ELMB} is a general-purpose plug-on board developed by the ATLAS collaboration for various detector control tasks, and used since the beginning of \RunOne.
 
\subsubsection{Services\label{muonSS:services}} 
The \glspl{NSW} re-use, as far as possible, the services installed for the \glspl{SW}, including the gas racks that supplied the \glspl{TGC}, which use the same n-pentane/CO$_2$ mixture as the \glspl{TGC}, \glspl{MDT} 
and \glspl{CSC}.
The \gls{MDT} gas system was modified to re-purpose the racks supplying the \gls{SW}. 
The former \gls{CSC} gas system was modified to supply the \gls{MM} detectors with a new gas mixture.
The \gls{TGC} gas system distribution racks were modified to allow the supply of the new chambers in the \gls{NSW}.
 
The \gls{LV} requirements of the \gls{NSW} are different from those of the \glspl{SW}, and required a new system, described in Section~\ref{muonParLV}.
Its greater power consumption required upgrades to the chilled water cooling systems of ATLAS, which now provide \SI{175}{\kilo\watt} to each side of ATLAS.
 
\paragraph{Gas Distribution}
\subparagraph{\glstext*{MM} Gas}
The \gls{MM} use a mixture of 93\% argon, 5\% CO$_2$ and 2\% isobutane (iC$_4$H$_{10}$). The mixer is required to supply a nominal flow of \SI{600}{\litre/\hour}, with \num{32} flow cells calibrated to receive the same mixture at a flow of \SIrange{0}{30}{\litre/\hour} per flow cell.
The \gls{MM} gas is supplied to each wheel through \num{16} input channels, each serving two wedges in the same plane of azimuthally adjacent Large or Small sectors.
A bundle of eight inlet pipes goes to the upper half of each wheel and another to the lower half.
Four of the eight pipes serve one quadrant,
each supplying two adjacent Large or two adjacent Small wedges.
Each pipe then splits into four at a manifold, with an input line going to each edge of each of the two wedges served.
Gas enters the \gls{MM} wedges at the outer rim.
Impedances in the input lines and manifolds respectively ensure uniform gas flow through every wedge of the same type and through every layer of a wedge.
The gas is provided to each detector layer in every sector via two intermediate buffer volumes.
The average gas pressure in the chambers is \SIrange{0.1}{0.2}{\kPa} 
above atmospheric pressure, depending on the height of each chamber.
Gas is provided simultaneously to both inlets of the outer module and distributed by two small pipes to the inner module. The outlet of the inner module goes to the return line.
Outlet lines serving the same groupings of detectors as the inlet lines are connected at the inner end of each wedge.
 
\subparagraph{sTGC Gas}
The \glspl{sTGC} use the same strongly quenching mixture of 45\% n-pentane and 55\% CO$_2$ as the existing \glspl{TGC}, also at atmospheric pressure.
The \gls{TGC} gas system was modified to add \num{12} channels for the n-pentane:CO$_2$ mixture and two channels for the CO$_2$ envelope.
The CO$_2$ channel surrounding the \gls{sTGC} gas volumes, described in Section~\ref{par:sTGCwedgestruct}, is exhausted to a line equipped with flammable gas sniffers, so that in case flammable gas should leak from the detectors, it would be detected in the output of the CO$_2$ channel, and not escape into the ATLAS cavern.
 
The maximum flow rate for the entire \gls{TGC} gas system for \RunThr, including the \gls{sTGC}, remains \SI{5000}{\litre/\hour}. This does not constrain the \gls{sTGC} flow rate, and will change after \gls{LS3}, when the gas-mixing system will be modified.
Gas is supplied to each wheel through two main gas input channels, each connected to a manifold distributing the gas to the eight sectors of the upper or lower half of the wheel.
Calibrated flow restrictors ensure that all wedges, Large or Small, receive the same number of volume exchanges per day, irrespective of their position on the wheel.
To ensure uniform flow and avoid damaging multiple layers in case of accidental contamination with dirt or debris, each layer of the wedge is separately supplied.
Gas enters each layer at one side of the small end of the innermost quadruplet and follows a serpentine path around the ends of the wire supports inside the gas gap, until it reaches the corner of the large end of the same side of the quadruplet, passes through a tube into the second quadruplet, and then passes in the same way to the outermost quadruplet, finally exiting at the outer radius of the wedge. There are thus four gas input channels entering the wedge at the inner radius and four gas outlet channels leaving it at the outer radius, all on the same edge of the wedge.
 
\paragraph{Grounding}
The grounding of the \gls{MS} was designed~\cite{NSWelx} to follow the overall ATLAS policy~\cite{Blanchot:1073170}: for the definition of the DC level, and for safety, all wedges are connected from one point to the ATLAS structure ground. This ``star'' grounding does not allow currents to circulate in ground loops. It is, however, difficult to adhere strictly to this ideal in practice, and additional grounding connections were added in many places during construction and commissioning of the \gls{sTGC} and \gls{MM} detectors to suppress noise.
 
\paragraph{High Voltage Distribution}
The \gls{sTGC} anode wires are biased at \SI{2.8}{\kilo\volt} during running for optimal efficiency. A single \gls{HV} is supplied to all the wire groups in a single gas volume of each quadruplet, except for the innermost quadruplets, where the wires are divided in two,
and the inner- and outer-radius ends of each gas volume receive separate \gls{HV}. Each of the 32 wedges in each \gls{NSW} thus requires sixteen \gls{HV} channels in total, connected to the wedge with a single, multi-channel Lemo-Redel connector.
The \gls{SHV} cables from the distributors connect to patch-panels on the rim of the wheel. 
Like all the \gls{HV} and \gls{LV} systems of the ATLAS \gls{MDT}, \gls{RPC} and \gls{TGC}, the \gls{NSW} \gls{HV} distribution uses the CAEN \gls{EASY} Crate system, with \gls{EASY}~3000 crates containing the power supplies and \gls{HV} modules placed in racks in UX15, and mainframes and branch controllers in the \gls{USA15} cavern. The AC-DC power supplies in the \gls{US15} service cavern are the same that were used for the \gls{TGC} of the original EI-wheel, and the original power cables are also reused.
The \gls{HV} distributors occupy the two racks in the main \gls{UX15} experimental cavern previously used for the \gls{SW} \glspl{TGC}.
The \gls{HV} modules that powered the \gls{SW} \glspl{TGC} are re-used, supplemented with new, additional radiation-hard modules.
Their cables were all replaced with new ones for \RunThr.

\paragraph{Low Voltage Distribution \label{muonParLV}}
The \num{7500} on-detector electronics boards of the \gls{NSW} draw about \SI{110}{\kilo\watt} of power in aggregate. 
The boards are all powered by \gls{FEAST} \glspl{ASIC}~\cite{FEAST}, which can operate from an input supply of at most \SI{12}{\volt}. As mechanical and thermal considerations limit the permissible cable bulk and require a supply of at least \SI{48}{\volt} DC,  a power conversion stage on the detector is therefore necessary: the \gls{ICS}.
 
The \gls{ICS} is required to withstand up to \SI{96}{\gray} of ionising radiation, up to \num{5.8e12} of \SI{1}{MeV}-equivalent \si{n\per\cm\squared}, 
and single-event hadron fluences exceeding \SI{e12}{p\per\cm\squared} (E>\SI{20}{\MeV}), and to function in a magnetic field of up to (or perhaps exceeding) \SI{0.5}{\tesla}. The \glspl{ICS} are installed near the rim of the \gls{NSW}, water-cooled, monitored and controlled remotely.
 
Each \gls{ICS} module provides eight \gls{LV} channels.
The \gls{MM} system requires \num{80} \glspl{ICS} (five per sector),
and the \gls{sTGC} system needs \num{48} \glspl{ICS} (three per sector),
for a total of \num{1024} \gls{LV} channels in all.
 
\paragraph{Cooling Water}
Numerous chips on the front-end boards of both \gls{MM} and \gls{sTGC} wedges require water cooling, as do the \glspl{L1DDC} of both technologies and the \glspl{ADDC} of the \gls{MM}, as well as the \gls{ICS} for the \gls{LV} power,
the repeaters, 
and the electronics in the Rim Crates at the edge of the wheel. 
For each wheel, two cooling loops supply the eight upper and lower sectors of the wheel respectively.
The maximum height difference between elements supplied by the same loop is therefore \SI{3.6}{\m}.
Cooling pipes from the cooling station enter each wheel via the flexible chains.
A manifold splits the input flow into twenty individual lines, eight for the  \gls{MM} sectors, eight for the \gls{sTGC} sectors, and four supplying the rim electronics.
In each line, flow can be switched off by means of a manual cut-off valve, and adjusted by means of a needle valve.
An identical manifold, with the cut-off valves but without needle valves, combines the return lines from the sectors and rim electronics.

The \gls{MM} wedges take advantage of the hollow spacer frame between them, which contains integrated stainless steel cooling channels.
The \gls{sTGC} have no access to the central spacer, so their cooling water circulates along the radii of the wedges: each \gls{sTGC} wedge has two cooling loops, one on each long edge, outside the main Faraday cage, with both the inlet and outlet at the outer radius. The cooling pipes are made of copper, bent to route them through the hot zones of all the \glspl{FEB}, and thermally coupled via conductive foam and metal connectors to the heat-emitting chips.  They then loop through the \gls{L1DDC} at the outer radius before exiting the wedge. To minimise the risk of leaks, the cooling loop on each side is a single continuous loop, with electrically insulating connectors placed as far as possible from the wedge electronics.
 
The \glspl{FEB} have built-in temperature monitoring, and temperature sensors are also mounted on the input and outputs of the cooling pipes and on the outer surfaces of the \gls{sTGC} and \gls{MM} wedges, as described in Section~\ref{sss:TandBsensors}, ensuring that a cooling failure would be rapidly detected.

\subsection{Upgrades to the Muon Spectrometer Barrel Coverage} 
This section describes all upgrades to the Barrel \gls{MS} since \RunOne.
\gls{MDT} chambers serve as the primary precision tracker for the entire barrel of the ATLAS \gls{MS}, as well as for the EM and EO stations of the endcaps; however, because of the relatively large \SI{30}{\mm} diameter of the individual drift tubes, they have long drift times, are subject to high occupancies in regions where background rates are high, and there are some regions of the detector where space is very tight.
The \glspl{RPC} that provide the barrel muon trigger are subject to similar constraints.
In response to these challenges, new, smaller-volume versions of both technologies have been developed to fill gaps in coverage in the original \gls{MS}, and improve trigger coverage in the complicated transition region, where tracks pass through both barrel and endcap chambers. The \gls{sMDT} and \gls{sRPC} technologies are described in Sections~\ref{muonSS:smdt} and~\ref{muonSS:srpc} respectively.
Chambers added during \gls{LS1} and \RunTwo are described in  Section~\ref{muonSS:BMEBOEBMG}, and those added in \gls{LS2} in Section~\ref{muonSS:BIS78}.
 
\subsubsection{Small-Diameter MDT Chambers \label{muonSS:smdt}}
For regions where high backgrounds or tight space constraints make the use of traditional \glspl{MDT} impractical, new \glsfirst{sMDT}~\cite{sMDT1,sMDT2} chambers have been developed
with a \SI{15}{\mm} tube diameter: half the diameter of the drift tubes of the \gls{MDT} chambers. Operated with the same mixture of 93\% argon and 7\% CO$_2$ gas at \SI{3}{\bar}, with the same gas gain of \num{20000}, and read out by the same eight-channel
ASD chips, the spatial resolution of the \gls{sMDT} as a function of the drift radius is similar to that of the \gls{MDT}~\cite{sMDT_res}
(see
Section~\ref{muonSS:BIS78}).
The average \gls{sMDT} drift tube resolution is expected to remain below or around \SI{110}{\micron}, even for the maximum background rates expected at the \gls{HL-LHC} in the inner barrel region of ATLAS.
 
The \gls{sMDT} spatial resolution and efficiency remain comparable to those obtained with \glspl{MDT} in the absence of any background rate, even when rates are increased by roughly an order of magnitude,
because the smaller tube diameter reduces the tube cross section exposed to background radiation and the maximum drift time from about \SI{720}{\ns} to \SI{175}{\ns} 
and leads to a strong suppression of the effects of
space charge caused by the photon and neutron background radiation in the cavern~\cite{sMDT1,sMDT2}. \gls{sMDT} chambers have shown no ageing up to a charge accumulation on the wire of \SI{9}{\coulomb/\cm}~\cite{sMDT1,sMDT2}, equivalent to several \gls{HL-LHC} lifetimes in all regions of the detector.
 
\subsubsection{New Thin-Gap Resistive Plate Chambers \label{muonSS:srpc}}
Like the \glspl{MDT}, the \glspl{RPC} used in the trigger can also be made more compact and robust, and faster, by shrinking the gas gap.
Each singlet of the ATLAS \glsfirst{sRPC}, consists of a gas gap of \SI{1}{\mm} between two \SI{1.2}{\mm}-thick graphite-coated \gls{HPL} electrodes,
with the precise gap distance maintained by a grid of insulating polycarbonate spacers placed at \SI{7}{\cm} intervals.
 
The new chambers are operated with the same gas mixture\footnote{Studies are in progress to find a suitable replacement to conform to environmental requirements at the \gls{HL-LHC}~\cite{Mandelli:2021zxi}.}
as the original ATLAS \glspl{RPC}, a mixture of C$_2$H$_2$F$_4$ (94.7\%)-C$_4$H$_{10}$(5\%)-SF$_6$(0.3\%), 
and reach full efficiency at a relatively modest \gls{HV} of \SI{5.6}{\kV}
with \numrange{7}{10} times smaller average avalanche charge than the
legacy \glspl{RPC}. 
The rate capability and lifetime are correspondingly enhanced by a factor 10 with respect to the legacy \glspl{RPC}~\cite{RPC_aging}, such that the chambers can sustain a \SI{1}{\kHz/\cm\squared} counting rate over the lifetime of the \gls{HL-LHC},
at least twice the requirement for the barrel inner layer~\cite{quanyin2018}. 
The time resolution of the \glspl{sRPC} is better than \SI{350}{\ps}  
due to the smaller avalanches, resulting in a triplet time resolution of about \SI{200}{\ps}.

\subsubsection{Improved Coverage in the Vicinity of Barrel Feet and Elevator Shafts\label{muonSS:BMEBOEBMG}}
In the original \RunOne configuration, there were acceptance holes in the bottom sector (Sector~13) 
of the muon spectrometer barrel middle and outer layers due to the gaps required to allow elevators to pass.
In 2013, during \gls{LS1}, these acceptance gaps in the outer barrel layer were projectively covered  by the ``BOE'' chambers, mounted below the outer barrel layer.
The BOE are standard \gls{MDT} and \gls{RPC} detectors of the original types used throughout the ATLAS \gls{MS}.
 
In addition, two ``BME'' chambers were installed in Sector~13 in 2014, also during \gls{LS1}, to cover the elevator holes in the barrel middle layer.
The BME stations use \gls{sMDT} (operated with spare electronics from the original \gls{MDT} construction) and \gls{sRPC}, with new \analog front-end electronics, similar to the \gls{BIS78} chambers described in Section~\ref{muonSS:BIS78}. They are mounted on rails, which allow them to move into a parking position when the elevators are in use.
The BME \gls{sMDT} chambers were operated in ATLAS throughout \RunTwo, improving the muon momentum resolution in these sectors~\cite{BMEBMG1,BMEBMG2} by providing a third track point measurement.
 
\begin{figure}[!h]
\centering
\includegraphics[width=0.99\textwidth]{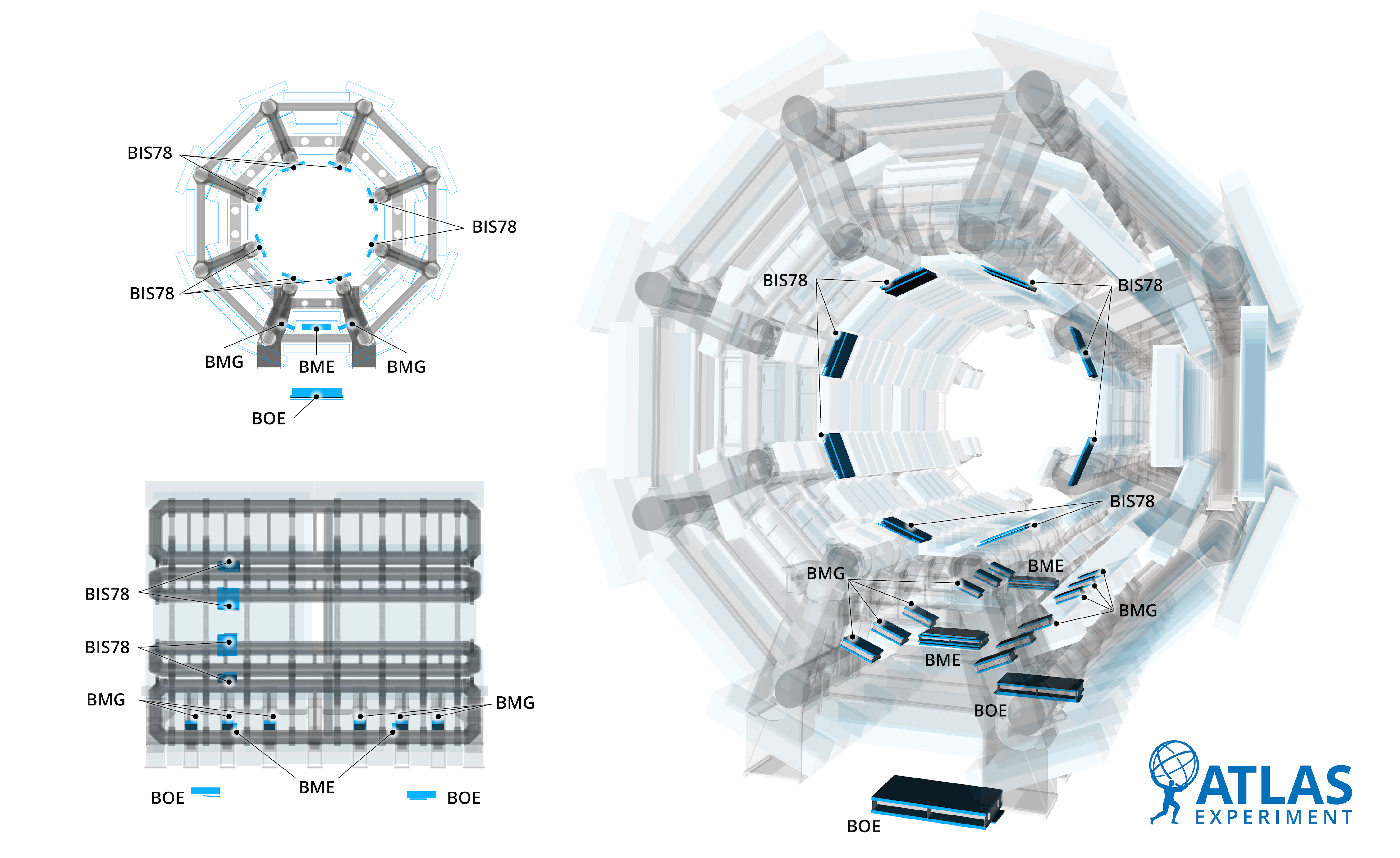}
\caption{Upgraded barrel muon chambers are shown in solid blue, and include the two BOE \gls{MDT} chambers and the two BME \gls{sMDT} chambers, covering the elevator shafts, the twelve BMG \gls{sMDT} chambers in the calorimeter feet, and the eight \gls{sMDT} and \gls{sRPC} integrated \gls{BIS78} modules on Side~A of the transition region between the barrel and endcap.}
\label{fig:muon_barrel_phase1}
\end{figure}

Twelve additional \gls{sMDT} chambers (BMG) were then installed in the barrel middle layer during the 2016/17 winter shutdown to fill acceptance gaps inside the detector feet in Sectors~12 and~14 that support the calorimeters, the \gls{NSW} and the endcap toroid cryostats. These were the first \gls{sMDT} chambers operated using the newer front-end electronics also used for the \gls{BIS78} upgrade, described in Section~\ref{muonSS:BIS78}.
The locations of all the chambers described in this section are shown in Figure~\ref{fig:muon_barrel_phase1}.

\subsubsection{New BIS78 Muon Chambers \label{muonSS:BIS78}}
While the requirement of a matching segment in the \gls{NSW} will reduce the \gls{L1} single-muon trigger rate in the $\abseta > 1.3$ region, a similar background exists in the region covered partly by barrel and partly by endcap chambers ($1.0<\abseta <1.3$).
In the original \RunOne configuration
the \gls{L1} trigger rate for muons with $\pT > \SI{20}{GeV}$ originating from this small region was about
20\% of the total \gls{L1} trigger rate. 
As in the region covered by the \gls{NSW}, this fake-muon background can also be reduced by requiring a coincidence between the trigger chambers of the EM wheels and an additional inner-layer chamber (see Figure~\ref{fig:TDAQL1MuonEndcapPerformance}).
In the Large sectors of the \gls{MS} transition region, the additional trigger inner layer is provided by the \gls{EIL4} \gls{TGC} chambers (mounted between the barrel toroid coils as shown in Figure~\ref{fig:MuonLargeQuadrant}), but in the Small sectors, the barrel toroid coils preclude the existence of ``EIS'' chambers.
 
\begin{figure}[!h]
\subfloat[]{\label{fig:BIS78position}\includegraphics[width=0.4\textwidth]{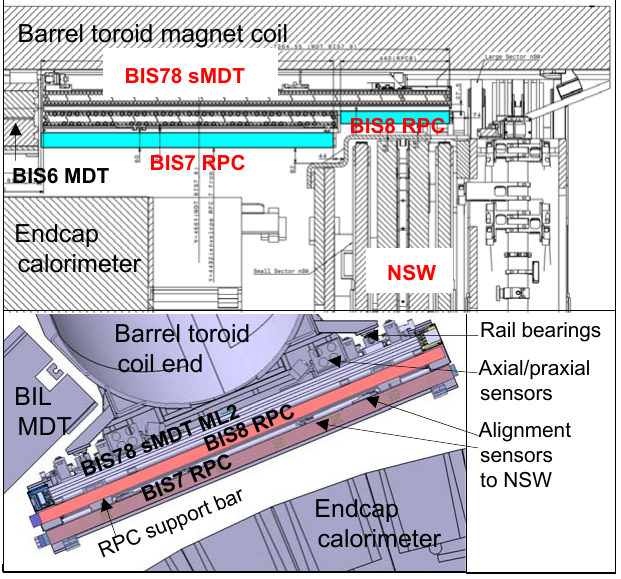}}
\subfloat[]{\label{fig:BIS783D}\includegraphics[width=0.6\textwidth]{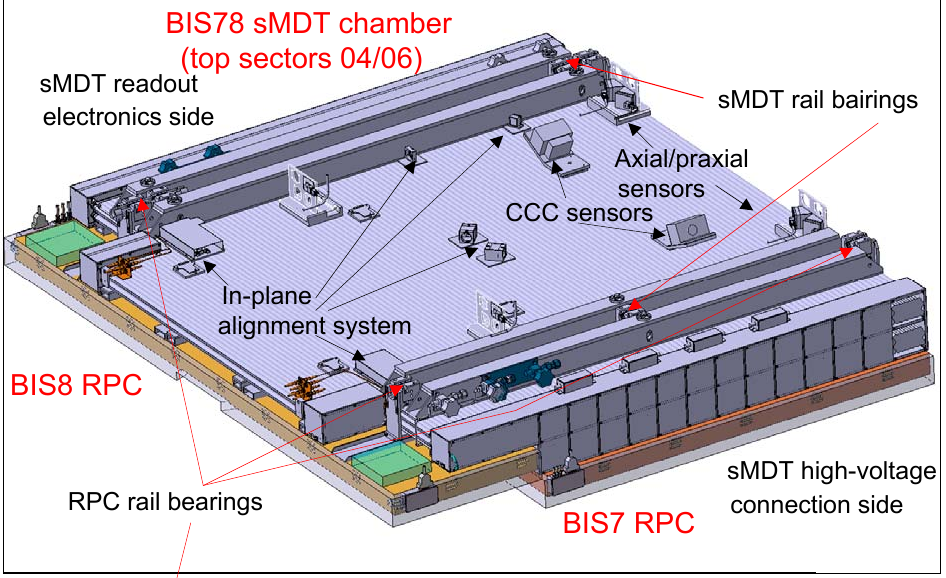}}
\caption{\protect\subref{fig:BIS78position} Placement of the \gls{BIS78} \gls{sMDT} and \gls{sRPC} chambers
in the Small sectors of the inner endcap layer at the interface with the \gls{NSW} and \protect\subref{fig:BIS783D} 3D model of a \gls{BIS78} module.}
\end{figure}

To fill these gaps in the Small sectors of the inner layer of the endcap, the barrel inner layer extends into the transition region, with BIS chambers installed on the surface of the barrel toroid magnet coils.
During \RunOneTwo, these BIS7 and BIS8 chambers served only for tracking, and were not part of the trigger: like the other barrel-inner chambers, they consisted only of \glspl{MDT}.
Because of the limited space between the \gls{SW} and the barrel toroid magnet coils (see Figure~\ref{fig:BIS78position}) the original BIS8 chambers comprised only a single multilayer of \glspl{MDT}.
 
In order to add triggering capabilities in this region and thus reduce the trigger rate in the Small sectors, the \glspl{MDT} had to be replaced by a more compact technology with space to add \glspl{RPC}.
Both the BIS7 and BIS8 \gls{MDT} chambers on side~A of ATLAS were removed during \gls{LS2}, and replaced by eight new integrated modules combining integrated  \gls{sMDT} BIS7/BIS8 chambers and separate BIS7 and BIS8 \glspl{sRPC} in a single mechanical structure~\cite{ATLAS-TDR-26} (see Figure~\ref{fig:BIS783D}).
Because the new integrated modules cover the areas formerly covered by the BIS7 and BIS8 chambers (see Figure~\ref{fig:muon_barrel_phase1}), they are called ``\gls{BIS78}''.
The corresponding BIS7 and BIS8 chambers on side~C will be replaced by \gls{BIS78} during \gls{LS3} for operation at \gls{HL-LHC}.
 
\begin{figure}[!h]
\centering
\includegraphics[width=0.475\textwidth]{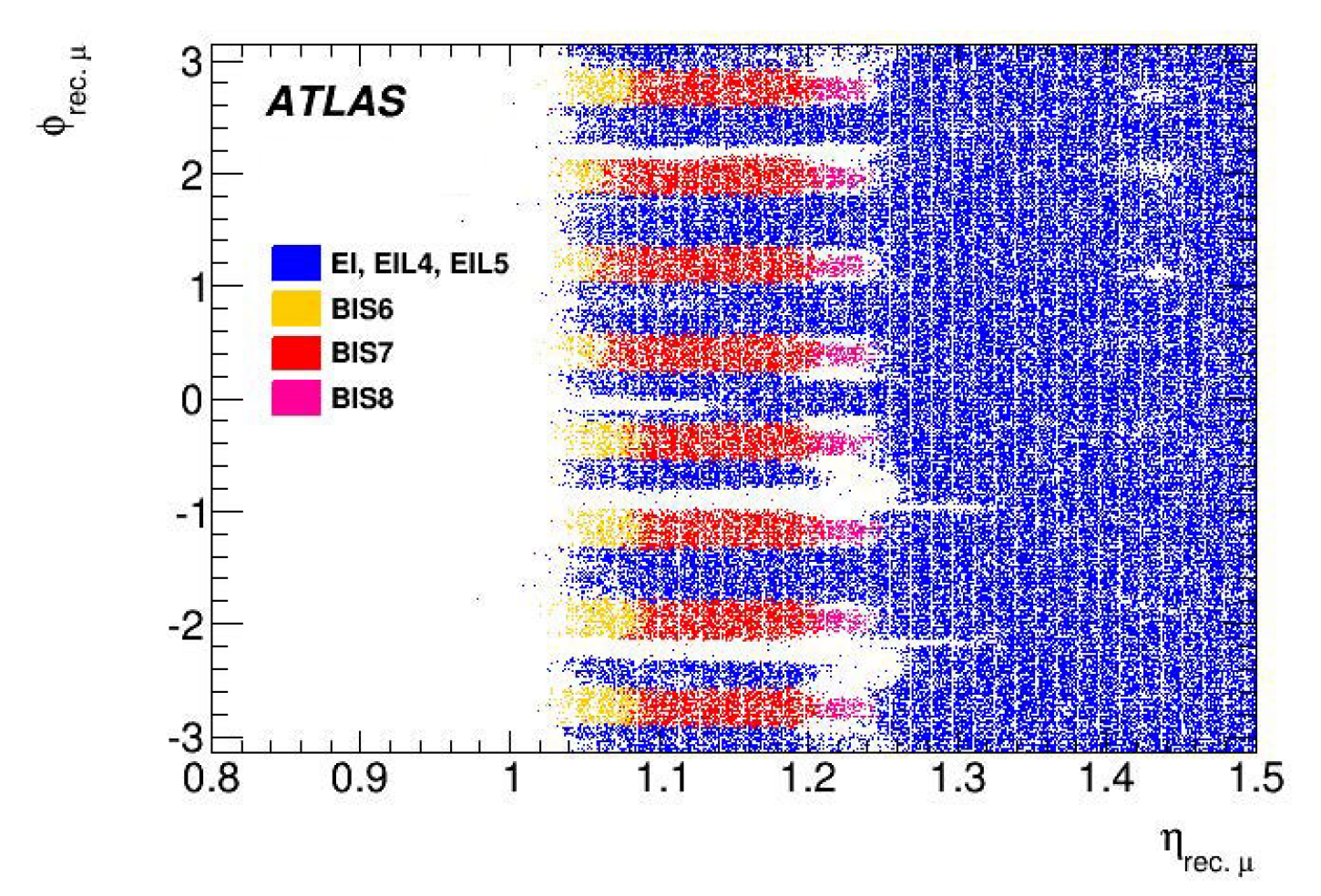}
\caption[Muon rates]{
The figure shows the $\eta-\phi$ distribution of reconstructed muons with $\pT >$ \SI{20}{\GeV} and associated to an endcap trigger (from the \glspl{TGC} of the EM wheels)
from track segments in the EI chambers (blue) and in the BIS chambers (all other colours). \label{fig:Muon_rate}}
\end{figure}

The coverage of the range $1< \eta <1.3$ 
with the \glspl{sRPC} in \gls{BIS78} is illustrated in Figure~\ref{fig:Muon_rate},
while the resulting anticipated rate reduction
is shown in Figure~\ref{fig:TDAQL1MuonEndcapPerformance}.
The new, smaller-volume technologies are also better able to withstand the high background rates anticipated in this region at the \gls{HL-LHC}.
The \gls{BIS78} chambers constitute a pilot project for the complete replacement of the BIS \gls{MDT} layer by the same type of integrated \gls{sMDT} and \gls{sRPC} chambers and the addition of \gls{sRPC} triplets to the BIL layer after \RunThr for operation at \gls{HL-LHC}~\cite{ATLAS-TDR-26}.
The impact of the improved time resolution that could be obtained at the \gls{HL-LHC} by using \glspl{sRPC} more extensively in the BI layer
will also be studied with the \gls{BIS78} chambers in \RunThr.

\begin{figure}[!h]
\centering
\includegraphics[width=0.95\textwidth]{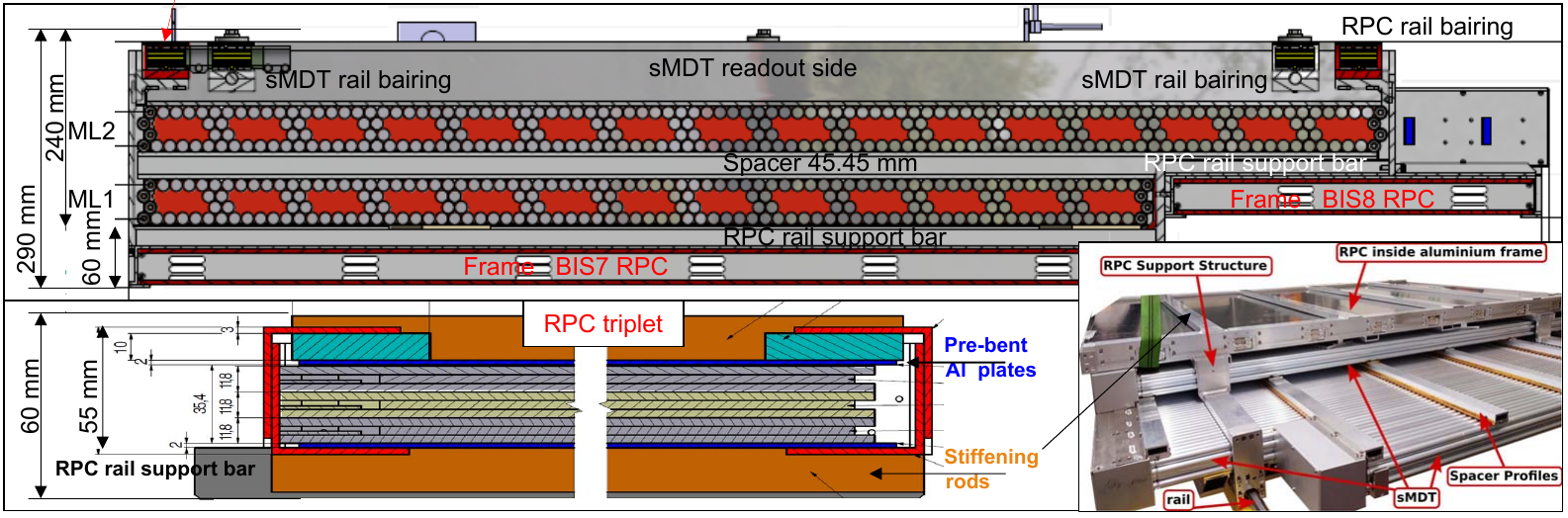}
\caption{Cross-sectional view of \gls{BIS78} module, with inset 3D view (bottom right) and zoomed view of the \gls{sRPC} triplet (bottom left).
\label{fig:BIS78Xsec}}
\end{figure}

Like the BIS7 \gls{MDT} chambers they replace, the
\gls{BIS78} \gls{sMDT} chambers consist of two four-layer multilayers of small-diameter aluminium drift tubes.
The two multilayers are separated by a \SI{45.5}{\mm}-high aluminium spacer frame, for an overall \gls{sMDT} module height of \SI{240}{\mm} (see  Figure~\ref{fig:BIS78Xsec}).
The radially outer multilayer, ML2, covers the full area formerly covered by the BIS7 and BIS8 chambers, while the inner multilayer, ML1, covers only the BIS7 area, and is thus always shorter than ML2 by \num{30} tubes per layer, to make space for the BIS8 \gls{sRPC} as shown in the upper part of Figure~\ref{fig:BIS78Xsec}.
The \gls{BIS78} outer multilayers, ML2, mounted directly against the barrel toroid coils, carry the rail supports,
the in-plane alignment monitoring system, the external 
optical alignment sensors connecting chambers to each other (which are the same as for the \RunOneTwo alignment configuration) and the magnetic field sensors.
The new light sources of optical monitoring systems connecting the barrel to the \glspl{NSW} are mounted on the inner multilayer, ML1, in the gap between the \gls{sMDT} and the \gls{sRPC} chambers.
The new chambers have twice as many electronics channels as the previous \glspl{MDT}, in total almost \num{11000}.
 
Most tubes in the BIS78 \gls{sMDT} chambers are \SI{1660}{\mm} long, the full azimuthal width of the chambers. The final \num{12} tubes per layer at the BIS8 end are in all cases shortened to \SI{1000}{\mm}, giving the chambers the "T"-shape seen in the lower right inset of Figure~\ref{fig:BIS78Xsec} and in Figure~\ref{fig:BIS783D} that allows them to interleave with the \gls{NSW}.
In order to accommodate \gls{NSW} support structures, however, chambers in Sectors~\numlist{2;12;14;16} have cutouts where \num{30} tubes per layer in the BIS7 region have a reduced length of \SI{1530}{\mm}.
Because the \gls{NSW} support structures are not radially symmetric, the lengths of the BIS78 chambers are also different from one sector to another, with \num{108} tubes per ML1 layer and \num{78} per ML2 layer for Sectors~\numlist{2;4;6} and, for all the other sectors, \num{96} per ML1 layer and \num{66} per ML2 layer.
The chambers in the uppermost Sectors, 4 and 6 are identical, as are those in Sectors~8 and 10; each of the four bottom chambers is unique.
There are thus six different \gls{sMDT} chamber types in the eight Small azimuthal sectors.
The chambers that will eventually be installed on Side~C of the detector (after \gls{LS3}) are mirror images of the corresponding Side~A chambers with respect to the mid-plane at $z=0$.
 
The BIS7 and BIS8 \gls{sRPC} each consist of three \gls{sRPC} detector singlets, as described in Section~\ref{muonSS:srpc}.
With the small gas gap and electrode thickness, these \gls{sRPC} singlets are just \SI{11.8}{\mm} thick.
Their readout panels consist of two \SI{0.3}{\mm}-thick \glspl{PCB} carrying the ground plane and the strip pattern, respectively, which are
glued to either side of a \SI{3}{\mm}-thick FOREX\textsuperscript{\textregistered} \gls{PVC} foam spacer.
The ground planes of the two panels constitute the singlet Faraday cage into which the frontend electronics are integrated, in order to fully exploit its high sensitivity.
The \gls{RPC} singlets are electrically completely decoupled from each other, and each \gls{sRPC} singlet is individually read out on each side of the detector plane with $\eta$ and $\phi$ strip panels, respectively.
Each of the eight \gls{BIS78} \gls{sRPC} stations has \num{544} $\eta$ and \num{544} $\phi$ strips,
so there are, in total, \num{8704} \gls{sRPC} readout channels.

The \gls{sRPC} in the \gls{BIS78} chambers are equipped with new highly sensitive readout electronics using silicon \gls{BJT} charge preamplifiers with highly increased sensitivity of better than \SI{4}{\milli\volt/\femto\coulomb} at \num{1700} electrons RMS noise amplitude,
less than \SI{600}{\ps} rise time and up to \SI{100}{\MHz} bandwidth~\cite{RPC_FE}. The frontend boards also contain new fast 4-channel discriminator \glspl{ASIC} in SiGe \gls{BiCMOS} technology with \SI{3}{\milli\volt} minimum threshold and \SI{500}{\MHz} bandwidth.
The discriminated \gls{LVDS} pulses from each \gls{BIS78} module are transmitted through a \SI{5}{\m}-long cable to a set of 18 external \gls{TDC} boards employing \gls{HPTDC} \glspl{ASIC}~\cite{HPTDC}.
The design of the front-end electronics has been made particularly robust, with low power consumption of less than \SI{20}{\milli\watt}/channel, radiation hardness, and \gls{ESD} protection, as they cannot be replaced after the chamber has been constructed.
 
The \gls{sRPC} triplets fit into aluminium support frames \SI{55}{\mm} high, (see bottom left image of Figure~\ref{fig:BIS78position})
and are pressed together by \SI{2}{\mm}-thick pre-bent aluminium plates at the top and the bottom, which distribute the force.
\gls{sMDT} chambers and \gls{sRPC} modules are mounted independently, and electrically insulated from each other on the
BIS \gls{MDT} rail system using support brackets passing between the chambers in the $z$ direction. They fit within the allowed height envelope of \SI{290}{\mm}.
The plastic gas inlets of the \gls{sRPC} are designed as part of the gas-gap frame, increasing their mechanical stability and reducing external stress in order to prevent cracking.
The BIS7 \glspl{sRPC} all have an azimuthal dimension of \SI{1840}{\mm}. Those in Sectors \numlist{2;4;6} have a $z$ dimension of \SI{1180}{\mm}, while for Sectors \numlist{8;10;12;14;16} the $z$ length is reduced to \SI{990}{\mm}.
The BIS8 \glspl{sRPC} all measure \SI{440}{\mm} in $z$. Those in Sectors \numlist{4;6;8;10} occupy the full \SI{1840}{\mm} azimuthally, while those in Sectors \numlist{2;12;14;16} are reduced in azimuth to \SI{1720}{\mm}.
 
\subsubsection{BIS78 Data Acquisition and Trigger System}
The \gls{BIS78} \gls{sMDT} are read out in a very similar way to the legacy \gls{MDT}, except that new \gls{HPTDC} \glspl{ASIC}~\cite{HPTDC} replaced the original \gls{TDC} (as for the BMG chambers described in Section~\ref{muonSS:BMEBOEBMG}).
The front-end boards 
of the \gls{sMDT} chambers each contain three eight-channel \gls{ASD} chips~\cite{ASD} to read out \num{24} tubes (as for the legacy \glspl{MDT}) and one \gls{HPTDC} \gls{ASIC}. A radiation-tolerant Actel ProASIC3 \gls{FPGA} is used on each board to configure the \gls{ASD} and \gls{HPTDC} chips.
The \glspl{HPTDC} send the hit information via twisted pair cables to an \gls{MDT} \gls{CSM}, of which there are two per chamber, mounted in accessible places on the barrel toroid magnet coils. The \glspl{CSM} transmit the serialised
data via optical fibres to the legacy \glspl{MROD}
and also send the ATLAS clock and trigger signals to the \glspl{HPTDC}.
 
The readout for the \glspl{RPC} is based on the new \gls{FELIX}~\cite{FELIX} system described in Section~\ref{subsubsec:tdaq_daqhlt_felixswrod}, with new electronics of the type that will be used in the Phase~II upgrades of the \gls{RPC} system.
The digitised hit times from the strips of a \gls{BIS78} \gls{sRPC} chamber are serialised and transmitted via a \gls{GOL}~\cite{GOL} to the trigger Pad board, which is mounted on the chamber.
The Pad board uses a Xilinx Kintex-7 \gls{FPGA}~\cite{RPC_FPGA} to determine the local trigger coincidences from the \gls{RPC} triplet hits.
Radiation-tolerant firmware design employing \gls{SEM} core and triple redundancy logic is used to ensure robustness against \glspl{SEU} and \gls{TID} radiation effects.
The \gls{RPC} data transmission to the \gls{FELIX} data acquisition system is handled by a \gls{GBTx} chip on the Pad board, with a \gls{GBT-SCA}
for monitoring and configuration, while the trigger candidate information is transmitted by the \gls{FPGA} via an optical fibre to the off-detector endcap Sector Logic board.
The endcap Sector Logic boards combine the trigger information of the \gls{BIS78} \glspl{RPC}, the \glspl{TGC} and the \glspl{NSW} to generate the \gls{L1} endcap muon trigger, as described in Chapter~\ref{sec:TDAQ}.



\clearpage
\newpage
 
\section{Forward Detectors} 
\label{sec:Forward}

\begin{figure}[hbtp]
\includegraphics[width=1.0\textwidth]{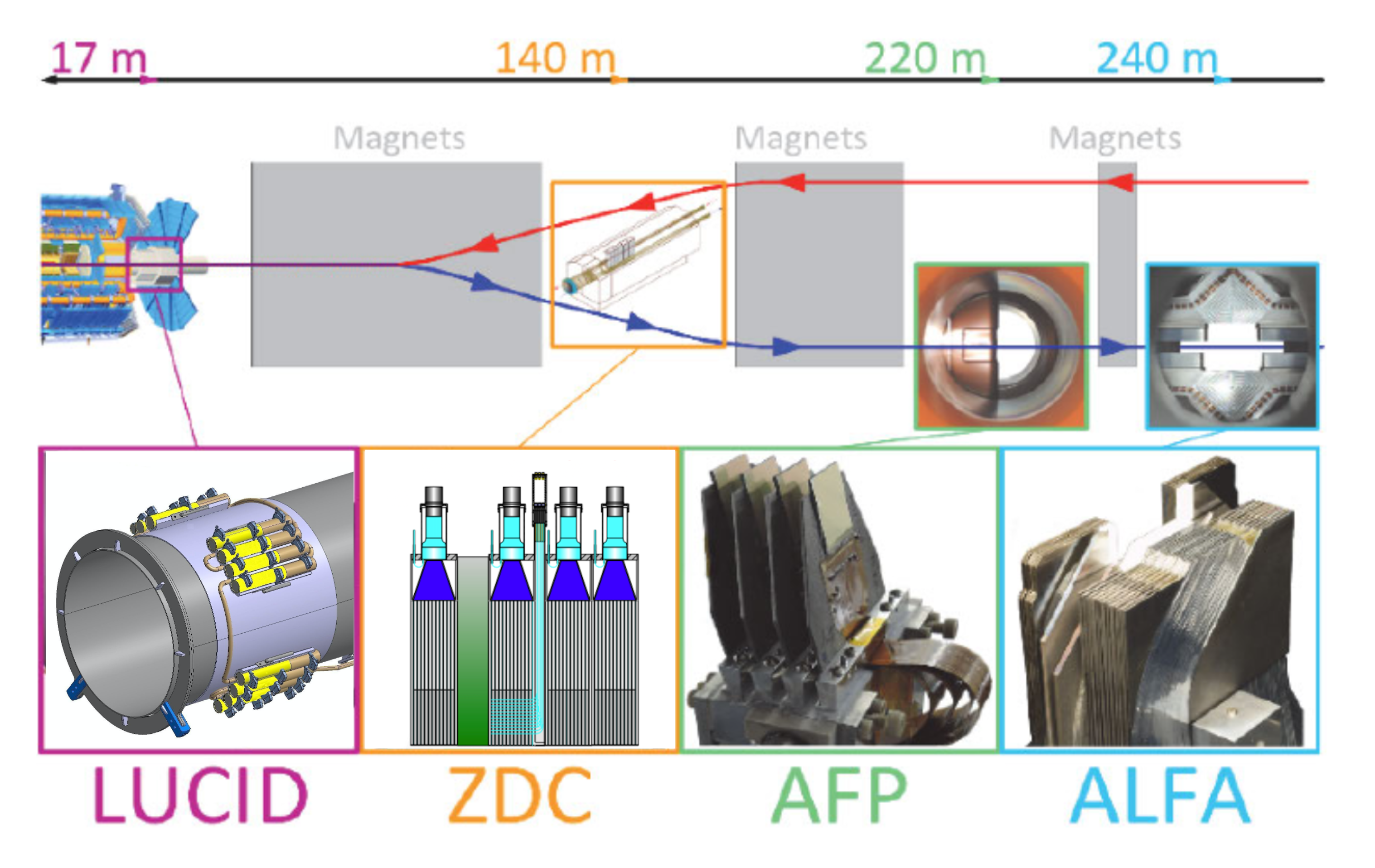}
\caption{Layout of the forward detectors.
The \glstext{LUCID} detector inside the main ATLAS volume provides a high-precision luminosity measurement; the \glstext{ZDC}, located just beyond the point where the two beams diverge, measure the energy of ``spectator'' neutrons in heavy-ion collisions; the \glstext{AFP} detectors measure time-of-flight and can trigger on bunch crossings producing forward protons in special low-pileup runs; and the \glstext{ALFA} Roman Pot detectors determine the total proton-proton cross-section by measuring elastic proton scattering at very small angles.
\label{fig:fwd}}
\end{figure}

The ATLAS Forward detectors are a set of four detectors installed along the \gls{LHC} \beampipe at different distances
from the \gls{IP}. These four detectors are used for different purposes, all of them using the information carried by particles whose detection is missed by the rapidity coverage of the ATLAS central detector.
The physics topics that can be studied with these detectors range from total cross section measurement (\gls{ALFA}, described in Section~\ref{alfa}),
to diffractive physics (\gls{AFP} --- Section~\ref{afp} --- and \gls{ALFA}), to heavy ion physics (\gls{ZDC}, Section~\ref{zdc}), to absolute luminosity determination (\gls{LUCID}, Section~\ref{lucid}).
The layout of the forward detectors is shown in Figure~\ref{fig:fwd}.

 
\subsection{\glstext*{LUCID} Detector}\label{lucid}
\glsfirst{LUCID} is designed for high-precision measurement of the luminosity delivered to the ATLAS experiment by the \gls{LHC}, both in real time and, with final calibrations and a more refined analysis, for offline reconstruction and physics analyses.
 
\gls{LUCID} is the only ATLAS luminometer capable of delivering per-bunch absolute-luminosity measurements under all the various \gls{LHC} beam conditions: high-luminosity \pp collisions at top energy, heavy-ion collisions, special-purpose \pp collisions at lower energies or very low luminosity conditions, and is therefore the reference luminosity detector for ATLAS. \gls{LUCID} is complemented by several luminometers with different capabilities and covering various luminosity regimes: the \gls{BCM}~\cite{BCM}, and the \gls{MBTS} detector with per-bunch absolute-luminosity measurement under specific low-luminosity conditions; the per-bunch relative-luminosity measurements based on counting charged-particle tracks reconstructed in the \gls{ID} at low and high luminosities, or based on counting pixel-clusters in the Pixel detector (under development); the bunch-integrated luminometers based on currents in the \gls{LAr} gaps of a subset of \gls{EMEC} and \gls{FCal} cells, and on integrated currents of the scintillator-\glspl{PMT} in a selection of \gls{Tile} extended-barrel modules, mostly for high luminosity conditions. For more details, see \Sect{\ref{lucid_control}}.
The luminosity reported by these luminometers is assumed to be proportional to the flux of charged particles striking the corresponding detector, integrated over a time interval ranging from one second to around one minute.
 
\emph{Online} \gls{LUCID} instantaneous luminosity measurements (bunch-integrated) are provided as fast feedback on a one- to two-second time scale to the ATLAS online-monitoring software and to the \gls{LHC} control system, as are luminosity measurements from the other  experiments (\acrshort{ALICE}, \acrshort{CMS} and \acrshort{LHCb}). This short latency and high sampling rate are of primary importance for the \gls{LHC} operators to monitor beam conditions and optimise the accelerator performance. The absolute accuracy of the reported luminosity is limited to about \SI{5}{\percent} (typically with larger uncertainties early in the running period, and improvements over the course of the data-taking year).
 
The \emph{offline} determination of the absolute luminosity scale is based on the \glsfirst{vdM} method~\cite{vdm}, and carried out in dedicated runs at low luminosity ($\mathcal{L} \sim 10^{31}\mathrm{cm^{-2}\,s^{-1}}$), and under accelerator conditions optimised to reduce the systematic uncertainties in the luminosity calibration~\cite{DAPR-2013-01, ATLAS-CONF-2019-021}. This calibration is extrapolated to high-luminosity physics conditions and monitored throughout the data-taking year using a methodology that involves most of the ATLAS luminometers, and that is outlined in Section~\ref{lucid_control}.
The overall uncertainty in the luminosity for proton collisions at $\rts=\SI{13}{\TeV}$  was determined to be \SI{0.83}{\percent}~\cite{DAPR-2021-01}.
Specialised \pp datasets and heavy-ion periods require their own calibrations, with uncertainties as low as \SI{1.5}{\percent} in \RunTwo.
 
The design of the \gls{LUCID} detector evolved over the course of \gls{LHC} \RunOneTwo. A first version of \gls{LUCID}, described in Ref.~\cite{PERF-2007-01}, was used during \RunOne; the \RunTwo version~\cite{lucid2-pmt,lucid2}, called \gls{LUCID}~2, will also be used for \RunThr. The changes in the detector design were dictated by the rapidly changing \gls{LHC} beam conditions, and by the increase in the annual radiation dose impinging on the detector. \gls{LUCID}~2 has been shown to perform very well at the highest instantaneous luminosity delivered by the \gls{LHC}. It was able to provide a highly accurate luminosity determination over the entire \RunTwo period, thanks in part to its built-in redundancy -- each \gls{LUCID} \gls{PMT} can provide a luminosity measurement either alone or in combination with other \glspl{PMT} -- and also to the extensive use of the numerous independent, complementary and well-understood luminosity measurements provided by \gls{ID}-, \gls{LAr}- and \gls{Tile}-based systems, that were used to cross-check or correct the \gls{LUCID} response.

\begin{figure}[hbtp]
\includegraphics[width=0.8\textwidth]{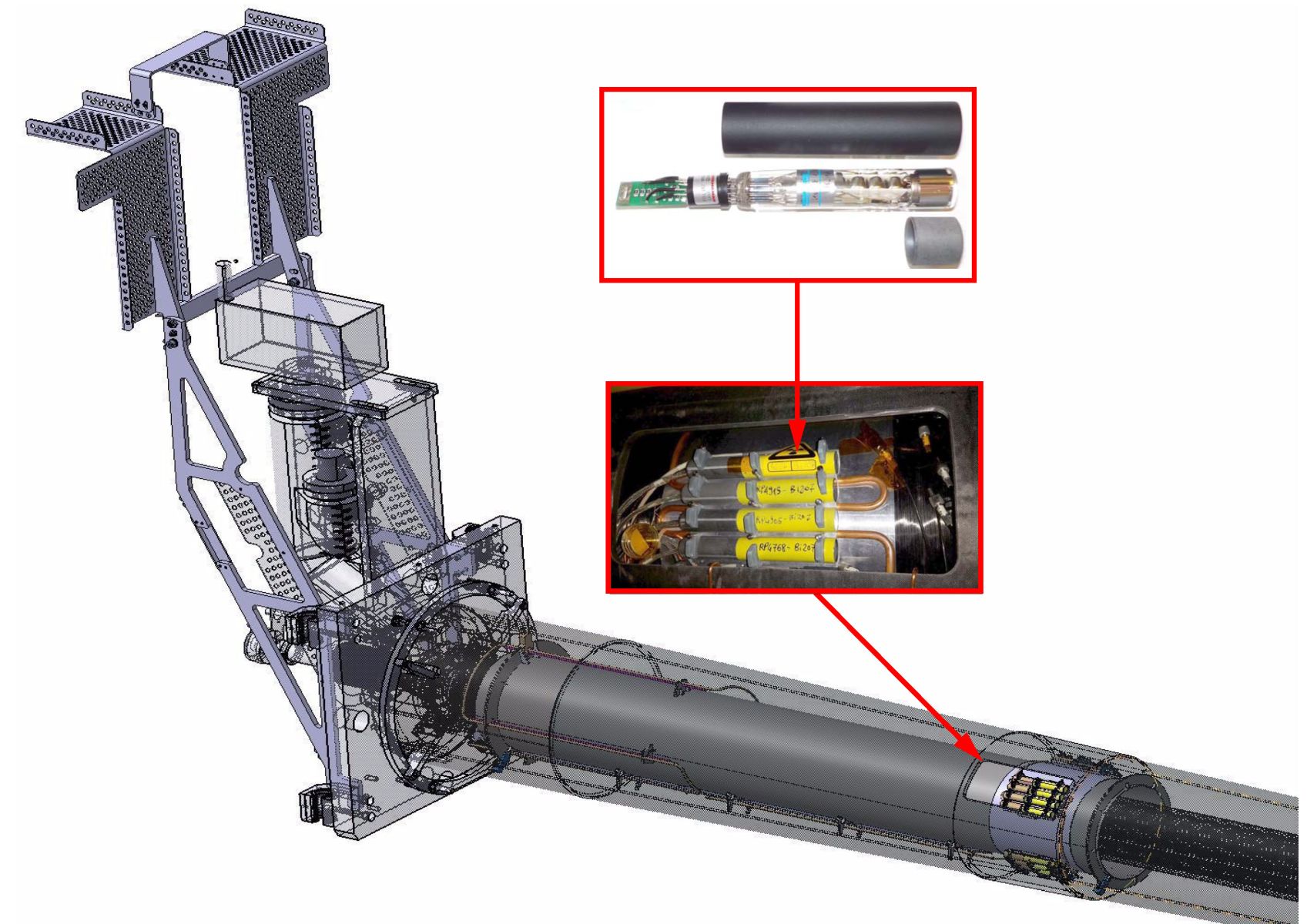}
\caption{Mechanical design of the \gls{LUCID}~2 detector. The left of the diagram is toward the ATLAS cavern wall, and the cone (referred to in the text as the ``VJ cone'') points toward the \gls{IP}, holding \gls{LUCID} inside the JF shielding, approximately midway between the EM and EO wheels of the \gls{MS}.  The insets show pictures of a group of \glspl{PMT} and details of one \gls{PMT} and its magnetic shield case. }
\label{fig:lucid_detector}
\end{figure}
 
\subsubsection{\glstext*{LUCID}~2 Detector Description}
The \gls{LUCID}~2 detector is composed of two arms placed symmetrically
at about \SI{\pm 17}{\m} on either side of the ATLAS \gls{IP}. The mechanical design of one arm is shown in Figure~\ref{fig:lucid_detector}.
 
The \gls{PMT} detector is formed by four groups of four \glspl{PMT} each, arranged symmetrically around the \beampipe. The \glspl{PMT} are of the R760 model manufactured by Hamamatsu. The quartz entrance window of each \gls{PMT} is used as a Cherenkov radiator. Charged particles with momentum above the Cherenkov threshold in the quartz window produce a number of photons which, on average, correspond to about 35 photoelectrons (p.e.). This signal is well above the noise threshold (1~p.e.) of the readout chain.
 
The \glspl{PMT} are mounted around an aluminium cooling cylinder, which is in thermal contact with a water cooling pipe, keeping the environmental temperature at an acceptable level for the \glspl{PMT} during the bake-out of the \gls{LHC} \beampipe. The \gls{PMT} detectors are surrounded by a carbon fibre support cylinder that can be opened at four access points to allow easy replacement of the \glspl{PMT} and their high voltage divider bases for each of the four \gls{PMT} groups.
The main conceptual difference in the \gls{LUCID}~2 detector for \RunThr with respect to the \RunTwo configuration is that the bases are now, like the \glspl{PMT}, completely replaceable during the winter shutdowns of the \gls{LHC} schedule.
The reason for this was that about half of the \gls{PMT} channels developed contact problems between the bases and the \gls{PMT} pins after three years of running.
It is possible that this connector problem was due to radiation, due to the location of the connectors;
however, radiation tests with a gamma source have not confirmed this hypothesis.
The possibility of completely replacing a \gls{PMT} and its base during a winter shutdown will help to maintain a high percentage of working readout channels during each one-year running period.
 
Each \gls{PMT} base is of the boosted type, with four \gls{HV} feeding points to ensure good linearity of the \gls{PMT} response up to about \SI{100}{\micro\ampere} of current. The whole \gls{LUCID} \gls{HV} system consists of 160 independent channels powered by a CAEN SY1527 module~\cite{caen_sy1527} housing
four \gls{HV} cards of type A1535N and three of type A1535SN.
The \gls{LUCID}~2 detector is fully integrated into the ATLAS \gls{DCS} (see Section~\ref{TDAQ_DCS_OPCUA}), which continuously monitors the \gls{HV} and current of every channel as
well as the temperature in various locations of the detector.
\subsubsection{Prototypes for HL-LHC}
\begin{figure}[hbtp]
\centerline{\includegraphics[width=0.8\textwidth]{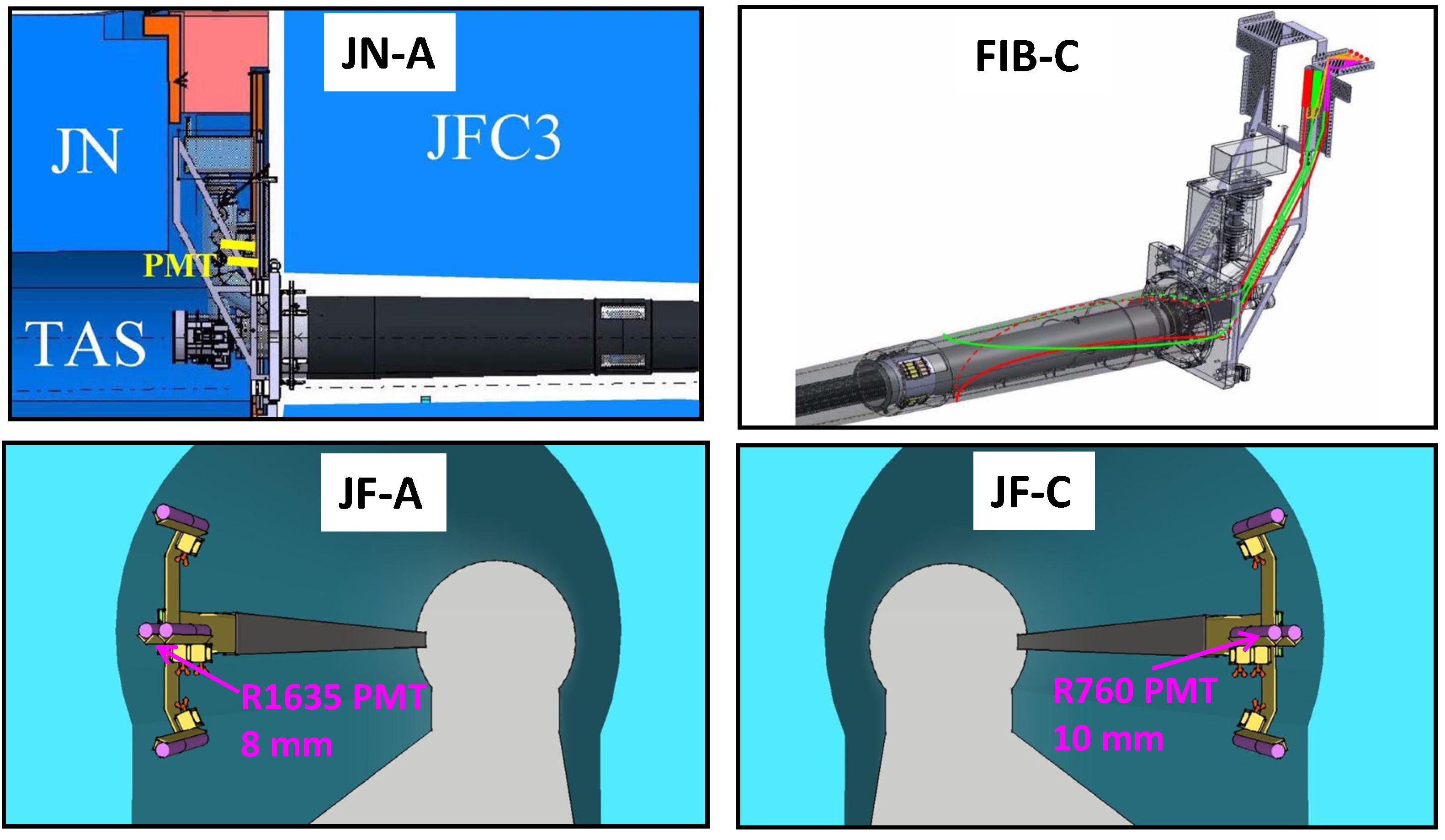}}
\caption{The three \gls{LUCID}~3 prototypes for \RunFour installed. From top left, clockwise: the JN \gls{PMT}, the side~C fibre detector, the side A/C JF \glspl{PMT}. JN in this chapter is the same as TX1S in Figure~\ref{fig:fw_shielding}.}
\label{fig:lucid_prototypes}
\end{figure}
In addition to the \gls{LUCID}~2 \gls{PMT} detector, three types of prototypes are present in the \gls{LUCID} \RunThr configuration, aiming to both measure luminosity in \RunThr
and to validate the \gls{LUCID} upgrade project for
\gls{HL-LHC} (called \gls{LUCID}~3).
Refer to Figure~\ref{fig:fw_shielding} (where JN is labelled as ``TX1S'') for the locations of the shielding components.
The three prototypes are:
\begin{itemize}
\item the JF prototype is a PMT-based detector located in the inner channel of the JF shielding at a larger distance from the \beampipe than \gls{LUCID}~2 .
\item the JN prototype is a low-rate PMT-based detector located in a shielded region behind the JN shielding surrounding the \gls{TAS};
\item the Fibre prototype is a fibre-based detector located around the VJ-cone
surrounding \gls{LUCID}~2
\end{itemize}
Figure~\ref{fig:lucid_prototypes} contains sketch of the three prototypes.
A short description of the prototypes follows, while an extensive description of
the \gls{LUCID}~3 upgrade project can be found in Ref.~\cite{LUCID-3_IDR}.
\subsubsection{JF detector}
Two main limitations will prevent \gls{LUCID}~2 from measuring the luminosity at the \gls{HL-LHC}: the increased pileup, which will lead to the saturation of the event/hit-counting algorithms, and the total radiation to which the detector will be exposed, and the resulting activation of material, which will make it unsafe to carry out the necessary annual maintenance of the \glspl{PMT}. In addition, the \gls{VAX} would strongly interfere with \gls{LUCID} if it remained in its present location. Both aspects can be mitigated by moving the detector further from the \beampipe. A possible location was identified in the inner channel of the JF shielding
hosting the \beampipe, as shown in Figure~\ref{fig:lucid_prototypes} (bottom images), where the \glspl{PMT} will be at a radius of about \SI{30}{\cm}, compared to about \SI{12}{\cm} for \gls{LUCID}~2. In the new location, both the radiation dose and the charged particle fluxes will be reduced by about \SI{30}{\percent} with respect to the \gls{LUCID}~2 location. Moreover, no interference with the \gls{VAX} will be present. Finally,
the JF shielding is removed and brought to the surface at each end-of-year shutdown, allowing for maintenance in a safer location than that of
\gls{LUCID}~2 (which cannot be removed from the \beampipe).
The prototype detectors consist of four \glspl{PMT} per side. Two \glspl{PMT} of each group will be read out by a \gls{LUCROD} card while the other two can be considered as spares, and read out in case faults develop in the initial two.
On Side~A, four standard R760 \glspl{PMT}
are installed, while on Side~C three R760 and one R1635 \gls{PMT} are installed.
The R1635 is a smaller acceptance \gls{PMT} (\SI{8}{\mm} diameter instead of \SI{10}{\mm} for
the R760) which has been custom-modified by Hamamatsu for \gls{LUCID}
to be equipped with a quartz window, needed both for the Cherenkov light production and for its radiation hardness.
The smaller acceptance will allow a further limitation of the hit-counting algorithm saturation, assuming the new \gls{PMT} proves suitable for \gls{LUCID} use.
\subsubsection{JN prototypes}
A further reduction of the particle flux and radiation exposure can be obtained
by placing \glspl{PMT} in the shadow of the forward shielding: here the levels are
reduced to about \SI{10}{\percent} with respect to the \gls{LUCID} 2 location, potentially
avoiding any hit-counting saturation and possibly
reducing or eliminating the non-linearity with the luminosity which has been
observed in \gls{LUCID}~2. The proposed location is shown in Figure~\ref{fig:lucid_prototypes} (top left).
The two R760 \glspl{PMT} are at different radii, therefore with different levels of
shadowing from the JN, and are expected to have quite different acceptances
despite the small distance between them, due to the predicted steep radial dependence of the
particle flux with the position behind the shielding. In this way it will be
possible to have a direct measurement of the relative particle fluxes and
the final \gls{PMT} position in \gls{LUCID}~3 can be optimised.
\subsubsection{Fibre detector}
During \RunTwo, four fibre bundles per side constituted the \gls{LUCID} Fibre detector. The Cherenkov light produced by charged particles along the fibres
is routed to \glspl{PMT} located in a low-radiation area in the
vertical channel behind the JN shielding. The fibre detector did not behave
as expected, in particular due to poor long term stability. The reason
for this poor behaviour is not entirely clear, but may be related both to
the monitoring system of the \gls{PMT} gain (based not on the Bismuth source, but
on \glspl{LED}) and to the opacification of the fibres, which was not
monitored.
On the other hand, the charge algorithms in general, and in particular
that of the fibre detector, showed remarkable linearity with the
luminosity, which was not the case for the hit-counting algorithms. For this
reason, a new attempt to exploit this technology is being made in \RunThr but
with major modifications aimed at solving the main problems of the former
fibre detector:
\begin{itemize}
\item{Quartz fibres of type UVNSS 600/624/660 are now used, motivated by the improved radiation hardness of their Fluorine-doped silica cladding compared to that of the silicon cladding of the original fibres. There are two bundles of fibres and each bundle is read out
at the end by a \gls{PMT};}
\item{A more effective calibration and monitoring system obtained by the combination of a Bismuth source deposited onto the window of the readout \gls{PMT} and an \gls{LED} system to monitor ageing of the fibres.}
\end{itemize}
The fibre detector is shown in Figure~\ref{fig:lucid_prototypes} (top right) in which one can see the routing of the two fibre bundles and the placement of the two readout \glspl{PMT}, on top of the support structure for the VJ cone (in black in the figure) in a region of strongly reduced radiation dose. Each fibre bundle starts from the \gls{PMT} window, reaches the \gls{LUCID}~2 detector area and then returns back, with an open end used to inject \gls{LED} light to monitor fibre ageing.
A set of six \glspl{LED} with wavelength variable from the green to the UV injects light both directly to the \gls{PMT} front window
(prompt signal) and to the opposite end of the fibre bundle (delayed signal).
The prompt and delayed signal have enough time separation to be clearly distinguished, allowing their relative amplitudes to be monitored. Changes in the relative amplitude of the two signals
are expected to provide an estimate of the decrease in the light
transmission of the fibre bundles due to radiation damage: this
information will
be used to make offline corrections to the luminosity measured during the run
to compensate for this effect.
The readout \gls{PMT} is a Hamamatsu R7459, which has a large area window, allowing it to accommodate both the fibre bundle end and the Bismuth radioactive source used for calibration purposes.

\subsubsection{\glstext*{LUCID}~2 Readout Electronics}
\begin{figure}[htbp]
\subfloat[]{\label{fig:lucroda}\includegraphics[width=0.45\textwidth,trim=0 0 350 0,clip]{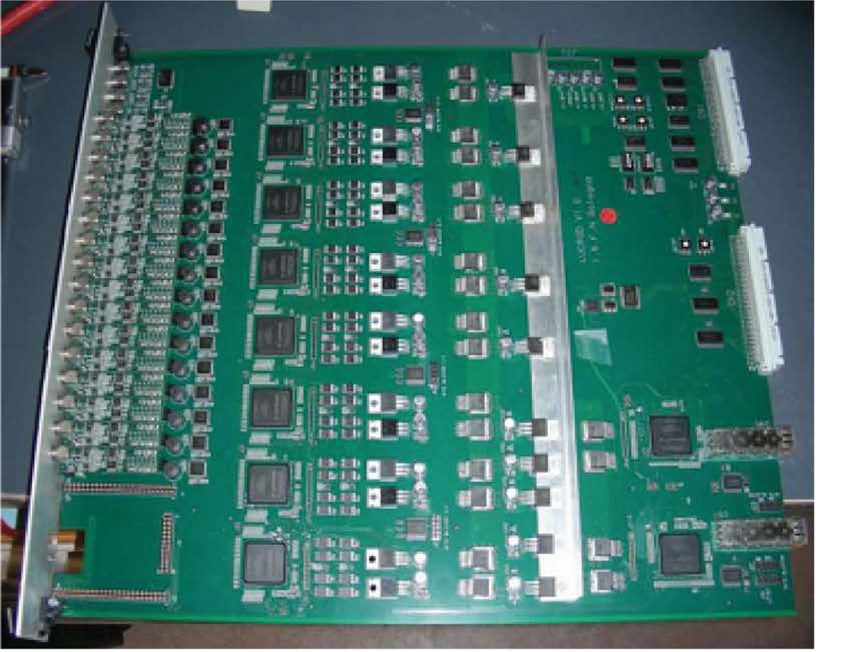}}
\subfloat[]{\label{fig:lucrodb}\includegraphics[width=0.45\textwidth,trim=350 0 0 0,clip]{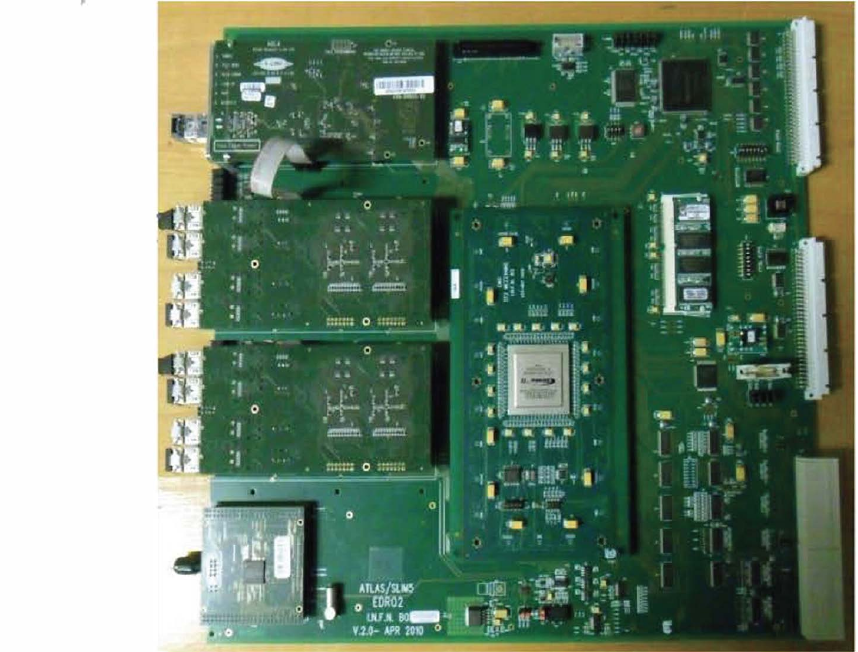}}
\caption{Photographs of the  \gls{LUCID}~2 readout cards: \protect\subref{fig:lucroda} \gls{LUCROD} and \protect\subref{fig:lucrodb} \gls{LUMAT}.
}
\end{figure}

\gls{LUCID}~2 is designed to measure the luminosity delivered in each of the \num{3564} bunch crossings during one \gls{LHC} orbit.
In order to achieve this, a fast \analog readout and signal processing chain has been designed.
The signals delivered by each \gls{PMT} are fed into a \SI{15}{\m} low-loss coaxial cable and then received by the powerful \gls{LUCROD} processing card
pictured in Figure~\ref{fig:lucroda}. The relatively short cable length does not significantly degrade the fast \gls{PMT} signal,
which can then be amplified by a \SI{350}{\MHz} bandwidth amplifier and fed to a fast 12-bit ADC, sampling at a rate of \SI{320}{\MHz} (eight samples
per \gls{LHC} machine \gls{BC}). Each digital waveform signal is processed in one FPGA providing two types of information:
\begin{itemize}
\item Signal amplitudes above a preset threshold, called {\textit \HITs} in the following;
\item Signal amplitudes integrated over the eight waveform samples, called {\textit \CHARGE} in the following.
\end{itemize}
The \HIT signals of each readout channel are then stored in the other \gls{FPGA} where they are delivered to a second processing card, the \gls{LUMAT}
, pictured in Figure~\ref{fig:lucrodb}, which counts the total number of \HITs per detector arm, runs several \HIT-based algorithms, and provides two \gls{L1} Trigger signals (the AND and OR of the two detector arms) to the ATLAS \gls{CTP}.

The \gls{LUCROD} and \gls{LUMAT} boards are both implemented in the \gls{VME} 9U standard. The \gls{LUCROD} card accepts \num{16} \analog input channels and the connection to the \gls{LUMAT} card is over a fast optical fibre (\gls{s-link}~\cite{slink}). Each detector arm is served by two \gls{LUCROD} cards. All \gls{LUCROD} cards continuously send \HITs to two \gls{LUMAT} cards which combine the information from the two detector arms. The \gls{LUCROD} cards sit in one \gls{VME} crate on each detector side, placed on a service platform close to the detector, in order to limit the length of the signal cables. The two \gls{LUMAT} cards sit in a \gls{VME} crate placed in the ATLAS \gls{USA15} service area.
 
Since \gls{LUCID} has to provide online luminosity measurements not only during ATLAS data-taking but also during \gls{LHC} machine tuning, it can run continuously in standalone mode, independent of the ATLAS \gls{TDAQ} system.
A fraction of the fully reconstructed waveforms is stored locally for possible cross-checks or to study detector performance.
 
\subsubsection{\glstext*{LUCID} absolute-luminosity calibration and luminosity-measurement methodology}\label{lucid_control}
 
A fundamental ingredient for a stable luminosity estimate based on \gls{LUCID} is the control, at the \SI{1}{\percent} level, of the stability of the \gls{LUCID} \gls{PMT} gains. This is accomplished by a very effective calibration system based on the radiation emitted by the (approximately)  \SI{40}{\kilo\becquerel} \Bisource source deposited on each \gls{PMT} window. This source emits, among other decay products, monochromatic electrons of about \SI{1}{\MeV} kinetic energy, which is enough for them to cross the \SI{1.2}{\mm} quartz window, and to produce roughly the same amount of light as a high-energy charged particle traversing that same window. At low luminosity, for example, during heavy-ion running or in the tails of \gls{vdM}-calibration scans, the $\mathcal{O}$\mbox{(\SI{10}{\kHz})} rate from this monochromatic electron source represents a non-negligible background that must be subtracted from the raw LUCID signal.
 
\begin{figure}[htbp]
\includegraphics[width=0.9\textwidth]{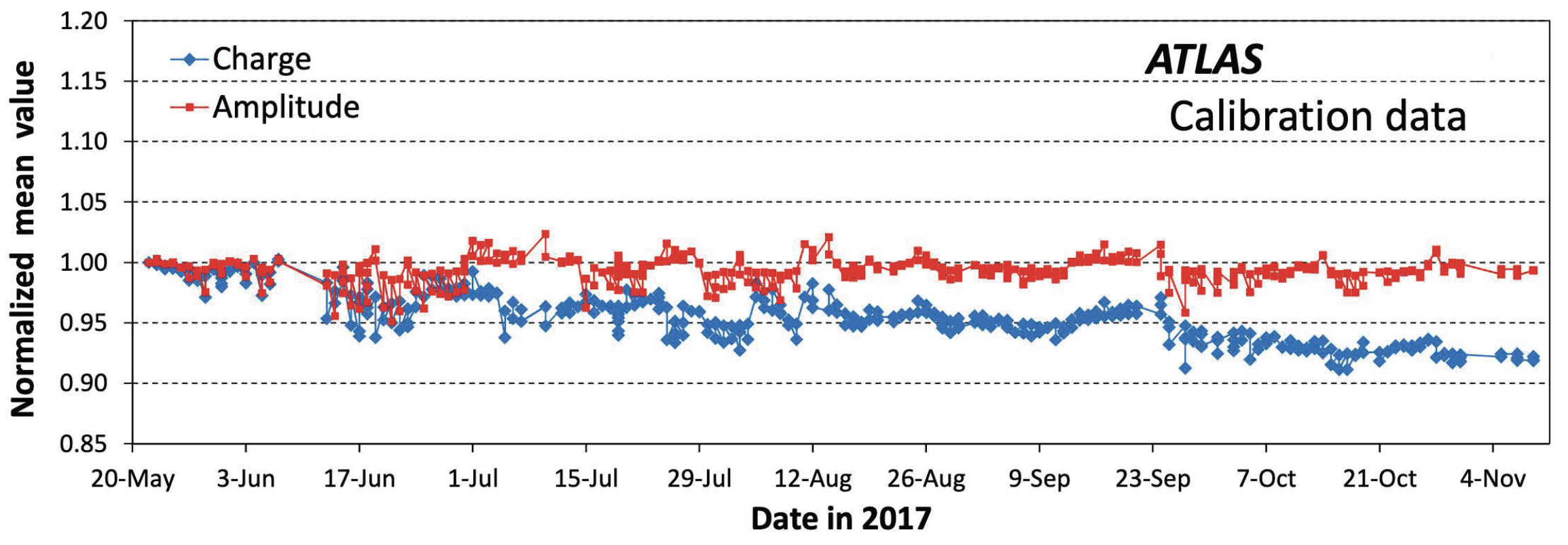}
\caption{Gain variation with respect to the first day of calibration in 2017. The \gls{PMT} \gls{HV} was corrected between collision runs aiming to keep \Bisource signals average amplitude constant (red line).}
\label{fig:lucid_calibration}
\end{figure}
 
The \gls{PMT} gains are monitored on an approximately daily basis by recording the amplitude spectrum of the \Bisource signal from each \gls{PMT} during periods without beam. The typical amplitude and charge inferred from these spectra, normalised to unity at the beginning of the running period, are shown in Figure~\ref{fig:lucid_calibration} as functions of time over the year 2017. The red points refer to the average amplitude, and the blue ones to the average charge. The average amplitudes of the calibration spectra are kept constant within a few percent throughout the entire period, while the average charge diverges systematically away from unity, decreasing by \SI{8}{\percent} over the six-month time span shown in the figure. Since the relationship between signal amplitude and charge changes when the \gls{HV} setting is increased to keep the gain constant~\cite{lucid2}, only one of the two can be stabilised throughout the data-taking year.
 
Keeping the amplitude constant is essential for the \HIT algorithms, since their luminosity measurement is based on the fraction of beam crossings in which this amplitude exceeds a given threshold (defining a ``\HIT''); charge algorithms, in contrast, rely on the stability of the charge measurement. In early \RunTwo, the charge algorithms were originally perceived as the most promising. In 2015 and 2016, therefore, the choice was made to keep the average charge constant. Experience, however, demonstrated the superior performance of the \HIT algorithms, so that in 2017 and 2018 the average amplitude was stabilised instead. Offline corrections to the hit-counting efficiency that had been introduced in 2015 and 2016 became obsolete.
 
In  \HIT-counting algorithms, the average fraction $f$ of \glspl{BC} containing a \HIT (typically a signal above a given threshold) is assumed to obey binomial statistics (while the pileup parameter $\mu$ is assumed to be Poisson-distributed). The luminosity per bunch crossing, $\mathcal{L}_{BC}$, is given by
\begin{equation}
\mathcal{L}_{BC}=\frac{-\ln(1-f)}{\sigma_{vis}}\cdot f_\text{rev},
\end{equation}
where $f_\text{rev}$ is the \gls{LHC} revolution frequency, and $\sigma_{vis}$ is a calibration constant called the visible cross section that is measured by the \gls{vdM} method. A \HIT-counting algorithm based on one of the two independent sets of eight \glspl{PMT} denoted as Bi1 or Bi2, is  typically chosen as the ``preferred'' algorithm both for online luminosity monitoring and for a first rough estimate of the offline luminosity. For the final offline luminosity, the preferred algorithm is chosen in each data-taking period (typically one year) based on a detailed analysis of its linearity and relative stability across that period. In \RunTwo, the preferred offline algorithm was based either on a subset of eight \glspl{PMT} (in 2016 and 2017), or on a single \gls{PMT} (in 2015 and 2018).
 
For \CHARGE algorithms, the average charge collected in a given \gls{BCID}, is intrinsically proportional to the bunch luminosity $\mathcal{L}_{BC}$ associated with that \gls{BCID}:
\begin{equation}
\mathcal{L}_{BC}=\frac{C}{K_{cal}}\, ,
\end{equation}
where $C$ is the charge averaged over a time interval, denoted as a \gls{LB}\footnote{\glspl{LB} have a typical duration of about one minute, within which the instantaneous luminosity and data-taking configuration are considered to be stable, and serve as an approximate ``time-unit'' in ATLAS luminosity measurement.}, collected in the \gls{BCID} considered\footnote{Care must be taken to exclude those bunch crossings in which the \gls{LUCID} \analog readout chain may be saturating.}, and $K_{cal}$ is a calibration constant to be determined by the  \gls{vdM} method. Because of their intrinsic linearity, \CHARGE algorithms are in principle better behaved than \HIT counting algorithms in high pileup conditions. The downside is that the instantaneous luminosity inferred from charge measurements is directly proportional to the \gls{PMT} gains, which have been observed to vary by up to a few percent over the course of a single \gls{LHC} fill. Solutions to compensate for this disadvantage are still under study.
 
The absolute luminosity scale of each \gls{LUCID} algorithm, or equivalently the corresponding visible cross-section $\sigma_{vis}$, is measured by the \gls{vdM} method under experimental conditions optimised to minimise systematic uncertainties: low pileup ($\mu \sim 0.5$) to eliminate \gls{LUCID} non-linearities, at most 150 isolated bunch pairs colliding in the \gls{LHC} to avoid long-range beam-beam crossings or out-of-time pileup in the luminometer electronics, and therefore low instantaneous luminosity ($\mathcal{L} \sim \SI{e31}{\per\cm\squared\per\s}$). The bunch intensity is lowered to around \num{0.8e11} protons/bunch, and the injected emittance increased to  \SIrange[range-phrase = --]{2.5}{3.5}{\micron\radian}, so as to minimise longitudinal charge leakage out of the nominally filled positions along the \gls{LHC}-ring, as well as beam-beam--induced distortions of the \gls{vdM}-scan curves. In addition, the crossing angle is set to zero to minimise orbit-drift and beam-beam correction uncertainties, and the $\beta$ function\footnote{The $\beta$ function describes the single-particle betatron motion around the central orbit, and in particular the variation of the transverse beam envelope along the beam trajectory. $\betastar$ is the value of the $\beta$ function at the \gls{IP}, and indicates how squeezed the beams are at the \gls{IP}, by giving the distance along the beam direction after which the $\beta$ function (the transverse beam size) has been multiplied by a factor of 2 ($\sqrt{2}$) compared to its value at the \gls{IP}. A large $\betastar$ indicates wide and almost parallel beams, a small $\betastar$ indicates narrow, but divergent beams.} at the \gls{IP} is increased to $\betastar = \SI{19}{\m}$ to widen the luminous region in order to facilitate non-factorisation corrections~\cite{DAPR-2013-01, ATLAS-CONF-2019-021}.
 
These luminosity calibrations, that are obtained in the so-called ``\gls{vdM} regime'', cannot be directly applied to physics data-taking conditions which exhibit non-zero crossing angle, with \betastar values in the range \SIrange{0.3}{0.6}{\m}, \SI{40}{\percent} lower emittance, \SI{50}{\percent} higher bunch intensity, two orders of magnitude higher pileup, up to around \num{2500}  bunches grouped in trains, as well as three orders of magnitude larger instantaneous luminosity. This is primarily because \gls{LUCID} suffers from significant pileup-dependent non-linearities that result in an overestimate of the luminosity of up to \SI{10}{\percent} at $\mu \sim 50$. These non-linearities are corrected, separately for each algorithm, using a calibration-transfer procedure based on counting custom-reconstructed charged-particle tracks emerging from inelastic collisions in randomly selected colliding-bunch crossings -- an observable proportional to the number of \pp interactions per \gls{BC} and thus to the instantaneous luminosity.
 
The custom track reconstruction uses only hits in the \gls{IBL}, the \gls{Pixel} and the \gls{SCT} detectors; the track-selection criteria are optimised for luminosity monitoring over the full range of pileup levels encountered during \pp running ($0.01 < \mu < 60$). During both \RunOne and \RunTwo, \gls{TC} demonstrated very good linearity with respect to $\mu$, as well as excellent long-term stability under evolving \gls{ID} conditions. Because the per-bunch \gls{TC}-based luminosity is statistically limited under the low-$\mu$ conditions of \gls{vdM} scans, this algorithm is not independently calibrated by the \gls{vdM} method, but cross-calibrated to \gls{LUCID} during an extended head-on collision period in the same \gls{LHC} fill in which the \gls{vdM} scans are performed. \gls{TC} is used to transfer the absolute-luminosity calibration of \gls{LUCID} to physics data-taking conditions, in a reference, high-luminosity \gls{LHC} fill in which the measured \gls{TC}/\gls{LUCID} luminosity ratio is used to parameterise the $\mu$ dependence of the uncorrected \gls{LUCID} response.
 
The linearity of the \gls{TC} algorithm is cross-validated against relative-luminosity measurements based on the \gls{Tile} and \gls{LAr} calorimeters. In the case of \gls{Tile}, the dynamic range of the cryostat scintillators E3 and E4 (see Figure~\ref{fig:TileE1-E4Counters}) provides sufficiently sensitive bunch-integrated luminosity measurements during head-on collisions all the way from the \gls{vdM} to the physics regime. For \gls{EMEC} and \gls{FCal}, specialised fills with isolated bunches that cover the full range of $\mu$ values are used to compare the \gls{TC}- and \gls{Tile}-based measurements with those inferred from \gls{LAr} energy-flow measurements.
 
During \RunTwo, the calibration-transfer procedure was applied from one to three times per year, and proved essential to ensure the year-long stability of the \gls{LUCID} response. This stability is quantified by comparing, over the course of each data-taking year, the \gls{TC}-corrected, bunch-integrated \gls{LUCID} luminosity with that inferred from the \gls{PMT} currents in the D6 cells of the \gls{Tile}, and from the \gls{LAr}-gap currents in the \gls{EMEC} and the \gls{FCal}. Since these luminometers are not sensitive enough to be calibrated in the low-luminosity \gls{LHC} fill in which \gls{vdM} scans are recorded, they are cross-calibrated to \gls{TC} in a small subset of high-luminosity fills close in time to the \gls{vdM}-calibration session. Additional luminometers, such as TimePix sensors~\cite{TimePix}, provide further cross-checks. Finally, monitoring the rate of reconstructed  $\Zboson \rightarrow \ell\ell$ decays, known as $\Zboson$-counting, allows a fully independent check of the relative long-term stability of the ATLAS luminosity measurements, both within each running year and across multiple data-taking years~\cite{ATL-DAPR-PUB-2021-001}. This technique, which does not rely on theoretical predictions of the absolute $\Zboson$ cross-section, is also used to compare the integrated luminosity delivered to the ATLAS and \acrshort{CMS} experiments.
 
The luminosity measurement methodology outlined above provided among the most accurate absolute-luminosity measurements at any hadron collider to date~\cite{DAPR-2021-01}, and serves as a starting point for high-precision luminosity determination at the ATLAS \gls{IP} in \RunThr.


 
\subsection{\glstext*{ALFA} Roman Pot Detector}\label{alfa} 
The \glsfirst{ALFA} consists of four stations, each equipped with two scintillator detectors in an upper and lower Roman Pot. Two stations are placed on each side of the ATLAS detector, about \SI{240}{\m} from the \gls{IP}, in the long straight sections of the \gls{LHC} tunnel.
\gls{ALFA} is described in Ref.~\cite{ALFA-Detector} as it was operated during special runs  in \RunTwo\ for elastic cross-section and diffraction studies at \betastar values from \SI{11}{\m} to \SI{2.5}{\km}.
The following paragraphs describe only the changes to allow operation of \gls{ALFA} in \RunThr.
 
The readout electronics of \gls{ALFA} have aged and some readout parts were already replaced during \RunTwo\ because of a water leak in the \gls{LHC} tunnel.
The scintillating fibre detectors and readout electronics are expected to age further during \RunThr. Moreover, the \gls{LHC} TCL6 collimator settings during insertions of the \gls{AFP} Roman pots in the last years of \RunTwo increased the total radiation dose of the \gls{ALFA} detectors significantly, by a factor of \numrange{5}{10} compared to the situation with no \gls{AFP} insertion (and TCL6 closure).
 
\begin{figure}[ht!]
\centering
\includegraphics[width=0.6\textwidth]{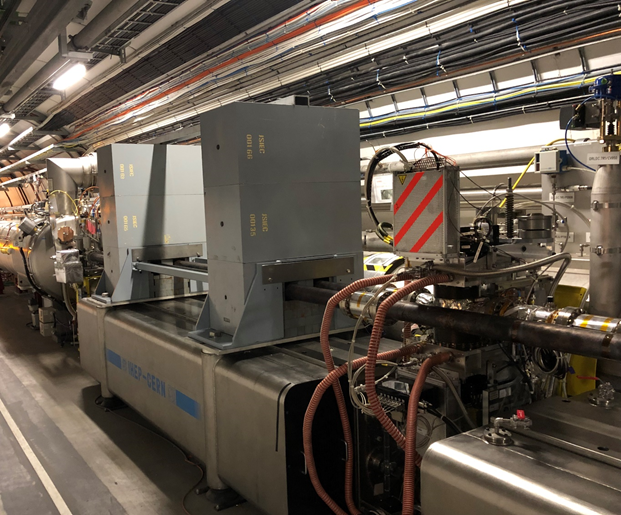}
\caption{The pair of shielding walls between the \gls{ALFA} far station (left, before Q7) and near (right) station in the \gls{LHC} tunnel in LSS1. Another pair is situated before the near station (not shown).}
\label{fig:AlfaShielding}
\end{figure}

Irradiation studies have shown that the readout motherboards and trigger boards start malfunctioning after about \SI{500}{\gray} and, in particular, that the embedded \glspl{ELMB} start failing after \SI{50}{\gray}~\cite{Almeghari_2018}. The Kuraray SCSF-78-SJ scintillating fibres only start to degrade above \SI{10}{\kilo\gray}, with an efficiency loss of \SI{20}{\percent} and higher. The dose accumulated in \RunTwo\ is about \SI{20}{\gray} for the motherboards, whereas the dose to the fibres, strongly dependent on the distance to the beam, was measured to be in the range of \SIrange{1}{3}{\kilo\gray} in the last year of \RunTwo.
To minimise radiation damage, it is therefore important that an \gls{ALFA} high-\betastar run occur early in \RunThr. To reduce the exposure of \gls{ALFA} in \RunThr, a \SI{40}{\cm} thick iron shielding wall was erected at \SI{221}{\m} after TCL6 and before Q6; and two pairs of similar walls between Q6 and each of the two \gls{ALFA} stations at \SI{235}{\m} and \SI{243}{\m}, as shown in Figure~\ref{fig:AlfaShielding}. The shielding reduces the radiation load to about half
, which should allow \gls{ALFA} to operate until the end of \RunThr.
 
Five new motherboards were produced to replace the most irradiated boards in the outer stations and to have spares in \RunThr.
 
During \gls{LS2} the Roman Pot movement system~\cite{ALFA-Detector} was refurbished and all the stopper and limit switches were readjusted. In \RunTwo, the \gls{ALFA} trigger, which operates outside the standard ATLAS latency, was connected directly to the ATLAS \gls{CTP} (via a specialised \gls{ALFA}-\gls{CTPIN} module) to minimise the trigger path length. The firmware of this \gls{CTPIN} module has been modified to provide a direct path to the \gls{CTP} for the \gls{L1Topo} trigger, and to be switched to the \gls{ALFA} trigger only for \gls{ALFA} special runs. The Readout system is unchanged.

\subsection{\glstext*{AFP} Detector}\label{afp} 
The \glsfirst{AFP} Phase I upgrade project
~\cite{ATLAS-TDR-24} 
consists of four detector stations, two on either side of the ATLAS \gls{IP} at \SI{\pm205}{\m} (near stations) and \SI{\pm217}{\m} (far stations).
Together, the near and far stations on one side constitute an Arm of the detector.
Each station contains a single horizontal ($x$) Roman pot housing the \gls{AFP} detectors. The pots present a \SI{300}{\micron}-thin window to the \gls{LHC} beam. Each near pot contains a four-plane \gls{SiT}, which is a 3D pixel detector similar to the \gls{IBL} except for a \SI{150}{\micron} thin edge parallel to the long pixel direction ($y$) on the side of the \gls{LHC} beam \cite{Lange_2015,Lange_2016}. Each far pot contains, in addition, a \gls{ToF} detector~\cite{bib:AFP-ECR2016} consisting of novel quartz Cherenkov hodoscopes with radiation-hard electronics behind the tracker to measure the vertex position of double-proton final states with \SIrange{3}{5}{\mm} precision.

During the 2015-2016 winter shutdown, the C-side arm of \gls{AFP} was installed, but without the \gls{ToF}. The \gls{AFP} detector was completed with the \gls{ToF} detectors during the 2016-2017 winter shutdown.
Over the course of 2017, ATLAS collected data where \gls{AFP} was read out corresponding to an integrated luminosity of \SI{20}{\ifb}.
From July 2017 onwards, the Roman pots were operated at a distance of $11.5\sigma_{\mathrm{beam}}+\SI{0.3}{\mm}$ from the \gls{LHC} beam, which is around \SI{1.7}{\mm} and \SI{2.8}{\mm} for the far and near stations respectively for $\betastar=\SI{40}{\cm}$.
 
In preparation for \RunThr, the \gls{AFP} detector was refurbished during \gls{LS2}: the movement mechanics were re-tuned and all switches were readjusted. The \gls{PXI} movement controller hardware and software were updated.
Because of radiation damage during \RunTwo, all silicon detectors were replaced by newly produced 3D silicon pixel tracker modules of the same type, as described in Section~\ref{ss:sit}. The \gls{ToF} detector underwent a design change to prevent corona discharge problems in vacuum, described in Section~\ref{ss:tof}.
The trigger and the readout electronics and software were updated, as discussed in Section~\ref{ss:afptrig}.
 
\subsubsection{Silicon 3D tracker \label{ss:sit}} 
The \gls{AFP} slim-edge 3D silicon detector planes are located in the near pots. During \gls{LS2}, all of them were replaced with new 3D modules of the same type.
The irradiation pattern on the \gls{SiT} planes is a characteristic narrow band that extends away from the \gls{LHC} beam and either diagonally up or down depending on the sign of the vertical crossing angle of the colliding beams. Thus, switching \gls{SiT} modules between the arms (or changing the sign of the crossing angle), exposes a fresh area of the sensor module to diffractive protons. It is expected that the new modules will operate efficiently for the first two full years of \RunThr.
After one year of running or an accumulated dose corresponding to about \SI{50}{\ifb}, a swap of detectors between the two detector arms is foreseen.
New \gls{SiT} detector modules may be required for the last years of \RunThr\ and different solutions (based on RD53~\cite{Garcia-Sciveres:2287593} or TimePix4~\cite{Llopart_2022} \glspl{ASIC}) are under consideration.
 
The cooling for the \gls{SiT} is based on a two-stage vortex tube system, using compressed air~\cite{Vacek_2013}. The 3D tracker planes are mounted in good thermal contact on top of a heat exchanger box, and cold air at \SI{-30}{\degreeCelsius} circulates through the box.
For \RunThr,
the heat exchanger efficiency was improved by about 30\% by filling it with open-cell metal foam.
 
\subsubsection{\glstext*{AFP} Time-of-Flight detector \label{ss:tof}} 
 
\begin{figure}[ht!]
\centering
\subfloat[]{\label{fig:afpSiTToF}\includegraphics[width=0.4\textwidth]{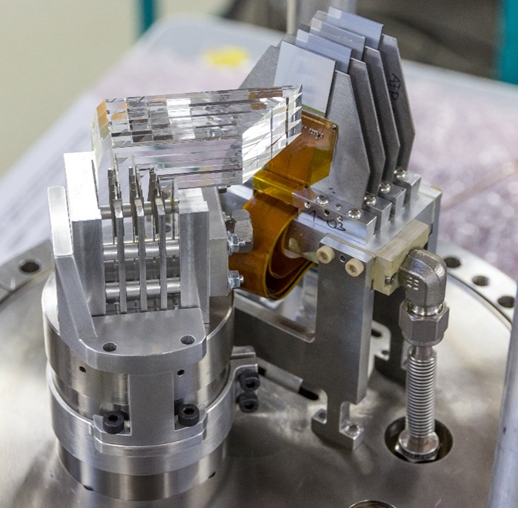}}
\subfloat[]{\label{fig:afpToFLQbarMatrix}\includegraphics[width=0.5\textwidth]{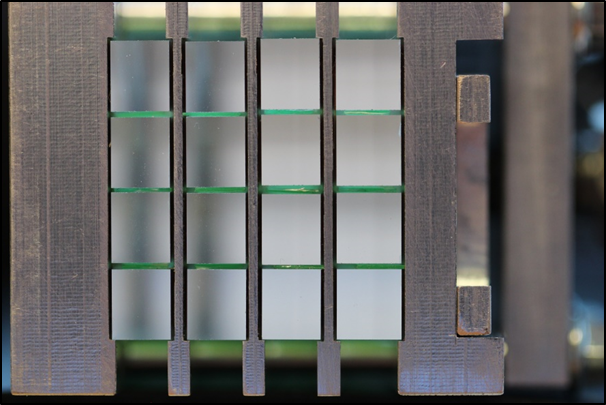}}
\caption[\gls{AFP} Silicon 3D Tracker and quartz radiators.]{\subref{fig:afpSiTToF} The \gls{AFP} far detector platform flange holding the 4-layer Silicon 3D Tracker (\gls{SiT}, top right) and the Time-of-Flight detector (\gls{ToF}, bottom left).
The \gls{SiT} is mounted on a cold-air heat exchanger. The normal to the \gls{SiT} planes makes a \ang{14} angle with the incoming protons to eliminate dead area between pixels. The \gls{ToF} multi-anode \gls{MCP-PMT} is mounted in air inside a tubular feedthrough housing. \subref{fig:afpToFLQbarMatrix} shows the ends of the 4$\times$4 L-shaped Suprasil (``quartz'') radiators (LQbars). Cherenkov light from the LQbars reaches the \gls{MCP-PMT} through a quartz window that separates the \gls{PMT} from the secondary safety vacuum of the detector environment.}
\label{fig:NewFlange}
\end{figure}

The \gls{AFP} \gls{ToF} detectors are located in the far Roman Pots.
The latest version consists of solid L-shaped fused silica bars and customized \glspl{MCP-PMT} with an extended lifetime which operate at low gains (order of 1000)~\cite{Nozka:23}. The improvements were aimed to increase the efficiency, the lifetime as well as the radiation hardness of the detector which has been designed to operate in high radiation areas (above 400 kGy/year).
They have a time resolution in the expected range of \SIrange{20}{26}{\ps} per proton~\cite{Cerny_2019}.
In 2017, the \gls{ToF} efficiency was very low (\SIrange{2}{9}{\percent}), and this was understood to be because the integrated charge received by the \glspl{PMT} during that first year exceeded the limits for which the \glspl{PMT} were intended.
In 2018, it was not possible to install the \gls{ToF} refurbished with long-life \glspl{PMT} because of unresolved
high-voltage breakdown when the \glspl{MCP-PMT} were operated in the secondary vacuum of the pot. In a change of approach, the new \gls{ToF} is now read through a quartz window separating the Cherenkov light guides (inside the pot) from the \gls{MCP-PMT} (outside the pot). This requires a ``Cherenkov light-feedthrough'' on the pot's flange (which serves as the detector platform) which is adjustable in depth to match the specific depth of the pot.
The design is shown in Figure~\ref{fig:NewFlange}.
In this view, forward protons enter the detectors from right-top and the Cherenkov radiators are oriented at \ang{48} with respect to the direction of the forward protons. Vacuum bellows seal the tube holding the square 4$\times$4  multi-anode \gls{MCP-PMT}~\cite{MiniPlanacon} to the flange, allowing for height adjustment of the \gls{ToF} structure.

\subsubsection{AFP Trigger \label{ss:afptrig}} 
While at high luminosity most bunch crossings in ATLAS produce at least one forward proton in one of the \gls{AFP} detectors (and it is therefore usually unnecessary to use this information in the trigger), for special low-pileup runs with $\langle\mu\rangle\sim 1$, it is useful to trigger on events with forward protons. For \RunTwo\ this was accomplished with a \gls{SiT}-plane hit-based majority trigger which had, however, a rather long dead time of about \SI{250}{\ns}. A similar, but \gls{ToF}-based, option was provided in the \RunTwo\ \gls{TDC} module, using the on-board \gls{FPGA} to select events where a majority of \gls{ToF} Cherenkov bars in a train of four successive bars fired. This option is also available in the new pico\gls{TDC} module.
 
In addition, a digital trigger module was added for use in \RunThr, inserted in the digital data path between the constant-fraction discriminator modules and the \gls{TDC}~\cite{Zich_2019}, improving the \gls{TDC} resolution from \SI{18}{\ps} to \SI{4}{\ps}.
This new trigger module provides programmable majority logic for each \gls{ToF} train, and outputs the observed trigger pattern as a sequence of 
\SIrange{1}{2}{\ns}-wide \glsentryshort{NIM} pulses (a start/trigger pulse followed by the trigger status of each of the four trains) on the fast air-core trigger cable to the \gls{CTP}.
 
\subsubsection{AFP Data Acquisition} 
New \gls{DAQ} boards~\cite{bib:SLAC-RCE} were designed for the \gls{AFP}. They are housed in an \gls{ATCA} crate.
The \gls{DCS} interface to the \gls{ATCA} system (see Section~\ref{sec:TDAQ_DCS_ATCA}) was implemented for control and
monitoring, and the \gls{AFP} \gls{DCS}~\cite{Banas_2017} was updated
to control and monitor the voltages, currents, and temperatures
of the new front-ends, digital Trigger modules, pico\glspl{TDC}, and \glspl{VLDB}.
The OptoBoards in the \gls{LHC} tunnel that were used for digital-to-optical conversion in \RunTwo\ were replaced by faster \glspl{VLDB} developed for \RunThr\ and beyond.

Front-end status information is collected by the \gls{DAQ}, for instance the number of parity errors or module busy signals. This information is monitored and in case the \gls{DAQ} decides that some \gls{SiT} or \gls{ToF} device is corrupted, e.g. by a single event upset, a request is sent to the \gls{DCS} for power-cycling the affected device.
 
A new package for \gls{AFP} data quality monitoring was written for \RunThr and operates within the general ATLAS multi-threaded data quality monitoring system and conforms to the ATLAS guidelines.


\subsection{The Zero Degree Calorimeters}\label{zdc}
 
The \glspl{ZDC} play a central role in the ATLAS heavy
ion physics program.  Their primary function is to measure the energy of
``spectator'' neutrons which do not participate in hadronic processes as
the nuclei collide, to determine impact parameters when the nuclei do not overlap completely.
The neutrons propagate in the original beam direction, with minimal deflection,
carrying on average the per-nucleon beam energy.
Because of its ability to observe neutrons from nuclear breakup,
the \gls{ZDC} plays several critical roles during heavy ion operations and data analysis:
\begin{itemize}
\item It is an integral component of the heavy ion triggering scheme and of some \gls{BSM} searches, as it allows the deliberate enhancement and suppression of electromagnetic processes, including photonuclear and elementary photon-photon processes such as light-by-light scattering.
\item It is the only part of the detector that records an unbiased selection of peripheral collisions at low multiplicities, essential for studies of angular correlations in heavy ion collisions.
\item Correlation of the \gls{ZDC} energy with transverse energy in the
central detector demonstrates the clear geometric nature of particle
production in heavy ion collisions. It also provides a well-tested
means, using two-dimensional cuts, to reject both in-time and
out-of-time pileup, which is key for precise physics measurements.
\end{itemize}
 
\subsubsection{Overview of the \glstext*{ZDC} upgrades}
\begin{figure}[t!]
\includegraphics[width=0.9\textwidth]{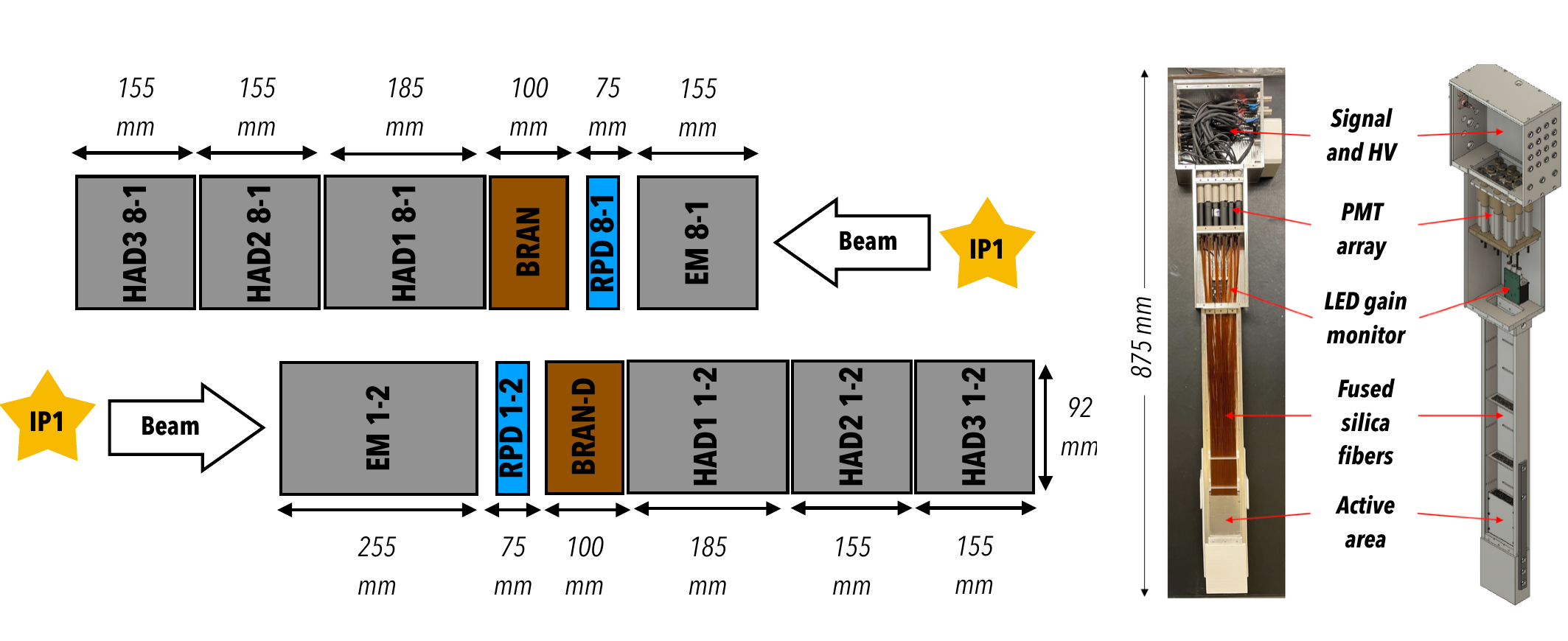}
\caption{The left-hand diagrame shows the arrangement of the ATLAS \glstext{ZDC} arms on Side~A (top) and Side~C (bottom), including the new \glstext{RPD} and indicating the different \glstext{BRAN} detectors being used in the first year of \RunThr. The new ATLAS \glstext{RPD} detector is shown on the right, with a built detector shown head-on, next to a CAD rendering showing the various cable connections on the side.}
\label{fig:zdc_layout}
\end{figure}

In \RunOneTwo, the \gls{ZDC} detector was designed to nearly fill the space
available in the \gls{TAN} region, which sits \SI{140}{\m} from the nominal
\gls{IP}, just after the beams diverge, and protects one
of the \gls{LHC} beamline dipoles.
As shown in Figure~\ref{fig:zdc_layout}, there are four modules on each side
(forward and backward), each of which consists of \num{1.1} nuclear
interaction lengths of tungsten plates (\num{11} plates, each of \SI{1}{\cm}
thickness) interleaved with a row of \SI{1.5}{\mm}-diameter quartz
rods.  Light produced by Cherenkov radiation from shower products is
transported up to the top of the detector, guided through a
trapezoidal prism light guide and into a Hamamatsu H6559 \gls{PMT}~\cite{zdc-pmt} assembly.
The signals from each \gls{PMT} are carried by cables of approximately
\SI{200}{\m} to \gls{USA15}, where they are digitised and included in the
\gls{ZDC} trigger.
\RunTwo began with a detector very similar to that used in \RunOne, after
periodic replacements of the commodity-grade quartz, which was
inexpensive but susceptible to radiation damage, particularly during
proton-proton operation.  However, the increased rates relative to \RunOne
led to large anode currents, and consequently to large changes in the
observed detector gain.  To mitigate these effects, the \glspl{PMT} were replaced
with a modified version (H6559 MOD) with high voltage boosters in the
last three dynode stages.
 
Several modifications to the detector were implemented for \RunThr:
\begin{itemize}
\item To minimise the impact of radiation damage on the signal
properties, the quartz rods were replaced with fused silica
rods of the same dimensions.  A combination of irradiation studies
and comparisons of fused silica samples placed in the \gls{TAN} (in a prototype of
the \gls{BRAN} detector utilised by the \gls{LHC} for luminosity measurements~\cite{MATIS2017114}),
were used to determine that high concentrations of OH and
H$_2$ doping provide additional radiation hardness.  Studies of the
expected dose in \RunThr have led to a final choice of fused silica rods doped with
H$_2$, which provides good stability in the ultraviolet range up to
several million \si{\gray}.
\item Digitisation is handled by the \glspl{PPM}, which run
at a sampling rate of either \num{40} or \SI{80}{\MHz}.  This is well matched to the
signal shape induced by the long coaxial cables, which transform a
fast signal into one with a rise time of \SI{4}{\ns} and an exponential decay
time of about \SI{25}{\ns}, for a total signal length of about \SI{70}{\ns}.
This was already just barely acceptable for the \SI{75}{\ns} bunch spacing
reached in 2018, but is too long for the \SI{50}{\ns} expected for heavy ion operations in
\RunThr.
To provide a faster signal with less attenuation, the coaxial
cables used in \RunTwo have been replaced by air-core coaxial cables similar to those
used in \RunTwo for the \gls{AFP} detector.  These reduce the attenuation
by a factor of nearly \num{10} and the signal shaping is substantially
reduced, such that nearly \SI{97}{\percent} of the charge is contained within a single bunch crossing.
\item The necessary updates to the digitisation scheme will be
discussed in the next section. The sampling rate is
increased by a factor of \numrange{4}{6}, making the after-pulsing of
the \glspl{PMT}, observed in test beam data with much shorter cables,
visible in typical events.  While it would be desirable to find a
photomultiplier tube specifically designed to reduce the impact of
after-pulsing, it turns out that the reduction in the effective
aperture of the photocathode is strong enough to make this
option unnecessary in the next \gls{LHC} run.
\item A new \gls{RPD} has been built, and
tested in an \gls{SPS} test beam, to determine the direction of the
event-by-event deflection of the cloud of spectator neutrons on each
side, and to measure the correlation of the directions between the
two sides.  The two sides are expected to show a clear
anti-correlation of the deflection direction, reflecting the angle
of the internuclear impact parameter, referred to as the
Reaction Plane.  The reaction plane angle can be utilised to
study the directed flow 
of emitted particles, and also
provides sensitivity to the presence of strong magnetic fields in
the initial state of the nuclear collision.  The \gls{RPD} consists of a $\SI{4}{\cm}\times\SI{4}{\cm}$ array of rad-hard optical fibres of different lengths read
out by Hamamatsu R1635 \glspl{PMT} at the top of the \gls{TAN}.  These provide an
effective $4\times $ array of $\SI{1}{\cm} \times \SI{1}{\cm}$ cells in two dimensions.  Since
the longer fibres overlap the shorter ones, machine learning
approaches are needed to disentangle the correlated signals between
the fibres.  In simulations, the resolution of the reaction plane
angle is found to be excellent and comparable to similar
detectors, for example in the STAR experiment at the RHIC accelerator.
 
\end{itemize}
\subsubsection{\glstext*{ZDC} Readout and Trigger Electronics}
The shorter signals from the air-core cables dramatically
mitigate the impact of both in-time and out-of-time pileup on the
measured \gls{ZDC} waveforms, nearly eliminating the overlapping signals
from adjacent filled bunch crossings and even collisions closer in
time due to debunched beam.  However, this requires a much faster
sampling rate than the \RunTwo readout system (based on
electronics originally developed for the \gls{L1} calorimeter system) can provide.
The \gls{LUCROD} card designed for \gls{LUCID}, as described in Section~\ref{lucid}, was adapted to rework the entire trigger and readout
approach for the \gls{ZDC} detectors.  The new card has modified
firmware and is referred to as the \gls{LUCROD}/\gls{ZDC}.
 
The \RunTwo \gls{ZDC} trigger
and readout scheme split the incoming signals into two
paths, one for digitising the signal and the other for providing
\gls{L1} triggers for ATLAS.  In the trigger path, the incoming \analog signals were summed
in hardware and a threshold was applied just below the single neutron peak for each side, using \glsentryshort{NIM}
discriminators.  The readout path involved a fourfold splitting of the
incoming signals into both high and low gain, to obtain an effective 12-bit dynamic range,
and applying a \SI{12.5}{\ns} delay to achieve an \SI{80}{\MHz} effective sampling rate even when using the \gls{L1Calo} \glspl{PPM},
which sampled at \SI{40}{\MHz}.
 
For \RunThr, the \gls{LUCROD} card, with special firmware for the \gls{ZDC},
combines digitisation, triggering and event logging into the same
card, with waveform sampling performed at \SI{320}{\MHz}.
Each \gls{LUCROD}/\gls{ZDC} card provides 8 channels, sampled with 12-bit flash \glspl{ADC}, with an \gls{FPGA} for
each channel (\gls{FPGACH}), one \gls{FPGA} (\gls{FPGAV}) which receives data from all 8
channels simultaneously to perform trigger calculations, and one
main \gls{FPGA} (\gls{FPGAM}) which serialises and transports the channel data to ATLAS.
One \gls{LUCROD} is used to digitise the signals without amplification,
while providing an \analog copy to a second card which amplifies the
signals by a factor of 16. Combined, the two cards provide an effective
16-bit dynamic range.  The high-gain \gls{LUCROD}/\gls{ZDC} also runs trigger
algorithms that provide the functionality previously provided by the
\gls{NIM} modules.  Peak finding is performed for each channel in the \gls{FPGACH}
using a threshold on the second derivative of the waveform, while
\gls{FPGAV} performs digital sums and uses two sets of lookup tables to form
a \gls{L1} decision for the ATLAS \gls{CTP}.  A lookup table for each side is used
to generate multiple energy thresholds -- specifically, distinguishing
one neutron from no activity, which tags nuclear breakup, and
distinguishing five neutrons or more, which distinguishes hadronic and
electromagnetic processes.  A combined lookup table then incorporates
the threshold level from the two sides, and provides a three-bit output
word to the ATLAS \gls{CTP} which reflects the correlations between the
energies of the two sides, distinguishing (for example) between events with
energy on only one side and those with substantial energy on both
sides.  Upon receipt of a \gls{L1A} from ATLAS, the \gls{FPGAM}
serialises the data from the channel \glspl{FPGACH}, prepares packets for
transmission over \gls{s-link} and manages \gls{s-link} transmission to the
ATLAS \gls{ROS}.  With this implementation, the \gls{LUCROD}s replaces essentially
all of the \gls{ZDC} electronics utilised for \RunTwo.
 
\subsubsection{\glstext*{ZDC} Detector control}
The \gls{ZDC} \glspl{PMT} are controlled by a CAEN mainframe
located in \gls{USA15}, with long \gls{HV} cables running to distribution
boxes located under the \gls{TAN}, and finally to the detectors, with
standard \gls{SHV} cables to drive the primary \gls{PMT} \gls{HV}, and with multicore
\gls{HV} cables to drive the three boosters per \gls{PMT}.  The ATLAS \gls{DCS} (as described in Section~\ref{TDAQ_DCS_OPCUA}) is used
to set and monitor the high voltage and current for each channel.
It is crucial to track the booster currents at higher rates,
to prevent damage to the \glspl{PMT}.
As the signals were calibrated in short intervals of several minutes throughout \RunTwo, based on a calibration
peak from events with one neutron -- at full beam energy -- in either
\gls{ZDC}, a stable response was not required over longer timescales, and the \RunTwo detector was simply continually monitored for degradation
of the observed energy of the one-neutron peak.  The \gls{HV} in one or
more \glspl{PMT} on each side was then increased when the fluctuations of that peak
began to approach the noise peak seen in empty events.  In \RunThr, the
radiation-hard fused silica rods replacing the quartz should
provide good stability throughout each heavy ion run period, and
minimise or eliminate any need for \gls{HV} adjustments.



\clearpage
\newpage
 
\section{Trigger and Data Acquisition System} 
\label{sec:TDAQ}

 
\subsection{System Overview}
\begin{figure}[b]
\centerline{\includegraphics[width=0.95\textwidth]{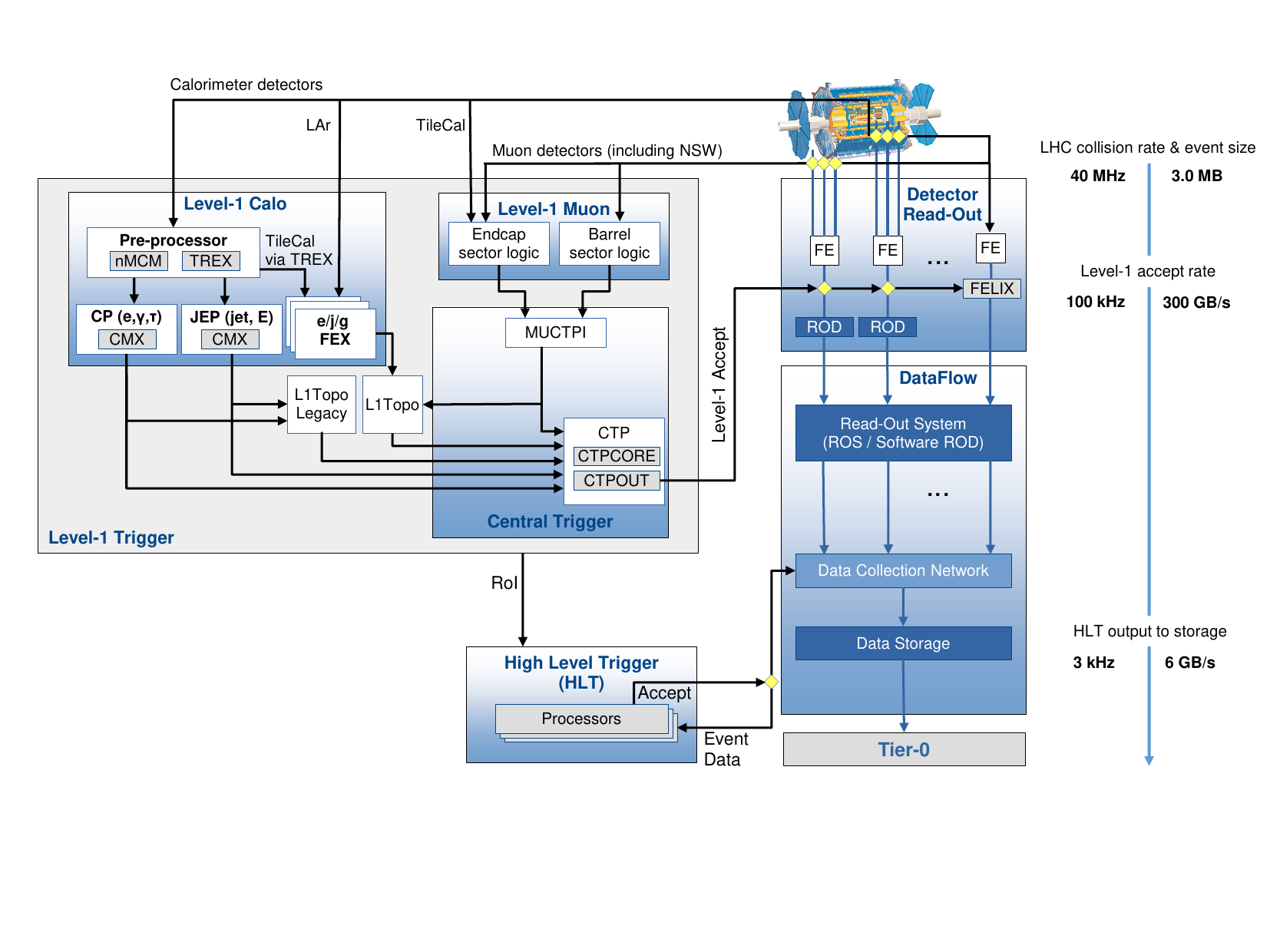}}
\caption{Schematic overview of the Trigger and DAQ system at the beginning of \RunThr.}
\label{fig:TDAQ_Overview}       
\end{figure}

As described in Section~\ref{subsec:OverviewTDAQ} and shown in Figure~\ref{fig:TDAQ_Overview}, the \glsfirst{TDAQ} system is based on a two-level trigger (event selection) system served by  the \glsfirst{DAQ} system that transports triggered data from custom subdetector electronics through to offline processing. The \glsfirst{L1} trigger system is based on custom-built electronics  and the \glsfirst{HLT}  is based on software implemented on commodity computers.
 
\begin{figure}[htbp]
\centerline{\includegraphics[width=0.95\textwidth]{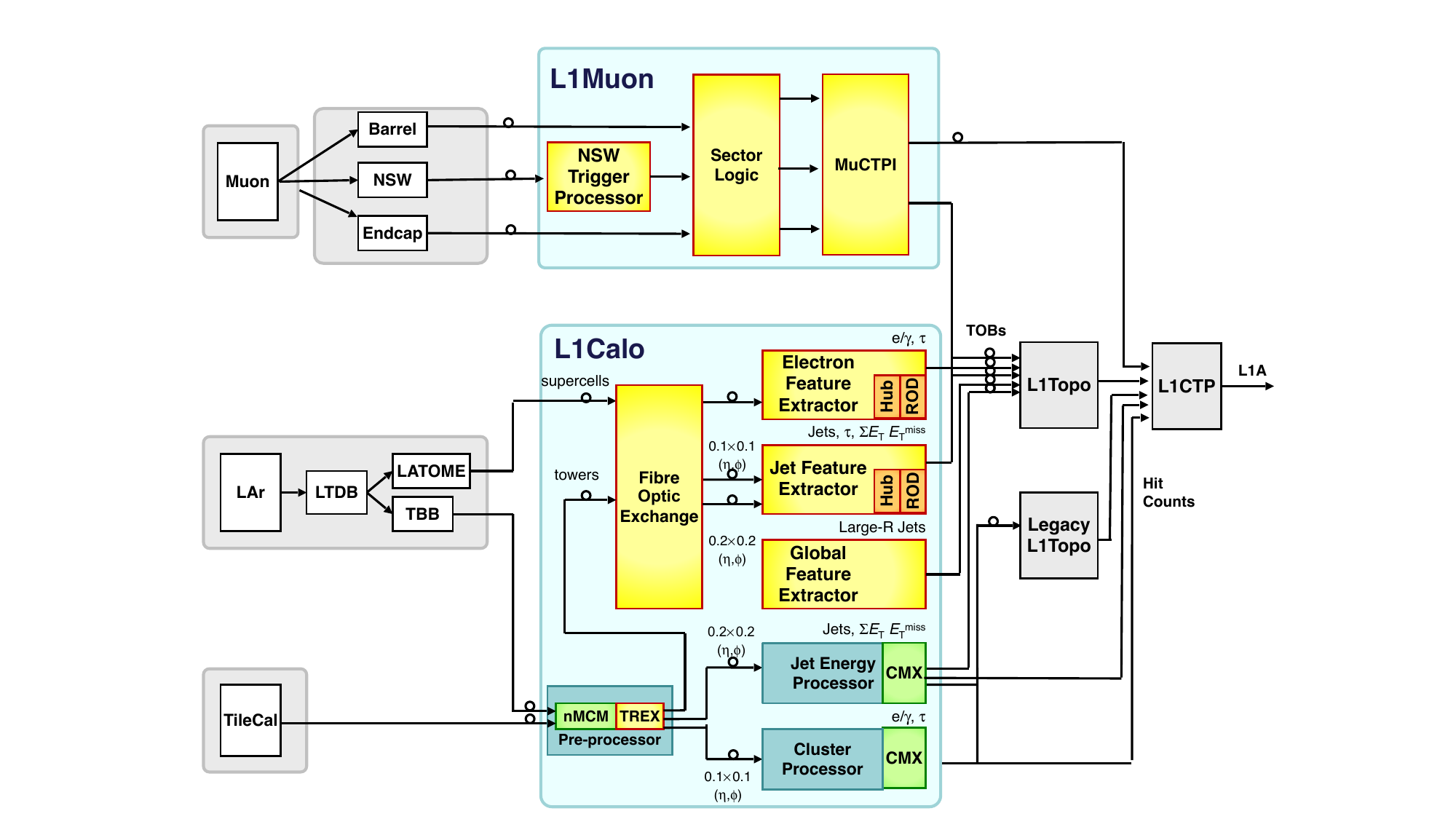}}
\caption{Schematic overview of the \gls{L1} trigger system in \RunThr.  The new and refurbished elements of the \gls{L1Muon} Phase-I trigger system are shown in yellow.  During the commissioning period, both the legacy (shown in green) and Phase-I \gls{L1Calo} (shown in yellow) modules will run in parallel.  Once the Phase-I system has been commissioned, the legacy \gls{L1Calo} \gls{JEP}, \gls{CP} and \gls{L1Topo} modules will be removed; the Phase-I \gls{L1Calo} system and upgraded \gls{L1Topo} module will provide triggers for physics in \RunThr.}
\label{fig:TDAQ_L1Overview}
\end{figure}

The \gls{L1} Trigger, shown in Figure~\ref{fig:TDAQ_L1Overview}, uses reduced-granularity information from the calorimeters and muon system to search for signatures from high-\pT muons, electrons, photons, jets, and $\tau$-leptons decaying into hadrons, as well as events with large missing transverse energy or large total transverse energy.  It consists of the \glsfirst{L1Calo} and \glsfirst{L1Muon} trigger systems, the \glsfirst{MuCTPI}, the \glsfirst{L1Topo} and the \glsfirst{CTP}. The \gls{L1Calo} trigger system (Section~\ref{sec:TDAQ_L1Calo}) identifies high-\ET objects such as electrons, photons, jets, and $\tau$-leptons.  Isolation requirements may also be applied to these objects.  It also selects events with large \MET or large total transverse energy.  The \gls{L1Muon} trigger system (Section~\ref{sec:TDAQ_L1Muon}) selects events containing high-\pT muons, based on inputs from the \glspl{RPC} in the barrel region and from the \glspl{TGC} and \glspl{NSW} in the endcaps, and then transmits data to the \gls{CTP} via the \gls{MuCTPI}. The \gls{L1Topo} trigger system (Section~\ref{sec:TDAQ_L1Topo}) takes input \glspl{TOB} containing kinematic information (e.g. \ET and $\eta-\phi$ coordinates) from the \gls{L1Calo} and \gls{L1Muon} systems and applies topological selections. As shown in the figure, both the legacy \gls{L1Topo} and upgraded \gls{L1Topo} systems are in use during the commissioning phase.
 
The final \gls{L1} trigger decision is made by the \gls{CTP} (Section~\ref{sec:TDAQ_CTP}).  The \gls{CTP} receives hit multiplicities from the \gls{L1Calo}, \gls{L1Muon}, and \gls{L1Topo} systems and accepts events satisfying requirements based on object type and threshold multiplicity.  Up to 512 distinct \gls{L1} trigger items may be configured in the \gls{CTP}.  The \gls{L1} trigger decision, as well as the 40.08~MHz \gls{LHC} bunch-crossing clock, is distributed to the detector front-end and readout systems via the \gls{TTC} system~\cite{bib:ttc}.
 
The parameters of the \gls{L1} trigger system depend on the pipeline memories, as specified and built for the original construction of the detector,  in custom electronics located on or near the detector to store information while the trigger decision is in progress.
This necessitates that the \RunThr \gls{L1} latency, the time from a given bunch crossing to the trigger decision, must adhere to the original specification of the detector.
The design of the trigger and detector front-end systems requires that the \gls{L1} latency be less than \SI{2.5}{\micro\s}.  The maximum \gls{L1A} rate supported by the detector readout systems is \SI{100}{\kHz}, and so the menu of the trigger selections is tuned to allocate a share of the rate budget to each underlying physics object according to ATLAS analysis goals.
 
If accepted by the \gls{L1} trigger, events are then sent to the \gls{HLT}. At this level, software processes reconstruct the event at higher levels of detail than \gls{L1}, seeded by \glspl{RoI}, which are regions of the detector in which possible trigger objects have been identified by the \gls{L1} trigger.
\gls{HLT} algorithms use full-granularity information from the calorimeters and the muon and tracking systems to provide better energy and momentum resolution for threshold selections, as well as precision tracking for particle identification. The overall selection criteria for the \gls{HLT} are once again provided by the trigger menu, tuned so as to maximise the data available for key analyses for the given resources.
 
The \gls{HLT} software is designed to reproduce the offline selection as closely as possible. On average, the event processing time at the \gls{HLT} in 2018 was approximately \SI{400}{\ms} for runs with a peak luminosity of \lumiruntwopeak. In \RunThr the \gls{HLT} reduces the event rate from \SI{100}{\kHz} after the \gls{L1} selection to approximately \SI{3}{\kHz} (averaged over the course of an \gls{LHC} fill); the data are then stored for offline analysis.
The \gls{HLT} software has undergone a significant redesign in order to take advantage of multithreaded \gls{CPU} architectures; this framework, which is also used in the offline software, is called \gls{AthenaMT} and is described in Section~\ref{subsubsec:tdaq_daq_hlt_athenamt}.
 
Underpinning the operation of the two trigger levels, the data acquisition system begins with detector-specific on- and off-detector electronics which perform a variety of data processing and monitoring features before passing events either to the \gls{L1} trigger or to the downstream system. In \RunOneTwo the off-detector stage was performed in (typically \gls{VME}-based) detector-specific custom hardware modules called \glspl{ROD}. On receipt of a \gls{L1A}, these \glspl{ROD} read out their data to the first common stage of the \gls{DAQ} system, the \gls{ROS}. The \gls{ROS} buffers event data and serves it to \gls{HLT} nodes on request to facilitate a trigger decision. As is described in Section~\ref{subsubsec:tdaq_daqhlt_felixswrod}, \RunThr will see the introduction of the new \gls{FELIX} and \gls{SW ROD}  systems into the readout path, with the goal to reduce the amount of custom hardware in the system and perform more processing in the software domain. In the new system the \gls{ROD} and \gls{ROS} are replaced by \gls{FELIX}, which receives readout data from \gls{L1} trigger processors  and detector front-end electronics via point-to-point links and then routes them to software applications running on commodity servers (the \glspl{SW ROD}) that perform the same tasks previously handled in the hardware \glspl{ROD}, plus the buffering function of the \gls{ROS}. By \RunFour, all of ATLAS readout will be via this new mechanism, but in \RunThr it has been rolled out for those systems with new or upgraded front-end electronics. The interface to the \gls{HLT} is identical between \gls{ROS} and \gls{SW ROD}, with events routed on demand to \gls{HLT} processing nodes in both cases.
 
The final stage of the \gls{DAQ} system, once \gls{HLT} processing has been completed, is for accepted events to be sent to a dedicated cluster of servers, known for historical reasons as \glspl{SFO}, for packing, compression, and finally transfer to offline storage.
 
\subsubsection{Motivation for upgrades}
 
Since \RunOne, the instantaneous luminosity and number of interactions per bunch crossing (pileup) have increased substantially, as illustrated in Figure~\ref{fig:Overview:lumi-pileup}.  In \RunThr, the improvements in the \gls{LHC} (see Section~\ref{ss:LHC}) and luminosity levelling will allow a much larger fraction of each run to be near the peak instantaneous luminosity of \lumirunthree, causing the average pileup to increase to \murunthree\ or beyond. Increased pileup degrades the calorimeter resolution and isolation of single particles, which leads to decreased trigger efficiency and necessitates the use of higher trigger thresholds to mitigate the resulting increase in rate.  One of the main goals of the ATLAS trigger upgrade is to reduce the impact of pileup in order to maintain low thresholds, especially at \gls{L1}, thereby maximising the dataset recorded for precision measurements (e.g.\ of the Higgs boson) as well as for searches for physics beyond the Standard Model.  Since the events passing the \gls{L1} selection will contain a larger number of pileup interactions, further upgrades at the \gls{HLT}, including improved precision tracking and the expansion of the computing farm, are intended to maintain a reasonable output rate within bandwidth and storage limitations.
 
\subsubsection{Improvements for \RunTwo}
 
Substantial upgrades to the ATLAS \gls{TDAQ} system were made prior to the beginning of \RunTwo in 2015.  These changes are described in detail in~\cite{TRIG-2016-01}, but are briefly summarised here for completeness.
 
During \gls{LS1}, the \gls{L1Muon} barrel system was equipped with additional trigger electronics to improve the acceptance in the feet and elevator regions (where new detectors were also installed, as discussed in Section~\ref{muonSS:BMEBOEBMG}).  In the endcap region, new trigger coincidence logic was deployed in order to improve the rejection of background originating outside the \gls{IP}.
This logic also made use of new signals derived from the outermost Tile calorimeter cells.
 
Several significant upgrades were made to the \gls{L1Calo} system as shown in Figure~\ref{fig:TDAQ_L1Overview}.   The \glspl{PPM}~\cite{TDAQ-2019-01} were upgraded to provide improved bunch-crossing identification and pedestal subtraction capabilities, resulting in better pileup suppression and lower \gls{L1} jet and \MET trigger rates.  The \gls{CP} and \gls{JEP} systems were upgraded to allow an increased data transmission rate. Additional improvements were made to the \gls{CP} to allow for energy-dependent isolation requirements to be applied on \gls{EM} object candidates.  New extended \glspl{CMX}~\cite{TRIG-2016-01} were added to replace the existing \glspl{CMM}, which transmitted the threshold multiplicities to the \gls{CTP}.  In addition to this functionality, the \gls{CMX} permits the transmission of \glspl{TOB}, containing such information as the position and energy of physics objects passing the \gls{L1} trigger, to the new \gls{L1Topo} system.
 
The \gls{L1} Topological trigger system (L1Topo)~\cite{TRIG-2019-02}, described in Section~\ref{sec:TDAQ_L1Topo}, was installed and commissioned during \RunTwo. The system uses \glspl{TOB} from \gls{L1Calo} and \gls{L1Muon} and applies topological selections in order to reduce background rates while preserving signal efficiency.
 
For the \gls{DAQ} system, \RunTwo saw the introduction of a newer version of the \gls{I/O} card housed in the \gls{ROS} servers which provides the interface to front-end optical links and buffers data throughout the \gls{HLT} selection process. The move to this new board, the RobinNP, resulted in a system up to 4 times denser than its predecessor, saving significant rack space, while also providing 10 times higher buffering capacity per data-link than before. The performance of the upgraded \gls{ROS} was also nearly 4 times that of its predecessor in terms of supported maximum data request rate from the trigger.
 
Alongside the new \gls{ROS}, the \gls{HLT} itself was significantly redesigned for \RunTwo. What was previously two separate farms, one focused on \gls{RoI}-driven selection and one studying full events, was merged into one with a more dynamic event-building approach. The combined farm eliminated the previously complex load balancing required to manage the two-level version, while also minimising any duplication of data requests to the \gls{ROS}. Thanks to this redesign, it was also possible to greatly simplify the data collection network connecting the \gls{HLT} to the \gls{ROS} and \gls{SFO}. What was previously two separate networks became one high-performance implementation, with 20 times higher bandwidth than before (40 \gls{GbE} vs 2 \gls{GbE}).
 
Finally, the \gls{SFO} systems were also upgraded for \RunTwo. While the functionality remained the same, the new servers provided significantly improved staging and streaming capability, able to buffer more than twice as much data as before, making it possible to stably process and write much higher data volumes to offline storage.
 
All of these upgrades combined allowed the
ATLAS trigger system
in \RunTwo to execute more complex algorithms with higher overall processing time and to writing out larger and richer datasets, making it possible to significantly improve the sophistication of the trigger menu and greatly enhance the ATLAS physics programme.
 
\subsection{Level-1 Calorimeter Trigger}\label{sec:TDAQ_L1Calo}
The \glsfirst{L1Calo} trigger, pictured in Figure~\ref{fig:TDAQ_L1Overview}, identifies events containing electrons, photons, jets, and $\tau$-leptons decaying into hadrons, as well as events with large missing transverse energy (\MET) or large total transverse energy (\sumET) using custom hardware processors based on \gls{FPGA} technology. To manage increasing rates in this regime, while preserving an effective and efficient calorimeter trigger, significant upgrades are necessary.  These upgrades include more refined processing of electromagnetic calorimeter information at higher granularity, which provides improved identification of isolated electrons and photons and improved rejection of jets, thereby maintaining low trigger thresholds, which are especially important for electroweak physics.  Improved processing at higher granularity and new algorithms will also enable better identification of jets and discrimination against pileup, which will benefit searches for physics involving boosted objects and \MET.
 
\begin{figure}[htbp]
\centering
\subfloat[]{
\includegraphics[width=0.3\textwidth]{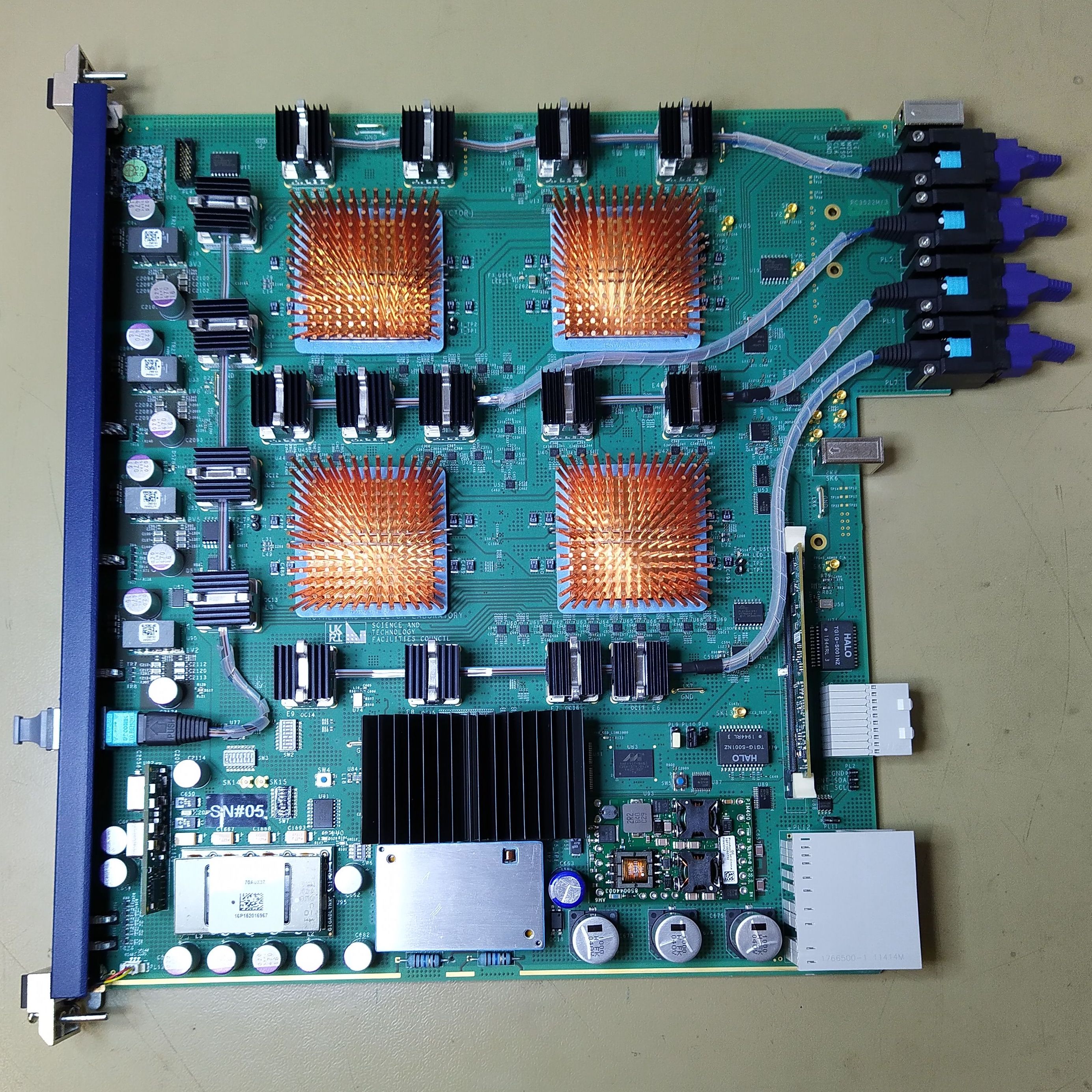}
\label{fig:TDAQL1CaloeFEXPhoto}
}
\subfloat[]{
\includegraphics[width=0.3\textwidth]{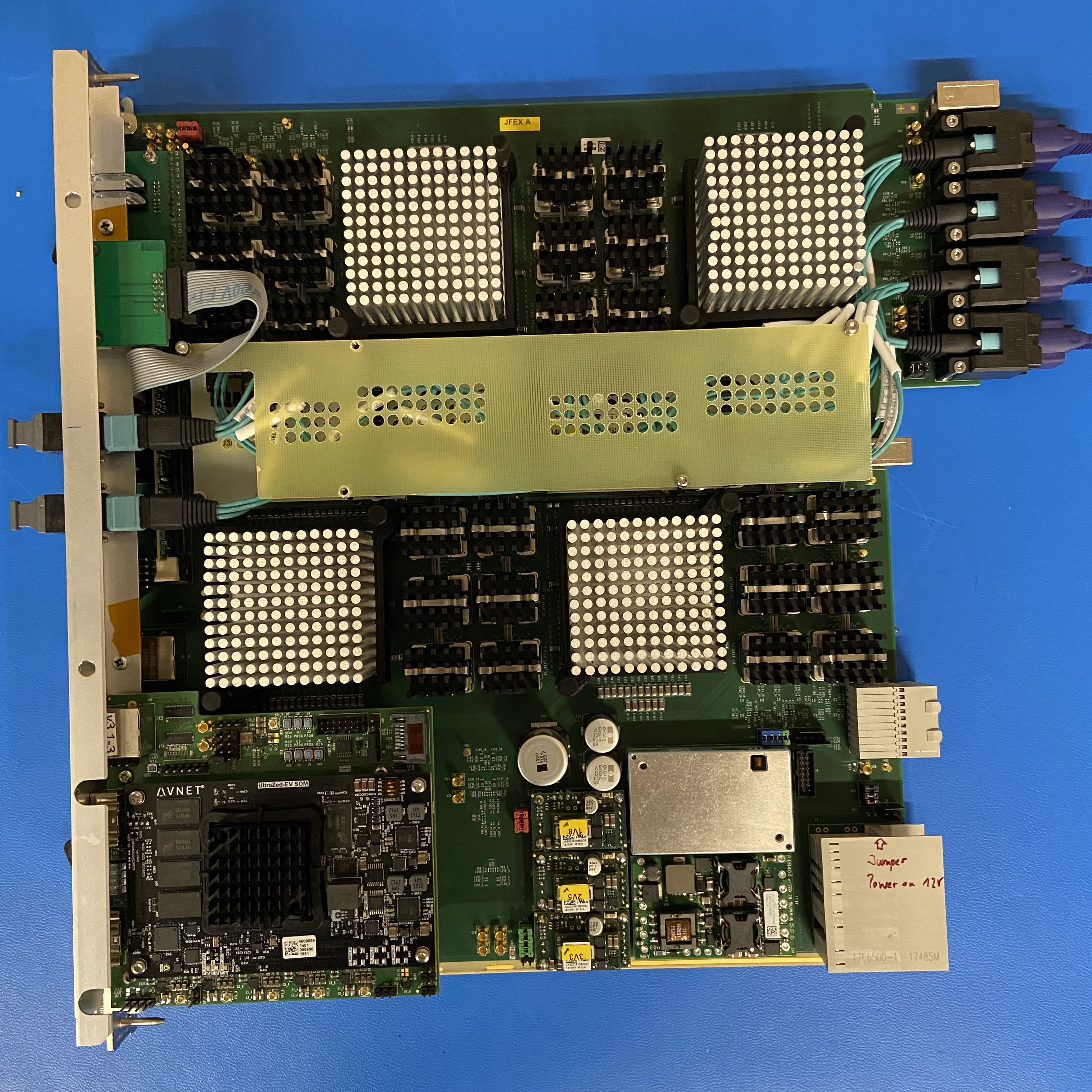}
\label{fig:TDAQL1CalojFEXPhoto}
}
\subfloat[]{
\includegraphics[width=0.3\textwidth]{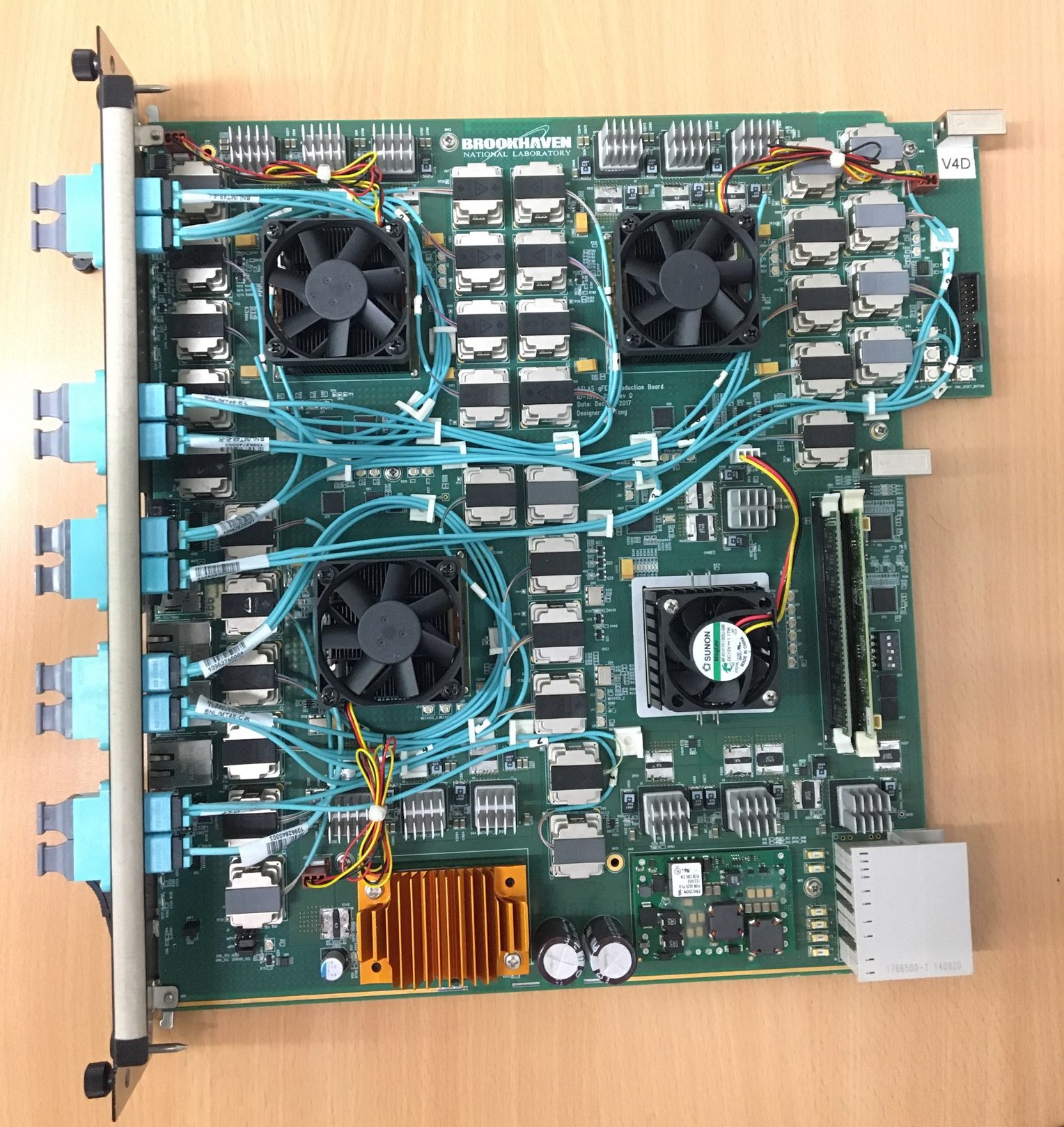}
\label{fig:TDAQL1CalogFEXPhoto}
}
\caption{
\protect\subref{fig:TDAQL1CaloeFEXPhoto} A production \gls{eFEX} module.
\protect\subref{fig:TDAQL1CalojFEXPhoto} A production \gls{jFEX} module.
\protect\subref{fig:TDAQL1CalogFEXPhoto} A production \gls{gFEX} module.
}
\label{fig:TDAQL1CaloFEXPhotos}
\end{figure}

The \RunThr system includes new electromagnetic and jet feature extractors: the \gls{eFEX} (Section~\ref{sec:TDAQ_eFEX}) identifies electron, photon, and $\tau$ objects, while the \gls{jFEX} (Section~\ref{sec:TDAQ_jFEX}) identifies jets, \MET, and $\tau$s.  A global feature extractor (\gls{gFEX}, Section~\ref{sec:TDAQ_gFEX}) identifies large-area jets, \MET, and \sumET.
The new \gls{FEX} modules are shown in Figure~\ref{fig:TDAQL1CaloFEXPhotos}.
 
\begin{figure}[htbp]
\centerline{\includegraphics[width=0.5\textwidth]{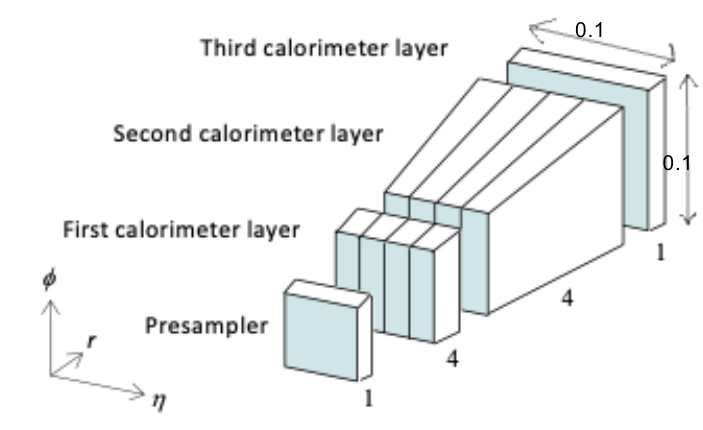}}
\caption{The finer granularity of the \RunThr \gls{L1Calo} trigger towers after the upgrade of the liquid argon calorimeter electronics.  Each tower is divided into ten \glspl{SC}, each providing an \ET\ value.}
\label{fig:TDAQ_L1CaloSuperCell}
\end{figure}

Compared to \RunOneTwo, \gls{L1Calo} receives finer-granularity input data from the \gls{LAr} calorimeter in \RunThr.  In \RunOneTwo, these inputs consisted of ``trigger towers'' spanning $0.1 \times 0.1$ in $\eta$ and $\phi$.  In the \RunThr system, the \gls{LAr} Digital Processing System (\gls{LDPS}; Section~\ref{sec:LArDigitalTrigger}) provides electromagnetic calorimeter information in the form of \glspl{SC} containing sums of four or eight calorimeter cells.  The detailed mapping is provided in Table~\ref{tab:LArSuperCells} and summarised here. Each trigger tower contains 10 \glspl{SC}, as shown in Figure~\ref{fig:TDAQ_L1CaloSuperCell}: one in the presampler, four in each of the first and second layers, where the majority of shower energy is deposited, and one in the third layer.  The full granularity, corresponding to $\Delta\eta \times \Delta\phi = 0.025 \times 0.1$ in the first and second layers, is available to the $e/\gamma$ and $\tau$ triggers, while the jet, \ET, and \MET triggers use coarser granularity, comparable to the size of the $0.1 \times 0.1$ trigger towers.  For comparison, in \RunTwo, \gls{L1} jet triggers were based on $0.2 \times 0.2$ jet elements.
 
In \RunOneTwo, the calorimeter data arrived at \gls{L1Calo} in \analog format, and was then digitised and calibrated by the pre-processors.  In \RunThr, the digitisation and calibration of the \gls{LAr} calorimeter data are now performed in the \gls{LAr} calorimeter electronics (\ref{sec:LArDigitalTrigger}).  The Tile calorimeter data are still received in \analog format and are digitised by the pre-processors, and then the new \gls{TREX} (Tile Rear EXtension; Section~\ref{sec:TDAQ_L1CaloInfrastructure}) modules transmit them to the \glspl{FEX} and the legacy system.  It is planned to upgrade the Tile calorimeter system to perform the signal digitisation for running at the \gls{HL-LHC}~\cite{ATLAS-TDR-28}.
 
The digitised and calibrated calorimeter inputs are sent to \gls{L1Calo} via optical fibres. Mapping and routing of the fibres to the \glspl{FEX} is performed by the \gls{FOX} optical plant (Section~\ref{sec:TDAQ_L1CaloInfrastructure}).
 
The output from \gls{L1Calo} consists of \glspl{TOB} which include the position, \ET, object type, and energy sum.  These are sent to \gls{L1Topo}, which computes the multiplicity of \glspl{TOB} passing a given threshold and sends the information to the \gls{CTP}.  Another optical plant (TopoFOX; Section~\ref{sec:TDAQ_L1CaloInfrastructure}) is used to map the output fibres from \gls{L1Calo} to L1Topo.  In addition, \gls{L1Calo} sends \gls{RoI} information to the \gls{HLT} and readout data to the \gls{FELIX} and \gls{SW ROD} (Section~\ref{subsubsec:tdaq_daqhlt_felixswrod}).
 
During the start-up and commissioning phase of \RunThr, triggers for physics will be provided by the legacy (\RunOneTwo) \gls{L1Calo} system.  Once the Phase-1 system has been commissioned and validated, it will be enabled to provide physics triggers and the legacy system will be removed.

\subsubsection{Electron Feature Extractor (eFEX)}\label{sec:TDAQ_eFEX}
 
\begin{figure}[htbp]
\centerline{\includegraphics[width=0.75\textwidth]{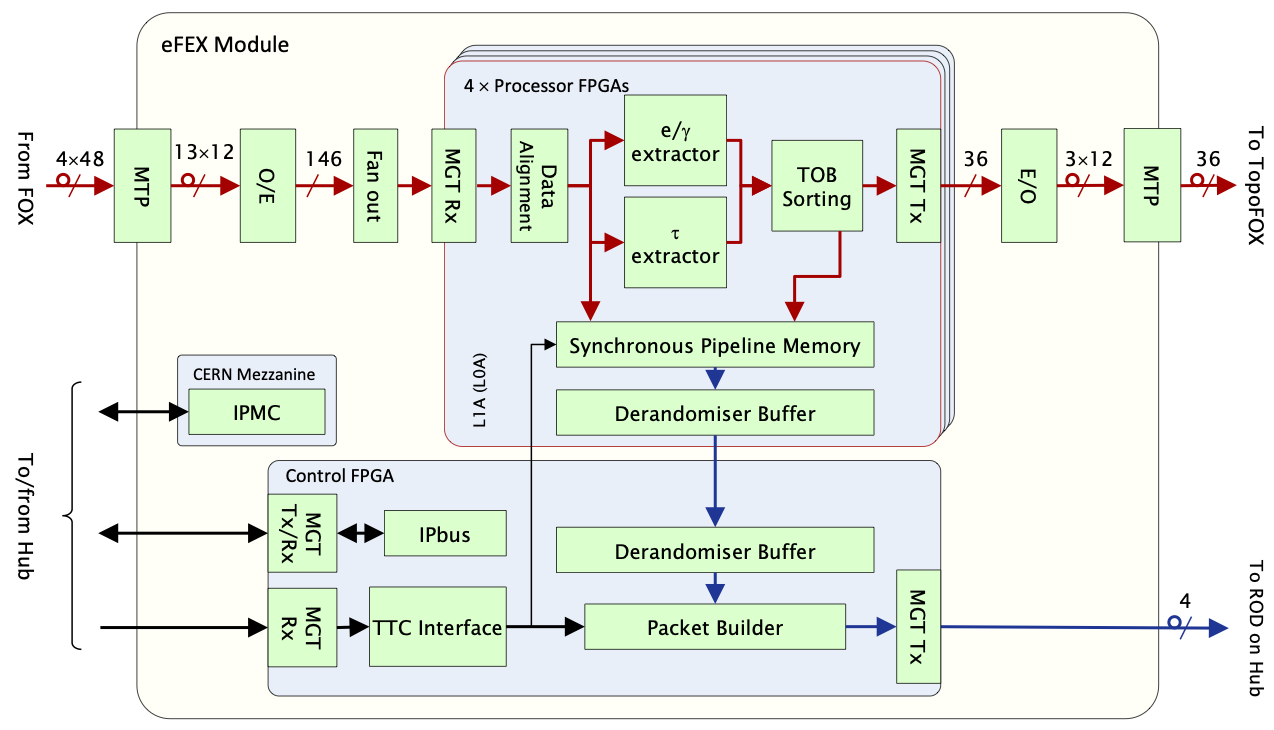}}
\caption{Block diagram of an \gls{eFEX} module, illustrating the real-time and readout paths.  Control and monitoring signals, except for the \gls{L1A}, are not shown.}
\label{fig:TDAQL1CaloeFEXModule}
\end{figure}

\paragraph{\glstext*{eFEX} module design}
The \gls{eFEX} system consists of 24 modules located within two \gls{ATCA} shelves, with each module containing four algorithm-processing \glspl{FPGA} and one control \gls{FPGA}.  A block diagram of an \gls{eFEX} module is shown in Figure~\ref{fig:TDAQL1CaloeFEXModule}, and a photograph in Figure~\ref{fig:TDAQL1CaloeFEXPhoto}.
 
The system covers a region of $|\eta|\le 2.5$ and the full $\phi$ range.  Each of the 24 modules receives data on up to 136 fibre links (11.2 Gb/s), covering a calorimeter area of up to $\Delta\eta\times\Delta\phi = 1.7 \times 1.0$.  Each \gls{FPGA} processes data for 32 algorithm cores of area 
$\Delta\eta\times\Delta\phi = 0.1\times 0.1$, for a total area of $\Delta\eta\times\Delta\phi = 0.4 \times 0.8$ per \gls{FPGA}.  On the borders, the total number of algorithm cores processed per \gls{FPGA} increases to 40.  The input to each algorithm core includes the surrounding environment, which measures $\Delta\eta\times\Delta\phi = 0.3 \times 0.3$.  Each \gls{FPGA} contains an algorithm module which produces $e/\gamma$ and $\tau$ \glspl{TOB}. The internal clock for processing is \SI{200}{\MHz} and the output rate is \SI{280}{\MHz}. \glspl{TOB} are first sorted locally on each \gls{FPGA} before being sent to dedicated $e/\gamma$ and $\tau$ sorter modules on two of the four \glspl{FPGA}; this sorting is also performed at \SI{280}{\MHz}.
 
The \glspl{TOB} computed by the \gls{eFEX} algorithms are merged and sorted across the entire system.  A maximum of six \glspl{TOB} per algorithm ($e/\gamma$ and $\tau$) per module may be sent to \gls{L1Topo}; in case more than six \glspl{TOB} are identified, the six with the highest transverse energy \glspl{TOB} are sent.
 
\textbf{Timing and readout}  \gls{TTC} signals are received on the backplane from the \gls{Hub} (see Section~\ref{sec:TDAQ_L1CaloHubROD}).  Upon receiving a \gls{L1A}, \glspl{TOB} and input data (prescaled to limit the bandwidth usage) are sent to the \gls{ROD} (described in Section~\ref{sec:TDAQ_L1CaloHubROD}).  Each \gls{eFEX} module sends its readout data over 8 electrical links (\SI{6.4}{Gb/\s}) on the \gls{ATCA} backplane using the multi-lane Aurora protocol~\cite{bib:Aurora}.  From the \gls{ROD}, the readout data are then sent to the \gls{FELIX} and \gls{SW ROD} (Section~\ref{subsubsec:tdaq_daqhlt_felixswrod}).
 
\textbf{Configuration, control, and monitoring} Configuration is performed by the control \gls{FPGA}.  Monitoring and control are handled by the IPbus~\cite{bib:IPbus} firmware implemented in the control \gls{FPGA}.  A CERN-standard \gls{IPMC} module~\cite{bib:CERN-IPMC} performs low-level control functions.
 
\begin{figure}[htbp]
\centering
\subfloat[]{
\includegraphics[width=0.35\textwidth]{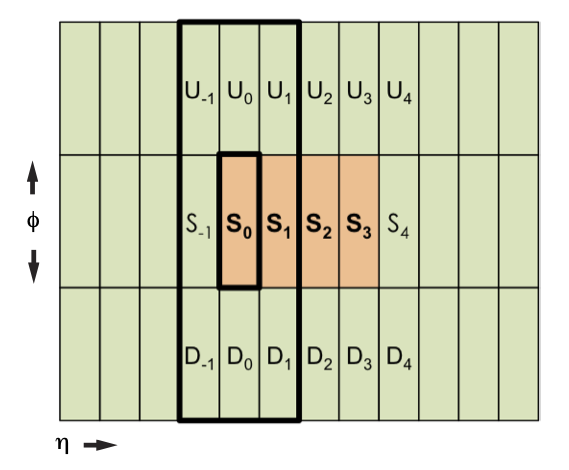}
\label{fig:TDAQL1CaloeFEXSeed}
}
\subfloat[]{
\includegraphics[width=0.35\textwidth]{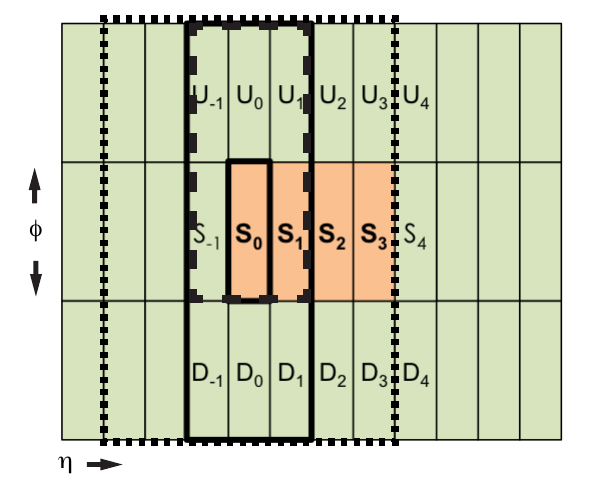}
\label{fig:TDAQL1CaloeFEXfield}
}
\caption{\protect\subref{fig:TDAQL1CaloeFEXSeed}  The \gls{L1Calo} \gls{eFEX} seed finder algorithm.  The four potential seeds (S) in the algorithm core are compared in order to determine the local energy maximum.  The directionality of the cluster, upward (U) or downward (D) is determined by finding the largest energy deposit among the \glspl{SC} surrounding the seed.
\protect\subref{fig:TDAQL1CaloeFEXfield} The dashed rectangle shows of $3 \times 2$ \glspl{SC} used to compute the cluster energy, assuming that the local energy maximum was at cell $S_0$ and that the most energetic directly neighbouring \glspl{SC} to $S_0$  was $U_0$.  The dotted rectangle shows the  $7 \times 3$ \glspl{SC}   area used for the calculation of the shower width in the second calorimeter layer.
}
\end{figure}

\paragraph{\glstext*{eFEX} algorithms}\mbox{}\\
\textbf{$e/\gamma$ algorithm}  The \gls{eFEX} $e/\gamma$ algorithm searches for a ``seed'', i.e. a local energy maximum, by comparing the four \glspl{SC} in a given algorithm core in the second layer of the electromagnetic calorimeter.  This process is illustrated in Figure~\ref{fig:TDAQL1CaloeFEXSeed}.  The $\phi$-direction of the cluster is established by finding the highest-energy
\gls{SC} adjacent to the seed.  A cluster of $3 \times 2$, \glspl{SC} (see Figure~\ref{fig:TDAQL1CaloeFEXfield}),  is formed around the seed, and the cluster energy is then computed by adding the energy of the corresponding \glspl{SC} in the presampler, first, and third layers of the calorimeter.
 
To improve background rejection, isolation requirements may be applied. These are generally defined as a ratio between the energy sum in the cluster and that in the surrounding area.  Such requirements are applied in the form of selections based on several electromagnetic shower observables; the following observables can be computed in the \gls{eFEX}:
 
\begin{itemize}
\item Shower width in the second calorimeter layer \begin{equation} R_{\eta} = 1 - \frac{E_{\mathrm{T}, 3 \times 2}}{E_{\mathrm{T}, 7 \times 3}},\end{equation}
 
where $E_{\mathrm{T}, 3 \times 2}$ is the transverse energy sum in an area of $3 \times 2$ \glspl{SC} in the second layer of the calorimeter centred on the local maximum and its highest-\ET neighbour in $\phi$, and $E_{\mathrm{T}, 7 \times 3}$ is the transverse energy sum in an area of $7 \times 3$ \glspl{SC} in the second layer of the calorimeter centred on the same locus.
 
\item Hadronic fraction \begin{equation} R_{\mathrm{had}} = \frac{E_{\mathrm{T,had}}}{E_{\mathrm{T,EM}}+E_{\mathrm{T,had}}} \end{equation}
 
where $E_{\mathrm{T,had}}$ is the transverse energy sum in a window of $0.3 \times 0.3$ in the hadronic calorimeter and $E_{\mathrm{T,EM}}$ is the transverse energy sum in the four layers of the electromagnetic calorimeter, computed in a window of $3 \times 2$ ($1 \times 3$) \glspl{SC} in the presampler and third layers, centred as in the case of $R_\eta$.
 
\item Shower width in the first calorimeter layer \begin{equation} w_{\mathrm{s, tot}} = \sqrt{\frac{\sum{E_{\mathrm{T},i} \times (i-i_{\mathrm{max}})}^2}{\sum{E_{\mathrm{T},i}}}},\end{equation}
 
where $i$ runs over five \glspl{SC} in $\eta$, centred on the most energetic \gls{SC} in the first layer of the calorimeter.

\end{itemize}

\begin{figure}[htbp]
\centering
\subfloat[]{
\includegraphics[width=0.2\textwidth]{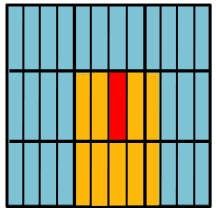}
\label{fig:TDAQL1CaloeFEXTauFine}
}
\subfloat[]{
\includegraphics[width=0.2\textwidth]{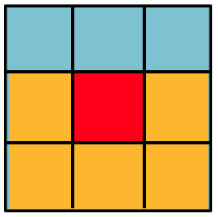}
\label{fig:TDAQL1CaloeFEXTauCoarse}
}
\caption{
\protect\subref{fig:TDAQL1CaloeFEXTauFine} The \gls{L1Calo} \gls{eFEX} tau cluster, as defined in the first and second \gls{EM} calorimeter layers.  The seed \gls{SC} is shown in red, while the other \glspl{SC} making up the cluster are shown in orange.  The surrounding environment, which is not included in the cluster energy computation, is shown in blue.
\protect\subref{fig:TDAQL1CaloeFEXTauCoarse} The \gls{L1Calo} \gls{eFEX} tau cluster, as defined in the presampler and third \gls{EM} calorimeter layers, as well as in the hadronic layer.
}
\label{fig:TDAQL1CaloeFEXTau}
\end{figure}

\textbf{$\tau$ algorithm} The \gls{eFEX} $\tau$ algorithm has been designed to identify prompt hadronic decays of $\tau$-leptons.
First, the cluster seed tower ($\Delta\eta \times \Delta\phi = 0.1 \times 0.1$) is chosen by identifying a local maximum in an environment of $3 \times 3$ towers.
From this seed tower, the seed SuperCell is then chosen by identifying its constituent SuperCell with the highest energy; this search is performed only in the second layer of the \gls{EM} calorimeter.
A cluster is then formed in the direction (up or down in $\phi$) of the highest-energy adjacent SuperCell.  In the fine-granularity layers of the \gls{EM} calorimeter (the first and second layers), the cluster spans a region of $5 \times 2$ \glspl{SC} ($\Delta\eta \times \Delta\phi = 0.125 \times 0.2$).
In the coarse-granularity layers (the presampler and the third layer of the \gls{EM} calorimeter, plus the hadronic calorimeter), the cluster comprises $3 \times 2$ \glspl{SC} ($\Delta\eta \times \Delta\phi = 0.3 \times 0.2$).  The energies in each of these regions are summed to compute the total cluster energy.
The $\tau$ clustering algorithm is illustrated in Figure~\ref{fig:TDAQL1CaloeFEXTau}.

The \gls{eFEX} $\tau$ algorithm can also apply an isolation criterion based on the shower shape in the \gls{EM} calorimeter.  The discriminating variable $F_{\mathrm{core}}$ is used; this variable is defined as the ratio of the transverse energy deposited in a region of $3 \times 2$ \glspl{SC} surrounding the seed ($E_{\mathrm{T}, 3\times2}$) to that deposited in a region of $9 \times 2$ \glspl{SC} surrounding the seed ($E_{\mathrm{T}, 9\times2}$):
 
\begin{equation} F_{\mathrm{core}} = \frac{E_{\mathrm{T}, 3\times2}}{E_{\mathrm{T}, 9\times2}}. \end{equation}

\subsubsection{Jet Feature Extractor (jFEX)}\label{sec:TDAQ_jFEX}
\begin{figure}[htbp]
\centerline{\includegraphics[width=0.75\textwidth]{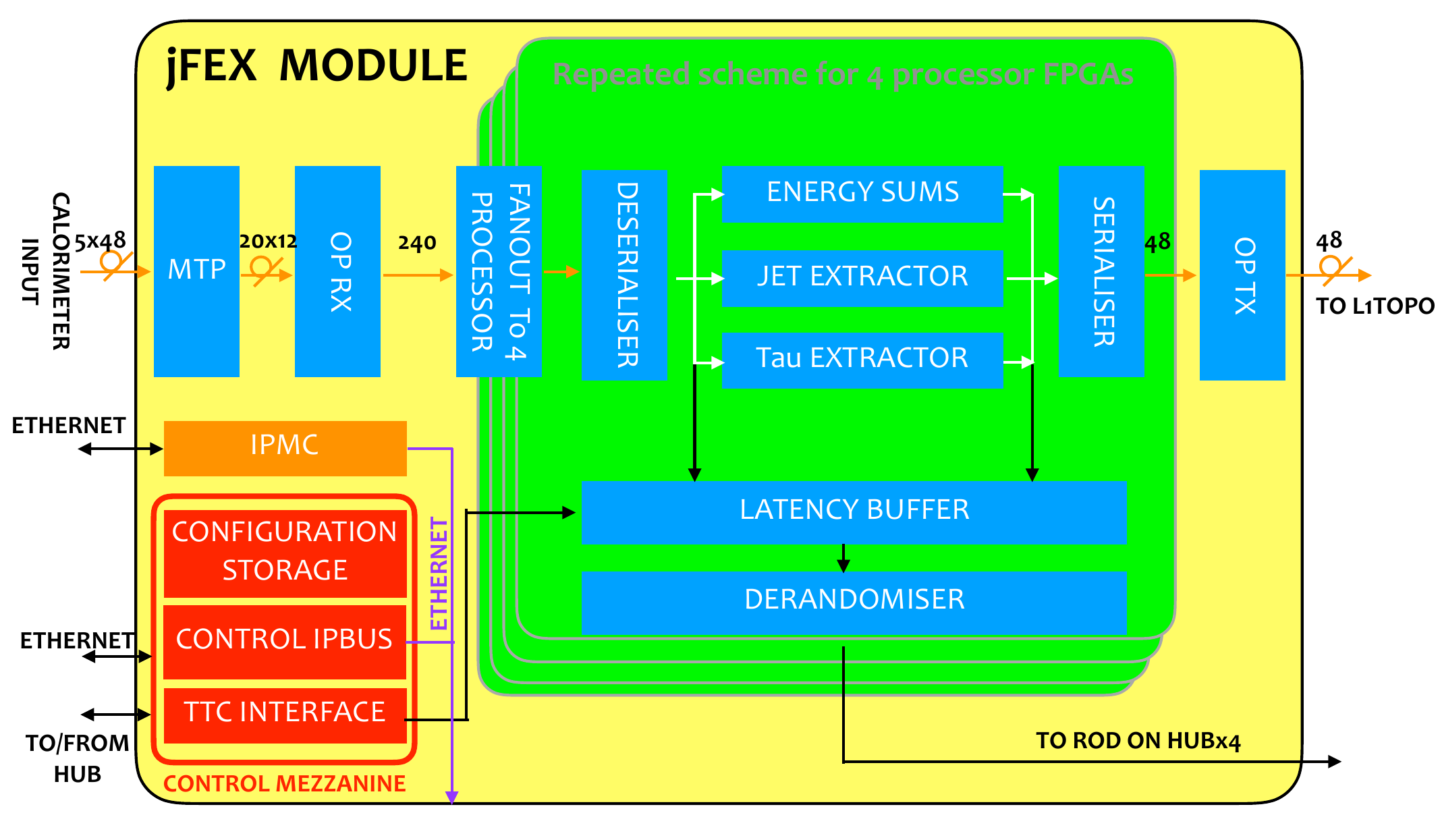}}
\caption{Block diagram of a \gls{jFEX} module, illustrating the real-time and readout paths.  Control and monitoring signals are only shown for the \gls{L1A}.}
\label{fig:TDAQL1CalojFEXBlockDiagram}
\end{figure}

\paragraph{\glstext*{jFEX} module design}
The \gls{jFEX} system identifies jets, hadronically-decaying $\tau$-leptons, \MET, and \sumET in the range $|\eta| \le 4.9$.  The system consists of one \gls{ATCA} shelf, equipped with six \gls{jFEX} modules.  Digitised electromagnetic and hadronic calorimeter data arrive at the \gls{jFEX} via the \gls{FOX} (Section~\ref{sec:TDAQ_L1CaloInfrastructure}) and are received by optoelectronic devices (MiniPODs).  Each module contains 24 MiniPODs (20 receivers and four transmitters) and four Xilinx\textregistered~Ultrascale+\texttrademark~processor \glspl{FPGA}.  Each processor \gls{FPGA} contains 120 \glspl{MGT} and can handle an input bandwidth of up to \SI{3.6}{Tb/\s}.  \glspl{TOB} identified by the \gls{jFEX} are sent to \gls{L1Topo}.  A block diagram of a \gls{jFEX} module is shown in Figure~\ref{fig:TDAQL1CalojFEXBlockDiagram} and a photograph in Figure~\ref{fig:TDAQL1CalojFEXPhoto}.
 
\textbf{Timing and readout}
Timing and \gls{TTC} signals are received on the backplane from the \gls{Hub} (Section~\ref{sec:TDAQ_L1CaloHubROD}). For each \gls{L1A}, a single readout data packet is constructed per event from the processor \gls{FPGA} data and is sent to the \gls{Hub} via high-speed links on the backplane.  Upon receiving a \gls{L1A}, \glspl{TOB} and input data (prescaled to limit the bandwidth usage) are sent to the \gls{ROD} (described in Section~\ref{sec:TDAQ_L1CaloHubROD}), which then sends the data to \gls{FELIX} and the \gls{SW ROD} (Section~\ref{subsubsec:tdaq_daqhlt_felixswrod}).
 
\textbf{Configuration, control, and monitoring}
The processor \glspl{FPGA} are configured by the control \gls{FPGA}.  Control and monitoring of the \glspl{FPGA} is done via IPbus~\cite{bib:IPbus}, which executes such functions as setting parameters for the algorithms, controlling high-speed links, access of playback and spy memories, and environmental monitoring.  Low-level control, compliant with \gls{ATCA} standards, is performed by an \gls{IPMC} module~\cite{bib:CERN-IPMC}.
 
\paragraph{\glstext*{jFEX} algorithms}\mbox{}\\
Each \gls{jFEX} module provides full $\phi = 2\pi$ coverage.  Four \gls{jFEX} modules cover the barrel region, spanning $|\eta| \le 1.6$, and two modules cover the endcap and forward regions corresponding to $1.6 < \abseta < 4.9$.  Each of the four processor \glspl{FPGA} in a given module covers a core of $\Delta\eta\times\Delta\phi = 0.8 \times 1.6$ in the barrel region or $3.3 \times 1.6$ in the endcap and forward regions.  In order to properly identify \glspl{TOB} on the edges of a single \gls{FPGA}, each \gls{FPGA} is supplied with a copy of the data from the surrounding environment, corresponding to a so-called ``overlap'' area of $2.4 \times 3.2$.
 
\begin{figure}[htbp]
\centerline{\includegraphics[width=0.5\textwidth]{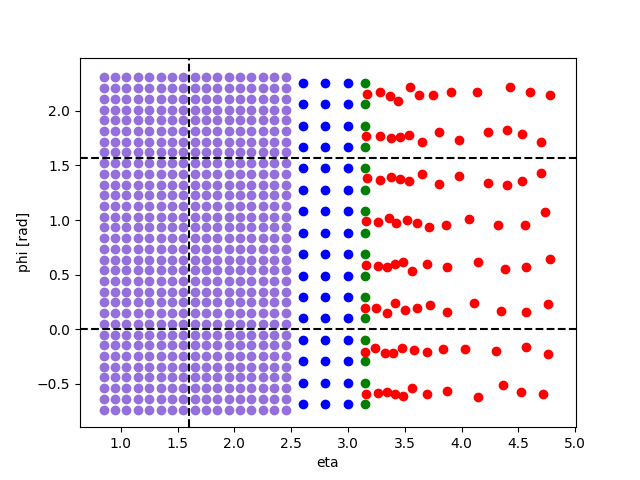}}
\caption{Granularity of trigger tower ($|\eta| < 3.1$) and \gls{SC} ($3.1 < |\eta| < 4.9$) inputs to the \gls{jFEX}.  The different colours serve to highlight the different granularities, as described in the text.  The dashed lines illustrate the boundaries between the core region of a particular \gls{FPGA} ($1.6 < |\eta| < 4.9$, $0 < \phi < \pi/2$) and the surrounding environment.}
\label{fig:TDAQL1CalojFEXGranularity}
\end{figure}

Calorimeter data inputs to the \gls{jFEX} consist of $0.1 \times 0.1$ trigger towers in the region $|\eta| < 2.5$.  The granularity is slightly coarser in the endcap and forward regions: $0.2 \times 0.2$ in the region $2.5 < |\eta| < 3.1$ and $0.1 \times 0.2$ in the region $3.1 < |\eta| < 3.2$.  In the forward region, $3.1 < |\eta| < 4.9$, the inputs consist of \glspl{SC};  in this region, the granularity is irregular, corresponding to the positions of the calorimeter electrodes; this is illustrated in Figure~\ref{fig:TDAQL1CalojFEXGranularity}.
 
\textbf{Pileup subtraction and noise cuts.} The pileup density $\rho$ is computed and subtracted from each trigger tower $\ET$.  Noise cuts may also be applied to individual trigger towers.  Separate noise cuts can be applied for the computation of $\MET$ versus all other objects.
 
\begin{figure}[htbp]
\centering
\subfloat[]{
\includegraphics[width=0.4\textwidth]{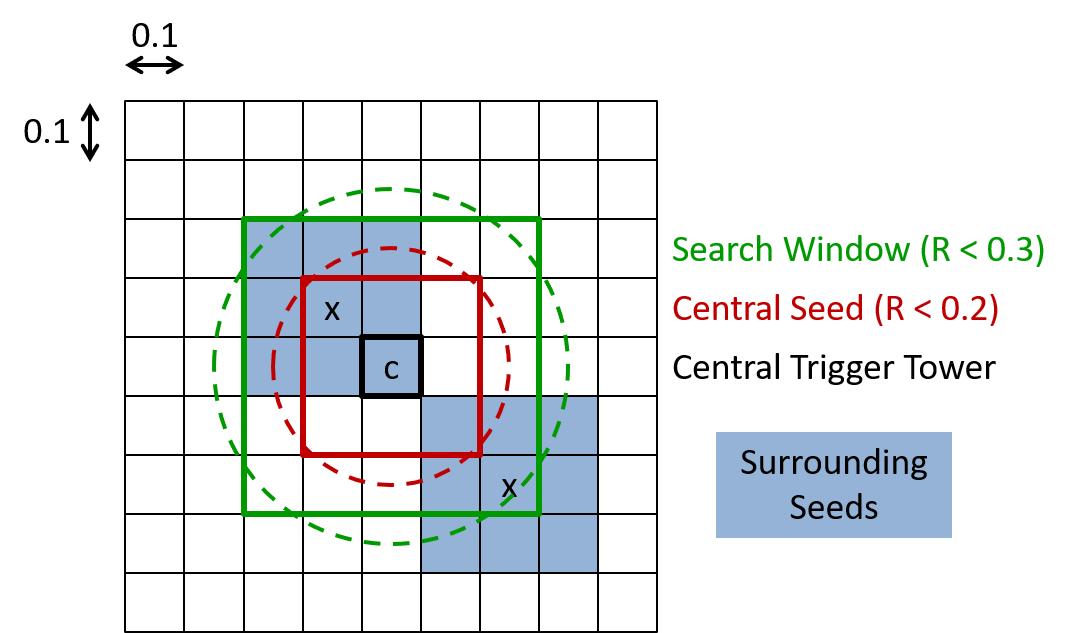}
\label{fig:TDAQL1CalojFEXJetSeeding1}
}
\subfloat[]{
\includegraphics[width=0.2\textwidth]{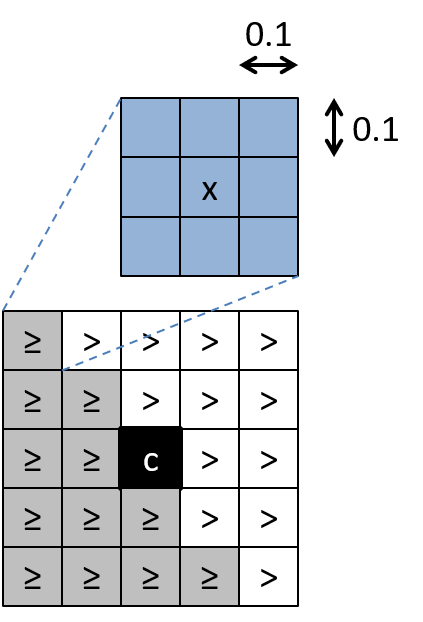}
\label{fig:TDAQL1CalojFEXJetSeeding2}
}
\subfloat[]{
\includegraphics[width=0.35\textwidth]{fig_83c.pdf}
\label{fig:TDAQL1CalojFEXSmallRJetEnergySum}
}
\caption{
\protect\subref{fig:TDAQL1CalojFEXJetSeeding1} The seeding process by which local maxima are identified in the \gls{jFEX}.
\protect\subref{fig:TDAQL1CalojFEXJetSeeding2} Comparative operators used to identify local maxima in a given search window.
\protect\subref{fig:TDAQL1CalojFEXSmallRJetEnergySum} A small-radius jet as defined by the \gls{jFEX}.
}
\label{fig:TDAQL1CalojFEXSmallRJet}
\end{figure}

\textbf{\gls{jFEX} small-radius jet algorithm.} The \gls{jFEX} small-radius jet algorithm is illustrated in Figure~\ref{fig:TDAQL1CalojFEXSmallRJet}.  Small-radius jets are identified using a sliding-window algorithm in a search window of $5 \times 5$ towers, spanning $\eta \times \phi = 0.5 \times 0.5$ (equivalent to $\Delta R < 0.3$).  ``Seeds'' of $3 \times 3$ towers (equivalent to $\Delta R < 0.2$) are constructed around each tower in the search window.
To identify local maxima without double counting and without missing jet objects,  the energy sums of the seeds in the search window are compared using either $\geq$ or $>$ operators.
The central tower of the seed with the largest energy sum in a given search window is then chosen as the centre of the jet.  The energy sum of the seed is added to the sum of the tower energies in a ring of radius $0.2 \leq R < 0.4$ to form a ``round'' jet object consisting of 45 towers. 
 
\textbf{\glstext*{jFEX} large-radius jet algorithm.} The large-radius jet algorithm also uses a sliding-window algorithm.  In order to reduce resource usage, it uses some elements of the small-radius jet algorithm.  The energy sum of a small-radius jet is added to the energy sum of a ring of radius $0.4 \leq R < 0.8$ to form a large-radius jet of $R < 0.8$ consisting of 148 trigger towers.
 
\textbf{\glstext*{jFEX} \MET and \sumET algorithms}
The total transverse energy sum (\sumET) is computed by adding the \gls{EM} and hadronic trigger tower \ET values in slices of constant $\eta$.  Each \sumET \gls{TOB} contains two sums corresponding to different regions in $|\eta|$.  The final energy summation over the full detector area is done by \gls{L1Topo}.
 
\MET is computed in a manner analogous to \sumET: transverse energy values are summed in slices of constant $\phi$ and then weighted by $\mathrm{cos}(\phi)$ and $\mathrm{sin}(\phi)$ in order to determine the $x$- and $y$-components, respectively.
 
\textbf{\glstext*{jFEX} $\tau$ algorithm}
\gls{jFEX} identifies hadronically-decaying $\tau$-leptons in the range $|\eta| \leq 2.5$.  Seeds are identified as for jets, using a search window of size $R < 0.2$.  In order to minimise resource usage, the algorithm reuses jet seeds ($R < 0.2$) as $\tau$ cluster energies.  The isolation is computed using the energy ring of $0.2 \leq R < 0.4$ used by the small-radius jet algorithm.
 
\textbf{\glstext*{jFEX} forward electron algorithm}
\gls{jFEX} identifies electrons in the region $2.3 \leq |\eta| < 4.9$.  This region extends beyond the acceptance of \gls{eFEX} whose coverage is limited to that of the tracking acceptance.   The seeding procedure, which uses only \gls{EM} trigger towers as inputs, is analogous to that used for jets and $\tau$-leptons.  The search window varies in size depending on the calorimeter granularity: $R < 0.2$ in the region $2.3 \leq |\eta| < 2.5$ and $R < 0.3$ in the region $2.5 \leq |\eta| < 4.9$.  The cluster energy is defined as the energy sum of the electron seed (a single \gls{EM} trigger tower) and the most energetic neighbouring tower.  The \gls{EM} isolation is defined as the transverse energy sum of all \gls{EM} towers within $R < 0.4$, excluding the electron cluster itself.
 
\subsubsection{Global Feature Extractor (gFEX)}\label{sec:TDAQ_gFEX}
\paragraph{\glstext*{gFEX} module design}\mbox{}\\
The \gls{gFEX} has been designed such that the data from the entire calorimeter can be processed on a single module, thus permitting the use of full-scan algorithms.  This functionality is intended to facilitate identification of boosted objects and global observables such as \MET, which are of particular interest in a number of searches for new physics. Jet substructure algorithms may be employed to discriminate between signals from boosted boson or top quark decay products and the \gls{QCD} multi-jet background.  The \gls{gFEX} also incorporates pileup subtraction capabilities, which provide robustness against pileup for \MET triggers.  \MET trigger rates are particularly sensitive to pileup and are of great importance for searches for \gls{SUSY}, dark matter, and Higgs bosons decaying to invisible final states.
 
\begin{figure}[htbp]
\centerline{\includegraphics[width=0.85\textwidth]{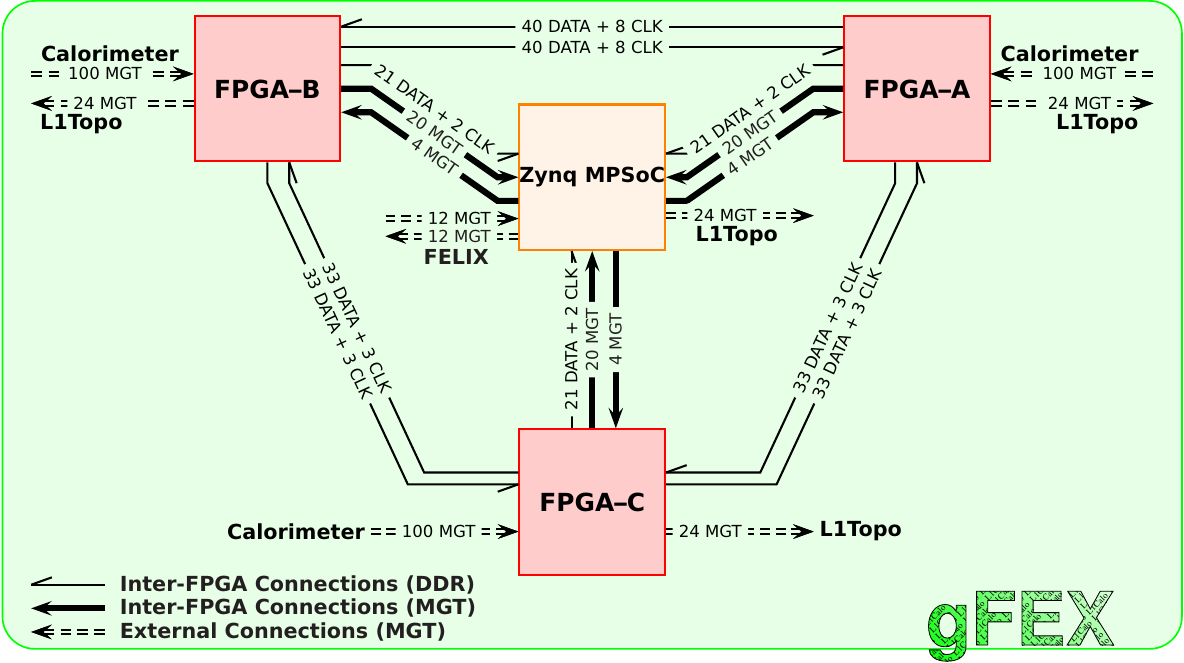}}
\caption{Block diagram of the \gls{L1Calo} global Feature EXtractor (gFEX).}
\label{fig:TDAQL1CalogFEX}
\end{figure}

The \gls{gFEX} module, illustrated as a block diagram in Figure~\ref{fig:TDAQL1CalogFEX} and with a photograph in Figure~\ref{fig:TDAQL1CalogFEXPhoto}, includes three processor \glspl{FPGA} and a Zynq+ \gls{SOC} which performs control and monitoring functions.  Each processor \gls{FPGA} covers the entire $\phi$ ring; two cover the barrel region $|\eta| < 2.5$ and the remaining \gls{FPGA} covers the endcap and forward regions.  Data arrive at the \gls{gFEX} from the \gls{LDPS} (LAr -- Section~\ref{sec:LArDigitalTrigger}) and \gls{TREX} (Tile) via the \gls{FOX} (Section~\ref{sec:TDAQ_L1CaloInfrastructure}) at a transmission speed of \SI{11.2}{Gb/\s}.
 
\textbf{Timing and readout}
\gls{TTC} information is transmitted to and from the Zynq+ \gls{FPGA} via \gls{FELIX} (described in Section~\ref{subsubsec:tdaq_daqhlt_felixswrod}).  Upon receipt of a \gls{L1A}, readout data and \glspl{TOB} are sent to \gls{FELIX}.  From the \gls{FELIX}, readout data are sent to the \gls{SW ROD}, which builds event fragments to be sent to the \gls{HLT} via the \gls{DAQ} system.
 
\textbf{Configuration, control, and monitoring}
The IPbus protocol~\cite{bib:IPbus} is used to control access to playback and spy memories on the board for diagnostic and commissioning purposes.  Monitoring and control of the module are performed via \gls{IPMC}~\cite{bib:CERN-IPMC}, as well as by \gls{DCS} (see Section~\ref{sec:DCS}).
 
\begin{figure}[htbp]
\centerline{\includegraphics[width=0.75\textwidth]{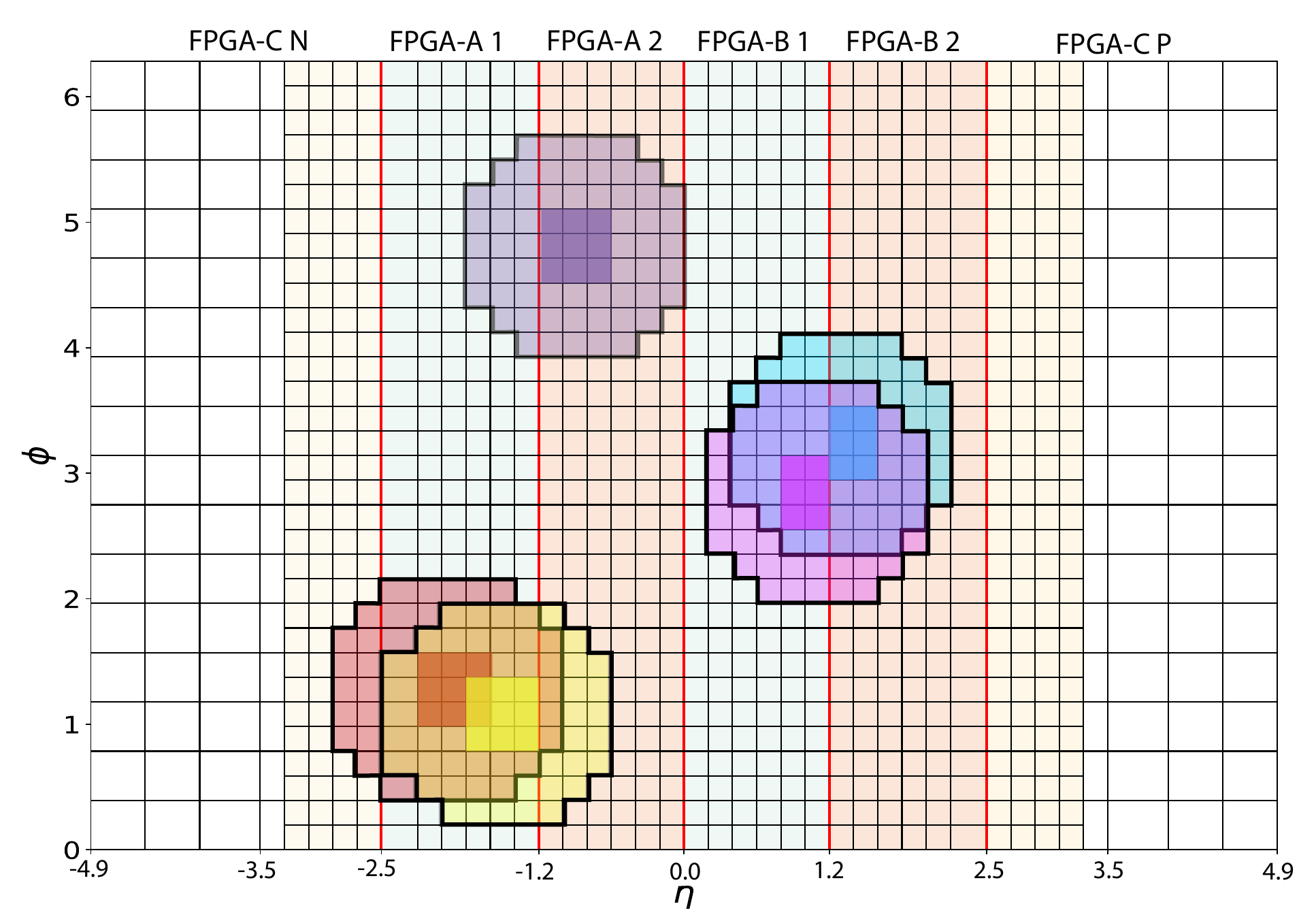}}
\caption{Schematic of the \gls{gFEX} processor \gls{FPGA} coverage, illustrating the granularity of the gTower inputs.  $3 \times 3$ ($2 \times 3$ on \gls{FPGA} region boundaries)  groups of gTowers make up gBlocks.  Note that both gBlocks and large-radius jets  are permitted to overlap.
All possible  large-radius ($R < 0.9$) jets are constructed and those passing an optional seed requirement are sorted by energy and transmitted separately to \gls{L1Topo} in two
$\eta$ regions per \gls{FPGA}.
}
\label{fig:TDAQL1CalogFEXgTowers}
\end{figure}

\paragraph{\glstext*{gFEX} algorithms}\mbox{}\\
Inputs to the \gls{gFEX} are combined into ``gTowers'' by summing  transverse energy from the electromagnetic and hadronic calorimeters and
subsequently applying a calibration with a lookup table.
The granularity in the barrel region is $0.2 \times 0.2$, increasing in the endcaps and forward region as illustrated in Figure~\ref{fig:TDAQL1CalogFEXgTowers}.  Contiguous groups of $3 \times 3$ gTowers ($2 \times 3$ gTowers on the boundaries between \glspl{FPGA} and \gls{FPGA} divisions) are called ``gBlocks''. gBlocks have configurable thresholds and can be used as seeds for large-radius jet identification as described below.

Two types of \gls{TOB} are identified by the \gls{gFEX}.  Jet \glspl{TOB} are produced on a single processor \gls{FPGA} and include large-radius jets, gBlocks, the local pileup energy density ($\rho$) and optionally substructure information.
Global \glspl{TOB} are produced on the Zynq \gls{FPGA} using information from all of the processor \glspl{FPGA} and can include such global observables as  \MET, \sumET, and \HT.  All \glspl{TOB} computed by the \gls{gFEX} are sent to \gls{L1Topo}.
 
\textbf{\gls{gFEX} pileup suppression algorithm.}  Pileup suppression is performed on an event-by-event basis in the firmware. The energy density $\rho$ is computed per processor \gls{FPGA} on an event-by-event basis, and then multiplied by a factor of 69 to take into account the 69 gTowers that make up a large-radius jet as defined below. This value is then subtracted from the \ET of the large-radius jet.
 
\textbf{\gls{gFEX} large-radius jet algorithm.}  Large-radius jets (``gJets'') are formed using a cone algorithm by summing the gTower \ET in a ``circular'' $1.8 \times 1.8$ area ($R < 0.9$) around each gTower.
A total of 69 gTowers are used for each large-radius jet as shown in Figure~\ref{fig:TDAQL1CalogFEXgTowers}.
Since gJets may span the area covered by two different processor \glspl{FPGA}, information is shared between processor \glspl{FPGA} in the form of partial energy sums. gJets are permitted to overlap.
Large-radius jets  and gBlocks \glspl{TOB} are separately sorted on both the right and left $\eta$ divisions of the A and B  processor  \glspl{FPGA} and sent to \gls{L1Topo} on separate fibres.
Initially, only the most energetic gJet in each division will be transmitted.
If latency allows, a sub-leading non-overlapping gJet will be transmitted as well.
A non-zero seed threshold can be applied to the central gBlock of the large-radius gJets before sorting.  For the gBlocks, the sub-leading gBlock is required to have its centre outside of the area of the leading gBlock.
 
\textbf{\gls{gFEX} ``jets-without-jets'' \MET algorithm.}  The ``jets without jets'' (JwoJ) algorithm~\cite{JetsWithoutJets} is based on the concept that jet observables may be transformed into global event-shape observables.
The JwoJ algorithm is motivated by the fact that
diffuse transverse energy  is likely to be associated with pileup interactions whereas clustered transverse energy is more likely to be associated with the interaction of interest.
The \gls{gFEX} computes \MET by separating the \ET sums into ``hard'' (MHT) and ``soft'' (MST) terms, where the hard term consists of the \ET sum of towers with the associated gBlock satisfying $\ET > 25$~GeV and the soft term consists of the \ET sum of the remaining towers.  The \MET is then computed as a linear combination of the hard and soft terms:
 
\begin{equation}E_{\mathrm{T} x,y}^{\mathrm{miss}} = a_{x,y}{\mathrm{MHT}}_{x,y} + b_{x,y}{\mathrm{MST}}_{x,y},\end{equation}
 
where the $a$ and $b$ coefficients can be optimised for resolution and overall performance.

\subsubsection{L1Calo infrastructure}\label{sec:TDAQ_L1CaloInfrastructure}
 
\paragraph{TREX} 
In \RunThr, the output signals from the Tile calorimeter will still be in \analog format (Section~\ref{sec:tile}) but inputs to the \glspl{FEX} must be in digital format.  The digitisation is performed in the \glspl{PPM} of the legacy \gls{L1Calo} system~\cite{TDAQ-2019-01}, and new \gls{TREX} modules, located on the \glspl{PPM}, process and duplicate the digitised Tile signals and send them to the Phase-1 \gls{L1Calo} system via optical links at 11.2 Gb/s.  For backwards compatibility with the legacy system, the Tile signals are sent electrically to the legacy \gls{CP} and \gls{JEP}.  The readout data are sent to both the \gls{FELIX} and the legacy \gls{ROD} via optical links (9.6 Gb/s and 960 Mb/s, respectively).  \gls{TTC} info is received from the \glspl{PPM} via a \gls{VME}-P2 connector and from the \gls{FELIX} via an optical link at 4.8 Gb/s.
 
\begin{figure}[htbp]
\centering
\subfloat[]{
\includegraphics[width=0.65\textwidth]{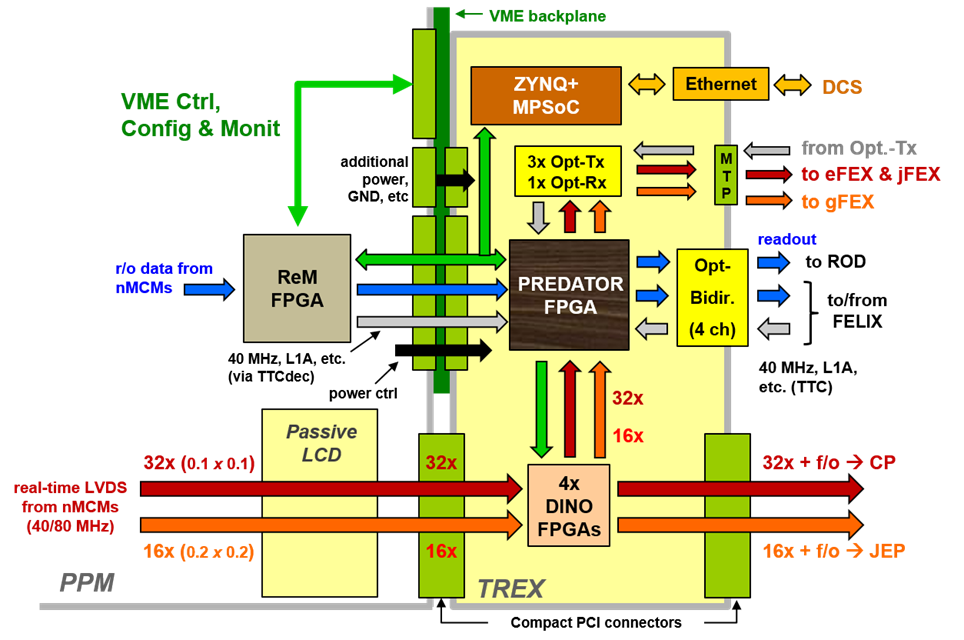}
\label{fig:TDAQL1CaloTREXDiagram}
}
\subfloat[]{
\includegraphics[width= 0.3\textwidth]{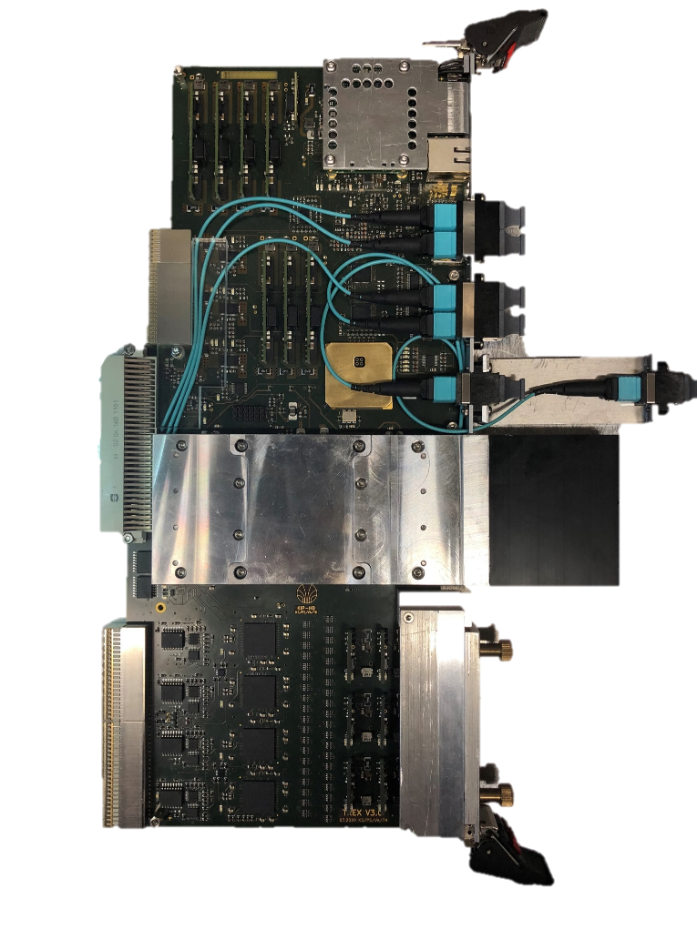}
\label{fig:TDAQL1CaloTREXPhoto}
}
\caption{
\protect\subref{fig:TDAQL1CaloTREXDiagram} Schematic diagram of the \gls{TREX}.
\protect\subref{fig:TDAQL1CaloTREXPhoto} A photo of a \gls{TREX} module.
}
\label{fig:TDAQL1CaloTREX}
\end{figure}

The \gls{TREX} module, pictured in Figure~\ref{fig:TDAQL1CaloTREX}, consists of a pre-processor data collector (PREDATOR) Xilinx\textregistered~Ultrascale\texttrademark~\gls{FPGA}, and four \gls{LVDS} data in-out (DINO) \glspl{FPGA} (Xilinx\textregistered~Artix\textregistered~-7).  It includes six Samtec FireFly\texttrademark~optical transceivers: four 12-channel transmitters to the \glspl{FEX} and two four-channel Duplex transceivers to handle \gls{TTC} and readout signals.  Monitoring of environmental conditions is performed using a Zynq\textregistered~Ultrascale+\texttrademark~MPSoC (multi-processor system-on-chip) device, and the data are sent to \gls{DCS} via Gigabit Ethernet interface.  Configuration, control, and monitoring of the module are performed via the \glspl{PPM} using interfaces to \gls{VME}.
 
\paragraph{FOX} 
The digitised inputs to \gls{L1Calo} from the \gls{LAr}  calorimeter and the \gls{TREX}  arrive via the \gls{FOX} optical plant.  The \gls{FOX}  separates the fibre bundles (approximately 7500 fibres in total) and re-routes them according to the necessary mapping required by the \gls{L1Calo} FEXes.  It also routes the input signals as required for data sharing between modules covering contiguous areas.  Duplicate output connectors are provided in order to allow for spares for the \gls{gFEX} board.
 
The \gls{FOX}  is internally divided into LArFOX and TileFOX components, which route the fibres from the \gls{LAr}  calorimeter and \gls{TREX}, respectively.  The LArFOX consists of four ``\gls{FOX}  boxes'', which route the \gls{EM} and hadronic inputs to the \gls{eFEX}, while the TileFOX consists of two \gls{FOX}  boxes, which route the \gls{EM} and hadronic inputs to the \gls{jFEX} and \gls{gFEX}.  Fibres may also be routed from one \gls{FOX}  box to another before finally being sent to a given \gls{FEX}.  A \gls{FOX}  box consists of a 19-inch rack-mount ``2U'' chassis containing internal shuffle modules specified by their input and output \gls{MTP} connectors and the input-to-output fibre mapping.
 
\paragraph{TopoFOX} 
Inputs to \gls{L1Topo} from the \gls{L1Calo} FEXes and \gls{MuCTPI} are handled by the TopoFOX, which performs fibre mapping and routing to the three \gls{L1Topo} processor modules in a manner analogous to the \gls{FOX}.  Each of the four processor \glspl{FPGA} in a given \gls{FEX} module sends a 12-fibre ribbon to the 48-fibre output connector of that module, while each of the side-A and side-C \glspl{FPGA} of the \gls{MuCTPI} send two 12-fibre ribbons to the 48-fibre muon output bundle.  The TopoFOX receives and sorts these fibre outputs, combining them into twelve 72-fibre bundles, two of which are sent to each of the six \glspl{FPGA} making up the three \gls{L1Topo} modules.
 
\paragraph{Hub \& ROD}\label{sec:TDAQ_L1CaloHubROD}
Common communications functionalities for the \gls{eFEX}, \gls{jFEX}, and \gls{L1Topo} \gls{ATCA} shelves are provided by the \gls{Hub} modules.  The \gls{Hub} supports the system readout via a \gls{ROD} daughter card
contained on a mezzanine on the module, provides switching functionality for module control and \gls{DCS}, and distributes timing and control signals to the modules.  Each \gls{ATCA} shelf contains two \gls{Hub} modules.  In total, the \gls{L1Calo} system contains seven \gls{Hub} modules: four for the \gls{eFEX} system, two for the \gls{jFEX} system, and one for \gls{L1Topo}.
 
In a given shelf, the \gls{Hub} module located in logical slot 1 provides switching capability for module control signals; it also receives the \gls{LHC} clock and \gls{TTC} information from the \gls{FELIX} via a 12-channel MiniPOD optical receiver.  This information is then fanned out to the \gls{ROD} daughter card contained on the Hub, the \gls{FEX} modules and the second \gls{Hub} in the shelf.  This second \gls{Hub} module, located in logical slot 2, provides switching for the \gls{DCS} network.  Readout control data from its \gls{ROD} is sent to the \gls{Hub} in slot 1 to be included in the combined readout data stream.  At the \gls{HL-LHC}, \gls{FEX} data will be multiplexed between the two \glspl{ROD} in the shelf in order to cope with the higher data rate.
 
High-level control of the \gls{Hub} is performed via an IPbus interface~\cite{bib:IPbus}.  The \gls{Hub} also connects to the \gls{IPMB}~\cite{bib:IPMI}) via an \gls{IPMC} card~\cite{bib:CERN-IPMC} located on the module.  An I$^2$C bus is used to manage communications on the \gls{Hub} itself.
 
Readout data from the \gls{eFEX}, \gls{jFEX}, and \gls{L1Topo} modules are sent to the new \gls{FELIX} and \gls{SW ROD} readout system (described in Section~\ref{subsubsec:tdaq_daqhlt_felixswrod}) by the \gls{ROD} daughter card mounted on the \gls{Hub} module.  The \gls{gFEX} sends its data directly to \gls{FELIX}.  The \gls{ROD} receives the readout data from all \gls{FEX} modules in a given shelf. The data are then decoded, the checksum evaluated, and finally merged into a single packet and buffered before being transmitted to \gls{FELIX}.
 
\begin{figure}[htbp]
\centerline{\includegraphics[width=0.45\textwidth]{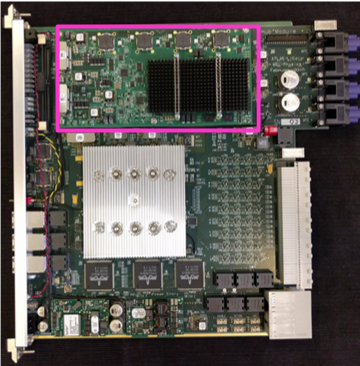}}
\caption{A \gls{Hub} module, with the \gls{ROD} mezzanine mounted (at the top of the module, outlined in pink).}
\label{fig:TDAQL1CaloHubROD}
\end{figure}

The \gls{ROD} includes a Xilinx\textregistered~Virtex\textregistered~-7 \gls{FPGA} and four MiniPOD optical transceivers, which transmit the readout data to \gls{FELIX}.  A photo of a \gls{Hub} module, with the \gls{ROD} mounted, is shown in ~\ref{fig:TDAQL1CaloHubROD}.
 
\subsection{L1 Muon Trigger}\label{sec:TDAQ_L1Muon}
 
In \RunOne, the \glsfirst{L1Muon} trigger decision in the endcap region ($1.05 < |\eta| < 2.4$) was based on the coincidence of hits in the \gls{TGC} stations of the endcap middle layer (\gls{EM-TGC}), called the Big Wheels.  The \gls{EM-TGC} have three stations (TGC-M1, TGC-M2, and TGC-M3) per side, with the M1 station consisting of three layers and the outer two stations (M2 and M3) each consisting of two layers, for a total of seven layers.
 
\begin{figure}[htbp!]
\centerline{\includegraphics[width=0.95\textwidth]{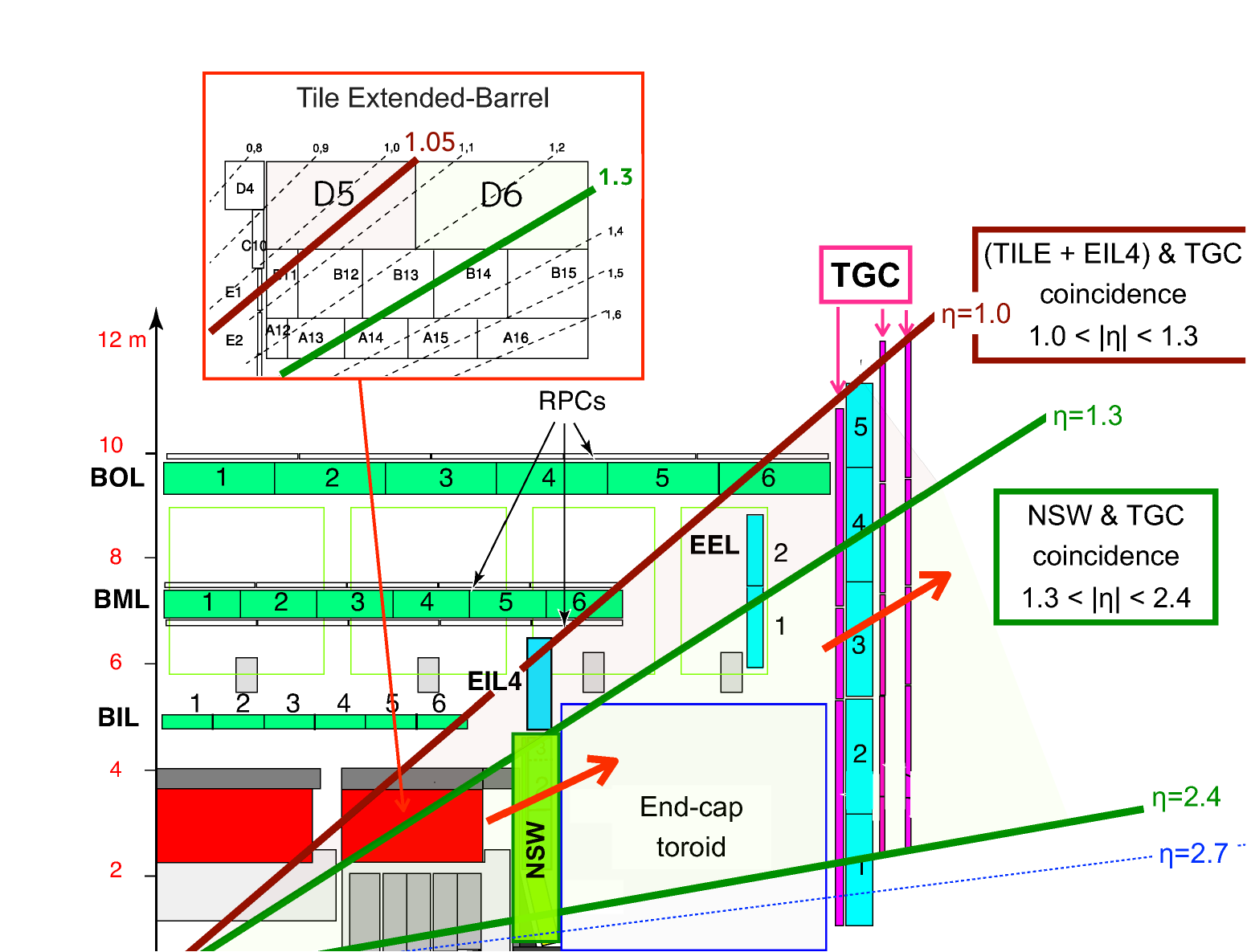}}
\caption{Schematic overview of the upgraded \gls{L1Muon} endcap system, illustrating the \gls{NSW} and Tile-Muon coincidence.}
\label{fig:TDAQL1MuonEndcapOverview}
\end{figure}

To improve the rejection of fake muons in the full $\eta$ range of the \gls{EM-TGC}, several upgrades have been performed, with the aim of reducing the \gls{L1} trigger rate while keeping the efficiency high~\cite{TRIG-2018-01}.  During \RunTwo, an additional requirement of a coincidence between the \gls{EM-TGC} and the \gls{D-layer} cells of the Tile calorimeter (see Figure~\ref{fig:TDAQL1MuonEndcapOverview}) in the range $1.05 < |\eta| < 1.3$, was introduced.
 
Further rate reduction (see Figure~\ref{fig:MuonTriggerPerformance} for the improvement in \gls{L1} Muon endcap trigger rate) in the range $1.3 < |\eta| < 2.4$ is achieved in \RunThr by replacing the existing muon endcap inner stations (the Small Wheels) by the \gls{NSW} described in Section~\ref{MuonSS:NSW}, comprising \gls{sTGC} and \gls{MM} detectors with high-rate tolerance and improved resolution. The upgrades to the \gls{L1Muon} endcap trigger are illustrated in Figure~\ref{fig:TDAQL1MuonEndcapOverview}. At \lumirunthree, the total \gls{L1} trigger rate for single muons with $\pT > \SI{20}{\GeV}$ was expected to be \SI{18}{\kHz} before these improvements; with these improvements, the expected rate is \SI{13}{\kHz}. The additional fake muon rejection provided by the upgrade will be even more important during \RunFour\ and beyond. 
The expected performance of the combined upgrades to the \gls{L1Muon} system is also shown in Figure~\ref{fig:TDAQL1MuonEndcapPerformance}.

In the \gls{L1Muon} trigger, muon candidates are identified by measuring the degree to which their paths deviate (in both $R$ and $\phi$) from the pattern of hits expected from a muon with infinite momentum; this deviation is inversely proportional to the \pT of the muon.
Using the deviations expected for different muon \pT, so-called ``muon roads'' can be defined for different \pT thresholds.
The \RunTwo trigger electronics permitted the use of six programmable \pT thresholds; in \RunThr,  the upgraded sector logic (see Section~\ref{sec:EndcapSL}) will allow 15 programmable thresholds in the endcap.
 
\subsubsection{TGC EI-FI Coincidence}
The main background source in the \gls{L1Muon} endcap trigger is low-momentum charged particles produced in the endcap toroid magnets and beam shielding.
In order to reject backgrounds due to these particles, a coincidence requirement between the Big Wheel and the \gls{TGC-FI} chambers of the old \gls{EI} wheels was introduced in 2015, in part as a proof of principle for the \gls{NSW}, which ultimately replaced the \gls{EI} wheels.
An additional coincidence requirement between the Big Wheel and the \gls{EIL4} \gls{TGC} chambers was introduced in 2016, and remains in effect for \RunThr and beyond. \RunTwo papers refer to the combination of these two vetoes as the ``TGC EI-FI coincidence''; however, for \RunThr only the ``TGC EIL4 coincidence'' remains.
 
\subsubsection{Tile-Muon Coincidence}
 
\begin{figure}[htbp!]
\centering
\subfloat[]{
\includegraphics[width=0.48\textwidth]{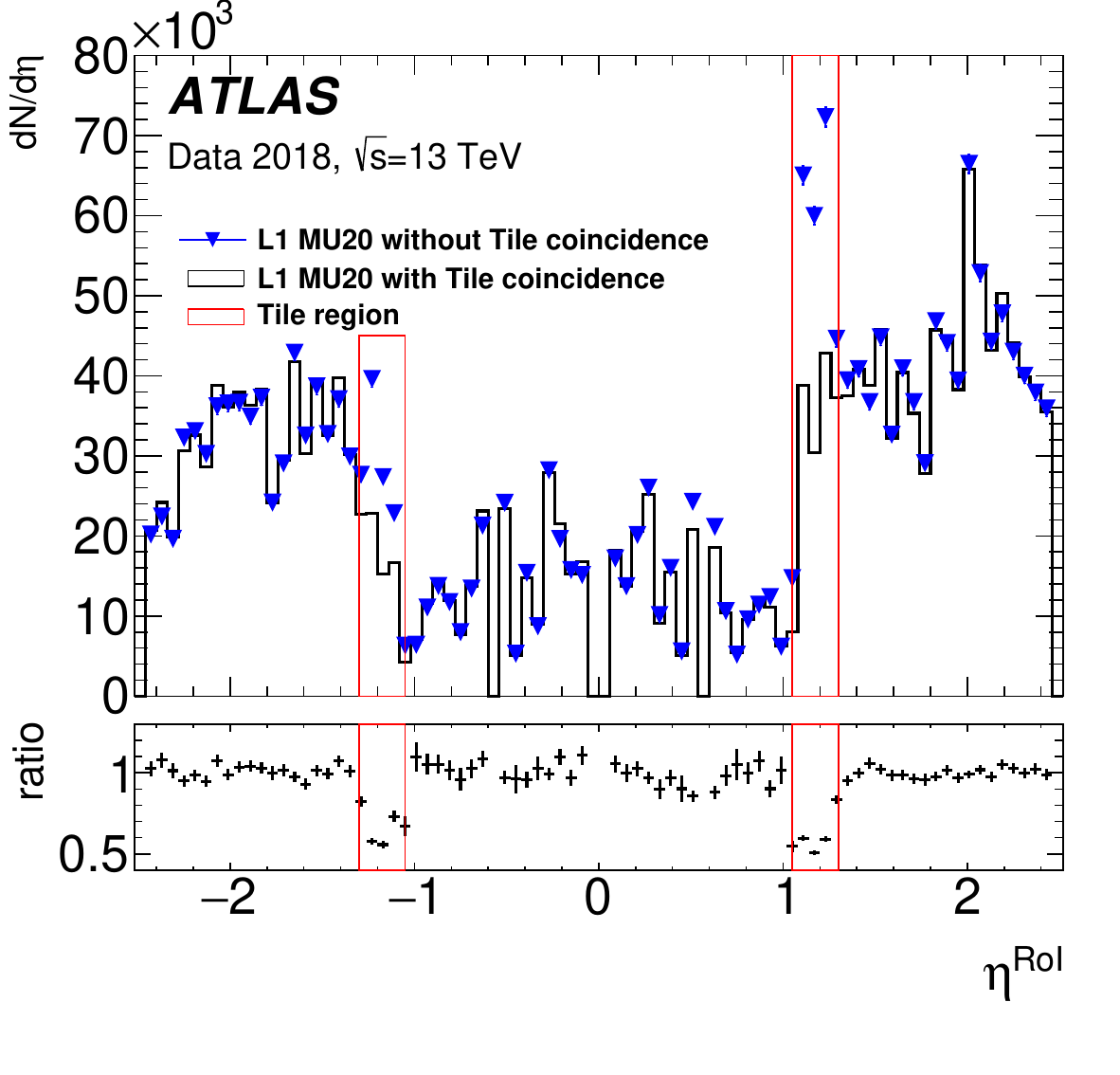}
\label{fig:TDAQL1MuonTileMuonRoIVsEta}
}
\subfloat[]{
\includegraphics[width= 0.48\textwidth]{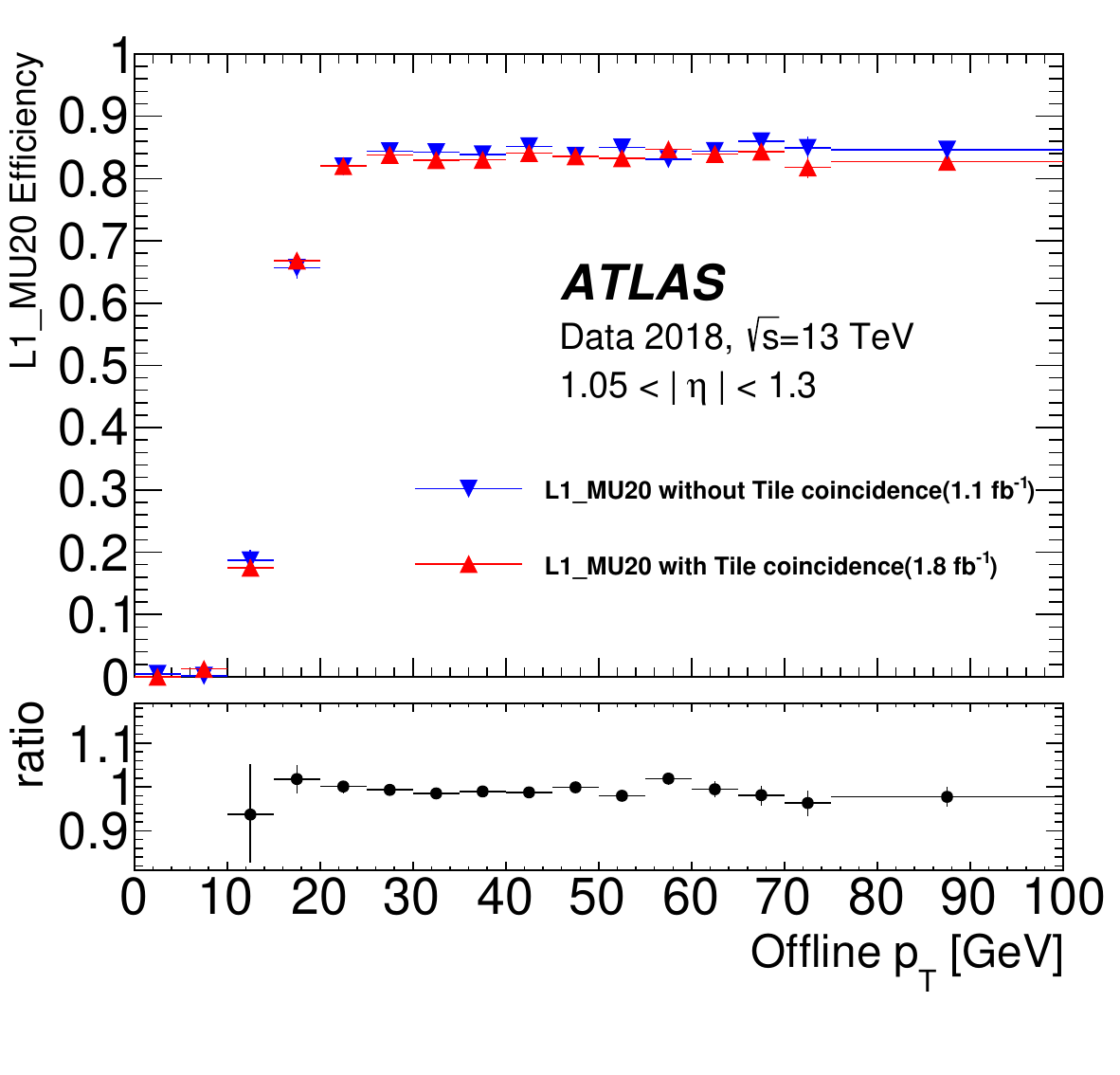}
\label{fig:TDAQL1MuonTileMuonEfficiency}
}
\caption{
\protect\subref{fig:TDAQL1MuonTileMuonRoIVsEta} The pseudorapidity distribution of \RunTwo \gls{L1Muon} \glspl{RoI} with $\pT > \SI{20}{\GeV}$ before and after the deployment of the Tile calorimeter coincidence requirement.  The reduction in the range $1.05 < |\eta_{\mathrm{RoI}}| < 1.3$, where the coincidence requirement is applied, is highlighted.
\protect\subref{fig:TDAQL1MuonTileMuonEfficiency} The efficiency of the \RunTwo \gls{L1} trigger selection for muon \glspl{RoI} with $\pT > \SI{20}{\GeV}$ in this pseudorapidity range.
}
\label{fig:TDAQL1MuonTileMuonPerformance}
\end{figure}
 
In the region $1.05 < |\eta| < 1.3$, where the inner layer of the muon system provides incomplete coverage due to the presence of the toroid magnets, a coincidence between the \gls{EM-TGC} chambers and the Tile calorimeter D-cell layers is required in order to improve the rejection of fake muons, which consist primarily of low-\pt protons.  The distribution of this background can be seen in Figure~\ref{fig:TDAQL1MuonTileMuonRoIVsEta}.
 
As shown in Figure~\ref{fig:TDAQL1MuonEndcapOverview}, a high-\pt muon originating from the \gls{IP} and passing through the endcap in the region $1.0 < |\eta| < 1.3$ can be expected to traverse the D5 or D6 cells of the Tile calorimeter extended barrel.  These modules have a granularity of $\eta \times \phi = 0.2 \times 0.1$, thus providing finely segmented energy measurements.  Muon candidates with $\pt > 20$~GeV are required to coincide with an energy deposit satisfying a pre-determined threshold in at least one of the corresponding Tile modules mapped in $\phi$ to the associated muon trigger sector.
 
New \glspl{TMDB} were installed in \RunTwo to perform matching in $\phi$ between the Tile extended barrel modules and the \gls{L1Muon} endcap sector logic.  Each extended barrel consists of 64 modules in $\phi$ (128 in total), while the \gls{L1Muon} endcap region consists of 48 trigger sectors.  Each \gls{TMDB} receives the D5 and D6 inputs from eight Tile modules and three \gls{L1Muon} endcap sector logic boards.  Hence, 16 \glspl{TMDB} in total are required.  The \glspl{TMDB} are 9U \gls{VME} modules situated in two crates in \gls{USA15}.  They perform the following tasks:
 
\begin{itemize}
\item Receive and digitise the \analog signals from the Tile D5 and D6 cells.
\item Provide calibration for the Tile signals.
\item Provide signal detection for each Tile cell.
\item Provide \gls{BCID} information using timing information from the \gls{TTC} receiver \cite{bib:ttc}.
\item Provide $\eta$, $\phi$, and \gls{BCID} information from cells in which a signal has been detected to the corresponding three muon sector logic boards.
\item Share information with neighbouring receiver boards to accommodate for non-perfect matching in $\phi$ between the eight Tile modules and three muon sector logic boards.
\item Provide readout data fragments to the \gls{DAQ} system.
\end{itemize}

For single muons with $\pt > \SI{20}{\GeV}$, the trigger rate was reduced by 6\% after this coincidence requirement was introduced, with only a 2.5\% efficiency loss in the range $1.05 < |\eta| < 1.3$, as shown in Figure~\ref{fig:TDAQL1MuonTileMuonPerformance}.
 
\subsubsection{NSW Trigger\label{sec:NSWTP}}

\begin{figure}[ht]
\centering
\includegraphics[width=0.7\textwidth]{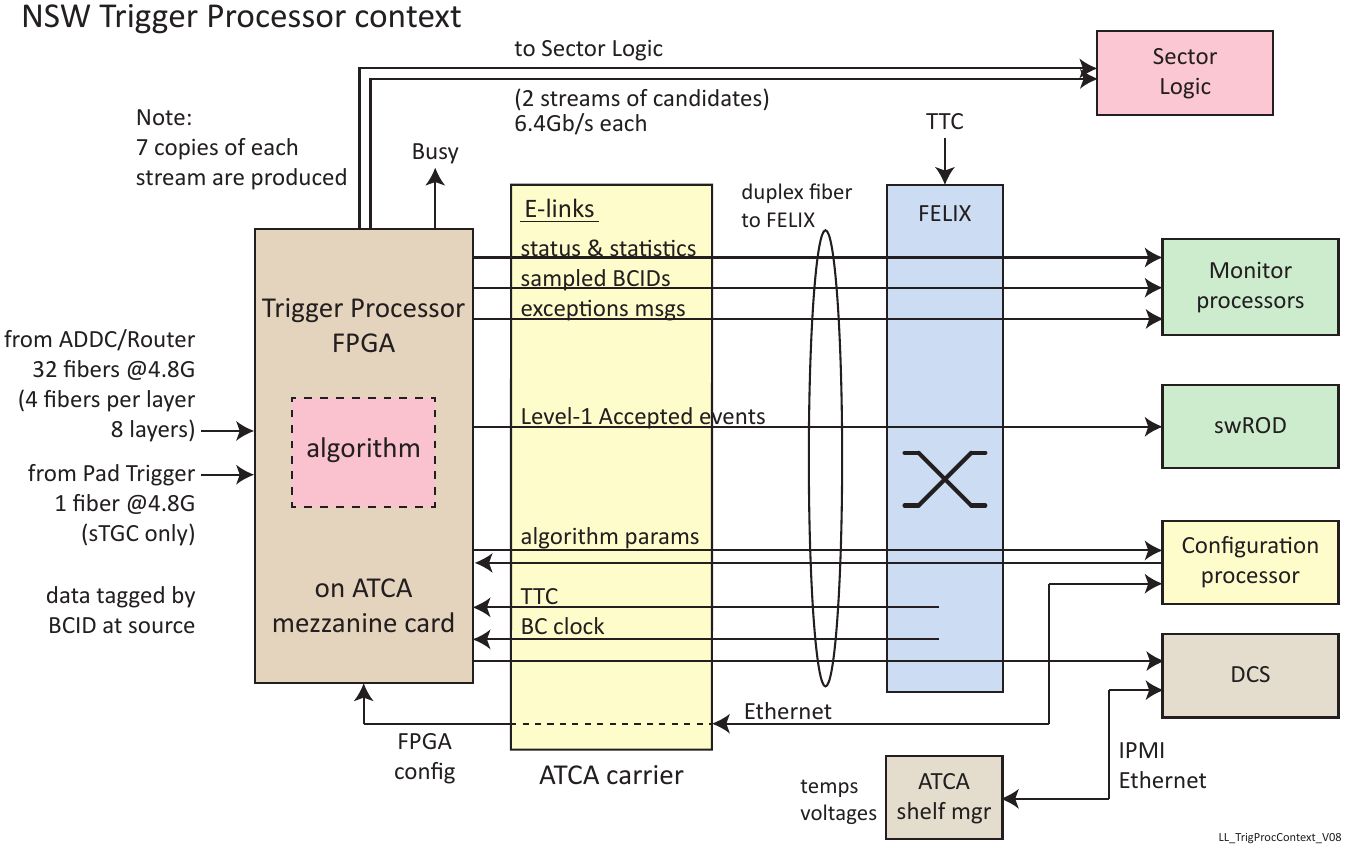}
\caption{Functional diagram of the \gls{NSW-TP}.}
\label{fig:TdaqNswTp}
\end{figure}

 
Trigger inputs from both the \gls{sTGC} and the \gls{MM} chambers in a sector are processed by algorithms running on the \gls{NSW-TP} cards (Figure~\ref{fig:TdaqNswTp}).
 
\paragraph{NSW Trigger Processor requirements}
The \gls{NSW-TP} provides track segments from the \gls{NSW} detectors (as described in Section\,\ref{muonSS:trigger}) to be matched in the \gls{SL} with coincidences found in the \gls{EM-TGC}.
Each track segment is characterised by its radial position $\eta$, its angle $\Delta\theta$ with respect to an ideal infinite-momentum track (a line from the \gls{IP} to the segment's radial position in the \gls{NSW}), measured by the precision strips of the \gls{sTGC} or \gls{MM} detectors, and by its azimuthal position, $\phi$, measured by the \gls{sTGC} pad towers (see also Section~\ref{muonSS:stgcPadTrigger}) or by the output of \gls{MM} ``diamond'' fitter (explained in Section~\ref{tdaq:MMtrigger}).
The required $\eta$-resolution of both the \gls{sTGC} and \gls{MM} strip triggers is \num{0.005}.
The \gls{sTGC} pad towers have an $r\phi$-resolution of \SIrange{7}{38}{\mm}, increasing with radius (based on the size of the logical pads), and the \gls{MM} trigger diamonds provide an $r\phi$-resolution of better than \SI{11}{\mm} at all radii.
The angular resolution of the \gls{NSW} segments required for Phase-II is \SI{1}{\milli\radian} in order to match the expected angular resolution of the \gls{EM-TGC}. 
For \RunThr,
$\Delta\theta$ is passed to the \gls{SL} but not used in the Phase-I trigger decision.
 
\paragraph{NSW-TP mezzanine cards and input interfaces}
The \gls{NSW-TP} is housed in two \gls{ATCA}~\cite{bib:ATCA-RCE} 
crates in \gls{USA15}.
The \gls{NSW-TP} processes information from the \gls{MM} and \gls{sTGC} layers in the sector in separate algorithms (one for the eight \gls{MM} layers, and one for the eight \gls{sTGC} layers) in separate Virtex-7 XC7V690 \glspl{FPGA} 
on a mezzanine card.
Each blade contains two mezzanine cards, to serve one \gls{NSW} octant (one Large sector and one adjacent Small sector); a crate of eight blades serves a full wheel.
 
One \gls{FPGA} on each mezzanine card takes \num{32} \gls{MM} fibre inputs (from the sixteen \glspl{ADDC} per sector).
Each \gls{ADDC} GigaBit transceiver packet can contain \gls{ART} data from up to eight triggered \glspl{VMM}.
When a packet is received, the \gls{ART} data are decoded, for each strip, into a strip number and the slope associated with a line from the hit strips to the \gls{IP}.
The decoded data are provided to the \gls{MM} \gls{TP} algorithm (Section~\ref{tdaq:MMtrigger}), which runs on the \gls{FPGA}.
 
The other \gls{FPGA} on the mezzanine takes \num{32} \gls{sTGC} fibre inputs from the Routers (four for each of the eight layers), plus two redundant fibres from the \gls{sTGC} Pad Trigger (see Section~\ref{muonSS:stgcPadTrigger}) for the sector.
Only data from strips passing through the tower selected by the \gls{sTGC} Pad Trigger (see Section~\ref{muonSS:stgcPadTrigger}) are transmitted to the \gls{NSW-TP}. A maximum of four such track segments per sector may be transmitted per \gls{BCID}.
The data from the Routers consist of a Band-ID identified by the Pad Trigger, the 6-bit \gls{ADC} values from \num{14} strips, a flag indicating whether the 14~high or low strips of the 17~strips in the band are transmitted, along with the six low bits of the \gls{BCID}, and the $\phi$-ID (see Section\,\ref{muonSS:stgcStripTrigger}). The decoded data are used by the \gls{sTGC} \gls{TP} algorithm (Section~\ref{tdaq:stgcStripTrigger}), which runs on the \gls{FPGA}.

\paragraph{NSW sTGC Strip Trigger Algorithm \label{tdaq:stgcStripTrigger}}
The \gls{sTGC} strip trigger uses the digitised peak data describing the quantity of charge deposited
on each active strip selected by the Pad Trigger to compute a centroid for each of the eight \gls{sTGC} layers and, from the centroids in the eight layers, track segments.
A configurable charge threshold is used to define the ``active'' strips used in the centroid calculation.
A configurable look-up table defines which patterns of active strips are accepted: wide clusters due to neutrons or $\delta$-rays can be rejected;
certain patterns with isolated hits near the cluster may be configured to be accepted.
An average centroid is calculated from the valid layer centroids in each of the two wedges.
An $R$-index, defined as the radial position of the trajectory projected onto a virtual plane passing through the nominal $z$-coordinate of the most downstream wire plane of the \gls{NSW}, and $\Delta\theta$ are calculated directly from the values of the two wedge-centroids, using \glspl{LUT}.
 
The \gls{sTGC} trigger produces at most four candidates per sector; the \gls{MM} trigger can transmit up to eight candidates. The FPGAs are connected by a high-speed, low-latency 64-bit \gls{LVDS} link. Since the \gls{MM} \gls{TP} algorithm results are available sooner, \gls{MM} candidates are sent to the \gls{sTGC} \gls{TP} for merging, with no impact on the latency due to the data transfer.
The merging algorithm limits the total number of merged candidates to at most eight, removing duplicates.
Since $\phi$ from the \gls{MM} is more precise and $R$ from the \gls{sTGC} is more precise (see Tables~\ref{table:Muon_MMparams} and~\ref{table:Muon_sTGCparams}), a duplicate segment can take its $\phi$ from the \gls{MM} and its $R$-index and $\Delta\theta$ from \gls{sTGC}.
Further details of the merging algorithm are given in Ref.~\cite{NSWelx}.

\paragraph{Micromegas Trigger Algorithm \label{tdaq:MMtrigger}}
The \MM trigger algorithm first converts incoming hits to slope values, based on the slope of the line between the hit strip and the ATLAS \gls{IP}.
Hits are stored in a buffer with a tuneable number of bunch crossings (currently five). A set of overlapping roads, eight strips wide, is formed for the $\eta$ strips;
the number of strip hits within a road required for further processing can be set between two and four (currently three).
 
\begin{figure}[!h]
\subfloat[]{\label{figMuon:oneDiamond}\includegraphics[width=0.45\textwidth]{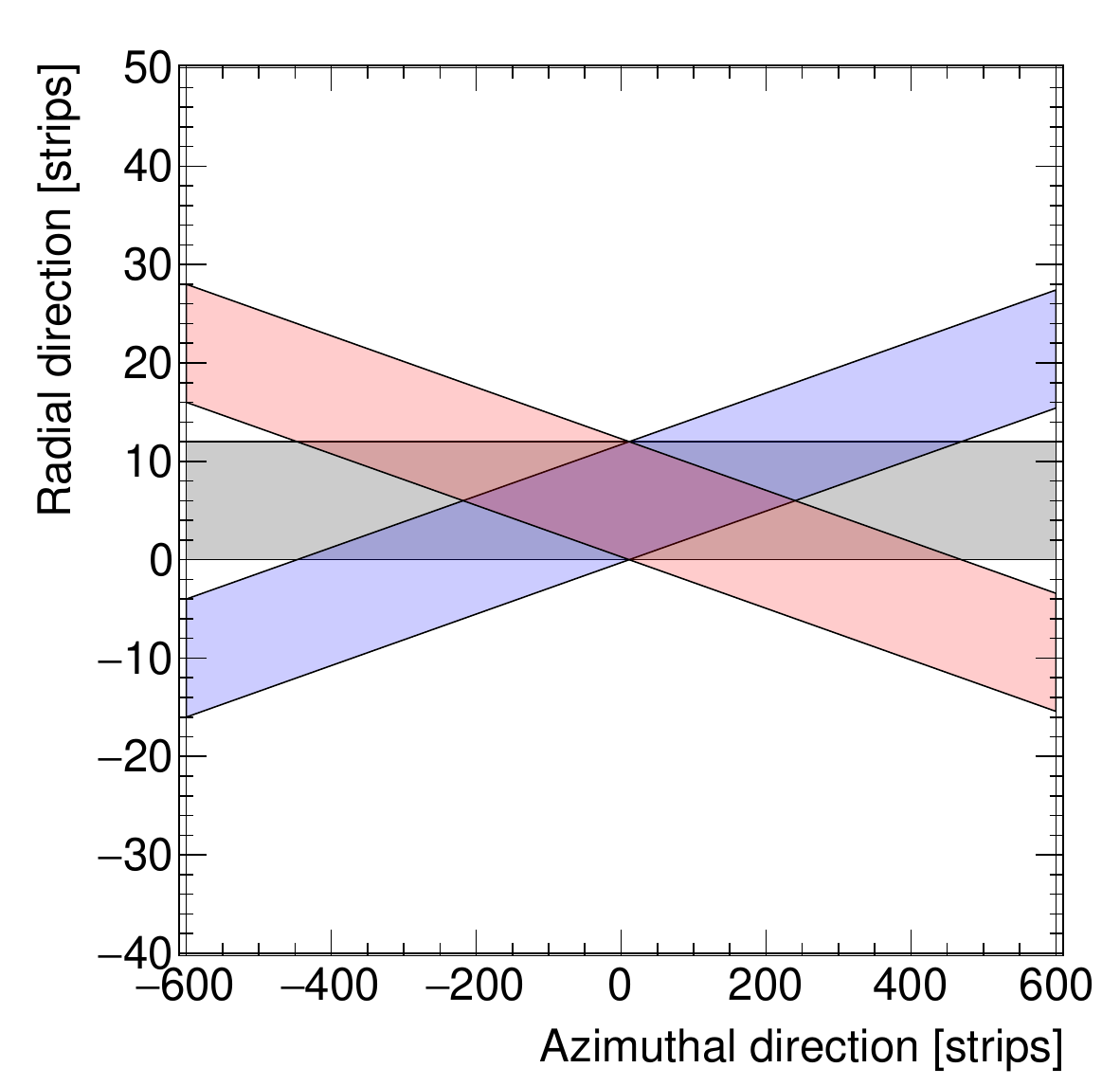}}
\subfloat[]{\label{figMuon:Ndiamonds}\includegraphics[width=0.45\textwidth]{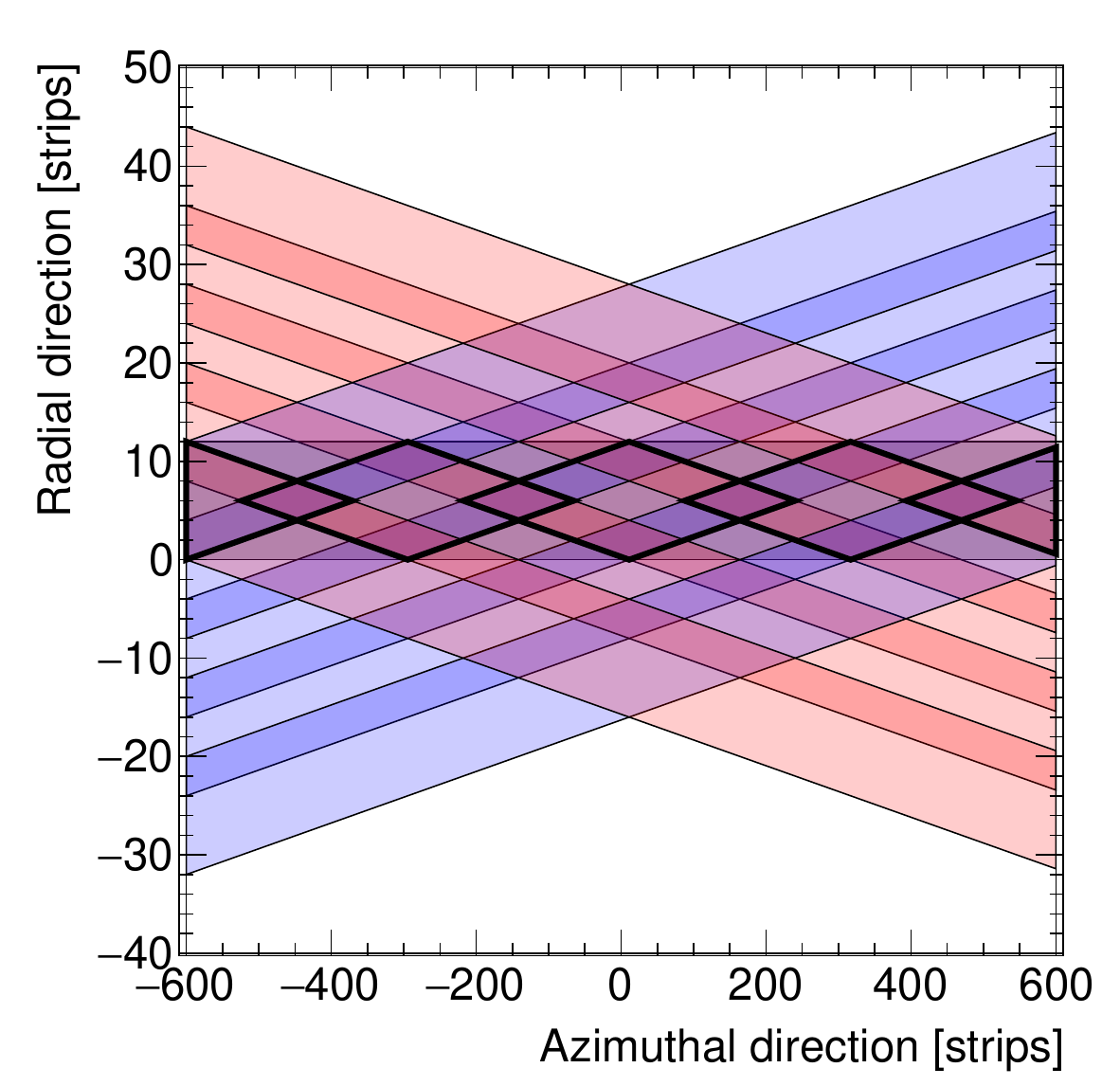}}
\caption{The \gls{MM} trigger is built from ``roads'' \num{12} strips wide. \protect\subref{figMuon:oneDiamond} shows an $\eta$-road (grey), and two opposite-angled stereo roads (red and blue) which intersect to form a diamond.
\protect\subref{figMuon:Ndiamonds} shows one $\eta$-road (grey), and five positive and five negative-angled overlapping stereo roads, forming five diamonds spanning the full length of the $\eta$-road. Both axes on each plot use units of strip pitch.
The $\eta$-road overlaps by four strip pitches with the $\eta$-roads immediately above and below it. The stereo roads also overlap by four strip pitches with their immediate neighbours, as shown.
The trigger diamonds are the intersections of twelve stereo strips in each direction. The diamonds overlap with their neighbours, as shown by the black outlines.
The \gls{MM} \gls{TP} \gls{FPGA} performs a fit to the strips in the roads within the selected diamonds. This fit determines the azimuthal resolution of the \gls{MM} trigger.}
\end{figure}

The \gls{MM} trigger identifies ``roads'': sets of strips in the eight \gls{MM} layers of a sector that are consistent with the path of a particle (with momentum above a given threshold) coming from the \gls{IP}.
In practice, \gls{MM} roads are twelve strips wide (\SI{5.1}{\mm} or \SI{5.4}{\mm}), with the four outer strips at each side overlapping the neighbouring roads.
Each wedge contains two layers of ``eta'' strips and two layers of ``stereo'' strips (one with a stereo angle of \ang{1.5} and the other \ang{-1.5} with respect to the eta strips).
When a coincidence of three $\eta$ strips is found (of the four $\eta$ layers in the sector), there is a range of possible stereo hits in both stereo projections.
The intersection of two opposite-angle roads of stereo strips (also \num{12} strips wide) forms a diamond (see Figure~\ref{figMuon:oneDiamond}).
The number of these diamonds required to span the full azimuthal extent of the chamber (see Figure~\ref{figMuon:Ndiamonds}) depends on the radius.
Seven diamonds are sufficient to span the full azimuth at the innermost radius of the Large sectors, while seventeen are needed at the outer radius.
For the $\eta$ road, all diamonds are sequentially parsed for possible stereo hits. The number of stereo hits required is also tuneable from two of opposite direction to a maximum of four.  Currently the default is three, but there must be at least two with opposite inclination angles. Once the coincidences in the $\eta$ roads and diamonds satisfy the coincidence requirements, the hit strips are then used to calculate slopes. See also Figures~\protect\subref{figMuon:oneDiamond} and \protect\subref{figMuon:Ndiamonds}.
Candidate diamonds are required to have hits in a configurable number of the eta layers in the sector and a configurable number of the corresponding stereo layers.
A high-efficiency configuration would be at least two out of four hits in the eta layers and at least two out of four in the stereo layers, whereas a high-rejection configuration would require three out of four in both.
Coincidences are considered within a configurable sliding window of time, in the range of four to eight \glspl{BC}. These configurable parameters are chosen based on detector and electronics performance, and on \gls{LHC} conditions.
 
A local slope is calculated from the hits on the horizontal $\eta$ strips and used to calculate $\Delta\theta$. The stereo hits are used to find the $\phi$ of the track segment. If a fit is found to be consistent with a projection back to the \gls{IP} using $\Delta\theta$, its $R$-index is determined using a lookup table and sent along with  $\Delta\theta$ for merging with the \gls{sTGC} segments.
 
\paragraph{NSW-TP output interfaces}
The outputs of both \gls{TP} algorithms are segment candidates.
The \gls{NSW-TP} requires a valid track segment to be within a configurable angle (up to \SI{\pm 15}{\milli\radian}) of the corresponding infinite momentum track through the \gls{IP}.
Each 24-bit segment includes a radial $R$-index, an azimuthal $\phi$-index, a track angle $\Delta\theta$, and resolution flags that indicate whether the segment is from the \gls{sTGC} (\gls{sTGC} segments have one bit lower $\phi$ resolution than \gls{MM}), or whether one quadruplet had only a 3-out-of-4 coincidence, which implies worse $\Delta\theta$ resolution.
The \gls{sTGC} \gls{TP} algorithm produces up to four segment candidates, and then accepts up to eight segment candidates from the output of the \gls{MM} \gls{TP} algorithm running on the same mezzanine card.
The \gls{TP} algorithm running on the \gls{sTGC} \gls{FPGA} then merges the two lists of segments.
The merging procedure removes duplicates and drops segments beyond the eight allowed.
Priority is currently given to \gls{sTGC} segments. There are options to ignore one or the other of \gls{MM} or \gls{sTGC} segments and in case of duplicates, to take the $\phi$-id from \gls{MM} and the other variables from the \gls{sTGC} segment.~\cite{NSWelx}
 
The surviving segments are then sent via fast serial links from the \gls{NSW-TP} to the \gls{SL} described in Section~\ref{sec:EndcapSL}, where endcap Muon candidates are formed by successfully matching \gls{NSW} track segments with coincidences found in the \gls{EM-TGC} (Big Wheels).
Each \gls{NSW} sector sends the track segment data to the \gls{SL} via optical fibres: up to eight candidates per sector per \gls{BCID} on two fibres, each running at \SI{6.4}{Gb/\s}, four candidates per fibre,
including the sector ID and \gls{BCID}.
Seven copies (14~links) are sent out from the \gls{NSW-TP}  \gls{sTGC} \gls{FPGA}
to up to seven different endcap \gls{SL} modules.
(A single \gls{SL} board receives data from at most six \gls{NSW} trigger sectors, but each \gls{NSW} trigger processor may need to deliver data to up to seven \gls{SL} boards.
This is needed to cover the overlap of \gls{NSW} sectors with \gls{EM-TGC} sectors, with multiple scattering, misalignments and magnetic field bending also taken into account.)
For an overview of the \gls{NSW} trigger-path electronics see Figure~\ref{fig:Muon_NSWelectronics}.
The whole process repeats for every \gls{BCID} with a fixed latency of \SI{1075}{\ns} 
from the time of collision to the time the signal reaches the \gls{SL}.

Via the \gls{ATCA} card's two \glspl{FPGA} and Rear Transition Module, each mezzanine has two fibre connections to \gls{FELIX}, one for \gls{sTGC} and one for \gls{MM}, carrying, on different \glspl{e-link}, the \gls{L1A} event readout, exception messages, statistics, sampled events for monitoring, algorithm parameters, and \gls{TTC} signals.
The \gls{BC} clock recovered from these links is distributed to the various \gls{FPGA} and transceiver clocks via jitter cleaners.
 
A Zynq \gls{SOC} \gls{FPGA} on the \gls{ATCA} card handles configuration of all the \glspl{FPGA} and communicates various temperatures and voltages from elements of the  \gls{NSW-TP} via Ethernet to monitoring applications.
An \gls{IPMC} card on the \gls{ATCA} card communicates critical temperatures and voltages to \gls{DCS} via the \gls{ATCA} Shelf Manager.
If any temperature exceeds a configurable threshold, the shelf manager can shutdown a complete \gls{ATCA} board or specific mezzanines.


\subsubsection{New Endcap Sector Logic}\label{sec:EndcapSL}
\begin{figure}[htbp!]
\centerline{\includegraphics[width=0.5\textwidth]{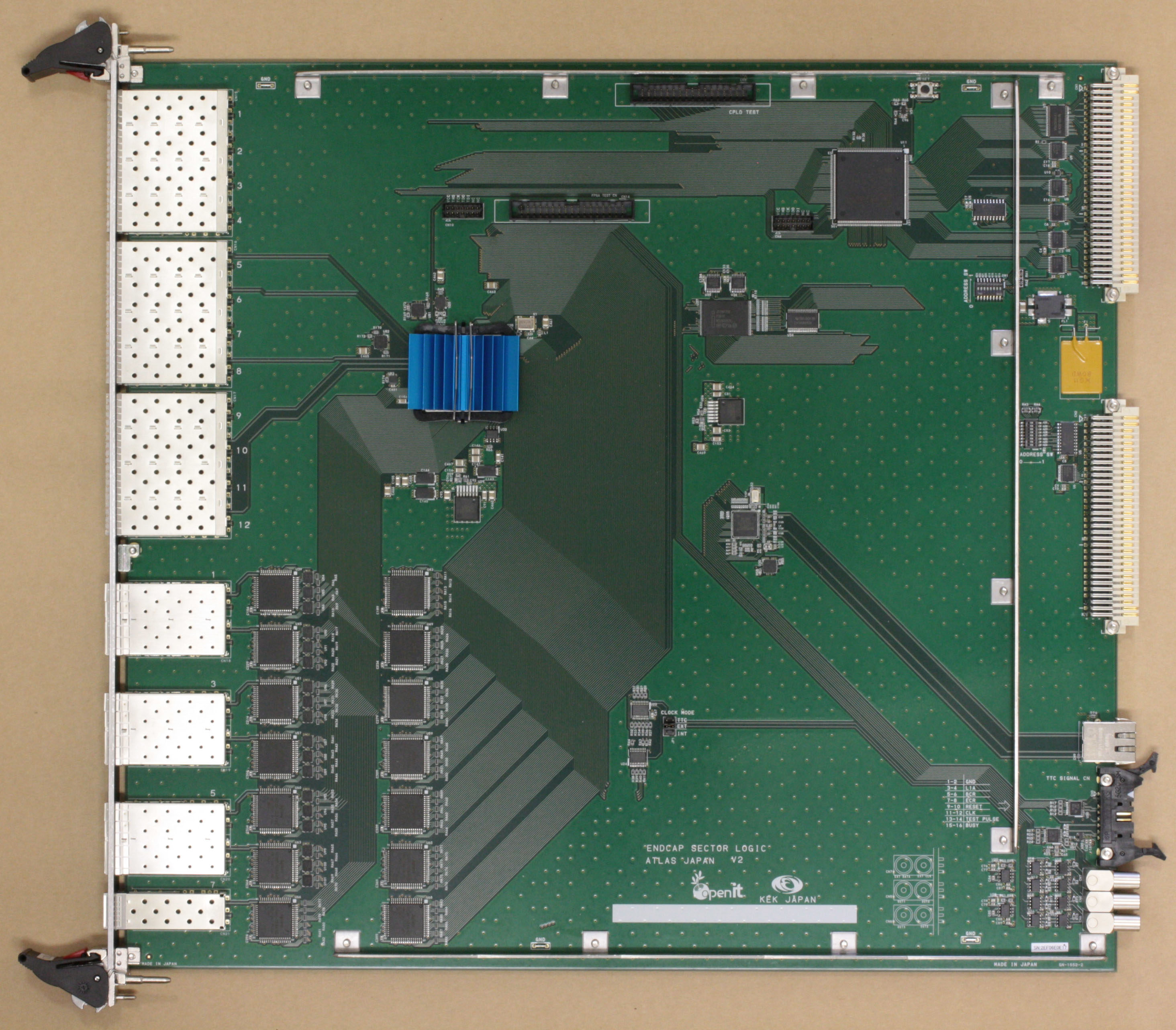}}
\caption{A \gls{L1Muon} sector logic board.}
\label{fig:TDAQL1MuonSectorLogicBoard}
\end{figure}

Muon track candidates are reconstructed based on coincidence logic and assigned to transverse momentum intervals by the
sector logic boards.  For \RunThr, new sector logic boards have been designed in order to receive the information from new detector components, and apply additional coincidence requirements.  Inputs to the new sector logic boards include information from the legacy \gls{EM-TGC}, \gls{EIL4} \gls{TGC}, and Tile calorimeters, as well as the new \gls{RPC-BIS78} and \gls{NSW} chambers.  The new sector logic boards utilise a Xilinx\textregistered~Kintex\textregistered~-7 \gls{FPGA}, featuring Multi-Gigabit Transceiver technology (\gls{GTX}); each board has 12 GTX interfaces and 14 G-Link receivers.  A sector logic board is shown in Figure~\ref{fig:TDAQL1MuonSectorLogicBoard}.
 
\begin{figure}[htbp!]
\centerline{\includegraphics[width=0.5\textwidth]{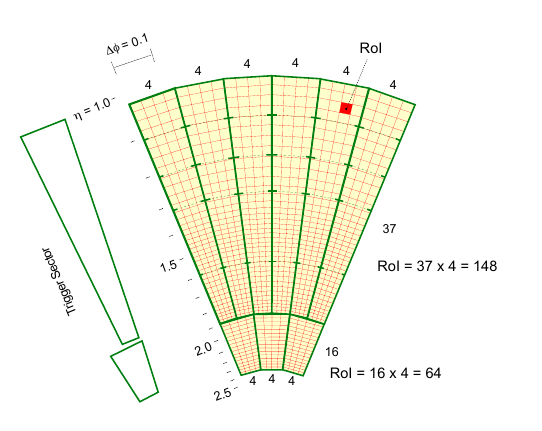}}
\caption{Schematic of the \gls{L1Muon} endcap trigger segmentation for one octant of the detector, showing independent trigger sectors outlined in green and \glspl{RoI} outlined in red.  An example of the grouping of the detector elements into a trigger sector is shown on the left side of the figure.
}
\label{fig:TDAQL1MuonEndcapRoI}
\end{figure}

The ``endcap'' region ($|\eta| < 1.9$) is subdivided into 48 independent trigger sectors per side  in $\phi$, as shown in Figure~\ref{fig:TDAQL1MuonEndcapRoI}, and 24 sectors per side in the ``forward'' ($|\eta| > 1.9$) region.  A sector logic board processes information from two adjacent sectors.  In total, 72 new sector logic boards are required to cover the entire \gls{L1Muon} endcap trigger system.
 
\begin{figure}[htbp!]
\centerline{\includegraphics[width=0.95\textwidth]{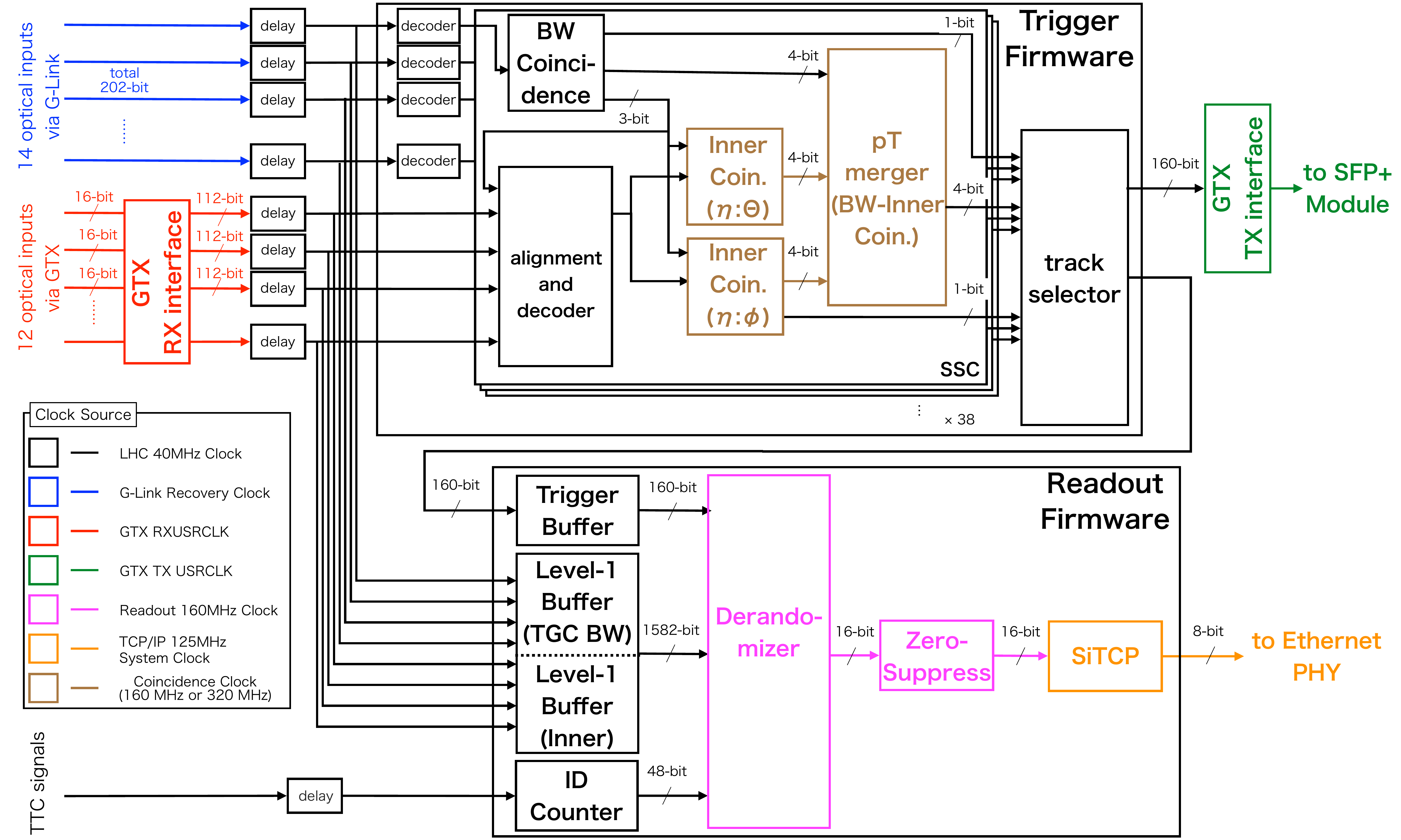}}
\caption{Block diagram of the \gls{L1Muon} sector logic.  The Big Wheel (G-Link) inputs are shown in blue and the \gls{NSW} (GTX) inputs are shown in red.}
\label{fig:TDAQL1MuonSectorLogic}
\end{figure}

\begin{figure}[htbp!]
\centerline{\includegraphics[width=0.95\textwidth]{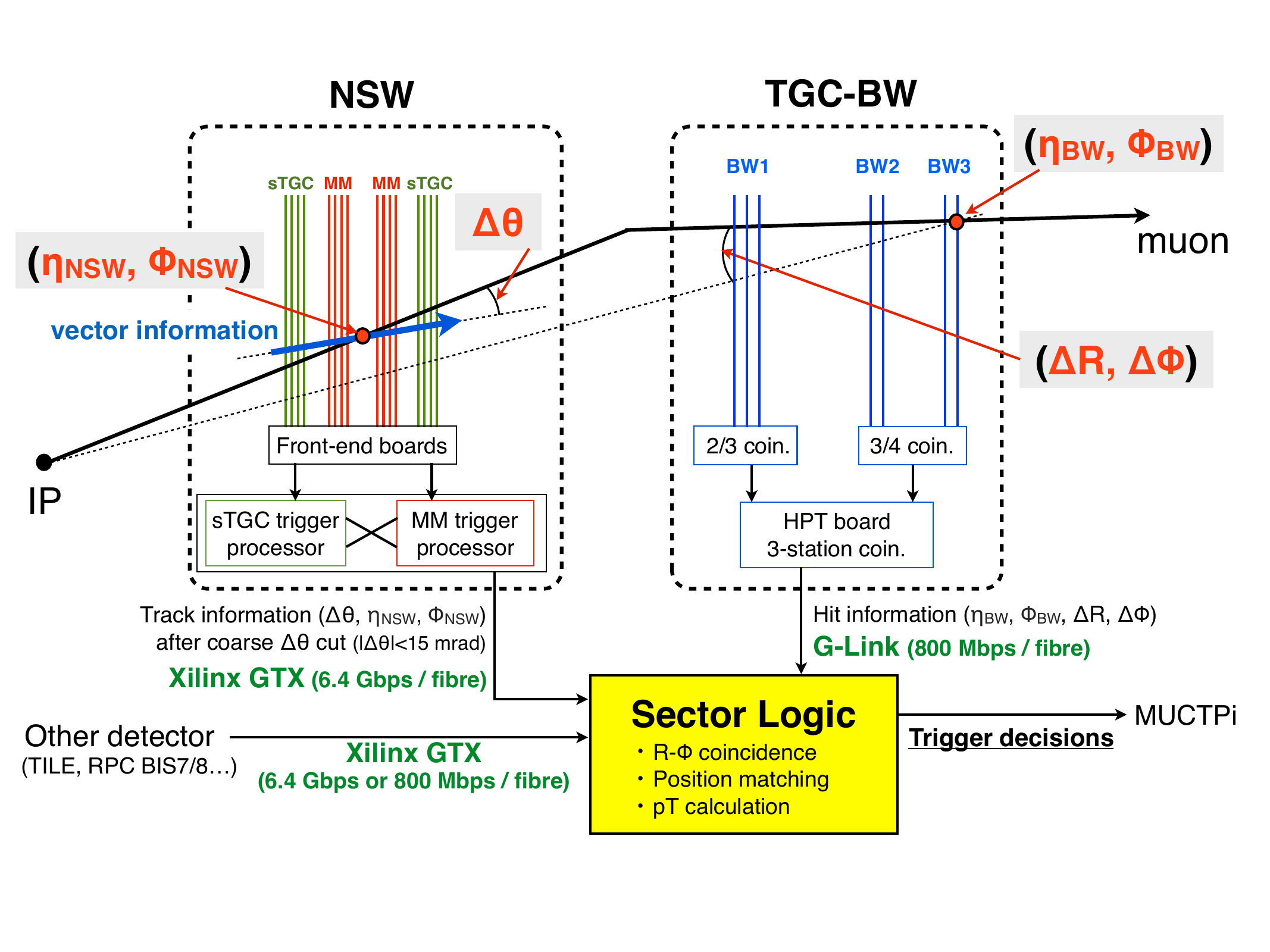}}
\caption{Overview of the \gls{L1Muon} endcap trigger system, showing the tracking information used to compute the trigger decision.
The solid line shows the trajectory of a muon with finite momentum, the dotted line shows the trajectory of a hypothetical infinite momentum muon used to define the centre of the trigger road.}
\label{fig:TDAQL1MuonEndcapTrigger}
\end{figure}

A block diagram of the \gls{L1Muon} trigger sector logic is shown in Figure~\ref{fig:TDAQL1MuonSectorLogic}.  Hit positions in the third station of the \gls{EM-TGC} ($\eta_{\mathrm{M3}}, \phi_{\mathrm{M3}}$) are extrapolated back to the \gls{IP}, as shown in Figure~\ref{fig:TDAQL1MuonEndcapTrigger}, thus determining the ``road'' corresponding to the straight track of a muon with infinite momentum.  Deviations from the centre of these straight roads ($\Delta R, \Delta\phi$) depend on the track momentum and are computed at the first \gls{TGC} station (\gls{TGC}-M1).
Inputs from the \gls{NSW} to the new sector logic include the position of the \gls{NSW} track segment ($\eta_{\mathrm{NSW}}, \phi_{\mathrm{NSW}}$) and the deviation of the reconstructed track segment with respect to the \gls{IP} ($\Delta\theta$).
 
Look-up tables are used to determine the $R-\phi$ coincidence between signals from the \gls{EM-TGC} and the \gls{NSW}, \gls{EIL4} \gls{TGC} chambers, Tile calorimeter D-layer cells, or \gls{RPC-BIS78} chambers. Tracks accepted by the sector logic are divided into intervals corresponding to the programmable \pT thresholds.  Up to four of the highest-\pT tracks per sector are sent to the \gls{MuCTPI}.  \glspl{RoI}, spanning an area of $\eta \times \phi = 0.025 \times 0.033$, are sent to the \gls{HLT}.
 
\subsubsection{RPC Feet and Elevator Chambers}
Muon triggers in the barrel region ($|\eta| < 1.05$) are provided by three concentric layers of \gls{RPC} doublets.  The \gls{L1} trigger decision in the barrel region relies on coincidence logic.  For the low-\pT thresholds, a coincidence of three out of four layers in the middle station is required.  The high-\pT trigger thresholds require the low-\pT trigger logic to be satisfied, as well as an additional hit in one of two layers in the outer barrel station.
 
During \RunOne, the sectors 12, 13 and 14 of the barrel spectrometer had a trigger coverage approximately 20\% lower than the other sectors due to the presence of the toroid feet support structure (sectors 12 and 14, corresponding to $-2.16 < \phi < -1.77$ and $-1.37 < \phi < -0.98$) and the elevator shaft ($|\eta| \simeq 0.7, \phi \simeq -1.57$).  In \gls{LS1}, additional chambers were installed in the toroid feet and elevator regions in order to increase the \gls{L1Muon} barrel trigger coverage by about 3\%, as described in Section~\ref{muonSS:BMEBOEBMG}.
 
\begin{figure}[htbp!]
\centering
\subfloat[]{
\includegraphics[width=0.48\textwidth]{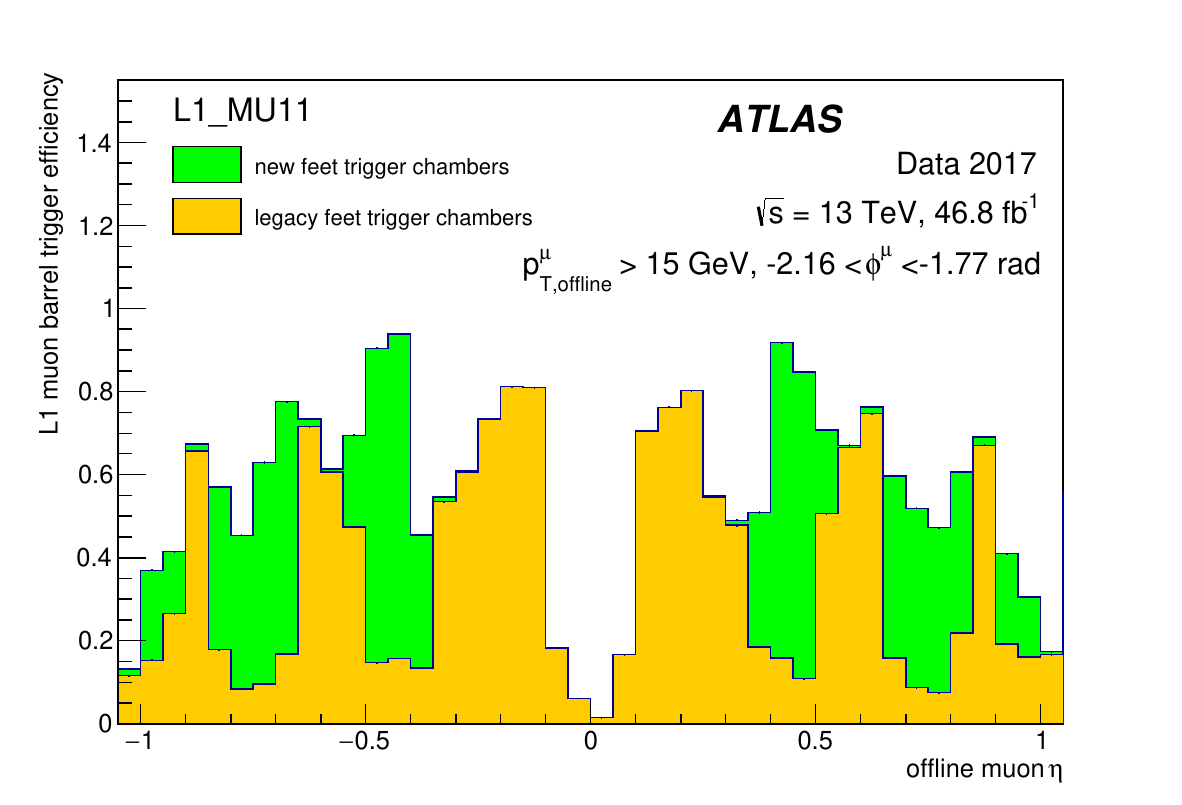}
\label{fig:TDAQL1MuonBarrelFeetSector12}
}
\subfloat[]{
\includegraphics[width= 0.48\textwidth]{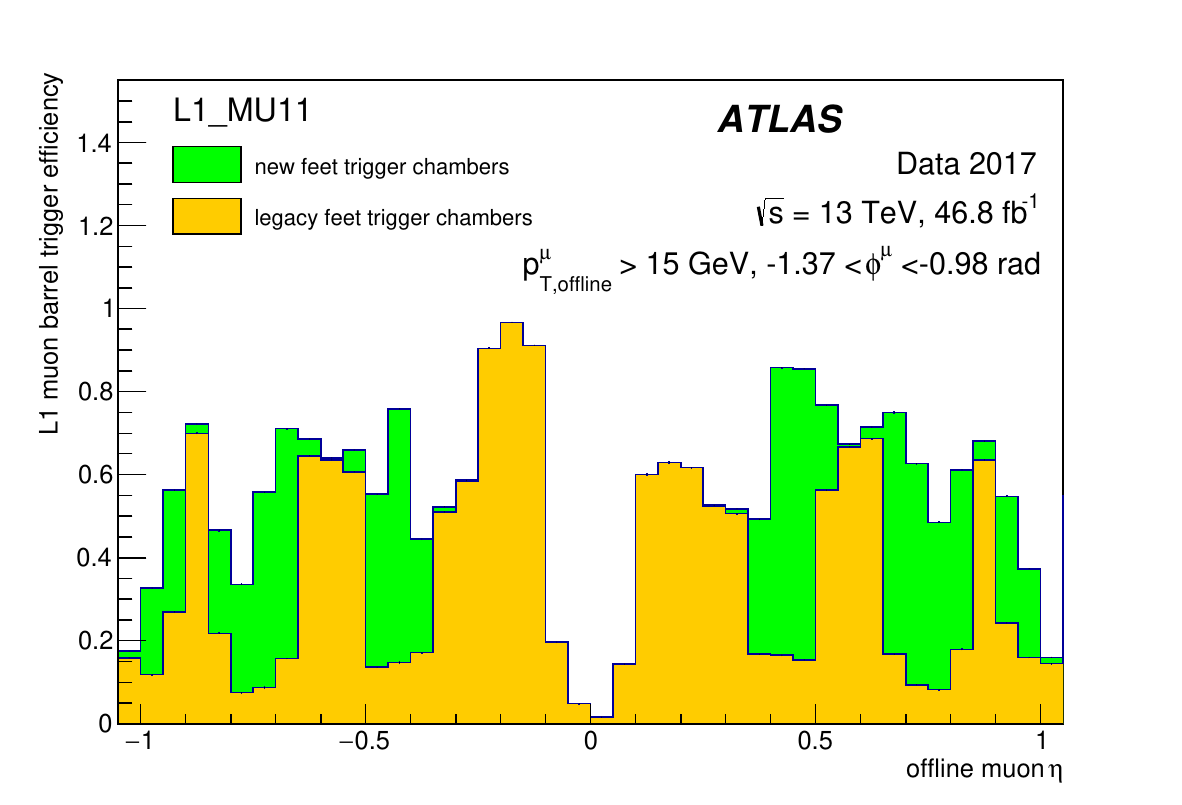}
\label{fig:TDAQL1MuonBarrelFeetSector14}
}
\caption{Trigger efficiency for muons with $\pT > \SI{15}{\GeV}$ and satisfying a three-station coincidence requirement in the barrel region.  The increased efficiency due to the additional chambers in the feet region can be seen for the two $\phi$-sectors where the chambers were added: \protect\subref{fig:TDAQL1MuonBarrelFeetSector12} Sector~12 and \protect\subref{fig:TDAQL1MuonBarrelFeetSector14} Sector~14.
}
\label{fig:TDAQL1MuonBarrelFeetChambers}
\end{figure}

In the feet region, the new chambers consist of two layers of \gls{RPC} doublets, instead of the usual three layers.  A two-layer coincidence is required for the high-\pT thresholds, but since there are fewer doublet layers in this region, the fraction of fake muon triggers in this region is higher than in the rest of the barrel.  The impact of the new chambers in the feet region can be seen in Figure~\ref{fig:TDAQL1MuonBarrelFeetChambers}.
 
\subsubsection{RPC BIS78}
 
In the regions of the inner barrel (BI) that receive the highest background rates, additional \gls{RPC} modules have been deployed, as discussed in Section~\ref{muonSS:BIS78}.  These new \gls{BIS78} (Barrel Inner Small~\cite{ATLAS-TDR-26}) chambers, covering $\eta$ stations 7 and 8, are located on the A-side of ATLAS, in the transition region between the barrel and endcap ($1.0 < \eta < 1.3$).  The high background rate in this region is due to secondary charged tracks originating from beam halo protons and a lack of detector shielding and instrumentation.  The \gls{BIS78} chambers provide stand-alone detection and localisation of charged tracks and have an angular resolution of \SI{3}{\milli\radian}, which provides improved trigger rejection of low-\pT muons.  These new chambers are a pilot project for the Phase-2 BI upgrade, which is planned to compensate for the reduced efficiency of the \glspl{RPC} in the \gls{HL-LHC} environment, and increase the geometrical acceptance of the muon trigger in the barrel with BI-BO combinations.
 
The trigger logic for the \gls{BIS78} chambers is performed on Pad boards~\cite{LoffredoBIS78}, which are installed on each station.  The Pad is an \gls{FPGA}-based board which collects the \gls{BIS78} \gls{RPC} hit data from the front-end electronics over 18 serial links at \SI{1.6}{Gb/\s}.  It selects muon trigger candidates by requiring a local 2/3 coincidence of the \gls{RPC} hits, applies a zero-suppression algorithm, and then sends the trigger information to the endcap sector logic board (described in Section~\ref{sec:EndcapSL}), located off the detector.  Readout data are sent to \gls{FELIX} (Section~\ref{subsubsec:tdaq_daqhlt_felixswrod}) through optical links via a gigabit transceiver (GBTx) chip~\cite{WYLLIE20121561}.
 
The Pad board utilises a Xilinx\textregistered~Kintex\textregistered~-7 \gls{FPGA}, which supports optical transmission with fixed latency and provides robustness against radiation, including both single-event upsets and total ionising dose effects.
 
The expected rate reduction expected from the \gls{BIS78} \glspl{RPC} is illustrated in Figure~\ref{fig:TDAQL1MuonEndcapPerformance}.
 
\subsection{L1 Topological Trigger}\label{sec:TDAQ_L1Topo}
 
The ATLAS physics programme relies significantly on electroweak-scale processes involving hadronically-decaying tau leptons, jets, and \MET, such as $H \rightarrow \tau\tau$ and $ZH \rightarrow \nu\overline{\nu}b\overline{b}$.  Also of importance are processes with unique topologies, such as $B$- or $J/\psi$-meson decays to low-\pT leptons or vector-boson fusion production of a Higgs boson, which then decays invisibly (\gls{VBF} $H \rightarrow \mathrm{invisible}$).  These signatures have large multi-jet backgrounds, and so the ability to reject the background and improve signal purity while still maintaining low trigger thresholds is of critical importance.
 
To accomplish this, a new \glsfirst{L1Topo} system was introduced in \RunTwo and commissioned in 2016.  It consisted of a single \gls{ATCA} shelf with two \gls{FPGA}-based processor modules which performed selections based on geometric or kinematic observables of \glspl{TOB} received from the \gls{L1Calo} and / or \gls{L1Muon} systems, e.g $\Delta\eta, \Delta\phi, \Delta R$, and invariant mass ($M_{\mathrm{inv}}$).  The details of the hardware design for the \RunTwo system may be found in~\cite{TRIG-2019-02}.
 
\begin{figure}[htbp]
\centerline{\includegraphics[width=0.45\textwidth]{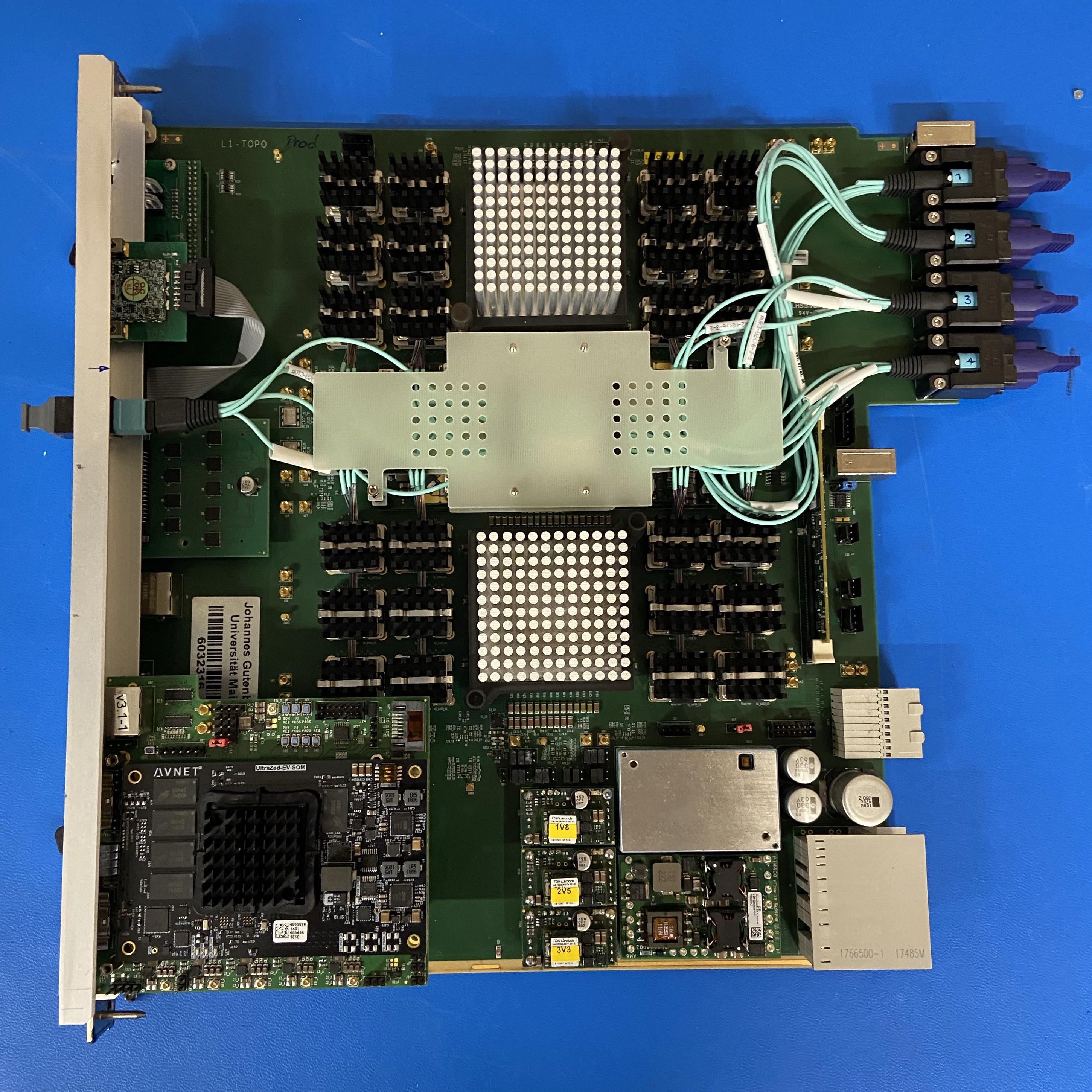}}
\caption{A production \gls{L1Topo} module.}
\label{fig:TDAQL1TopoPhoto}
\end{figure}

The \gls{L1Topo} system has been redesigned for \RunThr.
It consists of three \gls{ATCA} modules: one module computes \gls{L1Calo} trigger multiplicities (Topo1)  and the other two (Topo2 and Topo3)  apply topological selections as in \RunTwo.
(In the legacy system, the multiplicity task was performed by the \gls{L1Calo} \gls{CMX}~\cite{TRIG-2016-01}.)
Each module includes two Xilinx\textregistered~Ultrascale+ processor \glspl{FPGA} for algorithm computation, which provide increased processing power compared to the \RunTwo version.  Each \gls{FPGA} has 118 input and 24 output fibres.  Each module also contains two mezzanine cards: one for power and control, and another for communication with the \gls{CTP}.  An \gls{L1Topo} module is shown in Figure~\ref{fig:TDAQL1TopoPhoto}.
 
\gls{L1Topo} receives \glspl{TOB} for jets, \MET,  e/$\gamma$ clusters, and muons from the \gls{L1Calo} and \gls{L1Muon} systems via the TopoFOX optical plant (described in Section~\ref{sec:TDAQ_L1CaloInfrastructure}).  The data received include the object type, ($\eta, \phi$) coordinates, and transverse energy of the object, and its isolation (in the case of e/$\gamma$ objects).  Table~\ref{tab:TDAQL1TopoTOBs} summarises the number of \glspl{TOB} per object type sent to \gls{L1Topo}.
 
\begin{table}[htbp]
\begin{center}
\caption{A summary of the maximum number of \glspl{TOB} sent to \gls{L1Topo} for each object type.}\label{tab:TDAQL1TopoTOBs}
\begin{tabular}{|l|c|}
\hline
Object & Number of \glspl{TOB} \\
\hline
e/$\gamma$ (eFEX) & 144 \\
Tau (eFEX) & 144 \\
Forward e/$\gamma$ ( jFEX) &  5\\
Large Tau (jFEX) & 6 \\
Small-radius jet (jFEX) & 168 \\
Large-radius jet (jFEX) & 24  \\
\MET  (jFEX) & 7 \\
\ET     (jFEX) & 7\\
Small-radius jet (gFEX) & 6 \\
Large-radius jet (gFEX) & 3 \\
\MET  (gFEX) & 3 \\
\sumET     (gFEX) & 1\\
Muon & 32 \\
\hline
\end{tabular}
\end{center}
\end{table}
 
The \gls{L1Topo} algorithms are implemented in a pipelined, deadtime-free manner using a fixed latency.  Algorithms are distributed across the \glspl{FPGA} as evenly as possible such that optimal resource usage is achieved.  The configuration of algorithms, as well as their configurable parameters, are stored in the trigger menu.
 
The real-time output data sent by \gls{L1Topo} to the \gls{CTP} consist of individual bits indicating the algorithm decisions as well as an overflow bit for the topological algorithms.  The output of the multiplicity algorithms includes multiple bits indicating the number of \glspl{TOB} fulfilling the requirements of a given algorithm.  The outputs are sent to the \gls{CTP} via optical or electrical cables.
 
Upon receipt of an \gls{L1A}, the \gls{L1Topo} real-time output data are captured and sent to the \gls{DAQ} system.  Readout to \gls{FELIX} and the \gls{SW ROD} is done via the \gls{ROD} as described in Section~\ref{sec:TDAQ_L1CaloHubROD}.  The clock and \gls{TTC} signals are received from the \gls{Hub} module via the backplane.
 
Control and configuration are performed via an IPbus interface on every \gls{FPGA}, as well as on a control \gls{FPGA} which controls the entire module.  Voltage and temperature are monitored by the \gls{IPMC} via an \gls{I2C} bus and are provided to \gls{DCS} by the shelf manager.
 
\begin{figure}[htbp!]
\centering
\subfloat[]{
\includegraphics[width=0.44\textwidth]{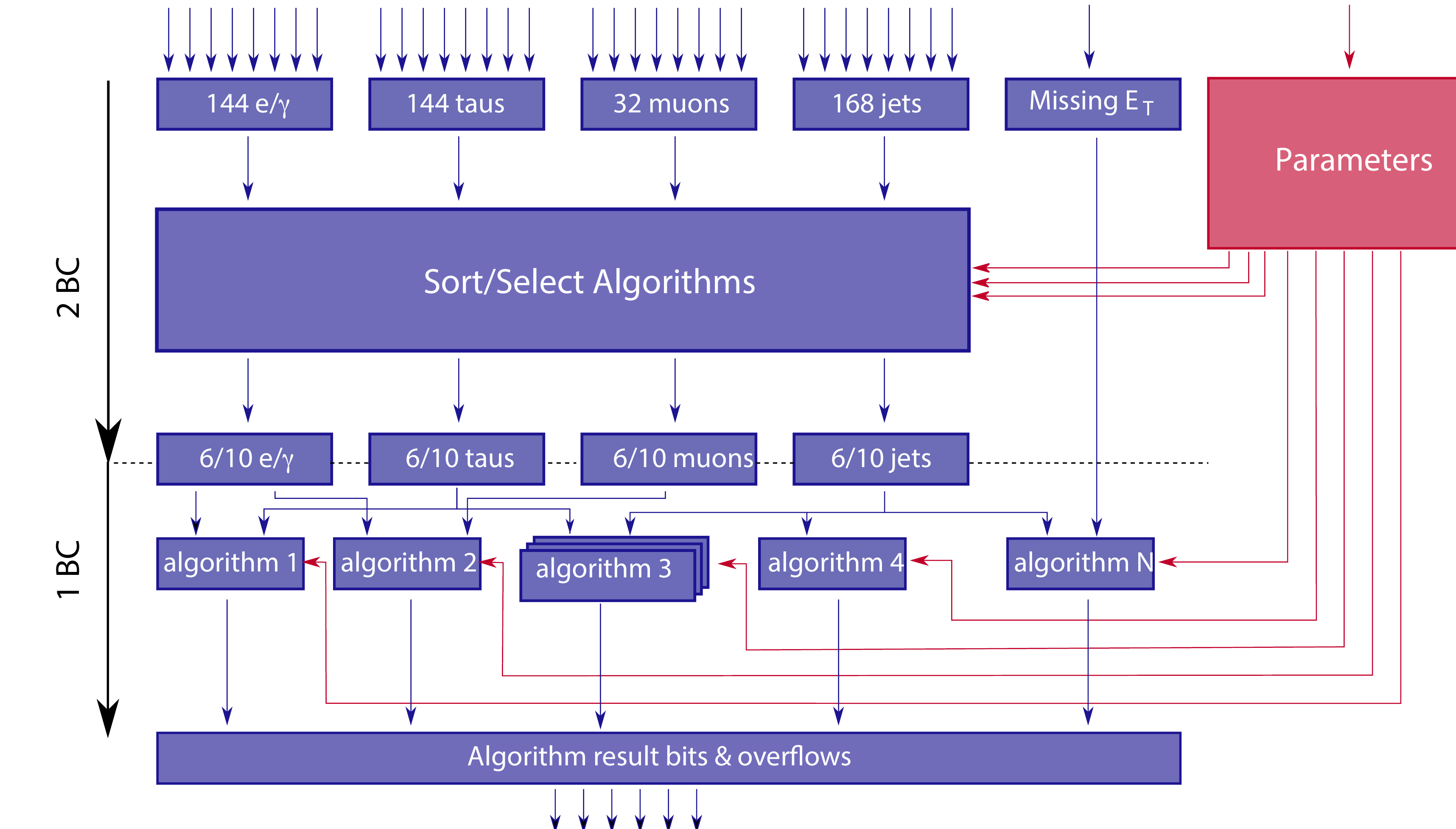}
\label{fig:TDAQL1TopoAlgoBlock1}
}
\subfloat[]{
\includegraphics[width= 0.54\textwidth]{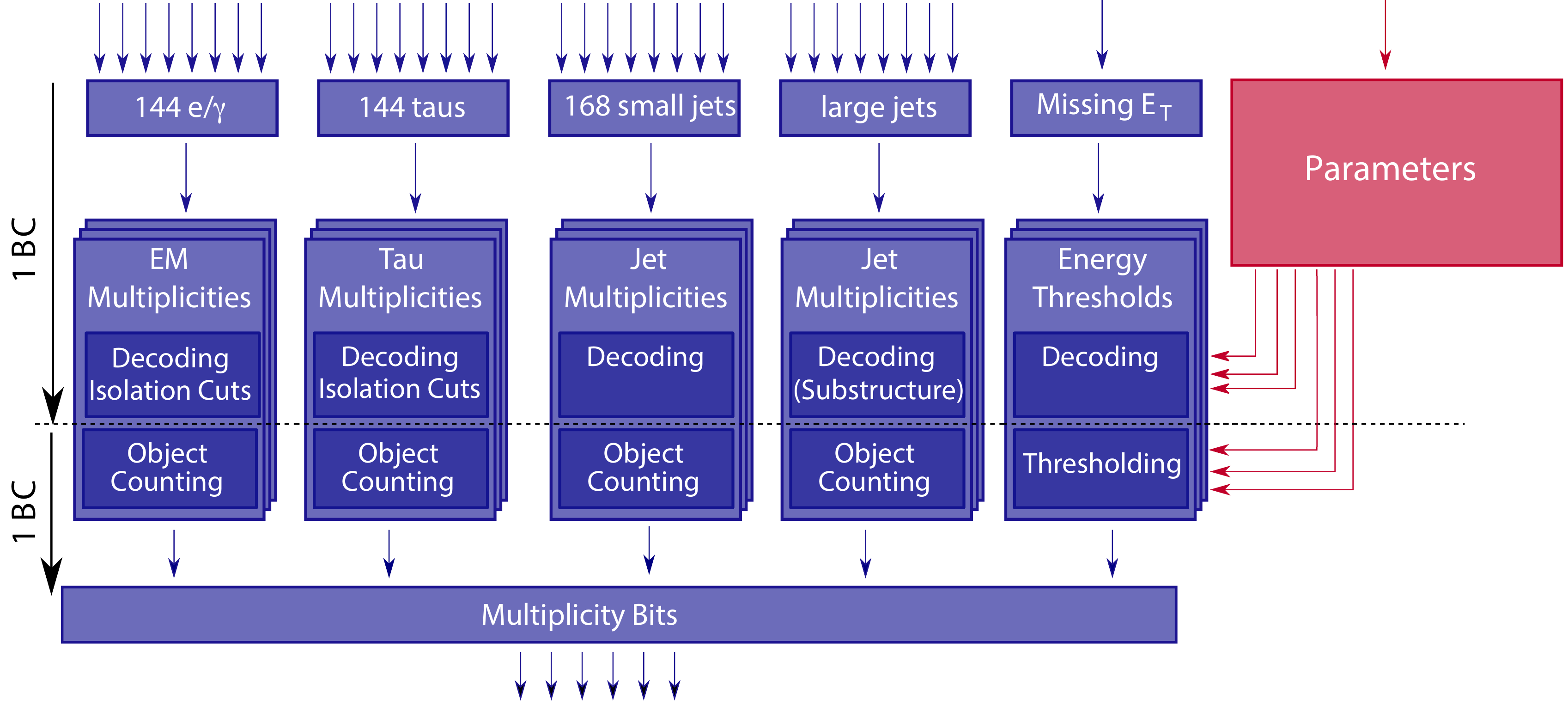}
\label{fig:TDAQL1TopoAlgoBlock2}
}
\caption{The \gls{L1Topo} algorithm structure is illustrated in \protect\subref{fig:TDAQL1TopoAlgoBlock1}. \gls{TOB} inputs are first passed to sort / select / no sort algorithms, which produce shortened lists of \glspl{TOB}, which are in turn passed to the decision algorithms. Subfigure~\protect\subref{fig:TDAQL1TopoAlgoBlock2} shows the multiplicity algorithm structure
which has as input all \glspl{TOB} produced by \gls{eFEX}, \gls{jFEX} and \gls{gFEX}.   These \gls{TOB} inputs are first passed to ``no-sort'' algorithms (labelled as ``Decoding'' in the figure).  The resulting lists of \glspl{TOB} in a global format are then passed to the multiplicity algorithms; the output multiplicity bits are sent to the \gls{CTP}.
}
\label{fig:TDAQL1TopoAlgoBlockDiagram}
\end{figure}

Three types of \gls{L1Topo} algorithms exist:
\begin{itemize}
\item \textbf{Sort / select / no-sort algorithms} These algorithms take all input \glspl{TOB} and convert the various \gls{TOB} formats to a single global data format.  They also sort the \glspl{TOB} by \ET or \pT,  or select all \glspl{TOB} with \ET or \pT exceeding a configurable threshold value, or satisfying object-specific criteria, such as isolation or substructure requirements.  The output consists of reduced lists of sorted \glspl{TOB}, which may then be used as inputs to the decision algorithms, described below.  This reduction is necessary to handle the otherwise excessively large combinatorics.  The final lists contain six \glspl{TOB} in the case of ``sort'' algorithms and 10 \glspl{TOB} for ``select'' algorithms.  ``No sort'' algorithms only perform the conversion to a global data format and are used as inputs to the multiplicity algorithms.  The latency of these algorithms is two bunch crossings (\SI{50}{\ns}).
\item \textbf{Decision algorithms} These algorithms determine whether a given trigger condition has been satisfied; examples include selections on $\Delta\eta$, $\Delta\phi$, and $M_{inv}$.  Their outputs include decision and overflow bits which are sent to the \gls{CTP}.  The latency of the decision algorithms is limited to one bunch crossing (\SI{25}{\ns}).
\item \textbf{Multiplicity algorithms} These algorithms count the number of objects passing a given threshold, e.g. \ET, or located within a given region in $\eta$.  The output multiplicity bits are transmitted to the \gls{CTP}.  The latency of the multiplicity algorithms is limited to three bunch crossings (\SI{75}{\ns}).
\end{itemize}
Configurable parameters (e.g. threshold values) of the algorithms are specified in the trigger menu.  The three algorithm types are illustrated in block format in Figure~\ref{fig:TDAQL1TopoAlgoBlockDiagram}.
 
\subsection{Central Trigger \& TTC} 
The \gls{L1} \gls{CTP} receives trigger information from the other \gls{L1} trigger systems and forward detectors and executes the final \gls{L1} trigger decision.  Muon trigger inputs are received via the \gls{MuCTPI}, which is described in Section~\ref{sec:TDAQ_MUCTPI}. The \gls{CTP} itself is described in Section~\ref{sec:TDAQ_CTP}.
 
\subsubsection{Muon to central trigger processor interface}\label{sec:TDAQ_MUCTPI}
\paragraph{Functional overview}
The muon barrel and endcap trigger processors send their results as input to the
\glsfirst{MuCTPI}, which has been redesigned and replaced during the Phase-I upgrade \cite{bib:muctpi-phase1}.  The replacement was necessary in order to provide full-granularity muon information at the \gls{BC} rate to the \gls{L1} topological processor and to be able to interface to the sector logic modules using high-speed optical links.
 
The \gls{MuCTPI} receives information for up to four muon track candidates per muon trigger sector from the endcap trigger sector logic modules, and up to two candidates from the barrel trigger sector logic modules.  The information includes the position and \pT threshold passed by the track candidates (15 \pT thresholds for the endcap, and six \pT thresholds for the barrel), along with additional information, such as track quality flags, geometrical flags, and in the case of the endcap, a flag indicating the electric charge of the muon candidate.
 
The \gls{MuCTPI} processes the information in three parallel paths:
 
\begin{itemize}
\item It combines the information from all trigger sectors to calculate the total multiplicity of muon candidates per muon threshold and sends the multiplicities to the \gls{CTP} for each bunch crossing.  Up to 64 bits of multiplicity information can be sent to the \gls{CTP} for each bunch crossing.  The multiplicity can either be a 2- or 3-bit value, and the maximum multiplicity value for a given \pT threshold includes the cases with even more muon candidates.
 
\item It sends muon position and transverse momentum information of selected muon candidates to the \gls{L1} topological trigger system, to be used in subsequent topological trigger algorithms in combination with calorimeter information.  The muon candidates sent are ordered according to decreasing \pT.
 
\item It can apply muon-only topological trigger algorithms to be sent to the \gls{CTP}.
\end{itemize}
 
Care is taken to avoid double-counting of muons which traverse more than one detector region due to geometrical overlap of the chambers and the deflection of the muons in the magnetic field.  Many cases of overlaps are resolved within the barrel and endcap muon trigger processors.  The remaining overlaps to be treated by the \gls{MuCTPI} are those between neighbouring trigger sectors.
 
The \gls{MuCTPI} also provides data to the \gls{HLT} and to the data acquisition system for events selected at \gls{L1}.
A subset of the muon candidate information, ordered by \pT is sent to the  \gls{HLT} to be used as \glspl{RoI} for further processing.  The \gls{DAQ} system records a more complete set of information, including the computed multiplicity values, which is used to monitor the functions of the  \gls{MuCTPI}.

\paragraph{System implementation}
 
\begin{figure}[htbp!]
\centering
\subfloat[]{
\includegraphics[width=0.65\textwidth]{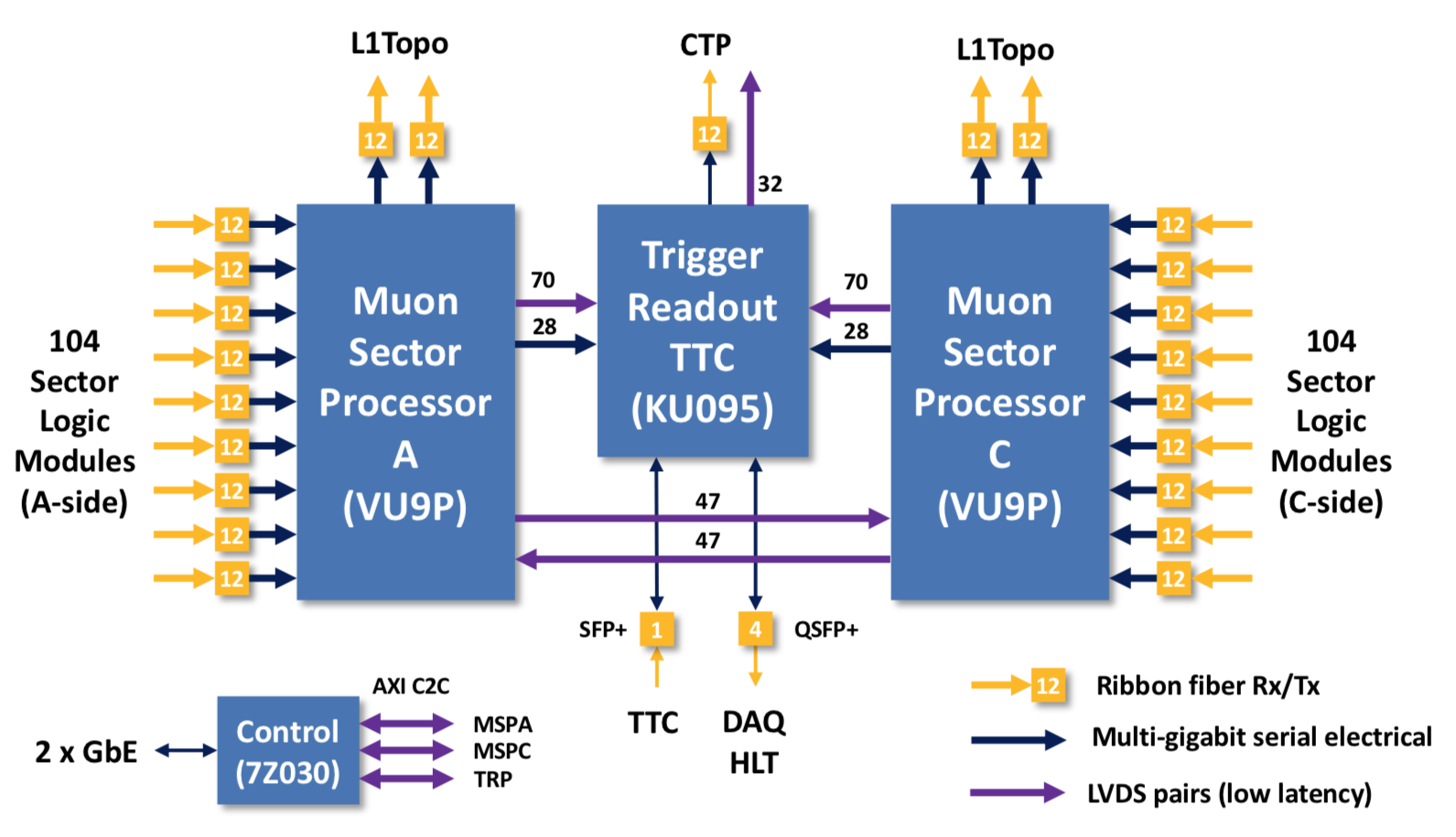}
\label{fig:TDAQMUCTPIBlock}
}
\subfloat[]{
\includegraphics[width= 0.33\textwidth]{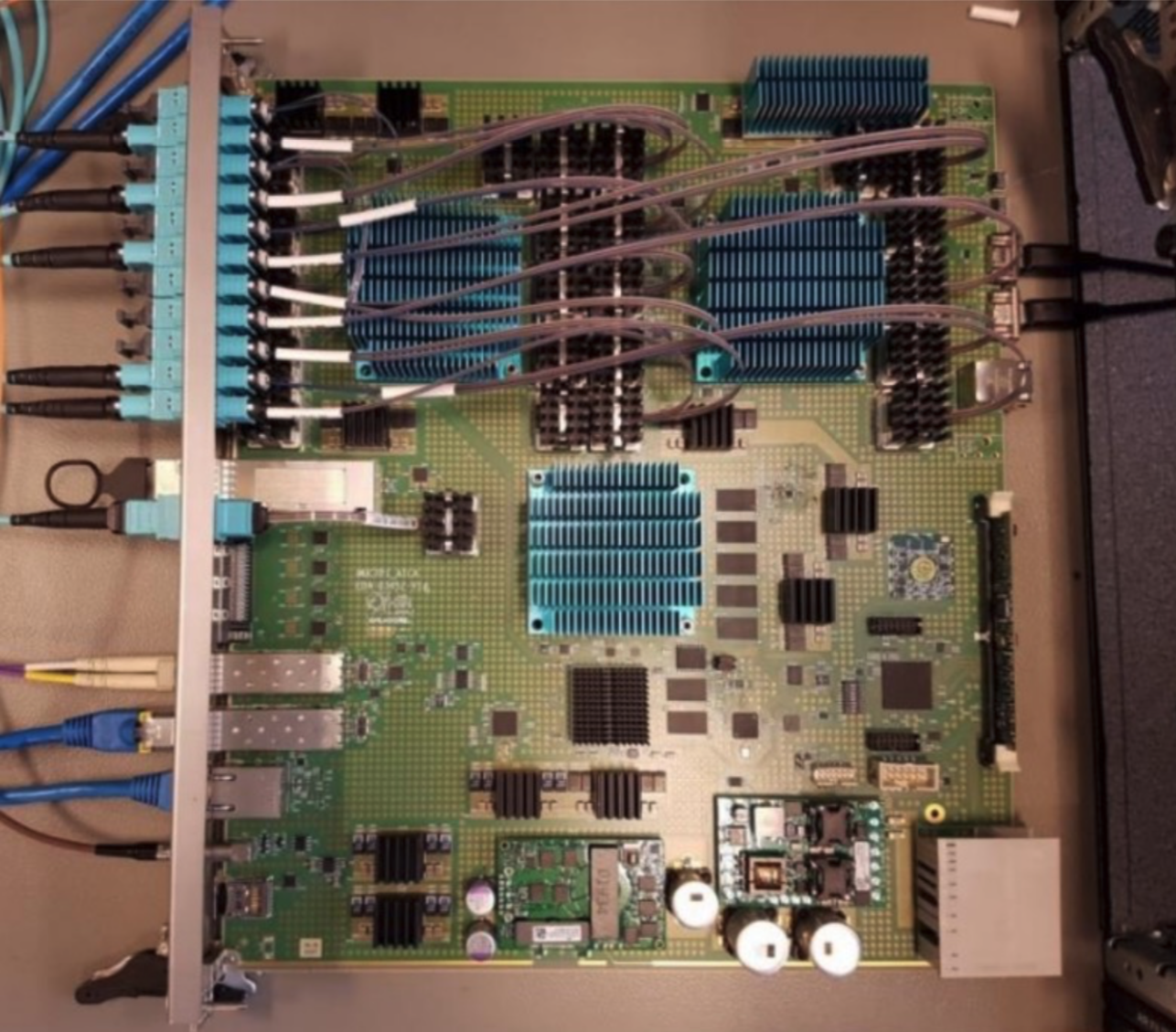}
\label{fig:TDAQMUCTPIPhoto}
}
\caption{The \gls{MuCTPI} module shown \protect\subref{fig:TDAQMUCTPIBlock} as a block diagram and \protect\subref{fig:TDAQMUCTPIPhoto} in a photograph.}
\label{fig:TDAQMUCTPI}
\end{figure}
 
The Phase-I \gls{MuCTPI}, shown in Figure~\ref{fig:TDAQMUCTPI}, is integrated on a single \gls{ATCA} blade, replacing the 18 9U \gls{VME} cards of the \RunTwo system.  The architecture is based on a highly integrated generation of \glspl{FPGA}, featuring a large number of on-chip \glspl{MGT} as well as 12-channel ribbon fibre optics receiver and transmitter modules (MiniPODs) for the data transfer.
 
The two Muon Sector Processor \glspl{FPGA} -- each taking care of one detector hemisphere -- receive and process muon trigger data from 208 inputs from sector logic modules connected through high-speed serial optical links using MiniPOD receiver modules.  The Muon Sector Processor \glspl{FPGA} also copy information on selected muon trigger objects to several \gls{L1Topo} modules using MiniPOD transmitter modules.
 
The Trigger and Readout Processor \gls{FPGA} merges the muon multiplicity information received from the two Muon Sector Processor \glspl{FPGA} and sends the results to the \gls{CTP}.  In addition, it can also implement muon topological trigger algorithms.  This is possible because all the trigger information is available in a single module with low latency.  The same \gls{FPGA} also receives, decodes and distributes the \gls{TTC} information.
 
A \gls{SOC} is used for configuration, control and monitoring of the module.  The device integrates a programmable logic part with a dual-core \gls{ARM} processor subsystem.  The processor subsystem runs the required software to interface the \gls{MuCTPI} to the ATLAS run control system through a \gls{GbE} interface.  It will also be used for environmental monitoring of components of the board such as the power supply, optical modules, and \glspl{FPGA}.  The values read include voltages, currents, temperatures, optical input power, clock status, etc.
 
\subsubsection{Central Trigger Processor}\label{sec:TDAQ_CTP}
 
\paragraph{Functional overview}
The \glsfirst{CTP} is the last stage of processing of the Level-1 trigger system.  It receives digital trigger information from the \gls{L1Topo}, legacy \gls{L1Calo}, \gls{MuCTPI} and \gls{ALFA} systems, and from various forward detectors.  The \gls{CTP} system used during \RunOne has been described in~\cite{PERF-2007-01} and has since been upgraded.
 
Table~\ref{tab:ctp-inputs} shows the various inputs of the \gls{CTP}.  Three different input paths are available:
\begin{itemize}
\item The traditional electrical trigger path via the \gls{CTPIN} modules is used for trigger signals coming from various forward detectors, for calibration signals from some sub-detectors, and for special triggers such as a filled-bunch trigger based on beam-pickup monitors, and a minimum-bias trigger based on scintillation counters.  It is also used for the legacy \gls{L1} calorimeter trigger system during the start-up phase of \RunThr and the commissioning of the \gls{L1} topological trigger system.
\item The electrical low-latency input via the \gls{CTPCORE+} module, which is used for latency-critical trigger signals, such as those from the \gls{L1} topological processor, muon-only topological triggers from the  \gls{MuCTPI} system, and signals from the \gls{ALFA} detector.  For the \gls{L1} topological processor, these are signals from the Topo2 and Topo3 modules: Topo2 provides topological algorithms that combine jets with one or more other objects, comprehensive combinations of  jet and $\tau$, and jet and
\gls{EM} clusters, along with muons and missing transverse energy where appropriate, while Topo3 is largely dedicated to topological triggers typically involving electrons, ranging from the standard model performance triggers, to exotics triggers involving many lepton flavours.
\item The optical input via the \gls{CTPCORE+} module is used for the muon threshold multiplicities from the \gls{MuCTPI} and for non-latency critical \gls{L1} topological algorithms, mainly coming from the Topo1 module.  These include simple multiplicity triggers on electromagnetic,
hadronic $\tau$, and jet objects, energy sums, missing transverse energy, and other fast topological algorithms.
\end{itemize}
 
\begin{table}[htbp]
\caption{\label{tab:ctp-inputs} Overview of the trigger inputs to the \gls{CTP} in \RunThr.}
\begin{center}
\begin{tabular}{|p{0.15\textwidth}|p{0.25\textwidth}|p{0.15\textwidth}|p{0.4\textwidth}|}
\hline
\textbf{\gls{CTP} input} & \textbf{Cable origin} & \textbf{Number of bits} & \textbf{Trigger information} \\
\hline
\gls{CTPIN}  & Various sources & 36 & Forward detectors, calibration, special triggers\\
\hline
\multirow{2}{0.15\textwidth}{CTPCORE electrical} & MUCTPI/ALFA & \multirow{2}{0.15\textwidth}{77$-$149} & Muon-only topological algorithms and ALFA\\ \cline{2-2}\cline{4-4}
& Topo2, Topo3& & Topological algorithms \\
\hline
CTPCORE optical & \gls{MuCTPI} & 64 & Muon threshold multiplicities \\ \cline{2-4}
& Topo1 & 263$-$335 & Simple multiplicity triggers, energy sums, missing energy\\
\hline
\end{tabular}
\end{center}
\end{table}
 
\begin{figure}[htbp!]
\centerline{\includegraphics[width=0.77777775\textwidth]{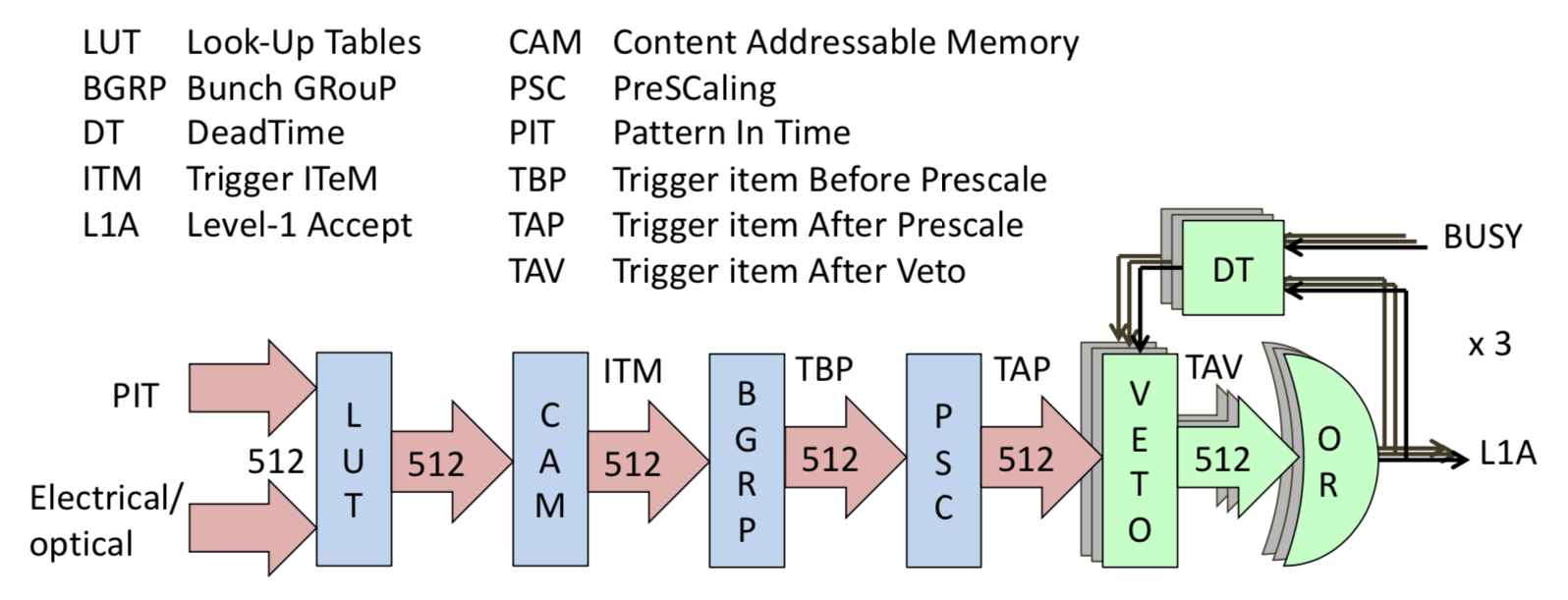}}
\caption{Block diagram of the trigger path in the CTP.}
\label{fig:TDAQ_CTPTriggerPath}
\end{figure}
 
The trigger path of the \gls{CTP} is shown in Figure~\ref{fig:TDAQ_CTPTriggerPath}.  Up to 512 trigger inputs can be used in programmable look-up tables to form trigger conditions from the input signals, as described in more detail below.  
 
Each trigger item can be put in coincidence with any combination of the 16 bunch groups, which are programmable collections of bunch-crossings.
 
In the next step, the rate of each trigger item is controlled by pre-scaling, where an algorithm rejects pseudo-randomly on average a certain fraction of the trigger item signals.  The fraction is called the pre-scale factor and can be individually set for each trigger item.  Each trigger item can subsequently be enabled or disabled to be used in the trigger decision.  The \gls{L1A} signal generated by the \gls{CTP} is the logical OR of all enabled trigger items.
 
For each \gls{L1A} signal, the \gls{CTP} provides an 8-bit trigger-type word signal, whose bits indicate certain categories of types of triggers and can be used to select options in the event data processing and monitoring in the sub-detector readout chain.  For each \gls{L1A} signal, the \gls{CTP} also sends information about the trigger decision for all trigger items to the \gls{HLT} and to the data acquisition system.  The \gls{CTP} can support up to three data-taking partitions for calibration and test runs.  The trigger items after pre-scaling are partitioned and assigned to the three partitions, each of which individually enables the trigger items of interest, gates them with the veto signal, forms the \gls{L1A} of the partition, and calculates the deadtime taking into account the \gls{L1A} and the BUSY signals of the corresponding partition.
 
In addition to its function as trigger processor, the \gls{CTP} is also the timing reference of the detector.  It receives the beam-synchronous clock and timing signals from the \gls{LHC} and distributes them together with the \gls{L1A} signal to the ATLAS sub-detectors, preserving their timing information.  It also interfaces with the \gls{LHC} \gls{GPS} timing system and attaches a nano-second precision absolute time-stamp to the read-out data of each \gls{L1}-accepted event.  In addition, it is in charge of generating luminosity blocks, which divide a data acquisition run into small time intervals of typically one minute.  These luminosity blocks are the shortest time interval for which the integrated luminosity, corrected for deadtime and pre-scale effects, can be determined.  They are also used as the time interval for online and offline data quality assessments.
 
The \gls{CTP} features many monitoring facilities that allow the monitoring of the rate of each incoming trigger signal, and of each trigger item at the different stages of trigger processing in the \gls{CTP}.  These rates are determined, published online, and preserved for offline usage, every few seconds and on luminosity block boundaries.  Some of the trigger rates, such as those before and after veto, are used to correct the integrated luminosity for deadtime losses.  The deadtime and its components are monitored directly by corresponding deadtime counters.  In addition, the \gls{CTP} has counter facilities that allow the monitoring of trigger rates and deadtime fractions per \gls{BCID} and thus the study of bunch-by-bunch effects of the trigger rates and deadtime.
 
\paragraph{System implementation}
\begin{figure}[htbp!]
\centerline{\includegraphics[width=0.5\textwidth]{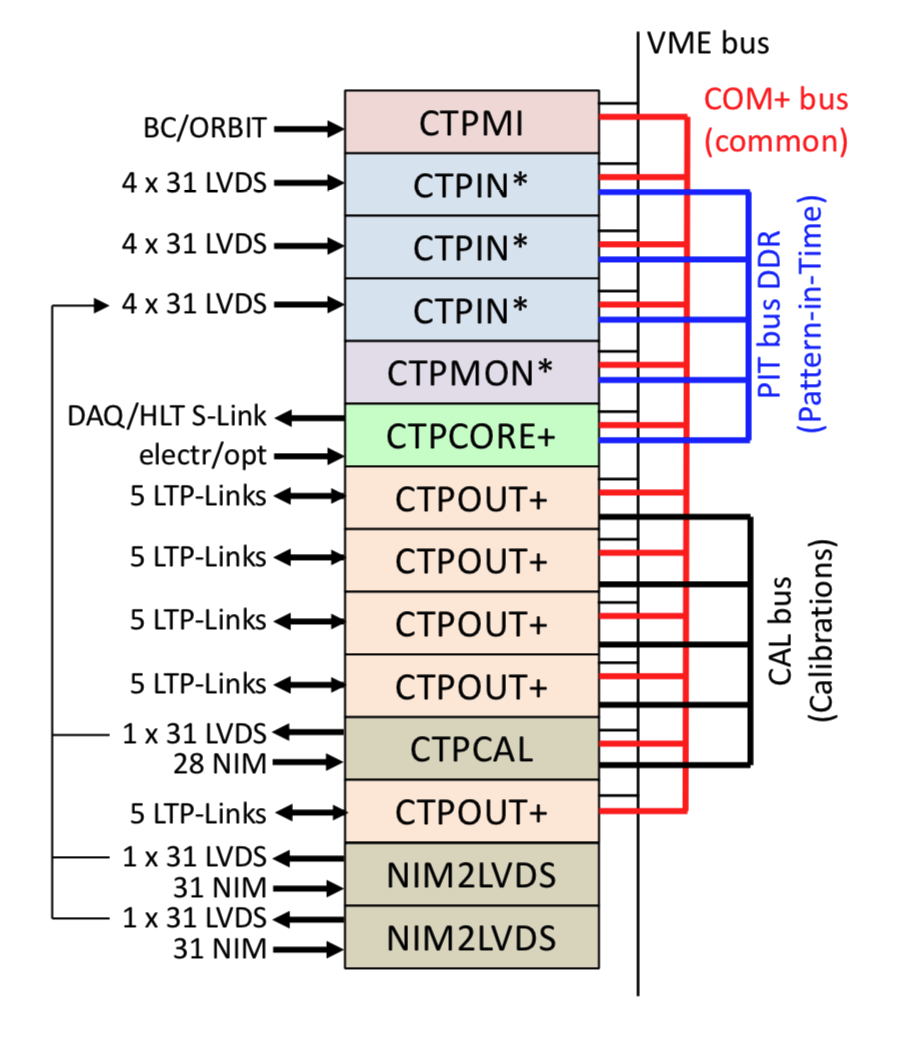}}
\caption{Block diagram of the various components of the \gls{CTP}. Components marked with $+$ were newly built during \gls{LS1}, while components marked with $*$ use the \RunOne hardware, but with upgraded firmware. The firmware of the \gls{CTPCORE+} was further upgraded during \gls{LS2}.}
\label{fig:TDAQ_CTPBoards}
\end{figure}
 
The \gls{CTP} for \RunOne has been described in detail in~\cite{PERF-2007-01} and has since been modified during \gls{LS1} and \gls{LS2}.  The upgraded \gls{CTP} system consists of six types of modules which are housed in a single 9U \gls{VME} crate, as shown in Figure~\ref{fig:TDAQ_CTPBoards}:
\begin{itemize}
\item The \gls{CTPMI} module as the machine interface module.
\item Three \gls{CTPIN} modules to receive trigger input signals.
\item The \gls{CTPCORE+} module to receive additional direct trigger input signals and implement the trigger logic.
\item The \gls{CTPMON} module for per-bunch-monitoring of trigger input signals from the \gls{CTPIN}.
\item Five \gls{CTPOUT+} modules as interfaces to the sub-detector \gls{TTC} distribution.
\item One \gls{CTPCAL} module to receive calibration request signals from sub-detectors and receive electrical signals.
\item Two \gls{NIM2LVDS} boards to receive further electrical signals.
\end{itemize}
 
The controller \gls{CPU} interacts with the modules via the \gls{VME} backplane.  Internal communication between the modules proceeds by custom bus systems implemented as custom-built backplanes installed in the \gls{CTP} crate: The \gls{PITbus} is used for routing synchronised and aligned trigger inputs from the \gls{CTPIN} boards to the \gls{CTPCORE+} module, the \gls{COMbus+} is for the interchange of common timing, trigger and control signals between all the modules, and the CALbus is used for sub-detector calibration requests.
 
For \RunTwo, the following components of the \gls{CTP} were upgraded:
\begin{itemize}
\item The firmware of the 3 \gls{CTPIN} modules was upgraded to drive trigger input signals to the \gls{PITbus} in double-data rate at \SI{80}{\MHz}, hence doubling the number of input signals available on the PITbus from 160 to 320.
\item The CTPCORE board was replaced by the newly designed \gls{CTPCORE+} board, which allows trigger logic to be applied to up to 512 trigger inputs (to be compared with 256 inputs in \RunOne).  The inputs are a programmable selection of signals from the PITbus and additional 192 direct electrical inputs from the \gls{L1Topo} trigger processor and the \gls{ALFA} sub-detector.  The \gls{CTPCORE+} can form up to 512 trigger items (256 in \RunOne) that can be put into coincidence with up to 16 bunch groups (eight in \RunOne).  It also features the possibility to support up to three data-taking partitions for calibration and test runs.  
\item The four CTPOUT boards have been replaced by five newly designed \gls{CTPOUT+} boards, each supporting running with up to three concurrent data-taking partitions, and featuring additional monitoring and testing facilities.
\item The COMbus backplane was replaced by the newly designed \gls{COMbus+} backplane, which allows the distribution of timing signals for up to three concurrent data-taking partitions, and includes the communication with a fifth \gls{CTPOUT+} board.
\item The \gls{CTPMON} firmware was updated to select up to 160 trigger inputs from the \gls{PITbus} for per-bunch monitoring.
\end{itemize}
 
For \RunThr, the \gls{CTPCORE+} firmware has been updated to receive optical trigger signals from \gls{L1Topo} via up to 12 optical fibres running at a link speed of \SI{6.4}{GBaud}, corresponding to 96 trigger bits per optical fibre.  A programmable switch matrix maps the incoming trigger signals to up to 512 usable trigger inputs.
 
The timing signals from the \gls{LHC} are received by the \gls{CTPMI}, which can also generate these signals internally for stand-alone running.  This board also controls and monitors the internal and external busy signals; for example, the busy signal transmitted from a sub-detector in case of overload on its data acquisition system.  The module sends the timing signals to the \gls{COMbus+}, thereby making them available to all of the other modules in the \gls{CTP}.
 
The \gls{CTPIN} modules receive trigger inputs from forward detectors, the legacy \gls{L1Calo} system, and from various other sources, such as calibration systems, the beam-pickup monitors, and a minimum-bias trigger based on scintillation counters.  The input boards select and route the trigger inputs to the \gls{PITbus}, after synchronising them to the clock signal and aligning them with respect to the bunch-crossing.  Three boards with four connectors of 31 trigger input signals each allow for a total of 372 input signals to be connected, of which up to 320 can be made available on the \gls{PITbus} at any given time, using double-data rate at \SI{80}{\MHz}.
 
The trigger decision module (\gls{CTPCORE+}) receives the trigger inputs from the \gls{PITbus} and from two external sources: via three electrical connectors of 64 trigger input signals each, and via up to 12 optical fibres containing up to 96 trigger bits each.  A selection of these trigger input signals is made using a programmable switch matrix.  The \gls{CTPCORE+} module combines the trigger input signals from the \gls{PITbus}, the electrical and the optical \gls{CTPCORE+} inputs, and internally generated signals, using several programmable look-up tables to form up to 512 trigger conditions.  In a further step the trigger conditions are combined using content-addressable memories to form up to 512 trigger items.  Any of the up to 512 trigger conditions may participate in any of the up to 512 trigger items.  The trigger items are put into coincidence with individual combinations of up to 16 bunch groups, subsequently undergo pseudo-random prescaling and deadtime gating, and can be individually enabled or disabled to take part in the final \gls{L1A} trigger decision.  The trigger results are transmitted to the \gls{COMbus+}.  The \gls{CTPCORE+} module also acts as the readout driver of the system, sending information to the \gls{HLT} trigger and the data acquisition for each accepted event.
 
The output module (\gls{CTPOUT+}) receives the timing and trigger signals from the \gls{COMbus+} and fans them out to the sub-detectors.  The module receives back from the sub-systems the busy signals, which are sent to the \gls{COMbus+}, and 3-bit calibration trigger requests, which are routed to the \gls{CALbus}.  The calibration module time-multiplexes the calibration requests on the \gls{CALbus} and sends them via a front-panel cable to one of the input modules.  The calibration module also has front-panel inputs for beam pick-up monitors, minimum-bias scintillators, other forward detectors, and test triggers.  In addition, two \gls{NIM2LVDS} modules provide further front-panel inputs for such trigger signals.
 
The \gls{TTC} signals of the \gls{CTP} are distributed electrically to the sub-detectors via the \gls{CTPOUT+} links, through which in return the sub-detector busy signals and some calibration request signals are received.  A local trigger system acts as an interface between the \gls{CTP} and the \gls{TTC} distribution of each sub-detector, and imitates the function of the \gls{CTP} during stand-alone data taking of the sub-detector.
 
During \RunOneTwo, the local trigger system consisted for all sub-detectors of a series of custom 6U \gls{VME} electronics boards \cite{bib:ttc}: an \gls{LTP} module as the \gls{TTC} switch board and generator of local signals, an \gls{LTPI} module for interconnections between sub-detector partitions, a TTCvi for the serialisation of the \gls{TTC} information, and a TTCex for the optical transmission to the sub-detector front- and back-end electronics, where \gls{TTC} receiver chips (TTCrx) decode the transmitted information and make it available as electrical signals for further use.
 
During \gls{LS2}, the \gls{ALTI} has been introduced as the local trigger system for new Phase-I sub-detector partitions and to replace the legacy set of \gls{TTC} modules (\gls{LTPI}, \gls{LTP}, TTCvi and TTCex) for all sub-detectors, except for the \gls{RPC}, \gls{TGC}, and \gls{LUCID} detectors.
 
The \gls{ALTI} module provides the same electrical and optical interfaces as the legacy set of \gls{TTC} modules in a single 6U \gls{VME} electronics board.  Using a single modern \gls{FPGA}, all the functions of the legacy set of \gls{TTC} modules are replicated.  New, useful functions were added, such as an optical input and memory to analyse optical \gls{TTC} signals, memories to store incoming electrical signals, additional monitoring features including per-bunch monitoring of trigger signals, and a range of specific functions serving to generate \gls{L1A} sequences similar to the ones from the \gls{CTP}.  These latter functions include a programmable look-up table for defining the trigger logic of the trigger inputs and generated trigger signals of the \gls{ALTI} board, pseudo-random trigger generators, pseudo-random pre-scaling, bunch group masking, and simple and complex deadtime generation.  Where possible, the same algorithms as in the \gls{CTP} were used.
 
From the \gls{ALTI} module or the legacy set of \gls{TTC} modules, the timing signals are distributed to the detector electronics using the \gls{TTC} system.  The implementation and use of the \gls{TTC} system is sub-system specific.  As an example, the muon trigger systems use \gls{TTC} standard components to transmit the timing signals all the way to the electronics mounted on the chambers, while in case of the inner tracking detector a custom-built distribution system is used to transmit the signals from the counting rooms to the cavern.
 
\subsection{DAQ/HLT}
 
\subsubsection{\glstext*{FELIX}/\glstext*{SW ROD}}
\label{subsubsec:tdaq_daqhlt_felixswrod}
\glsfirst{FELIX} and the \glsfirst{SW ROD} are new detector readout components introduced into the ATLAS \gls{DAQ} system for \RunThr. \gls{FELIX} is designed to act as a configurable data router, receiving packets from detector front-end electronics and transferring them to peers on a commodity high-bandwidth ethernet network. Whereas previous detector readout implementations relied on diverse custom hardware platforms, known as \glspl{ROD}, as the interface between the detector electronics and the common \gls{DAQ} chain, the motivation for \gls{FELIX} is to unify all readout across one well supported and flexible platform. As well as its readout function, \gls{FELIX} will also serve as a relay for trigger accept and clock information from the \gls{TTC} system to front-end electronics. It will also be possible to use \gls{FELIX} to send general purpose control data to front-end electronics to manage modules throughout data taking and calibration.
 
Another key aspect of the \gls{FELIX} concept is to take advantage of advances in technology, for example: larger and faster \glspl{FPGA}, the advent of multicore \glspl{CPU} and high-performance networking, to move tasks which were previously performed in customised hardware (the aforementioned \glspl{ROD}) into the more flexible firmware and software domains. In this new architecture, with \gls{FELIX} acting as a simple router, detector data processing, monitoring and control functions are instead implemented in software hosted by commodity server systems subscribed to \gls{FELIX} data. The primary peer on this network will be the \swrod, which will perform functions such as event fragment building, data formatting and other detector-specific processing to prepare data and facilitate online selection. The \swrod will also buffer event fragments and supply them on request to \gls{HLT} nodes, via an identical interface to the legacy \gls{ROS}.
 
\begin{figure}[htbp!]
\centerline{\includegraphics[width=0.8\textwidth]{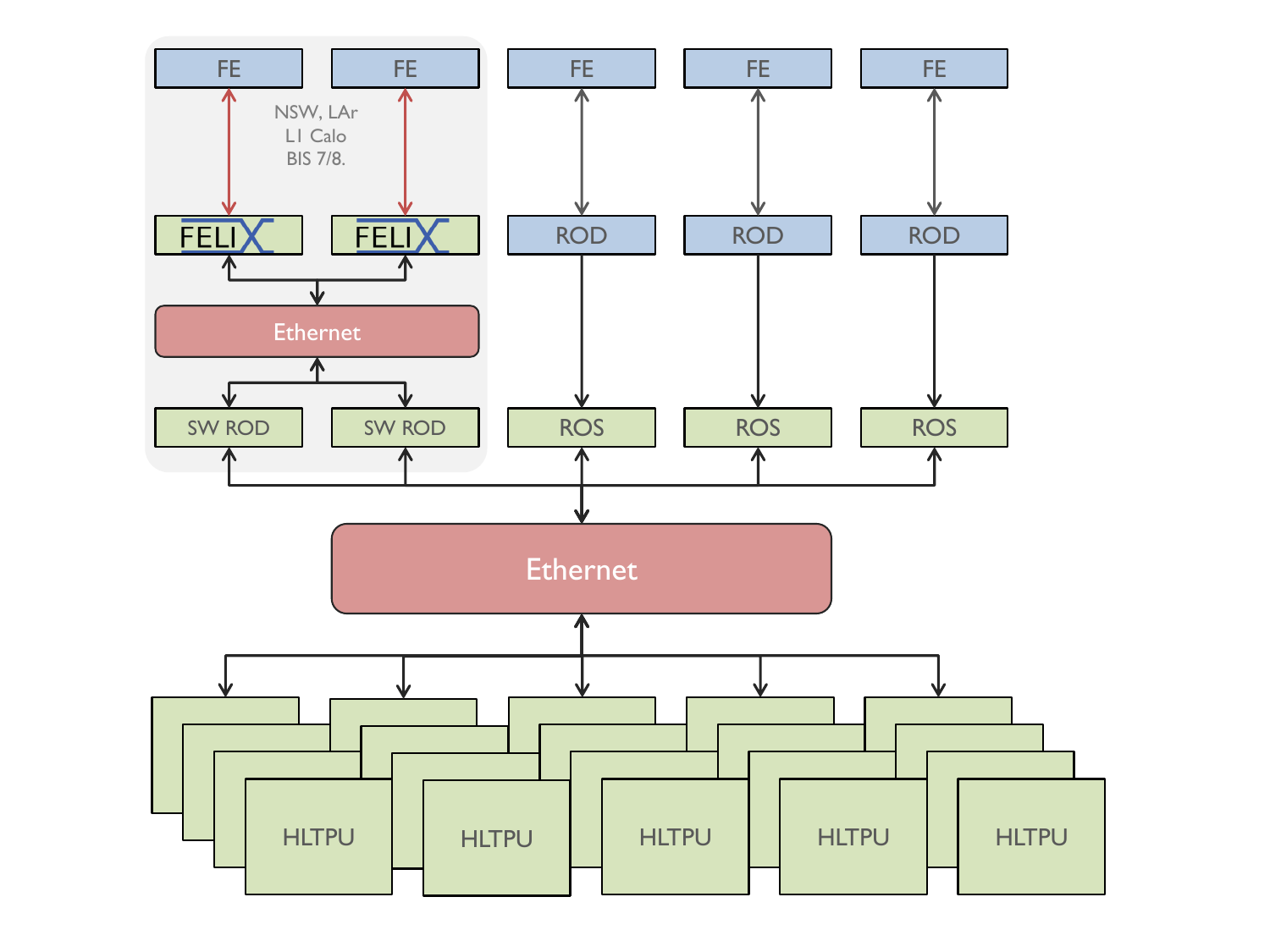}}
\caption{Comparison of Readout Architecture in \RunTwo (right-hand side, with hardware \glspl{ROD}) and \RunThr (pink box on left-hand side) showing new \gls{FELIX} readout paths. \glspl{HLTPU} are able to sample event data from both paths via an identical interface.}
\label{fig:TDAQ_DAQHLT_Run3Arch}
\end{figure}

\gls{FELIX} and \swrod-based readout paths will exist alongside the legacy \gls{ROS} system for the duration of \RunThr. An overview of the architecture is presented in Figure~\ref{fig:TDAQ_DAQHLT_Run3Arch}. \gls{FELIX} and \swrod installations were deployed during \gls{LS2} for systems undergoing significant detector or readout upgrades in preparation for \RunThr operation. These are the \gls{NSW}, \gls{LAr} digital readout and new \gls{L1Calo} systems described earlier in this chapter. Smaller scale demonstrators for upgraded Barrel \gls{RPC-BIS78} and the Tile Calorimeter were also installed during this shutdown. In total the \RunThr installation comprises about \num{60} \gls{FELIX} servers (hosting \num{100} I/O cards between them), with \num{30} software \gls{ROD} servers. This is similar in size to the legacy \gls{ROS} system in \RunThr, which consists of approximately \num{100} servers. The remaining ATLAS systems will then all be migrated to \gls{FELIX} 
during the next long shutdown before the \gls{HL-LHC}, resulting in the final decommissioning of the legacy \gls{ROS}.
 
\gls{FELIX} systems are able to interface with front-end electronics over one of two optical link protocols: GigaBit Transceiver (GBT~\cite{WYLLIE20121561}), a radiation-hard standard developed at CERN, where multiple lower speed links (E-links) from separate pieces of electronics can be aggregated into a single \SI{4.8}{Gb/\s} link; and FULL mode, an in-house design with no link substructure for higher bandwidth (\SI{9.6}{Gb/\s}) communication between \glspl{FPGA}. Data streams for either protocol can be configured to use different encoding, although 8b10b is typically used for normal dataflow.
 
\paragraph{\glstext*{FELIX} \& \glstext*{SW ROD} Server Hardware}
 
Each \gls{FELIX} server is 2U high and hosts custom \gls{I/O} cards (FLX-712, described in Section~\ref{sss:felixio}), with firmware to interface with either of the two link protocols. For the \gls{GBT} case each server hosts two cards, for the higher bandwidth FULL mode case each server hosts one card (driven primarily by the number of available \gls{PCIe} lanes). Each server also hosts high-bandwidth network interface cards (dual-port 25 \gls{GbE} for \gls{GBT}, dual-port 100 \gls{GbE} for FULL mode). Each \gls{FELIX} server has an Intel\textregistered~Xeon\textregistered~E5-1660 V4 CPU (8 cores \@ \SI{3.2}{\GHz}) and 32 GB of \gls{DDR}4 \gls{ECC} \gls{RAM}. Each \swrod server is 1U high, featuring dual Intel\textregistered~Xeon\textregistered~Gold 5218 CPUs (16 cores \@ \SI{2.3}{\GHz}) and 96 GB of \gls{DDR}4 \gls{ECC} \gls{RAM}. Each server also hosts a dual-port 100 \gls{GbE} network interface.
 
\paragraph{\glstext*{FELIX} I/O Card \label{sss:felixio}}
 
\begin{figure}[htbp!]
\centerline{\includegraphics[width=0.8\textwidth]{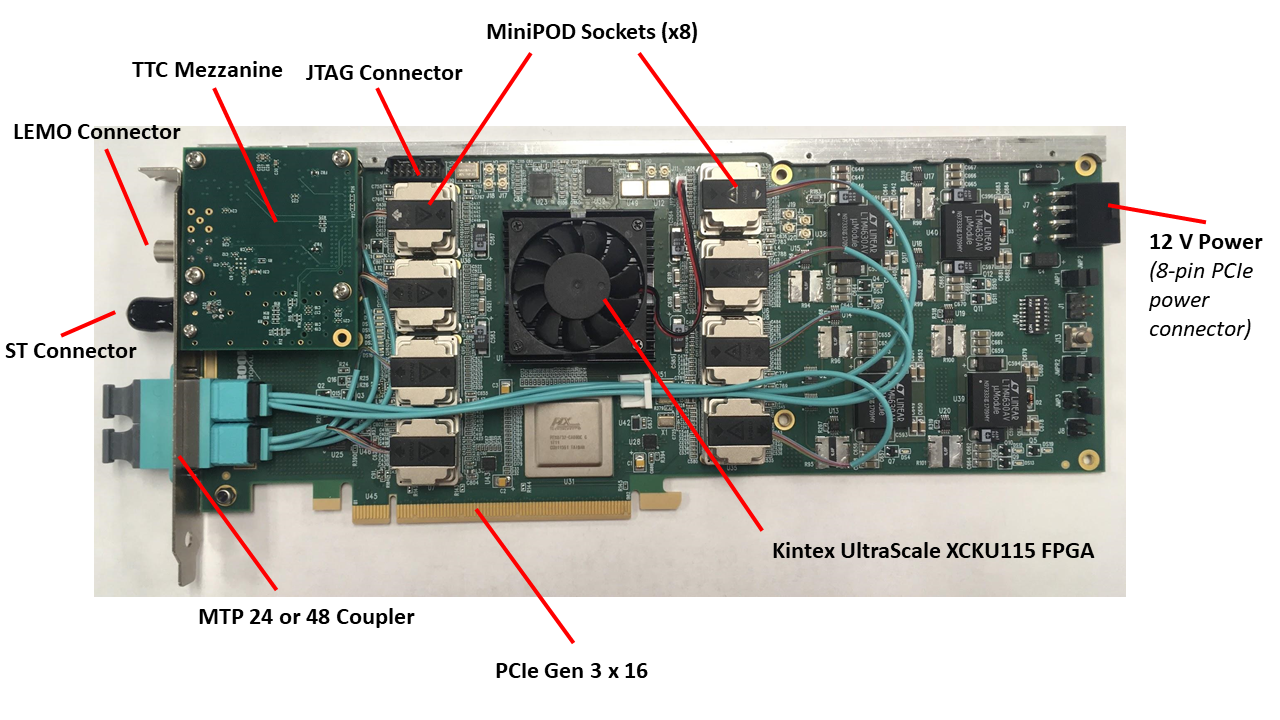}}
\caption{\RunThr \gls{FELIX} hardware platform (FLX-712), with key components labelled.}
\label{fig:TDAQ_DAQHLT_flx712}
\end{figure}

The FLX-712, shown in Figure~\ref{fig:TDAQ_DAQHLT_flx712}, is a \gls{PCIe}  card supporting a 16-lane Gen 3 interface, able to reach a throughput of up to 100 Gb/s. An \gls{MTP} 24 or 48 coupler provides the interface to external data fibres, after which the light is internally routed to one of eight MiniPOD transceivers (four for reception and four for transmission) handling 12 links each. A maximum of 48 bi-directional optical links can therefore be connected to each board. A Xilinx\textregistered~Kintex\textregistered~Ultrascale\texttrademark~(XCKU115) \gls{FPGA} provides the platform for all on-board firmware features. An on-board PEX8732 \gls{PCIe}  switch makes it possible to map two separate \gls{PCIe}  8-lane endpoints into one 16-lane bus. A \gls{JTAG} connector is provided to facilitate \gls{FPGA} configuration, though this may also be stored in an on-board FLASH chip. \gls{FPGA} programming and card health monitoring and control are also possible over \gls{PCIe}.
 
Finally, an interchangeable mezzanine card provides an interface for a number of timing and control systems. The ATLAS \gls{TTC} system in \RunThr connects to \gls{FELIX} via an optical fibre for the distribution trigger and clock information and a LEMO connector for the receipt of BUSY signals. An SI5345 jitter cleaner on-board the FLX-712 itself ensures a sufficiently good quality clock for all \gls{FELIX} use cases.
 
During \gls{LS2},
approximately 60 \gls{FELIX} servers hosting
a total of 100 FLX-712 cards
were
deployed, routing data to 30 \swrod systems. A
significantly larger number (of order 6 times more) will be deployed in \gls{LS3} to service all remaining ATLAS systems.
 
\paragraph{\glstext*{FELIX} Firmware}
 
\begin{figure}[htbp!]
\centering
\includegraphics[width=0.8\linewidth]{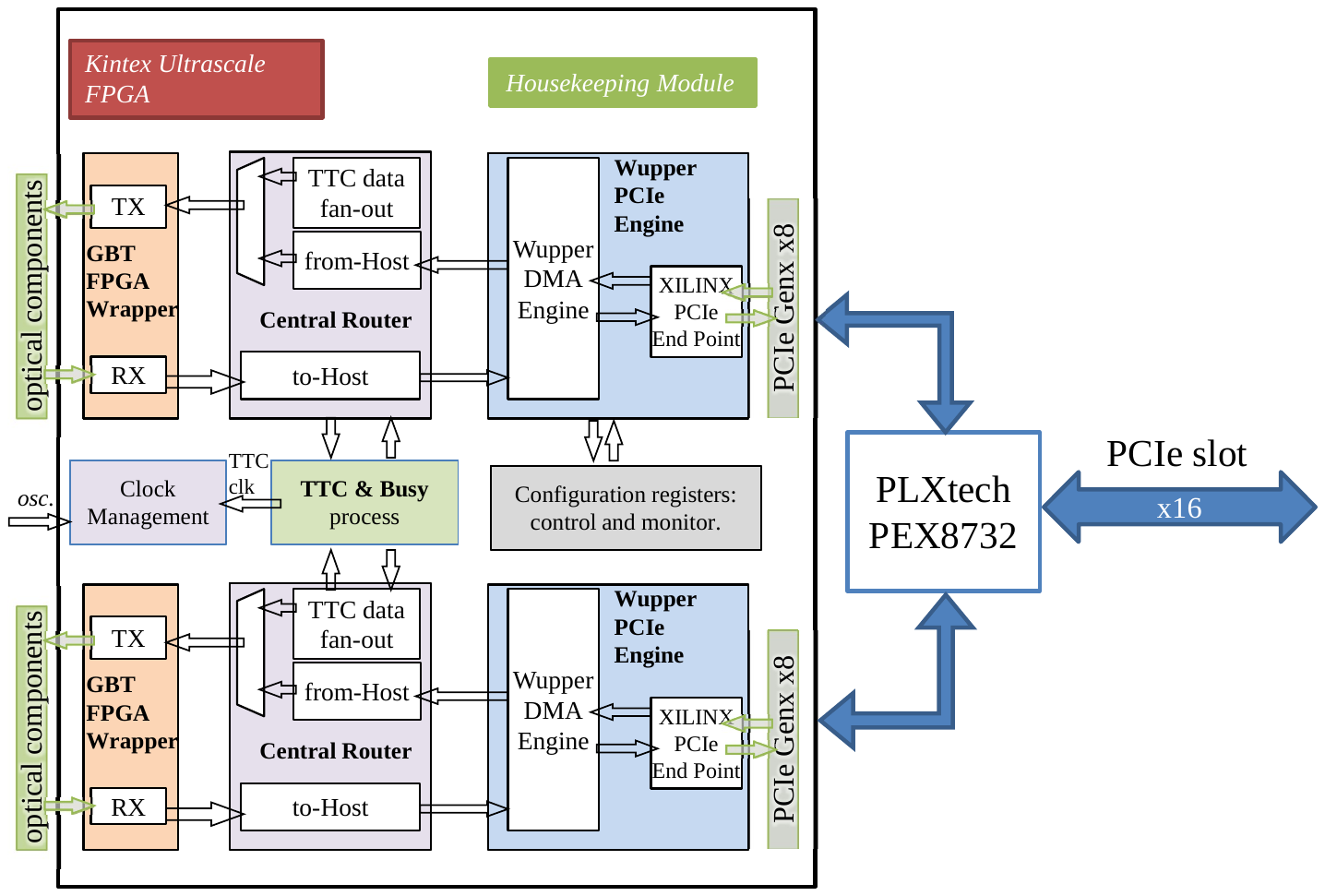}
\caption{Diagram of the firmware deployed on the \gls{FPGA} of the FLX-712 in GBT mode. In FULL mode the incoming (RX) GBT link wrapper is replaced with a dedicated FULL mode module. The Central Router is also much simpler in this case, as the data processing requirements are less severe.}
\label{fig:TDAQ_DAQHLT_flx-712-firmware}
\end{figure}

The \gls{FELIX} firmware, a diagram of which can be found in Figure~\ref{fig:TDAQ_DAQHLT_flx-712-firmware}, is designed to be modular and flexible. Separate components manage different key functions, such as the link wrapper (GBT or FULL mode) and the \gls{PCIe}  and \gls{DMA}~\cite{JEFFERS2013243} engines. Between these lies the Central Router module, which performs the most data intensive workload. Here data arriving over different links are decomposed according to protocol and converted into regular \SI{1}{kB} elements for optimal \gls{DMA} transfer to the host server's memory. In order to optimise \gls{FPGA} resource utilisation and timing, the \gls{FELIX} firmware deployed in the FLX-712 consists of two duplicate paths with identical modules, each servicing half the input links and reading out to an 8-lane \gls{PCIe}  interface. As such, each FLX-712 card appears to the host server as two 8-lane devices.
 
Alongside the primary dataflow path, a separate common module interacts with the \gls{TTC} system, injecting trigger and clock signals into the data path and relaying BUSY signals back to the central trigger should operating conditions require a pause in dataflow. The \gls{TTC} module is also responsible for generating an information packet for each trigger accept received on a special stream for downstream subscribers to use to facilitate event fragment building and synchronisation. Other common modules also manage configuration registers, clock control/distribution and general housekeeping.
 
By re-using the basic blocks above it is possible to flexibly produce firmware designs for different use cases. For ATLAS, separate designs are maintained for both \gls{GBT} and FULL mode, where the primary differences are the link wrapper and the complexity of the Central Router (which is significantly lower for FULL mode). Due to \gls{FPGA} resource utilisation constraints the maximum number of \gls{GBT} links which can be supported for primary dataflow is 24. For FULL mode the limitation comes from the \gls{PCIe}  bandwidth of the FLX-712, which can accommodate a maximum of 12 links. However, the standard approach is to build firmware to support up to 24 links, this giving the option to operate more than 12 at lower occupancy. Thus a higher link density can be provided within the same bus constraints.
 
\paragraph{\glstext*{FELIX} Software}
 
\begin{figure}[htbp!]
\centering
\includegraphics[width=0.75\linewidth]{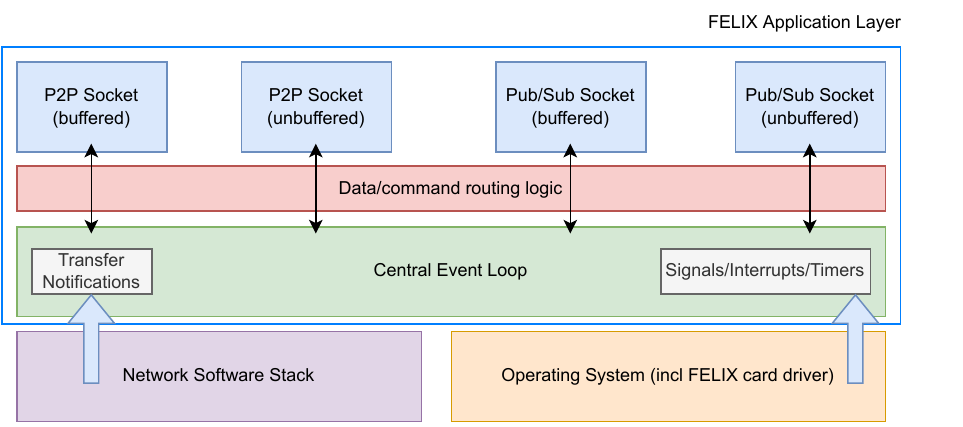}
\caption{Diagram showing the different layers of the \gls{FELIX} software architecture, with the central event loop (shown in green) able to interact with low level network software (supporting \gls{RDMA}) and operating system signals (including interrupts from the \gls{FELIX} \gls{I/O} card). The higher-level layer implements application-specific logic and handles subscriptions from external clients via dedicated sockets, available both in peer-to-peer or publisher/subscriber mode. Sockets can also either operate in ``buffered'' mode, whereby data are packed into an output buffer before transfer across the network (useful to optimise throughput for small packets) or ``unbuffered'' mode, where data are transferred immediately (useful for larger packets).}
\label{fig:TDAQ_DAQHLT_felix_software}
\end{figure}

The \gls{FELIX} software suite comprises high- and low-level components. Alongside a dedicated device driver, low-level tools make it possible to test all firmware features in a laboratory setting and debug any issues which may arise. At a higher level, a high-performance daemon operates in an ``always on'' fashion in order to receive data from the FLX-712 over \gls{DMA} and provide onward routing. \gls{DMA} transfers are received via a separate ring buffer for each \gls{PCIe}  device visible to the host server (hence two per FLX-712). The software daemon, based on the NetIO architecture~\cite{8071057} is designed to be event driven, able to react to hardware interrupts from the card, indicating incoming data, or signals from the network interface or operating system. The design is such that copies of the data in memory are kept to a bare minimum to maximise throughput. Finally, \glsfirst{RDMA} technology is used to transfer data to connected network peers without intermediate processing. An overview of the complete software stack is presented in Figure~\ref{fig:TDAQ_DAQHLT_felix_software}.
 
\paragraph{\swrod Design}
 
\begin{figure}[htbp!]
\centerline{\includegraphics[width=0.8\textwidth]{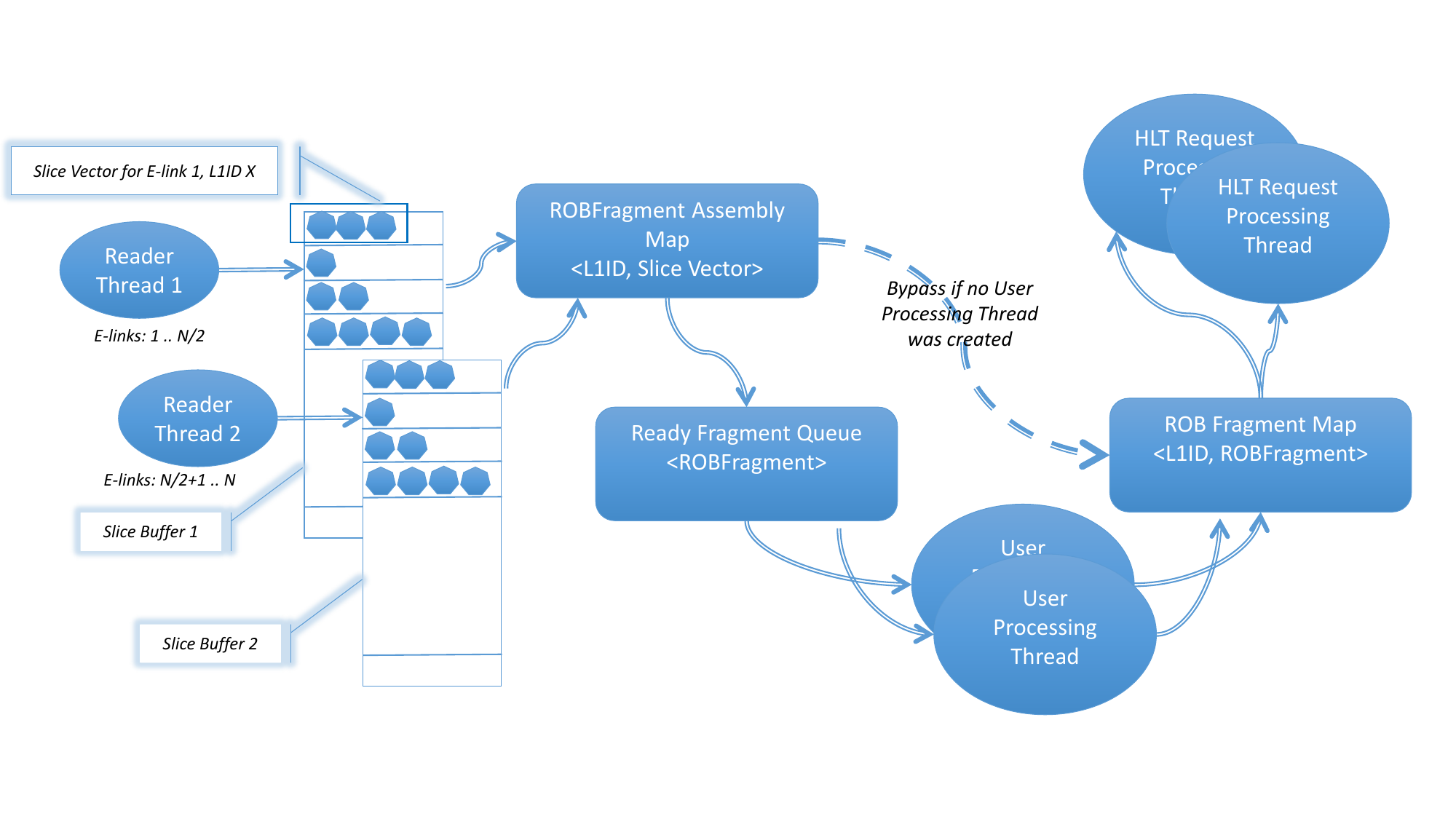}}
\caption{\swrod Architecture with subdetector plugins. Data are aggregated by common \gls{L1ID} in the input stage before being passed (if needed) to dedicated subdetector code, before being buffered and transferred onward as needed during \gls{HLT} processing. In this example only two Reader Threads are used for clarity, but in principle more can be configured as needed.}
\label{fig:TDAQ_DAQHLT_swrod}
\end{figure}

The \swrod is implemented as an ATLAS Run Control-aware application running on dedicated servers. The application is designed to be able to support flexible workloads, with configurable event fragment building and processing. The overall architecture of the \swrod application is shown in Figure~\ref{fig:TDAQ_DAQHLT_swrod}. At the input stage, dedicated ``Reader'' threads subscribe to \gls{FELIX} data streams and write incoming packets into ``slice'' buffers, with each buffer made up of ``vectors'' of data packets corresponding to a specific \gls{L1ID} for a specific input link. From here packets are placed into an associative ``\gls{ROB}Fragment assembly'' map structure, with all data in a previously assembled slice associated with their corresponding \gls{L1ID}. Thus the aggregation from many packets per \gls{L1ID} to one \gls{ROB}Fragment object happens in two steps. At this point, if required, fragments are placed into a ``Ready'' queue for handling by a subdetector-specific plugin, operating in a dedicated user processing thread. Within this thread, a variety of fragment analysis, error checking, monitoring and statistics gathering can take place, making it possible to replicate the actions of the old generation of hardware \glspl{ROD} which the \swrod replaces. Should such processing not be required fragments can be sent directly to the output stage. At this stage, fragments are held in a ``\gls{ROB}Fragments'' map while the \gls{HLT} performs selection operations on the events of which they form part. A pool of request processing threads transfers any requested fragment data from the map to the \gls{HLT} during this period. This final stage is designed to be functionally identical to the legacy \gls{ROS} buffers, making use of a common interface to the \gls{HLT}, such that any \gls{HLT} node requesting data makes no distinction as to whether it comes from a \gls{ROS} or a \swrod.
 
\subsubsection{Network}
 
As part of the preparations for \RunThr, the \gls{TDAQ} network is undergoing a major programme of updates. The overall design is based on two physical levels, with machines operating in the electronics cavern (\gls{FELIX}, \swrod, \gls{ROS} and various \gls{DCS} and detector infrastructure nodes) communicating with systems on the surface (\gls{HLT}, core infrastructure, control calibration and monitoring nodes) via a 40~\gls{GbE} backbone. Different communication workloads are supported, from bulk dataflow to diverse control and monitoring traffic.
 
\begin{figure}[htbp!]
\centerline{\includegraphics[width=0.8\textwidth]{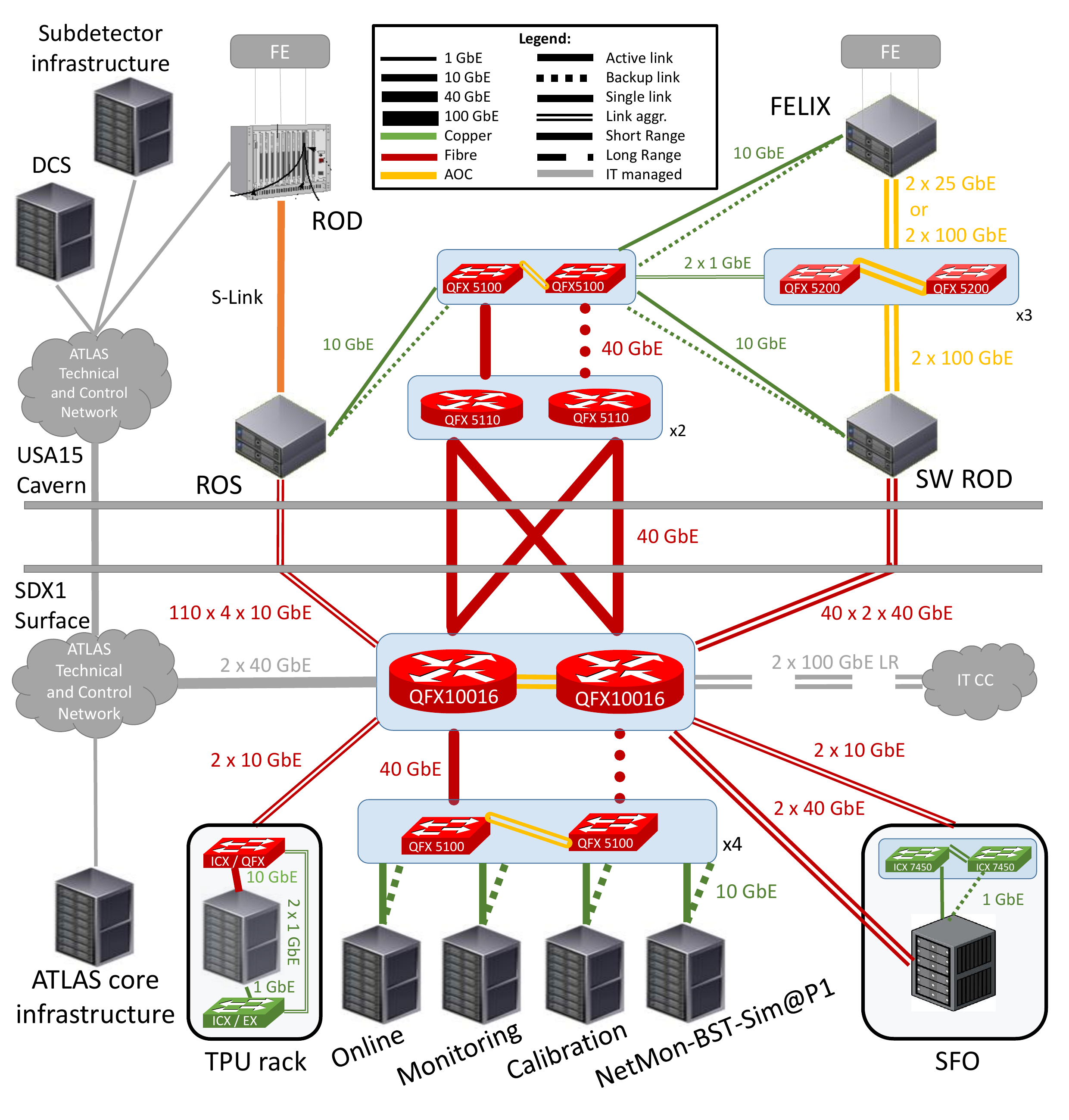}}
\caption{\gls{TDAQ} Network for \RunThr, showing upgraded core routers (QFX10016 in the centre) and new \gls{FELIX}/\swrod network (top right). Client routers (QFX5100) act as the interface between the core and the processing farms performing online control, monitoring and calibration functions. Nodes participating in primary dataflow (such as \gls{ROS}, \swrod, TPUs and \gls{SFO}) are connected directly to the core. In the case of \gls{ROS}, \gls{FELIX}, and \swrod, separate connections for the purposes of node control are also present. The \gls{ATCN} operates in parallel, connecting \gls{DCS} and ATLAS core infrastructure, and interfaces to the \gls{TDAQ} network via the core and via connections to hardware \gls{ROD} components.
}
\label{fig:TDAQ_DAQHLT_network}
\end{figure}

While the overall core throughput between the electronics cavern and the surface has not changed since \RunTwo (remaining at 40~\gls{GbE}), the overall number of nodes connected to the network has increased significantly, with more extensive deployment of virtual networks to serve different functions, eliminating the need for further physical interconnects. The network's router infrastructure has also been replaced, both at the core router level (now two Juniper QFX10016s) and all client switches at various levels (now made up of 50 Juniper QFX5100), with improved performance and the introduction of active-active redundancy (based on MC-LAG technology). All routers now make use of the next generation of \glspl{ASIC} (Broadcom Trident II in the client switches and Juniper Q5 in the core routers). A mixture of copper and optical fibre connectivity is used, so as to optimise the cost for each use case. The overall architecture of the network for \RunThr is shown in Figure~\ref{fig:TDAQ_DAQHLT_network}, along with a more detailed description of node connectivity.

The majority of new clients on the \gls{TDAQ} network belong to the new \gls{FELIX} and \swrod readout paths. Each \gls{FELIX} server hosts a dual 25 or 100 \gls{GbE} network interface card, with each \swrod hosting a dual 100 \gls{GbE} interface. The servers are connected via dedicated high-performance switches (a total of six Juniper QFX 5200), which feature lossless operation over \gls{RDMA} and active-active redundancy. Each \swrod then features a separate dual 40~\gls{GbE} interface to the common dataflow network, for control and bulk dataflow through to the \gls{HLT} and beyond. Further new nodes may also be added to the network in future as the \gls{HLT} farm continues to expand on the surface. As such, additional contingency has been factored into the specification of the core network to allow for such evolution to take place.
 
As mentioned before, the physical network is subdivided into separate \glspl{VLAN}, each with a dedicated quality of service policy, to enable efficient management of separate workloads without extra cabling. There are five separate \glspl{VLAN} in operation: Management, for network control and monitoring; Control, for server control (DHCP, NFS, DNS, etc.) and \gls{DAQ} control; Data, for \gls{DAQ} traffic; Monitoring, for Low-priority monitoring and backup (operated on a best-effort basis)
and finally Sim\@P1: for simulation jobs running in the \gls{HLT} farm. With this implementation, virtual router instances also provide traffic isolation and enhanced security (e.g. the ATLAS technical network is protected against traffic flooding).
 
Beyond the upgrades described above, significant effort has gone into reworking the cabling structure in the surface cavern and improving uninterruptible power supply coverage for the \gls{DAQ} network.
 
\subsubsection{HLT/AthenaMT}
\label{subsubsec:tdaq_daq_hlt_athenamt}
 
Event processing frameworks for most \gls{LHC} experiments have traditionally been designed to process single events serially on a single \gls{CPU} core, with events distributed between independent processing nodes. Unfortunately, for offline processing, such a model no longer matches trends in computer architecture. Crucially, while the number of \gls{CPU} cores available in a standard compute node has increased, the amount of memory per core is increasing at a lower rate. In order to effectively make use of the increasing core count, the average memory utilisation per job must therefore be reduced. The most effective way to achieve this is to share memory between cores. This can be achieved by sharing event processing between multiple threads, i.e. multi-threading, thus reducing the overall memory footprint per core. Memory has never been a limiting factor for the ATLAS \gls{HLT}, due to the different operational model, but the system can still benefit from any optimisation to gain additional margin for future evolution, and to aid with other workflows with differing resource limitations such as HLT simulation as part of MC event generation.
 
To tackle the memory issue, a significant redesign of the ATLAS software framework (Athena) was undertaken to allow it to process events across multiple threads. The new \gls{AthenaMT} framework to be used for \RunThr positions ATLAS software to optimally exploit future server technology evolution. An additional feature of \gls{AthenaMT} is its built-in hooks to facilitate potential future use of co-processors, such as \glspl{GPU} and \glspl{FPGA}. Asynchronously offloading compute-intensive tasks to these devices can free up \gls{CPU} cores for work better suited to the \gls{CPU}. While the use of co-processors is not planned for the start of \RunThr, the structure of \gls{AthenaMT} means they can be integrated in future without major architectural changes.
 
The Athena framework is mainly written in C++, with a configuration layer in Python.  The underlying Gaudi framework~\cite{gaudi} is shared with the \acrshort{LHCb} experiment. The Gaudi layer defines the basic classes used for event processing, and also provides a component known as the scheduler, which is responsible for optimal algorithm execution.  In \RunTwo, the Gaudi scheduler was used by offline reconstruction,  but the \gls{HLT} used a custom implementation: “Trigger Steering”. The steering layer was needed to implement additional \gls{HLT} functionality, to facilitate features such as RoI-based processing and the sequencing of trigger selection criteria to arrive at a decision for each event. The Trigger Steering layer fulfilled these requirements successfully during \RunOneTwo, but with a significant development and maintenance overhead.
 
\begin{figure}[htbp!]
\centerline{\includegraphics[width=0.8\textwidth]{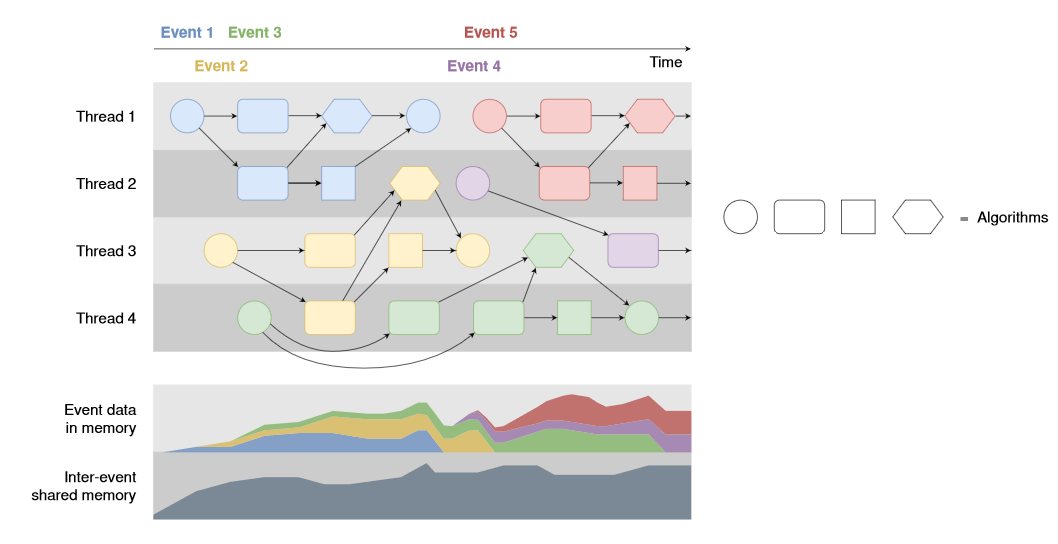}}
\caption{Diagram demonstrating the three different kinds of parallelism available with \gls{AthenaMT}: inter-event, where multiple events are processed in parallel; intra-event, where multiple algorithms can run in parallel for an event and in-algorithm, where algorithms can utilise multi-threading and vectorisation.}
\label{fig:TDAQ_DAQHLT_athenamt_parallel}
\end{figure}

Taking into consideration the above, the key drivers of the design of the new \gls{AthenaMT} framework were to reduce maintenance overhead and to make effective use of hardware. As already discussed, effective use of modern multi-core machines requires multi-threading. From the beginning, the new framework was designed to meet both offline and trigger requirements, eliminating the need for a custom trigger-specific layer. Data and control flow, as well as regional reconstruction, were designed to be part of the scheduler. \gls{AthenaMT} makes it possible to implement three different kinds of parallelism: inter-event, where multiple events are processed in parallel; intra-event, where multiple algorithms can run in parallel for an event and in-algorithm, where algorithms can utilise multi-threading and vectorisation. A pictorial representation of the different options is presented in Figure~\ref{fig:TDAQ_DAQHLT_athenamt_parallel}.
 
As described in Section~\ref{sec:Overview:PhysPerfGoals}, the trigger menu translates the physics priorities of the experiment into allocations of the total \gls{L1} and \gls{HLT} rates. Each unique selection combination is called a chain, with the set of chains making up the menu. During \RunTwo, more than 2000 unique combinations of trigger selection criteria were employed to distinguish interesting events.  The configuration for each selection was generated from a large 
Python ``TriggerMenu'' package. As it is important to know exactly which menu is used both for data-taking and production of simulated samples, the configuration is stored in a database.  As part of the software upgrade for \RunThr, both the actual configuration and its database storage mechanism have been redesigned. The key requirement was to reduce the complexity of the configuration and to ease maintenance and development overhead while also improving the performance of the configuration serialisation/deserialisation. The reduced complexity is the result of a redesign of the configuration (both in the trigger and across Athena as a whole), while the improved serialisation/deserialisation performance comes from storing the configuration in a JSON~\cite{JSON} representation rather than in a relational format.
 
\begin{figure}[htbp!]
\begin{minipage}{0.4\textwidth}
\centerline{\subfloat[]{\label{fig:hltmppu}\includegraphics[width=\textwidth]{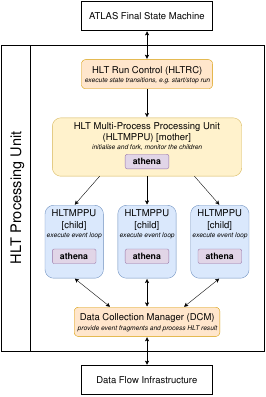}}}
\end{minipage}\hfill
\begin{minipage}{0.55\textwidth}
\centerline{\subfloat[]{\label{fig:hltputree}\includegraphics[width=\textwidth]{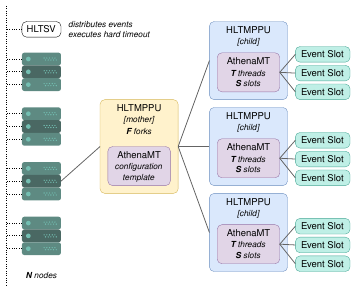}}}
\end{minipage}
\caption{\protect\subref{fig:hltmppu} Athena process operating within the wider \gls{HLT} processing framework in \RunTwo, where multiple forked instances performed all processing and \protect\subref{fig:hltputree} \RunThr equivalent, potentially with fewer forks, with each fork capable of spawning multiple processing threads\protect\footnotemark.}.
\label{fig:TDAQ_DAQHLT_hlpu}
\end{figure}

The \gls{AthenaMT} software framework must interact with the online trigger and data-acquisition infrastructure, as well as new hardware triggers installed for \RunThr. During \RunTwo, each node in the \gls{HLT} forked multiple sub-processes, thus sharing memory between them via the copy-on-write mechanism. In \RunThr, each forked sub-process will additionally run with multiple threads, and potentially process multiple events across these threads.  This design implies that tuning of the number of forks, threads and concurrent events will be necessary to ensure maximum performance. It also implies that if a sub-process crashes or takes too long, all concurrently processed events must be saved for offline reprocessing. A comparison of the \RunTwo and \RunThr architectures is presented in Figure~\ref{fig:TDAQ_DAQHLT_hlpu}.
 
\footnotetext{Image includes an icon made by Smashicons from \url{flaticon.com}.}
 
\subsubsection{Online Software \& Monitoring}
 
\begin{figure}[htbp!]
\centerline{\includegraphics[width=\textwidth]{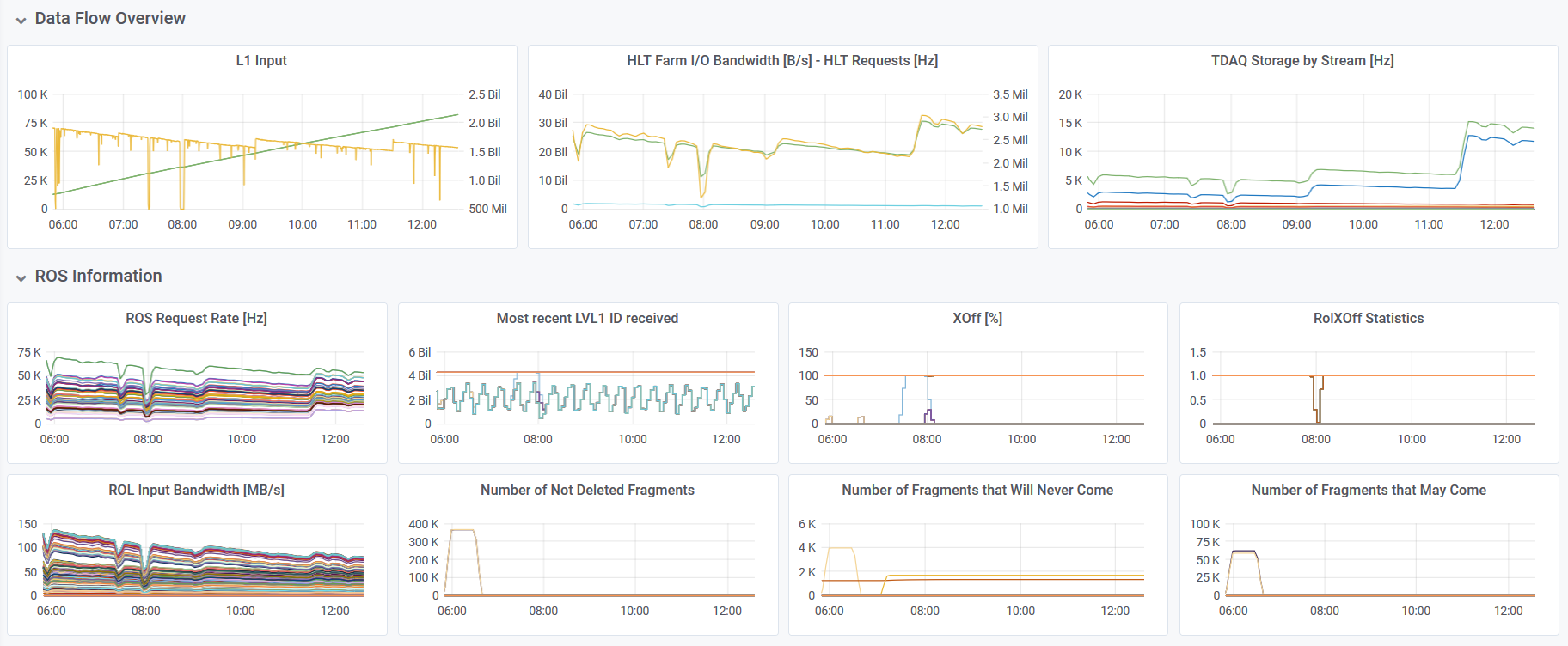}}
\caption{Example dashboard produced with the Grafana web interface.}
\label{fig:TDAQ_DAQHLT_grafana}
\end{figure}

The ATLAS Control \& Configuration software suite underwent a significant
overhaul for the start of \RunTwo. The features introduced have since been
refined throughout data taking; thus a modern and flexible system was in place at the
the start of \RunThr. The goals of the upgrade
were three-fold: first, to properly accommodate additional requirements that
could not be seamlessly included during data taking; to re-factor software that
had been repeatedly modified to include new features, and make it more
maintainable; and to take the opportunity to modernise the software base,
making the most of the rapid evolution in information technology during the lifetime
of the \gls{LHC}. This upgrade was carried out retaining the important constraint
to minimally impact the operational mode of the system or any public \glspl{API}, thus
making it as easy as possible for the large user community to integrate all the
changes. An example of the changes is the introduction of a completely redesigned
online monitoring archiving engine (P-BEAST) able to aggregate and serve data to
a number of clients (principally a Grafana web interface for the production
of time series displays). P-BEAST interfaces with the existing statistics distribution
system (IS) without \gls{API} changes in the latter, but provides enormous improvements
in user experience and monitoring capability. An example of a typical online monitoring
time series display with P-BEAST and Grafana is shown in Figure~\ref{fig:TDAQ_DAQHLT_grafana}.
 
\begin{figure}[htbp!]
\centerline{\includegraphics[width=0.8\textwidth]{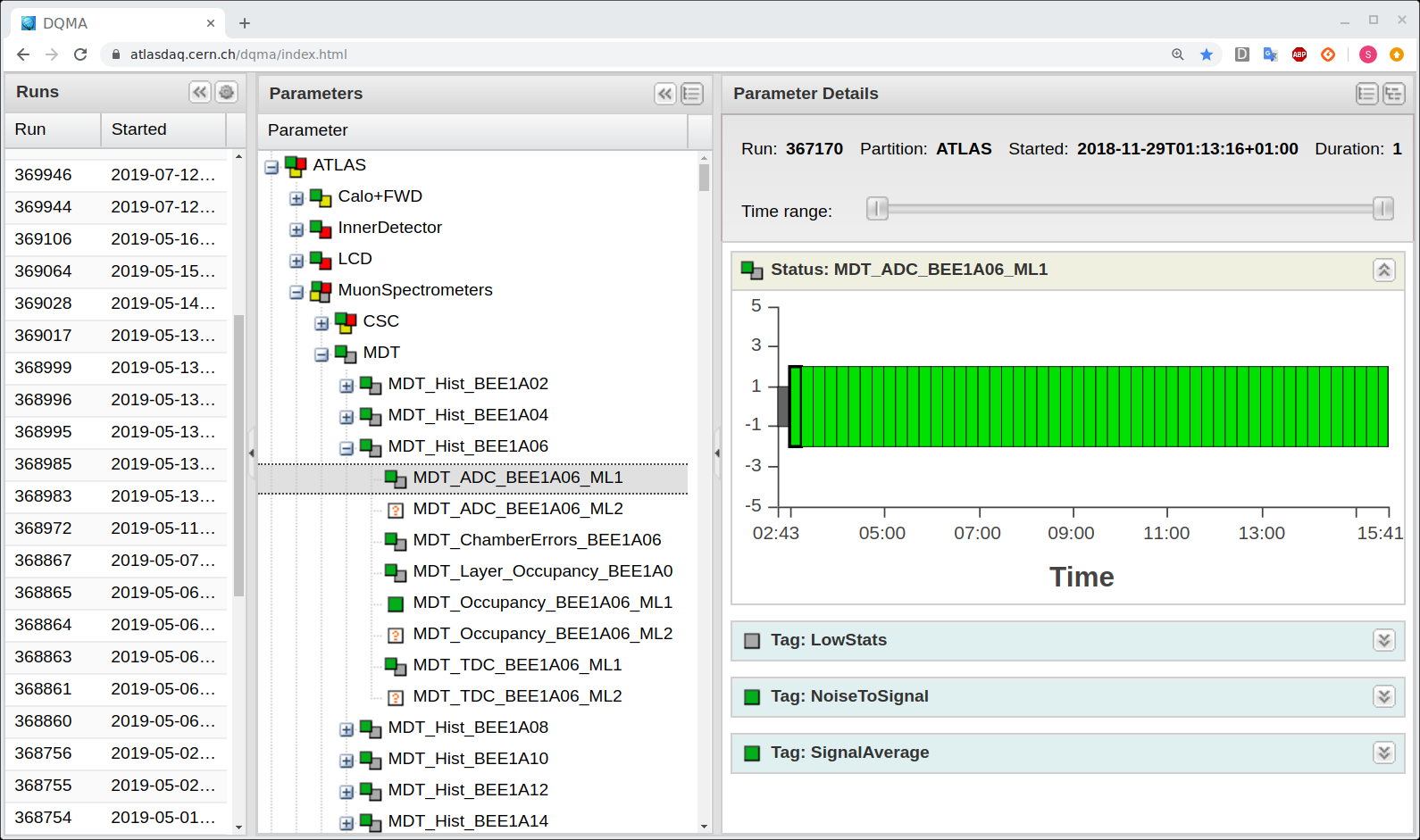}}
\caption{Example screenshot of the Data Quality Monitoring Application.}
\label{fig:TDAQ_DAQHLT_DQMA}
\end{figure}

Alongside the tools already mentioned, ATLAS also deploys specific Data Quality monitoring tools
to aid shifters in the task of monitoring the performance of a data-taking session. The Data Quality Monitoring Framework automates data quality assessment by applying data quality algorithms to the histograms produced by event selection applications running in the \gls{HLT}. The results produced by these algorithms are displayed to the shift crew in a hierarchical manner, so that the shift operator can quickly spot problems and easily identify their origin. This framework was further enhanced throughout \RunTwo with the addition the Data Quality Monitoring Archiver, a new feature to store results in ROOT format for offline analysis. An example of a web display produced by the tool's integrated archive viewer is shown in Figure~\ref{fig:TDAQ_DAQHLT_DQMA}.
 
\subsection{Detector Control System\label{sec:DCS}}

 
The purpose of the \glsfirst{DCS}~\cite{ATLAS_DCS_Jinst} is to ensure coherent and safe operation of ATLAS and to serve as a homogeneous interface to all subdetectors and to the technical infrastructure of the experiment.
The \gls{DCS} must be able to bring the
detector into any desired operational state, to continuously monitor
and archive the operational parameters, to signal any abnormal
behaviour to the operator, and to allow manual or automatic actions to
be taken.
It must also provide bi-directional
communication between \gls{DCS} and run control to synchronise the state of the detector with the
operation of the data acquisition system.
The \gls{DCS} also handles interactions between ATLAS
sub-detectors and other systems that are controlled externally,
such as the \gls{LHC} accelerator, CERN technical services, the ATLAS
magnets and the \gls{DSS}.
 
Apart from consolidation and modernisation efforts for the \gls{DCS} hardware and software components, several conceptual upgrades have been implemented since \RunOne. First, the standard middleware employed by all systems was migrated from \gls{OPC} Data Access to the \glsfirst{OPC UA} standard~\cite{OPCUA}. Middleware applications interface front-end components with the \gls{DCS} back-end software applications based on the \gls{WinCC OA} \gls{SCADA} package~\cite{PVSS} (formerly known as PVSS). Second, for the new detector systems introduced for \RunThr, \gls{DCS} functions of most of the new or upgraded front-end electronics components are covered by the \gls{GBT-SCA} \gls{ASIC}\cite{GBTSCA,GBTSCA2} which communicates with the back-end electronics using the same optical links as the readout or trigger data processed by the \gls{FELIX} system. These monitoring and control data are relayed to and from the \gls{DCS} back-end by the \gls{GBT-SCA} software suite, a dedicated set of middleware applications. Finally, a significant part of the new back-end electronics of the upgrade systems is based on crates and blades following the \gls{ATCA} standard for which a dedicated integration solution was conceived. The \gls{ATCA} equipment significantly extends the \gls{DCS} functionality compared to the previous implementation based on \gls{VME} crates. These three aspects are described in more detail in the following three sub-sections.
 
\subsubsection{Standard Middleware \glstext*{OPC UA} }
\label{TDAQ_DCS_OPCUA}
 
The middleware standard for \gls{DCS} systems is \gls{OPC UA}, an industry
standard for machine-to-machine communication in the controls domain, allowing independence from the operating system and development in various programming environments. Features like robust data modelling, 
custom hardware embedding and secure communications are among the advantages of this standard.

In order to reduce development and maintenance efforts, a framework
for \gls{OPC UA}  server creation is available -- the
\texttt{\gls{quasar}} \cite{bib:quasar, bib:quasar_CHEP}. Development starts with
creation of a design file, in \gls{XML} format, describing an
object-oriented information model of the target system or
device. Using this model, the framework generates an executable \gls{OPC UA}
server application, which exposes the per-design \gls{OPC UA}  address
space, without the developer writing a single line of
code. Furthermore, the framework generates skeleton code into which
the developer adds the required target device/system integration
logic.  This approach allows both developers unfamiliar with the \gls{OPC UA}  standard, and advanced \gls{OPC UA}  developers, to create servers
for the systems they are experts in while greatly reducing design and
development effort as compared to developments based purely on \gls{COTS}
\gls{OPC UA}  toolkits. Higher level software may further benefit from
the explicit device model by using the \gls{XML} design description as the
basis for generating client connectivity configuration and server data
representation. Moreover, having the \gls{XML} design description at hand
facilitates automatic generation of validation tools.
 
\begin{figure}[pht]
\begin{center}
\includegraphics[width=\linewidth]{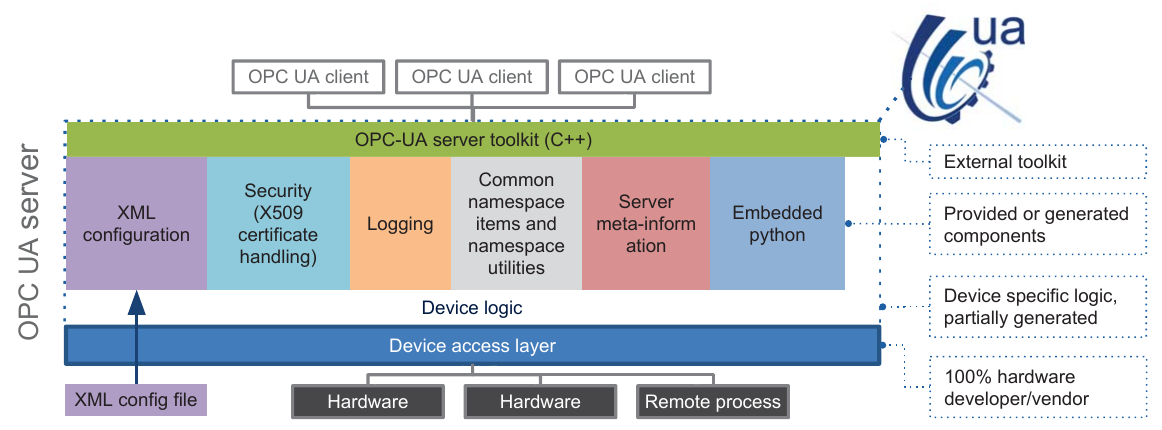}
\caption{\label{fig:DCS:quasar} \gls{OPC UA}  server architecture within the \texttt{\gls{quasar}} framework.}
\end{center}
\end{figure}
 
Figure~\ref{fig:DCS:quasar} gives an overview of the different layers
of \texttt{\gls{quasar}} put into context. Controllable devices or
systems are accessed using their specific access layer -- often
provided together with the specific device.  The device logic layer
functions as an interface with the high level layers of
\texttt{\gls{quasar}}, which comes in several modules covering
different functional aspects.  The address space module sits in the
\gls{OPC UA}  end of the server, exposing data towards \gls{OPC UA}
clients, and is implemented using an \gls{OPC UA}  back-end layer with
exchangeable back-end implementations. A configuration module
facilitates address space and device instantiation and the definition
of their relations.  \gls{XML} is used as the configuration format backed by
\gls{XML} schema definitions. A  \gls{XML} schema to C++\ mapping generator is used to build actual instances from configuration files.  An additional subsystem called Calculated Items, operating entirely in the address space, enables creation of new variables which are derived from existing ones using mathematical
functions. \texttt{\gls{quasar}} further comes with optional modules
such as component based logging, certificate handling, server
meta-data, embedded Python processing, \gls{WinCC OA} integration
tools and \gls{SQL}/\gls{NoSQL} archiving with historic data access. A
ready-to-use build system based on CMake along with pre-configured
tool-chains for several platforms such as \texttt{x86\char`_64} or
ARM-based Linux and Microsoft Windows are provided. Finally, an
\gls{OPC UA}  client generation facility called UaoForQuasar is
available for building C++\ clients for \texttt{\gls{quasar}}-based
servers.
 
By the start of \RunThr, all middleware applications for Java Card OpenPlatform-supported devices such as power supplies and the \gls{ELMB}~\cite{ELMB}, as well as their \gls{WinCC OA} integration components were migrated to the \gls{OPC UA}  standard. Furthermore, at the time of writing, more than 20 different \gls{OPC UA}  server implementations for numerous custom device types used in ATLAS have been developed and integrated, including servers running on embedded platforms such as system-on-chip devices.
 
\subsubsection{Controls Software for \glstext*{GBT-SCA}-based Front-End Electronics}
\label{TDAQ_DCS_GBT_SCA}

The \gls{GBT-SCA} is a radiation-tolerant \gls{ASIC} and part of a chip-set of the \gls{GBT} project, providing simultaneous transfer of readout data, timing and trigger signals as well as slow control and monitoring data, by multiplexing multiple logical electrical data links, onto a single optical link using the rad-tolerant \gls{GBTx} \gls{ASIC} on the front-end side.
The \gls{GBT-SCA} serves as an interface to the control and monitoring signals of front-end electronics on the detectors, using two redundant \glspl{e-link} to connect to a \gls{GBTx}.
 
The  \gls{GBT-SCA} software suite \cite{icalepcs2019-wepha102} provides a high level of abstraction and an interface to all communication channels of a \gls{GBT-SCA}, profiting from the hardware parallelism between independent channels. To ensure reliability, the software does the necessary bookkeeping for the synchronous communication and transaction tracking.
 
\begin{figure*}[!tbh]
\centering
\includegraphics[width=\textwidth]{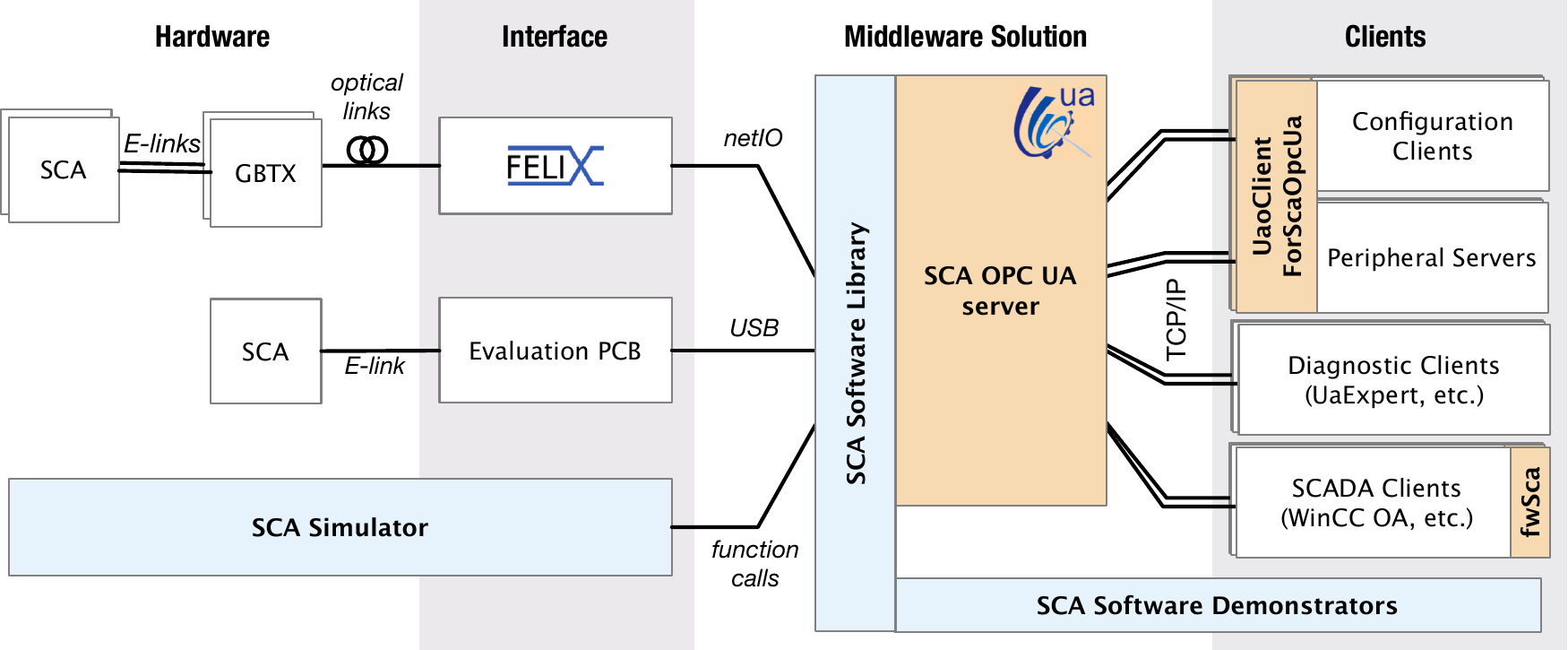}
\caption{Global picture of the software suite. The \gls{GBT-SCA} Software package, in light blue, comprises the \gls{GBT-SCA} Software \gls{API} to communicate with the  \gls{GBT-SCA} via different back-ends, the \gls{GBT-SCA} Simulator to emulate  \gls{GBT-SCA} traffic for testing and development, and the Demonstrator tools which are used for standalone operations. The \gls{GBT-SCA} \gls{OPC UA} server and its ecosystem, in orange, is the middleware of choice to exchange data with the front-ends. UaoClientForScaOpcUa is a library that clients use to communicate with the  \gls{GBT-SCA} server. Finally, the fwSca module automates the integration of the server data into  \gls{SCADA} systems.}
\label{fig:globalPicture}
\end{figure*}

Moreover, the software achieves high performance and low latency, including features such as grouping requests for lengthy operations requiring transfers of large amounts of data over \gls{JTAG}, such as \gls{FPGA} programming. Since thousands of  \glspl{GBT-SCA} are used in the detector systems, scalability is an important design aspect. At the same time, monitoring and control tasks require availability close to 100\%, implying the need for a high level of robustness. For the final production system, the \gls{GBT-SCA} software is interfaced with the optical link receiver system, \gls{FELIX}, via a dedicated communication link called netIO.
Figure~\ref{fig:globalPicture} shows an overview of the  \gls{GBT-SCA} integration chain, illustrating the interplay of the components of the \gls{GBT-SCA} software suite.
 
\paragraph{ \glstext*{GBT-SCA} \glstext*{OPC UA}  Ecosystem}
 
In the \gls{GBT-SCA} software package core there is a library that is structured in modules that implement the required functionality in various layers. The library was designed to be flexible and easily adaptable to the diverse systems intended to use it by its polymorphic high-level data link control back-end. Moreover, the \gls{GBT-SCA} software package contains the Demonstrators, which are tools that directly use the library and are used for testing and for low-level diagnostics. Finally, as part of the package, a \gls{GBT-SCA} Simulator was developed that is able to generate  \gls{GBT-SCA} traffic, simulating realistic  \gls{GBT-SCA} behaviour, in order to allow for development and testing without real hardware.
 
The \gls{GBT-SCA} \gls{OPC UA} server was implemented using the \texttt{\gls{quasar}} framework, taking advantage of \texttt{\gls{quasar}}'s built-in features such as calculated variables, threading, different types of variables and methods. The server architecture divides the  \gls{GBT-SCA} channels into device classes, according to their respective hardware functions. In addition, a Global Statistician module was developed to collect and measure general statistics across the setup and to expose the collected metrics to the clients. Finally, a \gls{GBT-SCA} Supervisor software module oversees the state of the system and provides supervisory functionality such as automatic recovery from communication loss with \glspl{GBT-SCA},  \gls{GBT-SCA} ID validation and other administrative tasks.
 
\begin{figure}[!htb]
\centering
\includegraphics[width=0.6\textwidth]{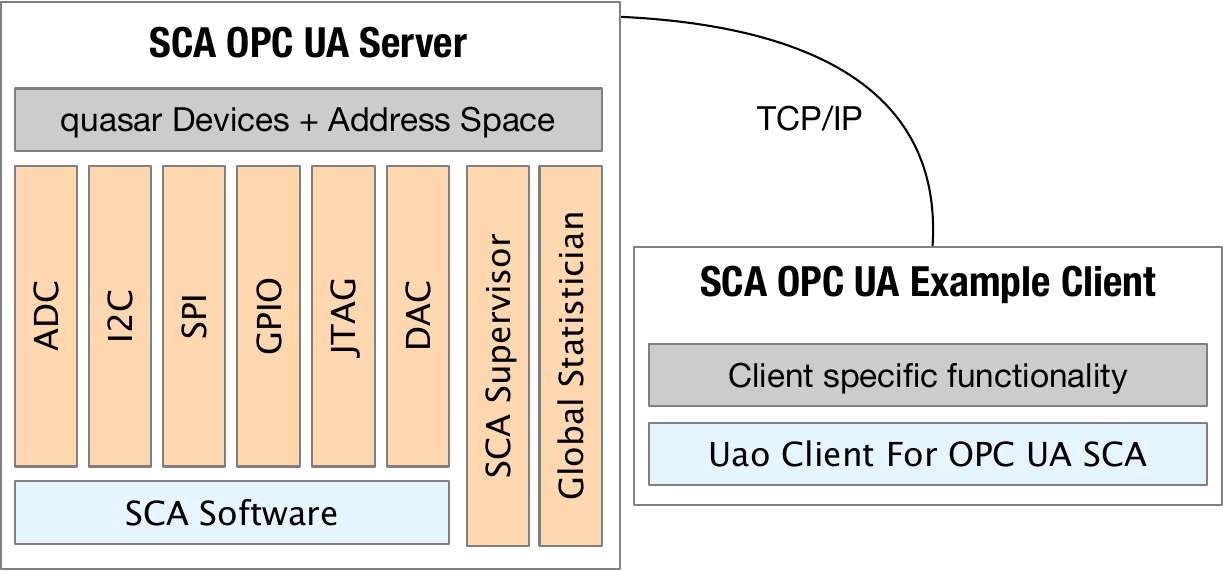}
\caption{\gls{GBT-SCA} \gls{OPC UA} stack.}
\label{fig:scaopcua-stack}
\end{figure}
 
For any sub-system application using the  \gls{GBT-SCA} functionality -- the  \gls{GBT-SCA} \gls{OPC UA}  Clients -- the \gls{GBT-SCA} \gls{OPC UA} server is used as the hub to transmit and synchronise the data, and a choice was made to decentralise all specialised applications (as shown in Figure~\ref{fig:globalPicture}). This choice facilitates maintenance, has the advantage of dividing responsibilities among different communities, allows for staging by creating higher-level wrapper applications, and allows for interoperability between diverse clients.
To support the concept, a \texttt{\gls{quasar}}-generated C++ library, namely UaoClientForOpcUaSca is provided for building ad-hoc \gls{OPC UA}  clients. This library supplies the interface to the \gls{GBT-SCA} \gls{OPC UA}  server and is created based on information sourced from the design of the server. Applications in ATLAS, that use the aforementioned library, are \gls{TDAQ} \gls{OPC UA}  clients used for configuration, or peripheral servers which perform sub-detector specific higher-level operations.
The simplified architecture of the \gls{OPC UA} \gls{GBT-SCA} server and an example of a \gls{GBT-SCA} \gls{OPC UA} client 
is depicted in Figure~\ref{fig:scaopcua-stack}. The server allows for the usage of any general-purpose \gls{OPC UA}  clients for diagnostic purposes such as the publicly available \textit{UaExpert} tool \cite{uaexpert}. Finally, the most common way that an \gls{OPC UA}  server is used by the \gls{DCS} is through  \gls{SCADA}  \gls{WinCC OA}-based systems. These systems employ \gls{OPC UA}  clients which connect to the servers to retrieve the data and visualise the information in a user interface, usually deployed in the control room where the system is monitored by a shifter. In the case of \gls{WinCC OA}, the \gls{OPC UA}  connectivity is realised via a software module, \textit{fwSca}, which is 
supplied by the \gls{GBT-SCA} software suite. This module allows for fast integration as it creates all the necessary configuration in the \gls{WinCC OA} application based on a priori information of the \gls{GBT-SCA} \gls{OPC UA} information schema.
 
\paragraph{ \glstext*{GBT-SCA} Software Performance}
 
The server has been designed to serve setups of various sizes and types. The biggest challenge is the  \gls{NSW} detector upgrade (see Section~\ref{MuonSS:NSW}). In this system, \num{6976} \glspl{GBT-SCA} are employed, distributed over different types of front-end electronics boards. The traffic of the \glspl{GBT-SCA} is handled by \num{30}  \gls{FELIX} hosts with \num{18} optical fibre connections each, and a corresponding number of \gls{GBT-SCA} \gls{OPC UA} servers.
 
In an early integration setup, the \gls{GBT-SCA} \gls{OPC UA} server was tested against a full-sector slice of the \gls{NSW} \gls{MM} sub-system with eight detector layers fully equipped with their front-end electronics. The slice serves \num{160} \glspl{GBT-SCA}, handled by a single server which was used in various realistic scenarios. The  \glspl{GBT-SCA} are separated in three categories/types of electronics with different functionality and interfaces as described in Table~\ref{tab:sca-setup}.
 
The setup used one \gls{FELIX} host equipped with an Intel(R) Xeon(R) CPU E5-1650 v4 @\SI{3.60}{\GHz}. The \gls{GBT-SCA} \gls{OPC UA} server runs in the  \gls{FELIX} host machine along with  \gls{FELIX} software. In a first constant-throughput scenario, a \gls{WinCC OA} \gls{SCADA} application monitored the \analog inputs from a separate host while three \gls{OPC UA}  clients were used for diagnostics. In the second burst traffic scenario, the server was used by 128 additional configuration clients.
 
\begin{table}[!htp]
\centering
\caption{ \gls{GBT-SCA} channel usage in the ATLAS  \gls{NSW} \gls{MM} full-sector slice. The setup was used to evaluate the performance of the server.}
\begin{tabular}{lccc}
\toprule
\textbf{Board Name}				& \textbf{\gls{MMFE8}}	& \textbf{\gls{ADDC}}						& \textbf{\gls{L1DDC}}\\
\midrule
\multirow{ 2}{*}{Functionality}	&\multirow{ 2}{*}{readout}		&trigger					&data \\
&											&aggregator				&aggregator\\
 
\gls{GBT-SCA} Numbers					&128								&16								&16\\
 
\gls{ADC} Inputs				&15								&10								&9\\
 
Calculated 				&\multirow{ 2}{*}{15}		&\multirow{ 2}{*}{10}		&\multirow{ 2}{*}{9}\\
variables					&									&									&	\\
 
\gls{I2C} Master 				&2								&6								&2\\
 
\gls{I2C} Slave 					&44+60						&6								&2\\
 
\gls{SPI} Slave					&8								&-									&-\\
 
\gls{GPIO}			&19								&18								&-\\
\bottomrule
\end{tabular}
 
\label{tab:sca-setup}
\end{table}
 
\textbf{Constant-throughput Monitoring Traffic:} Even when no configuration activities are performed, the server is used constantly to provide monitoring data from the detector electronics. These data, mostly from \analog inputs, correspond to the minimum possible activity of the server. In the \gls{MM} full-sector slice setup the global request rate was measured to be around \SI{7800}{requests/\s} for \num{2192} \gls{ADC} inputs (each \analog input consists of two \gls{GBT-SCA} requests). That resulted in an actual refresh rate of about \SI{2}{\Hz} per \analog input. The CPU usage of the server reached about 25\% on average and the share of available physical memory used was \SI{340}{MB}, a metric that is stable and not dependent on the usage.
 
\textbf{On-demand  \gls{GBT-SCA} Traffic - Front-end Configuration:} The most challenging aspect, in terms of open sessions and process complexity, is the configuration of the front-end boards. Emulating the cold start of the  \gls{NSW} \gls{MM} detector, a full-sector configuration was attempted in addition to the constant-throughput monitoring traffic as described above. During the process, up to \num{58} concurrent sessions were established from various \gls{OPC UA}  clients. The configuration clients were programming the front-end electronics using a combination of interleaving operations between \gls{GPIO}, \gls{I2C}, and \gls{SPI} totalling around \num{2700} requests for each \gls{GBT-SCA}. The global request rate reached about \SI{35000}{requests/\s}. The instantaneous CPU usage peaked at 218\%. The total time to initialise all the front-ends was measured to be \SI{10}{\s}. In ATLAS, each sector is independent, so these times and rates are also applicable to the full detector.
 
\subsubsection{Controls for \glstext*{ATCA} Back-End Electronics}
\label{sec:TDAQ_DCS_ATCA}
 
\begin{figure}[!htb]
\centering	\includegraphics[width=1.0\textwidth]{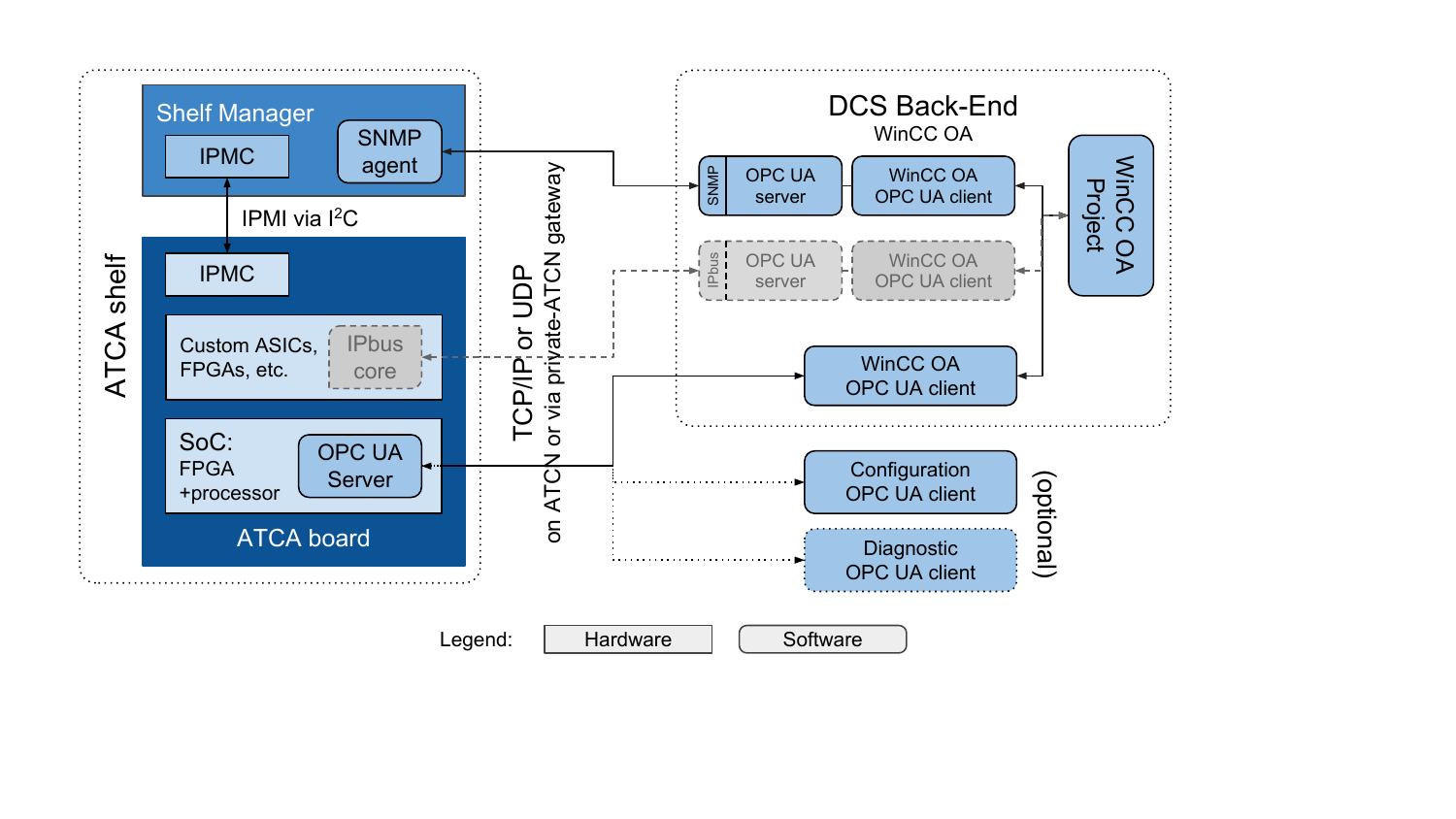}
\caption{Overview of the scheme based on \gls{OPC UA} to manage \gls{ATCA} shelves via their shelf manager interface. During \RunThr, some systems still use the IPbus-based pathway shown in grey, but this is deprecated and will be phased out by the end of \gls{LS3}.
\label{fig:TDAQ_DCS_ATCAserver_overview}}
\end{figure}

The \gls{ATCA} standard is employed as a back-end platform by many of the system upgrades for \RunThr (the \gls{NSW} Trigger Processor, the \gls{LAr} \gls{LATOME}, and the \gls{TDAQ} \gls{L1} \gls{gFEX}, \gls{eFEX}, \gls{MuCTPI}, \gls{jFEX}, and \gls{L1Topo} systems), replacing the \gls{VME} standard as the preferred back-end technology for new electronics systems. To allow the \gls{DCS} to work with \gls{ATCA}, an integration solution based on the common \gls{OPC UA}  toolset was developed, managing \gls{ATCA} shelves via their shelf manager \gls{SNMP} interface and providing control and monitoring of shelf and blade functions. The solution covers the “CERN-standard” Pigeon Point Shelf Managers (ShMM 500, ShMM 700R) and, while designed for \gls{ATCA}, is compatible with the broader device family. An overview of this scheme is illustrated in Figure~\ref{fig:TDAQ_DCS_ATCAserver_overview}.
 
The \gls{ATCA} \gls{OPC UA} server, a modularised software application
(see also \ref{TDAQ_DCS_OPCUA}), models selected parts of the \gls{ATCA} standard functionality for compliant devices in an object-oriented design 
and interfaces with the shelf manager \gls{SNMP} agent via the experiment controls network. Code generation techniques are used to implement the selected device functions based on the \gls{SNMP} Management Information Base provided by the shelf manufacturer. The \gls{SNMP} back-end within the server implementation is a C++ wrapper of the Net-\gls{SNMP} open-source library and provides a generic interface to any \gls{SNMP} device. In addition, the \gls{ATCA} \gls{OPC UA} server provides features such as automatic hardware discovery that queries a given set of shelf managers and automatically creates a map of all blades, fan trays, power supplies and other field-replaceable units, and makes their functionality available to \gls{OPC UA}  clients. The supervision of custom blade functions such as on-board sensors controlled by the blade \gls{IPMC} is also supported.
 
Finally, a set of associated tools, allowing for easy client integration with the  \gls{WinCC OA}\gls{DCS} applications, are part of the \gls{ATCA} software solution. These tools take care of datapoint creation, alarm handling, archiving, and Finite State Machine integration, and facilitate the creation of operator interfaces with pre-built generic user interface panels, making shelf integration into the ATLAS \gls{DCS} an easy and efficient task.
 
The \gls{LAr} Calorimeter back-end electronics (see also Section~\ref{sec:LArDigitalTrigger}) and the \gls{TDAQ} Level-1 Calorimeter Trigger electronics (see also Section~\ref{sec:TDAQ_L1Calo}) use this method to monitor onboard component parameters such as temperature and voltage.



\clearpage
\newpage
 
\section{Outlook} 
\label{sec:Outlook}

Extensive upgrades were performed to ready the ATLAS detector for \RunThr based on the experience gained from \RunOneTwo.
While the running parameters foreseen for \RunThr are not expected to subject the detector to more extreme conditions than it has experienced already, it is expected that with luminosity levelling, ATLAS will perpetually experience conditions close to the extremes of \RunOneTwo during \RunThr.
The emphasis of the Phase-I upgrades was therefore on making the detector and its trigger as robust as possible, so that ATLAS can run comfortably during \RunThr with un-prescaled single-lepton trigger thresholds comparable to those of \RunOne. The Phase-I upgrades were designed to last for the remaining lifetime of ATLAS, and constitute the first step in preparing ATLAS for the rigours of data-taking at the \gls{HL-LHC}, when it is expected that the instantaneous luminosity could rise as high as \lumihllhchigh, with an average pileup of \muhllhc.

The remaining upgrades required for ATLAS to run at the \gls{HL-LHC} are being prepared now, and will be installed during \gls{LS3}, at the end of \RunThr.
The largest of these is the complete replacement of the ATLAS \gls{ID} by the new, all-silicon, \gls{ITk}~\cite{ATLAS-TDR-25,ATLAS-TDR-30} that is designed for a similar or even improved tracking performance compared to the current ATLAS \gls{ID}, but in the challenging pileup conditions of the \gls{HL-LHC}. The \gls{ITk} pixel system, which builds on the technology used in the \gls{IBL}, will be composed of five barrel layers plus endcap rings covering up to $\abseta = 4$. It will be surrounded by a large-area strip detector composed of four barrel layers and six endcap discs, extending up to a maximum radius of \SI{1}{m} and and covering up to $\abseta < 2.7$.
There will be substantial upgrades to the electronics of numerous subsystems for Phase-II, with a focus on radiation-hard readout electronics for the \gls{LAr}~\cite{ATLAS-TDR-27} and Tile~\cite{ATLAS-TDR-28} Calorimeters that must operate at the trigger rates and latencies required for Phase-II luminosities.
The already excellent timing resolution of the calorimeters will be augmented with the addition of a dedicated \gls{HGTD}~\cite{ATLAS-TDR-31}.
The Muon Spectrometer~\cite{ATLAS-TDR-26} will see further improvements to trigger coverage and redundancy, with the replacement of on- and off-detector readout and trigger electronics, the replacement of the \gls{TGC} \gls{EIL4} doublet modules by more efficient triplets, and with the addition of \glspl{RPC} to the entire inner barrel layer, building on the \gls{BIS78} pilot project described in Section~\ref{muonSS:BIS78}.
Information from the \gls{MDT} detectors will be incorporated in the first level of the Muon trigger, increasing its granularity and improving the sharpness of the trigger transverse momentum thresholds.
The replacement of the front-end electronics of the existing ATLAS detector systems during the Phase-I and Phase-II upgrades allows for a substantially higher hardware-level trigger rate and a longer latency.
The \gls{TDAQ} system for ATLAS in the \gls{HL-LHC}~\cite{ATLAS-TDR-29,ATLAS-TDR-29-ADD-1} will consist of a single-level hardware trigger that analyses calorimeter and muon detector information at \SI{40}{\MHz} within \SI{10}{\micro\s} latency. After the hardware-based trigger decision, the resulting full detector and trigger data will be read out at a rate of \SI{1}{\MHz}. An upgraded high-level trigger system will be implemented using commodity hardware, refining the trigger objects in order to achieve a maximum output rate of \SI{10}{\kHz}.
 
\begin{figure}[t]
\centerline{\includegraphics[width=0.95\textwidth]{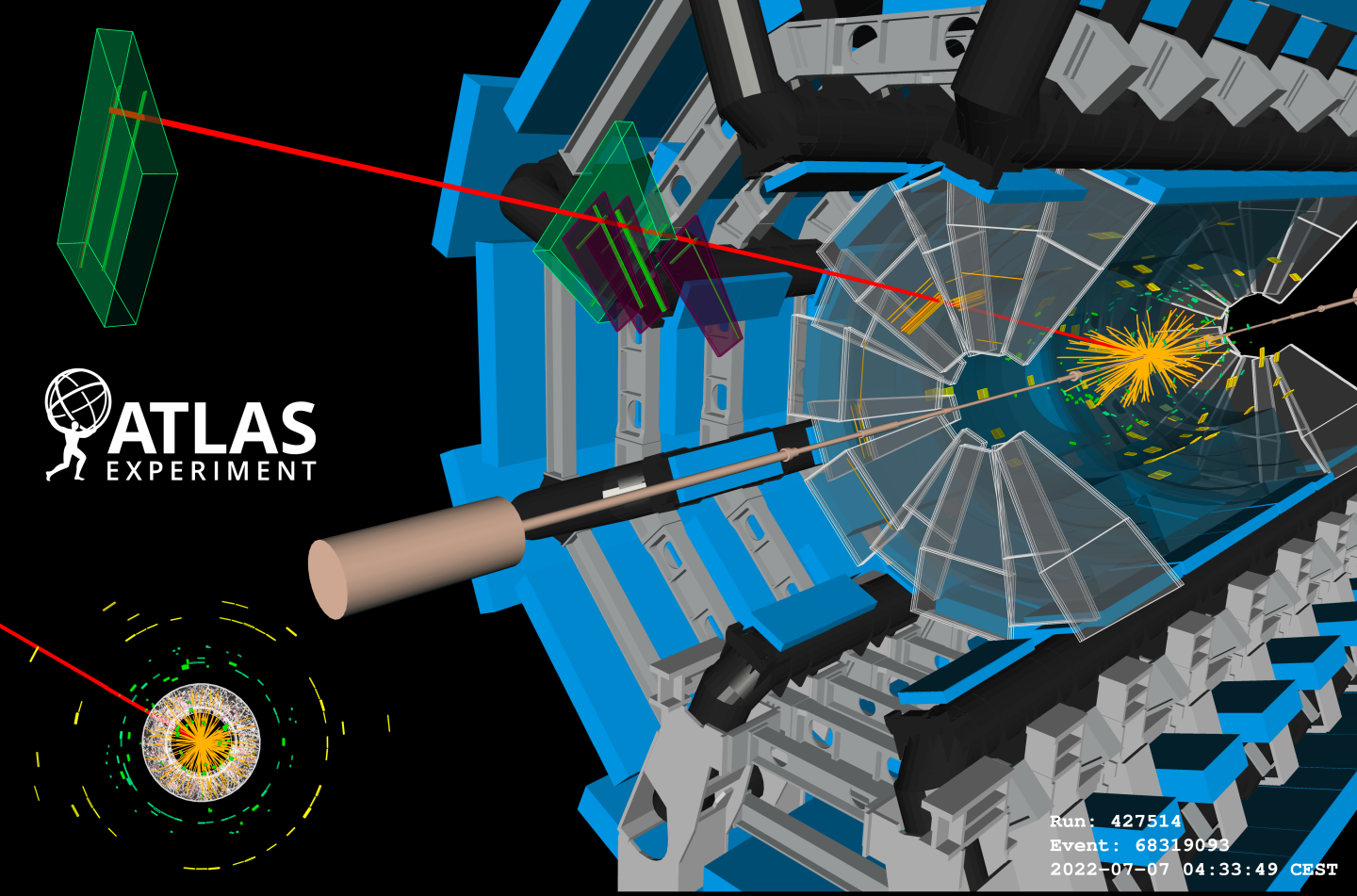}}
\caption{Event display (Run 427514, Event 68319093) of a collision event recorded in ATLAS on 7 July 2022, when stable beams of \SI{6.8}{\TeV} protons were delivered to ATLAS by the \gls{LHC}. The red line shows a muon candidate with a transverse momentum of \SI{15}{\GeV} reconstructed using information from the inner tracking detectors and the ATLAS \glsfirst{MS} endcap. The muon candidate was among the first reconstructed using hits in the \gls{MM} chambers of the \gls{NSW} on side C. The \gls{NSW} is outlined in white and the \gls{MM} hits are shown as orange lines. Additional muon chambers associated with the track are shown as green (\gls{MDT} endcap) and purple (\gls{TGC} endcap) boxes. 
}
\label{fig:Run3EventDisplay_NSW}
\end{figure}

The Phase-I upgrade project is an essential part of a broad upgrade program through the lifetime of the Large Hadron Collider and is fully compatible with the future Phase-II upgrade program of the ATLAS experiment. The collection of $\rts=$\SI{13.6}{\TeV} collisions in July 2022 is the realisation of the Run~3 configuration of the ATLAS detector (see Figure~\ref{fig:Run3EventDisplay_NSW}) and has marked the start of the vibrant physics programme planned for Run 3 and beyond.
 
\clearpage
\newpage
 
\section*{Acknowledgements}


We thank CERN for the very successful operation of the LHC, as well as the
support staff from our institutions without whom ATLAS could not be
operated efficiently.
 
We acknowledge the support of
ANPCyT, Argentina;
YerPhI, Armenia;
ARC, Australia;
BMWFW and FWF, Austria;
ANAS, Azerbaijan;
CNPq and FAPESP, Brazil;
NSERC, NRC and CFI, Canada;
CERN;
ANID, Chile;
CAS, MOST and NSFC, China;
Minciencias, Colombia;
MEYS CR, Czech Republic;
DNRF and DNSRC, Denmark;
IN2P3-CNRS and CEA-DRF/IRFU, France;
SRNSFG, Georgia;
BMBF, HGF and MPG, Germany;
GSRI, Greece;
RGC and Hong Kong SAR, China;
ISF and Benoziyo Center, Israel;
INFN, Italy;
MEXT and JSPS, Japan;
CNRST, Morocco;
NWO, Netherlands;
RCN, Norway;
MEiN, Poland;
FCT, Portugal;
MNE/IFA, Romania;
MESTD, Serbia;
MSSR, Slovakia;
ARRS and MIZ\v{S}, Slovenia;
DSI/NRF, South Africa;
MICINN, Spain;
SRC and Wallenberg Foundation, Sweden;
SERI, SNSF and Cantons of Bern and Geneva, Switzerland;
MOST, Taiwan;
TENMAK, T\"urkiye;
STFC, United Kingdom;
DOE and NSF, United States of America.
In addition, individual groups and members have received support from
BCKDF, CANARIE, Compute Canada and CRC, Canada;
PRIMUS 21/SCI/017 and UNCE SCI/013, Czech Republic;
COST, ERC, ERDF, Horizon 2020 and Marie Sk{\l}odowska-Curie Actions, European Union;
Investissements d'Avenir Labex, Investissements d'Avenir Idex and ANR, France;
DFG and AvH Foundation, Germany;
Herakleitos, Thales and Aristeia programmes co-financed by EU-ESF and the Greek NSRF, Greece;
BSF-NSF and MINERVA, Israel;
Norwegian Financial Mechanism 2014-2021, Norway;
NCN and NAWA, Poland;
La Caixa Banking Foundation, CERCA Programme Generalitat de Catalunya and PROMETEO and GenT Programmes Generalitat Valenciana, Spain;
G\"{o}ran Gustafssons Stiftelse, Sweden;
The Royal Society and Leverhulme Trust, United Kingdom.
 
The crucial computing support from all WLCG partners is acknowledged gratefully, in particular from CERN, the ATLAS Tier-1 facilities at TRIUMF (Canada), NDGF (Denmark, Norway, Sweden), CC-IN2P3 (France), KIT/GridKA (Germany), INFN-CNAF (Italy), NL-T1 (Netherlands), PIC (Spain), ASGC (Taiwan), RAL (UK) and BNL (USA), the Tier-2 facilities worldwide and large non-WLCG resource providers. Major contributors of computing resources are listed in Ref.~\cite{ATL-SOFT-PUB-2021-003}.


\addcontentsline{toc}{part}{Acknowledgements}

\appendix
\phantomsection
\addcontentsline{toc}{part}{Glossary}
\setglossarystyle{mylist}
\printnoidxglossaries
 
\phantomsection
\addcontentsline{toc}{part}{References}
\printbibliography

@article{ALFA-Detector,
  title = {The ALFA Roman Pot Detectors of ATLAS},
  collaboration ={ATLAS ALFA},
  author = {Abdel Khalek, S. and Allongue, B. and Anghinolfi, F. and others},
  journal = "JINST",
  volume = {\textbf 11},
  year = {2016},
  pages = {11013},
  eprint = {1609.00249},
  archivePrefix  = {arXiv},
  primaryClass   = {physics.ins-det}
}

@mastersthesis{Almeghari_2018,
      author       = "Almeghari, Abdelrahman",
      title        = "Radiation resistance of the ATLAS-ALFA electronics and trigger efficiency analysis at $\sqrt{s}=13$~TeV and $\beta$*=2.5~km",
      school       = "Islamic University of Gaza",
      address      = {Gaza City},
      year         = {2018},
      month  = {5},
      url = {https://library.iugaza.edu.ps/thesis/124502.pdf}
}

@techreport{bib:SLAC-RCE,
      title        = "Design of the SLAC RCE Platform: A general purpose ATCA based data acquisition system",
      author       = "Herbst, R. and others",
      institution  = "SLAC",
      address      = "Pasadena",
      number  = "SLAC-PUB-16182",
      year         = "2014",
      url          = {https://www.slac.stanford.edu/pubs/slacpubs/16000/slac-pub-16182.pdf}
}

@article{Lange_2015,
	doi = {10.1088/1748-0221/10/03/c03031},
	url = {https://doi.org/10.1088_1748-0221_10_03_c03031},
	year = 2015,
	month = {3},
	volume = {10},
	number = {03},
	pages = {C03031},
	author = {J. Lange and E. Cavallaro and S. Grinstein and I. L{\'{o}}pez Paz},
	title = {3D silicon pixel detectors for the {ATLAS} Forward Physics experiment},
	journal = "JINST"
}

@article{Lange_2016,
  author ={Lange, J. and others},
  title = {Beam tests of an integrated prototype of the ATLAS Forward Proton detector},
  journal = "JINST",
  volume = {11},
  year = {2016},
  month = {9},
  pages = {P09005}
}

@techreport{bib:AFP-ECR2016,
  author = {Ng, Christopher and Pinfold, J. and Rijssenbeek, M. and Sicho, P. and Sykora, T. and Trzebinski, M.},
  collaboration = {ATLAS},
  institution = {CERN},
  title = {Engineering Change Request - Installation of the ATLAS/AFP stations, Phase-2},
  numberl = {LHC-XAFP-EC-0003, EDMS 1705651},
  month = {8},
  year = {2016}
}

@techreport{Vacek_2013,
     title = {Commissioning of cooling system AIRCOOLER SPLIT before installation for TOTEM detectors},
     author = {Houška, D. and Vacek, V. and Doubek, Martin and Haubner, Michal},
     number = {internal CTU note},
     institution = {CTU, Prague},
     year = {2013},
     url = {https://pdfs.semanticscholar.org/59d0/a841d8f11ba217421c35296193de45491813.pdf}
}

@article{Cerny_2019,
   title = {Performance study of the ATLAS Forward Proton Time-of-Flight Detector System},
   author = {Cerny, K. and others},
   collaboration = {ATLAS AFP},
   journal = {Proceedings of Science},
   year = {2020},
   url = {https://cds.cern.ch/record/2703903}
}

@misc{MiniPlanacon,
   title = {Mini PLANACON XPM85112-S-R2D2: a long-life, high-rate, 4x4-channel multi-anode multi-channel plate photomultiplier },
   journal = {datasheet},
   note  =  {PHOTONIS Defense, Inc., 1000 New Holland Ave., Lancaster, PA 17601-5688, USA}, 
   url = {www.photonis.com}
}

@inproceedings{Zich_2019,
     title = {Multichannel Majority Coincidence Circuit for ToF AFP Detector},
     booktitle = "27th Telecommunications Forum TELFOR 2019",
     author = {Zich, Jan and Georgiev, Vjaceslav and Holik, Michael and Pavlicek, Vladimir and Vavroch, Ondrej},
     institution = {University of West Bohemia, Pilsen},
     year = {2019},
     address = "Belgrade, Serbia",
     month ="11",
     doi = {978-1-7281-4790-1/19/}
}

@article{Banas_2017,
   title = {Detector Control System for the AFP detector in ATLAS experiment at CERN},
   author = {Banaś, E. and Caforio, D. and Czekierda, S. and Z. Hajduk, Z. and Olszowska, J. and Seabra, L. and Šícho, P.},
   journal = {J. Phys. Conf. Ser.},
   volume = {898},
   year = {2017},
   pages = {032002},
   doi = {10.1088/1742-6596/898/3/032022},
   url = {http://iopscience.iop.org/article/10.1088/1742-6596/898/3/032022}
}

@article{HGTD,
 title = "ATLAS Phase-II Upgrade Scoping Document",
  author = "ATLAS Collaboration",
  journal = "CERN-LHCC-2015-20, LHCC-G-166, Sects. V.5. and XI.2.7.5.",
  month = "September",
  year = "2015",
  url  = {https://cds.cern.ch/record/2055248/}
}

@misc{bib:ATCA-RCE,
  title = "Advanced Telecom Computing Architecture",
  url = {http://www.picmg.org/openstandards/advancedtca/}
}

@article{MATIS2017114,
title = {The BRAN luminosity detectors for the LHC},
journal = {\NIMA},
volume = {848},
pages = {114-126},
year = {2017},
OPTissn = {0168-9002},
doi = {https://doi.org/10.1016/j.nima.2016.12.019},
url = {https://www.sciencedirect.com/science/article/pii/S0168900216312797},
author = {H.S. Matis and M. Placidi and A. Ratti and W.C. Turner and E. Bravin and R. Miyamoto}
}

@techreport{Garcia-Sciveres:2287593,
      author        = "Garcia-Sciveres, Maurice",
      collaboration = "RD53",
      title         = "{The RD53A Integrated Circuit}",
      institution   = "CERN",
      reportNumber  = "CERN-RD53-PUB-17-001",
      address       = "Geneva",
      year          = "2017",
      url           = "https://cds.cern.ch/record/2287593",
}

@article{Llopart_2022,
doi = {10.1088/1748-0221/17/01/C01044},
url = {https://dx.doi.org/10.1088/1748-0221/17/01/C01044},
year = {2022},
month = {1},
publisher = {IOP Publishing},
volume = {17},
number = {01},
pages = {C01044},
author = {X. Llopart and J. Alozy and R. Ballabriga and M. Campbell and R. Casanova and V. Gromov and E.H.M. Heijne and T. Poikela and E. Santin and V. Sriskaran and L. Tlustos and A. Vitkovskiy},
title = {Timepix4, a large area pixel detector readout chip which can be tiled on 4 sides providing sub-200 ps timestamp binning},
journal = {JINST}
}

@article{Nozka:23,
author = {Libor Nozka and Giulio Avoni and Elzbieta Banas and Andrew Brandt and Karel Cerny and Paul M. Davis and Serge Duarte Pinto and Vjaceslav Georgiev and Miroslav Hrabovsky and Tomas Komarek and Krzysztof Korcyl and Ivan Lopez-Paz and Marko Milovanovic and Goran Mladenovic and Dmitry A. Orlov and Michael Rijssenbeek and Petr Schovanek and Tomas Sykora and Maciej Trzebinski and Vladimir Urbasek and Jan Zich},
journal = {Opt. Express},
number = {3},
pages = {3998--4014},
publisher = {Optica Publishing Group},
title = {Upgraded Cherenkov time-of-flight detector for the AFP project},
volume = {31},
month = {1},
year = {2023},
url = {https://opg.optica.org/oe/abstract.cfm?URI=oe-31-3-3998},
doi = {10.1364/OE.480624}
}

@article{Sjostrand:2014zea,
      author         = "Sj{\"o}strand, Torbj{\"o}rn and Ask, Stefan and Christiansen,
                        Jesper R. and Corke, Richard and Desai, Nishita and Ilten,
                        Philip and Mrenna, Stephen and Prestel, Stefan and
                        Rasmussen, Christine O. and Skands, Peter Z.",
      title          = "{An introduction to PYTHIA 8.2}",
      journal        = "Comput. Phys. Commun.",
      volume         = "191",
      year           = "2015",
      pages          = "159",
      doi            = "10.1016/j.cpc.2015.01.024",
      eprint         = "1410.3012",
      archivePrefix  = "arXiv",
      primaryClass   = "hep-ph",
      reportNumber   = "LU-TP-14-36, MCNET-14-22, CERN-PH-TH-2014-190,
                        FERMILAB-PUB-14-316-CD, DESY-14-178, SLAC-PUB-16122,
                        --FERMILAB-PUB-14-316-CD",
      SLACcitation   = "%%CITATION = ARXIV:1410.3012;%%"
}

@article{Agostinelli:2002hh,
      author         = "{GEANT4 Collaboration} and Agostinelli, S. and others",
      title          = "{\textsc{Geant4} -- a simulation toolkit}",
      journal        = "Nucl. Instrum. Meth. A",
      volume         = "506",
      year           = "2003",
      pages          = "250",
      doi            = "10.1016/S0168-9002(03)01368-8",
      reportNumber   = "SLAC-PUB-9350, FERMILAB-PUB-03-339",
      SLACcitation   = "%%CITATION = NUIMA,A506,250;%%"
}

@Article{PERF-2007-01,
    author         = "{ATLAS Collaboration}",
    title          = "{The ATLAS Experiment at the CERN Large Hadron Collider}",
    journal        = "JINST",
    volume         = "3",
    year           = "2008",
    pages          = "S08003",
    doi            = "10.1088/1748-0221/3/08/S08003",
    primaryClass   = "hep-ex",
}

@Article{LARG-2009-01,
    author         = "{ATLAS Collaboration}",
    title          = "{Readiness of the ATLAS liquid argon calorimeter for LHC collisions}",
    journal        = "Eur. Phys. J. C",
    volume         = "70",
    year           = "2010",
    pages          = "723",
    doi            = "10.1140/epjc/s10052-010-1354-y",
    eprint         = "0912.2642",
    archivePrefix  = "arXiv",
    primaryClass   = "hep-ex",
}

@Article{HIGG-2012-27,
    author         = "{ATLAS Collaboration}",
    title          = "{Observation of a new particle in the search for the Standard Model Higgs boson with the ATLAS detector at the LHC}",
    journal        = "Phys. Lett. B",
    volume         = "716",
    year           = "2012",
    pages          = "1",
    doi            = "10.1016/j.physletb.2012.08.020",
    reportNumber   = "CERN-PH-EP-2012-218",
    eprint         = "1207.7214",
    archivePrefix  = "arXiv",
    primaryClass   = "hep-ex",
}

@Article{DAPR-2013-01,
    author         = "{ATLAS Collaboration}",
    title          = "{Luminosity determination in \(pp\) collisions at \(\sqrt{s} = 8\,\text{TeV}\) using the ATLAS detector at the LHC}",
    journal        = "Eur. Phys. J. C",
    volume         = "76",
    year           = "2016",
    pages          = "653",
    doi            = "10.1140/epjc/s10052-016-4466-1",
    eprint         = "1608.03953",
    archivePrefix  = "arXiv",
    primaryClass   = "hep-ex",
}

@Article{IDET-2013-01,
    author         = "{ATLAS Collaboration}",
    title          = "{Operation and performance of the ATLAS semiconductor tracker}",
    journal        = "JINST",
    volume         = "9",
    year           = "2014",
    pages          = "P08009",
    doi            = "10.1088/1748-0221/9/08/P08009",
    reportNumber   = "CERN-PH-EP-2014-049",
    eprint         = "1404.7473",
    archivePrefix  = "arXiv",
    primaryClass   = "hep-ex",
}

@Article{LARG-2013-01,
    author         = "{ATLAS Collaboration}",
    title          = "{Monitoring and data quality assessment of the ATLAS liquid argon calorimeter}",
    journal        = "JINST",
    volume         = "9",
    year           = "2014",
    pages          = "P07024",
    doi            = "10.1088/1748-0221/9/07/P07024",
    reportNumber   = "CERN-PH-EP-2014-045",
    eprint         = "1405.3768",
    archivePrefix  = "arXiv",
    primaryClass   = "hep-ex",
}

@Article{IDET-2015-01,
    author         = "{ATLAS Collaboration}",
    title          = "{Performance of the ATLAS Transition Radiation Tracker in Run~1 of the LHC: tracker properties}",
    journal        = "JINST",
    volume         = "12",
    year           = "2017",
    pages          = "P05002",
    doi            = "10.1088/1748-0221/12/05/P05002",
    reportNumber   = "CERN-EP-2016-311",
    eprint         = "1702.06473",
    archivePrefix  = "arXiv",
    primaryClass   = "hep-ex",
}

@Article{PERF-2015-07,
    author         = "{ATLAS Collaboration}",
    title          = "{Study of the material of the ATLAS inner detector for Run~2 of the LHC}",
    journal        = "JINST",
    volume         = "12",
    year           = "2017",
    pages          = "P12009",
    doi            = "10.1088/1748-0221/12/12/P12009",
    reportNumber   = "CERN-EP-2017-081",
    eprint         = "1707.02826",
    archivePrefix  = "arXiv",
    primaryClass   = "hep-ex",
}

@Article{TRIG-2016-01,
    author         = "{ATLAS Collaboration}",
    title          = "{Performance of the ATLAS trigger system in 2015}",
    journal        = "Eur. Phys. J. C",
    volume         = "77",
    year           = "2017",
    pages          = "317",
    doi            = "10.1140/epjc/s10052-017-4852-3",
    reportNumber   = "CERN-EP-2016-241",
    eprint         = "1611.09661",
    archivePrefix  = "arXiv",
    primaryClass   = "hep-ex",
}

@Article{IDET-2017-10,
    author         = "{ATLAS Collaboration}",
    title          = "{Modelling radiation damage to pixel sensors in the ATLAS detector}",
    journal        = "JINST",
    volume         = "14",
    year           = "2019",
    pages          = "P06012",
    doi            = "10.1088/1748-0221/14/06/P06012",
    reportNumber   = "CERN-EP-2019-061",
    eprint         = "1905.03739",
    archivePrefix  = "arXiv",
    primaryClass   = "hep-ex",
}

@Article{DAPR-2018-01,
    author         = "{ATLAS Collaboration}",
    title          = "{ATLAS data quality operations and performance for 2015--2018 data-taking}",
    journal        = "JINST",
    volume         = "15",
    year           = "2020",
    pages          = "P04003",
    doi            = "10.1088/1748-0221/15/04/P04003",
    reportNumber   = "CERN-EP-2019-207",
    eprint         = "1911.04632",
    archivePrefix  = "arXiv",
    primaryClass   = "physics.ins-det",
}

@Article{PIX-2018-001,
    author         = "Abbott, B. and others",
    title          = "{Production and integration of the ATLAS Insertable B-Layer}",
    journal        = "JINST",
    volume         = "13",
    year           = "2018",
    pages          = "T05008",
    doi            = "10.1088/1748-0221/13/05/T05008",
    eprint         = "1803.00844",
    archivePrefix  = "arXiv",
    primaryClass   = "physics.ins-det",
}

@Article{TRIG-2018-01,
    author         = "{ATLAS Collaboration}",
    title          = "{Performance of the ATLAS muon triggers in Run~2}",
    journal        = "JINST",
    volume         = "15",
    year           = "2020",
    pages          = "P09015",
    doi            = "10.1088/1748-0221/15/09/p09015",
    reportNumber   = "CERN-EP-2020-031",
    eprint         = "2004.13447",
    archivePrefix  = "arXiv",
    primaryClass   = "hep-ex",
}

@Article{IDTR-2019-05,
    author         = "{ATLAS Collaboration}",
    title          = "{Alignment of the ATLAS Inner Detector in Run-2}",
    journal        = "Eur. Phys. J. C",
    volume         = "80",
    year           = "2020",
    pages          = "1194",
    doi            = "10.1140/epjc/s10052-020-08700-6",
    reportNumber   = "CERN-EP-2020-108",
    eprint         = "2007.07624",
    archivePrefix  = "arXiv",
    primaryClass   = "hep-ex",
}

@Article{SCTD-2019-01,
    author         = "{ATLAS Collaboration}",
    title          = "{Operation and performance of the ATLAS semiconductor tracker in LHC Run~2}",
    journal        = "JINST",
    volume         = "17",
    year           = "2021",
    pages          = "P01013",
    doi            = "10.1088/1748-0221/17/01/P01013",
    reportNumber   = "CERN-EP-2021-178",
    eprint         = "2109.02591",
    archivePrefix  = "arXiv",
    primaryClass   = "physics.ins-det",
}

@Article{TDAQ-2019-01,
    author         = "{ATLAS Collaboration}",
    title          = "{Performance of the upgraded PreProcessor of the ATLAS Level-1 Calorimeter Trigger}",
    journal        = "JINST",
    volume         = "15",
    year           = "2020",
    pages          = "P11016",
    doi            = "10.1088/1748-0221/15/11/P11016",
    reportNumber   = "CERN-EP-2020-042",
    eprint         = "2005.04179",
    archivePrefix  = "arXiv",
    primaryClass   = "hep-ex",
}

@Article{TRIG-2019-02,
    author         = "{ATLAS Collaboration}",
    title          = "{Performance of the ATLAS Level-1 topological trigger in Run~2}",
    journal        = "Eur. Phys. J. C",
    volume         = "82",
    year           = "2021",
    pages          = "7",
    doi            = "10.1140/epjc/s10052-021-09807-0",
    reportNumber   = "CERN-EP-2021-040",
    eprint         = "2105.01416",
    archivePrefix  = "arXiv",
    primaryClass   = "hep-ex",
}

@Article{IDET-2020-01,
    author         = "{ATLAS Collaboration}",
    title          = "{Measurements of sensor radiation damage in the ATLAS inner detector using leakage currents}",
    journal        = "JINST",
    volume         = "16",
    year           = "2021",
    pages          = "P08025",
    doi            = "10.1088/1748-0221/16/08/P08025",
    reportNumber   = "CERN-EP-2021-055",
    eprint         = "2106.09287",
    archivePrefix  = "arXiv",
    primaryClass   = "hep-ex",
}

@Article{DAPR-2021-01,
    author         = "{ATLAS Collaboration}",
    title          = "{Luminosity determination in \(pp\) collisions at \(\sqrt{s} = 13\,\text{TeV}\) using the ATLAS detector at the LHC}",
    year           = "2022",
    reportNumber   = "CERN-EP-2022-281",
    eprint         = "2212.09379",
    archivePrefix  = "arXiv",
    primaryClass   = "hep-ex",
}

@Booklet{ATL-SOFT-PUB-2021-001,
    author         = "{ATLAS Collaboration}",
    title          = "{The ATLAS Collaboration Software and Firmware}",
    howpublished   = "{ATL-SOFT-PUB-2021-001}",
    url            = "https://cds.cern.ch/record/2767187",
    year           = "2021",
}

@Booklet{ATL-SOFT-PUB-2021-003,
    author         = "{ATLAS Collaboration}",
    title          = "{ATLAS Computing Acknowledgements}",
    howpublished   = "{ATL-SOFT-PUB-2021-003}",
    url            = "https://cds.cern.ch/record/2776662",
    year           = "2021",
}

@Report{ATLAS-TDR-02,
    author         = "{ATLAS Collaboration}",
    title          = "{ATLAS Liquid Argon Calorimeter: Technical Design Report}",
    type           = "ATLAS-TDR-2; CERN-LHCC-96-041",
    year           = "1996",
    url            = "https://cds.cern.ch/record/331061",
}

@Report{ATLAS-TDR-03,
    author         = "{ATLAS Collaboration}",
    title          = "{ATLAS Tile Calorimeter: Technical Design Report}",
    type           = "ATLAS-TDR-3; CERN-LHCC-96-042",
    year           = "1996",
    url            = "https://cds.cern.ch/record/331062",
}

@Report{ATLAS-TDR-19,
    author         = "{ATLAS Collaboration}",
    title          = "{ATLAS Insertable B-Layer: Technical Design Report}",
    type           = "ATLAS-TDR-19; CERN-LHCC-2010-013",
    year           = "2010",
    url            = "https://cds.cern.ch/record/1291633",
    related        = "ATLAS-TDR-19-addm",
    relatedstring  = "Addendum:",
}

@Report{ATLAS-TDR-19-addm,
    author         = "{ATLAS Collaboration}",
    type           = "ATLAS-TDR-19-ADD-1; CERN-LHCC-2012-009",
    year           = "2012",
    url            = "https://cds.cern.ch/record/1451888",
}

@Report{ATLAS-TDR-20,
    author         = "{ATLAS Collaboration}",
    title          = "{ATLAS New Small Wheel: Technical Design Report}",
    type           = "ATLAS-TDR-020; CERN-LHCC-2013-006",
    year           = "2013",
    url            = "https://cds.cern.ch/record/1552862",
}

@Report{ATLAS-TDR-22,
    author         = "{ATLAS Collaboration}",
    title          = "{ATLAS Liquid Argon Calorimeter Phase-I Upgrade: Technical Design Report}",
    type           = "ATLAS-TDR-022; CERN-LHCC-2013-017",
    year           = "2013",
    url            = "https://cds.cern.ch/record/1602230",
}

@Report{ATLAS-TDR-23,
    author         = "{ATLAS Collaboration}",
    title          = "{ATLAS TDAQ System Phase-I Upgrade: Technical Design Report}",
    type           = "ATLAS-TDR-023; CERN-LHCC-2013-018",
    year           = "2013",
    url            = "https://cds.cern.ch/record/1602235",
}

@Report{ATLAS-TDR-24,
    author         = "{ATLAS Collaboration}",
    title          = "{ATLAS Forward Proton Phase-I Upgrade: Technical Design Report}",
    type           = "ATLAS-TDR-024; CERN-LHCC-2015-009",
    year           = "2015",
    url            = "https://cds.cern.ch/record/2017378",
}

@Report{ATLAS-TDR-25,
    author         = "{ATLAS Collaboration}",
    title          = "{ATLAS Inner Tracker Strip Detector: Technical Design Report}",
    type           = "ATLAS-TDR-025; CERN-LHCC-2017-005",
    year           = "2017",
    url            = "https://cds.cern.ch/record/2257755",
}

@Report{ATLAS-TDR-26,
    author         = "{ATLAS Collaboration}",
    title          = "{ATLAS Muon Spectrometer Phase-II Upgrade: Technical Design Report}",
    type           = "ATLAS-TDR-026; CERN-LHCC-2017-017",
    year           = "2017",
    url            = "https://cds.cern.ch/record/2285580",
}

@Report{ATLAS-TDR-27,
    author         = "{ATLAS Collaboration}",
    title          = "{ATLAS LAr Calorimeter Phase-II Upgrade: Technical Design Report}",
    type           = "ATLAS-TDR-027; CERN-LHCC-2017-018",
    year           = "2017",
    url            = "https://cds.cern.ch/record/2285582",
}

@Report{ATLAS-TDR-28,
    author         = "{ATLAS Collaboration}",
    title          = "{ATLAS Tile Calorimeter Phase-II Upgrade: Technical Design Report}",
    type           = "ATLAS-TDR-028; CERN-LHCC-2017-019",
    year           = "2017",
    url            = "https://cds.cern.ch/record/2285583",
}

@Report{ATLAS-TDR-29,
    author         = "{ATLAS Collaboration}",
    title          = "{ATLAS TDAQ Phase-II Upgrade: Technical Design Report}",
    type           = "ATLAS-TDR-029; CERN-LHCC-2017-020",
    year           = "2017",
    url            = "https://cds.cern.ch/record/2285584",
}

@Report{ATLAS-TDR-30,
    author         = "{ATLAS Collaboration}",
    title          = "{ATLAS Inner Tracker Pixel Detector: Technical Design Report}",
    type           = "ATLAS-TDR-030; CERN-LHCC-2017-021",
    year           = "2017",
    url            = "https://cds.cern.ch/record/2285585",
}

@Report{ATLAS-TDR-31,
    author         = "{ATLAS Collaboration}",
    title          = "{A High-Granularity Timing Detector for the ATLAS Phase-II Upgrade: Technical Design Report}",
    type           = "ATLAS-TDR-031; CERN-LHCC-2020-007",
    year           = "2020",
    url            = "https://cds.cern.ch/record/2719855",
}

@Article{CMS-HIG-12-028,
    author         = "{CMS Collaboration}",
    title          = "{Observation of a new boson at a mass of 125 GeV with the CMS experiment at the LHC}",
    journal        = "Phys. Lett. B",
    volume         = "716",
    year           = "2012",
    pages          = "30",
    doi            = "10.1016/j.physletb.2012.08.021",
    reportNumber   = "CERN-PH-EP-2012-220",
    eprint         = "1207.7235",
    archivePrefix  = "arXiv",
    primaryClass   = "hep-ex",
}

@Article{LArPhaseIPaper,
  author = "Aad, G. and others",
    title = "{The Phase-I trigger readout electronics upgrade of the ATLAS Liquid Argon calorimeters}",
    eprint = "2202.07384",
    archivePrefix = "arXiv",
    primaryClass = "physics.ins-det",
    doi = "10.1088/1748-0221/17/05/P05024",
    journal = "JINST",
    volume = "17",
    number = "05",
    pages = "P05024",
    year = "2022"
}

@article{Jivan:2015mqk,
	doi = {10.1088/1742-6596/645/1/012019},
	url = {https://doi.org/10.1088/1742-6596/645/1/012019},
	year = 2015,
	month = {10},
	publisher = {{IOP} Publishing},
	volume = {645},
	pages = {012019},
	author = {H Jivan and E Sideras-Haddad and R Erasmus and S Liao and M Madhuku and G Peters and K Sekonya and O Solvyanov},
	title = {Radiation hardness of plastic scintillators for the Tile Calorimeter of the {ATLAS} detector},
	journal = {Journal of Physics: Conference Series},
	OPTnote={\textcolor{red}{Please check this is right reference!}}
}

@article{Liao:2015zsa,
	doi = {10.1088/1742-6596/645/1/012021},
	url = {https://doi.org/10.1088/1742-6596/645/1/012021},
	year = 2015,
	month = {10},
	publisher = {{IOP} Publishing},
	volume = {645},
	pages = {012021},
	author = {S Liao and R Erasmus and H Jivan and C Pelwan and G Peters and E Sideras-Haddad},
	title = {A comparative study of the radiation hardness of plastic scintillators for the upgrade of the Tile Calorimeter of the {ATLAS} detector},
	journal = {Journal of Physics: Conference Series},
	OPTnote={\textcolor{red}{Please check this is right reference!}}	
}

@article{cleland,
title = {Signal processing considerations for liquid ionization calorimeters in a high rate environment},
journal = {\NIMA},
volume = {338},
number = {2},
pages = {467-497},
year = {1994},
OPTissn = {0168-9002},
doi = {https://doi.org/10.1016/0168-9002(94)91332-3},
url = {https://www.sciencedirect.com/science/article/pii/0168900294913323},
author = {W.E. Cleland and E.G. Stern}
}

@Booklet{ATLAS-CONF-2019-021,
    author         = "{ATLAS Collaboration}",
    title          = "{Luminosity determination in \(pp\) collisions at \(\sqrt{s} = 13\,\text{TeV}\) using the ATLAS detector at the LHC}",
    howpublished   = "{ATLAS-CONF-2019-021}",
    url            = "https://cds.cern.ch/record/2677054",
    year           = "2019",
}

@article{FE-I4,
        Author = {M. Garcia-Sciveres and D. Arutinov and M. Barbero and R. Beccherle and S. Dube and D. Elledge and J. Fleury and D. Fougeron and F. Gensolen and D. Gnani and V. Gromov and T. Hemperek and M. Karagounis and R. Kluit and A. Kruth and A. Mekkaoui and M. Menouni and J.-D. Schipper},
	Journal = {\NIMA},
	Pages = {S155-S159},
	Title = {{The FE-I4 pixel readout integrated circuit}},
	Volume = {636},
        number = {1, Supplement },
        note = {7th International ""Hiroshima"" Symposium on the Development and Application of Semiconductor Tracking Detectors},
        issn = {0168-9002},
        doi = {https://doi.org/10.1016/j.nima.2010.04.101},
        url = {https://www.sciencedirect.com/science/article/pii/S0168900210009551},
	Year = {2011}}

@article{3D,
    title = {3D — A proposed new architecture for solid-state radiation detectors},
    journal = {\NIMA},
    volume = {395},
    number = {3},
    pages = {328-343},
    year = {1997},
    note = {Proceedings of the Third International Workshop on Semiconductor Pixel Detectors for Particles and X-rays},
    issn = {0168-9002},
    doi = {https://doi.org/10.1016/S0168-9002(97)00694-3},
    url = {https://www.sciencedirect.com/science/article/pii/S0168900297006943},
    author = {S.I. Parker and C.J. Kenney and J. Segal}
}

@Conference{ELMB,
         Author  = "{B. Hallgren, H. Boterenbrood, H.J. Burckhart and H. Kvedalen}",
         Title = {{The embedded local monitor board (ELMB) in the LHC front-end I/O control system}} ,
  booktitle = 	 {Proceedings of the 7th Workshop on Electronics for LHC Experiments:
                       Stockholm, Sweden 10 - 14 Sep 2001.},
  year = 	 {2001},
      url           = "https://cds.cern.ch/record/530675",
      doi           = "10.5170/CERN-2001-005.325",
  OPTkey = 	 {},
  editor =	 {C. Isabella},
  OPTvolume = 	 {},
  OPTnumber = 	 {},
  OPTseries = 	 {},
  pages =	 {325}, %means it is on p325 of the proceedings...
  OPTmonth = 	 {},
  address =	 {Geneva},
  organization = {CERN},
  publisher =	 {CERN},
  OPTnote = 	 {},
  OPTannote = 	 {}
}

@misc{PVSS,
          Author  = "{ETM professional control GmbH}", 
          Title = {{SIMATIC WinCC Open Architecture SCADA System}}, 
          url ={http://www.winccoa.com/},
         }

@article{Pixel,
        Author = {{ATLAS Collaboration}},
	xAuthor = {Aad, G. and Ackers, M. and Alberti, F.A. and Aleppo, M. and Alimonti, G. and others},
	Doi = {10.1088/1748-0221/3/07/P07007},
	Journal = {JINST},
	Pages = {P07007},
	Slaccitation = {%%CITATION = JINST,3,P07007;%%},
	Title = {{ATLAS pixel detector electronics and sensors}},
	Volume = {3},
	Year = {2008}}

@article{CYP,
               Author  = "{Ian Dawson (ed.)}",
               Journal = {CERN Yellow Reports: Monographs},
               Title      = {{Radiation effects in the LHC experiments: Impact on detector performance and operation}},
               url         = {https://doi.org/10.23731/CYRM-2021-001},
               Year      = {2021}
 }

@report{slink,
      author        = "van der Bij,Erik and Haas,Stefan",
      title         = "CERN S-LINK homepage",
      year           = "2003",
      url           = "http://hsi.web.cern.ch/HSI/s-link/"
}

@article{ROS,
      author        = "G. Crone and D. Della Volpe and B. Gorini and B. Green and M. Joos and G. Kieft and K. Kordas and A. Kugel and A. Misiejuk and N. Schroer and P. Teixeira-Dias and L. Tremblet and J. Vermeulen and F. Wickens and P. Werner",
      title         = "The ATLAS ReadOut System Performance with first data and perspective for the future",
      journal       = "\NIMA",
      volume        = "623",
      pages         = "534-536",
      year          = "2010",
      issn          = "0168-9002",
      note          = "1st International Conference on Technology and Instrumentation in Particle Physics",
      doi           = "https://doi.org/10.1016/j.nima.2010.03.061",
      url           = "http://www.sciencedirect.com/science/article/pii/S016890021000625X",
}

@report{QSFP,
      author         = "{SFF Committee}",
      title          = "Quad Small Form-factor Pluggable (QSFP) Transceiver Specification",
      year           = "2006",
      url            = "https://www.gigalight.com/downloads/standards/QSFP-MSA.pdf"
}

@report{HOLA,
      author         = "Aurelio Ruiz and Erik van der Bij and Stefan Haas",
      title          = "HOLA High-speed Optical Link for Atlas",
      year           = "2003",
      url            = "http://hsi.web.cern.ch/hsi/s-link/devices/hola/"
}

@article{Huffman,
  author={Huffman, David A.},
  journal={Proceedings of the IRE}, 
  title={A Method for the Construction of Minimum-Redundancy Codes}, 
  year={1952},
  volume={40},
  number={9},
  pages={1098-1101},
  doi={10.1109/JRPROC.1952.273898}}

@article{TRTElectronics,
      author        = {{The ATLAS TRT collaboration}},
      title = {The {ATLAS} {TRT} electronics},
	doi = {10.1088/1748-0221/3/06/p06007},
	year = 2008,
	month = {6},
        journal       = "JINST",
	publisher = {{IOP} Publishing},
	volume = {3},
	number = {06},
	pages = {P06007--P06007},
}

@article{TRT,
	author = {{The ATLAS TRT collaboration}},
	title = {The {ATLAS} Transition Radiation Tracker ({TRT}) proportional drift tube: design and performance},
	doi = {10.1088/1748-0221/3/02/p02013},
	year = 2008,
	month = {2},
        journal       = "JINST",
	publisher = {{IOP} Publishing},
	volume = {3},
	number = {02},
	pages = {P02013--P02013},
}

@article{TRTBarrel,
	author = {{The ATLAS TRT collaboration}},
	title = {The {ATLAS} {TRT} Barrel Detector},
	journal = {JINST},
	doi = {10.1088/1748-0221/3/02/p02014},
	year = 2008,
	month = {2},
	publisher = {{IOP} Publishing},
	volume = {3},
	number = {02},
	pages = {P02014--P02014},
}

@article{TRTEndcap,
	author = {{The ATLAS TRT collaboration}},
	title = {The {ATLAS} {TRT} end-cap detectors},
	doi = {10.1088/1748-0221/3/10/p10003},
	year = 2008,
	month = {10},
	journal = {JINST},
	publisher = {{IOP} Publishing},
	volume = {3},
	number = {10},
	pages = {P10003--P10003},
}

@article{DSS,
doi = {10.1088/1748-0221/4/09/P09012},
url = {https://dx.doi.org/10.1088/1748-0221/4/09/P09012},
year = {2009},
month = {9},
publisher = {},
volume = {4},
number = {09},
pages = {P09012},
author = {O Beltramello and  H J Burckhart and  S Franz and  M Jaekel and  M Jeckel and  S Lüders and  G Morpurgo and  F dos Santos Pedrosa and  K Pommes and  H Sandaker},
title = {The Detector Safety System of the ATLAS experiment},
journal = "JINST",
OPTabstract = {The ATLAS detector at the Large Hadron Collider at CERN is one of the most advanced detectors for High Energy Physics experiments ever built. It consists of the order of ten functionally independent sub-detectors, which all have dedicated services like power, cooling, gas supply. A Detector Safety System has been built to detect possible operational problems and abnormal and potentially dangerous situations at an early stage and, if needed, to bring the relevant part of ATLAS automatically into a safe state. The procedures and the configuration specific to ATLAS are described in detail and first operational experience is given.}
}

@article{lucid2,
    author = "Avoni, G. and others",
    title = "{The new LUCID-2 detector for luminosity measurement and monitoring in ATLAS}",
    doi = "10.1088/1748-0221/13/07/P07017",
    journal = "JINST",
    volume = "13",
    number = "07",
    pages = "P07017",
    year = "2018"
}

@article{lucid2-pmt,
    author = "Alberghi, G.L. and others",
    title = "{Choice and characterization of photomultipliers for the new ATLAS LUCID detector}",
    doi = "10.1088/1748-0221/11/05/P05014",
    journal = "JINST",
    volume = "11",
    number = "05",
    pages = "P05014",
    year = "2016"
}

@article{LUCID-3_IDR,
      author        = "{ATLAS Collaboration}",
      title         = "{The LUCID 3 detector for the ATLAS Phase-II Upgrade}",
      institution   = "CERN",
      reportNumber  = "CERN-LHCC-2021-016, LHCC-P-018",
      address       = "Geneva",
      year          = "2021",
      url           = "https://cds.cern.ch/record/2780604",
}

@article{BCM,
    author = "Cindro, V. and others",
    title = "{The ATLAS beam conditions monitor}",
    doi = "10.1088/1748-0221/3/02/P02004",
    journal = "JINST",
    volume = "3",
    pages = "P02004",
    year = "2008"
}

@Booklet{vdm,
   title = {Calibration of the effective beam height in the ISR},
   author = {S. van der Meer},
   collaboration = {},
   howpublished = {CERN-ISR-PO-68-31},
   year = {1968},
   url = {https://cds.cern.ch/record/296752}
}

@misc{caen_sy1527,
   title = {Caen SY 1527 Technical Information Manual},
   howpublished = {datasheet},
   url = {https://www.caen.it/documents/Events/4/opcrelease3x_rev10.pdf}
}

@misc{zdc-pmt,
   title = {H6559 Photomultiplier tube assembly},
   journal = {datasheet},
   note  =  {Hamamatsu, 325-6, Sunayama-cho, Naka-ku, Hamamatsu City,Shizuoka Pref., 430-8587, Japan}, 
   url = {www.hamamatsu.com/eu/en/product/type/H6559/index.html}
}

@article{TimePix,
    author = "Sopczak, Andr\'e and others",
    title = "{Precision Luminosity of LHC Proton\textendash{}Proton Collisions at 13 TeV Using Hit Counting With TPX Pixel Devices}",
    eprint = "1702.00711",
    archivePrefix = "arXiv",
    primaryClass = "physics.ins-det",
    doi = "10.1109/TNS.2017.2664664",
    journal = "IEEE Trans. Nucl. Sci.",
    volume = "64",
    number = "3",
    pages = "915--924",
    year = "2017"
}

@Booklet{ATLAS-Phase-I-LOI,
  title = 	 {{L}etter of {I}ntent for the {P}hase-{I} {U}pgrade of the {ATLAS} {E}xperiment},
  OPTkey = 	 {},
  author =	 {{ATLAS Collaboration}},
  howpublished = {{CERN-LHCC-2011-012. LHCC-I-020}},
  address =	 {Geneva},
  month =	 {11},
  year =	 {2011},
  OPTnote = 	 {},
  OPTannote = 	 {}
}

@Article{ref21TDR,
  title = {MICROMEGAS: a high-granularity position-sensitive gaseous detector for high particle-flux environments},
journal = {Nuclear Instruments and Methods in Physics Research Section A: Accelerators, Spectrometers, Detectors and Associated Equipment},
volume = {376},
number = {1},
pages = {29-35},
year = {1996},
OPTissn = {0168-9002},
doi = {https://doi.org/10.1016/0168-9002(96)00175-1},
url = {https://www.sciencedirect.com/science/article/pii/0168900296001751},
author = {Y. Giomataris and Ph. Rebourgeard and J.P. Robert and G. Charpak}
}

@Article{ref27TDR,
  author = 	 {V. Peskov and M. Cortesi and R. Chechik and A. Breskin},
  title = 	 {Further evaluation of a {THGEM} {UV}-photon detector for {RICH} - comparison with {MWPC}},
  journal = 	 {JINST},
  year = 	 {2010},
  OPTkey = 	 {},
  volume =	 {5},
  OPTnumber = 	 {},
  pages =	 {P11004},
  OPTmonth = 	 {},
  OPTnote = 	 {},
  eprint =       {1008.0151},
  archivePrefix ={arXiv},
  primaryClass = {physics.ins-det},
  OPTannote = 	 {}
}

@Article{ref26TDR,
  author = 	 {H. Raether},
  title =	 {Die Entwicklung der Elektronenlawine in den Funkenkanal},
  journal = 	 {Z. Phys.},
  year = 	 {1939},
  OPTkey = 	 {},
  volume =	 {112},
  OPTnumber = 	 {},
  pages =	 {464},
  OPTmonth = 	 {},
  OPTnote = 	 {},
  OPTannote = 	 {}
}

@article{Beker_2019,
	doi = {10.1088/1748-0221/14/08/p08010},
	url = {https://doi.org/10.1088%2F1748-0221%2F14%2F08%2Fp08010},
	year = 2019,
	month = {8},
	publisher = {{IOP} Publishing},
	volume = {14},
	number = {08},
	pages = {P08010--P08010},
	author = {M. Beker and G. Bobbink and B. Bouwens and N. Deelen and P. Duinker and J. van Eldik and N. de Gaay Fortman and R. van der Geer and H. van der Graaf and H. Groenstege and R. Hart and K. Hashemi and J. van Heijningen and M. Kea and J. Koopstra and X. Leijtens and F. Linde and J.A. Paradiso and H. Tolsma and M. Woudstra},
	title = {The Rasnik 3-point optical alignment system},
	journal = "JINST"
}

@article{ABUSLEME201685,
title = "Performance of a full-size small-strip thin gap chamber prototype for the ATLAS new small wheel muon upgrade",
journal = "\NIMA",
volume = "817",
pages = "85 - 92",
year = "2016",
OPTissn = "0168-9002",
doi = "https://doi.org/10.1016/j.nima.2016.01.087",
url = "http://www.sciencedirect.com/science/article/pii/S0168900216001285",
author = "A. Abusleme and C. Bélanger-Champagne and A. Bellerive and Y. Benhammou and J. Botte and H. Cohen and M. Davies and Y. Du and L. Gauthier and T. Koffas and S. Kuleshov and B. Lefebvre and C. Li and N. Lupu and G. Mikenberg and D. Mori and J.P. Ochoa-Ricoux and E. Perez Codina and S. Rettie and A. Robichaud-Véronneau and R. Rojas and M. Shoa and V. Smakhtin and B. Stelzer and O. Stelzer-Chilton and A. Toro and H. Torres and P. Ulloa and B. Vachon and G. Vasquez and A. Vdovin and S. Viel and P. Walker and S. Weber and C. Zhu"
}

@article{Aefsky_2008,
	doi = {10.1088/1748-0221/3/11/p11005},
	url = {https://doi.org/10.1088%2F1748-0221%2F3%2F11%2Fp11005},
	year = 2008,
	month = {11},
	publisher = {{IOP} Publishing},
	volume = {3},
	number = {11},
	pages = {P11005--P11005},
	author = {S Aefsky and C Amelung and J Bensinger and C Blocker and A Dushkin and M Gardner and K Hashemi and E Henry and B Kaplan and P Keselman and M Ketchum and U Landgraf and A Ostapchuk and J Rothberg and A Schricker and N Skvorodnev and H Wellenstein},
	title = {The Optical Alignment System of the {ATLAS} Muon Spectrometer Endcaps},
	journal = "JINST"
}

@ARTICLE{TDSref,
  author={J. {Wang} and L. {Guan} and J. W. {Chapman} and B. {Zhou} and J. {Zhu}},
  journal={IEEE Transactions on Nuclear Science}, 
 title={Design of a Trigger Data Serializer ASIC for the Upgrade of the ATLAS Forward Muon Spectrometer}, 
  year={2017},
  volume={64},
  number={12},
  pages={2958-2965},
  doi={10.1109/TNS.2017.2771266},
  OPTISSN={1558-1578},
  month={12},}

@article{ALEXOPOULOS2019125,
title = "Performance studies of resistive-strip bulk micromegas detectors in view of the ATLAS New Small Wheel upgrade",
journal = "\NIMA",
volume = "937",
pages = "125 - 140",
year = "2019",
OPTissn = "0168-9002",
doi = "https://doi.org/10.1016/j.nima.2019.04.050",
url = "http://www.sciencedirect.com/science/article/pii/S0168900219305194",
author = "T. Alexopoulos and M. Bianco and M. Biglietti and C. Bini and M. Byszewski and G. Iakovidis and P. Iengo and M. Iodice and E. Karentzos and S. Leontsinis and K. Ntekas and F. Petrucci and G. Sekhniaidze and O. Sidiropoulou and M. Vanadia and J. Wotschack"
}

@article{Kuger_2016,
	doi = {10.1088/1748-0221/11/11/c11010},
	url = {https://doi.org/10.1088%2F1748-0221%2F11%2F11%2Fc11010},
	year = 2016,
	month = {11},
	publisher = {{IOP} Publishing},
	volume = {11},
	number = {11},
	pages = {C11010--C11010},
	author = {F. Kuger},
	title = {Production and quality control of Micromegas anode {PCBs} for the {ATLAS} {NSW} upgrade},
	journal = "JINST"
}

@ARTICLE{router_7287804,
author={J. {Wang} and X. {Hu} and T. {Schwarz} and J. {Zhu} and J. W. {Chapman} and T. {Dai} and B. {Zhou}},
journal={IEEE Transactions on Nuclear Science},
title={FPGA Implementation of a Fixed Latency Scheme in a Signal Packet Router for the Upgrade of ATLAS Forward Muon Trigger Electronics},
year={2015},
volume={62},
number={5},
pages={2194-2201},
}

@article{INST20_MM,
doi = {10.1088/1748-0221/15/09/C09019},
url = {https://dx.doi.org/10.1088/1748-0221/15/09/C09019},
year = {2020},
month = {9},
publisher = {},
volume = {15},
number = {09},
pages = {C09019},
author = {I. Gnesi},
title = {Micromegas chambers for the ATLAS New Small Wheel upgrade},
journal = "JINST",
OPTseries =	 {Instrumentation for Colliding Beam Physics, 24-28 Feb 2020},
OPTaddress =	 {Novosibirsk, Russia},
OPTorganization = {INSTR2020},
OPTnote ={Proceedings from INSTR2020, Novosibirsk, Russia; also available as \href{https://cds.cern.ch/record/2718454}{ATL-MUON-PROC-2020-009}} ,
OPTannote = 	 { Should be identical to internal ATL-COM-MUON-2020-022 https://cds.cern.ch/record/2715768/files/ comes out.}
}

@article{Farina_LaRochelle,
  author = 	 {Edoardo Farina},
  title = 	 {{ATLAS} {NSW} {M}icromegas readout boards: Industrialisation and Quality Control and Quality Assurance},
  journal =	 {J. Phys.:Conf. Ser.},
doi = {10.1088/1742-6596/1498/1/012052},
url = {https://dx.doi.org/10.1088/1742-6596/1498/1/012052},
year = {2020},
month = {4},
publisher = {IOP Publishing},
volume = {1498},
number = {1},
pages = {012052},
author = {Edoardo Farina and on behalf of the ATLAS Muon Collaboration},
}

@InProceedings{Iakovidis_LaRochelle,
  author = 	 {George Iakovidis},
  title = 	 {VMM ASIC},
  OPTcrossref =  {},
  OPTkey = 	 {},
  booktitle =	 {J. Phys.:Conf. Ser.},
  year =	 {2020},
  OPTeditor = 	 {},
  volume =	 { 1498 },
  OPTnumber = 	 {},
  series =	 {MicroPattern Gaseous Detectors Conference 2019},
  pages =	 {012051},
  OPTmonth = 	 {},
  address =	 {La Rochelle, France},
  OPTorganization = {},
  OPTpublisher = {IOP Publishing Ltd},
  OPTnote =	 {ATL-MUON-PROC-2019-009},
  url  =         {https://doi.org/10.1088/1742-6596/1498/1/012051},
  OPTannote = 	 {}
}

@Article{sMDT1,
author = "Kroha, H. and others",
    editor = "Cervelli, Franco and Chiarelli, Giorgio and Forti, Francesco and Grassi, Marco and Scribano, Angelo",
    title = "{Construction and test of a full prototype drift-tube chamber for the upgrade of the ATLAS muon spectrometer at high LHC luminosities}",
    doi = "10.1016/j.nima.2012.08.055",
    journal = "Nucl. Instrum. Meth. A",
    volume = "718",
    pages = "427--428",
    year = "2013"
}

@Article{sMDT2,
doi = {10.1088/1748-0221/12/06/C06007},
url = {https://dx.doi.org/10.1088/1748-0221/12/06/C06007},
year = {2017},
month = {6},
volume = {12},
number = {06},
pages = {C06007},
author = {H. Kroha and R. Fakhrutdinov and A. Kozhin},
title = {New high-precision drift-tube detectors for the ATLAS muon spectrometer},
  author = 	 {H. Kroha and R. Fakhroutidinov and A. Kozhin}
}

@Article{sMDT_res,
    author = "Bittner, Bernhard and others",
    editor = "Bergauer, T. and Badurek, G. and Dragicevic, M. and Friedl, M. and Hrubec, J. and Jeitler, M. and Krammer, M.",
    title = "{Performance of drift-tube detectors at high counting rates for high-luminosity LHC upgrades}",
    eprint = "1603.09508",
    archivePrefix = "arXiv",
    primaryClass = "physics.ins-det",
    reportNumber = "MPP-2013-596",
    doi = "10.1016/j.nima.2013.07.076",
    journal = "Nucl. Instrum. Meth. A",
    volume = "732",
    pages = "250--254",
    year = "2013"
}

@Article{BMEBMG1,
  author = 	 {C. Ferretti and H. Kroha},
  title = 	 {Upgrades of the ATLAS Muon Spectrometer With sMDT Chambers},
  journal = 	 {Nucl.~Instr.~and Meth. A},
  year = 	 {2016},
  OPTkey = 	 {},
  volume =	 {824},
  OPTnumber = 	 {},
  pages =	 {538},
  OPTmonth = 	 {},
  OPTnote = 	 {},
  OPTannote = 	 {}
}

@Article{BMEBMG2,
  author = 	 {O. Kortner and others},
  title = 	 {Upgrades of the ATLAS Muon Spectrometer with New Small-Diameter Drift Tube Chambers},
  journal = 	 {Nucl.~Instr.~and Meth. A},
  year = 	 {2019},
  OPTkey = 	 {},
  volume =	 {936},
  OPTnumber = 	 {},
  pages =	 {509},
  OPTmonth = 	 {},
  OPTnote = 	 {},
  OPTannote = 	 {}
}

@Article{RPC_FE,
  author = 	 {L. Pizzimento and others},
  title = 	 {Development of a new Front End electronics in Silicon and Silicon-Germanium technology for the Resistive Plate Chamber detector for high rate experiments},
  journal = 	 {JINST},
  year = 	 {2019},
  OPTkey = 	 {},
  volume =	 {14},
  OPTnumber = 	 {},
  pages =	 {C10010},
  OPTmonth = 	 {},
  OPTnote = 	 {},
  OPTannote = 	 {}
}

@Misc{HPTDC,
  OPTkey = 	 {},
  author =	 {Jorgen Christiansen},
  title =	 {HPTDC -- High Performance Time to Digital Converter},
  url = {https://cds.cern.ch/record/1067476/files/cer-002723234.pdf},
  month =	 {3},
  year =	 {2004},
  OPTnote = 	 {The CERN HPTDC fast time digitizer chip, formerly documented at now-dead http://tdc.web.cern.ch/TDC/hptdc/hptdc.htm, info available at  https://kt.cern/technologies/high-performance-time-digital-converter},
  OPTannote = 	 {}
}

@Article{RPC_aging,
  author = 	 {G. Aielli and others},
  title = 	 {New results on ATLAS RPC's aging at CERN's GIF},
  journal = 	 {IEEE Trans.~Nucl.~Sci.},
  year = 	 {2006},
  OPTkey = 	 {},
  volume =	 {53},
  OPTnumber = 	 {},
  pages =	 {567},
  OPTmonth = 	 {},
  OPTnote = 	 {},
  OPTannote = 	 {}
}

@Article{quanyin2018,
  author = 	 {Q. Li and others},
  title = 	 {Performance study of HL-LHC ATLAS RPC prototype},
  journal = 	 {JINST},
  year = 	 {2019},
  OPTkey = 	 {},
  volume =	 {14},
  OPTnumber = 	 {},
  pages =	 {C09022},
  OPTmonth = 	 {},
  OPTnote = 	 {},
  OPTannote = 	 {}
}

@Article{ASD,
  author = 	 {Y. Arai and others},
  title = 	 {ATLAS Muon Drift Tube Electronics},
  journal = 	 {JINST},
  year = 	 {2008},
  OPTkey = 	 {},
  volume =	 {3},
  OPTnumber = 	 {},
  pages =	 {P09001},
  OPTmonth = 	 {},
  OPTnote = 	 {},
  OPTannote = 	 {}
}

@InProceedings{FELIX,
  author = 	 {J. Anderson and others},
  title = 	 {FELIX: The new approach for interfacing to front-end electronics for the ATLAS experiment},
  OPTcrossref =  {},
  OPTkey = 	 {},
  OPTbooktitle = {},
  year =	 {2016},
  OPTeditor = 	 {},
  OPTvolume = 	 {},
  OPTnumber = 	 {},
  series =	 {IEEE NSS Real Time Conference},
  OPTpages = 	 {},
  OPTmonth = 	 {},
  OPTaddress = 	 {},
  OPTorganization = {},
  OPTpublisher = {},
  note =	 {https://ieeexplore.ieee.org/document/7543142/},
  OPTannote = 	 {}
}

@article{GOL,
       author        = "Moreira, P and Cervelli, G and Christiansen, J and Faccio,
                       F and Kluge, A and Marchioro, A and Toifl, Thomas H and
                       Cachemiche, J P and Menouni, M",
      title         = "{A radiation tolerant gigabit serializer for LHC data
                       transmission}",
      year          = "2001",
      url           = "https://cds.cern.ch/record/588665",
      doi           = "10.5170/CERN-2001-005.145",
}

@Misc{RPC_FPGA,
  OPTkey = 	 {},
  OPTauthor = 	 {},
  title =	 {Xilinx Kintex-7 FPGA product information},
  OPThowpublished = {},
  OPTmonth = 	 {},
  OPTyear = 	 {},
  url =	 {https://www.xilinx.com/products/silicon-devices/fpga/kintex-7.html},
  OPTannote = 	 {}
}

@Misc{GBTx,
  OPTkey = 	 {},
  author =	 {P. Moreira},
  title =	 {GBTX MANUAL V0.15 },
  OPThowpublished = {},
  month =	 {10},
  year =	 {2018},
  url =	 {https://espace.cern.ch/GBT-Project/GBTX/Manuals/gbtxManual.pdf},
  OPTannote = 	 {}
}

@article{ROC,
    author = "Coliban, R.M. and Popa, S. and Tulbure, T. and Nicula, D. and Ivanovici, M. and Martoiu, S. and Levinson, L. and Vermeulen, J.",
    title = "{The Read Out Controller for the ATLAS New Small Wheel}",
    doi = "10.1088/1748-0221/11/02/C02069",
    journal = "JINST",
    volume = "11",
    number = "02",
    pages = "C02069",
    year = "2016"
}

@Misc{GBT,
  OPTkey = 	 {},
  author =	 {Moreira, P and Marchioro, A and Kloukinas},
  title =	 {The GBT: A proposed architecure for multi-Gb/s data
                       transmission in high energy physics},
  OPThowpublished = {},
  OPTmonth = 	 {},
  year =	 {2007},
  url           = "http://cds.cern.ch/record/1091474",
  doi           = "10.5170/CERN-2007-007.332",
  OPTnote = 	 {See also complete manual at https://espace.cern.ch/GBT-Project/GBTX/Manuals/gbtxManual.pdf},
  OPTannote = 	 {}
}

@Booklet{ATL-MUON-PUB-2022-003,
    author         = "{ATLAS Collaboration}",
    title          = "{The Serial and LVDS repeaters for the ATLAS New Small Wheel sTGC trigger}",
    howpublished   = "{ATL-MUON-PUB-2022-003}",
    url            = "https://cds.cern.ch/record/2841848",
    year           = "2022",
    OPTnote        = {Eventually this will be replaced by official entry in bib/PubNotes.bib}
}

@INPROCEEDINGS{ARTref,
author={Tang, S. and Chen, H. and Chen, K. and Dimitrios, M. and Polychronakos, V. and Yao, L.},
booktitle={2019 IEEE Nuclear Science Symposium and Medical Imaging Conference (NSS/MIC)},
title={The Development and Production of the ADDC for the Micromegas Detector of the ATLAS New Small Wheel Upgrade},
year={2019},
volume={},
number={},
pages={1-5},
OPTnote={This reference describes both the ADDC and the ART ASIC.},
doi={10.1109/NSS/MIC42101.2019.9060071}
}

@ARTICLE{FEAST,
author={Fuentes, C. and Allongue, B. and Blanchot, G. and Faccio, F. and Michelis, S. and Orlandi, S. and Pontt, J. and Rodriguez, J. and Kayal, M.},
journal={IEEE Transactions on Nuclear Science},
title={Optimization of DC-DC Converters for Improved Electromagnetic Compatibility With High Energy Physics Front-End Electronics},
year={2011},
volume={58},
number={4},
pages={2024-2031},
note = {For more details about the FEAST, see \url{https://espace.cern.ch/project-DCDC-new/}},
doi={10.1109/TNS.2011.2159395}
}

@article{ALLARD2022166143,
title = {The large inner Micromegas modules for the Atlas Muon Spectrometer upgrade: Construction, quality control and characterization},
journal = {\NIMA},
volume = {1026},
pages = {166143},
year = {2022},
OPTissn = {0168-9002},
doi = {https://doi.org/10.1016/j.nima.2021.166143},
url = {https://www.sciencedirect.com/science/article/pii/S0168900221010330},
author = {J. Allard and N. Andari and M. Anfreville and D. Attié and E. Aubernon and S. Aune and H. Bachacou and F. Balli and F. Bauer and J. Beltramelli and J. Bennet and T. Benoit and H. Bervas and T. Bey and S. Bouaziz and M. Boyer and G. Cara and T. Chaleil and T. Chevalérias and X. Coppollani and J. Costa and G. Decock and F. Deliot and D. Denysiuk and D. Desforge and G. Disset and G.A. Durand and R. Durand and J. Elman and E. Ferrer-Ribas and M. Fontaine and A. Formica and J. Galán and W. Gamache and A. Giganon and J. Giraud and P.F. Giraud and G. Glonti and C. Goblin and P. Graffin and J.C. Guillard and S. Hassani and S. Hervé and S. Javello and F. Jeanneau and D. Jourde and S. Jurie and M. Kebbiri and T. Kawamoto and C. Lampoudis and J.F. Laporte and D. Leboeuf and M. Lefèvre and M. Lohan and C. Loiseau and P. Magnier and I. Mandjavidze and J. Manjarrés and P. Mas and M. Mur and R. Nikolaidou and A. Peyaud and D. Pierrepont and Y. Piret and P. Ponsot and G. Prono and M. Riallot and F. Rossi and P. Schune and T. Vacher and M. Vandenbroucke and A. Vigier and C. Vuillemin and M. Usseglio and Z. Wu},
keywords = {MPGD, Micromegas, Resistive anode, High rate capability, HL-LHC, ATLAS, NSW, Cosmic bench, Validation tests},
%abstract = {The steadily increasing luminosity of the LHC requires an upgrade with high-rate and high-resolution detector technology for the inner end cap of the ATLAS muon spectrometer: the New Small Wheels (NSW). In order to achieve the goal of precision tracking at a hit rate of about 15 kHz/cm2 at the inner radius of the NSW, large area Micromegas quadruplets with 100µm spatial resolution per plane have been produced. IRFU, from the CEA research center of Saclay, is responsible for the production and validation of LM1 Micromegas modules. The construction, production, qualification and validation of the largest Micromegas detectors ever built are reported here. Performance results under cosmic muon characterization will also be discussed.}
}

@article{AGARWALA2022167285,
title = {Construction and test of the SM1 type Micromegas chambers for the upgrade of the ATLAS forward muon spectrometer},
journal = {\NIMA},
volume = {1040},
pages = {167285},
year = {2022},
OPTissn = {0168-9002},
doi = {https://doi.org/10.1016/j.nima.2022.167285},
url = {https://www.sciencedirect.com/science/article/pii/S0168900222006118},
author = {J. Agarwala and M.G. Alviggi and M. Antonelli and F. Anulli and C. Arcangeletti and S. Auricchio and P. Bagnaia and S. Bariani and A. Baroncelli and I. Bashta and M. Bauce and M. Beretta and C. Bini and D. Calabró and M.T. Camerlingo and V. Canale and E. Capitolo and M. Capponi and G. Capradossi and G. Carducci and A. Caserio and C. Cassese and S. Cerioni and V. Chiarella and G. Ciapetti and V. D’Amico and S. {De Cecco} and B. {De Fazio} and M. {Del Gaudio} and C. {Di Donato} and R. {Di Nardo} and D. D’Uffizi and A. Farilla and E. Farina and R. Ferrari and G. Fiore and G. Frattari and A. Freddi and G. Gaudio and P. Gauzzi and S. Giagu and S. Gigli and I. Gnesi and E. Gorini and F.G. Gravili and M. Greco and A. Iaciofano and P. Iengo and A. Innocente and G. Introzzi and M. Iodice and V. Ippolito and A. {Kourkoumeli Charalampidi} and F. Lacava and E. Lalli and A. Lanza and S. Lauciani and P. Laurelli and L. Luminari and G. Mancini and L. Martinelli and P. Massarotti and A. Miccoli and A. Mirto and A. Negri and D. Orestano and G. Paruzza and M. Pepe and M. Petruccetti and F. Petrucci and L. Pezzotti and G. Pileggi and M. Pirola and C. Piscitelli and G. Polesello and A. Policicchio and G. Pontoriere and B. Ponzio and M. Primavera and D. Rebuzzi and A. Rimoldi and E. Romano and L. Roscilli and G. Rovelli and V. Russo and A. Sansoni and F. {Safai Tehrani} and C. Scagliotti and M. Schioppa and G. Sekhniaidze and M. Sessa and S. Sottocornola and P. Trattino and E. Tskhadadze and P. Turco and D. Vannicola and L. Vannoli and T. Vassilieva and V. Vecchio and A. Ventura and F. Vercellati and E. Vilucchi and A. Zullo and G. Zunica},
OPTabstract = {Large-size Resistive Micromegas have been chosen for the upgrade of the forward muon spectrometer of the ATLAS experiment, the New Small Wheel project. These chambers, together with small-strip Thin Gap Chambers (sTGC), allow reconstruction of high-momentum muon tracks in a high-radiation environment and provide a robust low-threshold single-muon trigger. A collaboration of seven INFN units built 32 SM1 type chambers, corresponding to one fourth of the total number needed for this upgrade. Each SM1 chamber has a surface of approximately 2 m2 and four sensitive layers. The production was shared among five INFN construction sites and it was completed in fall 2020. The construction methods, as well as the results of the quality tests done on components of the detector and on the assembled chambers, are reported in the present paper.}
}

@article{MAJEWSKI1983265,
title = {A thin multiwire chamber operating in the high multiplication mode},
journal = {\NIM},
volume = {217},
number = {1},
pages = {265-271},
year = {1983},
OPTissn = {0167-5087},
doi = {https://doi.org/10.1016/0167-5087(83)90146-1},
url = {https://www.sciencedirect.com/science/article/pii/0167508783901461},
author = {S. Majewski and G. Charpak and A. Breskin and G. Mikenberg}
}

@PhdThesis{KNtekas,
  author = 	 {Konstantinos Ntekas},
  title = 	 {Performance characterization of the Micromegas detector for the New Small Wheel upgrade and Development and improvement of the Muon Spectrometer Detector Control System in the ATLAS experiment},
  school = 	 {National Technical University, Athens},
  year = 	 {2016},
  OPTkey = 	 {},
  OPTtype = 	 {},
  OPTaddress = 	 {},
  OPTmonth = 	 {},
  note =	 {(pp. 180--185)},
  url =          {https://cds.cern.ch/record/2143887?ln=en},
  OPTannote = 	 {}
}

@Booklet{Blanchot:1073170,
      author        = "Blanchot, G",
      title         = "{Grounding of the ATLAS experiment}",
      institution   = "CERN",
      address       = "Geneva",
      howpublished  = "ATL-ELEC-PUB-2007-002, ATL-COM-ELEC-2007-003",
      month         = "12",
      year          = "2007",
      url           = "https://cds.cern.ch/record/1073170",
}

@INPROCEEDINGS{4436512,
  author={Dubbert, J. and Horvat, S. and Kroha, H. and Legger, F. and Kortner, O. and Richter, R. and Rauscher, F.},
  booktitle={2007 IEEE Nuclear Science Symposium Conference Record}, 
  title={Development of precision drift tube detectors for very high background rates at the super-LHC}, 
  year={2007},
  volume={3},
  number={},
  pages={1822-1825},
  doi={10.1109/NSSMIC.2007.4436512}}

@article{Mandelli:2021zxi,
    author = "Mandelli, Beatrice and Guida, Roberto and Rigoletti, Gianluca",
    title = "{Performance studies of RPC detectors operated with new environmentally-friendly gas mixtures in presence of LHC-like radiation background}",
    doi = "10.22323/1.390.0857",
    journal = "PoS",
    volume = "ICHEP2020",
    pages = "857",
    year = "2021"
}

@article{Gkountoumis:2018usk,
    author = "Gkountoumis, Panagiotis",
    collaboration = "ATLAS Muon",
    title = "{LEVEL-1 DATA DRIVER CARD - A high bandwidth radiation tolerant aggregator board for detectors}",
    doi = "10.22323/1.322.0036",
    journal = "PoS",
    volume = "MPGD2017",
    pages = "036",
    year = "2019"
}

@Article{NSWelx,
doi = {10.1088/1748-0221/18/05/P05012},
url = {https://dx.doi.org/10.1088/1748-0221/18/05/P05012},
year = {2023},
month = {5},
publisher = {IOP Publishing},
volume = {18},
number = {05},
pages = {P05012},
  author = 	 {George Iakovidis and Lorne Levinson and others},
  title = 	 {The {N}ew {S}mall {W}heel electronics},
  journal = 	 {JINST},
  year = 	 {2023},
    eprint         = "2303.12571",
    archivePrefix  = "arXiv",
    primaryClass   = "hep-ex",
  OPTnote = 	 {should have a public CERN number...ATL-COM-MUON-2022-069 at https://cds.cern.ch/record/2845057 is internal},
  OPTannote = 	 {}
}

@article{Iodice:2017Kt,
  author = "Iodice, Mauro",
  title = "{Resistive Micromegas for the Muon Spectrometer Upgrade of the ATLAS Experiment}",
  doi = "10.22323/1.282.0275",
  journal = "PoS",
  year = 2017,
  volume = "ICHEP2016",
  pages = "275"
}

@article{ALEXOPOULOS2020162086,
title = {Construction techniques and performances of a full-size prototype Micromegas chamber for the ATLAS muon spectrometer upgrade},
journal = {Nuclear Instruments and Methods in Physics Research Section A: Accelerators, Spectrometers, Detectors and Associated Equipment},
volume = {955},
pages = {162086},
year = {2020},
OPTissn = {0168-9002},
doi = {https://doi.org/10.1016/j.nima.2019.04.040},
url = {https://www.sciencedirect.com/science/article/pii/S0168900219304966},
author = {T. Alexopoulos and M. Alviggi and M. Antonelli and F. Anulli and C. Arcangeletti and P. Bagnaia and A. Baroncelli and M. Beretta and C. Bini and J. Bortfeldt and D. Calabrò and V. Canale and F. Capocasa and G. Capradossi and G. Carducci and A. Caserio and C. Cassese and S. Cerioni and G. Ciapetti and V. D’Amico and B. {De Fazio} and M. {Del Gaudio} and P. Gkountoumis and C. {Di Donato} and R. {Di Nardo} and D. D’Uffizi and E. Farina and R. Ferrari and A. Freddi and C. Gatti and G. Gaudio and E. Gorini and F. Gravili and S. Guelfo Gigli and G. Iakovidis and P. Iengo and A. Innocente and G. Introzzi and M. Iodice and E. Karentzos and A. Koulouris and A. Kourkoumeli-Charalampidi and F. Lacava and A. Lanza and L. {La Rotonda} and S. Lauciani and L. Luminari and G. Maccarrone and S. Maltezos and G. Mancini and L. Martinelli and P. Massarotti and A. Miccoli and A. Mirto and P. Moschovakos and K. Ntekas and S. Palazzo and G. Paruzza and F. Petrucci and L. Pezzotti and G. Pileggi and A. Policicchio and G. Pontoriere and B. Ponzio and E. Romano and V. Romano and L. Roscilli and G. Rovelli and C. Scagliotti and M. Schioppa and G. Sekhniaidze and M. Sessa and S. Sottocornola and P. Turco and M. Vanadia and D. Vannicola and T. Vassileva and V. Vecchio and F. Vercellati and A. Zullo},
OPTabstract = {A full-size prototype of a Micromegas precision tracking chamber for the upgrade of the ATLAS detector at the LHC Collider has been built between October 2015 and April 2016. This paper describes in detail the procedures followed in the construction of the components of the chamber in various INFN laboratories and the final assembly in the Laboratori Nazionali di Frascati (LNF). In addition, the results of the chamber exposure to a particle beam at SPS/H8 at CERN in June 2016 are presented. The performances obtained in the construction and the results of the test beam are compared with the requirements set in order to sustain the high radiation levels expected during the data-taking of the LHC in the next years.}
}

@ARTICLE{VMMref,
  author={de Geronimo, Gianluigi and Iakovidis, George and Martoiu, Sorin and Polychronakos, Venetios},
  journal={IEEE Transactions on Nuclear Science}, 
  title={The VMM3a ASIC}, 
  year={2022},
  volume={69},
  number={4},
  pages={976-985},
  doi={10.1109/TNS.2022.3155818}}

@Article{LHCcontributionInThisJournal,
  author = 	 {G. Arduini and V. Baglin and L. Bottura and C. Bracco and B. Bradu and G. Bregliozzi and K. Brodzinski and R. Bruce and M.  Calviani and P. Chiggiato and P. Cruikshank and S. Claudet and D. Delikaris and S.Fartoukh and C. Garion and M. Himmerlich and M. Hostettler and G. Iadarola and S. Kostoglou and S. Le Naour and A. Lechner and T. Lefevre and L. Mether and Y. Papaphilippou and V. Petit and M. Pojer and A. Poyet and S. Redaelli and F. Rodriguez Mateos and G. Rumolo and B. Salvant and F. Sanchez Galan and A. Siemko and M. Solfaroli-Camillocci and G. Sterbini and M. Taborelli and L. Tavian and H. Timko and J.-Ph. Tock and A. Verweij and M. Wendt and J. Wenninger and D. Wollmann and Ch. Yin Vallgren},
  title = 	 {LHC Upgrades in preparation of Run 3},
  journal = 	 "JINST",
  year = 	 {2023},
  OPTkey = 	 {},
  OPTvolume = 	 {},
  OPTnumber = 	 {},
  OPTpages = 	 {},
  OPTmonth = 	 {},
  note =	 {To be published in the same volume as this paper.},
  OPTannote = 	 {}
}

@article{Steerenberg:2259071,
      author        = "Steerenberg, Rende",
      title         = "{Batch Compression Merging and Splitting (BCMS).}",
      month         = 4,
      year          = "2017",
      url           = "https://cds.cern.ch/record/2259071",
      note          = "CERN General Photo",
}

@Booklet{slipstacking,
    author = "Ankenbrandt, C.",
    title = "{Slip Stacking: A New Method of Momentum Stacking}",
    howpublished = "FERMILAB-FN-0352",
    month = "12",
    url = "https://inspirehep.net/literature/169350",
    year = "1981"
}

@Booklet{ATL-MUON-PUB-2015-001,
    author         = "{ATLAS Collaboration}",
    title          = "{Stereo Information in Micromegas Detectors}",
    howpublished   = "{ATL-MUON-PUB-2015-001}",
    url            = "https://cds.cern.ch/record/2052206",
    year           = "2015",
}

@Booklet{ATL-PHYS-PUB-2016-015,
    author         = "{ATLAS Collaboration}",
    title          = "{Electron and photon energy calibration with the ATLAS detector using data collected in 2015 at \(\sqrt{s} = 13~\text{TeV}\)}",
    howpublished   = "{ATL-PHYS-PUB-2016-015}",
    url            = "https://cds.cern.ch/record/2203514",
    year           = "2016",
}

@Booklet{ATL-DAPR-PUB-2021-001,
    author         = "{ATLAS Collaboration}",
    title          = "{Luminosity monitoring using \(Z\to{\ell^+\ell^-}\) events at \(\sqrt{s} = 13\,\text{TeV}\) with the ATLAS detector}",
    howpublished   = "{ATL-DAPR-PUB-2021-001}",
    url            = "https://cds.cern.ch/record/2752951",
    year           = "2021",
}

@Article{moll-TNS65:1561,
      author         = "Moll, Michael",
      title          = "Displacement Damage in Silicon Detectors for High Energy Physics",
      journal        = "IEEE Trans. Nucl. Sci.",
      volume         = "65",
      pages          = "1561",
      year           = "2018",
      doi            = "10.1109/TNS.2018.2819506",
}

@Article{huhtinen-nim491:194,
      author         = "Huhtinen, Mika",
      title          = "Simulation of non-ionising energy loss and defect formation in silicon",
      journal        = "Nucl. Instrum. Meth. A",
      volume         = "491",
      pages          = "194",
      year           = "2002",
      OPTissn = "0168-9002",
      doi = "https://doi.org/10.1016/S0168-9002(02)01227-5",
      url = "http://www.sciencedirect.com/science/article/pii/S0168900202012275",
}

@Article{huhtinen-nim450:155,
      author         = "Huhtinen, Mika and Faccio, Federico",
      title          = "Computational method to estimate Single Event Upset rates in an accelerator environment",
      journal        = "Nucl. Instrum. Meth. A",
      volume         = "450",
      pages          = "155",
      year           = "2000",
      OPTissn           = "0168-9002",
      doi            = "https://doi.org/10.1016/S0168-9002(00)00155-8",
      url            = "http://www.sciencedirect.com/science/article/pii/S0168900200001558",
}

@Article{huhtinen-nim335:580,
      author         = "Huhtinen, Mika and Aarnio, Pertti",
      title          = "Pion induced displacement damage in silicon devices",
      journal        = "Nucl. Instrum. Meth. A",
      volume         = "335",
      pages          = "580",
      year           = "1993",
      OPTissn           = "0168-9002",
      doi            = "https://doi.org/10.1016/0168-9002(93)91246-J",
      url            = "http://www.sciencedirect.com/science/article/pii/016890029391246J",
}

@Article{gcalorref,
      title = "The GEANT-CALOR interface and benchmark calculations of ZEUS test calorimeters",
      journal        = "Nucl. Instrum. Meth. A",
      author = "Zeitnitz, C. and Gabriel, T. A.",
      volume = "349",
      number = "1",
      pages = "106 - 111",
      year = "1994",
      OPTissn = "0168-9002",
      doi = "https://doi.org/10.1016/0168-9002(94)90613-0",
      url = "http://www.sciencedirect.com/science/article/pii/0168900294906130",
}

@Booklet{ flukaref1,
      author         = "Ferrari, Alfredo and Sala, Paola R. and Fasso, Alberto and Ranft, Johannes",
      title          = "{FLUKA: A multi-particle transport code}",
      year           = "2005",
      howpublished        = "CERN-2005-010",
      url            = "https://cds.cern.ch/record/898301",
}

@Article{ flukaref2,
      author         = "B{\"o}hlen, T T and others",
      title          = "{The FLUKA Code: Developments and Challenges for High Energy and Medical Applications}",
      journal        = "Nucl. Data Sheets",
      volume         = "120",
      pages          = "211",
      year           = "2014",
     doi            = "10.1016/j.nds.2014.07.049",
}

@Booklet{Ferrari:300336,
      author        = "Ferrari, A and Potter, K M and Rollet, S and Sala, P R",
      title         = "{Radiation calculations for the ATLAS detector and
                       experimental hall}",
      howpublished  = "CERN-EST-96-001",
      year          = "1996",
      url           = "https://cds.cern.ch/record/300336",
}

@PHDTHESIS{hamburgmodel,
      author       = {Moll, Michael},
      title        = {{R}adiation damage in silicon particle detectors:
                      {M}icroscopic defects and macroscopic properties},
      school       = {Universität Hamburg},
      type         = {Ph.D. Thesis},
      reportid     = {PUBDB-2016-02525, DESY-THESIS-1999-040},
      OPTpages        = {251},
      year         = {1999},
      OPTnote         = {Advisor: G. Lindstrom},
      OPTcin          = {L},
      OPTcid          = {I:(DE-H253)L-20120731},
      OPTpnm          = {899 - ohne Topic (POF3-899)},
      OPTpid          = {G:(DE-HGF)POF3-899},
      OPTexperiment   = {EXP:(DE-MLZ)NOSPEC-20140101},
      OPTtyp          = {PUB:(DE-HGF)11},
      doi          = {10.3204/PUBDB-2016-02525},
      url          = {https://bib-pubdb1.desy.de/record/300958},
}

@Article{TileDamage,
      author         = "R. {Pedro on behalf of the ATLAS Collaboration}",
      title          = "Optics robustness of the ATLAS Tile Calorimeter",
      journal        = "J. Phys.: Conf. Ser.",
      volume         = "1162",
      pages          = "012004",
      year           = "2019",
      url            = "https://iopscience.iop.org/article/10.1088/1742-6596/1162/1/012004",
}

@InProceedings{RadMonRef,
      author         = "J. Hartert and others",
      title          = "{The ATLAS radiation dose measurement system and its extension to SLHC experiments}",
      year           = "2008",
      booktitle      = {Proceedings of Topical Workshop on Electronics for Particle Physics, Naxos, Greece},
      url            = {http://cds.cern.ch/record/1158640?ln=en},
}

@Article{summers87,
   author={Summers, G. P. and Burke, E. A. and Dale, C. J. and Wolicki, E. A. and Marshall, P. W. and Gehlhausen, M. A.},
   journal={IEEE Transactions on Nuclear Science}, 
   title={Correlation of Particle-Induced Displacement Damage in Silicon}, 
   year={1987},
   volume={34},
   number={6},
   pages={1133-1139},
   url = "https://ieeexplore-ieee-org.ezproxy.cern.ch/document/4337442",
   doi={10.1109/TNS.1987.4337442}}

@article{ZHU1998297,
title = {Radiation damage in scintillating crystals},
journal = {Nuclear Instruments and Methods in Physics Research Section A: Accelerators, Spectrometers, Detectors and Associated Equipment},
volume = {413},
number = {2},
pages = {297-311},
year = {1998},
OPTissn = {0168-9002},
doi = {https://doi.org/10.1016/S0168-9002(98)00498-7},
url = {https://www.sciencedirect.com/science/article/pii/S0168900298004987},
author = {Ren-yuan Zhu}
}

@article{JetsWithoutJets,
    author         = "{D. Bertolini, T. Chan, and J. Thaler}",
    title          = "{Jet Observables Without Jet Algorithms}",
    journal        = "JHEP",
    volume         = "1404",
    year           = "2014",
    pages          = "013",
    doi            = "10.1007/JHEP04(2014)013",
    eprint         = "1310.7584",
    archivePrefix  = "arXiv",
    primaryClass   = "hep-ph",
}

@article{LoffredoBIS78, 
    author        = "{Salvatore Loffredo, ATLAS Muon Collaboration}",
    title         = "{The BIS78 Pad Trigger Board for the Phase-I Upgrade of the Level-1 Muon Trigger of the ATLAS Experiment at the LHC}",
    journal       = "2019 IEEE Nuclear Science Symposium and Medical Imaging Conference (NSS/MIC)",
    volume        = "",
    year          = "2019",
    pages         = "1",
    doi           = "10.1109/NSS/MIC42101.2019.9059637",
    url           = "https://ieeexplore.ieee.org/document/9059637",
}

@article{WYLLIE20121561,
title = "A Gigabit Transceiver for Data Transmission in Future High Energy Physics Experiments",
journal = "Physics Procedia",
volume = "37",
pages = "1561 - 1568",
year = "2012",
note = "Proceedings of the 2nd International Conference on Technology and Instrumentation in Particle Physics (TIPP 2011)",
issn = "1875-3892",
doi = "https://doi.org/10.1016/j.phpro.2012.02.487",
url = "http://www.sciencedirect.com/science/article/pii/S1875389212018706",
author = "K. Wyllie and S. Baron and S. Bonacini and Ö. Çobanoğlu and F. Faccio and S. Feger and R. Francisco and P. Gui and J. Li and A. Marchioro and P. Moreira and C. Paillard and D. Porret",
}

@misc{GBTSCA,
  title =       "{GBT-SCA User Manual V. 2019-002}",
  author =      "{S. Bonacini, A. Caratelli et al.}",
  howpublished = "(online)",
  year =        {2019},
  url =        "https://espace.cern.ch/GBT-Project/GBT-SCA/Manuals/GBT-SCA_Manual_2019.002.pdf",
}

@article{GBTSCA2,
	doi = {10.1088/1748-0221/10/03/c03034},
	url = {https://doi.org/10.1088/1748-0221/10/03/c03034},
	year = 2015,
	month = {3},
	publisher = {{IOP} Publishing},
	volume = {10},
	number = {03},
	pages = {C03034--C03034},
	author = {A. Caratelli and S. Bonacini and K. Kloukinas and A. Marchioro and P. Moreira and R. De Oliveira and C. Paillard},
	title = {The {GBT}-{SCA}, a radiation tolerant {ASIC} for detector control and monitoring applications in {HEP} experiments},
	journal = {JINST}
}

@article{icalepcs2019-wepha102,
  author       = {P. Moschovakos and H. Boterenbrood and A. Koulouris and P.P. Nikiel and S. Schlenker},
  title        = {{A Software Suite for the Radiation Tolerant Giga-bit Transceiver - Slow Control Adapter}},
  booktitle    = {Proc. ICALEPCS'19},
  pages        = {1333--1337},
  paper        = {WEPHA102},
  language     = {english},
  venue        = {New York, NY, USA},
  series       = {International Conference on Accelerator and Large Experimental Physics Control Systems},
  number       = {17},
  publisher    = {JACoW Publishing, Geneva, Switzerland},
  month        = {08},
  year         = {2020},
  OPTissn         = {2226-0358},
  isbn         = {978-3-95450-209-7},
  doi          = {10.18429/JACoW-ICALEPCS2019-WEPHA102},
  url          = {https://jacow.org/icalepcs2019/papers/wepha102.pdf},
  note         = {https://doi.org/10.18429/JACoW-ICALEPCS2019-WEPHA102},
}

@article{ATLAS_DCS_Jinst,
	author        = "{A.~Barriuso~Poy et al.}",
	title         = "{The detector control system of the ATLAS experiment}",
	journal        = "JINST",
	volume         = "3",
	pages          = "P0500",
	year           = "2008",
}

@inproceedings{bib:quasar,
	author         = "Schlenker, Stefan and Abalo Miron, Damian and Farnham,
	Ben and Filimonov, Viatcheslav and Nikiel, Piotr and
	Soare, Cristian-Valeriu",
	title          = "{quasar - A Generic Framework for Rapid Development of
	OPC UA Servers}",
	booktitle      = "{Proceedings, 15th International Conference on
	Accelerator and Large Experimental Physics Control Systems
	(ICALEPCS 2015): Melbourne, Australia, October 17-23,
	2015}",
	year           = "2015",
	url            = "http://jacow.org/icalepcs2015/papers/web3o02.pdf",
	pages          = "WEB3O02",
	doi            = "10.18429/JACoW-ICALEPCS2015-WEB3O02",
	SLACcitation   = "%%CITATION = INSPIRE-1481616;%%"
}

@article{bib:quasar_CHEP,
	doi = {10.1088/1742-6596/664/8/082039},
	url = {https://doi.org/10.1088%2F1742-6596%2F664%2F8%2F082039},
	year = 2015,
	month = {12},
	publisher = {{IOP} Publishing},
	volume = {664},
	number = {8},
	pages = {082039},
	author = {Piotr P Nikiel and Benjamin Farnham and Viatcheslav Filimonov and Stefan Schlenker},
	title = {Generic {OPC} {UA} Server Framework},
	journal = {Journal of Physics: Conference Series},
}

@misc{OPCUA,
	title         = "OPC UA Online Reference",
	howpublished  = "(online)",
	url           = "https://reference.opcfoundation.org/",
}

@misc{uaexpert,
	title         = "UaExpert—A Full-Featured OPC UA Client",
	howpublished  = "(online)",
	url           = "https://www.unified-automation.com/products/development-tools/uaexpert.html",
}

@article{bib:muctpi-phase1,
    author        = "{A. Armbruster et al.}", 
    title         = "{The ATLAS Muon to Central Trigger Processor Interface Upgrade for the Run 3 of the LHC}",
    journal       = "{2017 IEEE Nuclear Science Symposium and Medical Imaging Conference (NSS/MIC)}",
    volume        = "",
    year          = "2017",
}

@article{bib:ttc,
    author        = "{S. Ask et al.}",
    title         = "{The ATLAS central level-1 trigger logic and TTC system}",
    journal       = "{JINST}",
    volume        = "3",
    year          = "2008",
    pages         = "P08002"
}

@misc{JSON,
  title =       "{The JavaScript Object Notation (JSON) Data Interchange Format,  STD 90, RFC 8259}",
  editor =      "{Bray, T.}",
  howpublished = "(online)",
  year =        {2017},
  url =        "https://www.rfc-editor.org/info/std90"
  }

@misc{bib:Aurora,
    author        = "Xilinx\textregistered",
    title         = "{Aurora 8B/10B Protocol Specification}",
    howpublished  = {\url{https://docs.xilinx.com/v/u/en-US/aurora_8b10b_protocol_spec_sp002}},
    note          = {Accessed: 2023-08-09}
}

@misc{bib:CERN-IPMC,
    author        = "{CERN}",
    title         = "{CERN-IPMC}",
    howpublished  = {\url{https://cern-ipmc.web.cern.ch/doc}},
    note          = {Accessed: 2021-11-04},
}

@article{bib:IPbus,
    author        = "{C. Ghabrous Larrea, K. Harder, D. Newbold, D. Sankey, A. Rose, A. Thea and T. Williams}",
    title         = "{IPbus: a flexible Ethernet-based control system for xTCA hardware}", 
    journal       = "{JINST}",
    volume        = "10",
    year          = "2015",
    pages         = "C02019",
    doi           = "10.1088/1748-0221/10/02/C02019"
}

@misc{bib:IPMI,
    author        = "{Intel Corporation, Hewlett-Packard Company, NEC Corporation, Dell Inc.}",
    title         = "{Intelligent Platform Management Interface Specification, v. 2.0}",
    howpublished  = {\url{https://www.intel.com/content/dam/www/public/us/en/documents/product-briefs/ipmi-second-gen-interface-spec-v2-rev1-1.pdf}},
    year          = "2015",
    note          = {Accessed: 2023-08-09},

}

@incollection{JEFFERS2013243,
title = {Chapter 8 - Coprocessor Architecture},
editor = {Jim Jeffers and James Reinders},
booktitle = {Intel Xeon Phi Coprocessor High Performance Programming},
publisher = {Morgan Kaufmann},
address = {Boston},
pages = {243-268},
year = {2013},
isbn = {978-0-12-410414-3},
doi = {https://doi.org/10.1016/B978-0-12-410414-3.00008-6},
url = {https://www.sciencedirect.com/science/article/pii/B9780124104143000086},
author = {Jim Jeffers and James Reinders}
}

@INPROCEEDINGS{8071057,

  author={Schumacher, Jörn},

  booktitle={2017 IEEE 25th Annual Symposium on High-Performance Interconnects (HOTI)}, 

  title={Utilizing HPC Network Technologies in High Energy Physics Experiments}, 

  year={2017},

  volume={},

  number={},

  pages={57-64},

  doi={10.1109/HOTI.2017.25}}

@Report{ATLAS-TDR-29-ADD-1,
    author         = "{ATLAS Collaboration}",
    title          = "{Technical Design Report for the Phase-II Upgrade of the ATLAS Trigger and Data Acquisition System --- EF Tracking Amendment}",
    type           = "ATLAS-TDR-029; CERN-LHCC-2022-004",
    year           = "2022",
    url            = "http://cds.cern.ch/record/2802799",
}

@article{gaudi,
title = {GAUDI — A software architecture and framework for building HEP data processing applications},
journal = {Computer Physics Communications},
volume = {140},
number = {1},
pages = {45-55},
year = {2001},
note = {CHEP2000},
OPTissn = {0010-4655},
doi = {https://doi.org/10.1016/S0010-4655(01)00254-5},
url = {https://www.sciencedirect.com/science/article/pii/S0010465501002545},
author = {G. Barrand and I. Belyaev and P. Binko and M. Cattaneo and R. Chytracek and G. Corti and M. Frank and G. Gracia and J. Harvey and E.van Herwijnen and P. Maley and P. Mato and S. Probst and F. Ranjard},
OPTabstract = {We present a software architecture and framework that can be used to facilitate the development of data processing applications for High Energy Physics experiments. The development strategy follows an architecture-centric approach as a way of creating a resilient software framework that can withstand changes in requirements and technology over the long lifetimes of experiments. The software architecture, called GAUDI, supports event data processing applications that run in different processing environments, from the high level triggers in the on-line system to the final physics analysis. We present our major architectural design choices and outline the arguments that led to these choices. Several iterations of a software framework based on this architecture have been released and the framework is now being used by the physicists of the collaboration to facilitate the development of data processing algorithms. Object oriented technologies have been used throughout.}
}

\clearpage
 

\begin{flushleft} 
\hypersetup{urlcolor=black} 
{\Large The ATLAS Collaboration} 

\bigskip

\AtlasOrcid[0000-0002-6665-4934]{G.~Aad}$^\textrm{\scriptsize 102}$,
\AtlasOrcid[0000-0002-5888-2734]{B.~Abbott}$^\textrm{\scriptsize 120}$,
\AtlasOrcid[0000-0002-7248-3203]{D.C.~Abbott}$^\textrm{\scriptsize 103}$,
\AtlasOrcid{J.~Abdallah}$^\textrm{\scriptsize 8}$,
\AtlasOrcid[0000-0002-1002-1652]{K.~Abeling}$^\textrm{\scriptsize 55}$,
\AtlasOrcid[0000-0002-8496-9294]{S.H.~Abidi}$^\textrm{\scriptsize 29}$,
\AtlasOrcid[0000-0002-9987-2292]{A.~Aboulhorma}$^\textrm{\scriptsize 35e}$,
\AtlasOrcid{S.~Abovyan}$^\textrm{\scriptsize 110}$,
\AtlasOrcid[0000-0001-5329-6640]{H.~Abramowicz}$^\textrm{\scriptsize 151}$,
\AtlasOrcid[0000-0002-1599-2896]{H.~Abreu}$^\textrm{\scriptsize 150}$
% ...
\bigskip
\\

% ...
$^{8}$Department of Physics, University of Texas at Arlington, Arlington TX; United States of America.\\
% ...
$^{29}$Physics Department, Brookhaven National Laboratory, Upton NY; United States of America.\\
% ...
$^{35}$$^{(a)}$Facult\'e des Sciences Ain Chock, R\'eseau Universitaire de Physique des Hautes Energies - Universit\'e Hassan II, Casablanca;$^{(b)}$Facult\'{e} des Sciences, Universit\'{e} Ibn-Tofail, K\'{e}nitra;$^{(c)}$Facult\'e des Sciences Semlalia, Universit\'e Cadi Ayyad, LPHEA-Marrakech;$^{(d)}$LPMR, Facult\'e des Sciences, Universit\'e Mohamed Premier, Oujda;$^{(e)}$Facult\'e des sciences, Universit\'e Mohammed V, Rabat;$^{(f)}$Institute of Applied Physics, Mohammed VI Polytechnic University, Ben Guerir; Morocco.\\
% ...
$^{55}$II. Physikalisches Institut, Georg-August-Universit\"{a}t G\"ottingen, G\"ottingen; Germany.\\
% ...
$^{102}$CPPM, Aix-Marseille Universit\'e, CNRS/IN2P3, Marseille; France.\\
$^{103}$Department of Physics, University of Massachusetts, Amherst MA; United States of America.\\
% ...
$^{110}$Max-Planck-Institut f\"ur Physik (Werner-Heisenberg-Institut), M\"unchen; Germany.\\
% ...
$^{120}$Homer L. Dodge Department of Physics and Astronomy, University of Oklahoma, Norman OK; United States of America.\\
% ...
$^{150}$Department of Physics, Technion, Israel Institute of Technology, Haifa; Israel.\\
$^{151}$Raymond and Beverly Sackler School of Physics and Astronomy, Tel Aviv University, Tel Aviv; Israel.\\
% ...
$^{a}$ Also Affiliated with an institute covered by a cooperation agreement with CERN.\\
$^{b}$ Also at Borough of Manhattan Community College, City University of New York, New York NY; United States of America.\\
$^{c}$ Also at Bruno Kessler Foundation, Trento; Italy.\\
$^{d}$ Also at Center for High Energy Physics, Peking University; China.\\
$^{e}$ Also at Center for Interdisciplinary Research and Innovation (CIRI-AUTH), Thessaloniki ; Greece.\\
$^{f}$ Also at Centro Studi e Ricerche Enrico Fermi; Italy.\\
% ...

\end{flushleft}


\end{document}